\makeatletter
\def\input@path{{/Users/patrick/gitrepos/phd/DielectronAnalysis/docs_data/star-docs/thesis//}}
\makeatother
\documentclass[a4paper,11pt,twoside,openright]{report}
\IfFileExists{ifxetex.sty}{%
    \usepackage{ifxetex}%
  }{%
    \newif\ifxetex
    \xetexfalse
  }
  \ifxetex
\usepackage{fontspec}
\usepackage{xltxtra}
\defaultfontfeatures{Mapping=tex-text}
\else
\usepackage[T1]{fontenc}
\usepackage[latin1]{inputenc}
\fi
\usepackage{fancybox}
\usepackage{makeidx}
\def\hyperparam{draft}
\usepackage[hyperlink]{db2latex_huck}
\renewcommand{\DBKreleaseinfo}{}
\renewcommand{\DBKrevhistory}{}
\title{Beam Energy Dependence of Dielectron Production in Au+Au Collisions from STAR at RHIC}
\author{Patrick Huck}
\hypersetup{%
pdfcreator={DBLaTeX-0.3.5},%
pdftitle={Beam Energy Dependence of Dielectron Production in Au+Au Collisions from STAR at RHIC},%
pdfauthor={Patrick Huck}%
}
\renewcommand{\DBKindexation}{}
\makeindex
\makeglossary
\begin{document}
\lstsetup
\frontmatter
\maketitle
\tableofcontents
\setcounter{secnumdepth}{-1}
\addtocontents{toc}{\protect\setcounter{tocdepth}{-1}\ignorespaces}
\setcounter{tocdepth}{-1}

\chapter{Abstract}
\label{_abstract}\hyperlabel{_abstract}%

At sufficiently high temperatures and baryon densities, nuclear matter is
expected to undergo a transition into the Quark-{}Gluon-{}Plasma (QGP) consisting
of deconfined quarks and gluons and accompanied by chiral symmetry restoration.
Signals of these two fundamental characteristics of Quantum-{}Chromo-{}Dynamics
(QCD) can be studied in ultra-{}relativistic heavy-{}ion collisions producing a
relatively large volume of high energy and nucleon densities as existent in the
early universe. Dileptons are unique bulk-{}penetrating sources for this purpose
since they penetrate through the surrounding medium with negligible interaction
and are created throughout the entire evolution of the initially created
fireball. A multitude of experiments at SIS18, SPS and RHIC have taken on the
challenging task to measure these rare probes in a heavy-{}ion environment.
NA60's results from high-{}quality dimuon measurements have identified the
broadened \ensuremath{\rho} spectral function as favorable scenario to explain the low-{}mass
dilepton excess, and partonic sources as dominant at intermediate dilepton
masses.

Enabled by the addition of a TOF detector system in 2010, the first phase of
the Beam Energy Scan (BES-{}I) at RHIC allows STAR to conduct an unprecedented
energy-{}dependent study of dielectron production within a homogeneous
experimental environment, and hence close the wide gap in the QCD phase diagram
between SPS and top RHIC energies. This thesis concentrates on the
understanding of the LMR enhancement regarding its M$_{\text{ee}}$, p$_{\text{T}}$ and energy
dependence. It studies dielectron production in Au+Au collisions at beam energies
of 19.6, 27, 39, and 62.4 GeV with sufficient statistics. In conjunction with
the published STAR results at top RHIC energy, this thesis presents results on
the first comprehensive energy-{}dependent study of dielectron production.

This includes M$_{\text{ee}}$-{} and p$_{\text{T}}$-{}spectra for the four beam energies measured in
0-{}80\% minimum-{}bias Au+Au collisions with high statistics up to 3.5 GeV/c$^{\text{2}}$ and
2.2 GeV/c, respectively. Their comparison with cocktail simulations of hadronic
sources reveals a sizeable and steadily increasing excess yield in the LMR at
all beam energies. The scenario of broadened in-{}medium \ensuremath{\rho} spectral functions
proves to not only serve well as dominating underlying source but also to be
\emph{universal} in nature since it quantitatively and qualitatively explains the
LMR enhancements measured over the wide range from SPS to top RHIC energies.
It shows that most of the enhancement is governed by interactions of the \ensuremath{\rho}
meson with thermal resonance excitations in the late(r)-{}stage hot and dense
hadronic phase. This conclusion is supported by the energy-{}dependent
measurement of integrated LMR excess yields and enhancement factors. The former
do not exhibit a strong dependence on beam energy as expected from the
approximately constant total baryon density above 20 GeV, and the latter show
agreement with the CERES measurement at SPS energy. The consistency in excess
yields and agreement with model calculations over the wide RHIC energy regime
makes a strong case for LMR enhancements on the order of a factor 2 -{} 3.

The extent of the results presented here enables a more solid discussion of its
relation to chiral symmetry restoration from a theoretical point of view.
High-{}statistics measurements at BES-{}II hold the promise to confirm these
conclusions along with the LMR enhancment's relation to total baryon density
with decreasing beam energy.
\setcounter{secnumdepth}{3}
\addtocontents{toc}{\protect\setcounter{tocdepth}{5}\ignorespaces}
\setcounter{tocdepth}{5}
\setcounter{secnumdepth}{-1}
\addtocontents{toc}{\protect\setcounter{tocdepth}{-1}\ignorespaces}
\setcounter{tocdepth}{-1}

\chapter{Zusammenfassung}
\label{_zusammenfassung}\hyperlabel{_zusammenfassung}%

Die vorliegende Doktorarbeit besteht aus einem Hauptteil (Part~\ref{physics}) im
Themengebiet der hoch-{}energetischen Kernphysik und einem Nebenteil
(Part~\ref{software}) über Entwicklung und Beitrag von neuen Software Modulen zur
Kollaboration und der Öffentlichkeit. Der Hauptteil behandelt das Thema
der Energieabhängigkeit von Dielektronen-{}Produktion in
ultra-{}relativistischen Schwerionen-{}Kolli­sionen mithilfe des STAR Detektors
(Solenoidal Tracker at RHIC) am Brook­haven National Laboratory in Upton, NY. Im
Nebenteil werden zum Einen sieben C++ und Python Bibliotheken für
Datenanalyse und Dokumentation vorgestellt. Zum Anderen wird in diesem Teil der
Übergang der gesamten 15-{}jährigen Geschichte des STAR Softwarepakets
in ein modernes System der Versionskontrolle beschrieben. Beide Teile sind im
Folgenden ausführlich zusammengefasst.

\section{Messung der energieabhängigen Dielektronen-{}Produktion mit STAR}
\label{_messung_der_energieabh_auml_ngigen_dielektronen_produktion_mit_star}\hyperlabel{_messung_der_energieabh_auml_ngigen_dielektronen_produktion_mit_star}%

Der Phasenzustand von Kernmaterie ändert sich mit zunehmender Temperatur
und Baryonendichte (Section~\ref{intro_sec_qcd_phases}). Die Theorie der
Quanten-{}Chromo-{}Dynamik (QCD) hat sich sehr erfolg­reich darin erwiesen
[1], die Eigenschaften dieser Phasen vorherzusagen
und zu bestätigen. Im QCD Vakuum sind Quarks und Gluonen darauf
beschränkt als Bestandteile von farb-{}neutralen Hadronen vorzukommen.
Außerdem ist die globale chirale Symmetrie des QCD Langrangian im Vakuum
spontan gebrochen [2]. Mit diesen beiden Charakteristiken ist die
QCD in der Lage, die grundlegenden Eigenschaften des experimentellen
Hadronenspektrums zu beschreiben [3].\newline

Bei ausreichend hohen Temperaturen und Baryonendichten durchläuft
Kernmaterie den Phasen­über­gang ins Quark-{}Gluon-{}Plasma (QGP), in dem freie
Quarks und Gluonen die Freiheitsgrade bestimmen (\emph{Deconfinement}) und chirale Symmetrie
wiederhergestellt ist [4].  Diese
fundamentalen Eigenschaften der QCD Materie können in
ultra-{}relativistischen Schwerionen-{}Kollisionen untersucht werden
[5, 6]. In den Kernstößen wird ein
verhältnismäßig großes Vo­lu­men mit hohen Energie-{} und
Nukleonendichten produziert, die vergleichbar zu den Be­ding­ungen im frühen
Universum sind (Section~\ref{intro_sec_hics}). Während der darauffolgenden
Ausdehnung und Abkühlung durchschreitet die heiße und dichte Materie
mehrere QCD Phasen, bevor sie in die beobachtbaren Teilchen des QCD Vakuums
hadronisiert und ausfriert [7].\newline

Messungen des elliptischen Flusses von Mesonen und Baryonen, zum Beispiel,
geben Aufschlüsse sowohl über die Zeitskala der Thermalisierung des
Mediums als auch über den Grad der kollektiven Expansion
[8, 9]. Es ist allerdings wichtig, (eindeutige)
Signale für die beiden fundamentalen QCD Charakteristiken des
Deconfinements und der Chiralen Symmetrie Restauration zu beobachten und
umfassend zu messen: Wenn Quark-{}Antiquark ($q\bar{q}$) Paare in
einem thermisch ausgeglichenen \emph{deconfined} QCD Medium annihilieren, sollte man
thermische elektromagnetische Strahlung direkt vom QGP beobachten können;
Änderungen der Spektralfunktion von Vektormesonen im Medium könnten
Vorläufereffekte eines Verschwindens der QCD Vakuumstruktur sein
[10, 11]. Die Diskussion des letzteren Zusammenhangs
(Section~\ref{intro_sec_dielec}) legt dar, dass eine koordinierte Anstrengung von
theoretischer und experimenteller Seite notwendig ist, um die
Schlussfolgerungen aus den Dielektronenmessungen der \ensuremath{\rho}-{}Spek­tral­funk­tion
bezüglich (teilweiser) Chiraler Symmetrie Restauration zu stärken.\newline

In diesem Bezug sind Dileptonen ein einzigartiges Werkzeug. Sie durchdringen
das um­ge­ben­de Medium mit vernachlässigbarer
Wechselwirkung und entstammen
der gesamten Evolution des Systems (Figure~\ref{intro_fig_dielec_prod}). Die
Verteilungen von invarianter Masse (M$_{\text{ee}}$) und Transversalimpuls (p$_{\text{T}}$)
widerspiegeln ihre zeitlich geordnete Emission und sind daher empfindlich auf
die Evolutionsdynamik des Systems. Insbesondere die \emph{Low-{}Mass}-{} (LMR) und
\emph{Intermediate-{}Mass}-{}Regionen der Dielektronenspektren versprechen jeweils
Zugang zur in-{}Medium modifizierten \ensuremath{\rho}-{}Spektral­funk­tion und zur effektiven
QGP Temperatur. Für die Extraktion der QGP Temperatur muss allerdings der
Beitrag eines möglicherweise Medium-{}modifizierten Charm Kontinuums bekannt
sein [12].\newline

Eine Vielzahl von Experimenten an SIS18 (Schwerionen-{}Synchrotron), SPS (Super
Proton Synchrotron), und RHIC (Relativistic Heavy Ion Collider)
haben die schwere Aufgabe auf sich genommen, diese seltenen Proben in einer
Schwerionen-{}Umgebung zu messen (Section~\ref{intro_subsec_dielec_meas}). CERES
beobachtete als erstes Experiment einen beträchtlichen und
unerklärten LMR Produktionsüberschuss im Vergleich zu einem Cocktail
aus erwarteten hadronischen Quellen [13]. Die NA60 Resultate
von hoch-{}qualitativen Dimuonen-{}Messungen haben die verbreiterte
\ensuremath{\rho}-{}Spektralfunktion als bevorzugtes Szenario für den LMR
Überschuss und einen partonischen Ursprung für die dominanten Quellen
im IMR identifiziert [14]. PHENIX folgte mit
Di­elek­tro­nen-{}Messungen bei höchsten RHIC Energien, die eine bedeutende
Verstärkung des LMR Signals und gute Übereinstimmung der IMR
Produktion mit einem Charm Kontinuum von N$_{\text{bin}}$-{}ska­lier­ten p+p Kollisionen
aufweisen [15].\newline

Im Gegensatz zu den NA60 Ergebnissen, konnte die Größe der LMR
Signalverstärkung nicht durch existierende Modellrechnungen erklärt
werden -{} selbst mit Ausnutzung des ge­sam­ten verfügbaren Parameterraums.
Messungen der Dielektronen-{}Produktion bei iden­ti­schen Energien mit STAR am RHIC
führten zu einer moderateren Verstärkung, die im Einklang ist mit den
NA60 Resultaten und mit Modellrechnungen basierend auf in-{}Medium
ver­brei­ter­ten
\ensuremath{\rho}-{}Spektralfunktionen [16]. Es sei an dieser Stelle
angemerkt, dass HADES für niedrige Temperaturen und hohe Baryonendichten
das \emph{DLS Rätsel} mit einem LMR Verstärkungsfaktor von
\ensuremath{\sim}2-{}3 bestätigt hat [17, 18].\newline

Diese Vorgeschichte schafft die Voraussetzungen für die umfassende
Untersuchung der Dielektronenproduktion in dieser Doktorarbeit. Ermöglicht
durch die Hinzufügung des Time-{}Of-{}Flight (TOF) Detektors im Jahre 2010
liefert STAR ausgezeichnete Teilchenidentifikation, niedriges Materialbudget,
volle azimuthale Akzeptanz bei mittleren Rapiditäten, und breite
p$_{\text{T}}$-{}Abdeckung [19]. Insbesondere mit der ersten Phase des \emph{Beam
Energy Scans} (BES-{}I) [20] bie­tet STAR die noch nie da
gewesene Gelegenheit, eine energie-{}abhängige Untersuchung der
Dielektronenproduktion innerhalb eines homogenen experimentellen Umfelds
durchzuführen und damit die große Lücke im QCD Phasendiagram
zwischen SPS und höchsten RHIC Energien zu schliessen
(Section~\ref{intro_sec_thesis}).\newline

Die vorliegende Arbeit konzentriert sich auf die Analyse von
Au+Au Kollisionen bei Strahl­energien von 19.6, 27, 39, und 62.4 GeV, die
während BES-{}I mit ausreichender Statistik für eine solche
Untersuchung aufgenommen wurden (Section~\ref{ana_sec_exp}). In Verbindung mit den
ver­öffent­lichten STAR Ergebnissen bei höchsten RHIC Energien stellt
dies den ersten umfassenden und energie-{}abhängigen Datensatz bereit.
Angesichts des beträchtlichen Aufwands wird dieser Datensatz der wohl
einzig verfügbare bei diesen Energien in naher Zukunft
bleiben.\footnote{
STAR wird weiterhin Daten mit hoher Statistik bei
\ensuremath{\surd}s$_{\text{NN}}$ <{} 20 GeV während BES-{}II aufnehmen.
}\newline

Zu Beginn der Analyse werden geeignete Auswahlkriterien angewandt, um den
Datensatz auf hoch-{}qualitative Kollisionsereignisse (\emph{Events}) und
Teilchenspuren zu reduzieren (Section~\ref{ana_sec_dsets_evttrk}). Die zusätzliche
Rekonstruktion der Eventebene (Section~\ref{ana_subsec_evtplane}) ist Voraussetzung
für die statistische Generierung von Untergrund-{}Paarverteilungen.
Elektronen und Positronen werden durch Kombination von Impuls, Geschwindigkeit,
und Energieverlust gemessen im TOF Detektor und der Time-{}Projection-{}Chamber
(TPC) sauber identifiziert (Section~\ref{ana_sec_pid}). Dielektronen (e$^{\text{+}}$/e$^{\text{-{}}}$ Paare)
werden aus der resultierenden Auswahl von Elektronen und Positronen kinetisch
rekonstruiert und Untergrund-{}Beiträge statistisch abgezogen
(Section~\ref{ana_sec_pairrec}), indem von \emph{Same-{}} und \emph{Mixed-{}Event} Me­tho­den auf
p$_{\text{T}}$-{}integrierte und -{}differentierte Weise Gebrauch gemacht wird.\newline

Für den Vergleich zu theoretischen Rechnungen und zwischen Energien
müssen die rohen M$_{\text{ee}}$-{} und p$_{\text{T}}$-{}Verteilungen der Dielektronen korrigiert
werden (Chapter~\ref{effcorr}), um Verluste durch Detektor-{}Ineffizienzen und Auswahlkriterien zu
beheben. Zu diesem Zweck werden simulierte Teilchenspuren in reale Events
eingebettet (\emph{Embedding}) und mit denselben Mitteln wie für die
experimentelle Eventrekonstruktion durch den Detektor verfolgt
(Section~\ref{effcorr_sec_samples}). Die Abschätzung von systematischen
Unsicherheiten erfordert den Vergleich von simulierten zu sauberen
experimentellen Verteilungen basierend auf geeigneter Teilchenauswahl. Die
Berechnung von Gesamt-{}Effizienzen der Spurrekonstruktion und -{}verfolgung folgt
dem Vorgehen der experimentellen Datenanalyse und benutzt die dedizierte
Teilchenauswahl von Embedding und Experiment (Section~\ref{effcorr_sec_single}). Die
Spureffizienzen werden schließlich mithilfe einer einfachen
Zwei-{}Körper Monte-{}Carlo Simulation in Paareffizienzen übertragen
(Section~\ref{effcorr_sec_paireffs}).\newline

Die Physik-{}Interpretation der experimentellen Ergebnisse in dieser Arbeit
(Chapter~\ref{results}) ba­siert auf \emph{(i)} der Simulation hadronischer
Zerfallskanäle, die bekanntermaßen zur Dielektronenproduktion in
elementaren Kollisionen beitragen, und \emph{(ii)} effektive Modellrechnungen, die
die in-{}Medium Effekte beschreiben. Ersteres stellt den sogenannten
\emph{hadronischen Cocktail} dar, dessen Quellen und charakteristischen
Massenregionen in der Einleitung im Zusammenhang mit p+p Kollisionen
vorgestellt und mit früheren Dielektronen-{}Messungen in
Schwerionen-{}Kollisionen verglichen werden (Section~\ref{intro_sec_dielec}). Die
Einzelheiten der Cocktail Simulationen, wie sie für die
energie-{}abhängige Dielektronen-{}Produktion in Au+Au Kollisionen bei BES-{}I
Energien verwendet werden, sind ausführlich beschrieben
(Section~\ref{sim_sec_cocktail}). In der Einleitung (Section~\ref{intro_sec_dielec}) wird auch der
gemessene LMR Produktionüberschuss über dem Cocktail und dessen
mögliche Verbindung zu Chiraler Symmetrie Restaurierung mittels \emph{Hadronic
Many-{}Body Theory} (HMBT) behandelt. Im Simulationskapitel (Section~\ref{sim_sec_model})
werden einige veranschauliche Einzelheiten und die Ergebnisse dieser
Berechnungen für BES-{}I Energien diskutiert.\newline

Für die genannten vier Strahlenergien, legt diese Dissertation M$_{\text{ee}}$-{} and
p$_{\text{T}}$-{}Spektren vor (Chapter~\ref{results}), gemessen in 0-{}80\% \emph{minimum-{}bias} Au+Au
Kollisionen mit hoher Statistik bis zu jeweils 3.5 GeV/c$^{\text{2}}$ und 2.2 GeV/c.
Deren Vergleich zu Cocktail Simulationen aus hadronischen Quellen deckt einen
beträchtigen und stetig zunehmenden LMR Überschuss bei allen Energien
auf. Dieser kann weder durch Vakuum \ensuremath{\rho}/\ensuremath{\omega}-{}Spektralfunktionen, die durch
die Feuerball-{}Expansion propagiert wurden, noch durch die entsprechenden
bloßen Ausfrierungsbeiträge beschrieben werden. Modellrechnungen mit
einer in-{}Medium verbreiterten \ensuremath{\rho}-{}Spektralfunktion auf Grundlage von
HMBT erreichen stattdessen gute
Übereinstimmung sowohl über den ge­sam­ten Energiebereich als auch in
der Form der M$_{\text{ee}}$-{} und p$_{\text{T}}$-{} Verteilungen. Dieser Beitrag dominiert im LMR
über die $q\bar{q}$ Beiträge vom QGP und dient daher im
Allgemeinen gut als zugrunde liegende Ursache des gemessen Überschusses im
gesamten RHIC Energiebereich. Die Beobachtung bedeutet auch, dass der Großteil
der Signalverstärkung durch Wechselwir­kungen des \ensuremath{\rho}-{}Mesons mit
thermischen Resonanzanregungen in der spät(er)en heißen und dichten
hadronischen Phase geregelt wird.\newline

Die Energie-{}Abhängigkeit sowohl der M$_{\text{ee}}$-{}integrierten und auf invariante
Pionen-{}Pro­duk­tion normierten LMR Überschüsse als auch der LMR
Signalverstärkung unterstützen diese Schlussfolgerung. Im Einklang
mit der etwa konstanten totalen Baryonendichte überhalb von 20 GeV weisen
die absoluten Überschüsse keine starke Abhängigkeit von der
Strahlenergie auf.  Die Verstärkungsfaktoren des LMR Signals stimmen mit
CERES Messungen bei SPS Energien trotz unterschiedlicher Akzeptanzen
überein -{} die STAR Daten sind jedoch von höherer Qualität. Die
von STAR gemessene Beständigkeit des LMR Überschusses und dessen
Übereinstimmung mit Modellrechnungen über den breiten RHIC
Energiebereich liefern überzeugende Argumente für
Verstärkungsfaktoren eher im Bereich 2 -{} 3 als 5 oder mehr.\newline

Die Energie-{}Abhängigkeit der integrierten LMR Überschüsse
ermöglicht eine interessante Prognose für die zweite Phase des \emph{Beam
Energy Scans} (BES-{}II). Im Energiebereich unterhalb von 20 GeV weisen sowohl
Messungen der totalen Baryonendichten als auch \ensuremath{\rho}-{}Meson-{}basierte PHSD
Rechnungen auf eine ungefähre Verdopplung des LMR Überschusses hin
[21, 22]. Messungen während BES-{}II mit hoher
Statistik sollten ausreichend Genauigkeit bieten, um diese Vorhersagen zu
testen und unser Verständnis der LMR Signalverstärkung sowie dessen
zugrunde liegende Ursache weiter zu kräftigen [23]. Außer
gesteigerter Statistik wird BES-{}II mit der beabsichtigten iTPC Aufrüstung
feinere Spurverfolgung und verbesserte Fähigkeiten zur Dimuonen-{}Messung
liefern.\newline

In der gegenwärtigen Untersuchung sind verlässliche
Rückschlüsse auf QGP Strahlung im IMR leider beschränkt durch a)
die vorhandene Statistik, b) die Unsicherheiten in den $c\bar{c}$ Wirkungsquerschnitten in Nukleon-{}Nukleon Kollisionen, und c) die möglichen
Medium-{}Modifikationen des Charm Kontinuums durch Entkorrelation der
$e^+/e^-$ Zerfallsprodukte. Mit­hil­fe der kürzlich
abgeschlossenen Aufrüstungen der HFT (Heavy Flavor Tracker) und MTD (Muon
Telescope Detector) Detektor-{}Teilsysteme werden
es BES-{}II Messungen aber er­mög­lichen, das Charm Kontinuum und damit
indirekt QGP Strahlung zu untersuchen. Dies ist insbesondere wichtig für
den LMR, da der $c\bar{c}$ Beitrag zum Gesamt-{}Cocktail im 0.4 -{} 0.7
GeV/c$^{\text{2}}$ Massenbereich von etwa 20\% bei 19.6 GeV auf etwa 60\% bei 200 GeV
steigt.\newline

Zusammenfassend stellen die STAR Messungen während BES-{}I hoch-{}qualitative
Datensätze zur Verfügung, die wesentlich für das
Verständnis der LMR Signalverstärkung bezüglich ihrer M$_{\text{ee}}$-{}, p$_{\text{T}}$-{}
und Energie-{}Abhängigkeit sind. Die Untersuchung der zugehörigen
Zen­tra­litätsabhängigkeit in der nahen Zukunft würde die
extrahierbaren Ergebnisse vervollständigen und könnte neue
Informationen über die Energie-{}Abhängigkeit der Feuerball-{}Lebensdauer
preisgeben. Das Szenario der in-{}Medium verbreiterten \ensuremath{\rho}-{}Spektralfunktionen
erklärt quantitativ und qualitativ die gemessene LMR
Signalverstärkung über den breiten Bereich von SPS bis zu höchsten
RHIC Energien. Diese Erklärung erweist sich daher nicht nur dazu in der
Lage als dominante zugrunde liegende Ursache zu dienen sondern auch als
\emph{universell} in ihrem Wesen.

Der Umfang der dargelegten Ergebnisse erlaubt eine solidere Diskussion im
Hinblick auf deren Beziehung zu Chiraler Symmetrie Restauration vom
theoretischen Standpunkt aus. Messungen mit hoher Statistik während BES-{}II
verprechen, diese Folgerungen sowie die Relation von LMR
Signalverstärkung zu totaler Baryonendichte mit abnehmender Strahlenergie
zu bestätigen.

\pagebreak[4]

\section{Software Projekte für STAR und Open-{}Source}
\label{_software_projekte_f_uuml_r_star_und_open_source}\hyperlabel{_software_projekte_f_uuml_r_star_und_open_source}%

Die C++ Softwareklasse \emph{StV0TofCorrection} (Section~\ref{stv0tofcorr}) implementiert
Funktionen, um die Flug­zeit von Teilchen aus V$_{\text{0}}$ (\emph{Off-{}Vertex}) Zerfällen
zu korrigieren. Der zugrunde liegende Algorithmus basiert auf der berichtigten
Rekonstruktion der Mutterteilchen. Die Anwendung auf \ensuremath{\Lambda}-{} und
\ensuremath{\Omega}-{}Zerfälle legt dessen Fähigkeit dar, zusätzlich
Untergrund von anderen Zerfällen zu verwerfen. Außerdem erlaubt die
Klasse dem Benutzer, einen Standardfilter auf die korrigierten Teilchenmassen
anzulegen und damit das Signal-{}zu-{}Untergrund Verhätnis zu verbessern. Der
Quellcode dieser Arbeit ist seit Januar 2011 Teil des \emph{StBTofUtil} [24] Moduls in der offiziellen STAR Code Repositorie. Die neueste
Revision der Implementation ist von Dezember 2013. Ein interner Report wurde
der STAR Kollaboration zugänglich gemacht
[25] \footnote{
Dieser Teil der Dissertation wurde
vom DAAD durch ein Auslandsstipendium unterstützt.
}.\newline

Die C++ Softwareklasse \emph{StBadRdosDb} (Section~\ref{stbadrdos}) wurde im Rahmen dieser
Arbeit ent­wickelt, um im Laufe einer Analyse bequem Informationen über die
Anzahl der fehlenden Auslese-{}Boards der TPC zu erfragen. Die Daten der 2010 und
2011 Strahlzeiten werden in zeitliche Perioden (\emph{run ranges}) mit gleicher
Anzahl fehlender Boards unterteilt und in einem passenden Datenbankformat
gespeichert.  \emph{StBadRdosDb} kann dann benutzt werden, um die zugehörige
Periode für einen spezifischen \emph{Run} zu erhalten. Dies ist insbesondere
hilfreich für Analysen, die sich auf stabile Detektorbedingungen
während der Strahlzeit verlassen und daher empfindlich auf
\emph{run-{}range}-{}abhängige Effizienzkorrekturen sind. \emph{StBadRdosDb} wurde der
gesamten STAR Kollaboration bekannt gemacht [26] und
ebenfalls in der STAR CVS Repositorie zur Verfügung gestellt
[27].\newline

Die C++ Softwareklasse \emph{StRunIdEventsDb} (Section~\ref{strunideventsdb}) wurde
entwickelt, um die Handhabung von \emph{Run}-{}abhängigen Untersuchungen
über mehrere Strahlenergien (Part~\ref{physics}) zu er­leich­tern. Für alle
BES-{}I Energien der Jahre 2010 und 2011 stellt die Implementation eine Abbildung
von einer spezifischen \emph{Run} Identifikationsnummer zum zugehörigen
Listenindex bereit. Für eine Kombination von \emph{Run} und Trigger kann
desweiteren die Anzahl an Events für alle Energien abgefragt werden.\newline

Ein weiteres Kapitel im zweiten Teil dieser Dissertation (Chapter~\ref{cvs2git})
beschäftigt sich mit der Ausführung des Übergangs der zentralen
STAR Code Repositorie von CVS zu \emph{git}. Es wird vorgeführt, wie die
gesamte Geschichte aller STAR Module mithilfe von \emph{cvs2git} konvertiert werden
kann. Außerdem wird \emph{gerrit} als neues Review-{}System für
zukünftige Code-{}Änderungen und als Gateway für alle
Interaktionen mit der zentralen Repositorie eingeführt. Dies bezieht die
Be­reit­stellung der notwendigen Skripten und Anweisungen sowie die
De­mons­tra­tion
eines Einsatzes des Review-{}System auf einem entfernten Server mit ein. Als
Ersatz für \emph{cvsweb} zum Browsen der neuen Repositorien wird \emph{cgit} vorgeschlagen.\newline

Die Größe und projektinternen Bibliotheksabhängigkeiten machen
die Organisation von Software entwickelt für Physik-{}Analysen (Part~\ref{physics})
mühsam [28]. Oft werden die ROOT-{}basierten \emph{shared
libraries} von länglichen Makros anstatt von kompilierten Programmen aus
auf­ge­ru­fen, obwohl es die Qualität des Softwareprojekts
erheblich verbessern würde. Das C++ Programm \emph{ckon} (Section~\ref{ckon}) wurde daher
entwickelt, um den Arbeitsablauf der Kompilation und Ausführung von ROOT
Analysen zu automatisieren. Mit den Befehlen \texttt{cko\penalty5000 n s\penalty5000 e\penalty5000 t\penalty5000 u\penalty5000 p \&\penalty5000 \& c\penalty5000 kon} erlaubt
\emph{ckon} dem Benutzer, den Erstellungsprozess eines gesamten Softwareprojekts
automatisiert zu durchlaufen, aber gleichzeitig über dessen Ordner-{}Layout
zu entscheiden. Der Benutzer ist nicht nur vom \emph{Makefile}-{}Schreiben befreit
sondern kann auch auf einfache Weise fremden Code oder Dritt-{}Bibliotheken sowie
eigene Hilfs-{} und Datenbankklassen mithilfe einer YAML Konfigurationsdatei
einbauen.\newline

Das graphische Darstellungswerkzeug \emph{ccsgp} (Section~\ref{get-started-with-ccsgp}) wurde
entwickelt, um Ver­öffent­lich­ungs-{}fertige
wissenschaftliche Abbildungen mit Gnuplot und Python zu produzieren. Die
Repositorie \emph{wp-{}pdf} (Section~\ref{wp-pdf}) umfasst eine Ansammlung an Anweisungen und
Software, um\hspace{0.167em}\textemdash{}\hspace{0.167em}basierend auf denselben AsciiDoc Dateien\hspace{0.167em}\textemdash{}\hspace{0.167em}simultan eine
Website und einen gedruckten Report für den STAR-{}internen Review-{}Prozess
zu generieren. Das Kommandozeilen-{}Program \emph{rmrg} (Section~\ref{rmrg}) erweitert ROOTs
\texttt{hadd} Programm [29] mit fehlenden Features und besserer Kontrolle, um ROOT
Bäume oder Histogramme zusammenzufügen.
\setcounter{secnumdepth}{3}
\addtocontents{toc}{\protect\setcounter{tocdepth}{5}\ignorespaces}
\setcounter{tocdepth}{5}
\mainmatter
%
%

\part{Physics: Energy-{}Dependence of Dielectron Production in STAR}
\label{physics}\hyperlabel{physics}%


\chapter{Introduction}
\label{intro}\hyperlabel{intro}%
\begin{quote}

«We should investigate [novel] phenomena by distributing high energy or
high nucleon density over a relatively large volume.»

\hspace*\fill--- T.D. Lee (1974 [5])
\end{quote}

\section{The Phases of QCD}
\label{intro_sec_qcd_phases}\hyperlabel{intro_sec_qcd_phases}%

The Standard Model of particle physics [2] comprises a collection
of quantum field theories describing the electromagnetic, weak, and strong
interactions of subatomic particles with relative strengths at the size of a
nucleon according to their coupling constants:
\[ \alpha_\mathrm{s}\sim 1,\, \alpha_\mathrm{em}\sim 1/137,\,\mbox{ and } \alpha_\mathrm{w}\sim 10^{-6}. \]
\emph{Quantum-{}Chromo-{}Dynamics} (QCD)
[30-{}34] has
emerged [1] as the underlying theory of strong
interactions describing the fundamental properties of quarks and their
color-{}exchanging interactions via gluons. Even though QCD has been a very
successful theory to predict and guide experimental observations (e.g.
\ensuremath{\alpha}$_{\text{s}}$ [35], jets [36]), some fundamental
questions remain unresolved:
For instance, the nuclear constituents of ordinary matter are~known but it is
yet to be understood how they arrange themselves into the composite structures
observed in nature. The running coupling constant \ensuremath{\alpha}$_{\text{s}}$ of QCD separates the
low-{}energy regime (low momentum transfers \emph{Q}, large distances) from the
high-{}energy regime in which \ensuremath{\alpha}$_{\text{s}}$ is small enough for QCD to be treated
perturbatively, i.e. by Taylor expansion in orders of \ensuremath{\alpha}$_{\text{s}}$ [37]. This expansion is applicable down to \emph{Q} \ensuremath{\sim} 1 GeV/c or
above distances of about 0.2 fm which is still about an order of magnitude
smaller than the nucleon radius. Hence, the solution of the QCD equations and
therefore the behavior of the nuclear force on energy scales relevant for the
description of stable atomic nuclei remain unsettled.

Significant progress in this non-{}perturbative regime has most successfully
been made within the computational framework of \emph{Lattice QCD} as major
mathematical constraints
\footnote{
in form of the numerical sign problem impeding importance sampling.
} can be avoided at zero net baryon densities and finite temperatures
[38, 39]. However, on scales relevant to the internals of the
nucleons, i.e. for the study of \emph{quark matter} phases with quarks and gluons as
degrees of freedom, perturbative QCD suggests the existence of asymptotically
free (deconfined) quarks and gluons allowing them to escape their \emph{confinement} into color-{}neutral hadrons [37]. Hence, under high temperatures
and energy densities, the equations of QCD predict that nuclear matter
undergoes a phase transition into the so-{}called Quark-{}Gluon-{}Plasma (QGP)
[4]\hspace{0.167em}\textemdash{}\hspace{0.167em}a "soup" in which quarks and gluons constitute the
relevant degrees of freedom.

\pagebreak[4]

In 2005, theorists and experimentalists [40] claimed the discovery of a strongly
coupled QGP (sQGP) sourced in a so-{}called Color Glass Condensate (CGC) which
was critically discussed at the time by the experiments at the Relativistic
Heavy-{}Ion Collider (RHIC, [41]) emphasizing the need for additional
measurements of various other variables
[42-{}45]. It turns out that this sQGP
behaves much like a \emph{perfect fluid} with a viscosity near the quantum limit
[46] and unlike a weakly interacting plasma as would
intuitively be suggested by the small \ensuremath{\alpha}$_{\text{s}}$ in the perturbative regime of
QCD. In recent years, the study of the properties of this dense
nuclear matter has been the main focus of the \emph{High Energy Nuclear Physics} community [47-{}49].

\wrapifneeded{0.50}{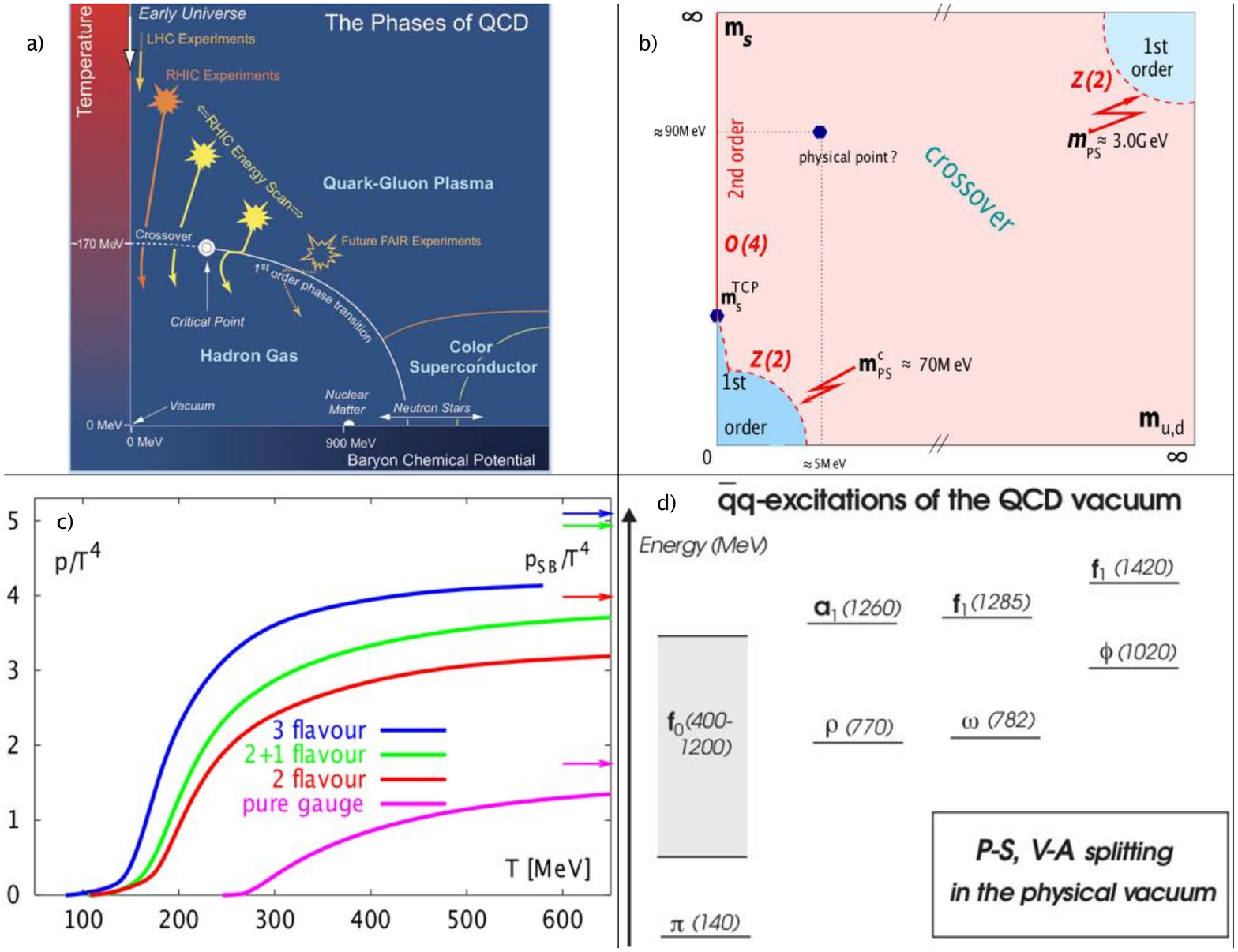}{(Details see text) a) Diagram highlighting the current conception on the possible phases of QCD versus temperature and baryon chemical potential. b) Columbia phase diagram from Lattice QCD as function of up/down (\emph{m}$_{\text{u,d}}$) and strange (\emph{m}$_{\text{s}}$) quark masses [38]. The lower left and upper right regions depict the first-{}order chiral and deconfinement transitions, respectively. The calculations reveal the point of the transition corresponding to physical quark masses to be located in the large crossover regime. c) Temperature dependence of QCD pressure for different degrees of freedom released during the transition from hadron gas to QGP [50]. Arrows indicate the respective Stefan-{}Boltzmann limit valid for vanishing coupling, i.e. for \emph{T}\ensuremath{\rightarrow}\ensuremath{\infty}. d) Measured quark-{}antiquark excitation spectrum showing the splitting of pseudoscalar-{}scalar (P-{}S) and vector-{}axialvector (V-{}A) chiral partners in the QCD vacuum [3]. Note that, Section~\ref{intro_sec_dielec} later refers to the \ensuremath{\rho}-{}a$_{\text{1}}$ partners as particularly interesting for the study of chiral symmetry breaking.}{intro_fig_phdiag_cross_eos_qqbar}{1} %

\pagebreak[4]

The QCD phase diagram shown in Figure~\ref{intro_fig_phdiag_cross_eos_qqbar}a represents
the current conception how strongly interacting nuclear matter evolves into
other phases when varying the thermodynamical environment, namely temperature
\emph{T} and baryon-{}chemical potential \ensuremath{\mu}$_{\text{B}}$ which is linked to net baryon
density. Reaching a better understanding of the QCD phase diagram has direct
implications on the role of nuclear matter in the cosmos. The part of the phase
diagram at low temperatures and high baryon densities, on the one hand, is
closely related to the dense nuclear matter at the core of neutron stars
[51]. Due to the tiny ratio of baryon to photon number of about
10$^{\text{-{}9}}$ (\ensuremath{\propto}\ensuremath{\mu}$_{\text{B}}$/\emph{T}) [2], the part at high temperatures and
low densities, on the other hand, corresponds to the thermodynamical path that
the expanding and cooling early universe took microseconds after the big bang.\newline

At moderate temperatures and baryon densities, nuclear matter exists in form of
a Hadron Gas [52] as dictated by the confining long-{}range nature
of QCD. Around temperatures of about 170 MeV, Lattice QCD predicts the
liberation of the confined quarks to the QGP [53] resulting in a
notable rise in pressure with increasing temperature due to the release of many
additional degrees of freedom (see Figure~\ref{intro_fig_phdiag_cross_eos_qqbar}c). The
calculation for 2+1 flavors using light up/down quarks and a heavy strange
quark corresponds to the most realistic case for the equation of state
[50]. The calculations in Figure~\ref{intro_fig_phdiag_cross_eos_qqbar}b at
zero net baryon density [38] also identify the transition to be
of rapid analytical crossover type given the location of the physical quark
masses in the broad range of possible m$_{\text{u,d}}$ and m$_{\text{s}}$ configurations in the
so-{}called Columbia phase diagram [54].\newline

Even though Taylor expansion in \ensuremath{\mu}$_{\text{B}}$ allows Lattice QCD to be extended into
the small non-{}zero \ensuremath{\mu}$_{\text{B}}$ regime, effective models like Nambu-{}Jona-{}Lasinio
[55, 56] need to be employed instead to cover predictions
beyond \ensuremath{\mu}$_{\text{B}}$ \ensuremath{\sim} \emph{T}. However, such models also need to incorporate the
second fundamental QGP characteristics after \emph{deconfinement}, namely \emph{chiral
symmetry breaking} (\ensuremath{\chi}SB). Chiral symmetry is evident in the zero quark mass
limit of the QCD Lagrangian in form of a re-{}organization of the QCD vacuum in
left and right handed contributions with respect to the orientation of spin versus
parity. It is spontaneously broken by the QCD vacuum exhibiting a non-{}zero
quark-{}antiquark condensate ($\langle\bar{q}q\rangle$)
[57], and hence expected to be restored at high temperatures
[58]. Spontaneous \ensuremath{\chi}SB explains the basic features of the
experimental hadron spectrum [3], i.e. the manifestation of eight (nearly) massless
Goldstone bosons (\ensuremath{\pi}, \emph{K}, \ensuremath{\eta}) as well as non-{}degenerated opposite parity
isospin multiplets as observed in Figure~\ref{intro_fig_phdiag_cross_eos_qqbar}d. The
effective models used at \ensuremath{\mu}$_{\text{B}}$ >{} \emph{T} suggest this chiral transition to be of
first order for realistic light quark masses [59, 60].
Together with the Lattice QCD predictions at zero \ensuremath{\mu}$_{\text{B}}$, it results in the
appearance of a critical point between the crossover and first-{}order regions.
The latter might end in another critical point at low temperatures due to an
unclear border between superfluid nuclear matter and superconducting quark
matter [61].

The occurrence of the deconfinement and chiral transitions can intuitively be
understood using the MIT bag model [62, 50] which treats
hadrons as objects confining the valence quarks within a finite size of
vanishing $\langle\bar{q}q\rangle$: At high temperatures, the
individual bags merge into a single large bag allowing quarks and gluons to
move freely (see percolation models for QCD [63]). Also, the
hadronic bag is larger than the constituent quark bag defined by current quarks
covered in a cloud of virtual quarks and gluons.  This has lead to the
speculation of distinct deconfinement and chiral phase transitions with
$T_d\leq T_\chi$ [64, 65].\newline

The preceding discussion only outlines the most prominent features of the rich
QCD phase diagram bearing relevance to the contents and objective of this
thesis (Section~\ref{intro_sec_thesis}). The publications [61] and
[66] including the references therein review the phase diagram
in dense QCD and at high baryon densities giving rise to additional partially
conjectured phase structures like \emph{quarkyonic matter}, for instance, which
shares properties from baryonic and quark matter phases.

\section{Ultra-{}relativistic Heavy-{}Ion Collisions}
\label{intro_sec_hics}\hyperlabel{intro_sec_hics}%

The quote at the beginning of Chapter~\ref{intro} [67], recalls T.D. Lee's
suggestion to try restoring the broken symmetries of the physical vacuum
temporarily by studying extended environments of high energy or nucleon
densities. It was subsequently demonstrated [5, 6]
that the latter could be achieved via relativistic heavy-{}ion collisions which
are still to date the only means to recreate nuclear matter in the laboratory
as it existed during the early universe. The phase diagram in
Figure~\ref{intro_fig_phdiag_cross_eos_qqbar}a also depicts \ensuremath{\mu}$_{\text{B}}$-{}\emph{T}-{}regions
currently reachable by experiments at the Large Hadron Collider (LHC), in
RHIC's top energy and Beam Energy Scan (BES) programs as well as at the future
Facility for Antiproton and Ion Research (FAIR)\hspace{0.167em}\textemdash{}\hspace{0.167em}to name a selected few.
The initial phase of the BES
program in the years 2010/11 (BES-{}I) serves the purpose of scanning the phase
diagram of QCD, first and foremost, for evidence of the critical point and the
first order phase transition. To locate these characteristics of nuclear
matter, the considerable capabilities of the \emph{Solenoidal Tracker at RHIC} (STAR) allow for a multitude of observables like particle production,
collective flow, fluctuations, cumulants, jets and others to be measured and
studied in BES-{}I [20]. Consistently combining the resulting
various signatures over a wide range of beam energies makes the BES program a
powerful tool to explore the \ensuremath{\surd}s-{}dependence as well as possible "turn
on/off" effects of partonic media.

Figure~\ref{intro_fig_phdiag_cross_eos_qqbar}a also indicates the according evolution
trajectories such a medium takes through the QCD phase diagram while expanding
and cooling. To provide the reader with better insights regarding the evolution
of a heavy-{}ion collision, the space-{}time picture of the subsequential stages
are depicted in Figure~\ref{intro_fig_hic_stages} and schematically discussed in the
following [68] in terms of the invariant \emph{proper time} $\tau=\sqrt{t^2-z^2}$.

\wrapifneeded{0.50}{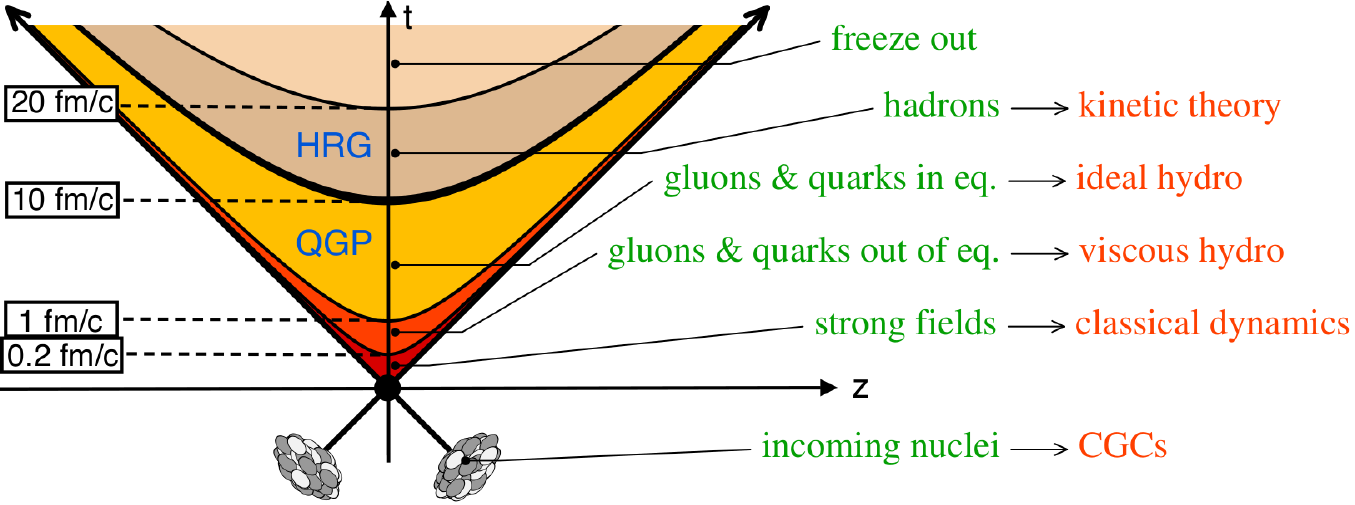}{Stages of a heavy-{}ion collision versus time \emph{t} and collision axis \emph{z} [68]. The invariant \emph{proper time} \ensuremath{\tau} is constant along the solid hyperbolic curves and approximate values corresponding to the various transitions are given in the boxes on the left. The annotations on the right denote some of the stages' features and theoretical treatments. See text for more details.}{intro_fig_hic_stages}{0.85} %

Before the collision (\ensuremath{\tau} <{} 0), the incoming nuclei travel at
ultra-{}relativistic speeds and time dilation causes them to appear as
pancake-{}like objects to an observer in the center-{}of-{}mass frame. The
nuclear constituents within these thin disks are forced by the Lorentz
contraction on very small distances at which (color-{}carrying) gluons have
taken over as the dominating content of hadrons [69]. The gluons can then
self-{}generate very high gluon densities via gluon-{}splitting (condensate) which
inversely corresponds to very high transverse momenta making the incoming
nuclei effectively behave like gluonic walls. Such a solid-{}type behavior on
short as opposed to liquid-{}type on long time scales coined the term for this
other new form of nuclear matter: \emph{color glass condensate} (CGC)
[70]. The stochastic source of these self-{}generating
high-{}density gluon fields is hence of basically static nature which allows the
parton distributions to be replaced by an ensemble of coherent classical fields
for the CGC's theoretical treatment [71].

The initial scattering at \ensuremath{\tau} = 0 is followed by so-{}called \emph{hard} processes
with momentum transfers of typically \emph{Q} >{} 10 GeV/c at LHC energies resulting
in the characteristic di-{}jet events [72, 73], for
instance. Around \ensuremath{\tau} \ensuremath{\sim} 0.2 fm/c, the transverse momentum scale becomes
\emph{semi-{}hard} with \emph{Q} \ensuremath{\sim} 1 GeV/c and the gluons composed in the CGC are
liberated generating most of the final-{}state multiplicity via fragmentation and
hadronization of these initial-{}state gluons.  The liberated partons form the
non-{}equilibrium state of partonic matter called \emph{Glasma} until it thermalizes
into the QGP at \ensuremath{\tau} \ensuremath{\sim} 1 fm/c [74]. Naively speaking,
thermalization occurs when scattering times characterized by small interaction
lengths are small compared to expansion times. Such a quick approach to
thermalization at about 1 fm/c requires coherent and strong self-{}interaction of
the liberated partons to compete the medium expansion [75].
However, redistribution of energy in spite of being rapidly driven apart only
works to a certain degree since the QGP keeps on expanding and cooling. The
result is a space-{}time dependent temperature profile which only allows for
local thermal equilibrium to be reached during expansion. The development of
thermal equilibrium distinguishes the respective theoretical treatment of
Glasma and QGP via viscous and ideal hydrodynamics [76] which
describe the medium's expansion and flow, thus referring to the fluid's
long-{}range behavior.

At \ensuremath{\tau} \ensuremath{\sim} 10 fm/c, hadronization traps quarks and gluons inside color-{}less
hadrons corresponding to the previously discussed phase transition into a
Hadron Resonance Gas (HRG). The continuously expanding and cooling
HRG subsequently undergoes two conceptually different so-{}called \emph{freeze-{}out} transitions [7]: (\emph{i}) a chemical freeze-{}out stopping inelastic
processes that change hadronic species into each other and hence freezing
hadronic abundances, and (\emph{ii}) a thermal freeze-{}out at which all
momentum-{}changing interactions, i.e. also all elastic collisions, cease. The
cross section of the abundance-{}changing inelastic hadron-{}hadron reactions is
(usually much) smaller than the total cross-{}section which drives all
momentum-{}changing and hence thermally equilibrating reactions. This makes it
increasingly harder for inelastic reactions to keep up with the continuously
decreasing density. They hence become unlikely earlier during the medium
expansion than elastic reactions, and thus
cause the chemical freeze-{}out to occur at higher temperatures than the thermal
freeze-{}out [77]. The latter happens at \ensuremath{\tau} \ensuremath{\sim} 20 fm/c when the
HRG cannot keep the local thermal equilibrium intact anymore. Freely streaming
particles are then allowed to escape to the detector measuring momentum
distributions that are expected to be the same as the thermal distribution in
the very late-{}stage expansion [78]. Note that, for RHIC (7.7 -{}
200 GeV) and SPS (160 -{} 200 AGeV) energy regimes, the chemical freeze-{}out
temperature is close to the critical temperature of the deconfinement
transition. At energies covered at the Alternating Gradient Synchrotron (AGS, <{}
11 AGeV), however, the chemical and thermal freeze-{}out temperatures approach
each other and finally coincide at energies measured at the Heavy-{}Ion
Synchrotron (SIS18, <{} 2 AGeV) [79].\newline

An important concept in the evolution of heavy-{}ion collisions is the
aforementioned development of a collective motion called \emph{flow} [80]. In the simplified case of a (strictly) central collision,
the interaction zone is circular when viewed along the incident beam
directions. The constituents rescattering in the expanding nuclear liquid move
outward collectively in a radially isotropic manner typically referred to as
\emph{radial flow} [81]. The according radial expansion velocity
\ensuremath{\beta} can be extracted from the paramaterization of the Blast-{}Wave model to
final-{}state hadron p$_{\text{T}}$ spectra [82] which are a result of
inherently isotropic thermal emission at temperature \emph{T} and radial flow.  The
latter is reflected in the definition of the initial effective temperature of
the medium [83, 84]:
\[T_\mathrm{eff} \approx T\sqrt{\frac{1+\beta}{1-\beta}}.\]
The initial overlap zone of two non-{}centrally colliding nuclei, illustrated in
Figure~\ref{intro_fig_flow} (left) [85], is usually of elliptic shape
which results in a higher pressure gradient along the shorter direction than
along the longer direction. Rescattering thus causes the constituents of the
liquid to collectively flow faster in the shorter direction effectively
converting the initial spatial into a momentum anisotropy. The azimuthal
anisotropy in flow velocity also reveals the self-{}quenching nature of this
so-{}called \emph{elliptic flow} as it drives the matter toward a spherical shape.
Elliptic flow is characterized by the second Fourier coefficient v$_{\text{2}}$ in
azimuthal multiplicity distributions of final state hadrons
[8, 9]:
\[ \frac{1}{N}\frac{\mathrm{d}N}{\mathrm{d}\phi} = v_0 + 2v_1\cos\phi + 2v_2\cos(2\phi) + \cdots \]
The chronological build-{}up of elliptic flow during the expansion is shown in
Figure~\ref{intro_fig_flow} (right) using the v$_{\text{2}}$ measure. The full magnitude of the
elliptic flow only develops late(r) in the evolution with the specific timing
depending on the effective parton scattering cross sections. Hence, particles
created early in the evolution and embedded in this anisotropically expanding
medium experience no significant flow, yet. The early saturation of elliptic
flow well before the hadronization transition is also indicative of the
thermalization to happen on the quark rather than the hadronic level [86].

\wrapifneeded{0.50}{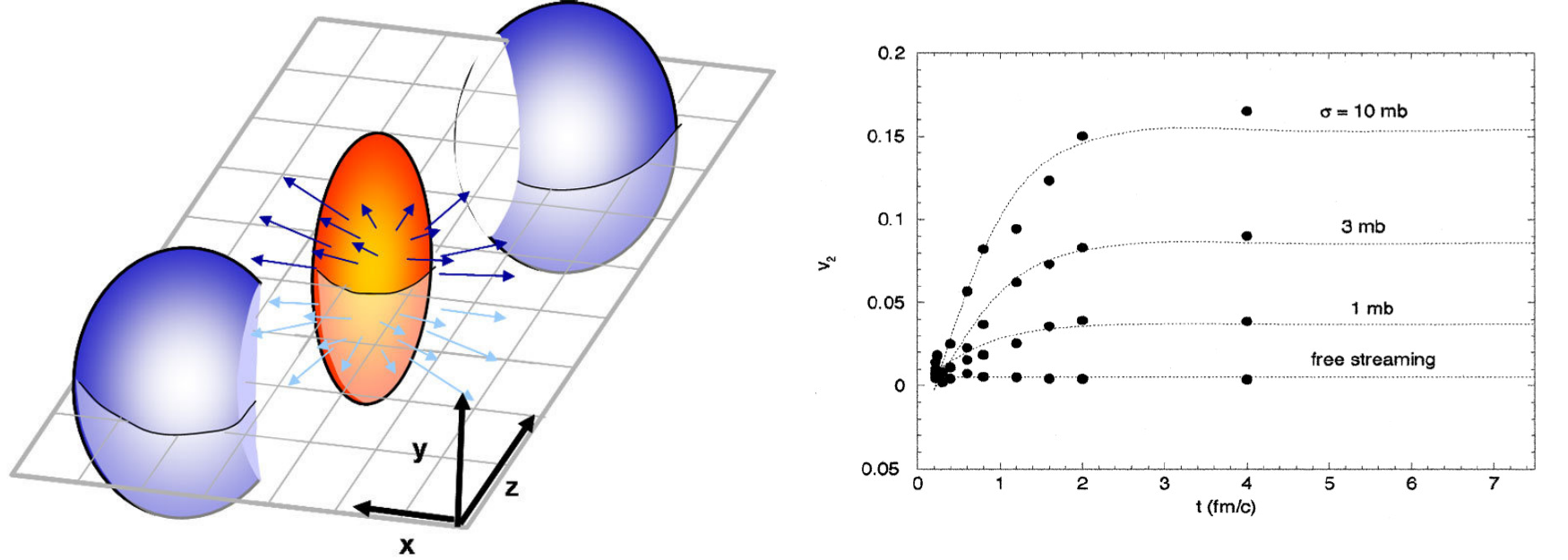}{(left) Illustration of the development of flow in a non-{}central heavy-{}ion collision caused by the asymmetric spatial shape of the collisional overlap zone. (right) Time-{}dependence of elliptic flow v$_{\text{2}}$ for different partonic scattering cross sections [86]. See text for discussion.}{intro_fig_flow}{0.85} %

In summary, Figure~\ref{intro_fig_hic_stages} reiterates the similarity of the
space-{}time picture of a heavy-{}ion collision to Big Bang cosmology and the
related discussion identifies a few fundamental questions that need particular
attention in the context of heavy-{}ion collisions\hspace{0.167em}\textemdash{}\hspace{0.167em}especially if one wants to
prove the creation of a QGP [87, 44]. First, whether thermal
equilibrium is reached through sufficient rescattering during the initial stage
of the collision. Measurements of elliptic flow at RHIC reveal a magnitude of
collective expansion that indicates thermalization time scales of less than 1-{}2
fm/c and hence, justify the characterization of the medium using bulk
thermodynamic variables [88]. Second, whether a distinctive
footprint for individual partons can be identified.  Number-{}of-{}constituent
(NCQ) scaling holding up for mesons and baryons suggests a collectively
expanding partonic source [89, 90]. Third, signals of the
two fundamental QCD characteristics, chiral symmetry restoration and
deconfinement, need to be comprehensively studied. For the latter
characteristic, detection of direct \emph{thermal electromagnetic radiation} as
expected from $q\bar{q}$ annihilation in a thermally equilibrated
deconfined QCD medium, would provide one of the desired \emph{smoking guns}. Another
would be to unequivocally measure the disappearance of the vacuum consequences
of spontaneous \ensuremath{\chi}SB in the QCD medium (see
Figure~\ref{intro_fig_phdiag_cross_eos_qqbar}d and Section~\ref{intro_sec_qcd_phases}).

As will be discussed in the remainder of this introductory chapter, both
signatures might be accessible via \emph{dileptons} which are regarded as \emph{golden
probes} of the early stages due to their bulk-{}penetrating electromagnetic
nature. To cover the various aspects of these striking final-{}state ingredients
of a heavy-{}ion collision, Section~\ref{intro_sec_dielec} further motivates the study of
dilepton production including a recapitulation of its experimental and
theoretical status. Section~\ref{intro_sec_thesis} concludes the introduction with a
brief overview of the thesis and its objective.

\section{Dileptons as Bulk-{}Penetrating Electromagnetic Probes}
\label{intro_sec_dielec}\hyperlabel{intro_sec_dielec}%

The precise measurement of the \ensuremath{\rho}-{} and a$_{\text{1}}$-{}mesons' spectral functions in
hadronic \ensuremath{\tau} decays at the Large Electron-{}Positron collider (LEP)
[91] is shown in Figure~\ref{intro_fig_dm_mr} (left), underlying the
vector-{}axialvector (V-{}A) splitting of about 500 MeV in
Figure~\ref{intro_fig_phdiag_cross_eos_qqbar}d. It exhibits a clear separation of the
respective pole masses as expected from spontaneous \ensuremath{\chi}SB in the QCD vacuum.
In a simplified picture, two scenarios can be distinguished for the
degeneration of the chiral partners in a quark matter environment
(Figure~\ref{intro_fig_dm_mr} right, [87]): (\emph{i}) the a$_{\text{1}}$ pole mass drops to
the \ensuremath{\rho} mass and/or they approach each other, hence keeping the resonant
strucure intact, or (\emph{ii}) the resonances "melt" down to approach the according
production rates from perturbative QCD.  The clarifying study of the fate of
chiral symmetry and its manner of possible restoration in dense states of
nuclear matter, thus requires experimental access to the so-{}called \emph{in-{}medium} spectral functions of the chiral partners.

\wrapifneeded{0.50}{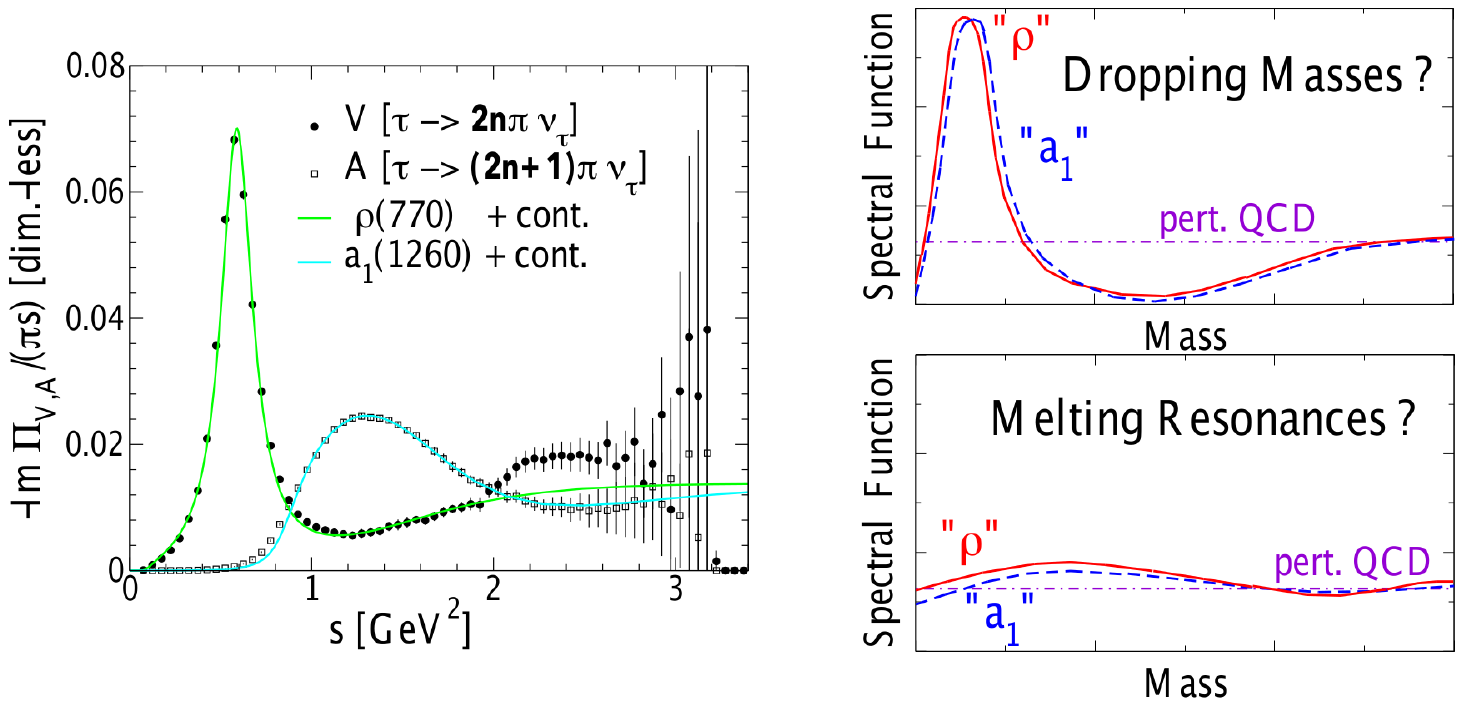}{(left) Measurement of the chiral partners \ensuremath{\rho}-{} and a$_{\text{1}}$-{}meson in hadronic decays at LEP. (right) Simplified scenarios of dropping pole masses or melting resonances for the degeneration of the chiral partners in an environment of hot and dense nuclear matter. Above an invariant mass of about 1.5 GeV/c$^{\text{2}}$ (\ensuremath{\surd}s = M$_{\text{ee}}$) called the \emph{duality threshold}, the production rates are expected to resemble the ones from perturbative QCD. [87]}{intro_fig_dm_mr}{0.75} %

However, the relation of medium-{}modifications of vector meson spectral
functions to chiral symmetry restoration (\ensuremath{\chi}SR) is not straight-{}forward. Spectral
functions need to be connected, on the one hand, to the order parameter of the
chiral transition, namely (a reduction in) the $q\bar{q}$ condensate, and, on the other hand, to the measured dilepton yields emanating
from the entire fireball evolution. This paragraph follows the arguments in
[10] (Section 2.7.1) and [11] in an effort to make the
connection that allows dilepton production measured in heavy-{}ion collisions to
be interpreted with respect to \ensuremath{\chi}SR.

\pagebreak[4]

\wrapifneeded{0.50}{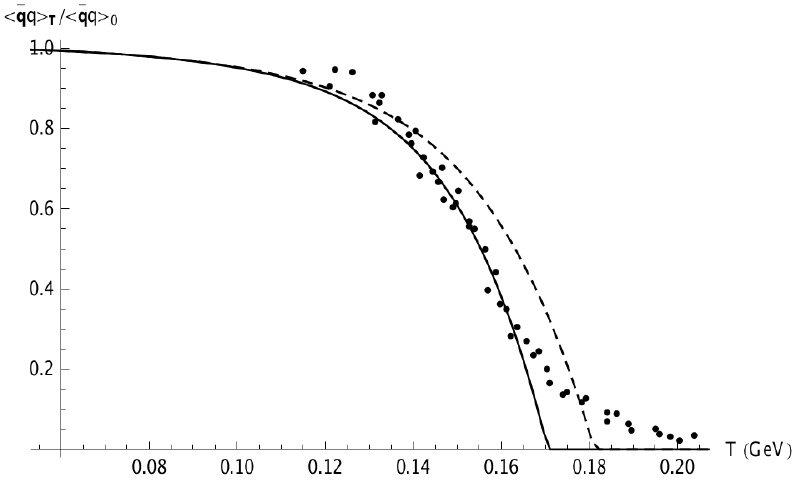}{Temperature dependence of the chiral quark condensate from Lattice QCD calculations (solid circles) compared to the condensate-{}diminishing effect of thermal excitations in the HRG (dashed line). A high-{}order temperature-{}dependent correction improves the agreement above 140 MeV (solid line). [11]}{intro_fig_qq_T}{0.49} %

Qualitatively, the relation can be retraced through the discussion of the
model-{}independent (\emph{T},\ensuremath{\mu}$_{\text{B}}$)-{}dependence of the chiral condensate in a pion
or nucleon gas at low \ensuremath{\mu}$_{\text{B}}$ and \emph{T} [92-{}94]:
\[\frac{\langle\bar{q}q\rangle(T,\mu_B)}{\langle\bar{q}q\rangle_0} = 1 - \sum_h\frac{\rho_h^s\Sigma_h}{f_\pi^2m_\pi^2}\]
with $f_\pi$ the pion decay constant, $\rho_h^s$ the
(scalar) density and $\Sigma_h=m_q\langle h\vert\bar{q}q\vert h\rangle$ the sigma commutator of hadron \emph{h}. In the non-{}relativistic limit,
$\Sigma_h/m_q$ denotes the scalar quark density in the
hadron. The expression illustrates that \ensuremath{\chi}SR is an additive effect since
every hadron contributes a certain decrease to the quark condensate according
to its $\Sigma_h$, with $\rho_h^s$ being produced
through thermally excited hadrons [94]. The expression has been
generalized for a hadron resonance gas in leading order of density, also taking
care of hadronic interactions in a manner consistent with chiral symmetry. In
other words, thermal resonance excitations in the hadron gas give rise to the
so-{}called "vacuum cleaner" effect since the resonances' quark content "sweeps
up" the chiral condensate of the vacuum [11]. Together with
Figure~\ref{intro_fig_qq_T}, this shows that presence of hadrons alone already results
in the reduction of the chiral quark condensate consistent with ab-{}initio
Lattice QCD calculations. The connection to spectral functions becomes clear
when studying the mechanisms of the effective hadronic model in
Section~\ref{sim_sec_model} and [95] that are required to comply with the
results of the dilepton measurements introduced in Section~\ref{intro_subsec_dielec_meas} and especially with the results of this thesis (Chapter~\ref{results}). In short, the
components that underlie the preferred scenario of a broadening of the
\ensuremath{\rho}-{}meson's spectral function, are (\emph{i}) the modified coupling of its pion
cloud to the medium ($\rho\pi\pi$), and (\emph{ii}) its direct
interactions with the mesons and baryons in the heat bath exciting further
resonances (\ensuremath{\rho}MB).  Hence, thermal resonance excitations driving the
reduction of the chiral condensate are actually one of the processes causing
the \ensuremath{\rho}-{}broadening (in addition to chiral mixing through the pion cloud).
This close relation of (partial) \ensuremath{\chi}SR and \ensuremath{\rho} spectral function is further
strengthened by the fact that the $\rho\pi\pi$ and \ensuremath{\rho}MB medium
effects find their counterparts in the $\Sigma_h$ term since it
decomposes into short-{} and long-{}distance contributions from the hadron's quark
core and pion cloud, respectively [96].

Taking the direct approach and measuring the a$_{\text{1}}$ meson in the medium is
experimentally difficult due to its low production rates and broad width.
Instead, a coordinated procedure has to be followed to quantitatively connect
the chiral quark condensate to measured dilepton yields through the \ensuremath{\rho}
meson's spectral function [10]: On the theoretical side, the vector
and axial-{}vector spectral functions are calculated within a chirally invariant
model carefully constrained by basic symmetry principles using effective
vertices in agreement with available data. Using Weinberg sum rules,
temperature-{}dependent pion decay constants and quark condensates can be deduced
and compared for consistency with Lattice QCD results. The spectral functions
are propagated through a realistic model for the fireball expansion taking into
account measured hadronic observables (e.g. v$_{\text{2}}$). On the experimental side,
this enables the detailed comparison to data with respect to centrality-{},
$\sqrt{s_\mathrm{NN}}$-{}, mass-{} and p$_{\text{T}}$-{}dependence. Besides
centrality, this thesis covers these pivotal ingredients and
hence tightens the connection for dilepton data to be interpreted in terms of
\ensuremath{\chi}SR. Since spectral functions are difficult to obtain
in Lattice QCD in the near future, this approach is the only way to enable
another leap forward in dilepton physics.

\pagebreak[4]

Vector mesons are generally a good choice for the \emph{in-{}medium} spectroscopy of a
quickly expanding and cooling fireball. As indicated by the life times listed
below, short-{}lived resonances such as the \ensuremath{\rho}-{}, \ensuremath{\omega}-{} and
$\phi$-{}mesons decay on time scales very similar to the evolution of
a heavy-{}ion collision (cf. Figure~\ref{intro_fig_hic_stages}). The long-{}lived
pseudo-{}scalar \ensuremath{\pi}-{} and \ensuremath{\eta}-{}mesons, on the other hand, decay long after
freeze-{}out and hence, are not suitable for the study of the early-{}stage QCD
medium.

{\centering\savetablecounter \begingroup%
\setlength{\newtblsparewidth}{1\linewidth-2\tabcolsep-2\tabcolsep-2\tabcolsep-2\tabcolsep-2\tabcolsep-2\tabcolsep-2\tabcolsep}%
\setlength{\newtblstarfactor}{\newtblsparewidth / \real{426}}%
\begin{longtable}{llllll}\hline
\multicolumn{1}{m{71\newtblstarfactor}|}{\centering%
\ensuremath{\rho}
}&\multicolumn{1}{m{71\newtblstarfactor}|}{\centering%
\ensuremath{\Delta}
}&\multicolumn{1}{m{71\newtblstarfactor}|}{\centering%
\ensuremath{\omega}
}&\multicolumn{1}{m{71\newtblstarfactor}|}{\centering%
$\phi$
}&\multicolumn{1}{m{71\newtblstarfactor}|}{\centering%
\ensuremath{\eta}
}&\multicolumn{1}{m{71\newtblstarfactor+\arrayrulewidth}}{\centering%
\ensuremath{\pi}$^{\text{0}}$
}\tabularnewline
\multicolumn{1}{m{71\newtblstarfactor}|}{\centering%
1.3 fm/c
}&\multicolumn{1}{m{71\newtblstarfactor}|}{\centering%
1.7 fm/c
}&\multicolumn{1}{m{71\newtblstarfactor}|}{\centering%
23 fm/c
}&\multicolumn{1}{m{71\newtblstarfactor}|}{\centering%
46 fm/c
}&\multicolumn{1}{m{71\newtblstarfactor}|}{\centering%
151 pm/c
}&\multicolumn{1}{m{71\newtblstarfactor+\arrayrulewidth}}{\centering%
26 nm/c
}\tabularnewline
\hline
\end{longtable}\endgroup%
\restoretablecounter%
}

Due to the matching quantum numbers $J^{PC}=1^{-\,-}$,
Vector Meson Dominance [97] also allows vector mesons to couple to virtual
photons which in turn directly decay into \emph{dielectrons} ($e^+e^-$ pairs) and \emph{dimuons} ($\mu^+\mu^-$ pairs) collectively called
\emph{dileptons}.  The electromagnetic nature of their interactions causes
interaction lengths \ensuremath{\lambda}$_{\text{mfp}}$ much larger than the fireball life time
rendering final-{}state interactions negligible.  Hence, \emph{dileptons} can serve as
pristine probes allowing for direct and undistorted access to the properties of
\emph{in-{}medium} spectral functions which makes them the most promising candidates
to conduct these studies.  Note that, even though \ensuremath{\Delta}-{}baryons exhibit life
times comparable to the \ensuremath{\rho}-{}meson, such in-{}medium information would be lost
in the respective hadronic decay channels $\Delta\to\pi\mathrm{N}$ and $\rho\to\pi\pi$.  The decay products suffer rescattering in the
fireball which basically restricts their sensitivity to the break-{}up stages of
the collision.

\wrapifneeded{0.50}{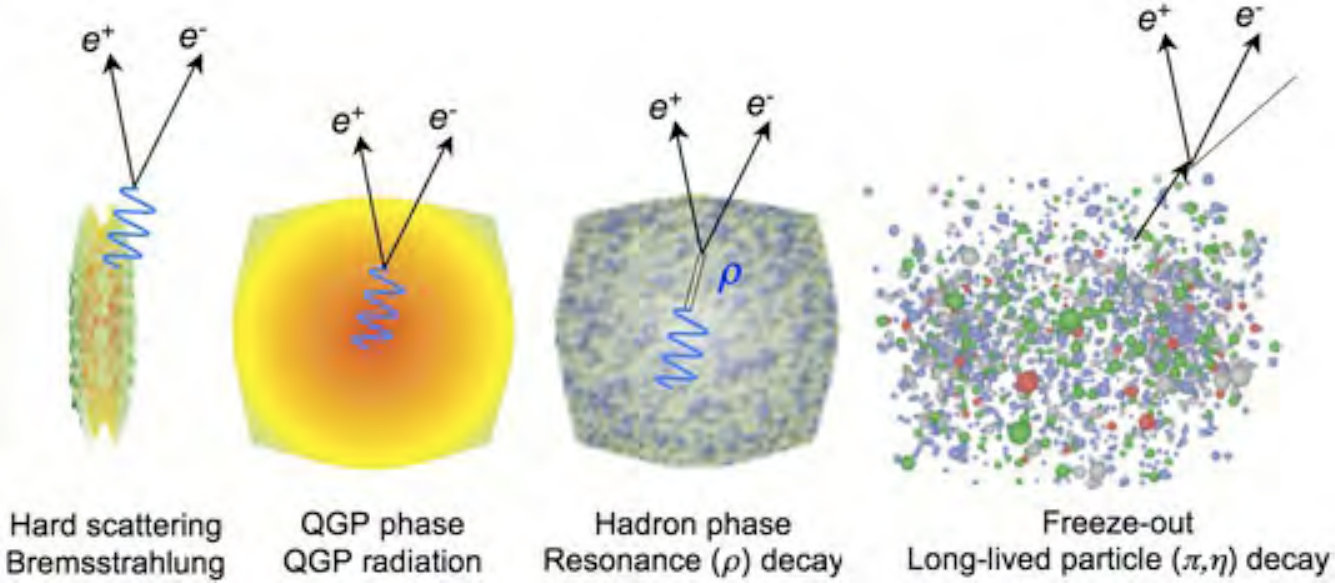}{Sources of dielectron production during the subsequential stages of a heavy-{}ion collision. Dielectrons can be referred to as \emph{bulk-{}penetrating} probes since they penetrate through the surrounding medium without noticeable interaction and also emanate from the earliest (bulk) stages.}{intro_fig_dielec_prod}{0.75} %

As visually depicted in Figure~\ref{intro_fig_dielec_prod}, dielectrons also emanate
from the initial hard scattering via Bremsstrahlung and from the QGP phase via
electromagnetic (thermal) radiation in addition to the decays during hadron gas
and freeze-{}out phases. Being thus created throughout the entire evolution of
the hot dense medium, dielectrons can be considered \emph{bulk-{}penetrating} probes
and provide dynamic as well as direct information about their original
heavy-{}ion collision stage encoded in their invariant mass (M$_{\text{ee}}$) and
transverse momentum (p$_{\text{T}}$).  At this point, it is illustrative to discuss the
prevalent features of a simulated dielectron invariant mass spectrum in p+p
collisions as shown in Figure~\ref{intro_fig_invmass} and compare to the expected
sources of dielectron production in heavy-{}ion collisions. There are multiple
categories of electromagnetic radiation and \emph{background} sources at play
contributing to different invariant mass regions of the "physical" spectrum. It
can generally be concluded that in heavy-{}ion collisions, earlier creation times
correspond to higher temperatures and thus to higher dielectron invariant
masses and vice versa.

\wrapifneeded{0.50}{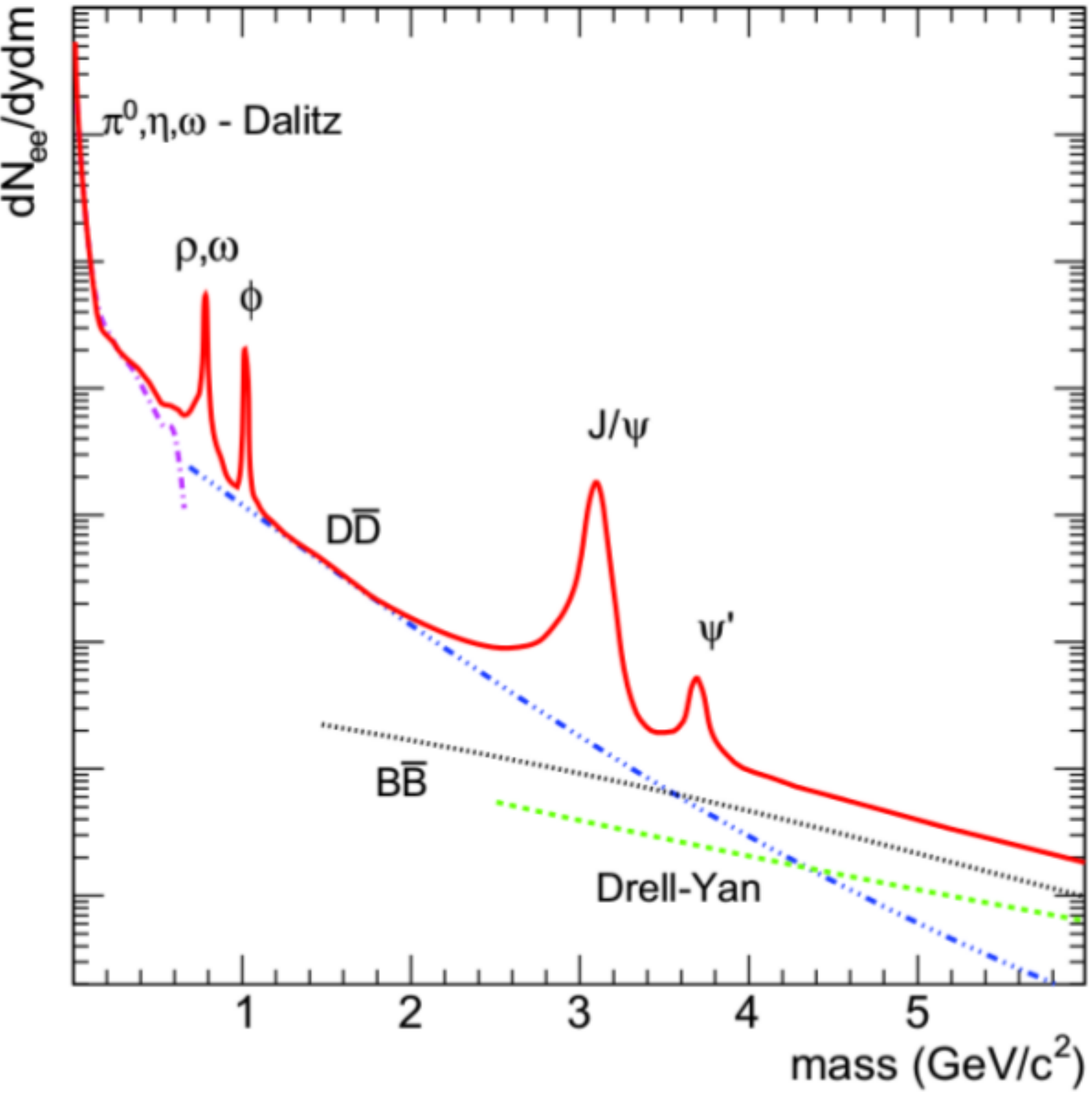}{Simulated dielectron invariant mass spectrum in p+p collisions at 200 GeV [12]. Details on the simulation at BES energies in STAR can be found in Section~\ref{sim_sec_cocktail}.}{intro_fig_invmass}{0.47} %

The High-{}Mass Region (HMR, M$_{\text{ee}}$ >{} 3 GeV/c$^{\text{2}}$) is primarily occupied by yields
from the prompt production during the primordial nucleon-{}nucleon collision.
Drell-{}Yan annihilation of quark and antiquark of two passing hadrons dominates
over thermal emission in this region.  Hidden heavy flavor vector mesons
(J/\ensuremath{\psi}, \ensuremath{\psi}') appear on top of this continuum that also has contributions
from semi-{}leptonic weak decays of B-{}mesons and \ensuremath{\Lambda}$_{\text{b}}$-{}baryons originating
from back-{}to-{}back $b\bar{b}$ pairs and hence, resulting in
correlated dileptons.

The Intermediate-{}Mass Region (IMR, 1.1 <{} M$_{\text{ee}}$ <{} 3 GeV/c$^{\text{2}}$) also exhibits a
continuous yield of correlated $e^+e^-$ pairs from semi-{}leptonic
decays of charmed mesons competing in a heavy-{}ion environment with thermal
radiation via $q\bar{q}$ annihilation from the QGP. With
contributions from possibly medium-{}modified correlated charmed decays known, an
initial effective temperature of the fireball can be deduced in this region
from the slope of dielectron m$_{\text{T}}$ spectra given sufficient statistics. Due to
the scarcity of electromagnetic probes, the latter impedes such a measurement
even at SPS or lowest RHIC energies where the $c\bar{c}$ cross
sections fall off steeply.

In the Low-{}Mass Region (LMR, M$_{\text{ee}}$ <{} 1.1 GeV/c$^{\text{2}}$), dielectron production is
dominated by the decay products from final-{}state hadronic sources culminating
in a \emph{cocktail} of light pseudo-{}scalar and vector mesons
(\ensuremath{\pi},\ensuremath{\eta},\ensuremath{\eta}',\ensuremath{\rho},\ensuremath{\omega},\ensuremath{\phi}). Electromagnetic \emph{radiation} due to
(multiple) reinteractions in the hot and dense HRG populates this region and
could provide information about in-{}medium modifications of the vector mesons'
spectral function properties.  Modifications of the dielectron rates and their
integration over the fireball evolution as well as consequences on the
invariant mass spectrum are addressed in the context of this thesis within an
effective multi-{}body hadronic theory in Section~\ref{sim_sec_model}.

\subsection{Measurements of Dilepton Production}
\label{intro_subsec_dielec_meas}\hyperlabel{intro_subsec_dielec_meas}%

The challenging measurement of dilepton production has been an active field of
research in experimental high energy nuclear physics through multiple
generations of detectors at colliders and fixed-{}target machines since the
1990's. The reader is referred to the main part of the thesis for details on
the general analysis and simulation techniques employed to arrive at the
spectra presented in the following. Here, we concentrate on the results of
selected influential measurements and their current interpretations and
conclusions leading up to the important contributions of this thesis to the
progress of dielectron physics.\newline

The CERES and NA45 collaborations
[13, 98-{}100] first reported
an unexplained LMR excess yield in S+Au collisions while achieving excellent
agreement of cocktail simulations and data in p+N collisions
(Figure~\ref{intro_fig_ceres_na45_na60}a). However, follow-{}up comparisons of measured
LMR invariant mass spectra with calculated dropping-{}mass and broadening
spectral functions of the \ensuremath{\rho}-{}meson in Pb+Au collisions, did not allow for
the distinction of the two scenarios. The emergence of high-{}quality dimuon
measurements in In+In collisions by the NA60 collaboration [14]
significantly accelerated the progress by clearly favoring the broadened
spectral function scenario for the modifications of the \ensuremath{\rho}-{}meson driven by
(multiple) baryonic interactions in the hot hadronic phase
(Figure~\ref{intro_fig_ceres_na45_na60}b). The tail above the \ensuremath{\rho}-{}mass can be
described by 4\ensuremath{\pi}-{}mixing, correlated charm and QGP contributions
[101]. Also particularly interesting are the conclusions that can be
drawn from NA60's measurement of the effective temperature T$_{\text{eff}}$ and its mass
dependence in Figure~\ref{intro_fig_ceres_na45_na60}c. The steady rise up to M$_{\text{\ensuremath{\mu}\ensuremath{\mu}}}$ of about 1 GeV/c$^{\text{2}}$ is consistent with the radial flow expected from the
hadronic emission source $\pi\pi\to\rho\to\mu\mu$. The sudden drop
and hence absence of significant flow above 1 GeV/c$^{\text{2}}$, however, suggests the
partonic emmission via $q\bar{q}\to\mu\mu$ as dominant source.
Accumulation of further evidence as discussed in [14], is even
consistent with the interpretation of the excess dimuons as thermal radiation.
This not only triggered renewed attention to dilepton physics by the theory
community, but also independent and dedicated high-{}accuracy measurements of the
dielectron decay channels at top RHIC energy\hspace{0.167em}\textemdash{}\hspace{0.167em}first by PHENIX and recently
complemented by STAR.

\wrapifneeded{0.50}{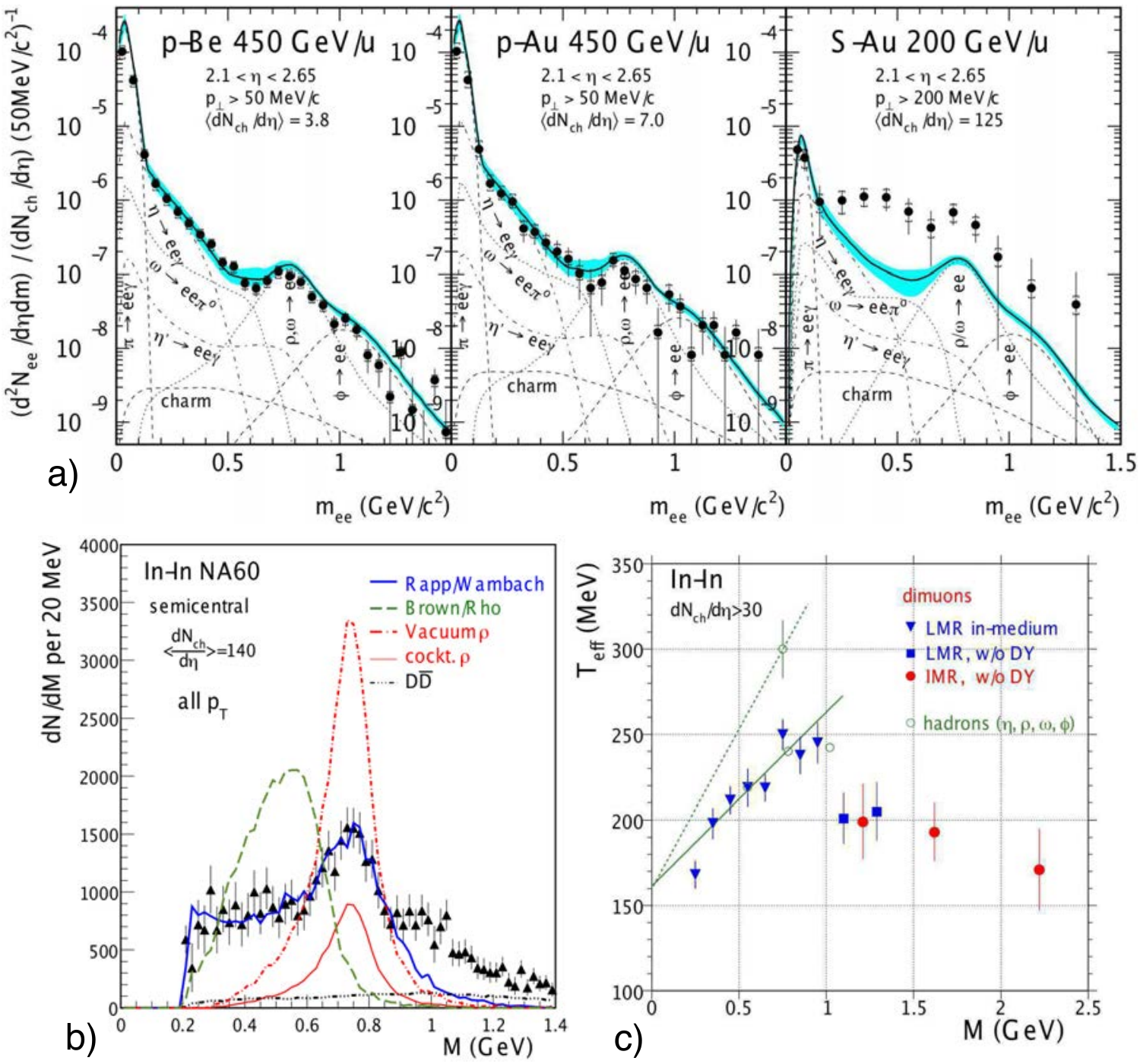}{a) CERES/NA45 dielectron measurements in p+N and N+N collisions reporting a LMR enhancement for the first time but failing to distinguish possible vector meson in-{}medium scenarios [13, 98-{}100]. b,c) NA60 dimuon measurement establishing \ensuremath{\rho} in-{}medium broadening and identifying partonic sources as dominant in IMR [14].}{intro_fig_ceres_na45_na60}{0.9} %

\pagebreak[4]

\wrapifneeded{0.50}{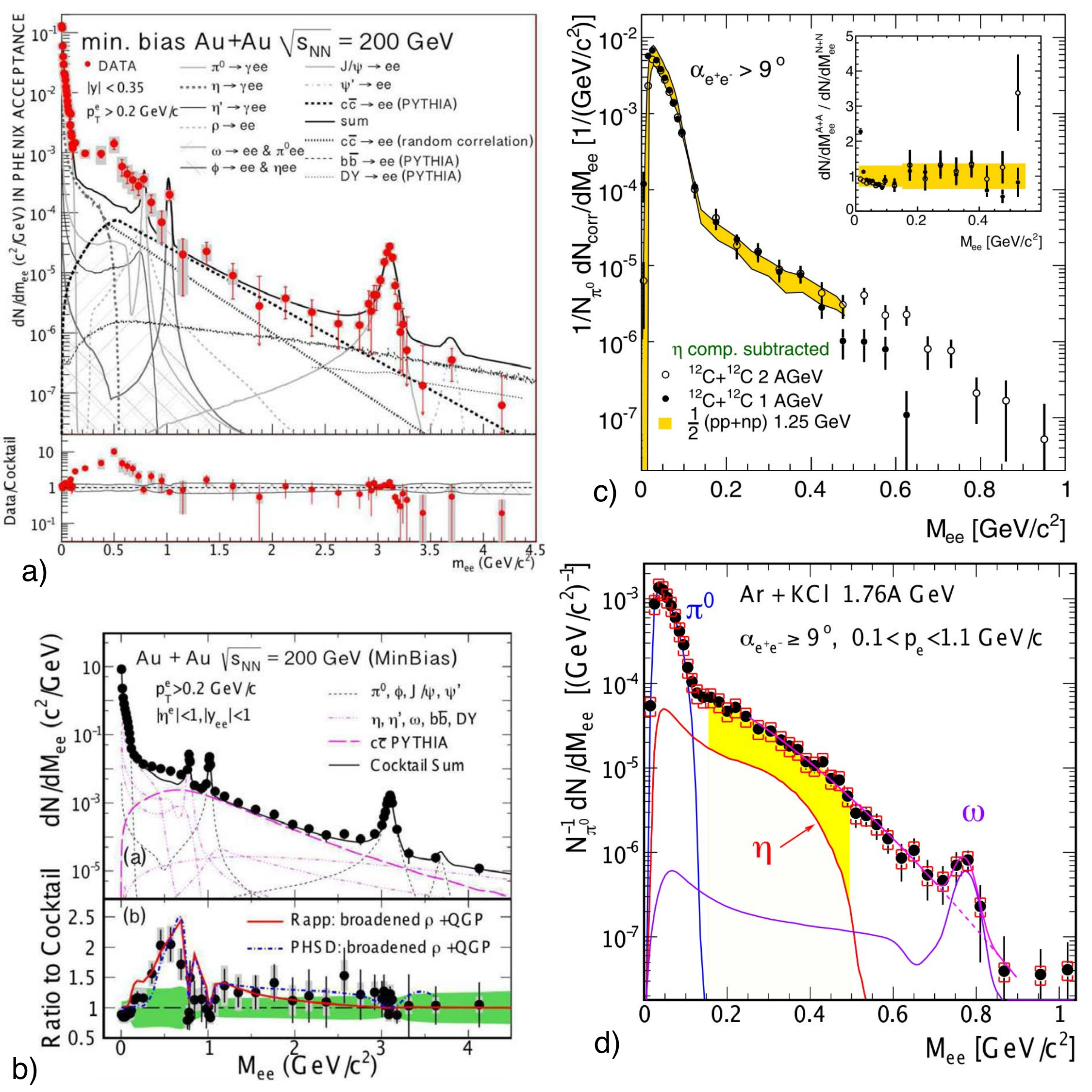}{a) PHENIX dielectron measurement at top RHIC energy exhibiting significant and yet insufficiently explained LMR excess [15]. b) STAR dielectron measurements at top RHIC energy exhibiting reasonable agreement with broadened \ensuremath{\rho} calculations for the LMR excess, also in its p$_{\text{T}}$ and N$_{\text{part}}$ dependence. c) HADES measurement in C+C collisions described by superposition of elementary reactions [102]. d) HADES measurement in Ar+KCl collisions at 1.76 AGeV also showing unexplained enhancement by a factor 2.5-{}3 [103].}{intro_fig_phenix_star_hades}{1} %

The PHENIX detector system has measured invariant mass and
transverse momentum spectra in d+Au collisions and seen very good agreement of
the data with next-{}to-{}leading order perturbative QCD calculations over the
entire accessible phase space [104]. Hence, any enhancement seen in the heavy-{}ion
collisions is not due to cold nuclear matter effects. More importantly, PHENIX
has also published dielectron invariant mass spectra in Au+Au collisions
[15] exhibiting a significant LMR enhancement of more than a factor
of 5 over an expected cocktail of hadronic sources
(Figure~\ref{intro_fig_phenix_star_hades}a).  None of the scenarios for in-{}medium
spectral functions of the vector mesons could satisfactorily explain the
magnitude of the enhancement, even by exhausting the full parameter space
allowed.  Same is true for PHSD calculations in which neither the incorporated
hadronic nor partonic sources were able to account for the enhancement observed
by PHENIX. It is interesting to note, that the latest dielectron measurements
by PHENIX using the novel Hadron-{}Blind Detector (HBD) show a more moderate
enhancement factor of 2-{}3 for the 20-{}40\% centrality bin [105]. It
remains to be seen whether the large enhancement of the original measurement
bears up in the future since the most central bin of the HBD measurement, to
which most of the enhancement has been attributed, has not been shown to date.\newline

As detailed in Section~\ref{intro_sec_thesis}, STAR has in recent years entered the
landscape of dielectron measurements enabled by major upgrades to this large
acceptance mid-{}rapidity (multi-{}purpose) detector.
Figure~\ref{intro_fig_phenix_star_hades}b shows STAR's main results of the very recently
completed analysis of dielectron production in Au+Au collisions at top RHIC
energy [16]. The statistical quality of the measurement allowed
STAR to observe an enhancement factor of about 1.8 in the mass region 0.30-{}0.76
GeV/c$^{\text{2}}$ which is significantly lower than reported by PHENIX. Neither
comparisons of cocktail simulations nor of production yields within PHENIX
acceptance revealed any explanation for the discrepancy in the magnitude of the
LMR excess.  A detailed comparison of the LMR invariant mass shapes shows that
the data disfavors a pure vacuum-{}\ensuremath{\rho} spectral function for the excess but is
in reasonable agreement with broadened \ensuremath{\rho} and QGP contributions calculated
within hadronic multi-{}body as well as PHSD models. This conclusion further
supports the NA60 findings in size and shape. When split in \ensuremath{\rho}-{}, \ensuremath{\omega}-{}
and \ensuremath{\phi}-{}like mass regions, the data over cocktail ratios in the latter two
can be reproduced by the cocktail calculations as opposed to the ratio in the
\ensuremath{\rho}-{}like region which is reasonably well described by the aforementioned
models and weakly dependent on the number of participants (N$_{\text{part}}$, centrality)
and transverse momentum. A power-{}law fit to the N$_{\text{part}}$-{}scaled \ensuremath{\rho}-{}like
dielectron excess results in an exponent of 0.54 \ensuremath{\pm} 0.18 possibly indicating
sensitivity to QCD medium dynamics like the lifetime of the HRG phase (see
[106]).\newline

For completeness, we also need to discuss measurements of dielectron production
in the energy range of normal to moderate temperatures and baryon densities.
One of the early measurements in 1.04 AGeV Ca+Ca collisions by the DiLepton
Spectrometer (DLS) at the BEVALAC [107] compares dielectron
invariant mass distributions to freeze-{}out decays as well as in-{}medium modified
\ensuremath{\omega}-{} and \ensuremath{\rho}-{}mesons. While reasonably described by at SPS energies
[101], neither the dropping mass nor the broadened in-{}medium spectral
function scenario can account for the considerable excess yield in the low mass
region [108, 109] which led to the so-{}called
"DLS-{}puzzle". The second-{}generation High Acceptance Di-{}Electron Spectrometer
(HADES) was subsequently built in an effort to systematically study this excess
with high precision in the SIS18 energy regime. Taking into account the
different acceptances, HADES was able to confirm the DLS measurements in its
invariant mass and transverse momentum dependence for 1 and 2 AGeV C+C
collisions [17, 18]. The HADES findings further
show that such light collision systems can in first order be described as an
isospin-{}averaged (incoherent) superposition of elementary N+N processes
(Figure~\ref{intro_fig_phenix_star_hades}c). However, in the heavier collision system of
Ar+KCl at 1.76 AGeV the \ensuremath{\eta}-{}subtracted invariant mass spectrum exceeds this
N+N reference by a factor of about 2.5 to 3 between 0.15 and 0.5 GeV/c$^{\text{2}}$ (Figure~\ref{intro_fig_phenix_star_hades}d). The excitation function covered by the DLS
and HADES experiments indicates that the excess closer resembles the scaling of
pion rather than \ensuremath{\eta} production. Hence, the interpretation of dielectron
production in the regime of 1-{}2 AGeV bombarding energies requires a better
theoretical understanding of the relevant N$^{\text{*}}$ resonances and their \ensuremath{\rho}-{}meson
coupling.

\subsection{Underlying source of Dilepton Enhancement}
\label{intro_subsec_theo_thermal}\hyperlabel{intro_subsec_theo_thermal}%

The final part of this section is dedicated to the empirical discussion of a
few theoretical, partly data-{}driven observations that collectively support the
interpretation of the experimental results as thermally emitted dileptons
[110]. The purpose is not to provide a complete and resilient
overview but rather briefly describe the interplay of a few theoretical points
which pinpoint the observed dilepton emission to temperatures around the chiral
transition. It provides an instructive exercise since the main conclusion of
this thesis plays an integral part in the argumentation.

\wrapifneeded{0.50}{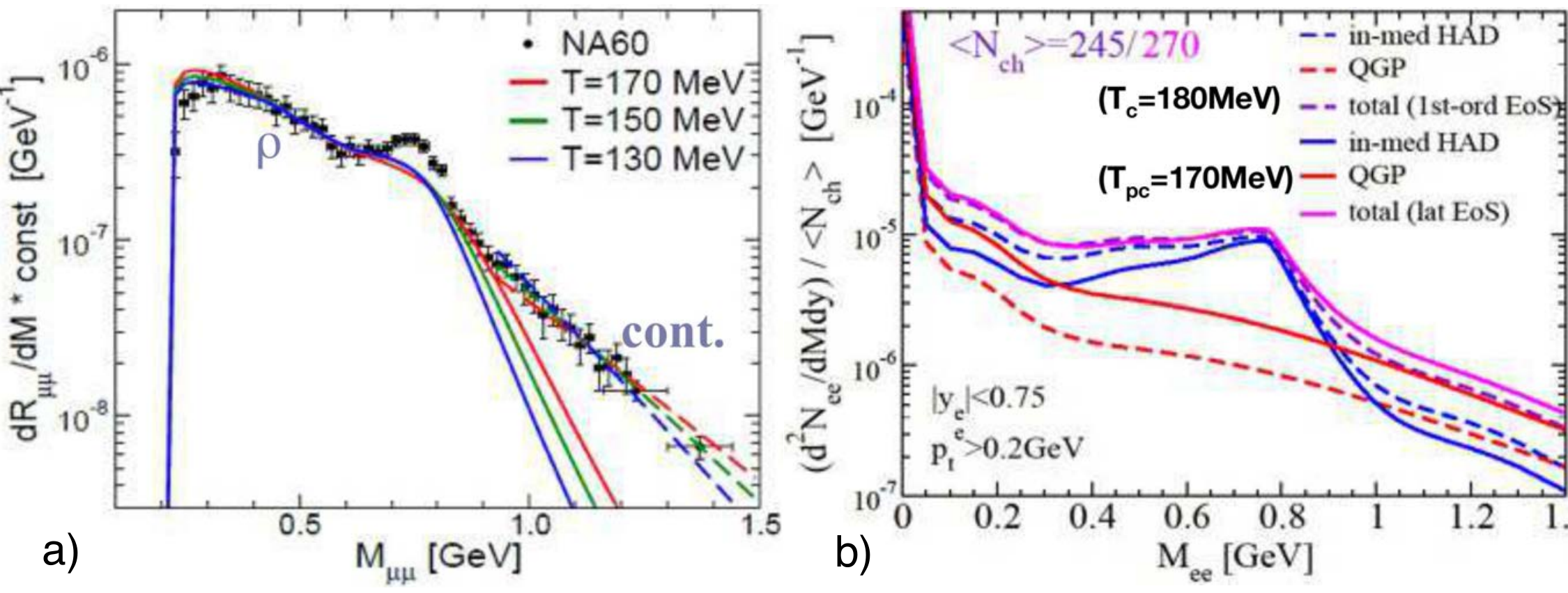}{Theoretical observations supporting the interpretation of the measured LMR excess as emanating from a thermal emission source around the chiral transition temperature T$_{\text{pc}}$$^{\text{\ensuremath{\chi}}}$ [110]. a) Indication of dimuon continuum radiation from T \ensuremath{\sim} 150 MeV. b) First-{}order and Lattice QCD equation-{}of-{}states lead to identical dilepton yields in the 150 MeV temperature range.}{intro_fig_thermal_rapp}{1} %

Due to its unique connection to theory, dilepton invariant mass spectra
directly reflect the thermal emission rate (see Section~\ref{sim_subsec_prorates}). The slopes of the dashed lines in
the continuum region of the NA60 dimuon measurements in
Figure~\ref{intro_fig_thermal_rapp}a reveal an emission temperature of about 150 MeV or
at least between 130 and 170 MeV. This temperature range happens to correspond
to the temperature of the chiral transition T$_{\text{pc}}$$^{\text{\ensuremath{\chi}}}$.
Figure~\ref{intro_fig_thermal_rapp}b studies in-{}medium hadronic and QGP contributions
for an equation of state (EoS) with a first-{}order phase transition at T$_{\text{c}}$ \ensuremath{\sim} 180 MeV with respect to a Lattice EoS with T$_{\text{pc}}$ at 170 MeV. The lower
T$_{\text{pc}}$ is balanced by enhanced QGP and depleted hadronic contributions resulting
in almost identical low-{}mass yields and corroborating a predominant emission
source around T$_{\text{pc}}$. Also, an effective excess temperature of 220 MeV as well
as rather large flow expansion velocities (\ensuremath{\beta}\ensuremath{\sim}0.3) both extracted from
direct photon spectra measured by PHENIX, suggest a "later" emission at T \ensuremath{\sim}
220 MeV / 1.35 \ensuremath{\sim} 160 MeV \ensuremath{\sim} T$_{\text{pc}}$. Theoretically, one observes that the
in-{}medium \ensuremath{\rho}-{}meson rates smoothly "melt" into the according rates from QGP
continuum radiation. Experimentally, the NA60 data has quantitatively confirmed
the broadened \ensuremath{\rho} spectral function scenario as explanation for the low-{}mass
excess yields.
Since, as previously discussed, the hadronic in-{}medium spectral functions
are related to the $q\bar{q}$ condensate, the development of a pion
cloud and excitation of baryonic as well as mesonic resonances as mechanisms
underlying this "\ensuremath{\rho}-{}melting", are possibly precursor to the dynamic
restoration of chiral symmetry. The preceding theoretical and experimental
observations, as a whole, allow the authors of [110] to speculate
about a \emph{three-{}fold degeneracy} of contributions from in-{}medium HRG,
perturbative QCD and Lattice QCD convening at the chiral phase
transition temperature.\newline

It is crucial, however, to study the dilepton LMR enhancements for various
regions of the QCD phase diagram and establish the degree of universality of
in-{}medium vector meson modifications as underlying source. In anticipation of
the results, this thesis reports on the extensive energy-{}dependent measurements
during STAR's BES-{}I which indeed indicate the universal character of this
scenario (see Chapter~\ref{results}). Hence, the study of emission temperature and
in-{}medium scenarios results in the important observation that spectral slopes,
yields and flow of low-{}mass dileptons at SPS and RHIC point toward a
\emph{universal} underlying source with average emission temperatures of about 150
MeV coinciding with T$_{\text{pc}}$$^{\text{\ensuremath{\chi}}}$.

However, there are caveats that need to be kept in mind. It is important to
also identify conditions in a heavy-{}ion environment under which these medium
effects turn off. While peripheral collisions are challenging in this regard
since the dense hadronic phase still persists, high(er) p$_{\text{T}}$ data like the NA60
dimuon measurements or elementary projectiles on cold nuclei like
\ensuremath{\gamma}+A\ensuremath{\rightarrow}eeX by CLAS are promising candidates [111]. Last
but not least, a thorough theoretical framework needs to relate the measured
\ensuremath{\rho} spectral functions to its chiral partner a$_{\text{1}}$ to convincingly address the
long-{}standing question of chiral symmetry restoration. Progress in this
direction has recently been made in [112].

\section{Thesis Objective and Overview}
\label{intro_sec_thesis}\hyperlabel{intro_sec_thesis}%

As later outlined in Section~\ref{sim_sec_model}, the most distinct features of
electromagnetic radiation from the hot hadronic phase are unfortunately hidden
by the according hadronic freeze-{}out decays.  However, model calculations still
predict about 50\% reduction in the \ensuremath{\rho}/\ensuremath{\omega} region and more importantly,
about a factor of two enhancement at 0.5 GeV/c$^{\text{2}}$ when comparing a broadened to
a vacuum-{}like spectral function for the dielectron spectrum from direct decays
of \ensuremath{\rho}, \ensuremath{\omega} and \ensuremath{\phi} over the fireball evolution
[113, 87]. The \ensuremath{\omega}/\ensuremath{\phi} resonances appear to be less
susceptible to the different scenarios.  The preceding introduction has argued
that theory can connect these modifications to chiral symmetry restoration with
sufficient experimental data at its disposal. However, the accuracy required to
test the respective differences from various scenarios in the (\ensuremath{\rho}) spectral
function, is experimentally very challenging.\newline

The completion of the Barrel Time-{}of-{}Flight detector (TOF) in 2010 has allowed
STAR to play an important role in the study of dielectron production
[16, 114, 115] with excellent particle
identification, low material budget, full azimuthal acceptance at mid-{}rapidity,
and a wide momentum coverage [19, 116, 117].  STAR
is thus one of the most versatile mid-{}rapidity detectors currently available to
the High Energy Nuclear Physics community. For instance, TOF efficiently
rejects slow hadrons and provides pure electron identification together with
the energy loss measured in STAR's Time-{}Projection-{}Chamber (TPC), which makes
the two detectors the primary subsystems employed in dielectron analyses at
STAR.  In particular, combined with the first phase of the Beam Energy Scan
Program (BES-{}I, [20]), STAR presents the unprecedented
opportunity to map out a significant portion of the QCD phase diagram within a
homogeneous experimental environment.\newline

The objective of this thesis in the main Part~\ref{physics} is to present STAR's novel
and successful endeavor extending the dielectron measurements from top RHIC
energies down to the SPS energy regime. The results of this thesis therefore
not only close a wide gap in the QCD phase diagram, but also provide the first
comprehensive dataset of dielectron measurements with respect to energy
dependence and experimental environment. In the particular context of the
dielectron analyses presented in this thesis, STAR aims to look for
energy-{}dependent changes in the in-{}medium spectral function modifications and
QGP radiation, possibly identifying the underlying universal nature of the
respective dielectron emission source(s). Only by varying beam energy and
accessing different regions of the phase diagram, we can objectively verify
conclusions reached for energies at which a QGP is or is not expected to form.
This fundamentally distinguishes the efforts in this thesis from dielectron
measurements previously undertaken. The discussion in Section~\ref{intro_sec_dielec} in
particular, has identified such measurements as crucial experimental input to
explore the connection between measured dilepton yields and chiral symmetry
restoration.\newline

The thesis is outlined as follows. Chapter~\ref{data_analysis} starts out by briefly
introducing the concepts of RHIC as well as important aspects of the STAR
detector setup relevant to dielectron analyses. It continues with the key
elements of the experimental raw data analyses ranging from event and track
selection including particle identification to pair reconstruction and
background subtraction. Chapter~\ref{effcorr} presents the procedure employed to derive
and apply corrections due to detector inefficiencies. Chapter~\ref{results} discusses
the obtained results including new and energy-{}dependent invariant mass spectra
as well as p$_{\text{T}}$ spectra in comparison to the simulations and model
calculations of Chapter~\ref{sim}. In the end, Chapter~\ref{summary} summarizes the
main results of the thesis and gives an outlook on the role of dielectron
measurements in the current STAR upgrades, the future second phase of the Beam
Energy Scan program (BES-{}II), and CERN's LHC. Preliminary versions of the
results in Chapter~\ref{results} have been presented at various major conferences in
recent years, and published in the respective proceedings [118].
Papers based on the results of this thesis are in preparation to be published
in the journals \emph{Physical Review Letters} and \emph{Physical Review C}.

Additional parts on software (Part~\ref{software}) and hardware (Chapter~\ref{hardware})
projects developed within the framework of this thesis contain a multitude of
contributions to the STAR \footnote{
a selection will possibly be published as
STAR notes in the future
} and Open-{}Source community as well as the inital
development of a Xe-{}Excimer Lamp designed to enable pad-{}by-{}pad efficiency
corrections for the RICH detector of the HADES experiment.


\chapter{Data Acquisition and Analysis}
\label{data_analysis}\hyperlabel{data_analysis}%

This chapter discusses the complex but integral part of gathering and analyzing
the (raw) data required to perform an energy-{}dependent
measurement of dielectron production at the Relativistic Heavy Ion Collider
(RHIC). Section~\ref{ana_sec_exp} briefly presents relevant aspects of experimental
setup and data acquisition by means of the Solenoidal Tracker at RHIC (STAR).
In Section~\ref{ana_sec_dsets_evttrk}, the datasets used in this thesis are introduced
and suitable selection criteria are discussed to reduce the data to
high-{}quality collision events and particle tracks. The separate reconstruction
and analysis of the event plane
described in Section~\ref{ana_subsec_evtplane} is a prerequisite for the statistical
generation of background pair distributions. The procedure to cleanly identify
electrons and positrons using STAR's detector subsystems Time-{}Of-{}Flight (TOF)
and Time Projection Chamber (TPC) is discussed in Section~\ref{ana_sec_pid} along with
the resulting sample purities.  Dielectrons (e$^{\text{+}}$/e$^{\text{-{}}}$ pairs) are reconstructed
and background contributions substracted statistically in Section~\ref{ana_sec_pairrec} by employing same and mixed event techniques in p$_{\text{T}}$-{}integrated and
-{}differential manner.  Note that corrections to account for efficiency losses
in the detector and to be applied on the raw spectra are derived in
Chapter~\ref{effcorr}.

\section{Experiment: RHIC and STAR}
\label{ana_sec_exp}\hyperlabel{ana_sec_exp}%

\wrapifneeded{0.50}{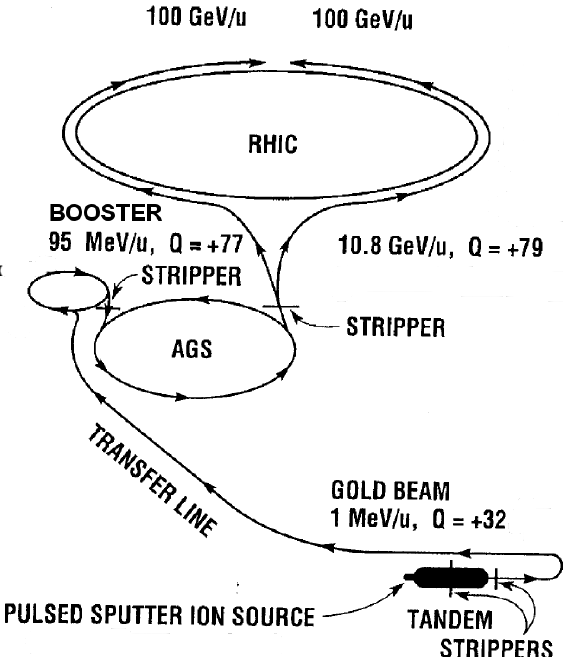}{RHIC acceleration scenario for gold beams [119].}{exp_fig_rhic}{0.36} %

It is beyond the scope of this thesis to give a complete introduction to
neither the functionality of heavy-{}ion accelerators nor the technology used to
build the detectors around the collision zone. Since excellent reviews about
RHIC and STAR already exist [120], this section gives overviews merely
outlining the general concepts of the two and concentrates on some additional
details relevant to dielectron analyses. Where appropriate the reader is
referred to references containing in-{}depth information.

The scenario to feed Au ions into RHIC using an injector chain of accelerator
rings is sketched out in Figure~\ref{exp_fig_rhic} [119].  The Tandem Van
de Graaff as well as the Booster and the Alternating Gradient (AGS)
synchrotrons are pre-{}existing hadron accelerators at the Brookhaven National
Laboratory (BNL) [121] that provide three different stages of increasing
beam energy after each of which the Au ions are partially stripped of their
electrons to adjust their charge to the according next-{}stage ring diameters and
intended energies. The Tandem [122] uses its characteristic
two-{}stage acceleration mode including mid-{}way electron stripping to boost Au$^{\text{-{}}}$ ions from the sputter ion source [123] to an energy of 1 MeV/u. A
charge state of +32 is subsequently selected by bending magnets for the
delivery via the Heavy Ion Transfer Line [124] to the Booster
Synchrotron [125, 126] in which the Au ions are accelerated
to 95 MeV/u. Au$^{\text{+77}}$ ions exit the Booster for the final step of acceleration
in the AGS [127] to an energy of 10.8 GeV/u before bunch-{}wise
injection into RHIC as Au$^{\text{+79}}$ via the AGS-{}to-{}RHIC transfer line
[128]. RHIC consists of a pair of superconducting rings with 3.8 km
circumference capable of accelerating and storing two counter-{}circulating gold
beams with energies up to 100 GeV/u each [129].  The rings
intersect at four points along the circumference allowing for a complementary
set of detectors [130] to conduct experiments using relativistic
heavy ion collisions. Each experiment operates a pair of Zero-{}Degree
Calorimeters (ZDCs) [131, 132] as a common detector subsystem
on either side of and close to the respective interaction zones. The ZDCs
detect neutron multiplicities from the collisions and enable the accelerator
staff to monitor luminosity and locate collision vertices ("steering"). For
more details on RHIC, the reader may want to use [133] as an entry
point.

The data for the energy-{}dependent analysis of dielectron production presented
here has been obtained by means of the STAR detector
[134, 135]. The STAR detector serves the study of a very
broad physics program ranging from high-{}density QCD, over the proton's spin
structure to ultra-{}peripheral collisions via electromagnetic interactions of
passing ions. Such a program requires a suite of multi-{}purpose and
complementary detector subsystems. Figure~\ref{exp_fig_star_setup} shows a graphical
rendering of the STAR detector with its subsystems Endcap [136] \&
Barrel [137] Electromagnetic Calorimeters (EEMC \& BEMC), Beam-{}Beam
Counter [138] (BBC), upgraded pseudo Vertex Position Detector
[139, 140] (upVPD), and Magnet [141]. Also depicted
are TPC and TOF, the combination of which is particularly important for the
study of dielectrons and discussed to some detail in Section~\ref{ana_subsec_tpctof} followed by STAR's Data Acquisition (DAQ) and Trigger systems in
Section~\ref{ana_subsec_trigger_daq}.

\wrapifneeded{0.50}{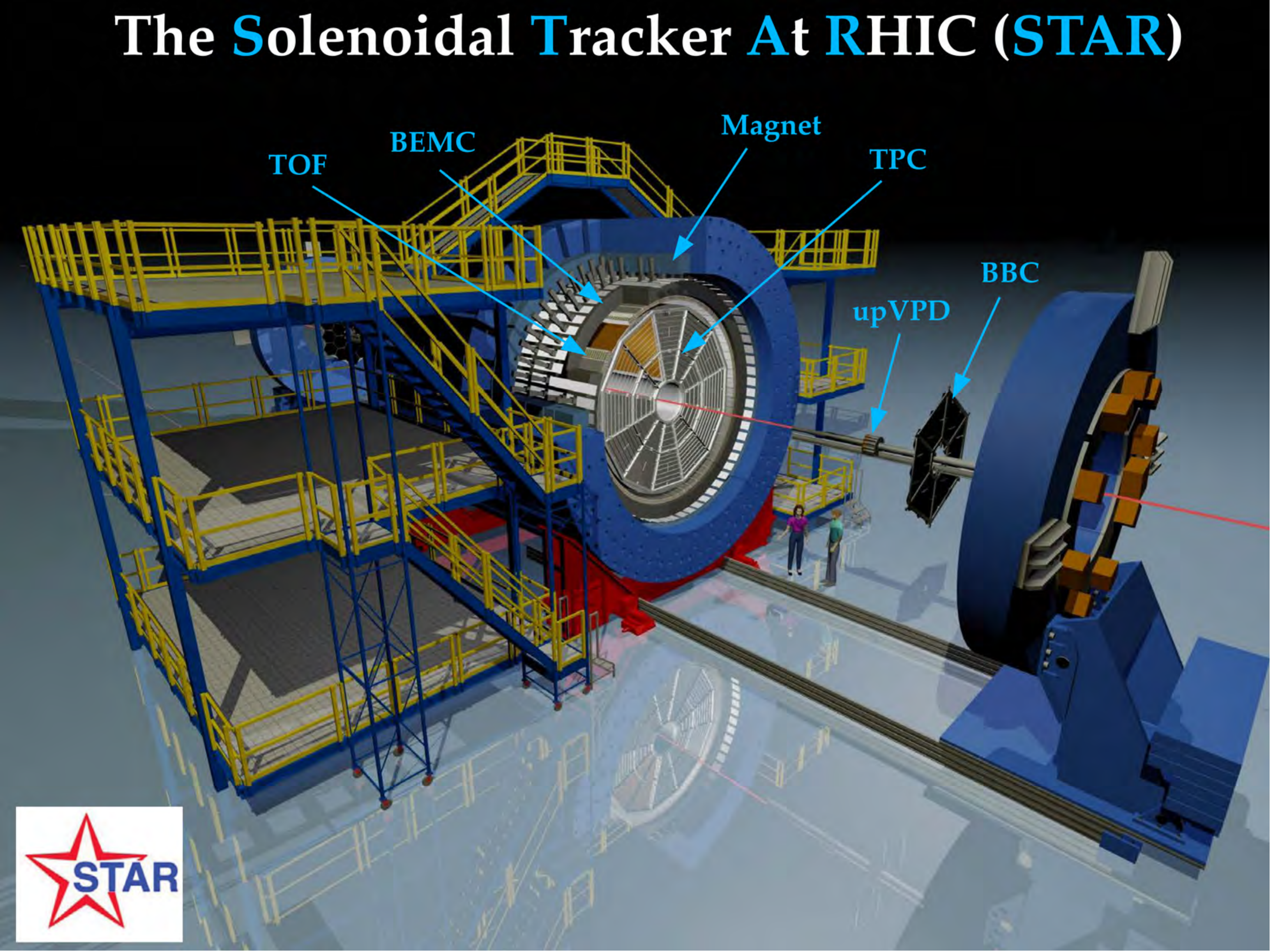}{Graphical rendering of the STAR detector system with subsystems reflecting the status in the years 2010/11 [142]. Details and references see text.}{exp_fig_star_setup}{0.78} %

\subsection{TPC and TOF}
\label{ana_subsec_tpctof}\hyperlabel{ana_subsec_tpctof}%

The cylindrical Time Projection Chamber (TPC) [143] constitutes
the very heart of the STAR detector system measuring 4 m in diameter as well as
length, and stretching from 50 to 200 cm in radial distance from the beam axis.
Stable and unstable particles emanate from the collision zone located
approximately in the center of the TPC and traverse through the covered volume
filled with P10 gas (10\% methane, 90\% argon). Along their path, the particles
ionize the P10 gas leaving behind a track of released secondary electrons. A
thin conductive membrane at the center of the TPC, the concentric field-{}cage
cylinders, and the end caps define a uniform electric field of \ensuremath{\sim}135 V/cm
which induces a drift of the electrons along its field lines toward the readout
caps at both ends of the chamber. P10 gas is a well-{}established choice for this
purpose since it exhibits fast drift velocities at comparably low electric
fields. The charge of the drifting electrons, registered when they reach the
TPC's readout plane, allows for the measurement of a track segment's x/y and z
coordinates through its projection into two dimensions and the elapsed drift
time, respectively.

The TPC is furthermore installed inside a large solenoidal, room-{}temperature
magnet providing a uniform field of up to 0.5 T along the z-{}direction (beam
axis). In addition to the spatial reconstruction, the charged particle's
momentum can hence be derived from the track's curvature in the magnetic field
enabling high-{}resolution 3D-{}imaging of the heavy-{}ion collision.  Electric field
uniformity is critical in this context to achieve a track reconstruction
precision of less than 1 mm with electron drift paths of up to 2.1 m. A
relative momentum resolution of ${\delta}p/p$ = 0.02 is achieved
which improves with the number of hits induced by a track's secondary electrons
in the TPC's readout end caps and with decreasing particle momentum.
Particularly noteworthy is the TPC's resulting large acceptance of close to
4\ensuremath{\pi} in azimuth and $\vert\eta\vert$ <{} 1 in pseudo-{}rapidity for
charged particle tracking and the full range of multiplicities.

The reconstructed momentum does not suffice to determine the particle species.
However, in every ionizing collision with the atoms along its path through the
gas, the particle releases energy from $\mathcal{O}$(10 eV)
frequently up to $\mathcal{O}$(100 eV) rarely. The momentum
dependence of this energy loss dE/dx per length dx of the ionization cluster is
specific to the particle species (c.f. Bethe formula [144]) and
its measurement hence allows for a very capable particle
identification below the relativistic rise (see Section~\ref{ana_sec_pid}). The
deposition of collisional energy in the ionization clusters along the track is
a stochastic process which impedes the accurate measurement of an average dE/dx
since the according track segments are too short to neglect the resulting
fluctuations. Instead, dE/dx distributions can satisfactorily be parameterized
by varying the most probable energy loss with the length of the ionization
cluster [145, 146]. With an energy loss resolution of 8\%,
pions and kaons can be identified cleanly up to a transverse momentum of 0.7
GeV/c and protons up to 1 GeV/c. The challenge in achieving the target dE/dx
resolution lies in TPC calibration and understanding of signal as well as gain
variations during TPC readout.

The TPC readout system [147] is based on Multi-{}Wire Proportional
Chambers (MWPC) with readout anode planes partitioned into \ensuremath{\sim}140k pads in
total. The high electric fields at the MWPC's anode wires cause an avalanche of
liberated electrons initiated by an ionizing collision of the drifting
electrons which amplifies the electronic signal by about three orders of
magnitude. The positive ions left behind by the avalanche electrons induce a
temporary image charge on the pad plane distributed over multiple neighboring
pads. The resulting charge distribution is accurate enough to determine the
original track segment's x/y-{}position within 20\% of the pad width. The readout
systems on either side of the TPC are composed of 12 sectors using different
pad geometries for inner and outer subsectors due to the varying track density.
The inner subsectors use small pad sizes arranged in widely spaced rows as
opposed to the outer subsectors with larger but densely packed pads. This setup
allows for good two-{}track separation and optimized dE/dx resolution using a
moderate number of pads at small and large radii, respectively. A future
upgrade of the inner TPC sectors [23] will cover the small radii with
densely packed pads, too. The electronic readout of the TPC pad planes consists
of small Front-{}End Electronics (FEE) cards responsible for amplification,
shaping and analog-{}to-{}digital conversion of the avalanche signal. Each readout
board (RDO) controls 36 FEE cards, provides power and sends the processed
signals to the DAQ system. See Section~\ref{stbadrdos} for details on so-{}called "bad"
RDOs missing during the time when the data sets used in this thesis were taken.

Finding and reconstructing particle tracks from the ionization clusters
recorded on adjacent pads of the TPC readout plane (so-{}called \emph{tracking} via
pattern recognition), is an integral ingredient of raw data processing to
enable event classification and successful physics analyses. The fast tracking
algorithm [148] employed in STAR utilizes the fact that the
solenoidal magnetic field forces the particles on helical trajectories. In the
plane perpendicular to the field (e.g. TPC readout plane), the helical paths
are projected into circles. These circular paths transform into straight lines
when applying a conformal mapping from cartesian coordinates to the
(\emph{s},\emph{z})-{}plane with \emph{s} the trajectory length. This procedure simplifies the
design of a fast algorithm by linking closest space points into track segments
and extending them with the unused space point closest to a straight line fit.
The algorithm proceeds inward due to the smaller track densities (less
ambiguity) on outer layers of the TPC and accepts or rejects track candidates
based on the \ensuremath{\chi}$^{\text{2}}$-{}quality of the fit. The initialization of this
\emph{nose-{}following} method requires a space point on the track to be known which
in the general, unbiased case of \emph{global} tracks, simply is the first point
associated with the track. For so-{}called \emph{primary} tracks that match well with
the interaction point, however, the primary vertex is included in the fit to
effectively constrain the track's origin to the event vertex. After a full
track is found it can be parameterized anew using a helix model in
three-{}dimensional space.  The reconstruction of secondary and tertiary vertices
for K$_{\text{0}}$, \ensuremath{\Lambda}, \ensuremath{\Xi} and \ensuremath{\Omega} decays, for instance, can then be achieved
by calculating the Distance-{}of-{}Closest-{}Approach (DCA) of two global helices via
a numerical minimization algorithm (c.f. Section~\ref{effcorr_sec_samples}). A difficulty
requiring special attention in this context, is the reconstruction of the
particle's momentum vector. For its physical determination the according track
needs to be correctly assigned to the particle's true vertex of origin.
However, in addition to the main interaction and secondary vertices, the latter
can also be a spurious interaction vertex caused by event pile-{}up or merely a
scattering center. The assignment of a track to a particular vertex is an
integral part of STAR's event reconstruction chain (see
Section~\ref{ana_sec_dsets_evttrk}) [149].

The level-{}3 trigger explained in Section~\ref{ana_subsec_trigger_daq} performs the above
tracking procedure in each of the TPC sectors separately and in real-{}time.
Tracks that cross sector boundaries will hence be split. This phenomenon also
increasingly occurs the further away the two incoming nuclei collide from the
TPC's central membrane. A track crossing the membrane will have two parts of
its ionization electrons drifting toward opposite sides of the TPC readout
system (end caps). To remove such split tracks in the data analysis, the
selection criteria in Table~\ref{ana_tab_dsets_evttrk} (Section~\ref{ana_sec_dsets_evttrk})
require more than half the possible number of ionization clusters (i.e.
electronic \emph{hits} in the pad plane) to be used in the reconstruction of a
track.\newline

The particle identification (PID) abilities of the TPC are limited to hadrons
with momenta up to the relativistic rise above which the particle separation
power ceases due the merging energy loss bands. Extracting the maximum physics
information contained in hadron spectra, for instance, hence makes it
imperative to extend the measurable hadron momentum range up to 2-{}3 GeV/c. This
has been achieved in STAR by the installation of an additional detector
subsystem in 2010 that allows for the measurement of a particle's velocity
\ensuremath{\beta} and its characteristic momentum dependence. The barrel Time-{}Of-{}Flight
(TOF) detector [150] is based on the Multi-{}gap
Resistive Plate Chamber (MRPC) technology and covers a pseudo-{}rapidity range of
$\vert\eta\vert$ <{} 1. Within the acceptance of the TOF detector,
the \ensuremath{\beta} selection on average doubles the fraction of kaons and protons
identified successfully with a separation up to momenta of \ensuremath{\sim}1.7 and
\ensuremath{\sim}3.0 GeV/c, respectively.

The basic quantity underlying the \ensuremath{\beta} measurement is the time interval
\ensuremath{\Delta}t elapsed between the initial nuclear collision and registration of a
particle in a TOF tray at a certain (\emph{r},\ensuremath{\eta},\ensuremath{\phi}) spatial position. Each
signal in the TOF's trays can be associated with a track reconstructed in the
TPC by extrapolation (\emph{TOF matching}) which additionally provides the track's
momentum \emph{p} and path length \emph{s}. The interval \ensuremath{\Delta}t hence suffices to
identify the particle species through its mass \emph{m} via
\[\beta=\left.s\middle/c{\Delta}t\right.\hspace{5mm}\mbox{and}\hspace{5mm}m^2=p^2\left(1\big/\beta^2-1\right).\]
Particle identification via \ensuremath{\beta} consequently requires the measurement of a
global \emph{start time} and many particle-{}specific \emph{stop times} in each event. The
separate detector subsystem upVPD [151, 140] mounted close to
the beam pipe at \emph{z} = \ensuremath{\pm}5 m is used to obtain the start time. It is,
however, unfeasible to distribute the start time from the upVPD to each TOF
tray taking the stop time and to install bulky electronics for the digitization
of each TOF tray logic signal with respect to the start time.  Instead, the
upVPD and TOF tray signals can be digitized in the same electronics versus a
common clock synchronized within 10-{}20 ps which does not change the relative
timing information required to determine \ensuremath{\beta}. All successful TOF detector
systems share this approach.\newline

The combination of information from TOF and TPC dE/dx does not only improve
STAR's hadron PID capabilities but also allows it to cleanly identify electrons
over a wide momentum range and with large acceptance coverage for the first
time (see Section~\ref{ana_sec_pid}). Physics-{}wise, it enables STAR to effectively
reconstruct and study the properties of resonances that are sensitive to the
dense medium in which they are created. Section~\ref{intro_sec_dielec} has extensively
discussed dielectrons as promising probes for these resonances which is why the
installation of TOF has been crucial in opening up an entirely new realm of
physics analyses to STAR (c.f. Section~\ref{intro_sec_thesis}).

\subsection{Trigger and DAQ}
\label{ana_subsec_trigger_daq}\hyperlabel{ana_subsec_trigger_daq}%

\wrapifneeded{0.50}{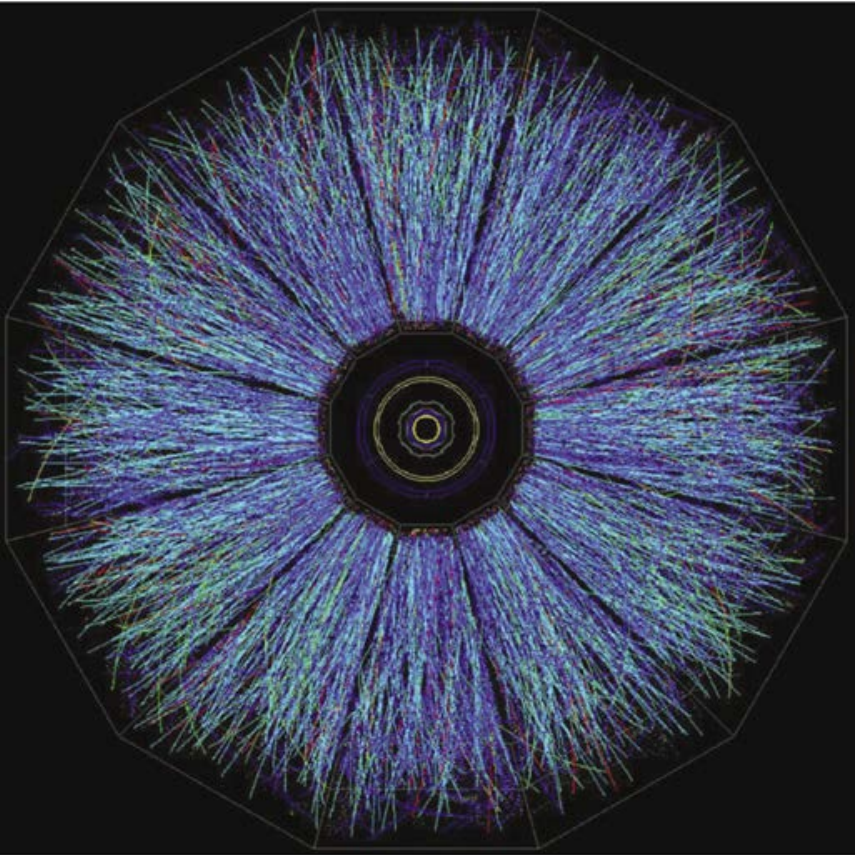}{TPC end view of a single Au+Au event with particle tracks reconstructed in real-{}time ("online") during the second stage of the STAR trigger system (Level-{}3) [135].}{exp_fig_star_event}{0.44} %

In the year 2000, RHIC collided Au ions with a bunch crossing rate of about 10
MHz which natively separates STAR's detector system into two classes. The
\emph{fast} subdetectors, on the one hand, can keep up with the beam crossing rate
and measure global event properties like the charged particle multiplicity at
mid-{}rapidity (Central Trigger Barrel, CTB), the neutron multiplicity at
forward/backward-{}rapidities (ZDCs), and energy localization for
electrons/photons (BEMC). The \emph{slow} subdetectors (primarily the TPC), on the
other hand, require additional time to conduct a track-{}based analysis for a
more sophisticated event selection resulting in a data rate of about 100 Hz.
STAR hence implements a multi-{}level trigger system [152] using the
fast detectors to reduce the data rate by about 5 orders of magnitude.

Its first stage consists of three trigger levels called \emph{Level-{}0}, \emph{-{}1}, and
\emph{-{}2} based on the signal from the fast detectors. These levels are fully
pipelined since the fast detectors do not need to process multiple events at
the same time. Within about 1.5 \ensuremath{\mu}s after each bunch crossing, Level-{}0
processes the measured global multiplicities and energies and signals the slow
detectors to initiate their cycle of amplification, digitization and
acquisition (ADA). During the Level-{}1 trigger phase of about 100 \ensuremath{\mu}s,
characteristics in the particle distributions are used to identify background
from beam-{}gas collisions and abort the ADA cycle during the 40 ms required for
the readout of the TPC. Signal digitization on its FEE takes about 10 ms and is
the reason for the above limit of 100 Hz on the TPC's readout rate. About 5 ms
elapse after the collision until the raw data in a fine pixel array is analyzed
on the Level-{}2 trigger CPU farm to decide whether to abort the event during the
TPC digitization phase and before transmission to the DAQ. The pixel array is
well-{}suited to isolate jets or to refine particle topologies which constrain
the event selection to specific interaction mechanisms. A minimal constraint on
the latter is achieved by selecting events with at least one neutron measured
coincidentally in each of the ZDCs. This corresponds to 95\% of the geometrical
cross-{}section and defines the \emph{minimum bias trigger} (MinBias).

The Level-{}2 trigger phase only provides a prototype event to be fed into the
second stage of STAR's trigger system. This so-{}called \emph{Level-{}3} trigger
[153] is responsible to reach a final decision whether or not to
store the full event information. For this purpose, it analyzes all available
data within 200 ms after the collision via a complete online event
reconstruction in a dedicated CPU farm. Single particle properties like
momentum and energy loss can subsequently be used to select J/\ensuremath{\psi} and
anti-{}nuclei candidates as well as back-{}to-{}back tracks in ultra-{}peripheral
collisions, for instance. An online display for the visual inspection of
individual events in real time is included in the Level-{}3 trigger
[154]. Figure~\ref{exp_fig_star_event} shows a screenshot of the online
display with the detected tracks emerging from a representative Au+Au collision
and demonstrates that the TPC is capable of reconstructing >{}1000 tracks per
pseudo-{}rapidity unit.

The STAR Data Acquisition System (DAQ) [155] receives data from
the many detector subsystems at input rates up to 100 Hz (c.f. Level-{}3 data
rate), reduces it to 20-{}50 MB/s via zero suppression (\ensuremath{\sim}2 events/s for
central collisions), and sends built events via Gigabit Ethernet to the RHIC
Computing Facility (RCF) [156] where they are stored to tape using
HPSS [157]. The duration of the Level-{}3 trigger phase limits the
processing time available for the DAQ and hence requires a fully integrated
farm of \ensuremath{\sim}50 CPUs [158] to parallelize the tracking. The DAQ also
needs to handle many events at the same time and in different stages of
completion since the decision from the Level-{}3 trigger to build an event
arrives long after the event has been received. The data stored on tape in a
unified format provides the basis for STAR's offline event reconstruction
chain. The latter is introduced in Section~\ref{ana_sec_dsets_evttrk} and the resulting
compact data files analyzed through-{}out the remainder of this chapter.

\section{Datasets and Event/Track Selection}
\label{ana_sec_dsets_evttrk}\hyperlabel{ana_sec_dsets_evttrk}%

The zero-{}suppression performed in the STAR DAQ system [155] only
does a minimum to reduce the size of the recorded raw data by discarding
signals that fall within the expected electronic noise. The task of further
data reduction becomes the duty and responsibility of the STAR event
reconstruction chain [149, 159, 160]. This broad first
stage of STAR data analysis combines raw digitized detector data into detector
hits (e.g. TPC cluster finder), integrates all detector components during
tracking to obtain physical particle properties (e.g. calibration and TOF
matching), and performs various post-{}tracking procedures such as vertex and
possibly secondary/tertiary vertex (V$_{\text{0}}$) finding. Each of these specific data
reduction tasks is implemented in a dedicated software module as part of the
so-{}called \emph{Big Full Chain} (BFC, see \texttt{Sta\penalty5000 r\penalty5000 C\penalty5000 l\penalty5000 a\penalty5000 s\penalty5000 s\penalty5000 L\penalty5000 i\penalty5000 b\penalty5000 r\penalty5000 ary} documentation in
[161]). As a result, the BFC produces the more compact \emph{Data
Summary Tapes} (DSTs) containing the "real data" in form of a list of events
with associated particles and their physical properties (see \texttt{StE\penalty5000 v\penalty5000 ent} and
\texttt{StT\penalty5000 r\penalty5000 ack} documentation in [161]). While being taken during
the running experiment, the raw data is run through parts of the reconstruction
chain to control and monitor its quality "online". In addition to the DST
production phase, the original raw detector data is needed offline only for
detector calibration purposes. Once calibration is completed the final set of
DST files is produced along with so-{}called \emph{micro-{}DST} files (MuDSTs) and both
committed to long-{}term archival storage at RCF. MuDSTs [162]
further reduce stored event sizes from 2-{}3 MB to 200-{}300 KB and obey a
standardized file format organized in ROOT [163] branches suitable for
physics analyses across all STAR working groups. Note that global and primary
tracks matching the event vertex are stored twice in the dual list of tracks in
the MuDST files. These derived data sets are redistributed to STAR member
institutions across the globe to be readily available for analysis by
collaborators.

The second stage of the STAR data analysis concentrates on the study of the
reconstructed collisions with respect to a specific physics phenomenon. It
builds on the MuDST files which already provide reconstructed tracks from the
best primary vertex matched to the according TOF hits. Using all global tracks
and established basic event-{} and track-{}wise selection criteria, so-{}called
\emph{PicoDSTs} are produced to further reduce the data needed for a common STAR
physics analysis [164]. Table~\ref{ana_tab_dsets_evttrk} summarizes the main
properties of the four datasets relevant to this thesis along with the event
and track selection criteria used in the production of derived ROOT trees
containing electron and positron candidates (also see Section~\ref{ana_sec_pid}). The
data taken in the years 2010/11 hold excellent premises for a successful study
of dielectron production. STAR's unique and versatile capabilities come
combined during these runs with particularly low material budget (due to the
retired Silicon Vertex Tracker) and high statistics both of which
are favorable for the difficult extraction of physics via the rare
electromagnetic probes. Of the energies recorded during this first phase of
the RHIC Beam Energy Scan (BES-{}I), $\sqrt{s_{NN}}$ = 19.6, 27, 39
and 62.4 GeV have been selected as very promising for the statistics-{}hungry
analysis of dielectron production.

\wrapifneeded{0.50}{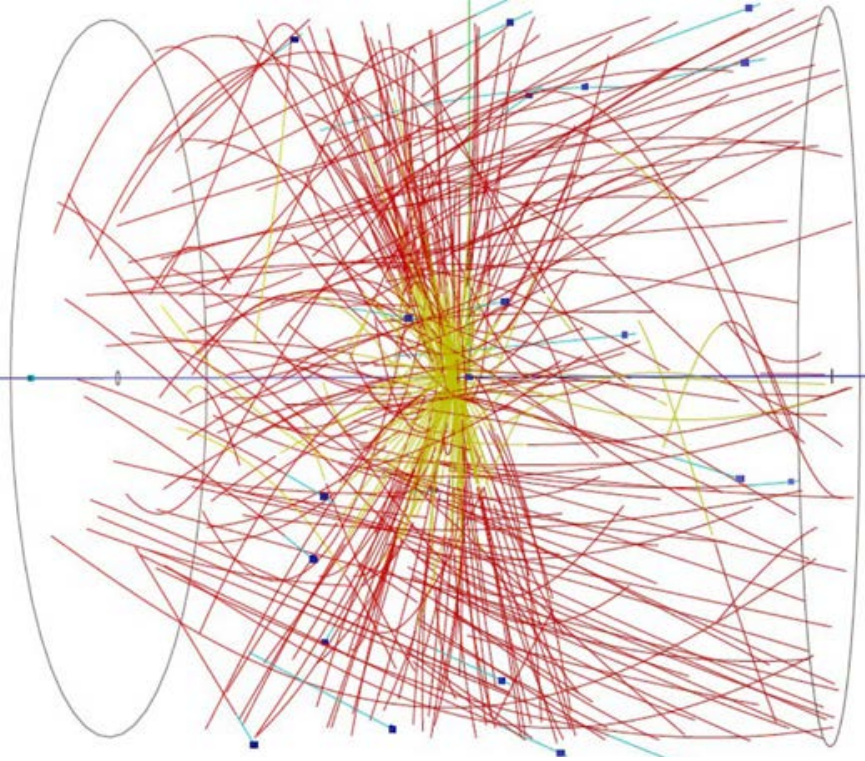}{Au+Au event with good primary vertex but triggered by a beam-{}pipe event just outside the TPC [165].}{dsets_evttrk_fig_beampipe_event}{0.41} %

Most of the criteria used in Table~\ref{ana_tab_dsets_evttrk} for data reduction are
considered established in STAR and ensure a minimum quality that defines a
"good" event or track, for instance. Their notation and meaning are briefly
explained in the according footnotes. Two less common choices for the criteria
should be noted, though: \emph{i}) The VPD cut used at 200 GeV [16]
is not employed at BES energies due to the VPD's significantly reduced
efficiency. \emph{ii}) An event like in Figure~\ref{dsets_evttrk_fig_beampipe_event} triggered by a beam-{}pipe collision with a simultaneous Au+Au event in the
center of the TPC, is not rejected by the selection of the 0-{}80\% most central
events. Unlike other beam-{}pipe events it also exhibits a (reference)
multiplicity appropriate for a Au+Au collision but most of the primary tracks
are not matched with TOF due to the wrong timing information (c.f.
Section~\ref{ana_subsec_tpctof}). Hence, an additional requirement of a minimum number of
TOF-{}matched primary tracks is used to remove such events [165].
Table~\ref{ana_tab_dsets_evttrk} also contains a detailed list of number of events per
analysis stage as well as the final number of events with which to normalize
the efficiency-{}corrected dielectron spectra (Chapter~\ref{results}). Further details in
Section~\ref{ana_subsec_refmult_runqa} accompany the analysis stages involving the
concept of reference multiplicity (RefMult) and refined run-{}by-{}run event
quality assurance (RunQA). Section~\ref{ana_subsec_evtplane} explains the procedure to
obtain uniform distributions of the event plane angle used for event
classification in Section~\ref{ana_sec_pairrec}.

\pagebreak[4] 

\begin{center}
\begingroup%
\setlength{\newtblsparewidth}{1\linewidth-2\tabcolsep-2\tabcolsep-2\tabcolsep-2\tabcolsep-2\tabcolsep-2\tabcolsep}%
\setlength{\newtblstarfactor}{\newtblsparewidth / \real{425}}%

\begin{longtable}{lllll}\caption[{Properties of the datasets analyzed in this thesis (PicoDST level) and the criteria used for event/track selection. Also included is the reduction in number of events with each analysis cut (in order of application). The footnotes give brief explanations of the selection criteria/stages.}]{Properties of the datasets analyzed in this thesis (PicoDST level) and the criteria used for event/track selection. Also included is the reduction in number of events with each analysis cut (in order of application). The footnotes give brief explanations of the selection criteria/stages.\label{ana_tab_dsets_evttrk}\hyperlabel{ana_tab_dsets_evttrk}%
}\tabularnewline
\hline
\multicolumn{1}{m{101\newtblstarfactor}|}{\raggedright\bfseries%
 \ensuremath{\surd}s$_{\text{NN}}$ (GeV) %
}&\multicolumn{1}{m{81\newtblstarfactor}|}{\centering\bfseries%
 19.6 %
}&\multicolumn{1}{m{81\newtblstarfactor}|}{\centering\bfseries%
 27 %
}&\multicolumn{1}{m{81\newtblstarfactor}|}{\centering\bfseries%
 39 %
}&\multicolumn{1}{m{81\newtblstarfactor+\arrayrulewidth}}{\centering\bfseries%
 62.4%
}\tabularnewline
\endfirsthead
\caption[]{(continued)}\tabularnewline
\hline
\multicolumn{1}{m{101\newtblstarfactor}|}{\raggedright\bfseries%
 \ensuremath{\surd}s$_{\text{NN}}$ (GeV) %
}&\multicolumn{1}{m{81\newtblstarfactor}|}{\centering\bfseries%
 19.6 %
}&\multicolumn{1}{m{81\newtblstarfactor}|}{\centering\bfseries%
 27 %
}&\multicolumn{1}{m{81\newtblstarfactor}|}{\centering\bfseries%
 39 %
}&\multicolumn{1}{m{81\newtblstarfactor+\arrayrulewidth}}{\centering\bfseries%
 62.4%
}\tabularnewline
\endhead
\multicolumn{1}{m{101\newtblstarfactor}|}{\raggedright%
Year
}&\multicolumn{1}{m{81\newtblstarfactor}|}{\centering%
2011
}&\multicolumn{1}{m{81\newtblstarfactor}|}{\centering%
2011
}&\multicolumn{1}{m{81\newtblstarfactor}|}{\centering%
2010
}&\multicolumn{1}{m{81\newtblstarfactor+\arrayrulewidth}}{\centering%
2010
}\tabularnewline
\multicolumn{1}{m{101\newtblstarfactor}|}{\raggedright%
Days
}&\multicolumn{1}{m{81\newtblstarfactor}|}{\centering%
112 -{} 122
}&\multicolumn{1}{m{81\newtblstarfactor}|}{\centering%
172 -{} 179
}&\multicolumn{1}{m{81\newtblstarfactor}|}{\centering%
99 -{} 112
}&\multicolumn{1}{m{81\newtblstarfactor+\arrayrulewidth}}{\centering%
80 -{} 82, 84 -{} 98
}\tabularnewline
\multicolumn{1}{m{101\newtblstarfactor}|}{\raggedright%
\# Runs
}&\multicolumn{1}{m{81\newtblstarfactor}|}{\centering%
483
}&\multicolumn{1}{m{81\newtblstarfactor}|}{\centering%
382
}&\multicolumn{1}{m{81\newtblstarfactor}|}{\centering%
615
}&\multicolumn{1}{m{81\newtblstarfactor+\arrayrulewidth}}{\centering%
735
}\tabularnewline
\multicolumn{1}{m{101\newtblstarfactor}|}{\raggedright%
\# Files
}&\multicolumn{1}{m{81\newtblstarfactor}|}{\centering%
9757
}&\multicolumn{1}{m{81\newtblstarfactor}|}{\centering%
13896
}&\multicolumn{1}{m{81\newtblstarfactor}|}{\centering%
24922
}&\multicolumn{1}{m{81\newtblstarfactor+\arrayrulewidth}}{\centering%
25569
}\tabularnewline
\multicolumn{1}{m{101\newtblstarfactor}|}{\raggedright%
\# Events (M)
}&\multicolumn{1}{m{81\newtblstarfactor}|}{\centering%
48.7024
}&\multicolumn{1}{m{81\newtblstarfactor}|}{\centering%
95.4347
}&\multicolumn{1}{m{81\newtblstarfactor}|}{\centering%
170.859
}&\multicolumn{1}{m{81\newtblstarfactor+\arrayrulewidth}}{\centering%
120.528
}\tabularnewline
\multicolumn{1}{m{101\newtblstarfactor}|}{\raggedright%
Size (TB)
}&\multicolumn{1}{m{81\newtblstarfactor}|}{\centering%
1.1
}&\multicolumn{1}{m{81\newtblstarfactor}|}{\centering%
1.9
}&\multicolumn{1}{m{81\newtblstarfactor}|}{\centering%
3.8
}&\multicolumn{1}{m{81\newtblstarfactor+\arrayrulewidth}}{\centering%
4.9
}\tabularnewline
\multicolumn{5}{m{101\newtblstarfactor+2\tabcolsep+\arrayrulewidth+81\newtblstarfactor+2\tabcolsep+\arrayrulewidth+81\newtblstarfactor+2\tabcolsep+\arrayrulewidth+81\newtblstarfactor+2\tabcolsep+\arrayrulewidth+81\newtblstarfactor+\arrayrulewidth}}{\centering%
\textbf{Event Selection}
}\tabularnewline
\multicolumn{1}{m{101\newtblstarfactor}|}{\raggedright%
MinBias Trigger \footnotemark{}
}&\multicolumn{1}{m{81\newtblstarfactor}|}{\centering%
3400 01/11/21
}&\multicolumn{1}{m{81\newtblstarfactor}|}{\centering%
360001
}&\multicolumn{1}{m{81\newtblstarfactor}|}{\centering%
280001
}&\multicolumn{1}{m{81\newtblstarfactor+\arrayrulewidth}}{\centering%
2700 01/11/21
}\tabularnewline
\multicolumn{1}{m{101\newtblstarfactor}|}{\raggedright%
Event Vertex \footnotemark{}
}&\multicolumn{2}{m{81\newtblstarfactor+2\tabcolsep+\arrayrulewidth+81\newtblstarfactor}|}{\centering%
V$_{\text{r}}$ <{} 2 cm \&
    V$_{\text{z}}$ <{} 70 cm
}&\multicolumn{2}{m{81\newtblstarfactor+2\tabcolsep+\arrayrulewidth+81\newtblstarfactor+\arrayrulewidth}}{\centering%
V$_{\text{r}}$ <{} 2 cm \& V$_{\text{z}}$ <{} 40 cm
}\tabularnewline
\multicolumn{1}{m{101\newtblstarfactor}|}{\raggedright%
Min. Ref. Mult. \footnotemark{}
}&\multicolumn{2}{m{81\newtblstarfactor+2\tabcolsep+\arrayrulewidth+81\newtblstarfactor}|}{\centering%
6
}&\multicolumn{2}{m{81\newtblstarfactor+2\tabcolsep+\arrayrulewidth+81\newtblstarfactor+\arrayrulewidth}}{\centering%
7
}\tabularnewline
\multicolumn{1}{m{101\newtblstarfactor}|}{\raggedright%
Others
}&\multicolumn{4}{m{81\newtblstarfactor+2\tabcolsep+\arrayrulewidth+81\newtblstarfactor+2\tabcolsep+\arrayrulewidth+81\newtblstarfactor+2\tabcolsep+\arrayrulewidth+81\newtblstarfactor+\arrayrulewidth}}{\centering%
\#GlobTracks <{} 3000 \footnotemark{} \&
   \#TofMatchedPrims \ensuremath{\geq} 2 \footnotemark{} \&
   \#TracksEP  >{} 0 \footnotemark{}
}\tabularnewline
\multicolumn{5}{m{101\newtblstarfactor+2\tabcolsep+\arrayrulewidth+81\newtblstarfactor+2\tabcolsep+\arrayrulewidth+81\newtblstarfactor+2\tabcolsep+\arrayrulewidth+81\newtblstarfactor+2\tabcolsep+\arrayrulewidth+81\newtblstarfactor+\arrayrulewidth}}{\centering%
\textbf{Primary Track Selection}
}\tabularnewline
\multicolumn{1}{m{101\newtblstarfactor}|}{\raggedright%
Momentum
}&\multicolumn{4}{m{81\newtblstarfactor+2\tabcolsep+\arrayrulewidth+81\newtblstarfactor+2\tabcolsep+\arrayrulewidth+81\newtblstarfactor+2\tabcolsep+\arrayrulewidth+81\newtblstarfactor+\arrayrulewidth}}{\centering%
p$_{\text{T}}$ >{} 0.2 \& p <{} 10 GeV/c
}\tabularnewline
\multicolumn{1}{m{101\newtblstarfactor}|}{\raggedright%
Track Quality
}&\multicolumn{4}{m{81\newtblstarfactor+2\tabcolsep+\arrayrulewidth+81\newtblstarfactor+2\tabcolsep+\arrayrulewidth+81\newtblstarfactor+2\tabcolsep+\arrayrulewidth+81\newtblstarfactor+\arrayrulewidth}}{\centering%
nHitsFit \ensuremath{\geq} 15 \footnotemark{} \&
   nHitsFit/nHitsPoss >{} 0.52 \footnotemark{} \&
   nHitsDedx \ensuremath{\geq} 15 \footnotemark{}
}\tabularnewline
\multicolumn{1}{m{101\newtblstarfactor}|}{\raggedright%
TOF Matching
}&\multicolumn{4}{m{81\newtblstarfactor+2\tabcolsep+\arrayrulewidth+81\newtblstarfactor+2\tabcolsep+\arrayrulewidth+81\newtblstarfactor+2\tabcolsep+\arrayrulewidth+81\newtblstarfactor+\arrayrulewidth}}{\centering%
TofMatchFlag >{} 0 \footnotemark{} \&
   TofYLocal <{} 1.8 cm \footnotemark{}
}\tabularnewline
\multicolumn{1}{m{101\newtblstarfactor}|}{\raggedright%
Others
}&\multicolumn{4}{m{81\newtblstarfactor+2\tabcolsep+\arrayrulewidth+81\newtblstarfactor+2\tabcolsep+\arrayrulewidth+81\newtblstarfactor+2\tabcolsep+\arrayrulewidth+81\newtblstarfactor+\arrayrulewidth}}{\centering%
$\vert\eta\vert$ <{} 1 \&
   glDCA <{} 1 cm \footnotemark{}
}\tabularnewline
\multicolumn{5}{m{101\newtblstarfactor+2\tabcolsep+\arrayrulewidth+81\newtblstarfactor+2\tabcolsep+\arrayrulewidth+81\newtblstarfactor+2\tabcolsep+\arrayrulewidth+81\newtblstarfactor+2\tabcolsep+\arrayrulewidth+81\newtblstarfactor+\arrayrulewidth}}{\centering%
\textbf{Number of Events (in Millions) by Analysis Stage}
}\tabularnewline
\multicolumn{1}{m{101\newtblstarfactor}|}{\raggedright%
STAR RunQA \footnotemark{}
}&\multicolumn{1}{m{81\newtblstarfactor}|}{\centering%
48.1469
}&\multicolumn{1}{m{81\newtblstarfactor}|}{\centering%
91.2154
}&\multicolumn{1}{m{81\newtblstarfactor}|}{\centering%
164.116
}&\multicolumn{1}{m{81\newtblstarfactor+\arrayrulewidth}}{\centering%
109.177
}\tabularnewline
\multicolumn{1}{m{101\newtblstarfactor}|}{\raggedright%
Minimum Bias
}&\multicolumn{1}{m{81\newtblstarfactor}|}{\centering%
46.4914
}&\multicolumn{1}{m{81\newtblstarfactor}|}{\centering%
90.4904
}&\multicolumn{1}{m{81\newtblstarfactor}|}{\centering%
164.116
}&\multicolumn{1}{m{81\newtblstarfactor+\arrayrulewidth}}{\centering%
84.2632
}\tabularnewline
\multicolumn{1}{m{101\newtblstarfactor}|}{\raggedright%
\#GlobTracks
}&\multicolumn{1}{m{81\newtblstarfactor}|}{\centering%
46.4909
}&\multicolumn{1}{m{81\newtblstarfactor}|}{\centering%
90.4904
}&\multicolumn{1}{m{81\newtblstarfactor}|}{\centering%
164.116
}&\multicolumn{1}{m{81\newtblstarfactor+\arrayrulewidth}}{\centering%
84.2618
}\tabularnewline
\multicolumn{1}{m{101\newtblstarfactor}|}{\raggedright%
Event Vertex
}&\multicolumn{1}{m{81\newtblstarfactor}|}{\centering%
37.0639
}&\multicolumn{1}{m{81\newtblstarfactor}|}{\centering%
71.3360
}&\multicolumn{1}{m{81\newtblstarfactor}|}{\centering%
133.213
}&\multicolumn{1}{m{81\newtblstarfactor+\arrayrulewidth}}{\centering%
70.7260
}\tabularnewline
\multicolumn{1}{m{101\newtblstarfactor}|}{\raggedright%
\#TofMatchedPrims
}&\multicolumn{1}{m{81\newtblstarfactor}|}{\centering%
36.1075
}&\multicolumn{1}{m{81\newtblstarfactor}|}{\centering%
70.8353
}&\multicolumn{1}{m{81\newtblstarfactor}|}{\centering%
132.195
}&\multicolumn{1}{m{81\newtblstarfactor+\arrayrulewidth}}{\centering%
68.2627
}\tabularnewline
\multicolumn{1}{m{101\newtblstarfactor}|}{\raggedright%
Refined RunQA \footnotemark{}
}&\multicolumn{1}{m{81\newtblstarfactor}|}{\centering%
33.6811
}&\multicolumn{1}{m{81\newtblstarfactor}|}{\centering%
67.2199
}&\multicolumn{1}{m{81\newtblstarfactor}|}{\centering%
131.324
}&\multicolumn{1}{m{81\newtblstarfactor+\arrayrulewidth}}{\centering%
62.6191
}\tabularnewline
\multicolumn{1}{m{101\newtblstarfactor}|}{\raggedright%
Min. Ref. Mult.
}&\multicolumn{1}{m{81\newtblstarfactor}|}{\centering%
30.8784
}&\multicolumn{1}{m{81\newtblstarfactor}|}{\centering%
61.7139
}&\multicolumn{1}{m{81\newtblstarfactor}|}{\centering%
118.704
}&\multicolumn{1}{m{81\newtblstarfactor+\arrayrulewidth}}{\centering%
56.7663
}\tabularnewline
\multicolumn{1}{m{101\newtblstarfactor}|}{\raggedright%
\#TracksEP
}&\multicolumn{1}{m{81\newtblstarfactor}|}{\centering%
30.8783
}&\multicolumn{1}{m{81\newtblstarfactor}|}{\centering%
61.7137
}&\multicolumn{1}{m{81\newtblstarfactor}|}{\centering%
118.691
}&\multicolumn{1}{m{81\newtblstarfactor+\arrayrulewidth}}{\centering%
56.7082
}\tabularnewline
\multicolumn{1}{m{101\newtblstarfactor}|}{\raggedright%
\emph{Final (weighted) \footnotemark{}}
}&\multicolumn{1}{m{81\newtblstarfactor}|}{\centering%
\emph{32.2307}
}&\multicolumn{1}{m{81\newtblstarfactor}|}{\centering%
\emph{63.6828}
}&\multicolumn{1}{m{81\newtblstarfactor}|}{\centering%
\emph{122.390}
}&\multicolumn{1}{m{81\newtblstarfactor+\arrayrulewidth}}{\centering%
\emph{59.4631}
}\tabularnewline
\hline
\end{longtable}\endgroup%

\end{center}
\addtocounter{footnote}{-15}\stepcounter{footnote}
\footnotetext{
STAR-{}internal IDs to select minimum bias trigger events.
}\stepcounter{footnote}
\footnotetext{
fiducial event vertex volume to remove beam on beam-{}pipe events.
}\stepcounter{footnote}
\footnotetext{
minimum reference multiplicity to select the 0-{}80\%
   most central events (see Section~\ref{ana_subsec_refmult_runqa}).
}\stepcounter{footnote}
\footnotetext{
total number of global tracks in an event.
}\stepcounter{footnote}
\footnotetext{
number of primary tracks matched to TOF hit(s) in an event.
}\stepcounter{footnote}
\footnotetext{
number of global tracks in an event used for its
   event plane reconstruction (see Section~\ref{ana_subsec_evtplane}).
}\stepcounter{footnote}
\footnotetext{
number of TPC dE/dx cluster hits used for particle helix fit.
}\stepcounter{footnote}
\footnotetext{
remove split tracks (see Section~\ref{ana_subsec_tpctof}).
}\stepcounter{footnote}
\footnotetext{
total number of TPC dE/dx cluster hits in an event.
}\stepcounter{footnote}
\footnotetext{
flag determining whether a track is matched to TOF
   hit(s): 0 -{} no match, 1/2 -{} one-{}to-{}one/multiple.
}\stepcounter{footnote}
\footnotetext{
constrain distance btw. TPC track
           extrapolation endpoint and TOF pad center to remove false TOF
           matches.
}\stepcounter{footnote}
\footnotetext{
distance-{}of-{}closest-{}approach of global track to event vertex.
}\stepcounter{footnote}
\footnotetext{
collaboration-{}wide list for bad run rejection via
 \texttt{StR\penalty5000 e\penalty5000 f\penalty5000 M\penalty5000 u\penalty5000 l\penalty5000 t\penalty5000 C\penalty5000 orr} (see Section~\ref{ana_subsec_refmult_runqa}).
}\stepcounter{footnote}
\footnotetext{
additional run rejection specific to this physics
           analysis (see Section~\ref{ana_subsec_refmult_runqa}).
}\stepcounter{footnote}
\footnotetext{
corrected for trigger inefficiencies (see
 Section~\ref{ana_subsec_refmult_runqa}).
}
\pagebreak[4]

\subsection{Reference Multiplicity and Quality Assurance}
\label{ana_subsec_refmult_runqa}\hyperlabel{ana_subsec_refmult_runqa}%

In heavy-{}ion collisions, the impact parameter of the two incoming nuclei cannot
be observed directly due to the femtoscopic length scales at play. However,
related geometric parameters such as the number of
participating nucleons (N$_{\text{part}}$) or the number of binary nucleon-{}nucleon
collisions (N$_{\text{coll}}$) are typically needed to classify the centrality of an
event during a physics analysis. For instance, in Section~\ref{ana_sec_pairrec} an
event's centrality class is used to group it with other similar events for the
generation of background distributions using the event mixing technique, and in
Section~\ref{sim_sec_cocktail} N$_{\text{coll}}$ constitutes the factor required to scale
simulations of charm continuum distributions from p+p to Au+Au collisions
(quoted in Table~\ref{sim_tab_pars}). This predicament is commonly addressed nowadays
using \emph{Glauber} Monte Carlo simulations [166] which provide a
theoretical mechanism to estimate the necessary quantities from experimental
data by modeling the multiple scattering of the traversing nucleons inside the
colliding nuclei. As illustrated in Figure~\ref{dsets_evttrk_fig_glauber} (left and
middle), the latter are assembled in-{}silico using suitable nuclear density
distributions and brought to collision with an impact parameter \emph{b} chosen
randomly according to $\mathrm{d}\sigma/\mathrm{d}b=2{\pi}b$. The
collision subsequently proceeds via a sequence of independent (inelastic)
binary nucleon-{}nucleon collisions with the participants traveling along
straight lines inbetween collisions. In very simple terms, a nucleon-{}nucleon
collision occurs when the criterion
$d^2\leq\sigma_\mathrm{inel}^\mathrm{NN}/\pi$ is satisfied for
their radial distance \emph{d} (\emph{black disk},
$\sigma_\mathrm{inel}^\mathrm{NN}$: total inelastic nucleon-{}nucleon
cross-{}section).

\wrapifneeded{0.50}{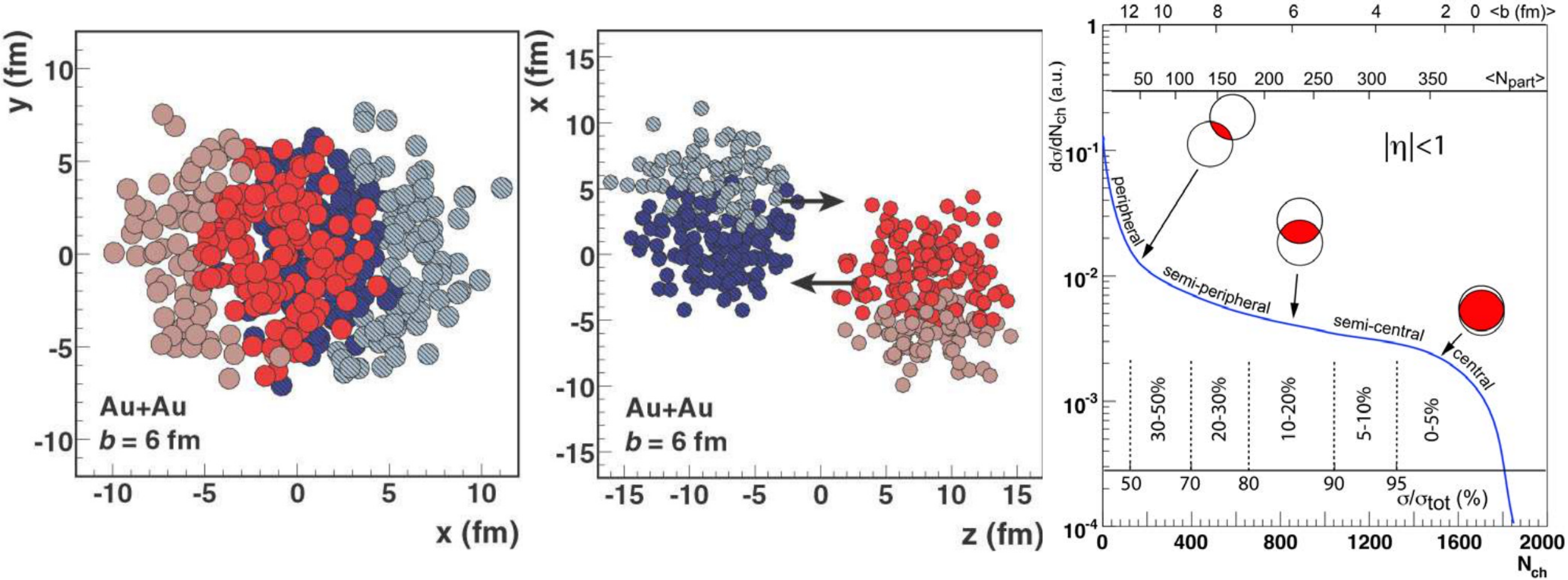}{Glauber Monte Carlo event viewed in transverse plane (left) and along the beam axis (middle). Nucleons are depicted with diameter \emph{d} according to the black disk collision criterion (see text). Darker disks represent participating nucleons. (right) Cartoon depicting the experimentally measurable charged particle multiplicity (N$_{\text{ch}}$) distribution along with mean impact parameter and N$_{\text{part}}$ calculated from Glauber simulations for the centrality classes. [166]}{dsets_evttrk_fig_glauber}{1} %

One of the advantages of Glauber Monte Carlo models is their capability to
simulate experimental final-{}stage observables like the charged particle
multiplicity N$_{\text{ch}}$. The cartoon in Figure~\ref{dsets_evttrk_fig_glauber} (right), for
instance, connects the N$_{\text{ch}}$ distribution d\ensuremath{\sigma}/dN$_{\text{ch}}$ to the fundamental
geometry-{}related parameters \emph{b} and N$_{\text{part}}$. The d\ensuremath{\sigma}/dN$_{\text{ch}}$ distribution
is thus a good experimental measure to indirectly determine a set of centrality
classes ranging from central to peripheral collisions by calculating fractions
of the total reaction cross section (\ensuremath{\sigma}/\ensuremath{\sigma}$_{\text{tot}}$). The N$_{\text{ch}}$ multiplicities then serve as a reference for the centrality of a collision
which is why a single measured N$_{\text{ch}}$ is commonly referred to as \emph{reference
multiplicity} (RefMult) of an event. In Table~\ref{ana_tab_dsets_evttrk}, a minimum
reference multiplicity has been used to select the 0-{}80\% most central events,
i.e. events with impact parameters approximately up to 80\% of the maximum
radial distance between the incoming nuclei
($\langle\mathrm{N}_\mathrm{part}\rangle$ >{} 0).

Figure~\ref{dsets_evttrk_fig_refmult} shows the (raw) reference multiplicity
distributions of charged particle tracks reconstructed by STAR within
$\vert\eta\vert$ <{} 0.5 at BES-{}I energies [167]. All
experimental distributions are accurately described by Glauber Monte Carlo
simulations over the full N$_{\text{ch}}$ range except for some of the most peripheral
events suffering trigger inefficiencies. The effect is corrected by using the
N$_{\text{ch}}$-{}dependent ratio of simulation over data as a weighting factor on event-{}
but also track-{}wise observables (Section~\ref{ana_sec_pairrec}). The final number of
events listed in Table~\ref{ana_tab_dsets_evttrk}, for instance, is obtained by
calculating the according weighted sums of events. Note that the correction is
as large as 30\% for the peripheral centrality bin whereas negligible in more
central collisions. The z-{}vertex dependence of other sources of inefficiencies
due to changes in the trigger configuration throughout the run are also
included in the correctional weighting factors.

\wrapifneeded{0.50}{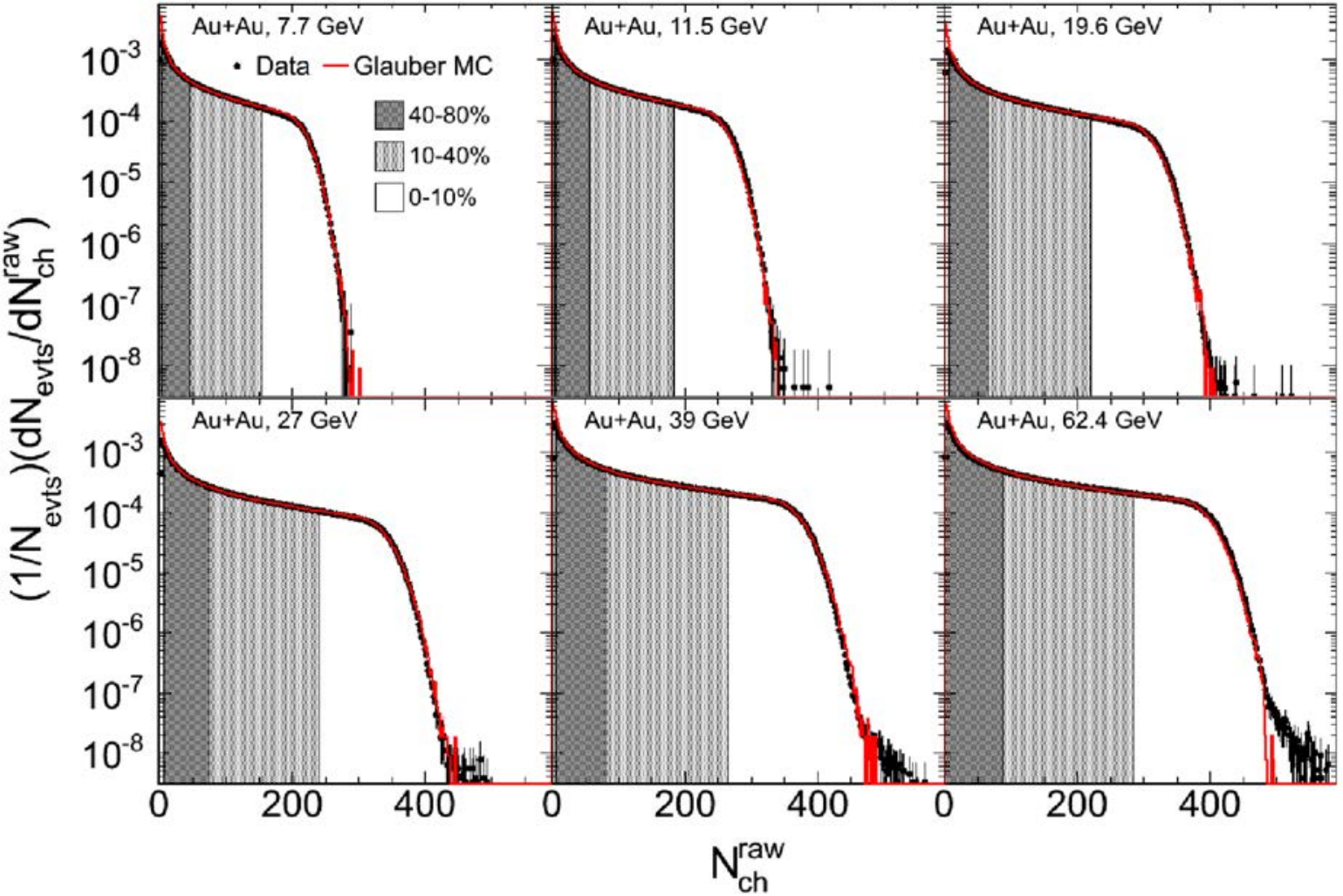}{Raw multiplicity distributions measured in STAR at $\vert\eta\vert$ <{} 0.5 for BES-{}I energies (black points). The data is compared to a Glauber Monte Carlo simulation (red solid line) and three different centrality classes are indicated (grey shaded regions). [167]}{dsets_evttrk_fig_refmult}{0.9} %
\begin{center}
\begingroup%
\setlength{\newtblsparewidth}{1\linewidth-2\tabcolsep-2\tabcolsep-2\tabcolsep-2\tabcolsep-2\tabcolsep-2\tabcolsep-2\tabcolsep-2\tabcolsep-2\tabcolsep-2\tabcolsep-2\tabcolsep}%
\setlength{\newtblstarfactor}{\newtblsparewidth / \real{422}}%

\begin{longtable}{llllllllll}\caption[{Reference multiplicities at BES-{}I energies used in this dielectron analysis to define the centrality classes [168]. An event belongs to a certain centrality class (in \%) if its reference multiplicity is larger than the value listed but smaller than the one for the next more central class.}]{Reference multiplicities at BES-{}I energies used in this dielectron analysis to define the centrality classes [168]. An event belongs to a certain centrality class (in \%) if its reference multiplicity is larger than the value listed but smaller than the one for the next more central class.\label{dsets_evttrk_tab_refmult}\hyperlabel{dsets_evttrk_tab_refmult}%
}\tabularnewline
\hline
\multicolumn{1}{m{31\newtblstarfactor}|}{\centering\bfseries%
 \ensuremath{\surd}s$_{\text{NN}}$ %
}&\multicolumn{1}{m{47\newtblstarfactor}|}{\centering\bfseries%
 70-{}80 %
}&\multicolumn{1}{m{47\newtblstarfactor}|}{\centering\bfseries%
 60-{}70 %
}&\multicolumn{1}{m{47\newtblstarfactor}|}{\centering\bfseries%
 50-{}60 %
}&\multicolumn{1}{m{47\newtblstarfactor}|}{\centering\bfseries%
 40-{}50 %
}&\multicolumn{1}{m{47\newtblstarfactor}|}{\centering\bfseries%
 30-{}40 %
}&\multicolumn{1}{m{47\newtblstarfactor}|}{\centering\bfseries%
 20-{}30 %
}&\multicolumn{1}{m{47\newtblstarfactor}|}{\centering\bfseries%
 10-{}20 %
}&\multicolumn{1}{m{31\newtblstarfactor}|}{\centering\bfseries%
 5-{}10 %
}&\multicolumn{1}{m{31\newtblstarfactor+\arrayrulewidth}}{\centering\bfseries%
 0-{}5%
}\tabularnewline
\endfirsthead
\caption[]{(continued)}\tabularnewline
\hline
\multicolumn{1}{m{31\newtblstarfactor}|}{\centering\bfseries%
 \ensuremath{\surd}s$_{\text{NN}}$ %
}&\multicolumn{1}{m{47\newtblstarfactor}|}{\centering\bfseries%
 70-{}80 %
}&\multicolumn{1}{m{47\newtblstarfactor}|}{\centering\bfseries%
 60-{}70 %
}&\multicolumn{1}{m{47\newtblstarfactor}|}{\centering\bfseries%
 50-{}60 %
}&\multicolumn{1}{m{47\newtblstarfactor}|}{\centering\bfseries%
 40-{}50 %
}&\multicolumn{1}{m{47\newtblstarfactor}|}{\centering\bfseries%
 30-{}40 %
}&\multicolumn{1}{m{47\newtblstarfactor}|}{\centering\bfseries%
 20-{}30 %
}&\multicolumn{1}{m{47\newtblstarfactor}|}{\centering\bfseries%
 10-{}20 %
}&\multicolumn{1}{m{31\newtblstarfactor}|}{\centering\bfseries%
 5-{}10 %
}&\multicolumn{1}{m{31\newtblstarfactor+\arrayrulewidth}}{\centering\bfseries%
 0-{}5%
}\tabularnewline
\endhead
\multicolumn{1}{m{31\newtblstarfactor}|}{\centering%
19.6
}&\multicolumn{1}{m{47\newtblstarfactor}|}{\centering%
6
}&\multicolumn{1}{m{47\newtblstarfactor}|}{\centering%
12
}&\multicolumn{1}{m{47\newtblstarfactor}|}{\centering%
23
}&\multicolumn{1}{m{47\newtblstarfactor}|}{\centering%
40
}&\multicolumn{1}{m{47\newtblstarfactor}|}{\centering%
66
}&\multicolumn{1}{m{47\newtblstarfactor}|}{\centering%
102
}&\multicolumn{1}{m{47\newtblstarfactor}|}{\centering%
152
}&\multicolumn{1}{m{31\newtblstarfactor}|}{\centering%
220
}&\multicolumn{1}{m{31\newtblstarfactor+\arrayrulewidth}}{\centering%
263
}\tabularnewline
\multicolumn{1}{m{31\newtblstarfactor}|}{\centering%
27
}&\multicolumn{1}{m{47\newtblstarfactor}|}{\centering%
6
}&\multicolumn{1}{m{47\newtblstarfactor}|}{\centering%
13
}&\multicolumn{1}{m{47\newtblstarfactor}|}{\centering%
26
}&\multicolumn{1}{m{47\newtblstarfactor}|}{\centering%
45
}&\multicolumn{1}{m{47\newtblstarfactor}|}{\centering%
74
}&\multicolumn{1}{m{47\newtblstarfactor}|}{\centering%
114
}&\multicolumn{1}{m{47\newtblstarfactor}|}{\centering%
168
}&\multicolumn{1}{m{31\newtblstarfactor}|}{\centering%
241
}&\multicolumn{1}{m{31\newtblstarfactor+\arrayrulewidth}}{\centering%
288
}\tabularnewline
\multicolumn{1}{m{31\newtblstarfactor}|}{\centering%
39
}&\multicolumn{1}{m{47\newtblstarfactor}|}{\centering%
7
}&\multicolumn{1}{m{47\newtblstarfactor}|}{\centering%
15
}&\multicolumn{1}{m{47\newtblstarfactor}|}{\centering%
28
}&\multicolumn{1}{m{47\newtblstarfactor}|}{\centering%
50
}&\multicolumn{1}{m{47\newtblstarfactor}|}{\centering%
81
}&\multicolumn{1}{m{47\newtblstarfactor}|}{\centering%
125
}&\multicolumn{1}{m{47\newtblstarfactor}|}{\centering%
185
}&\multicolumn{1}{m{31\newtblstarfactor}|}{\centering%
265
}&\multicolumn{1}{m{31\newtblstarfactor+\arrayrulewidth}}{\centering%
316
}\tabularnewline
\multicolumn{1}{m{31\newtblstarfactor}|}{\centering%
62.4
}&\multicolumn{1}{m{47\newtblstarfactor}|}{\centering%
7
}&\multicolumn{1}{m{47\newtblstarfactor}|}{\centering%
16
}&\multicolumn{1}{m{47\newtblstarfactor}|}{\centering%
30
}&\multicolumn{1}{m{47\newtblstarfactor}|}{\centering%
54
}&\multicolumn{1}{m{47\newtblstarfactor}|}{\centering%
88
}&\multicolumn{1}{m{47\newtblstarfactor}|}{\centering%
135
}&\multicolumn{1}{m{47\newtblstarfactor}|}{\centering%
199
}&\multicolumn{1}{m{31\newtblstarfactor}|}{\centering%
285
}&\multicolumn{1}{m{31\newtblstarfactor+\arrayrulewidth}}{\centering%
339
}\tabularnewline
\hline
\end{longtable}\endgroup%

\end{center}

The preceding procedure culminated in the development of a STAR software
module called \texttt{StR\penalty5000 e\penalty5000 f\penalty5000 M\penalty5000 u\penalty5000 l\penalty5000 t\penalty5000 C\penalty5000 orr} [168, 169] which not only
provides the centrality classes listed in Table~\ref{dsets_evttrk_tab_refmult} but also
includes basic quality assurance (QA) to reject bad runs based on their average
RefMult. The according list can be found in Table~\ref{app_tab_badruns} and is refined
for this dielectron analysis as follows.

\wrapifneeded{0.50}{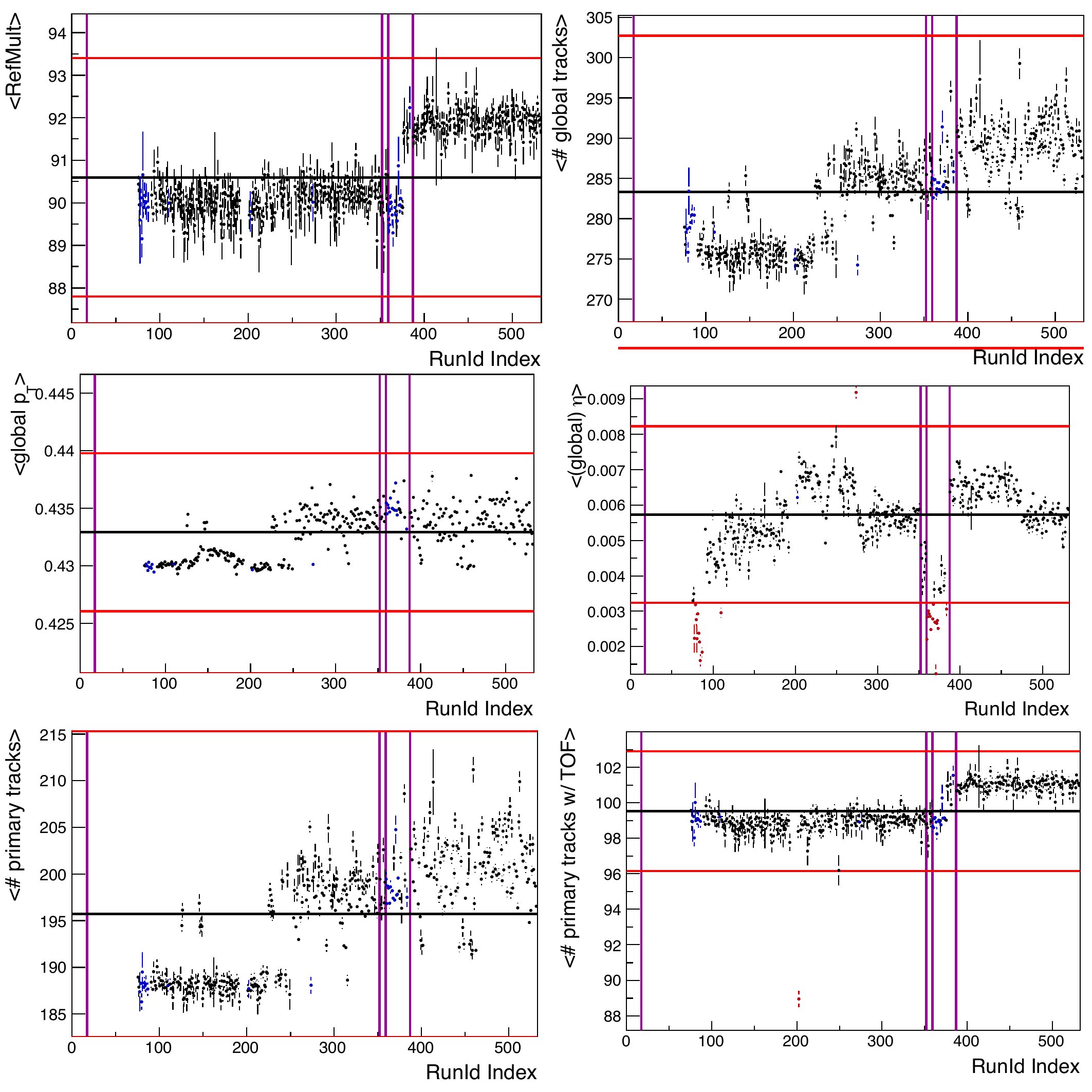}{Set of event-{}wise observables used for run-{}by-{}run quality assurance (Run QA) at 19.6 GeV, representatively. Figures for the Run QA at all BES-{}I energies used here are appended in Section~\ref{app_runqa}. Solid black points denote the experimental data for the observable averaged over the events within a run whereas black horizontal lines reflect the observable's average over the entire beam time. The runs are identified using the index obtained via the dedicated library \texttt{StR\penalty5000 u\penalty5000 n\penalty5000 I\penalty5000 d\penalty5000 E\penalty5000 v\penalty5000 e\penalty5000 n\penalty5000 t\penalty5000 sDb} in Section~\ref{strunideventsdb}. Vertical magenta lines indicate run ranges with the same number of missing RDOs (see Section~\ref{ana_subsec_tpctof} and the library \texttt{StB\penalty5000 a\penalty5000 d\penalty5000 R\penalty5000 d\penalty5000 o\penalty5000 sDb} in Section~\ref{stbadrdos}). Horizontal red lines result from the Grubbs test for outliers [170] and separate accepted runs (black) from rejected (red) runs. Blue points depict runs that have been rejected based on one of the other observables.}{dsets_evttrk_fig_runqa}{0.97} %

A so-{}called \emph{run} compiles a set of events taken during a time with steady
detector conditions and, depending on the beam energy, lasts for about 30-{}60 minutes with a few
million events. The choice to split the data acquisition into chunks called
\emph{runs} is rather arbitrary but simplifies time-{}dependent studies like the
quality assurance in Figure~\ref{dsets_evttrk_fig_runqa}. For the latter, the
event-{}by-{}event observables reference multiplicity (RefMult), number of
global/primary tracks with and without TOF match, global momentum and
pseudo-{}rapidity are averaged over the events contained in the run with an index
specific to the corresponding RunID (see Section~\ref{strunideventsdb}).  To obtain a
refined list of "bad" runs in addition to the one provided by \texttt{StR\penalty5000 e\penalty5000 f\penalty5000 M\penalty5000 u\penalty5000 l\penalty5000 t\penalty5000 C\penalty5000 orr},
outliers from the general trend of these observables across all runs are
detected using the Grubbs test [170]. In its two-{}sided form, this
test continuously recalculates the statistic
$G=\mathrm{max}\vert{Y_i-\bar{Y}}\vert/\sigma$ for the observable
\emph{Y}, its sample mean $\bar{Y}$ and standard deviation \ensuremath{\sigma}.
Using this statistic, the iterative algorithm [171] intrinsically
starts testing for outliers with the data point currently farthest away from
the sample mean when properly weighted with its statistical uncertainty. This
data point is marked as outlier at a significance level \ensuremath{\alpha} and removed
from the sample before the next iteration if the statistic satisfies
\[G>\frac{N-1}{\sqrt{N}}\sqrt{\frac{t^2}{N-2+t^2}}\]
with \emph{N} the sample size and \emph{t} the quantile of the Student's t-{}distribution
for probability \ensuremath{\alpha}/2N and N-{}2 degrees of freedom. The iteration is aborted
and no further RunIDs rejected once no outliers are found anymore according to
the above criterion. Note that normal distribution of the data technically is a
requirement for the validity of the Grubbs test. For the observables used for
the Run QA in Figure~\ref{dsets_evttrk_fig_runqa} and Section~\ref{app_runqa}, this requirement
could be satisfied by partitioning the data into suitable run ranges (e.g. the
ones provided by \texttt{StB\penalty5000 a\penalty5000 d\penalty5000 R\penalty5000 d\penalty5000 o\penalty5000 sDb} in Section~\ref{stbadrdos}) and applying the algorithm
on the subsets separately. However, this would have complicated the analysis
unnecessarily since the simplified implementation chosen in [171]
with \ensuremath{\alpha} = 0.95 has proven to be effective enough in rejecting the worst
outliers and can also be applied across all energies without modifications. The
resulting refined list of rejected runs is also included in
Table~\ref{app_tab_badruns}.

\subsection{Event Plane Reconstruction}
\label{ana_subsec_evtplane}\hyperlabel{ana_subsec_evtplane}%

The event plane method [9, 172] is a concept integral
to many physics analyses trying to measure elliptic flow \emph{v$_{\text{2}}$} in a heavy-{}ion
collision (see Section~\ref{intro_sec_hics}). In such analyses, an identified particle's
\emph{v$_{\text{2}}$} signal can be extracted from a fit to its yield distribution in the form of
\[\mathrm{d}N/\mathrm{d}(\phi-\Psi_2)\propto 1+2v_2\cos\left(2(\phi-\Psi_2)\right)\]
as function of its azimuthal angle $\phi$ with respect to the second harmonic event
plane angle \ensuremath{\Psi}$_{\text{2}}$ [167]. The measurement of dielectron
elliptic flow [173] is beyond the scope of this thesis and the v$_{\text{2}}$ signal extraction requiring corrections for self-{}correlations and calculation
of event plane resolutions, for instance, thus not part of this analysis.
However, an event's \ensuremath{\Psi}$_{\text{2}}$ and hence the full reconstruction of the event
plane are still needed to later allow for the correct statistical subtraction
of background from combinatorial $e^+e^-$ pairs. The latter is an
important step to obtain the dielectron spectra from physical sources. The
study of event mixing in Section~\ref{ana_sec_pairrec} further reveals that event
classification taking the event plane angle into account is imperative for the
correct background shape.

The reconstruction of the event plane relies on the particles' observed
azimuthal angles and hence the event's anisotropic flow itself to arrive at a
reasonable estimate for the more fundamental plane defined by the participating
nucleons. Direct experimental access to the participant plane is not possible
but an estimate for it is important since theoretical models commonly use it to
describe flow measurements. The initial elliptical overlap zone containing the
participant nucleons (see Figure~\ref{intro_fig_flow}) is oriented isotropically in
space due to the stochastic nature of a nucleus-{}nucleus collision. The reaction
plane defined by the incoming nuclei is hence distributed uniformly regarding
its azimuthal angle with respect to the beam axis. This uniformity in the initial
reaction consequently demands a flat distribution of the event plane angle
$\Psi_2=0.5\tan^{-1}\left(Q_y/Q_x\right)$ with
$\vec{Q}=(Q_x,Q_y)$ the event flow vector in the laboratory
determined by the measured azimuthal angles $\phi_i$ of the
reconstructed particles in an event:
\[\vec{Q}_\mathrm{raw} = \frac{1}{N} \sum\limits_{i=0}^N \binom{w_i\cos(2\phi_i)}{w_i\sin(2\phi_i)} \mbox{ with } w_i= \begin{cases} p_T & \mbox{if } p_T \leq 2\,GeV/c \\ 2 & \mbox{if } p_T > 2\,GeV/c.\end{cases}\]
where \emph{N} denotes the number of tracks with global DCA <{} 1 cm,
$\vert\eta\vert$ <{} 1 (full TPC), and global momenta of 0.15 -{} 5
GeV/c. In the optimal case, the weights \emph{w$_{\text{i}}$} in the above raw event flow
vector are chosen such that they reflect the elliptic flow \emph{v$_{\text{2}}$} which is
often linearly dependent on the particle's transverse momentum \emph{p$_{\text{T}}$} up to
about 2 GeV/c.

\wrapifneeded{0.50}{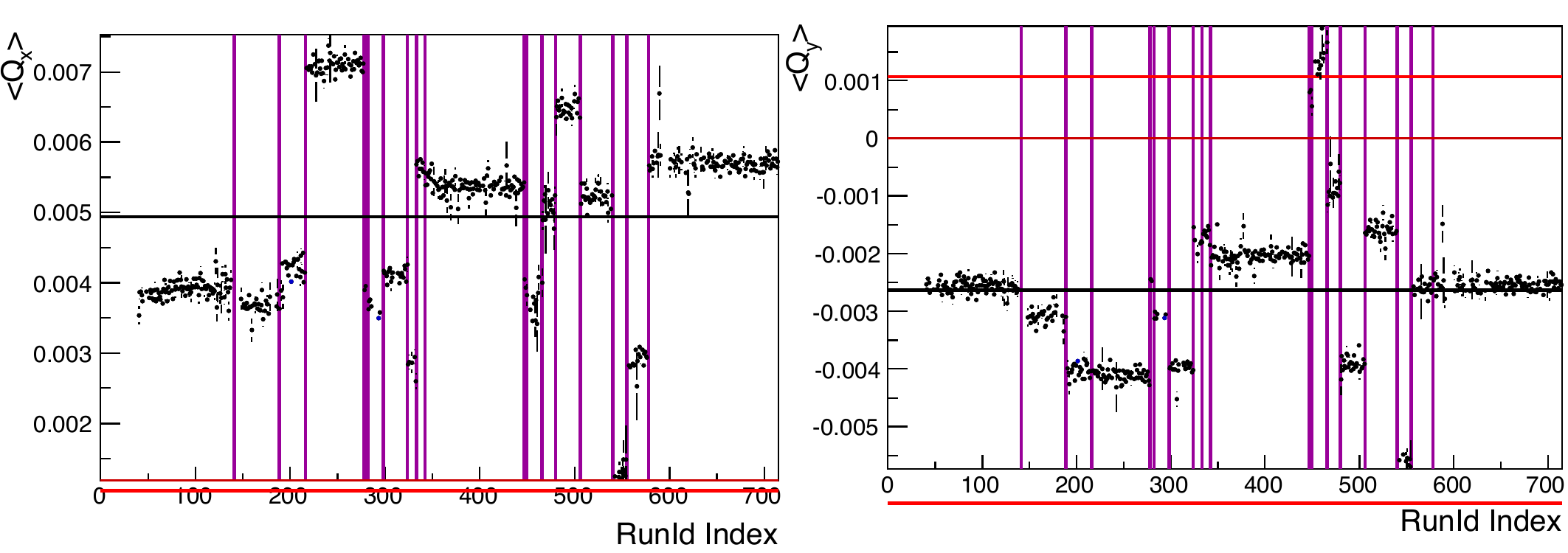}{RunID-{}dependence of the raw event flow vector at 39 GeV averaged over all events within an experimental run. Color coding and run indexing are identical to Figure~\ref{dsets_evttrk_fig_runqa}. Note however, that horizontal red lines do not bear any meaning here since the event flow vector is not used as an observable for Run QA. Figures for all BES-{}I energies are appended in Section~\ref{app_runqa}.}{evtplane_fig_QxQy39}{1} %

As a representative for all BES-{}I energies (see Section~\ref{app_runqa}),
Figure~\ref{evtplane_fig_QxQy39} depicts the run-{}by-{}run averaged raw event flow vectors
at $\sqrt{s_\mathrm{NN}}$ = 39 GeV. The averages clearly correlate
with the number of missing RDOs for the different run ranges indicated by the
magenta vertical lines (see Section~\ref{stbadrdos}). Within each run range, the average
Q-{}vectors are approximately constant when taking statistical uncertainties into
account. These observations are in agreement with expectations since such
inefficiencies in the detector response immediately destroy isotropy and induce
a preferential direction on the average Q-{}vector. Detector imperfections, in
general, hence introduce acceptance correlations on the \ensuremath{\Psi}$_{\text{2}}$ distribution.
This necessitates the application of analysis methods to correct for the
resulting effects and flatten the \ensuremath{\Psi}$_{\text{2}}$ distribution. The intuitive
procedure called \emph{Phi-{}Weighting} is to weight each contributing particle with
the inverse of an azimuthal angle distribution averaged over a sufficiently
large number of events. But strong variations in acceptance or entirely missing
sectors are difficult to compensate using Phi-{}Weighting.  Alternatively, an
average raw event flow vector $\langle\vec{Q}_\mathrm{raw}\rangle$ can be derived from a large enough selection of events and subtracted from
$\vec{Q}_\mathrm{raw}$ on an event-{}by-{}event basis to obtain the
\emph{recentered} event flow vector
$\vec{Q}_\mathrm{rc}=\vec{Q}_\mathrm{raw}-\langle\vec{Q}_\mathrm{raw}\rangle$ [9, 172].
This \emph{Recentering} method is not only more practical to implement but also
guarantees a zero $\langle\vec{Q}_\mathrm{rc}\rangle$. The
corrected event plane angle \ensuremath{\Psi}$_{\text{2,rc}}$ is subsequently recalculated using the
recentered Q-{}vector $\vec{Q}_\mathrm{rc}$.

The event samples used for the purpose of obtaining an average raw event flow
vector should be small enough to register changes in detector response and beam
performance during the experiment but also large enough to ensure statistically
significant averages. In this analysis [174, 175],
the datasets at each energy are therefore divided into 10 z-{}vertex (V$_{\text{z}}$)
intervals and raw Q-{}vector averages are calculated for each run separately.
Samples with less than four events for a specific V$_{\text{z}}$ bin and RunID are
ignored in the subsequent analysis steps. Instead of using equi-{}distant
intervals, the V$_{\text{z}}$ distributions for each energy are parameterized using the
functional forms listed in Table~\ref{evtplane_tab_vz} and partitioned such that each
bin contains the approximately equal number of events. The resulting V$_{\text{z}}$ intervals are appended in Table~\ref{evtplane_tab_vz} and also used for event
classification during event mixing in Section~\ref{ana_sec_pairrec}.  Exemplatory
$\langle\vec{Q}_\mathrm{raw}\rangle$ coordinates for all BES-{}I
energies are shown in Figure~\ref{evtplane_fig_recenter} as a function of reference
multiplicity. The dependence is parameterized with a straight line to smoothen
the statistical fluctuations and the resulting fit parameters are used to
obtain event-{}by-{}event $\vec{Q}_\mathrm{rc}$ vectors according to
the event's V$_{\text{z}}$, RunID, and RefMult.
\begin{center}
\begingroup%
\setlength{\newtblsparewidth}{1\linewidth-2\tabcolsep-2\tabcolsep-2\tabcolsep-2\tabcolsep-2\tabcolsep-2\tabcolsep-2\tabcolsep-2\tabcolsep-2\tabcolsep-2\tabcolsep-2\tabcolsep-2\tabcolsep-2\tabcolsep}%
\setlength{\newtblstarfactor}{\newtblsparewidth / \real{421}}%

\begin{longtable}{llllllllllll}\caption[{Parameterization of V$_{\text{z}}$ distributions for binning with equal number of events.}]{Parameterization of V$_{\text{z}}$ distributions for binning with equal number of events.\label{evtplane_tab_vz}\hyperlabel{evtplane_tab_vz}%
}\tabularnewline
\hline
\multicolumn{1}{m{31\newtblstarfactor}|}{\centering\bfseries%
 \ensuremath{\surd}s$_{\text{NN}}$ %
}&\multicolumn{5}{m{31\newtblstarfactor+2\tabcolsep+\arrayrulewidth+47\newtblstarfactor+2\tabcolsep+\arrayrulewidth+47\newtblstarfactor+2\tabcolsep+\arrayrulewidth+47\newtblstarfactor+2\tabcolsep+\arrayrulewidth+47\newtblstarfactor}|}{\centering\bfseries%
 Form %
}&\multicolumn{2}{m{31\newtblstarfactor+2\tabcolsep+\arrayrulewidth+31\newtblstarfactor}|}{\centering\bfseries%
 a %
}&\multicolumn{2}{m{31\newtblstarfactor+2\tabcolsep+\arrayrulewidth+31\newtblstarfactor}|}{\centering\bfseries%
 b %
}&\multicolumn{2}{m{31\newtblstarfactor+2\tabcolsep+\arrayrulewidth+16\newtblstarfactor+\arrayrulewidth}}{\centering\bfseries%
 c%
}\tabularnewline
\endfirsthead
\caption[]{(continued)}\tabularnewline
\hline
\multicolumn{1}{m{31\newtblstarfactor}|}{\centering\bfseries%
 \ensuremath{\surd}s$_{\text{NN}}$ %
}&\multicolumn{5}{m{31\newtblstarfactor+2\tabcolsep+\arrayrulewidth+47\newtblstarfactor+2\tabcolsep+\arrayrulewidth+47\newtblstarfactor+2\tabcolsep+\arrayrulewidth+47\newtblstarfactor+2\tabcolsep+\arrayrulewidth+47\newtblstarfactor}|}{\centering\bfseries%
 Form %
}&\multicolumn{2}{m{31\newtblstarfactor+2\tabcolsep+\arrayrulewidth+31\newtblstarfactor}|}{\centering\bfseries%
 a %
}&\multicolumn{2}{m{31\newtblstarfactor+2\tabcolsep+\arrayrulewidth+31\newtblstarfactor}|}{\centering\bfseries%
 b %
}&\multicolumn{2}{m{31\newtblstarfactor+2\tabcolsep+\arrayrulewidth+16\newtblstarfactor+\arrayrulewidth}}{\centering\bfseries%
 c%
}\tabularnewline
\endhead
\multicolumn{1}{m{31\newtblstarfactor}|}{\centering%
19.6
}&\multicolumn{5}{m{31\newtblstarfactor+2\tabcolsep+\arrayrulewidth+47\newtblstarfactor+2\tabcolsep+\arrayrulewidth+47\newtblstarfactor+2\tabcolsep+\arrayrulewidth+47\newtblstarfactor+2\tabcolsep+\arrayrulewidth+47\newtblstarfactor}|}{\setlength{\newtblcolwidth}{31\newtblstarfactor+2\tabcolsep+\arrayrulewidth+47\newtblstarfactor+2\tabcolsep+\arrayrulewidth+47\newtblstarfactor+2\tabcolsep+\arrayrulewidth+47\newtblstarfactor+2\tabcolsep+\arrayrulewidth+47\newtblstarfactor}\multirowii[m]{2}{\newtblcolwidth}{\centering%
$a-b\,V_z-c\,V_z^2$
}}&\multicolumn{2}{m{31\newtblstarfactor+2\tabcolsep+\arrayrulewidth+31\newtblstarfactor}|}{\centering%
6631.45
}&\multicolumn{2}{m{31\newtblstarfactor+2\tabcolsep+\arrayrulewidth+31\newtblstarfactor}|}{\centering%
7.6321
}&\multicolumn{2}{m{31\newtblstarfactor+2\tabcolsep+\arrayrulewidth+16\newtblstarfactor+\arrayrulewidth}}{\centering%
0.372181
}\tabularnewline
\multicolumn{1}{m{31\newtblstarfactor}|}{\centering%
27
}&\multicolumn{5}{m{31\newtblstarfactor+2\tabcolsep+\arrayrulewidth+47\newtblstarfactor+2\tabcolsep+\arrayrulewidth+47\newtblstarfactor+2\tabcolsep+\arrayrulewidth+47\newtblstarfactor+2\tabcolsep+\arrayrulewidth+47\newtblstarfactor}|}{\setlength{\newtblcolwidth}{31\newtblstarfactor+2\tabcolsep+\arrayrulewidth+47\newtblstarfactor+2\tabcolsep+\arrayrulewidth+47\newtblstarfactor+2\tabcolsep+\arrayrulewidth+47\newtblstarfactor+2\tabcolsep+\arrayrulewidth+47\newtblstarfactor}\multirowii[m]{2}{-\newtblcolwidth}{\centering%
$a-b\,V_z-c\,V_z^2$
}}&\multicolumn{2}{m{31\newtblstarfactor+2\tabcolsep+\arrayrulewidth+31\newtblstarfactor}|}{\centering%
48240.6
}&\multicolumn{2}{m{31\newtblstarfactor+2\tabcolsep+\arrayrulewidth+31\newtblstarfactor}|}{\centering%
8.16007
}&\multicolumn{2}{m{31\newtblstarfactor+2\tabcolsep+\arrayrulewidth+16\newtblstarfactor+\arrayrulewidth}}{\centering%
2.5618
}\tabularnewline
\multicolumn{1}{m{31\newtblstarfactor}|}{\centering%
39
}&\multicolumn{5}{m{31\newtblstarfactor+2\tabcolsep+\arrayrulewidth+47\newtblstarfactor+2\tabcolsep+\arrayrulewidth+47\newtblstarfactor+2\tabcolsep+\arrayrulewidth+47\newtblstarfactor+2\tabcolsep+\arrayrulewidth+47\newtblstarfactor}|}{\setlength{\newtblcolwidth}{31\newtblstarfactor+2\tabcolsep+\arrayrulewidth+47\newtblstarfactor+2\tabcolsep+\arrayrulewidth+47\newtblstarfactor+2\tabcolsep+\arrayrulewidth+47\newtblstarfactor+2\tabcolsep+\arrayrulewidth+47\newtblstarfactor}\multirowii[m]{2}{\newtblcolwidth}{\centering%
$a\big/\left(1+b(V_z+c)^2\right)$
}}&\multicolumn{2}{m{31\newtblstarfactor+2\tabcolsep+\arrayrulewidth+31\newtblstarfactor}|}{\centering%
1.67147e-{}3
}&\multicolumn{2}{m{31\newtblstarfactor+2\tabcolsep+\arrayrulewidth+31\newtblstarfactor}|}{\centering%
8.05854e-{}4
}&\multicolumn{2}{m{31\newtblstarfactor+2\tabcolsep+\arrayrulewidth+16\newtblstarfactor+\arrayrulewidth}}{\centering%
-{}2.00757
}\tabularnewline
\multicolumn{1}{m{31\newtblstarfactor}|}{\centering%
62.4
}&\multicolumn{5}{m{31\newtblstarfactor+2\tabcolsep+\arrayrulewidth+47\newtblstarfactor+2\tabcolsep+\arrayrulewidth+47\newtblstarfactor+2\tabcolsep+\arrayrulewidth+47\newtblstarfactor+2\tabcolsep+\arrayrulewidth+47\newtblstarfactor}|}{\setlength{\newtblcolwidth}{31\newtblstarfactor+2\tabcolsep+\arrayrulewidth+47\newtblstarfactor+2\tabcolsep+\arrayrulewidth+47\newtblstarfactor+2\tabcolsep+\arrayrulewidth+47\newtblstarfactor+2\tabcolsep+\arrayrulewidth+47\newtblstarfactor}\multirowii[m]{2}{-\newtblcolwidth}{\centering%
$a\big/\left(1+b(V_z+c)^2\right)$
}}&\multicolumn{2}{m{31\newtblstarfactor+2\tabcolsep+\arrayrulewidth+31\newtblstarfactor}|}{\centering%
1.77048e-{}3
}&\multicolumn{2}{m{31\newtblstarfactor+2\tabcolsep+\arrayrulewidth+31\newtblstarfactor}|}{\centering%
1.04252e-{}3
}&\multicolumn{2}{m{31\newtblstarfactor+2\tabcolsep+\arrayrulewidth+16\newtblstarfactor+\arrayrulewidth}}{\centering%
1.36677
}\tabularnewline
\multicolumn{12}{m{31\newtblstarfactor+2\tabcolsep+\arrayrulewidth+31\newtblstarfactor+2\tabcolsep+\arrayrulewidth+47\newtblstarfactor+2\tabcolsep+\arrayrulewidth+47\newtblstarfactor+2\tabcolsep+\arrayrulewidth+47\newtblstarfactor+2\tabcolsep+\arrayrulewidth+47\newtblstarfactor+2\tabcolsep+\arrayrulewidth+31\newtblstarfactor+2\tabcolsep+\arrayrulewidth+31\newtblstarfactor+2\tabcolsep+\arrayrulewidth+31\newtblstarfactor+2\tabcolsep+\arrayrulewidth+31\newtblstarfactor+2\tabcolsep+\arrayrulewidth+31\newtblstarfactor+2\tabcolsep+\arrayrulewidth+16\newtblstarfactor+\arrayrulewidth}}{\centering%
\textbf{V$_{\text{z}}$ intervals (in cm)}
}\tabularnewline
\multicolumn{1}{m{31\newtblstarfactor}|}{\centering%
19.6
}&\multicolumn{1}{m{31\newtblstarfactor}|}{\centering%
-{}70
}&\multicolumn{1}{m{47\newtblstarfactor}|}{\centering%
-{}54.88
}&\multicolumn{1}{m{47\newtblstarfactor}|}{\centering%
-{}40.88
}&\multicolumn{1}{m{47\newtblstarfactor}|}{\centering%
-{}27.72
}&\multicolumn{1}{m{47\newtblstarfactor}|}{\centering%
-{}14.84
}&\multicolumn{1}{m{31\newtblstarfactor}|}{\centering%
-{}1.96
}&\multicolumn{1}{m{31\newtblstarfactor}|}{\centering%
10.92
}&\multicolumn{1}{m{31\newtblstarfactor}|}{\centering%
24.36
}&\multicolumn{1}{m{31\newtblstarfactor}|}{\centering%
38.64
}&\multicolumn{1}{m{31\newtblstarfactor}|}{\centering%
54.32
}&\multicolumn{1}{m{16\newtblstarfactor+\arrayrulewidth}}{\centering%
70
}\tabularnewline
\multicolumn{1}{m{31\newtblstarfactor}|}{\centering%
27
}&\multicolumn{1}{m{31\newtblstarfactor}|}{\centering%
-{}70
}&\multicolumn{1}{m{47\newtblstarfactor}|}{\centering%
-{}54.04
}&\multicolumn{1}{m{47\newtblstarfactor}|}{\centering%
-{}39.48
}&\multicolumn{1}{m{47\newtblstarfactor}|}{\centering%
-{}25.76
}&\multicolumn{1}{m{47\newtblstarfactor}|}{\centering%
-{}12.6
}&\multicolumn{1}{m{31\newtblstarfactor}|}{\centering%
0.28
}&\multicolumn{1}{m{31\newtblstarfactor}|}{\centering%
13.16
}&\multicolumn{1}{m{31\newtblstarfactor}|}{\centering%
26.32
}&\multicolumn{1}{m{31\newtblstarfactor}|}{\centering%
40.04
}&\multicolumn{1}{m{31\newtblstarfactor}|}{\centering%
54.88
}&\multicolumn{1}{m{16\newtblstarfactor+\arrayrulewidth}}{\centering%
70
}\tabularnewline
\multicolumn{1}{m{31\newtblstarfactor}|}{\centering%
39
}&\multicolumn{1}{m{31\newtblstarfactor}|}{\centering%
-{}40
}&\multicolumn{1}{m{47\newtblstarfactor}|}{\centering%
-{}27.84
}&\multicolumn{1}{m{47\newtblstarfactor}|}{\centering%
-{}18.72
}&\multicolumn{1}{m{47\newtblstarfactor}|}{\centering%
-{}11.2
}&\multicolumn{1}{m{47\newtblstarfactor}|}{\centering%
-{}4.64
}&\multicolumn{1}{m{31\newtblstarfactor}|}{\centering%
1.44
}&\multicolumn{1}{m{31\newtblstarfactor}|}{\centering%
7.52
}&\multicolumn{1}{m{31\newtblstarfactor}|}{\centering%
13.92
}&\multicolumn{1}{m{31\newtblstarfactor}|}{\centering%
21.12
}&\multicolumn{1}{m{31\newtblstarfactor}|}{\centering%
29.76
}&\multicolumn{1}{m{16\newtblstarfactor+\arrayrulewidth}}{\centering%
40
}\tabularnewline
\multicolumn{1}{m{31\newtblstarfactor}|}{\centering%
62.4
}&\multicolumn{1}{m{31\newtblstarfactor}|}{\centering%
-{}40
}&\multicolumn{1}{m{47\newtblstarfactor}|}{\centering%
-{}28
}&\multicolumn{1}{m{47\newtblstarfactor}|}{\centering%
-{}19.36
}&\multicolumn{1}{m{47\newtblstarfactor}|}{\centering%
-{}12.32
}&\multicolumn{1}{m{47\newtblstarfactor}|}{\centering%
-{}6.24
}&\multicolumn{1}{m{31\newtblstarfactor}|}{\centering%
-{}0.48
}&\multicolumn{1}{m{31\newtblstarfactor}|}{\centering%
5.28
}&\multicolumn{1}{m{31\newtblstarfactor}|}{\centering%
11.52
}&\multicolumn{1}{m{31\newtblstarfactor}|}{\centering%
18.88
}&\multicolumn{1}{m{31\newtblstarfactor}|}{\centering%
28.16
}&\multicolumn{1}{m{16\newtblstarfactor+\arrayrulewidth}}{\centering%
40
}\tabularnewline
\hline
\end{longtable}\endgroup%

\end{center}

\wrapifneeded{0.50}{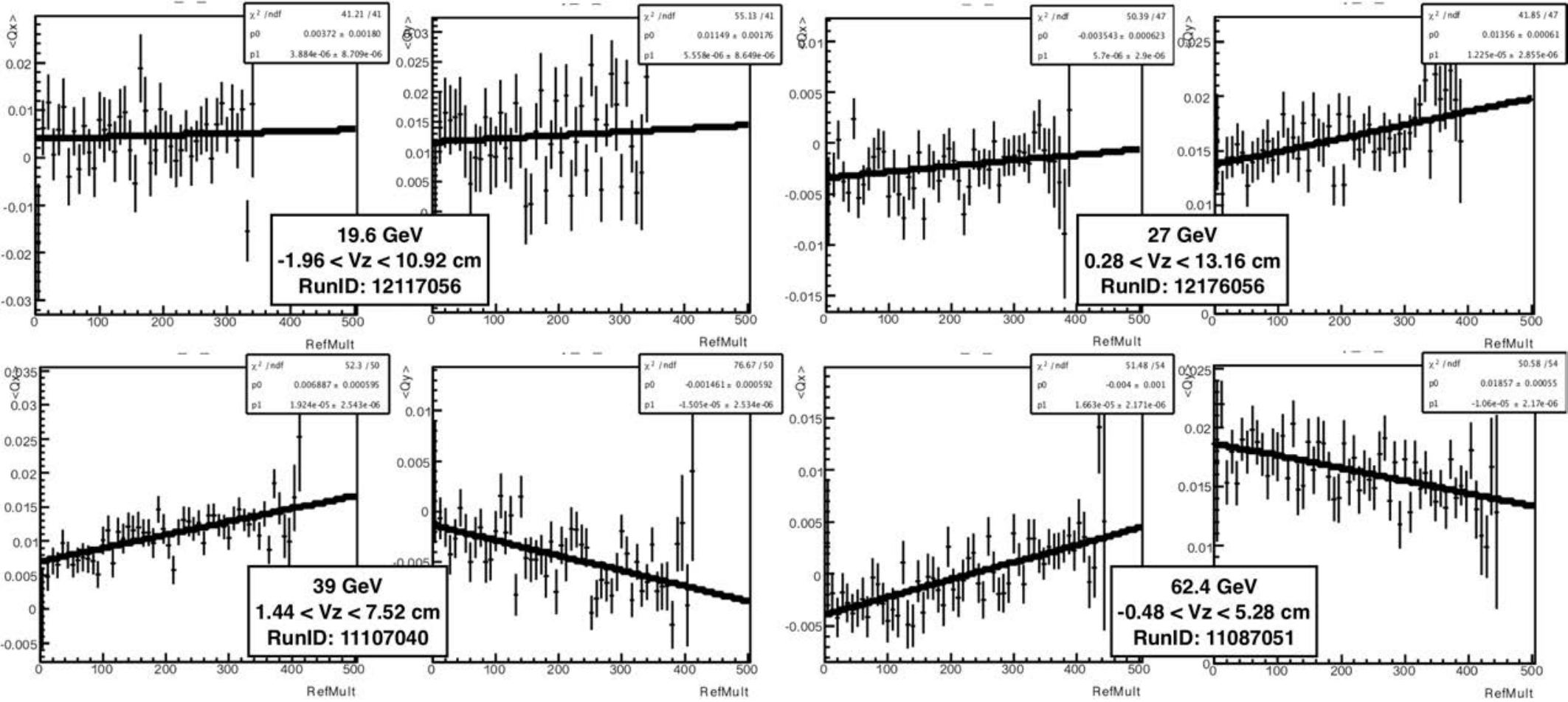}{Typical reference multiplicity dependencies of average raw event flow vector coordinates \emph{Q$_{\text{x,y}}$} at all BES-{}I energies for specific V$_{\text{z}}$ intervals and RunIDs. Also shown are linear fits and their parameters which are used for the recentering correction of the raw Q-{}vector.}{evtplane_fig_recenter}{1} %

Finally, any residual correlations remaining in the recentered \ensuremath{\Psi}$_{\text{2}}$ distribution can be forced flat on average via the \emph{Shift Correction} method [9, 172]: The \ensuremath{\Psi}$_{\text{2,rc}}$ distribution averaged over the events in each
combination of V$_{\text{z}}$ interval and RunID is fit with a second order Fourier
expansion
\[\mathrm{d}N/\mathrm{d}\Psi_\mathrm{2,rc}\propto 1+2\left( p_1\cos(2\Psi_\mathrm{2,rc}) + p_2\cos(4\Psi_\mathrm{2,rc}) +p_3\sin(2\Psi_\mathrm{2,rc}) + p_4\sin(4\Psi_\mathrm{2,rc})\right)\]
from which the fully corrected event plane angle
$\Psi_\mathrm{2,corr}=\Psi_\mathrm{2,rc}+\Psi_\mathrm{2,shift}$ can
be obtained using
\[\Psi_\mathrm{2,shift} = -p_3\cos(2\Psi_\mathrm{2,rc}) + p_1\sin(2\Psi_\mathrm{2,rc}) +0.5\left(-p_4\cos(4\Psi_\mathrm{2,rc})+p_2\sin(4\Psi_\mathrm{2,rc})\right).\]
Figure~\ref{evtplane_fig_shift} depicts typical second order Fourier expansion
parameterizations of the \ensuremath{\Psi}$_{\text{2,rc}}$ distributions at all BES-{}I energies.
Figure~\ref{evtplane_fig_check} compares the according raw event plane angle
distributions to the recentered and shift-{}corrected ones proving the successful
flattening of the averaged \ensuremath{\Psi}$_{\text{2}}$ distributions. The latter can hence be
split into several \ensuremath{\Psi}$_{\text{2}}$ intervals and the corrections used on the raw event
plane angles in Section~\ref{ana_sec_pairrec} to categorize similar events in the event
mixing technique.

\wrapifneeded{0.50}{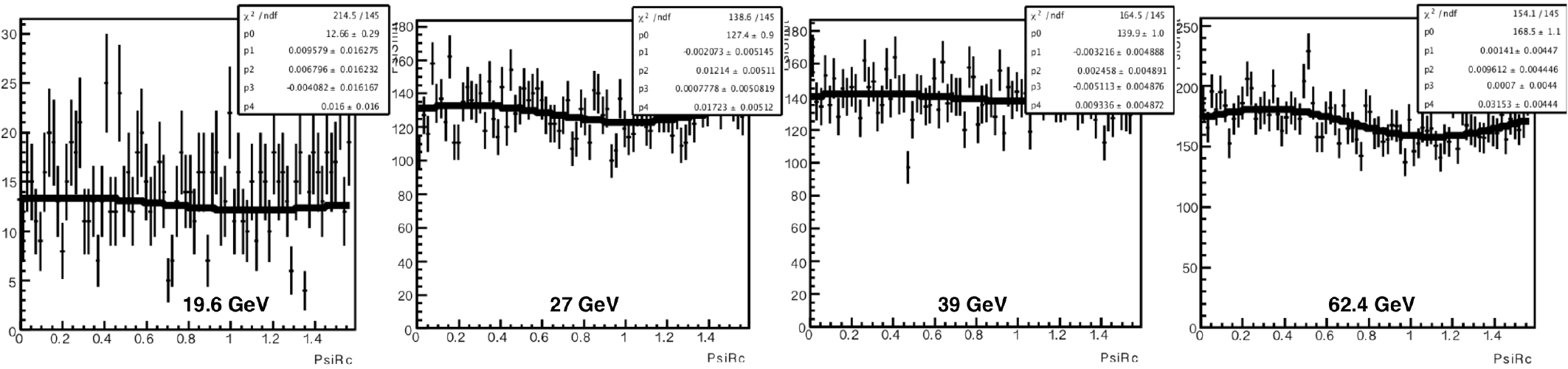}{Typical second order Fourier expansion parameterizations of the \ensuremath{\Psi}$_{\text{2,rc}}$ (PsiRc) distributions at all BES-{}I energies for the same V$_{\text{z}}$ intervals and RunIDs as in Figure~\ref{evtplane_fig_recenter}.}{evtplane_fig_shift}{1} %

\wrapifneeded{0.50}{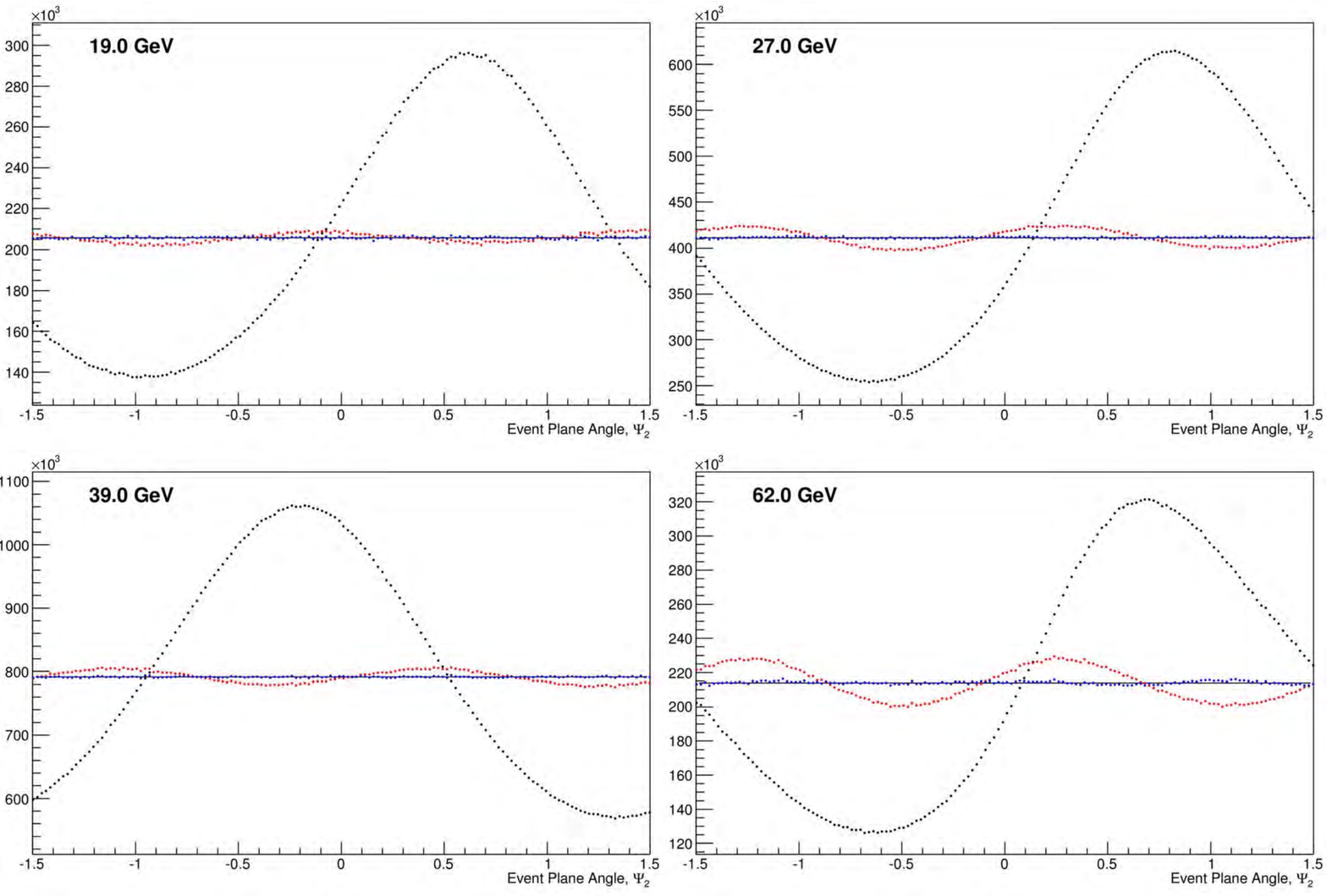}{Comparison of raw (black), recentered (red), and shift-{}corrected (blue) \ensuremath{\Psi}$_{\text{2}}$ distributions at all BES-{}I energies. The black solid line represents a constant fit to guide the reader's eye.}{evtplane_fig_check}{0.8} %

\section{Single-{}Electron/Positron Samples}
\label{ana_sec_pid}\hyperlabel{ana_sec_pid}%

So far, suitable event and track selection criteria (Table~\ref{ana_tab_dsets_evttrk})
have been applied on the BES-{}I datasets derived from the according STAR DSTs
[174]. This first raw data analysis step has also included
run-{}by-{}run quality assurance (Figure~\ref{dsets_evttrk_fig_runqa}) and event plane
reconstruction to provide improved event classification in Section~\ref{ana_sec_pairrec} via a uniformly distributed event plane angle \ensuremath{\Psi}$_{\text{2}}$ (Figure~\ref{evtplane_fig_check}). Next, the resulting \emph{good} tracks in an accepted
event need to be reduced to the particle species of interest in the particular
data analysis. In the context of this thesis and its intended physics results
(Chapter~\ref{results}), the suppression of hadrons facilitated through the usage of the
TOF detector is of crucial importance for the identification of electrons and
positrons.  Section~\ref{ana_subsec_pid_toftpc} describes the procedure with which clean
electron and positron samples are obtained in combination with the particles'
specific energy loss in the TPC. Section~\ref{ana_subsec_pid_purity} studies the samples
in detail revealing their high degree of purity which makes them ideally suited
for the study of dielectron production. Note that this section only includes
selected figures at a single energy and particle charge in order to support the
text whereas the complete set can be found in Section~\ref{app_pid_figures}.

\subsection{Electron/Positron Identification via TOF and TPC}
\label{ana_subsec_pid_toftpc}\hyperlabel{ana_subsec_pid_toftpc}%

In dependence of primary momentum and representatively for
$\sqrt{s_\mathrm{NN}}$ = 39 GeV, Figure~\ref{ana_fig_pid_beta39} shows the
distribution of the inverse velocity for positive particles measured with
respect to the expected positron velocity of about 1. In this representation
the positrons appear as an approximately constant band around zero which allows
for the determination of simplified criteria to select the region containing a
certain fraction of all reconstructed positrons.

\wrapifneeded{0.50}{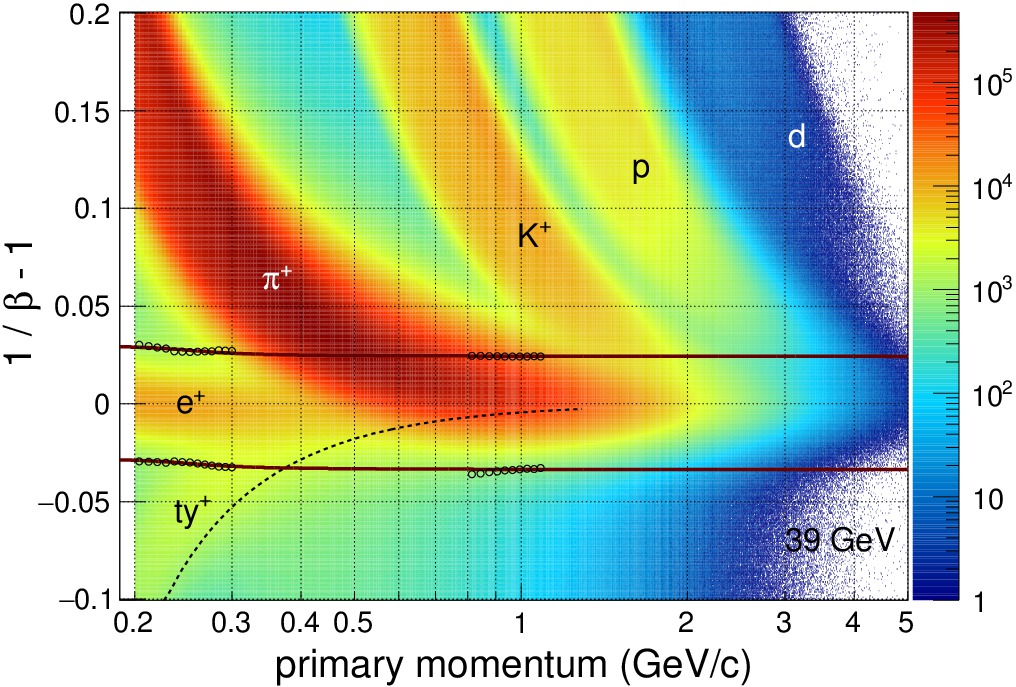}{Measured inverse velocity 1/\ensuremath{\beta} for positive particles as a function of primary momentum at \ensuremath{\surd}s$_{\text{NN}}$ = 39 GeV. Superposed are the velocity limits used to select positrons (solid lines) derived from the band's 3\ensuremath{\sigma} windows below 320 MeV/c and from the according distribution using pions as reference at higher momenta (open black circles). The expected correlation for the tachyonic artefacts (ty$^{\text{+}}$, dashed line) is depicted as well. Details see text. All remaining combinations of beam energy and particle charge are appended in Figure~\ref{app_pid_fig_beta}.}{ana_fig_pid_beta39}{0.49} %

A clear separation of the positrons from slow hadrons up to momenta of about
320 MeV/c is observed beyond which the positrons' band starts merging with the
one from the abundant pions. Together with a separate pion sample at 1 GeV/c,
this region is hence used in the following as part of the procedure to
determine a 3\ensuremath{\sigma} selection criterion which efficiently rejects the dominant
contamination from pions, kaons, protons and deuterons.  When applied on the
inverse velocity distribution, this criterion prepares a set of track
candidates from which the final positron sample can subsequently be extracted
via the TPC energy loss (see Figure~\ref{fig_pidQA}).

Also note the additional band of tracks that appear to be reconstructed in
Figure~\ref{ana_fig_pid_beta39} with \ensuremath{\beta} >{} 1.  This~non-{}physical band of seemingly
faster-{}than-{}light ("tachyonic") particles is observed even though
electrons/positrons must reach the TOF detector before all other detectable
particles. Photons, however, can convert into electron/positron pairs in the
outer TPC field cage or inside the TOF detector itself causing the according
TOF hit to have a photon timing. Due to the occupancy in the detector these TOF
hits are~accidentally~matched to hadron tracks in the TPC resulting in an
artefact of apparently tachyonic particles.

\wrapifneeded{0.50}{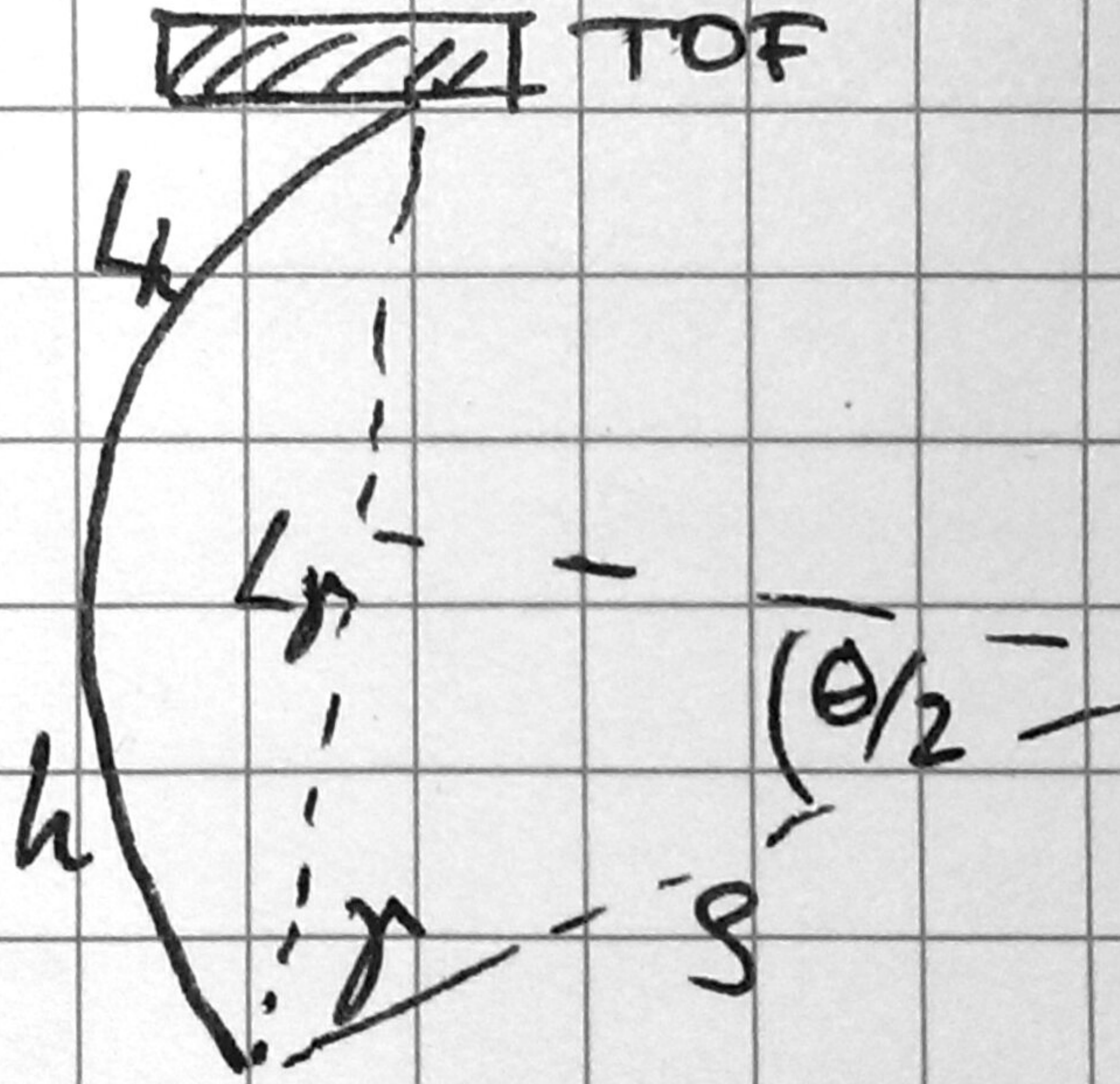}{TOF Tachyon}{ana_fig_pid_tachyon}{0.27} %

The momentum dependence of this
detector~peculiarity can~be derived through simple geometrical considerations
outlined briefly in the following and depicted in Figure~\ref{ana_fig_pid_beta39} as a
dashed line to guide the reader's eye.  For the definition of the variables see
Figure~\ref{ana_fig_pid_tachyon}.
The paths of a photon \ensuremath{\gamma} with length \emph{L$_{\text{\ensuremath{\gamma}}}$} and a hadron \emph{h} with
length \emph{L$_{\text{h}}$} meet in the same TOF hit with a timing corresponding to \emph{L$_{\text{h}}$}.
With the photon time-{}of-{}flight \emph{t$_{\text{\ensuremath{\gamma}}}$} = \emph{L$_{\text{\ensuremath{\gamma}}}$/c}, the tachyonic
velocity amounts to $\beta_\mathrm{ty}=L_h/ct_\gamma=L_h/L_\gamma$ where \emph{L$_{\text{h}}$} is related to \emph{L$_{\text{\ensuremath{\gamma}}}$} via
$L_h=\rho\theta=2\rho\arcsin(L_\gamma/2\rho)$ with \ensuremath{\rho} the radius
of the helix's projected circle and $\theta$ the according central
angle of the circular segment. Embedded in a magnetic field \emph{B}, the track's
transverse momentum is $p_T=0.3B\rho$ resulting in
$\beta_\mathrm{ty}=\arcsin(p_T^\ast)/p_T^\ast$ with
$p_T^\ast=0.3BL_\gamma/2p_T$.

The functional form of the 3\ensuremath{\sigma} TOF selection criterion is derived
programmatically [176] through a series of analysis steps
based on the momentum-{}dependent inverse velocity distribution. First,
one-{}dimensional 1/\ensuremath{\beta} distributions are obtained from the projection in
momentum intervals small enough to approximate a gaussian distribution for the
pions. Second, the 1/\ensuremath{\beta} interval with the most entries is identified as the
pion peak in the respective momentum bin and the distribution parameterized
with a gaussian in the peak's vicinity. Third, the resulting gaussian is
subtracted from the experimental 1/\ensuremath{\beta} distribution to remove contributions
from the pion tail to the positron band.  Fourth, the slopes of the entire
remaining distribution are determined bin-{}by-{}bin through linear interpolation
using five points at a time. Fifth, a sign change in the slopes and exception
of the tachyonic artefact then reveals the positron peak, the position of which
is used to determine the range limits of its gaussian parameterization. The
calculation of the two 3\ensuremath{\sigma} endpoints for the latter results in the
positions of the solid open circles below 320 MeV/c in Figure~\ref{ana_fig_pid_beta39}.
For the determination of the positron band's gaussian widths around 1 GeV/c the
1/\ensuremath{\beta}-{}1/\ensuremath{\beta}$_{\text{\ensuremath{\pi}}}$~representation is used assuming particle species
independence in the TOF measurement. Here 1/\ensuremath{\beta}$_{\text{\ensuremath{\pi}}}$ denotes the inverse
velocity expected for a particle with pion mass at the respective primary
momentum. The two regions are combined using a \ensuremath{\chi}$^{\text{2}}$ minimization of the
momentum-{}dependent selection functional
\[\frac{1}{\beta(p)}-1= A\cdot\left(\frac{\pi}{2}-\arctan\big(B\cdot(p-C)\big)\right)+ \begin{cases}D,&\mbox{upper limit}\\d,&\mbox{lower limit}\end{cases}\]
to the data points of the experimental 3\ensuremath{\sigma} windows. This allows for a
constant 3\ensuremath{\sigma}~selection around the positron distribution across all
momenta. Table~\ref{ana_tab_tofcut} compiles the parameters resulting from the above
fit of the TOF selection functional. The preceding arguments, conclusions and
procedure to extract positrons via TOF with a probability of 99.73\%\footnote{
at
the time of this writing, systematic uncertainties on the TOF selection have
not been taken into account, yet.
} are also valid and hence employed for other
particles and energies as apparent from Figure~\ref{app_pid_fig_beta}.
\begin{center}
\begingroup%
\setlength{\newtblsparewidth}{1\linewidth-2\tabcolsep-2\tabcolsep-2\tabcolsep-2\tabcolsep-2\tabcolsep-2\tabcolsep-2\tabcolsep}%
\setlength{\newtblstarfactor}{\newtblsparewidth / \real{426}}%

\begin{longtable}{llllll}\caption[{Parameters of electron/positron TOF selection functional.}]{Parameters of electron/positron TOF selection functional.\label{ana_tab_tofcut}\hyperlabel{ana_tab_tofcut}%
}\tabularnewline
\endfirsthead
\caption[]{(continued)}\tabularnewline
\endhead
\hline
\multicolumn{1}{m{71\newtblstarfactor}|}{\centering%
}&\multicolumn{1}{m{71\newtblstarfactor}|}{\centering%
A
}&\multicolumn{1}{m{71\newtblstarfactor}|}{\centering%
B
}&\multicolumn{1}{m{71\newtblstarfactor}|}{\centering%
C
}&\multicolumn{1}{m{71\newtblstarfactor}|}{\centering%
D
}&\multicolumn{1}{m{71\newtblstarfactor+\arrayrulewidth}}{\centering%
d
}\tabularnewline
\multicolumn{1}{m{71\newtblstarfactor}|}{\centering%
19.6 GeV
}&\multicolumn{1}{m{71\newtblstarfactor}|}{\centering%
2.322e-{}3
}&\multicolumn{1}{m{71\newtblstarfactor}|}{\centering%
16.2
}&\multicolumn{1}{m{71\newtblstarfactor}|}{\centering%
0.2458
}&\multicolumn{1}{m{71\newtblstarfactor}|}{\centering%
2.459e-{}2
}&\multicolumn{1}{m{71\newtblstarfactor+\arrayrulewidth}}{\centering%
-{}3.426e-{}2
}\tabularnewline
\multicolumn{1}{m{71\newtblstarfactor}|}{\centering%
27 GeV
}&\multicolumn{1}{m{71\newtblstarfactor}|}{\centering%
3.078e-{}3
}&\multicolumn{1}{m{71\newtblstarfactor}|}{\centering%
17.04
}&\multicolumn{1}{m{71\newtblstarfactor}|}{\centering%
0.2383
}&\multicolumn{1}{m{71\newtblstarfactor}|}{\centering%
1.693e-{}2
}&\multicolumn{1}{m{71\newtblstarfactor+\arrayrulewidth}}{\centering%
-{}2.627e-{}2
}\tabularnewline
\multicolumn{1}{m{71\newtblstarfactor}|}{\centering%
39 GeV
}&\multicolumn{1}{m{71\newtblstarfactor}|}{\centering%
2.0785e-{}3
}&\multicolumn{1}{m{71\newtblstarfactor}|}{\centering%
18.3
}&\multicolumn{1}{m{71\newtblstarfactor}|}{\centering%
0.2478
}&\multicolumn{1}{m{71\newtblstarfactor}|}{\centering%
2.429e-{}2
}&\multicolumn{1}{m{71\newtblstarfactor+\arrayrulewidth}}{\centering%
-{}3.361e-{}2
}\tabularnewline
\multicolumn{1}{m{71\newtblstarfactor}|}{\centering%
62.4 GeV
}&\multicolumn{1}{m{71\newtblstarfactor}|}{\centering%
2.158e-{}3
}&\multicolumn{1}{m{71\newtblstarfactor}|}{\centering%
12.56
}&\multicolumn{1}{m{71\newtblstarfactor}|}{\centering%
0.2265
}&\multicolumn{1}{m{71\newtblstarfactor}|}{\centering%
1.739e-{}2
}&\multicolumn{1}{m{71\newtblstarfactor+\arrayrulewidth}}{\centering%
-{}4.174e-{}2
}\tabularnewline
\hline
\end{longtable}\endgroup%

\end{center}

To conclude the identification of electrons and positrons, a momentum-{}dependent
selection criterion to be subsequently applied on the energy loss measured in
the TPC is developed in the following. For better handling of dE/dx selection
windows, the energy loss is transformed into
\[n\sigma_\mathrm{el}\propto\ln\left(\left.\frac{dE}{dx}\right|_\mathrm{meas} \Bigg/\left.\frac{dE}{dx}\right|_\mathrm{el}\right)\]
describing the logarithmic deviation of the measured from the expected dE/dx of
electrons/positrons in units of the dE/dx distribution's width \footnote{
due to
TPC mis-{}calibration at 27 GeV, a n\ensuremath{\sigma}$_{\text{el}}$ correction factor of 1.9 is taken
into account in Figure~\ref{app_pid_fig_dedx1}.
}. By means of this
representation, STAR's new particle identification (PID) capabilities are
illustrated in Figure~\ref{fig_pidQA} by comparing the momentum dependence of
n\ensuremath{\sigma}$_{\text{el}}$ distributions for positive particles at
$\sqrt{s_\mathrm{NN}}$ = 39 GeV with and without using the TOF
selection functional of Figure~\ref{ana_fig_pid_beta39}. Without employing TOF to reduce
hadron contamination, the positron band is effectively buried underneath the
substantial overlaps of pion, kaon, and proton energy losses below 1.5 GeV/c.
Combining TOF and TPC for particle identification purposes, however, now also
allows STAR to cleanly identify electrons and positrons down to momenta as low
as 0.2 GeV/c.  The combined PID continues to work well up to the high momenta
at which the magnetic calorimeters provide good energy resolution for
$e^+/e^-$ identification. Unlike elementary collisions (see Figure
47 in [150]), the well separated pions and positrons/electrons are
not the only bands visible in the n\ensuremath{\sigma}$_{\text{el}}$ distributions. Due to the high
occupancy in heavy-{}ion collisions, hadrons in the tachyon band above about 280
MeV/c survive the TOF selection in Figure~\ref{ana_fig_pid_beta39} and cause residual
contamination from kaons and protons as well as so-{}called \emph{merged} pions in
Figure~\ref{fig_pidQA} (right). The latter is caused by close-{}by pion pairs, the tracks
of which are not resolved in the TPC and thus result in approximately double
the energy loss with a momentum dependence resembling single pions.

\wrapifneeded{0.50}{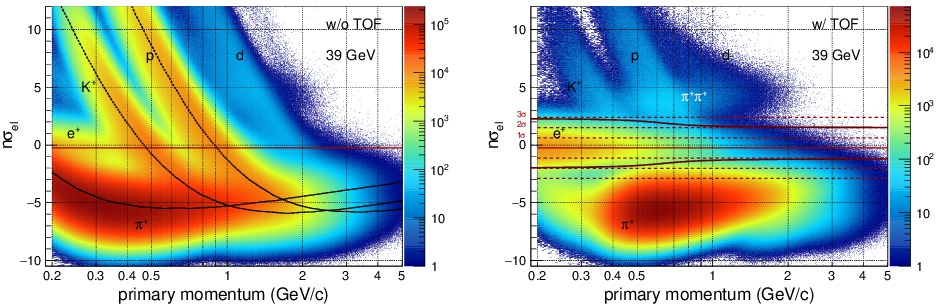}{Distributions of TPC energy loss relative to the expected positron dE/dx for positive particles at \ensuremath{\surd}s$_{\text{NN}}$ = 39 GeV in dependence of primary momentum and in units of the dE/dx width. The two versions shown are n\ensuremath{\sigma}$_{\text{el}}$ distributions without (left) and with (right) a TOF requirement selecting the positron 1/\ensuremath{\beta} band (see Figure~\ref{ana_fig_pid_beta39}). Horizontal red solid lines correspond to the positrons' momentum-{}integrated (global) gaussian mean \ensuremath{\mu}$_{\text{gl}}$ (Table~\ref{effcorr_tab_tpcselect}) derived from the separate positron sample of Section~\ref{effcorr_sec_samples}. Dashed horizontal lines denote 1/2/3\ensuremath{\sigma}-{}windows based on the \ensuremath{\sigma}$_{\text{opt}}$ widths in Table~\ref{effcorr_tab_tpcselect}. Also depicted are the adjusted Bichsel functions of Section~\ref{ana_subsec_pid_purity} (black lines in left figure) and the TPC selection functional (red solid lines in right figure). The procedure to select e$^{\text{+}}$/e$^{\text{-{}}}$ via their TPC energy loss is very similar for all BES-{}I energies (see Figure~\ref{app_pid_fig_dedx1} and Figure~\ref{app_pid_fig_dedx2}).}{fig_pidQA}{1} %

Horizontal solid and dashed lines in Figure~\ref{fig_pidQA} indicate gaussian mean as
well as 1\ensuremath{\sigma}, 2\ensuremath{\sigma} and 3\ensuremath{\sigma} windows. Mean and width are based on the separate analysis
of pure $e^+/e^-$ samples from photon conversions and Dalitz decays in
Section~\ref{effcorr_sec_samples} and Table~\ref{effcorr_tab_tpcselect}. The windows are used
as guidelines for the definition of the momentum-{}dependent criterion on the
TPC energy loss to select electrons and positrons. Similar to the TOF selection
functional, the resulting TPC dE/dx selection functional basically consist of a
common bare part plus an offset for the lower and upper limits, respectively:
\[n\sigma_\mathrm{el}(p)=\pm\frac{A}{\pi}\arctan(5p-3)+ \begin{cases}D,&\mbox{lower limit}\\d+A/2,&\mbox{upper limit}\end{cases}\]
with \emph{A} = 0.9 and the values of the parameters \emph{D} and \emph{d} listed in
Table~\ref{ana_tab_tpccut}. The upper limit at every BES-{}I energy is tuned to select
about 3\ensuremath{\sigma} of the electrons/positrons at very low momenta followed by a
smooth transition down to about 2\ensuremath{\sigma} across the region in which the
residual kaons and proton traverse the $e^+/e^-$ band. However, the
multiples of dE/dx width for the lower limit are chosen to transition from 2
down to 1 due to the significant potential of pion contamination into the
n\ensuremath{\sigma}$_{\text{el}}$ <{} 0 tail of the $e^+/e^-$ distributions. The figures
containing n\ensuremath{\sigma}$_{\text{el}}$ distributions and the resulting TPC selection functions
are included for all BES-{}I energies in Figure~\ref{app_pid_fig_dedx1} and
Figure~\ref{app_pid_fig_dedx2}.

\wrapifneeded{0.50}{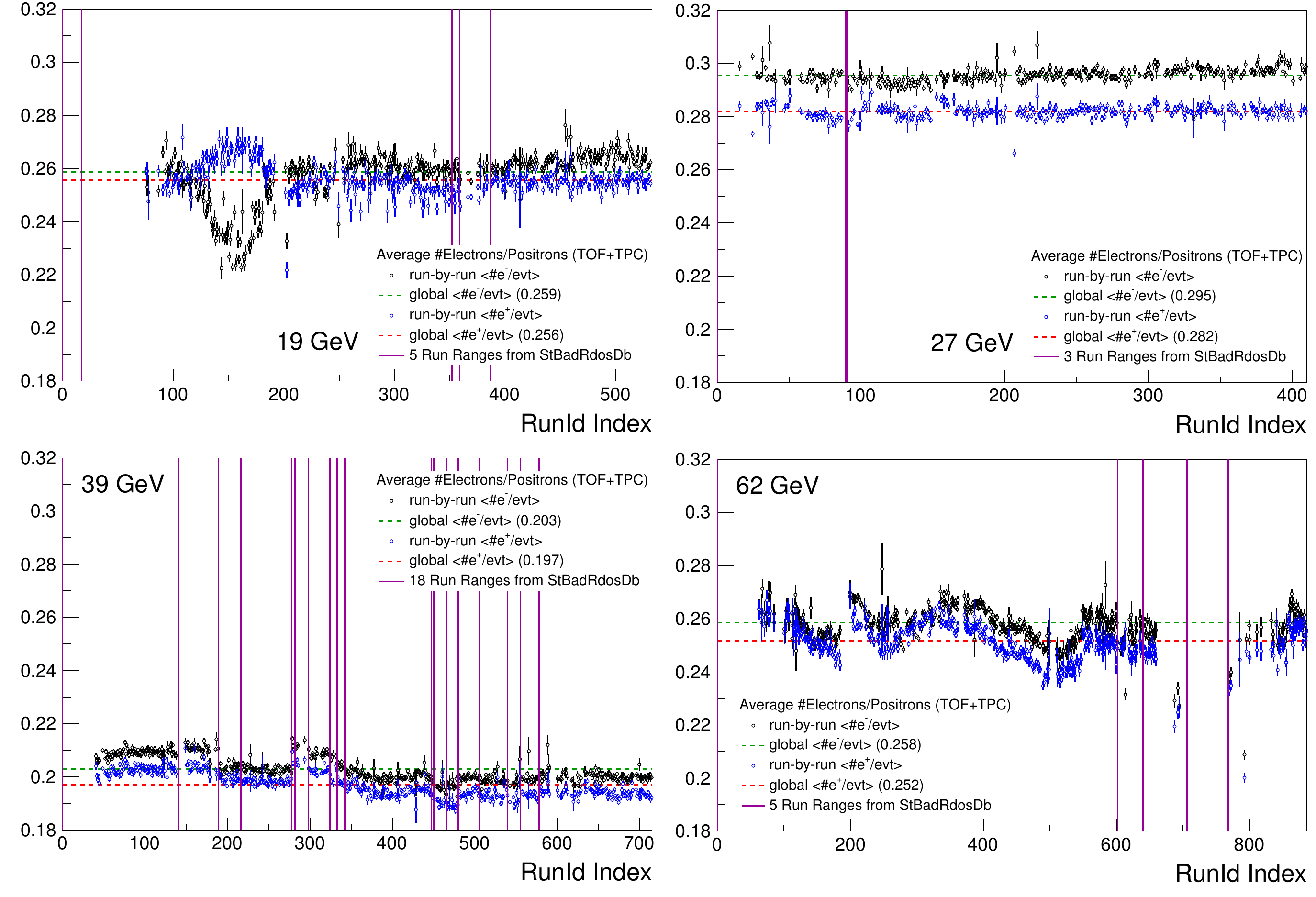}{Run-{}by-{}run averaged single electron (black open circles) and positron (blue open circles) yields per event in dependence of RunId index (Section~\ref{strunideventsdb}) for all BES-{}I energies. Also depicted are number of e$^{\text{-{}}}$ (green dashed line) and e$^{\text{+}}$ (red dashed line) averaged over the entire run. Vertical magenta lines indicate run ranges with the same number of missing RDOs (Section~\ref{stbadrdos}).}{ana_fig_pid_nepem}{1} %
\begin{center}
\begingroup%
\setlength{\newtblsparewidth}{1\linewidth-2\tabcolsep-2\tabcolsep-2\tabcolsep-2\tabcolsep-2\tabcolsep-2\tabcolsep-2\tabcolsep-2\tabcolsep-2\tabcolsep}%
\setlength{\newtblstarfactor}{\newtblsparewidth / \real{426}}%

\begin{longtable}{llllllll}\caption[{Parameters of electron/positron TPC dE/dx selection functional.}]{Parameters of electron/positron TPC dE/dx selection functional.\label{ana_tab_tpccut}\hyperlabel{ana_tab_tpccut}%
}\tabularnewline
\endfirsthead
\caption[]{(continued)}\tabularnewline
\endhead
\hline
\multicolumn{1}{m{24\newtblstarfactor+\arrayrulewidth}}{\centering%
}&\multicolumn{1}{m{71\newtblstarfactor+\arrayrulewidth}}{\centering%
19.6/39 GeV
}&\multicolumn{1}{m{47\newtblstarfactor+\arrayrulewidth}}{\centering%
27 GeV
}&\multicolumn{1}{m{71\newtblstarfactor+\arrayrulewidth}}{\centering%
62.4 GeV
}&\multicolumn{1}{m{24\newtblstarfactor+\arrayrulewidth}}{\centering%
}&\multicolumn{1}{m{71\newtblstarfactor+\arrayrulewidth}}{\centering%
19.6/39 GeV
}&\multicolumn{1}{m{47\newtblstarfactor+\arrayrulewidth}}{\centering%
27 GeV
}&\multicolumn{1}{m{71\newtblstarfactor+\arrayrulewidth}}{\centering%
62.4 GeV
}\tabularnewline
\multicolumn{1}{m{24\newtblstarfactor+\arrayrulewidth}}{\centering%
\emph{D}
}&\multicolumn{1}{m{71\newtblstarfactor+\arrayrulewidth}}{\centering%
-{}1.65
}&\multicolumn{1}{m{47\newtblstarfactor+\arrayrulewidth}}{\centering%
-{}1
}&\multicolumn{1}{m{71\newtblstarfactor+\arrayrulewidth}}{\centering%
-{}1.3
}&\multicolumn{1}{m{24\newtblstarfactor+\arrayrulewidth}}{\centering%
\emph{d}
}&\multicolumn{1}{m{71\newtblstarfactor+\arrayrulewidth}}{\centering%
1.5
}&\multicolumn{1}{m{47\newtblstarfactor+\arrayrulewidth}}{\centering%
0.5
}&\multicolumn{1}{m{71\newtblstarfactor+\arrayrulewidth}}{\centering%
1.7
}\tabularnewline
\hline
\end{longtable}\endgroup%

\end{center}

The electron and positron samples resulting from the combined TOF/TPC PID are
characterized in Figure~\ref{ana_fig_pid_nepem} at all energies by means of their
respective run-{}by-{}run dependent average number per event.
Some of the observed structures at 39 GeV, for instance,
can be linked to the different phases of data taking using the same set of
missing RDOs as indicated by vertical magenta lines. Others, however, e.g. at
19.6 GeV, are related to other possible acceptance effects of the detector. The
number of electrons and positrons averaged over the entire beam time ranges
from 0.2 at 39 GeV, and 0.26 at 19.6/62.4 GeV, up to 0.29 at 27 GeV. Not only
does 27 GeV deliver the highest average number of electrons/positrons per event
but it also appears to be the most stable beam time in terms of changing
detector conditions. Figure~\ref{app_pid_fig_nepem} in the appendix shows the number of
events for a given combination of electrons and positrons reiterating that the
bulk of the events (91-{}96\%) does not contain any electrons/positrons at all,
only one of each or at most a single unlike-{}sign pair. Further, more detailed
studies of samples quality follow in Section~\ref{ana_subsec_pid_purity}.

\subsection{Sample Purities}
\label{ana_subsec_pid_purity}\hyperlabel{ana_subsec_pid_purity}%

The study of hadron contamination levels in the electron/positron samples
generated in Section~\ref{ana_subsec_pid_toftpc} is especially important for the analysis
of dielectron production if a detailed understanding of the sample qualities in
terms of purity is to be obtained. Due to scarcity of these electromagnetic
probes, however, the accurate momentum-{}dependent determination of single hadron
contributions to the electron/positron samples is challenging and complex. A
generic functional form that works well across beam energies and momentum
ranges needs to be derived to parameterize the n\ensuremath{\sigma}$_{\text{el}}$ distributions in
Figure~\ref{fig_pidQA} resulting from the application of electron/positron TOF selection
criteria (see Figure~\ref{ana_fig_pid_beta39}). It is immediately evident, though, that
the multitude of possible contaminations by pions, kaons, protons, and merged
pions causes a high number of floating parameters for the \ensuremath{\chi}$^{\text{2}}$ minimization
of n\ensuremath{\sigma}$_{\text{el}}$ parameterizations. A fully automated procedure to successfully
perform such \emph{multi-{}particle} fits hence necessitates a preceding analysis of
pure hadron samples in the n\ensuremath{\sigma}$_{\text{el}}$-{}vs-{}momentum space that provides a good
handle on their respective functional shapes and allows for most of the
parameters to be either fixed or constrained within reasonable limits. As a
result, an elaborate two-{}prong approach consisting of (\emph{i}) \emph{multi-{}gaussian} fits to pure hadron n\ensuremath{\sigma}$_{\text{el}}$ distributions, and (\emph{ii}) simultaneous
\emph{multi-{}particle} fits to n\ensuremath{\sigma}$_{\text{el}}$ distributions of electron/positron
candidates has been developed. It is presented in detail in the remainder of this subsection
which concludes with momentum-{}dependent purities calculated at all BES
energies. The purities serve as a basis for the important estimate of
systematic uncertainties on dielectron invariant mass spectra caused by hadron
contaminations (see Section~\ref{ana_sec_pairrec}).\newline

{\bf Analysis of pure hadron distributions.} 
The same functional form and parameters as for the electron/positron selection
via TOF (see Table~\ref{ana_tab_tofcut} and accompanying text) are used for the
generation of pure \ensuremath{\pi}/K/p samples. Merely the selection window is reduced to
1\ensuremath{\sigma} around the 1/\ensuremath{\beta}-{}1/\ensuremath{\beta}$_{\text{POI}}$ band (\ensuremath{\beta}$_{\text{POI}}$ = velocity of the
particle of interest) using the offset parameters \emph{d} and \emph{D} at the respective
energy. As discussed in Section~\ref{ana_subsec_pid_toftpc}, a qualitatively sufficient
sample of merged pions (\ensuremath{\pi}\ensuremath{\pi}) appears as part of the pure pion sample at
about twice the single pion energy loss. In Figure~\ref{ana_fig_pid_puresamp}, the
resulting n\ensuremath{\sigma}$_{\text{el}}$ distributions of the pure hadron samples are
subsequently analyzed in dependence of particle momentum. Before starting their
iterative parameterization via \emph{multi-{}gaussian} fits, the initial values for
the first gaussian means and a suitable n\ensuremath{\sigma}$_{\text{el}}$ parameterization range are
determined by studying the momentum dependence of the most probable energy loss
using Bichsel functions [145, 146]
(Figure~\ref{ana_fig_pid_puresamp} left column). The peak of the projected n\ensuremath{\sigma}$_{\text{el}}$ distribution in each momentum bin is taken as the most probable energy loss
(solid crosses) and its momentum dependence parameterized (solid red line) via

\[n\sigma_\mathrm{el}^\mathrm{POI}=a\ln\left[ \hspace{1mm}dE/dx(m_\mathrm{POI})\hspace{1mm}/ \hspace{1mm}dE/dx(m_\mathrm{e})\hspace{1mm} \right]+b\]
with \emph{a} the proportionality factor and \emph{dE/dx} the Bichsel function depending
on e$^{\text{+}}$/e$^{\text{-{}}}$ mass \emph{m}$_{\text{e}}$ and hadron mass \emph{m}$_{\text{POI}}$. The shift parameter \emph{b} is
used to set the upper and lower range limits (dashed red lines). For merged
pions, \emph{a} and \emph{m}$_{\text{POI}}$ are fixed to the final parameters of the pion fit with
the offset initiated to $b=a\ln{2}$. Table~\ref{ana_tab_bichsel_fit} lists
the initial and final parameters of all n\ensuremath{\sigma}$_{\text{el}}$ functions.

\wrapifneeded{0.50}{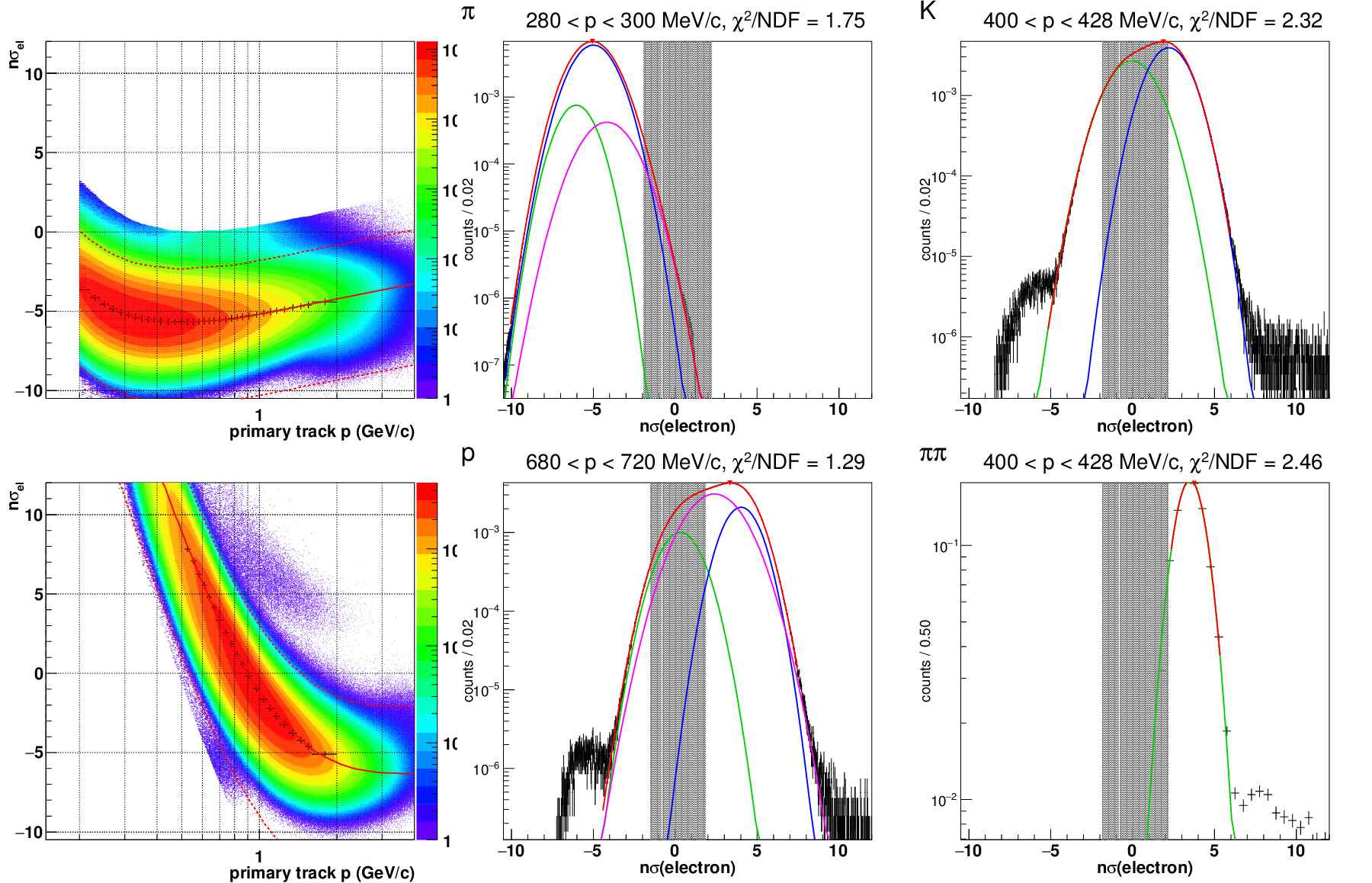}{(left column) Momentum-{}dependent n\ensuremath{\sigma}$_{\text{el}}$ distributions for pure pion (top) and proton (bottom) samples obtained through the according 1\ensuremath{\sigma} TOF selection on 1/\ensuremath{\beta}-{}1/\ensuremath{\beta}$_{\text{POI}}$. The peak positions in each momentum bin (black crosses) are parameterized with Bichsel energy loss functions [145, 146] (red solid line) which in turn are offset to obtain reasonable parameter limits for the \emph{multi-{}gaussian} fits. (right two columns) Representative selection of \emph{multi-{}gaussian} fits to n\ensuremath{\sigma}$_{\text{el}}$ distributions of pure \ensuremath{\pi}, K, p, and merged-{}\ensuremath{\pi} (\ensuremath{\pi}\ensuremath{\pi}) samples for different momentum ranges at \ensuremath{\surd}s$_{\text{NN}}$ = 39 GeV. The pure hadron samples are parameterized using a sum of multiple gaussians (green, blue, magenta) to describe the overall shape (red) sufficiently well (note the indicated reduced \ensuremath{\chi}$^{\text{2}}$). The grey dotted areas denote the TPC selection window defining the electrons/positrons samples at the corresponding momentum (c.f. Figure~\ref{fig_pidQA} right). The full momentum-{}dependent set of \emph{multi-{}gaussian} fits for all hadrons at all BES energies is appended in Section~\ref{app_pid_quality_figures}.}{ana_fig_pid_puresamp}{1.03} %

\pagebreak[4] 

\begin{center}
\begingroup%
\setlength{\newtblsparewidth}{1\linewidth-2\tabcolsep-2\tabcolsep-2\tabcolsep-2\tabcolsep-2\tabcolsep-2\tabcolsep-2\tabcolsep}%
\setlength{\newtblstarfactor}{\newtblsparewidth / \real{424}}%

\begin{longtable}{llllll}\caption[{Initial and final fit parameters of n\ensuremath{\sigma}$_{\text{el}}$ functions. Arrows denote the change in value.}]{Initial and final fit parameters of n\ensuremath{\sigma}$_{\text{el}}$ functions. Arrows denote the change in value.\label{ana_tab_bichsel_fit}\hyperlabel{ana_tab_bichsel_fit}%
}\tabularnewline
\endfirsthead
\caption[]{(continued)}\tabularnewline
\endhead
\hline
\multicolumn{2}{m{14\newtblstarfactor+2\tabcolsep+\arrayrulewidth+14\newtblstarfactor}|}{\centering%
\ensuremath{\surd}s$_{\text{NN}}$
}&\multicolumn{1}{m{99\newtblstarfactor}|}{\centering%
19.6 GeV
}&\multicolumn{1}{m{99\newtblstarfactor}|}{\centering%
27 GeV
}&\multicolumn{1}{m{99\newtblstarfactor}|}{\centering%
39 GeV
}&\multicolumn{1}{m{99\newtblstarfactor+\arrayrulewidth}}{\centering%
62.4 GeV
}\tabularnewline
\multicolumn{1}{m{14\newtblstarfactor}|}{\setlength{\newtblcolwidth}{14\newtblstarfactor}\multirowii[m]{3}{\newtblcolwidth}{\centering%
\ensuremath{\pi}
}}&\multicolumn{1}{m{14\newtblstarfactor}|}{\centering%
\emph{a}
}&\multicolumn{1}{m{99\newtblstarfactor}|}{\centering%
12.5 \ensuremath{\rightarrow} 12.5051
}&\multicolumn{1}{m{99\newtblstarfactor}|}{\centering%
6.579 \ensuremath{\rightarrow} 6.199
}&\multicolumn{1}{m{99\newtblstarfactor}|}{\centering%
12.5 \ensuremath{\rightarrow} 12.2
}&\multicolumn{1}{m{99\newtblstarfactor+\arrayrulewidth}}{\centering%
12.5 \ensuremath{\rightarrow} 12.421
}\tabularnewline
\multicolumn{1}{m{14\newtblstarfactor}|}{\setlength{\newtblcolwidth}{14\newtblstarfactor}\multirowii[m]{3}{-\newtblcolwidth}{\centering%
\ensuremath{\pi}
}}&\multicolumn{1}{m{14\newtblstarfactor}|}{\centering%
\emph{m}
}&\multicolumn{1}{m{99\newtblstarfactor}|}{\centering%
0.13957 \ensuremath{\rightarrow} 0.1235
}&\multicolumn{1}{m{99\newtblstarfactor}|}{\centering%
0.13957 \ensuremath{\rightarrow} 0.12298
}&\multicolumn{1}{m{99\newtblstarfactor}|}{\centering%
0.13957 \ensuremath{\rightarrow} 0.12346
}&\multicolumn{1}{m{99\newtblstarfactor+\arrayrulewidth}}{\centering%
0.13957 \ensuremath{\rightarrow} 0.12379
}\tabularnewline
\multicolumn{1}{m{14\newtblstarfactor}|}{\setlength{\newtblcolwidth}{14\newtblstarfactor}\multirowii[m]{3}{-\newtblcolwidth}{\centering%
\ensuremath{\pi}
}}&\multicolumn{1}{m{14\newtblstarfactor}|}{\centering%
\emph{b}
}&\multicolumn{1}{m{99\newtblstarfactor}|}{\centering%
0 \ensuremath{\rightarrow} -{}0.4446
}&\multicolumn{1}{m{99\newtblstarfactor}|}{\centering%
0 \ensuremath{\rightarrow} -{}0.22654
}&\multicolumn{1}{m{99\newtblstarfactor}|}{\centering%
0 \ensuremath{\rightarrow} -{}0.38356
}&\multicolumn{1}{m{99\newtblstarfactor+\arrayrulewidth}}{\centering%
0 \ensuremath{\rightarrow} -{}0.10062
}\tabularnewline
\multicolumn{1}{m{14\newtblstarfactor}|}{\setlength{\newtblcolwidth}{14\newtblstarfactor}\multirowii[m]{3}{\newtblcolwidth}{\centering%
K
}}&\multicolumn{1}{m{14\newtblstarfactor}|}{\centering%
\emph{a}
}&\multicolumn{1}{m{99\newtblstarfactor}|}{\centering%
12.5 \ensuremath{\rightarrow} 13.1061
}&\multicolumn{1}{m{99\newtblstarfactor}|}{\centering%
6.579 \ensuremath{\rightarrow} 6.5254
}&\multicolumn{1}{m{99\newtblstarfactor}|}{\centering%
12.5 \ensuremath{\rightarrow} 12.3772
}&\multicolumn{1}{m{99\newtblstarfactor+\arrayrulewidth}}{\centering%
12.5 \ensuremath{\rightarrow} 12.7731
}\tabularnewline
\multicolumn{1}{m{14\newtblstarfactor}|}{\setlength{\newtblcolwidth}{14\newtblstarfactor}\multirowii[m]{3}{-\newtblcolwidth}{\centering%
K
}}&\multicolumn{1}{m{14\newtblstarfactor}|}{\centering%
\emph{m}
}&\multicolumn{1}{m{99\newtblstarfactor}|}{\centering%
0.49368 \ensuremath{\rightarrow} 0.45557
}&\multicolumn{1}{m{99\newtblstarfactor}|}{\centering%
0.49368 \ensuremath{\rightarrow} 0.45243
}&\multicolumn{1}{m{99\newtblstarfactor}|}{\centering%
0.49368 \ensuremath{\rightarrow} 0.4542
}&\multicolumn{1}{m{99\newtblstarfactor+\arrayrulewidth}}{\centering%
0.49368 \ensuremath{\rightarrow} 0.45213
}\tabularnewline
\multicolumn{1}{m{14\newtblstarfactor}|}{\setlength{\newtblcolwidth}{14\newtblstarfactor}\multirowii[m]{3}{-\newtblcolwidth}{\centering%
K
}}&\multicolumn{1}{m{14\newtblstarfactor}|}{\centering%
\emph{b}
}&\multicolumn{1}{m{99\newtblstarfactor}|}{\centering%
0 \ensuremath{\rightarrow} -{}0.307
}&\multicolumn{1}{m{99\newtblstarfactor}|}{\centering%
0 \ensuremath{\rightarrow} -{}0.1462
}&\multicolumn{1}{m{99\newtblstarfactor}|}{\centering%
0 \ensuremath{\rightarrow} -{}0.4072
}&\multicolumn{1}{m{99\newtblstarfactor+\arrayrulewidth}}{\centering%
0 \ensuremath{\rightarrow} -{}0.0601
}\tabularnewline
\multicolumn{1}{m{14\newtblstarfactor}|}{\setlength{\newtblcolwidth}{14\newtblstarfactor}\multirowii[m]{3}{\newtblcolwidth}{\centering%
p
}}&\multicolumn{1}{m{14\newtblstarfactor}|}{\centering%
\emph{a}
}&\multicolumn{1}{m{99\newtblstarfactor}|}{\centering%
12.5 \ensuremath{\rightarrow} 11.7627
}&\multicolumn{1}{m{99\newtblstarfactor}|}{\centering%
6.579 \ensuremath{\rightarrow} 5.8754
}&\multicolumn{1}{m{99\newtblstarfactor}|}{\centering%
12.5 \ensuremath{\rightarrow} 11.299
}&\multicolumn{1}{m{99\newtblstarfactor+\arrayrulewidth}}{\centering%
12.5 \ensuremath{\rightarrow} 11.2449
}\tabularnewline
\multicolumn{1}{m{14\newtblstarfactor}|}{\setlength{\newtblcolwidth}{14\newtblstarfactor}\multirowii[m]{3}{-\newtblcolwidth}{\centering%
p
}}&\multicolumn{1}{m{14\newtblstarfactor}|}{\centering%
\emph{m}
}&\multicolumn{1}{m{99\newtblstarfactor}|}{\centering%
0.93827 \ensuremath{\rightarrow} 0.93989
}&\multicolumn{1}{m{99\newtblstarfactor}|}{\centering%
0.93827 \ensuremath{\rightarrow} 0.93538
}&\multicolumn{1}{m{99\newtblstarfactor}|}{\centering%
0.93827 \ensuremath{\rightarrow} 0.92899
}&\multicolumn{1}{m{99\newtblstarfactor+\arrayrulewidth}}{\centering%
0.93827 \ensuremath{\rightarrow}  0.94254
}\tabularnewline
\multicolumn{1}{m{14\newtblstarfactor}|}{\setlength{\newtblcolwidth}{14\newtblstarfactor}\multirowii[m]{3}{-\newtblcolwidth}{\centering%
p
}}&\multicolumn{1}{m{14\newtblstarfactor}|}{\centering%
\emph{b}
}&\multicolumn{1}{m{99\newtblstarfactor}|}{\centering%
0 \ensuremath{\rightarrow} -{}1.2483
}&\multicolumn{1}{m{99\newtblstarfactor}|}{\centering%
0 \ensuremath{\rightarrow} -{}0.6287
}&\multicolumn{1}{m{99\newtblstarfactor}|}{\centering%
0 \ensuremath{\rightarrow} -{}1.1916
}&\multicolumn{1}{m{99\newtblstarfactor+\arrayrulewidth}}{\centering%
0 \ensuremath{\rightarrow} -{}1.0686
}\tabularnewline
\multicolumn{1}{m{14\newtblstarfactor}|}{\setlength{\newtblcolwidth}{14\newtblstarfactor}\multirowii[m]{3}{\newtblcolwidth}{\centering%
\ensuremath{\pi}\ensuremath{\pi}
}}&\multicolumn{1}{m{14\newtblstarfactor}|}{\centering%
\emph{a}
}&\multicolumn{1}{m{99\newtblstarfactor}|}{\centering%
12.5051
}&\multicolumn{1}{m{99\newtblstarfactor}|}{\centering%
6.199
}&\multicolumn{1}{m{99\newtblstarfactor}|}{\centering%
12.2
}&\multicolumn{1}{m{99\newtblstarfactor+\arrayrulewidth}}{\centering%
12.421
}\tabularnewline
\multicolumn{1}{m{14\newtblstarfactor}|}{\setlength{\newtblcolwidth}{14\newtblstarfactor}\multirowii[m]{3}{-\newtblcolwidth}{\centering%
\ensuremath{\pi}\ensuremath{\pi}
}}&\multicolumn{1}{m{14\newtblstarfactor}|}{\centering%
\emph{m}
}&\multicolumn{1}{m{99\newtblstarfactor}|}{\centering%
0.1235
}&\multicolumn{1}{m{99\newtblstarfactor}|}{\centering%
0.12298
}&\multicolumn{1}{m{99\newtblstarfactor}|}{\centering%
0.12346
}&\multicolumn{1}{m{99\newtblstarfactor+\arrayrulewidth}}{\centering%
0.12379
}\tabularnewline
\multicolumn{1}{m{14\newtblstarfactor}|}{\setlength{\newtblcolwidth}{14\newtblstarfactor}\multirowii[m]{3}{-\newtblcolwidth}{\centering%
\ensuremath{\pi}\ensuremath{\pi}
}}&\multicolumn{1}{m{14\newtblstarfactor}|}{\centering%
\emph{b}
}&\multicolumn{1}{m{99\newtblstarfactor}|}{\centering%
8.66784 \ensuremath{\rightarrow} 9.05812
}&\multicolumn{1}{m{99\newtblstarfactor}|}{\centering%
4.29681 \ensuremath{\rightarrow} 4.51977
}&\multicolumn{1}{m{99\newtblstarfactor}|}{\centering%
8.45682 \ensuremath{\rightarrow} 8.82378
}&\multicolumn{1}{m{99\newtblstarfactor+\arrayrulewidth}}{\centering%
8.60922 \ensuremath{\rightarrow} 9.11032
}\tabularnewline
\hline
\end{longtable}\endgroup%

\end{center}

By extrapolation, this initialization phase not only allows access to reliable
estimates of the most probable energy loss as well as reasonable fit ranges
across the entire momentum range. It also ensures sanitized starting parameters
for the following iteration procedure to accurately describe the momentum
dependence of hadron n\ensuremath{\sigma}$_{\text{el}}$ distributions (Figure~\ref{ana_fig_pid_puresamp} right
two columns).

A particle's energy loss measured in the TPC is distributed normal around its
most probably energy loss which results in a gaussian function as valid
description of the rather momentum-{}independent n\ensuremath{\sigma}$_{\text{POI}}$ distributions with
normalized widths of \ensuremath{\sim}1. However, in the representation of a hadron's
energy loss in the n\ensuremath{\sigma}$_{\text{el}}$-{}vs-{}p plane, i.e. with respect to the energy
loss expected from electrons/positrons, its momentum dependence becomes
non-{}constant as illustrated for the protons in Figure~\ref{ana_fig_pid_puresamp} (lower
left). Especially in the regions where n\ensuremath{\sigma}$_{\text{el}}$ falls off steeply with
increasing momentum, several gaussians start contributing to the overall shape
unless momentum intervals are kept unmanagably small. The multiple overlapping
gaussians with approximately equal widths are slightly shifted towards each
other and contribute with different yields to the measured n\ensuremath{\sigma}$_{\text{el}}$ distribution of the hadrons in the respective momentum bin.  Iteratively
increasing the number of gaussians used to fit the pure hadron samples'
n\ensuremath{\sigma}$_{\text{el}}$ distributions until satisfactory agreement is achieved, thus makes
for an intuitive approach to solve this issue and cope with finite momentum bin
widths. The underlying algorithm reiterates a \ensuremath{\chi}$^{\text{2}}$ minimization of an
increasing sum of gaussians until either the reduced \ensuremath{\chi}$^{\text{2}}$ is less than 3.5
(or alternatively, the minimum \ensuremath{\chi}$^{\text{2}}$/NDF is reached), or the number of
gaussians used exceeds 6. The right two columns of Figure~\ref{ana_fig_pid_puresamp} depict representative examples of this procedure's end results for the \ensuremath{\pi}, K,
p and \ensuremath{\pi}\ensuremath{\pi}'s n\ensuremath{\sigma}$_{\text{el}}$ distributions at selected momenta. The full list
of these \emph{multi-{}gaussian} parameterizations at all BES energies are attached in
Section~\ref{app_pid_quality_figures} demonstrating the excellent level of control over
the distributions' shapes which is an essential requirement for the success of
the following simultaneous \emph{multi-{}particle} fits.\newline

{\bf Analysis of electron/positron candidate distributions.} 
The momentum dependence of the experimental n\ensuremath{\sigma}$_{\text{el}}$ distributions for
electron/positron candidates resulting from the 3\ensuremath{\sigma} TOF selection in
Figure~\ref{ana_fig_pid_beta39} is now parameterized using the functional representation
of the hadron n\ensuremath{\sigma}$_{\text{el}}$ distributions extracted in the previous paragraph.
Figure~\ref{ana_fig_nsigmael_fits} depicts a selection of distributions in momentum
ranges that illustrate how the \ensuremath{\pi}, K, p, and \ensuremath{\pi}\ensuremath{\pi} bands are
traversing and overlapping the electron/positron band. Each of the graphs shows
the total parameterization of the full experimental n\ensuremath{\sigma}$_{\text{el}}$ distribution
along with the respective single contributions, all of which are the outcome of
the minimization procedure discussed in the following.

\wrapifneeded{0.50}{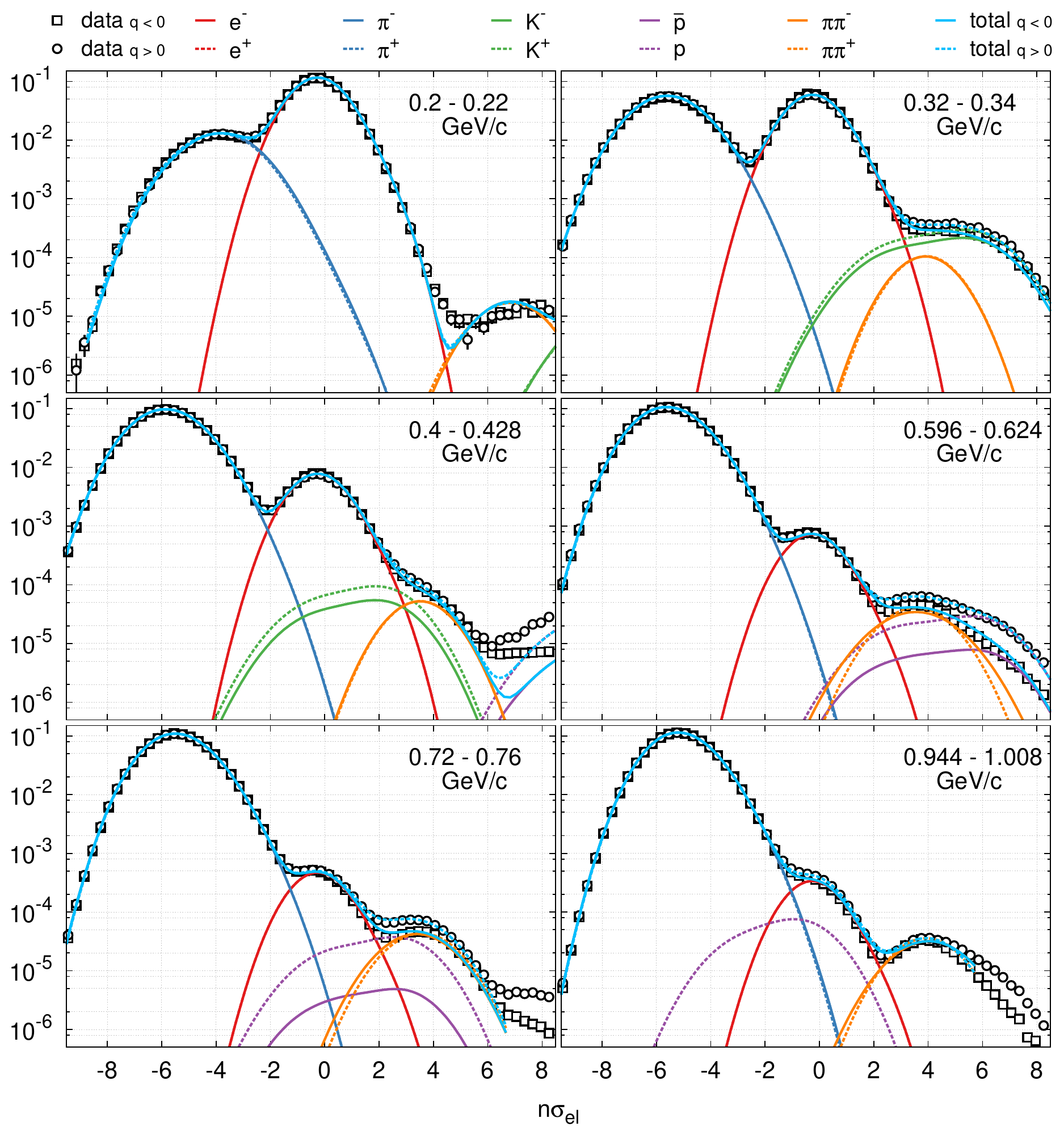}{Selected momentum ranges for \emph{multi-{}particle} fits to the experimental n\ensuremath{\sigma}$_{\text{el}}$ distributions for electron (open squares) and positron (open circles) candidates at \ensuremath{\surd}s$_{\text{NN}}$ = 39 GeV obtained after application of the 3\ensuremath{\sigma} TOF selection criterion from Figure~\ref{ana_fig_pid_beta39}. Residual contaminations caused by positive and negative hadrons as well as the resulting total parameterizations are depicted in solid and dashed lines, respectively. The selected momentum bins up to 1 GeV/c show the dominant overlap regions of \ensuremath{\pi}, K, and p bands with the electrons/positrons at n\ensuremath{\sigma}$_{\text{el}}$\ensuremath{\sim}0, and cover 99\% of the available total electron/positron statistics (c.f. Figure~\ref{ana_fig_purity}). The full momentum-{}dependent set of \emph{multi-{}particle} fits at all BES energies including the final values for the floating parameters is appended in Section~\ref{app_pid_quality_figures}. The single particle parameterizations provide the basis for the calculation of purity and contamination levels in Figure~\ref{ana_fig_purity}.}{ana_fig_nsigmael_fits}{1} %

The full momentum-{}dependent parameterization of electron/positron candidate
n\ensuremath{\sigma}$_{\text{el}}$ distributions involves a large number of parameters to be
optimized. The extraction of accurate shapes and positions for the according
contaminating hadron distributions through the analysis of pure hadron samples
allows for a considerable reduction of the floating parameter space. The
parameters of the resulting \emph{multi-{}gaussian} functions can be kept fixed and
symmetric with respect to particle charge. This leaves (at most) the
$4\cdot 2=8$ yields for the hadron contaminations as floating
parameters plus the yields as well as gaussian means and widths for electrons
and positrons. The latter can confidently be fixed to the charge-{}averaged mean
\ensuremath{\mu}$_{\text{gl}}$ and width \ensuremath{\sigma}$_{\text{opt}}$ (Table~\ref{effcorr_tab_tpcselect}) deduced from the
experimental electron/positron samples of photon conversion and Dalitz decays
in Section~\ref{effcorr_sec_samples}. However, a closer look at the lowest momentum range
0.2 -{} 0.22 GeV/c, in which hadron contaminations are small, reveals that a
secondary gaussian is required to satisfactorily describe the high-{}end tail of
the electron/positron band. In an initialization fit, this gaussian is tuned to
the tail in the lowest momentum bin and subsequently used in identical form
throughout all higher momenta. To utilitize the symmetries between electron and
positron candidate distributions and further reduce the parameter space by
linking most of the 10 remaining floating yields, a \emph{simultaneous} parameterization of the electron and positron candidate n\ensuremath{\sigma}$_{\text{el}}$ distributions has been implemented using the ROOT extension library \texttt{Roo\penalty5000 Fit} [177]. This allows for the $e^+/e^-$,
$\pi^+/\pi^-$, and $\pi\pi^+/\pi\pi^-$ yields to be
forced the same while $K^+/K^-$ and especially
$p/\bar{p}$ yields are allowed to vary separately. A successful fit
minimization further requires sensible initial values for the yields of the
single contributions to be set in each momentum bin. Reasonable estimates for
the current momentum bin are achieved by extrapolating the optimized yields of
the previous four momentum ranges via a power law (ax$^{\text{b}}$), and imposing 20-{}70\%
parameter limits based on the resulting central value (see last page of
Section~\ref{app_pid_quality_figures}). The n\ensuremath{\sigma}$_{\text{el}}$ range used for the
simultaneous fit corresponds to the lowest and highest limits given by the
underlying hadron n\ensuremath{\sigma}$_{\text{el}}$ parameterization ranges. Since kaons are skipped
between 0.48 -{} 1.2 GeV/c, (anti-{})protons below 0.4 and above 1.2 GeV/c, and
anti-{}protons above 0.8 GeV/c where their respective contributions are
negligible, most momentum bins only require five floating parameters for a
successful description of the electron/positron candidate n\ensuremath{\sigma}$_{\text{el}}$ distributions.  As demonstrated in Figure~\ref{ana_fig_nsigmael_fits} and
Section~\ref{app_pid_quality_figures}, the above procedure works very well across the
entire momentum range and most importantly across all energies without
modifications. The only minor adjustment happens at small momenta below 0.6 and
0.3 GeV/c where correctional shifts for \ensuremath{\pi} and \ensuremath{\pi}\ensuremath{\pi}, respectively, are
required in form of two additional floating parameters shifting the full
\emph{multi-{}gaussian} functions. The functional representation of the
electron/positron candidate n\ensuremath{\sigma}$_{\text{el}}$ distributions now allows for purity
and contamination levels to be calculated for the electron and positron samples
based on the TPC selection window applied in each momentum bin (see
Figure~\ref{fig_pidQA} right).\newline

{\bf Electron/positron sample purities and contamination levels.} 
For the purpose of quantifying sample qualities, the ultimate goal of this
section is to derive purities and contamination levels of the electron/positron
samples resulting from the application of TOF and TPC selection criteria.
Figure~\ref{ana_fig_purity} gives an overview of their momentum dependence at all BES
energies exhibiting characteristic patterns common to all energies. For 73\% of
the full electron/positron sample or up to 0.4 GeV/c, excellent purities very
close to 1 are achieved which drive the -{} for electrons and positrons
exceptional -{} total sample purities of \ensuremath{\sim}97-{}98.5\% listed in
Table~\ref{ana_tab_purity}. At higher and with increasing momentum the pion tail and
the traversing proton band begin to dominate the contamination of the
electron/positron samples causing a continuous reduction in purity down to
about 40\% at 2 GeV/c. Anti-{}protons are produced in heavy-{}ion collisions with
significantly smaller cross sections than protons, the effect of which is
observed at 0.7 -{} 1.2 GeV/c through an estimated purity decrease
of 10-{}15\% for the positron samples. It affects, however, only 5\% of the total
electron/positron statistics with momenta above 0.7 GeV/c. Contaminations by
kaons at 0.4 -{} 0.5 GeV/c and by merged pions predominantly at 0.7 -{} 0.8 GeV/c
are barely visible and on the sub-{}0.5\% level. During the pair reconstruction
procedure in Section~\ref{ana_sec_pairrec}, these impurities (as deviations from 100\%
pure electron/positron samples) are considered part of the systematic
uncertainties on the respective invariant mass spectra \footnote{
the
propagation of impurity levels to systematic uncertainties on M$_{\text{ee}}$ spectra is
in progress.
}.\newline
 As a final note on the side: The momentum dependence of upper and lower TPC
selection limits has been varied to optimize the window for best sample
significance. It turns out that only a minor improvement of the sample
significances could be achieved by updating the current TPC selection
functional with parameter values resulting from a fit to the result of the
significance tuning.

\begin{center}
\begingroup%
\setlength{\newtblsparewidth}{1\linewidth-2\tabcolsep-2\tabcolsep-2\tabcolsep-2\tabcolsep-2\tabcolsep-2\tabcolsep-2\tabcolsep-2\tabcolsep-2\tabcolsep-2\tabcolsep-2\tabcolsep-2\tabcolsep}%
\setlength{\newtblstarfactor}{\newtblsparewidth / \real{422}}%

\begin{longtable}{lllllllllll}\caption[{Total purities of electron and positron samples for all BES energies (in \%).}]{Total purities of electron and positron samples for all BES energies (in \%).\label{ana_tab_purity}\hyperlabel{ana_tab_purity}%
}\tabularnewline
\endfirsthead
\caption[]{(continued)}\tabularnewline
\endhead
\hline
\multicolumn{1}{m{85\newtblstarfactor}|}{\centering%
\ensuremath{\surd}s$_{\text{NN}}$ (GeV)
}&\multicolumn{1}{m{28\newtblstarfactor}|}{\centering%
19.6
}&\multicolumn{1}{m{28\newtblstarfactor}|}{\centering%
27
}&\multicolumn{1}{m{28\newtblstarfactor}|}{\centering%
39
}&\multicolumn{1}{m{28\newtblstarfactor}|}{\centering%
62.4
}&\multicolumn{1}{m{28\newtblstarfactor}|}{\centering%
}&\multicolumn{1}{m{85\newtblstarfactor}|}{\centering%
\ensuremath{\surd}s$_{\text{NN}}$ (GeV)
}&\multicolumn{1}{m{28\newtblstarfactor}|}{\centering%
19.6
}&\multicolumn{1}{m{28\newtblstarfactor}|}{\centering%
27
}&\multicolumn{1}{m{28\newtblstarfactor}|}{\centering%
39
}&\multicolumn{1}{m{28\newtblstarfactor+\arrayrulewidth}}{\centering%
62.4
}\tabularnewline
\multicolumn{1}{m{85\newtblstarfactor}|}{\centering%
e$^{\text{-{}}}$
}&\multicolumn{1}{m{28\newtblstarfactor}|}{\centering%
98.5
}&\multicolumn{1}{m{28\newtblstarfactor}|}{\centering%
97.8
}&\multicolumn{1}{m{28\newtblstarfactor}|}{\centering%
97.0
}&\multicolumn{1}{m{28\newtblstarfactor}|}{\centering%
98.2
}&\multicolumn{1}{m{28\newtblstarfactor}|}{\centering%
}&\multicolumn{1}{m{85\newtblstarfactor}|}{\centering%
e$^{\text{+}}$
}&\multicolumn{1}{m{28\newtblstarfactor}|}{\centering%
98.1
}&\multicolumn{1}{m{28\newtblstarfactor}|}{\centering%
97.8
}&\multicolumn{1}{m{28\newtblstarfactor}|}{\centering%
96.8
}&\multicolumn{1}{m{28\newtblstarfactor+\arrayrulewidth}}{\centering%
98.1
}\tabularnewline
\hline
\end{longtable}\endgroup%

\end{center}

\wrapifneeded{0.50}{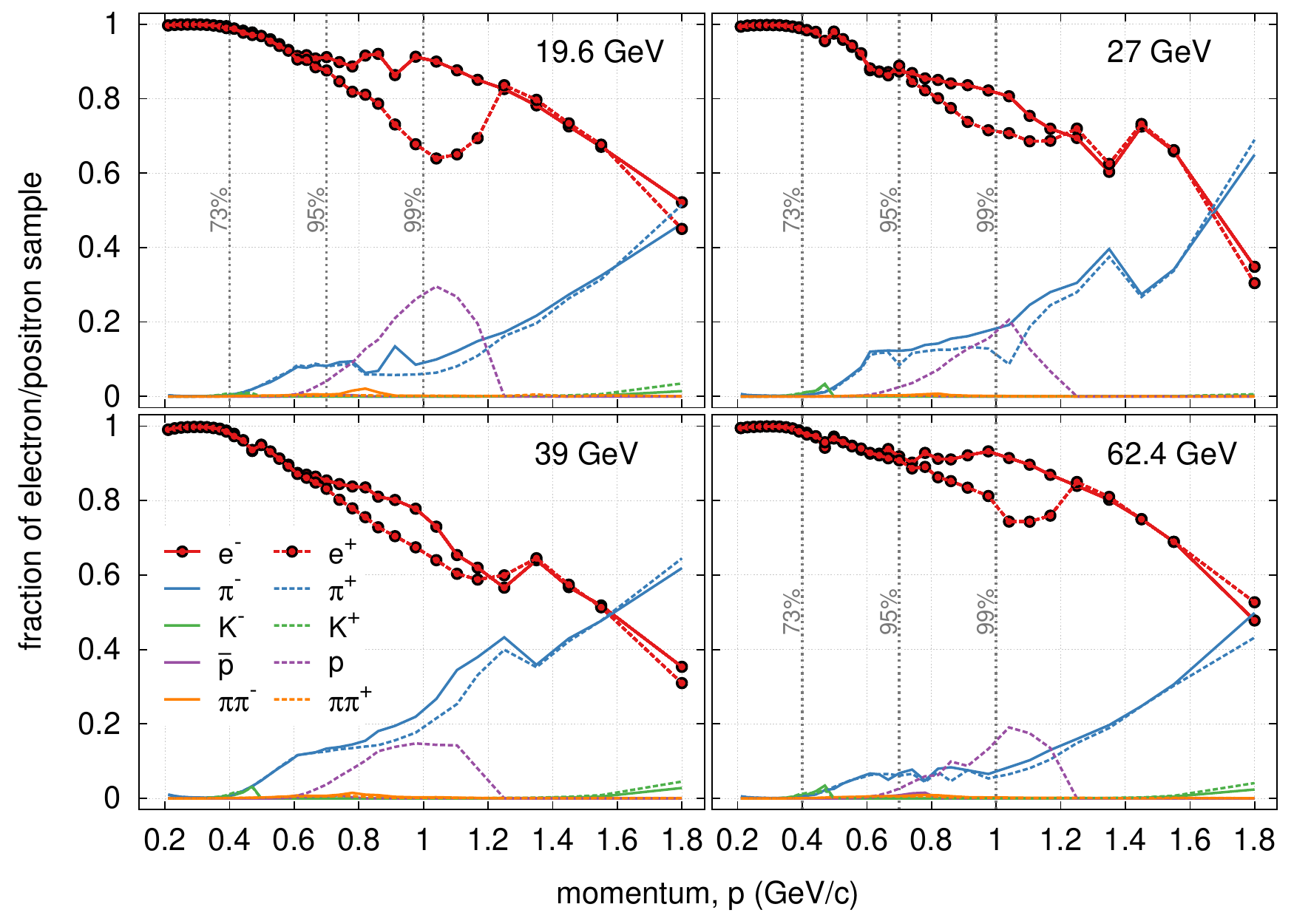}{Momentum dependence of purity (red) and hadron contamination (\ensuremath{\pi}: blue, K: green, p: purple, merged-{}\ensuremath{\pi}: orange) levels of electron/positron samples identified via the TOF \& TPC selection criteria discussed in Section~\ref{ana_subsec_pid_toftpc} at all BES energies. The dominating contributions to the electron/positron samples by the pion tail and the band-{}traversing proton are clearly visible. To guide the reader's eye, vertical grey dotted lines are depicted at momenta of 0.4, 0.7, and 1 GeV/c respectively corresponding to \ensuremath{\sim}73\%, 95\%, and 99\% of the available electron/positron statistics. The relation of purity to these sample fractions results in the high total purities listed in Table~\ref{ana_tab_purity}.}{ana_fig_purity}{0.95} %

\pagebreak[4]

\section{Pair Reconstruction}
\label{ana_sec_pairrec}\hyperlabel{ana_sec_pairrec}%

The single electrons and positrons identified with high purities in
Section~\ref{ana_sec_pid} on an event-{}by-{}event basis constitute a list of particles
originating from various, a priori unknown sources. Combining the four-{}momenta
of all electrons and positrons in an event to form dielectrons with invariant
masses M$_{\text{ee}}$ and transverse momenta p$_{\text{T}}$ results in the so-{}called unlike-{}sign
(opposite charge) foreground spectra. These contain the actual dielectron
signal plus background contributions from purely combinatorial or correlated
sources. The former consist of $e^+e^-$ pairs formed from disparate
decays and the latter of cross pairs caused by the combination of
$e^+e^-$ pairs within the same \ensuremath{\pi}$^{\text{0}}$ Dalitz or 2\ensuremath{\gamma} decay
(c.f. Figure~\ref{ana_fig_pair_normrange_accfac} left).
To identify the signal emanating from the interesting physical sources (see
Table~\ref{sim_tab_srcs}) by disentangling it from the background, two
well-{}established and complementary statistical techniques are employed: \emph{i})
the \emph{like-{}sign same-{}event method} which reproduces both combinatorial and
correlated background but is statistically limited at higher M$_{\text{ee}}$, and \emph{ii})
the \emph{mixed-{}event method} which exclusively provides the combinatorial
background but instead with effectively unlimited statistical accuracy. Since
the \ensuremath{\rho}/\ensuremath{\omega} region of STAR dielectron spectra typically exhibits
signal-{}to-{}background ratios of only 1/100 -{} 1/250 at BES-{}I energies, accurate
background subtraction is crucial to enable the successful reconstruction of
the dielectron signal and its comparison to the cocktail simulations and model
calculations of Chapter~\ref{sim}.

In Section~\ref{ana_subsec_pair_distros} [175], the prerequisite
p$_{\text{T}}$-{}dependent dielectron M$_{\text{ee}}$-{}distributions are prepared for all like-{} and
unlike-{}sign charge combinations of electrons and positrons from within the same
event and from different events via \emph{event-{}mixing}. Quality assurance of the
mixed events is presented and pair-{}based selection criteria applied on all
distributions to reject photon conversions. Section~\ref{ana_subsec_pair_signal} [178] explains the statistical techniques in detail and
performs the background subtraction. At the end of this chapter, raw dielectron
M$_{\text{ee}}$-{} and p$_{\text{T}}$-{}distributions are readily available for the application of
Chapter~\ref{effcorr}'s efficiency corrections which ultimately enables the spectra's
physics interpretation in Chapter~\ref{results}.

\subsection{Generation of Pair Distributions}
\label{ana_subsec_pair_distros}\hyperlabel{ana_subsec_pair_distros}%

The reconstruction of all $e^+e^-$ pairs for an event in a
so-{}called \emph{same-{}event analysis} and the subsequent generation of unlike-{}sign
foreground distributions by means of the particles' Lorentz vectors is a
straight-{}forward task in combinatorics. Special attention only has to be paid
in the case of like-{}sign combinations to avoid double counting of pairs (i.e.
start secondary iteration at \texttt{j=\penalty0 i+1}). In the \emph{like-{}sign same-{}event} background subtraction method discussed in Section~\ref{ana_subsec_pair_signal}, the two
resulting like-{}sign background distributions are averaged and corrected for the
acceptance difference between like-{} and unlike-{}sign pairs which is obtained
from mixed-{}event distributions.

The \emph{mixed-{}event technique} enables the population of unlike-{} and like-{}sign
background distributions with pairs from uncorrelated sources by combining
electrons and positrons from two different events within the same event class.
Here, an event class is defined by an event's primary z-{}vertex V$_{\text{z}}$ (Table~\ref{evtplane_tab_vz}), its reference multiplicity RefMult
(Table~\ref{dsets_evttrk_tab_refmult}), and its event plane angle \ensuremath{\Psi}$_{\text{2}}$ (Figure~\ref{evtplane_fig_check}). The space of event classes is discretized into 10, 9,
and 10 bins for the V$_{\text{z}}$-{}, RefMult-{}, and \ensuremath{\Psi}$_{\text{2}}$-{}dimension, respectively.
This ensures the quick aggregation of a sufficient amount of similar events
from which to generate high-{}statistics unlike-{} and like-{}sign distributions
while avoiding changes in shapes due to a too coarse binning. The particular
configuration choice is supported by the study of different event class
discretization levels in Figure~\ref{ana_fig_pair_evtclass}. The ratios of unlike-{}sign
distributions and acceptance correction factors -{} both characteristic for the
mixed-{}event technique (see Section~\ref{ana_subsec_pair_signal}) -{} with respect to the
respective fine-{}grained discretization reveal that coarser binnings for
\ensuremath{\Psi}$_{\text{2}}$ and V$_{\text{z}}$ affect the underlying distributions' shapes at least up to
M$_{\text{ee}}$ \ensuremath{\sim} 1 GeV/c$^{\text{2}}$. With finer binning the shape effects quickly diminish
making the above configuration choice sufficiently accurate within 0.2\%. These
observations motivate the use of \ensuremath{\Psi}$_{\text{2}}$ as additional dimension in the event
class definition which entails the separate reconstruction of the event plane
and the event-{}by-{}event correction of event flow vectors in Section~\ref{ana_subsec_evtplane}.

\wrapifneeded{0.50}{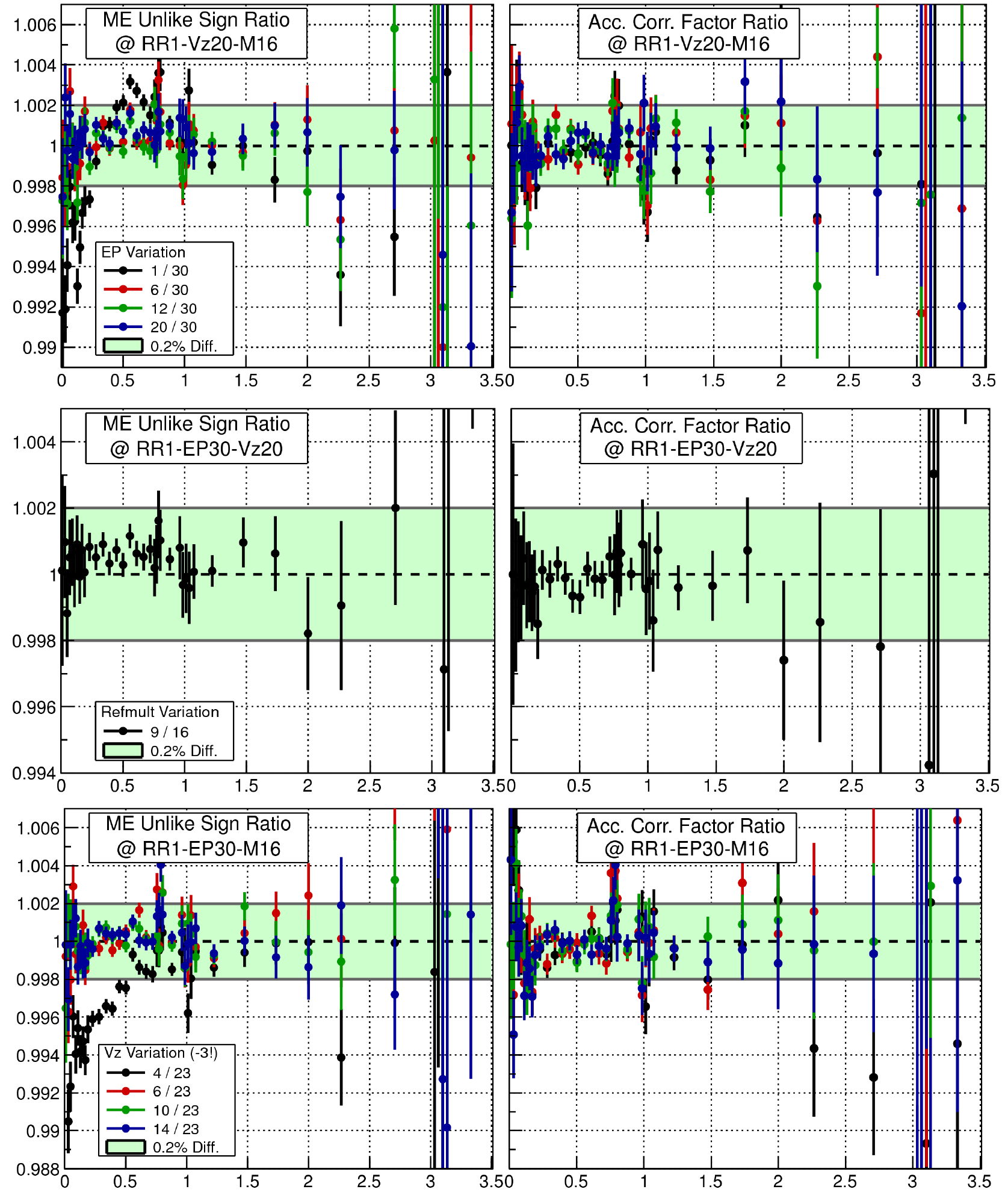}{M$_{\text{ee}}$-{}dependent ratios of mixed-{}event unlike-{}sign distributions (left column) and acceptance correction factors (right column) for varying configurations of event buffers (top to bottom) at \ensuremath{\surd}s$_{\text{NN}}$ = 39 GeV. For each dimension of an event class (\ensuremath{\Psi}$_{\text{2}}$, RefMult, and V$_{\text{z}}$), different discretization levels (black, red, green, and blue points) are compared to a reasonably fine-{}grained level to identify potential shape changes induced by event-{}mixing. The green band indicates 0.2\% deviations supporting the choice of 10 \ensuremath{\Psi}$_{\text{2}}$, 10 V$_{\text{z}}$, and 9 RefMult bins as sensible event buffer configuration.}{ana_fig_pair_evtclass}{0.88} %

On a technical level, the following strategy is pursued to obtain like-{} and
unlike-{}sign mixed-{}event distributions. During the (same-{})event loop, the
current event is appended to an in-{}memory \emph{event buffer} of size 20 pertaining
to the respective event class. When an event buffer at full capacity is
encountered, the (same-{})event loop is put on halt and the \emph{event-{}mixing} started using the list of events in the buffer. For each combination of two
unequal events A and B in the buffer, the respective particle lists are
iterated such that every $e^-$ ($e^+$) in A is
partnered (\emph{i}) with every $e^-$ ($e^+$) in B to form
like-{}sign pairs in which case no double counting occurs since the particle
lists originate from different events, and (\emph{ii}) with every $e^+$ ($e^-$) in B to form unlike-{}sign pairs. Once completed, the event
buffer is cleared and the (same-{})event loop resumed until the next full event
buffer is found and the procedure repeated.

An alternative strategy would be to trigger event-{}mixing during the
(same-{})event loop every time an event is attributed to the specific buffer.
When a full event buffer is encountered, one event is selected randomly and
removed to make space for the incoming event.  This method would result in more
statistics for the same buffer size but the two strategies are not expected to
cause significant discrepancies in the mixed-{}event distributions.\newline

Note that events without any electrons or positrons are also included in the
buffers and subsequently the even-{}mixing. If such empty events were skipped,
most event buffers would preferably be filled with more central events which
possibly changes the shape of the like-{} and unlike-{}sign distributions. Instead
of accounting for this effect by applying correction factors on the more
peripheral events, the loss in produced statistics is recovered by using an
increased buffer size.\newline

For the event buffer configuration chosen here, event-{}mixing requires that a
total of $9\cdot 10\cdot 10\cdot 20 = 18,000$ events are
simultaneously kept in memory. To keep the memory footprint to a minimum, a
dedicated event class is used containing only minimal event and track
information. Furthermore, storing events for re-{}usage during event-{}mixing
requires an explicit copy as opposed to a same-{}event analysis in which the same
allocated memory (via \texttt{TCl\penalty5000 o\penalty5000 n\penalty5000 e\penalty5000 s\penalty5000 A\penalty5000 r\penalty5000 ray}) can be used repeatedly for event storage.
A custom copy constructor has hence been implemented as part of the dedicated
event class which omits the event header and only clones the track arrays of
non-{}zero length when transfering the current event to the event buffer's
reusable STL container.\newline

In Section~\ref{ana_subsec_pair_signal}, the resulting mixed-{}event spectra are normalized
to the corresponding same-{}event distributions and used in the mass region where
combinatorial contributions account for the uncorrelated background.
Particularly the intermediate mass region of the dielectron spectrum (IMR)
requires the large statistics provided by event-{}mixing to contain the
statistical errors to the ones dictated by the unlike-{}sign foreground.\newline

Finally, an important question to pose is whether all event buffers are
sufficiently filled up to their intended size before performing event-{}mixing.
In Figure~\ref{ana_fig_pair_evtmixbuf}, an insufficient configuration for the setup of
event buffers and for the partitioning of the full statistics (left) is
compared to the optimal case (right). In the scenario on the left, the
available event statistics has been processed in parallel by dividing it into
smaller sub-{}samples and attributed to event buffers based on equal-{}length
segments in V$_{\text{z}}$. This approach evidently causes event-{}mixing based on
"truncated" event buffers observed in the reduced buffer utilization, both
overall as 95\% of buffers are actually full when event-{}mixing starts, as well
as at both ends of the V$_{\text{z}}$ selection window. These issues are resolved in the
scenario on the right, by partitioning the V$_{\text{z}}$ distribution into intervals
with approximately equal number of events (see Table~\ref{evtplane_tab_vz}) and by
processing the full event statistics sequentially.

\wrapifneeded{0.50}{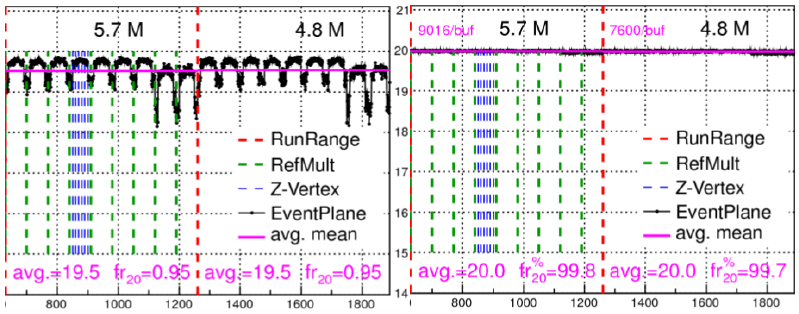}{Average sizes/utilization of event buffers defined by reference multiplicity (Table~\ref{dsets_evttrk_tab_refmult}), primary z-{}vertex (V$_{\text{z}}$, Table~\ref{evtplane_tab_vz}), and event plane angle (\ensuremath{\Psi}$_{\text{2,corr}}$, Figure~\ref{evtplane_fig_check}) used during event-{}mixing. Two setup configurations are compared for the same two RunID ranges (RunRange, Section~\ref{stbadrdos}): Partitioning of the available statistics and equi-{}distant V$_{\text{z}}$ binning (left) lead to only partially filled event buffers (horizontal magenta lines) as opposed to purely sequential processing and equi-{}event-{}number V$_{\text{z}}$ binning (right). See text for more details.}{ana_fig_pair_evtmixbuf}{0.9} %

Based on the like-{} and unlike-{}sign pairs extracted from the above same-{} and
mixed-{}event analyses, six p$_{\text{T}}$-{}dependent fore-{} and background pair invariant
mass distributions are generated in M$_{\text{ee}}$ and p$_{\text{T}}$ intervals of 1 MeV/c$^{\text{2}}$ and 10 MeV/c,
respectively. However, before the according histograms can be used as basis for
the statistical background subtraction and extraction of raw dielectron signal
spectra in Section~\ref{ana_subsec_pair_signal}, a few pair-{}based selections and
corrections need to be applied. First, only pairs with rapidities
$\vert y\vert<1$ are considered. Second, the event class dependent
weighting factors which correct the raw reference multiplicity distributions
(see Figure~\ref{dsets_evttrk_fig_refmult} and accompanying text) for trigger
inefficiencies and configuration changes, are applied on a pair-{}by-{}pair basis.\newline

\wrapifneeded{0.50}{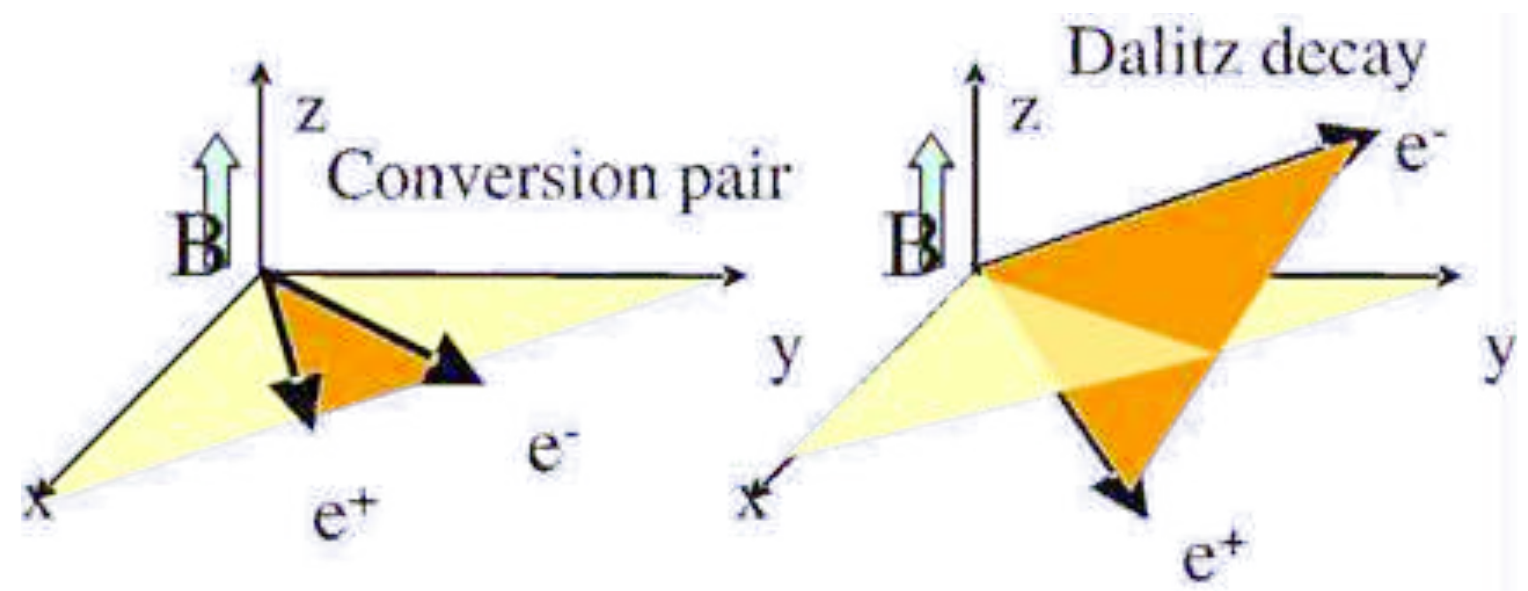}{Orientation of the dielectron decay plane in a B-{}field for background pairs from photon conversions (left) and signal pairs from \ensuremath{\pi}$^{\text{0}}$ Dalitz decays (right).}{ana_fig_pair_phiV}{0.45} %

Last, pairs emanating from photon conversions are rejected by reducing the
selection for M$_{\text{ee}}$ <{} 0.2 GeV/c$^{\text{2}}$ to pairs with a large enough so-{}called
\ensuremath{\phi}$_{\text{V}}$ angle. Figure~\ref{ana_fig_pair_phiV} highlights the different orientations of
the $e^+e^-$ decay planes with respect to the magnetic field for
background conversion pairs and signal pairs from \ensuremath{\pi}$^{\text{0}}$ Dalitz decays. The
bulk of the photons radiating outward from the primary vertex hit the radially
symmetric detector material close to the beam axis approximately orthogonally.
The created conversion pairs have no intrinsic opening angle causing the
involved electrons and positrons to only be deflected within the azimuthal
plane by the magnetic field. Electrons and positrons from \ensuremath{\pi}$^{\text{0}}$ Dalitz
decays, however, exhibit a non-{}zero initial opening angle and hence do not show
such a preferred orientation of their decay plane. These distinguishing in-{} and
out-{}of-{}plane orientations lie at the heart of the mathematical definition of
the \ensuremath{\phi}$_{\text{V}}$ angle found in [15]. The study of simulated versus
experimental \ensuremath{\phi}$_{\text{V}}$ distributions in Figure~\ref{ana_fig_pair_phiVsim} further
strengthens the use of \ensuremath{\phi}$_{\text{V}}$ as a potential means to separate photon
conversions from physical sources.

\pagebreak[4]

\wrapifneeded{0.50}{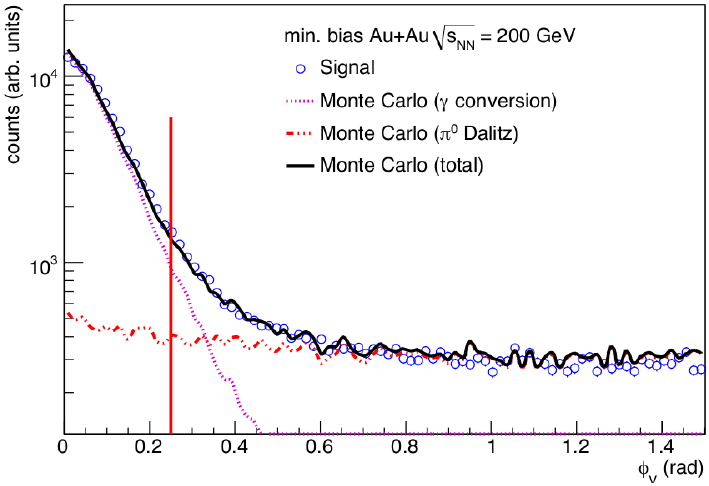}{Demonstration of \ensuremath{\phi}$_{\text{V}}$'s capabilities of \ensuremath{\gamma} conversion rejection [15]. Details see text.}{ana_fig_pair_phiVsim}{0.49} %

The sum of the Monte Carlo simulations for photon conversions and \ensuremath{\pi}$^{\text{0}}$ Dalitz decays describes the measured \ensuremath{\phi}$_{\text{V}}$ distribution very well. As
expected from a proper definition of \ensuremath{\phi}$_{\text{V}}$ reflecting the differences in
the respective decay plane orientations, the \ensuremath{\pi}$^{\text{0}}$ Dalitz distribution is
mostly uniform whereas the \ensuremath{\gamma} conversion distribution peaks at zero when
consistently ordering the charges of the pair for the \ensuremath{\phi}$_{\text{V}}$ calculation. Based on the simulations, PHENIX chose to reject all pairs with
\ensuremath{\phi}$_{\text{V}}$ <{} 0.25 (vertical red line) in Figure~\ref{ana_fig_pair_phiVsim} [15]. In contrast, for the
dielectron analyses at top RHIC energy [16], an M$_{\text{ee}}$-{}dependent
\ensuremath{\phi}$_{\text{V}}$ selection criterion has been employed to efficiently reject photon
conversions while still maintaining as much signal from the physical sources as
possible:
\[\phi_V(M_\mathrm{ee}) > 0.84326\,\exp(-49.4819\,M_\mathrm{ee}) - 0.996609\,M_\mathrm{ee} + 0.19801.\]
In Figure~\ref{ana_fig_pair_phivrej} (left), the above functional form is obtained from
a sample of $e^+e^-$ pairs produced via a photon embedding run
through the full STAR reconstruction chain (c.f. Section~\ref{effcorr_sec_samples}).
Applying the \ensuremath{\phi}$_{\text{V}}$ selection criterion on all like-{} and unlike-{}sign pairs
in same-{} and mixed-{}event analyses rejects more than 98\% of photon conversions.

\wrapifneeded{0.50}{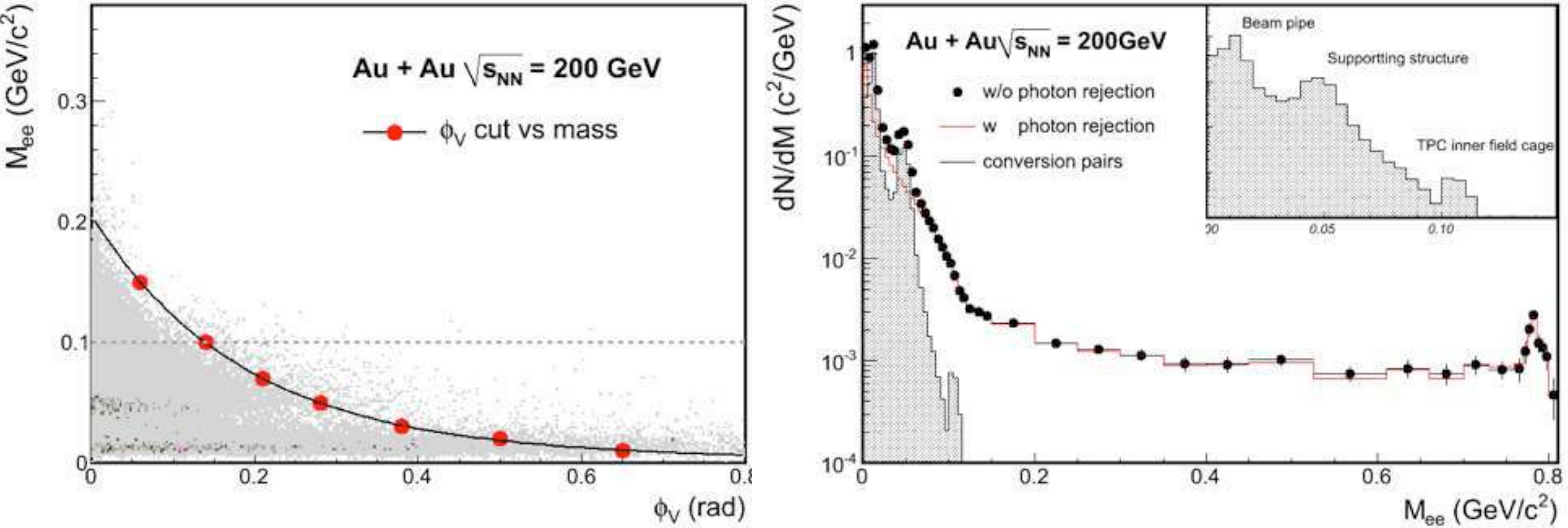}{Rejection of e$^{\text{+}}$e$^{\text{-{}}}$ pairs from photon conversions via \ensuremath{\phi}$_{\text{V}}$ at \ensuremath{\surd}s$_{\text{NN}}$ = 200 GeV [16]. (left) M$_{\text{ee}}$-{}dependence of \ensuremath{\phi}$_{\text{V}}$ along with the imposed selection functional (black line) based on a photon embedding driven through the full STAR reconstruction chain. (right) Raw dielectron invariant mass signal spectrum up to the \ensuremath{\omega} peak with (red line) and without (black points) conversion rejection. The distribution of the rejected conversion pairs exhibits features characteristic for the radial positions of inner STAR detector material (grey histogram and inlet).}{ana_fig_pair_phivrej}{1} %

Figure~\ref{ana_fig_pair_phivrej} (right) depicts the (raw) background-{}subtracted
dielectron invariant mass spectrum at $\sqrt{s_\mathrm{NN}}$ = 200
GeV from STAR [16] with and without the above \ensuremath{\phi}$_{\text{V}}$ selection applied. Peak-{}like structures caused by STAR's beam pipe, SVT support
cones and inner TPC field cage can be clearly identified in the distribution of
the rejected pairs. Their successive appearance with increasing invariant mass
is rooted in the tracking algorithm's assumption (and definition) that primary
tracks originate from the collision vertex. This introduces an artificial
opening angle in the reconstruction of pairs from off-{}vertex converting photons
which increases with the radial distance of the pair's point of origin. For
reasons of consistency and simplicity, the analyses of dielectron production at
BES energies presented in this thesis, use the same functional form for the
\ensuremath{\phi}$_{\text{V}}$ selection as derived at 200 GeV. Significant variations in the
\ensuremath{\phi}$_{\text{V}}$ selection's photon conversion rejection power are not expected due to
its inherent geometric nature. Still, individual \ensuremath{\phi}$_{\text{V}}$ selection criteria
for the BES energies as well as a study of its possible dependence on the V$_{\text{z}}$ selection are in preparation.

\subsection{Extraction of raw Signal Spectra}
\label{ana_subsec_pair_signal}\hyperlabel{ana_subsec_pair_signal}%

In the following, the background distributions obtained from the same-{} and
mixed-{}event analyses in Section~\ref{ana_subsec_pair_distros} are combined, scaled and
corrected to construct unlike-{}sign distributions suitable for subtraction from
the foreground. As argued in [15], since electrons and positrons
are always created pair-{}wise in heavy-{}ion collisions, the unlike-{}sign
background can in general be constructed as the geometric mean of the same-{} or
mixed-{}event like-{}sign background distributions regardless of single-{}particle
efficiencies or acceptances:
$\langle\mathrm{BG}_{+-}\rangle = 2\sqrt{ \langle\mathrm{BG}_{++}\rangle\langle\mathrm{BG}_{--}\rangle}$. The resulting
background-{}subtracted invariant mass (M$_{\text{ee}}$) and transverse momentum (p$_{\text{T}}$)
spectra then contain the dielectron signal steming from physical sources only
(c.f. Chapter~\ref{sim}). For the remainder of this section, the subscripts \texttt{+-{}\penalty0 }, \texttt{++},
and \texttt{-{}\penalty0 -{}\penalty0 } refer to  unlike-{}sign, positive like-{}sign, and negative like-{}sign
distributions, respectively, whereas \texttt{SE} and \texttt{ME} (\texttt{ME$^{\text{N}}$}) denote bin
contents of the respective same-{} and (normalized) mixed-{}event histograms.\newline

{\bf Mixed-{}Event Normalization.} 
As a consequence of the above argument, the unlike-{}sign mixed-{}event
distributions have to be normalized by scaling their integrated yield
($\sum_\mathrm{PS}\mathrm{ME}_{+-}$, PS: M$_{\text{ee}}$-{}p$_{\text{T}}$-{}space) to the
according yield obtained from the geometric mean of the positive and negative
like-{}sign same-{}event distributions. However, the latter two also contain
correlated $e^+e^+/e^-e^-$ pairs which are not accounted for in
distributions produced via event-{}mixing.  Hence, the proper yield to use for
the normalization factor \emph{A}$_{\text{\ensuremath{\pm}}}$ is reproduced by the geometric mean of
the normalized positive and negative like-{}sign \emph{mixed-{}event} distributions:

\[A_\pm = 2\sqrt{\left(\sum\limits_\mathrm{PS}\mathrm{ME}_{--}^N\right) \left(\sum\limits_\mathrm{PS}\mathrm{ME}_{++}^N\right)} \left/\sum\limits_\mathrm{PS}\mathrm{ME}_{+-}\right.\]
where $\mathrm{ME}_{++(--,+-)}^N=A_{+(-,\pm)}\cdot\mathrm{ME}_{++(--,+-)}$.
For the separate calculation of the required positive (\emph{A}$_{\text{+}}$) and negative
(\emph{A}$_{\text{-{}}}$) like-{}sign normalization factors,
\[A_{+(-)}=\sum_\mathrm{NR}\mathrm{SE}_{++(--)}\left/\sum_\mathrm{NR}\mathrm{ME}_{++(--)}\right.,\]
an adequate normalization region NR is chosen as M$_{\text{ee}}$ >{} 0.9 GeV/c$^{\text{2}}$ in
Figure~\ref{ana_fig_pair_normrange_accfac} (left) by comparing same-{} and mixed-{}event
like-{}sign spectra and identifying an invariant mass region in which their
respective shapes agree well within statistical errors. Table~\ref{ana_tab_pair_norms} lists the normalization factors for all BES energies including their
statistical and systematic uncertainties. An estimate for the systematic
uncertainties is determined by recalculating the normalization factors with
M$_{\text{ee}}$ = 1.3 GeV/c$^{\text{2}}$ as the lower limit of the normalization region and
subtracting the result from the values at M$_{\text{ee}}$ = 0.9 GeV/c$^{\text{2}}$.  Both,
statistical and systematic uncertainties on the normalization factors are
separately propagated to the final efficiency-{}corrected spectra presented in
Chapter~\ref{results}. The total relative uncertainties on \emph{A}$_{\text{\ensuremath{\pm}}}$ are also quoted
in Table~\ref{ana_tab_pair_norms} which can contribute substantially to the final spectra uncertainties in
invariant mass regions with the lowest signal-{}to-{}background ratios. In the BES
dielectron analyses, these can be as low \ensuremath{\sim}1/200 requiring relative errors
on the mixed-{}event normalizations of less than 0.5\% to avoid 100\% uncertainties
on the data points in the final spectra. All BES energies analyzed here achieve
an uncertainty level of about 0.5\% or less. Finally note that calculating
normalization factors for the mixed-{}event distributions in each p$_{\text{T}}$ interval
individually would only introduce larger statistical errors and hence
complicate the identification of adequate normalization regions.
\begin{center}
\begingroup%
\setlength{\newtblsparewidth}{1\linewidth-2\tabcolsep-2\tabcolsep-2\tabcolsep-2\tabcolsep-2\tabcolsep-2\tabcolsep}%
\setlength{\newtblstarfactor}{\newtblsparewidth / \real{424}}%

\begin{longtable}{lllll}\caption[{Normalization factors calculated at all BES energies for positive/negative like-{} and unlike-{}sign mixed-{}event distributions relative to their same-{}event counterparts (see text). The numbers given in parenthesis are statistical and systematic uncertainties on the last digits, respectively. Also listed are the total relative errors on \emph{A}$_{\text{\ensuremath{\pm}}}$ which folded with the signal-{}to-{}background ratio directly give the uncertainty on the signal spectra caused by mixed-{}event normalization.}]{Normalization factors calculated at all BES energies for positive/negative like-{} and unlike-{}sign mixed-{}event distributions relative to their same-{}event counterparts (see text). The numbers given in parenthesis are statistical and systematic uncertainties on the last digits, respectively. Also listed are the total relative errors on \emph{A}$_{\text{\ensuremath{\pm}}}$ which folded with the signal-{}to-{}background ratio directly give the uncertainty on the signal spectra caused by mixed-{}event normalization.\label{ana_tab_pair_norms}\hyperlabel{ana_tab_pair_norms}%
}\tabularnewline
\endfirsthead
\caption[]{(continued)}\tabularnewline
\endhead
\hline
\multicolumn{1}{m{20\newtblstarfactor}|}{\centering%
}&\multicolumn{1}{m{101\newtblstarfactor}|}{\centering%
19.6 GeV
}&\multicolumn{1}{m{101\newtblstarfactor}|}{\centering%
27 GeV
}&\multicolumn{1}{m{101\newtblstarfactor}|}{\centering%
39 GeV
}&\multicolumn{1}{m{101\newtblstarfactor+\arrayrulewidth}}{\centering%
62.4 GeV
}\tabularnewline
\multicolumn{1}{m{20\newtblstarfactor}|}{\centering%
$A_-$
}&\multicolumn{1}{m{101\newtblstarfactor}|}{\centering%
0.026629(29)(266)
}&\multicolumn{1}{m{101\newtblstarfactor}|}{\centering%
0.026533(18)(139)
}&\multicolumn{1}{m{101\newtblstarfactor}|}{\centering%
0.026780(18)(56)
}&\multicolumn{1}{m{101\newtblstarfactor+\arrayrulewidth}}{\centering%
0.026714(20)(142)
}\tabularnewline
\multicolumn{1}{m{20\newtblstarfactor}|}{\centering%
$A_+$
}&\multicolumn{1}{m{101\newtblstarfactor}|}{\centering%
0.026635(28)(23)
}&\multicolumn{1}{m{101\newtblstarfactor}|}{\centering%
0.026549(18)(136)
}&\multicolumn{1}{m{101\newtblstarfactor}|}{\centering%
0.026859(18)(253)
}&\multicolumn{1}{m{101\newtblstarfactor+\arrayrulewidth}}{\centering%
0.026737(20)(10)
}\tabularnewline
\multicolumn{1}{m{20\newtblstarfactor}|}{\centering%
$A_\pm$
}&\multicolumn{1}{m{101\newtblstarfactor}|}{\centering%
0.026668(21)(134)
}&\multicolumn{1}{m{101\newtblstarfactor}|}{\centering%
0.026554(13)(97)
}&\multicolumn{1}{m{101\newtblstarfactor}|}{\centering%
0.026816(13)(129)
}&\multicolumn{1}{m{101\newtblstarfactor+\arrayrulewidth}}{\centering%
0.026726(14)(71)
}\tabularnewline
\multicolumn{1}{m{20\newtblstarfactor}|}{\centering%
$\Delta A_\pm^\mathrm{tot}$
}&\multicolumn{1}{m{101\newtblstarfactor}|}{\centering%
0.51\%
}&\multicolumn{1}{m{101\newtblstarfactor}|}{\centering%
0.37\%
}&\multicolumn{1}{m{101\newtblstarfactor}|}{\centering%
0.48\%
}&\multicolumn{1}{m{101\newtblstarfactor+\arrayrulewidth}}{\centering%
0.27\%
}\tabularnewline
\hline
\end{longtable}\endgroup%

\end{center}

\wrapifneeded{0.50}{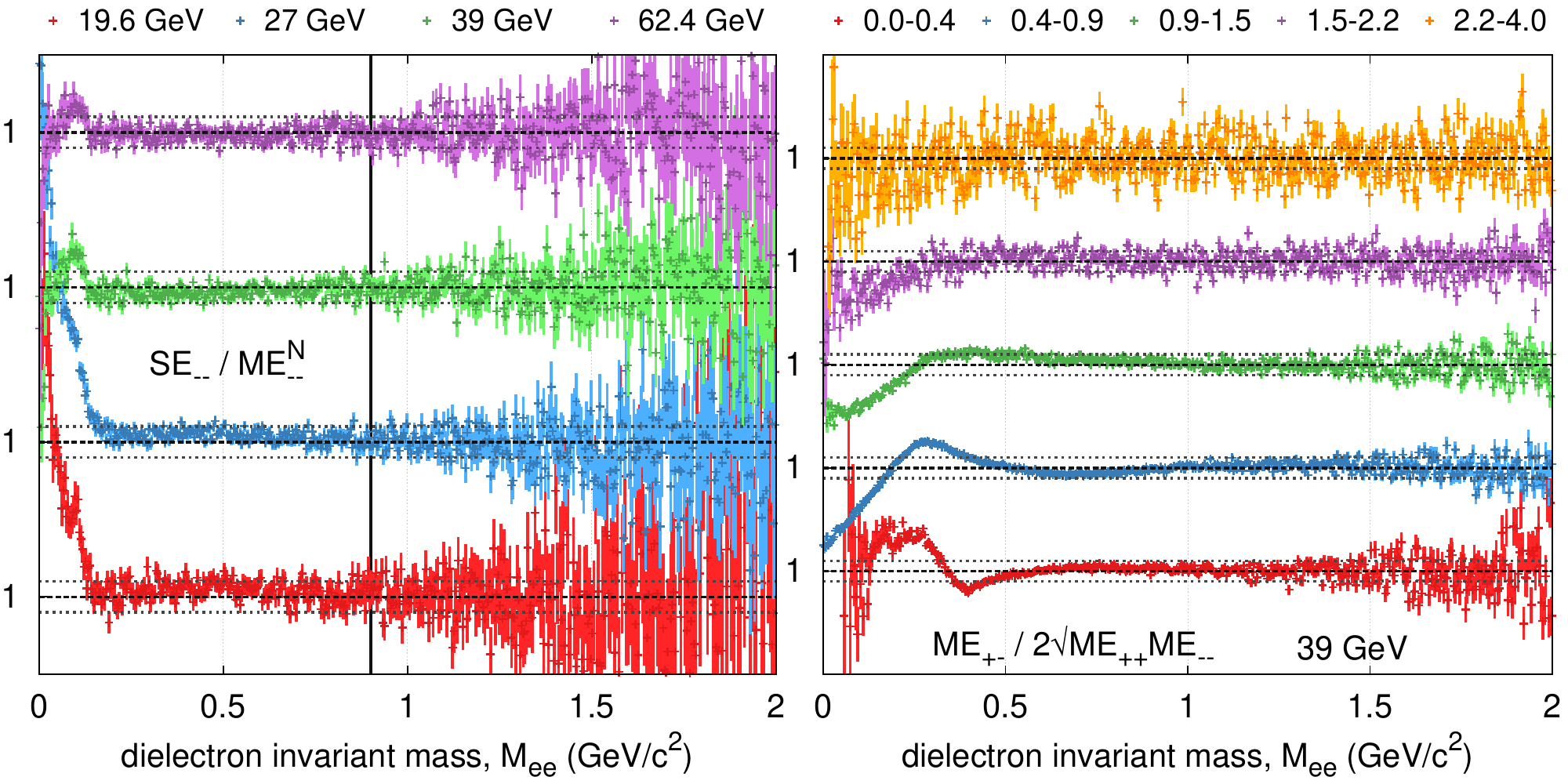}{(left) Identification of mixed-{}event normalization range M$_{\text{ee}}$ >{} 0.9 GeV/c$^{\text{2}}$ (vertical line) using the negative like-{}sign \texttt{SE/\penalty0 ME$^{\text{N}}$} ratios. (right) p$_{\text{T}}$-{}dependence of acceptance correction factor at \ensuremath{\surd}s$_{\text{NN}}$ = 39 GeV. Dotted lines denote a \ensuremath{\pm}2\% envelope. All graphs for the like-{}sign ratios and acceptance factors are collected in Figure~\ref{app_pair_fig_normrange_accfac_all}.}{ana_fig_pair_normrange_accfac}{1} %

{\bf Same-{}Event Acceptance Correction.} 
Below the normalization region identified in
Figure~\ref{ana_fig_pair_normrange_accfac} (left), correlated pairs play a role in the
shape of the same-{}event like-{} and unlike-{}sign distributions. Therefore, the
same-{}event like-{}sign distribution
$\mathrm{SE}_{\pm\pm}=2\sqrt{\mathrm{SE}_{++}\mathrm{SE}_{--}}$ needs to be subtracted from the foreground instead of the normalized
unlike-{}sign mixed-{}event distribution $\mathrm{ME}_{+-}^N$ to
correctly extract the dielectron signal. The like-{}sign pairs generating the
$\mathrm{SE}_{\pm\pm}$ distribution, however, are exposed to
different acceptance losses than the unlike-{}sign pairs of the foreground. The
effect can be accounted for by bin-{}wise applying the acceptance correction
factor \emph{f}$_{\text{acc}}$ deduced from the according mixed-{}event distributions on the
$\mathrm{SE}_{\pm\pm}$ distribution:

\[\mathrm{SE}_{\pm\pm}^\mathrm{corr} = \mathrm{SE}_{\pm\pm} \cdot f_\mathrm{acc} = \mathrm{SE}_{\pm\pm} \cdot \mathrm{ME}_{+-}\Big/2\sqrt{\mathrm{ME}_{++}\mathrm{ME}_{--}}.\]
To avoid subtraction artefacts due to limited statistics at high invariant
masses, the arithmetic instead of the geometric mean is used in the calculation
of $\mathrm{SE}_{\pm\pm}^\mathrm{corr}$ for invariant mass bins
with zero bin content in any of the involved like-{}sign same-{}event or
mixed-{}event distributions. The acceptance correction factors
(Figure~\ref{ana_fig_pair_normrange_accfac} right) are applied and
$\mathrm{SE}_{\pm\pm}^\mathrm{corr}$ calculated using the p$_{\text{T}}$ intervals 0.-{}0.4, 0.4-{}0.9, 0.9-{}1.5, 1.5-{}2.2, and 2.2-{}4 GeV/c which coincide
with the p$_{\text{T}}$ edges employed in the determination of pair efficiencies (c.f.
Figure~\ref{effcorr_fig_paireffs}). The lowest transverse momentum bin requires
special attention since it is significantly impacted by STAR's acceptance hole
for single track transverse momenta below 0.2 GeV/c. For the current version of
the efficiency-{}corrected dielectron p$_{\text{T}}$-{}spectra in Figure~\ref{results_fig_pT}, the
data point for the lowest p$_{\text{T}}$ interval is thus omitted but yet shown for
completeness as part of the raw data presented in this section.

\wrapifneeded{0.50}{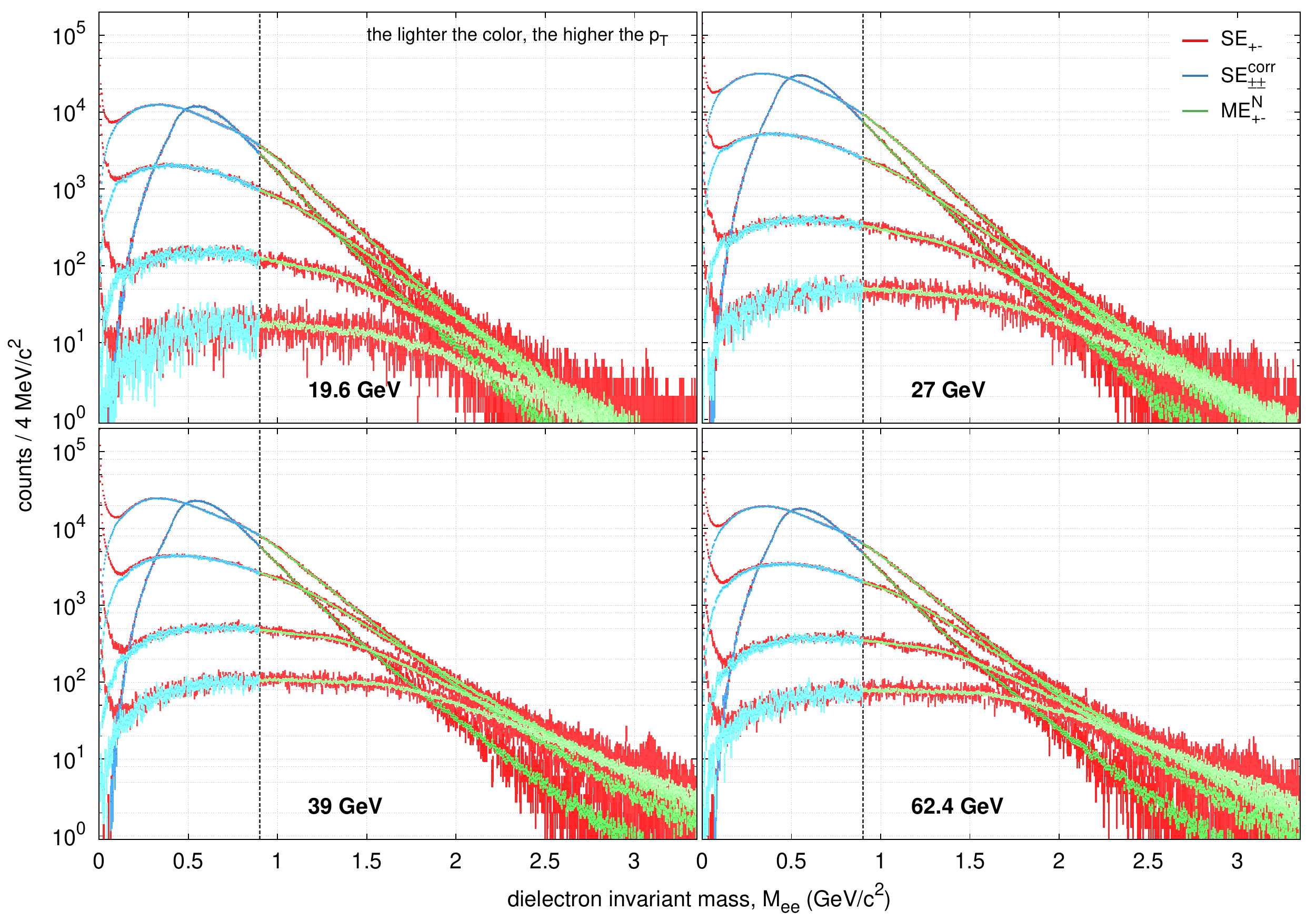}{p$_{\text{T}}$-{}dependent invariant mass distributions at all BES energies used to generate dielectron signal spectra via subtraction: the unlike-{}sign foreground (red) as well as the two backgrounds of acceptance-{}corrected like-{}sign same-{}event (blue) and normalized unlike-{}sign mixed-{}event (green). The lighter the respective colors, the higher the pair transverse momentum corresponding to the p$_{\text{T}}$ intervals 0.-{}0.4, 0.4-{}0.9, 0.9-{}1.5, 1.5-{}2.2, and 2.2-{}4 GeV/c. The vertical dashed lines indicate the transition from same-{} to mixed-{}event background subtraction at 0.9 GeV/c$^{\text{2}}$.}{ana_fig_pair_fg_bg}{1} %

{\bf Background Subtraction and Dielectron Signal.} 
Figure~\ref{ana_fig_pair_fg_bg} shows all resulting invariant mass distributions for the
unlike-{}sign foreground along with the corresponding background which are used
to generate dielectron signal spectra $\mathcal{S}$ in the
different p$_{\text{T}}$ intervals separately:

\[\mathcal{S} = \begin{cases} \mathrm{SE}_{+-} - \mathrm{SE}_{\pm\pm}^\mathrm{corr} & \mbox{if } M_{ee} \leq 0.9\,\mathrm{GeV}/c^2 \\ \mathrm{SE}_{+-} - \mathrm{ME}_{+-}^\mathrm{N} & \mbox{if } M_{ee} > 0.9\,\mathrm{GeV}/c^2 \end{cases}\]
The subtraction is based on the original distributions' 4 MeV/c$^{\text{2}}$ invariant
mass binning. In this representation, the raw dielectron signal also contains
negative entries due to statistical fluctuations in fore-{} and background. A
better representation of the signal spectra thus needs to be found in form of a
variable (invariant mass dependent) binning which widens the bin widths enough
to reduce statistical errors but still catches the important resonance as well
as continuum features of the distributions. In Figure~\ref{ana_fig_pair_rebin}, the
invariant mass spectra from the p$_{\text{T}}$-{}integrated analyses at all BES energies,
i.e. constructed based on p$_{\text{T}}$-{}integrated background generation and
subtraction, are used to establish a variable binning suited to the available
statistics at each individual energy. For this purpose, an automated rebinning
algorithm is employed with the following strategy.
\begin{itemize}[itemsep=0pt]

\item{} Start with the bin at the upper invariant mass limit.

\item{} Sum up bin contents until a positive number of entries is aggregated, and

\item{} the relative statistical error is smaller than a given mass-{}range-{}dependent limit, and

\item{} (optionally) a minimum, also mass-{}range-{}dependent bin width is reached.

\item{} When all criteria are satisfied, define the covered invariant mass range as
  bin for the rebinned distribution, reset the bin entries counter, and reiterate.

\end{itemize}

This strategy makes the prolonged trial-{}and-{}error approach of manually finding
a suitable variable binning obsolete. It is replaced by an algorithm-{}based
approach which is easily controllable via a small set of parameters, i.e. a
handful of statistical error limits per beam energy. Such a reproducible and
dynamic environment for the rebinning of dielectron signal spectra is
especially important when analyzing dielectron production at multiple beam
energies. The optimized bin definitions of Figure~\ref{ana_fig_pair_rebin} are
subsequently applied to the signal spectra obtained via the above subtraction
method from the p$_{\text{T}}$-{}differential analyses in Figure~\ref{ana_fig_pair_fg_bg}.\newline

These p$_{\text{T}}$-{}dependent rebinned invariant mass spectra form the basis for all
physics results discussed in Chapter~\ref{results}: The pair efficiencies of
Figure~\ref{effcorr_fig_paireffs} are applied in dependence of M$_{\text{ee}}$ and p$_{\text{T}}$ on the
differential spectra which are subsequently summed up to arrive at the
efficiency-{}corrected and p$_{\text{T}}$-{}integrated dielectron invariant mass spectra
shown in Figure~\ref{results_fig_stack} and Figure~\ref{results_fig_panel}. As can be seen in
Figure~\ref{ana_fig_pair_peaks}, the statistics in the signal spectra yields pronounced
resonance peaks for \ensuremath{\omega}, \ensuremath{\phi}, and J/\ensuremath{\Psi} but is not sufficient for
fully differential M$_{\text{ee}}$-{} and p$_{\text{T}}$-{}dependent analyses. The invariant mass
distributions in the different p$_{\text{T}}$ intervals are hence separately integrated
over the characteristic mass regions to obtain the efficiency-{}corrected
dielectron p$_{\text{T}}$-{}spectra depicted in Figure~\ref{results_fig_pT}.

\wrapifneeded{0.50}{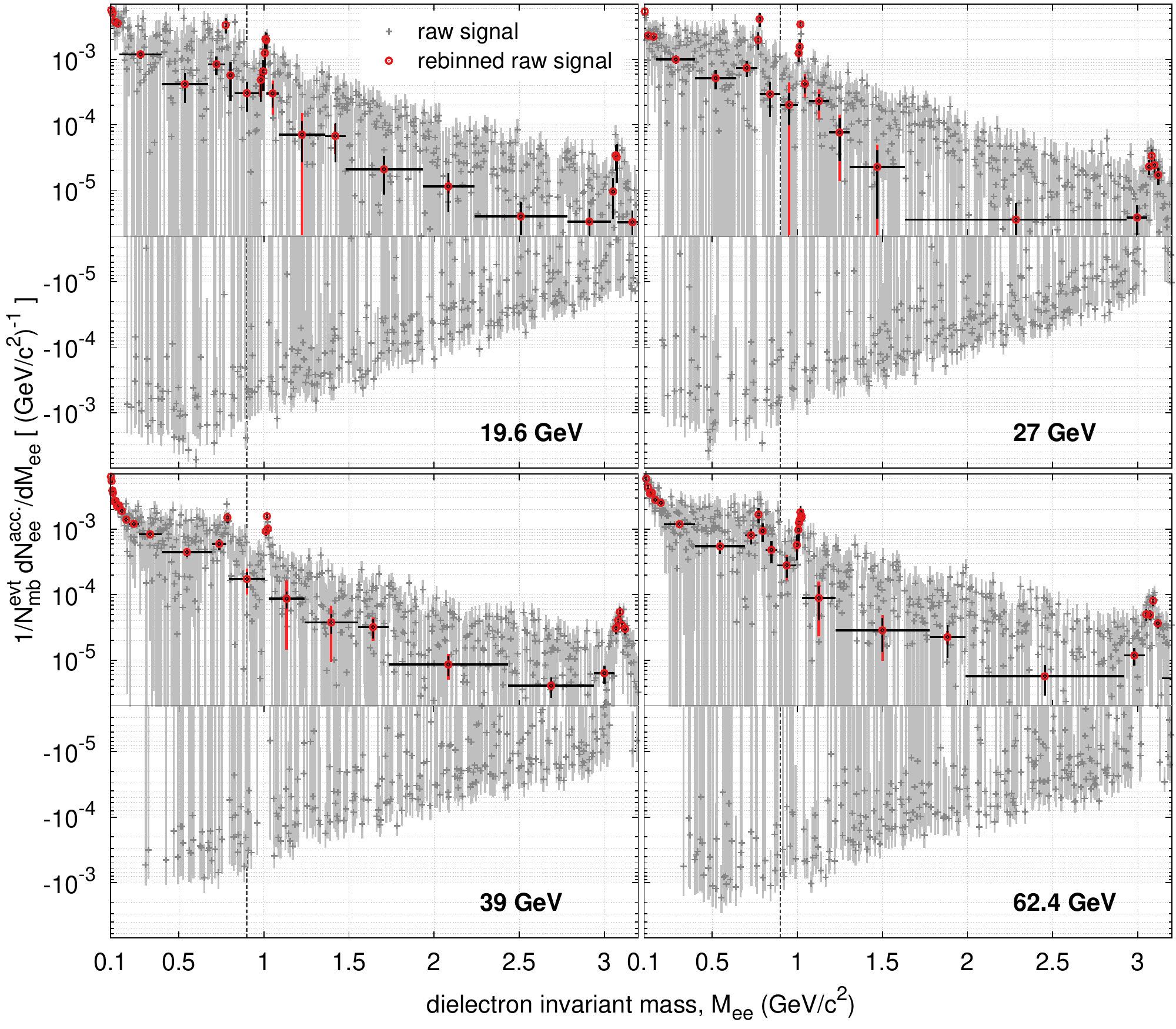}{Rebinning of raw signal spectra from p$_{\text{T}}$-{}integrated analyses at all BES energies. Only the mass region where actual rebinning takes place is shown (M$_{\text{ee}}$ >{} 0.1 GeV/c$^{\text{2}}$). All distributions are divided by the measured number of minimum bias events listed in Table~\ref{ana_tab_dsets_evttrk}. Grey and black error bars denote statistical errors and bin widths for signal distributions with equi-{}width and variable binning, respectively. Systematic uncertainties on the rebinned signal caused by the uncertainty in mixed-{}event normalization (Table~\ref{ana_tab_pair_norms}) are also included (red error bars). Note that the lower half of each energy contains the negative entries on a reversed logarithmic scale which are part of the background-{}subtracted raw signal distribution and hence have to be taken into account during the automated rebinning procedure. The vertical dashed line indicates the invariant mass of 0.9 GeV/c$^{\text{2}}$ at which same-{}event-{} is replaced by mixed-{}event-{}based background subtraction.}{ana_fig_pair_rebin}{1} %

\wrapifneeded{0.50}{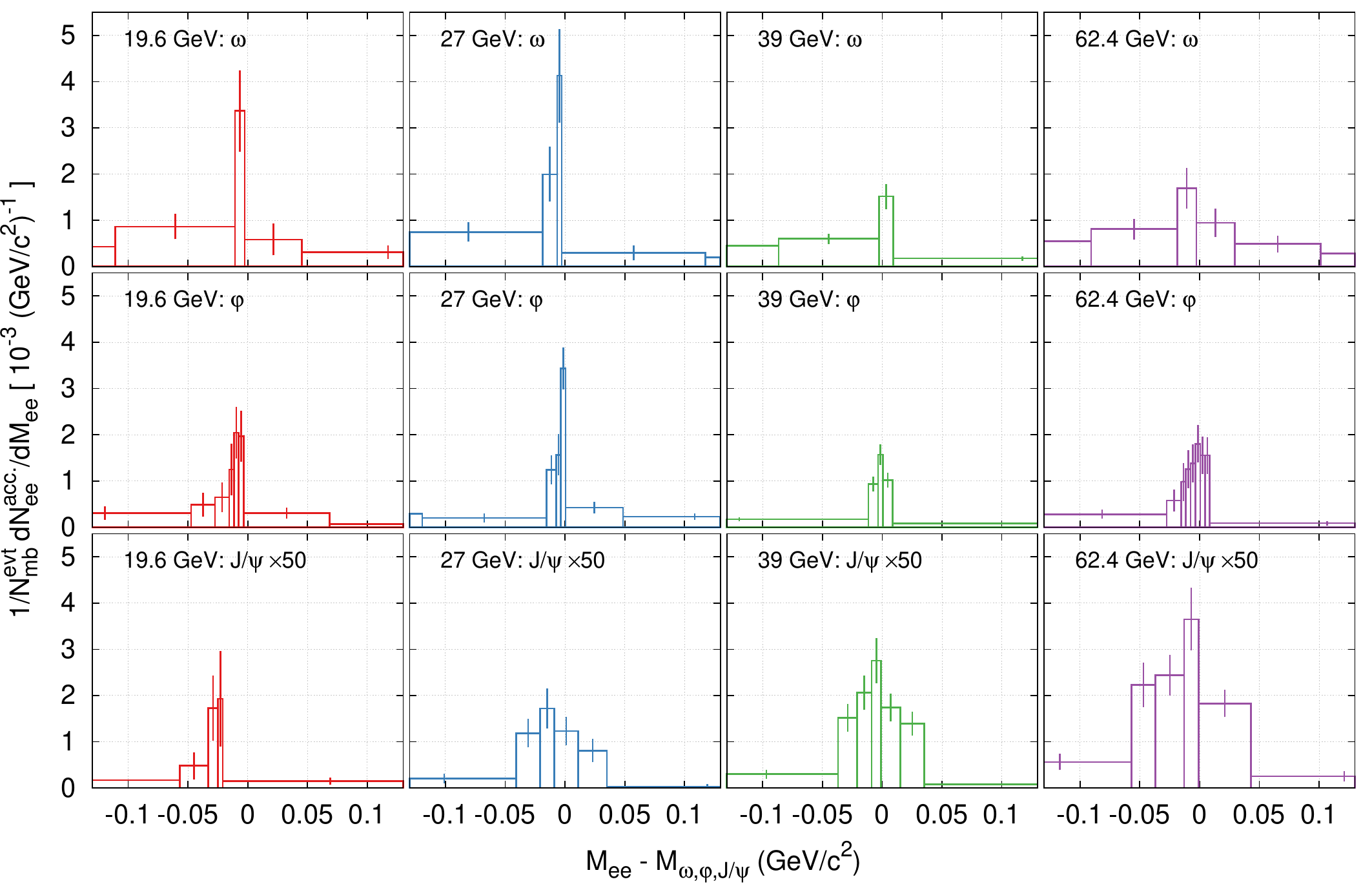}{Total raw dielectron invariant mass spectra for all BES energies after p$_{\text{T}}$-{}differential rebinning and summation concentrating on the \ensuremath{\omega}, \ensuremath{\phi}, and J/\ensuremath{\psi} mass regions on a linear scale.}{ana_fig_pair_peaks}{1} %


\chapter{Efficiency Correction}
\label{effcorr}\hyperlabel{effcorr}%

For the comparison to theoretical calculations and amongst energies
(Chapter~\ref{results}) the raw dielectron invariant mass and \emph{p$_{\text{T}}$} distributions
resulting from the data analysis in Chapter~\ref{data_analysis} have to be corrected for
losses caused by detector inefficiencies and by analysis selection cuts (c.f.
Table~\ref{ana_tab_dsets_evttrk}). For this purpose, simulated tracks are embedded into
real events and tracked through the detector using the same tools as for real
data event reconstruction (Section~\ref{effcorr_sec_samples}). The estimation of
systematic uncertainties requires the comparison of simulated to clean
experimental distributions of suitable particle samples as also described in
Section~\ref{effcorr_sec_samples}. Section~\ref{effcorr_sec_single} follows the analysis procedure
of Chapter~\ref{data_analysis} to calculate total single track efficiencies using the
particle samples from embedding and experiment.  Finally in
Section~\ref{effcorr_sec_paireffs}, the single track efficiencies are propagated into
pair efficiencies using a simple two-{}body Monte-{}Carlo simulation.

\section{Particle Samples}
\label{effcorr_sec_samples}\hyperlabel{effcorr_sec_samples}%

The procedure employed in Section~\ref{effcorr_sec_single} to obtain total single track
efficiencies is mostly based on electron and positron samples generated through
the \emph{embedding} technique. 5\% of the collision's multiplicity is included
within a measured Au+Au event as simulated tracks with known kinematics and of
known type. The percentage of embedded tracks is chosen such that the
additional multiplicity does not change the real event's centrality class.
Generation of events with the appropriate simulated multiplicity is achieved by
means of the software package GSTAR [179] which extends GEANT
[180, 181] with STAR-{}specific geometry definitions of
detailed detector material information. It also provides a plugin-{}like
framework for the development of external software modules to read event
generator output and define signals (\emph{hits}) induced by particles traversing
the sensitive detector elements. In the case of the TPC, for instance, GSTAR
produces electron clusters along the charged particle tracks provided by
GEANT's event generator. The clusters are used by the dedicated \emph{tss} plugin
library to realistically simulate their transport and diffusion in the drift
field to the TPC's wire plane followed by electron avalanche, spread over
read-{}out pads, and pixel-{}by-{}pixel digitization to ADC counts
[182]. The resulting output of this TPC response simulation is
subsequently propagated through the TPC cluster/hit finder of STAR's standard
event reconstruction software [159] which also includes, amongst
others, track and vertex reconstruction, and provides the usual data summary
files (DSTs) for further physics analysis. See Section~\ref{ana_sec_dsets_evttrk} for
details on this first analysis stage of event reconstruction from raw detector
data. As a result, the embedding procedure allows for the determination of
single track efficiencies by calculating the fraction of simulated tracks still
available at different stages of the dielectron analysis chain
(Chapter~\ref{data_analysis}).

\emph{Embedding} has been established as a request based STAR-{}wide procedure with
dedicated teams incorporated in the organizational structure
[183].  Table~\ref{effcorr_tab_requests} hence summarizes general
information on the charge-{}separated embedding requests needed to access the
simulation results for the minimum-{}bias dielectron analyses at BES-{}I energies
and analyzed in Section~\ref{effcorr_sec_single}. All single-{}electron embedding requests
have the uniform simulation of rapidity $y\in(-1.2,1.2)$ and
azimuthal angle $\varphi\in(0,2\pi)$ in common. All of them use a
subset of events ensured to cover the entire beam time.
\begin{center}
\begingroup%
\setlength{\newtblsparewidth}{\linewidth-2\tabcolsep-2\tabcolsep-2\tabcolsep-2\tabcolsep-2\tabcolsep-2\tabcolsep-2\tabcolsep-2\tabcolsep-2\tabcolsep-2\tabcolsep}%
\setlength{\newtblstarfactor}{\newtblsparewidth / \real{99}}%

\begin{longtable}{lllllllll}\caption[{STAR e+/e-{} embedding requests for dielectron analyses at BES-{}I energies [183] with MC and Matched the total number of tracks simulated and reconstructed, respectively. FSET denotes a STAR-{}internal suffix for the set of simulation output files used for analysis in Section~\ref*{effcorr_sec_single}.}]{STAR e$^{\text{+}}$/e$^{\text{-{}}}$ embedding requests for dielectron analyses at BES-{}I energies [183] with \emph{MC} and \emph{Matched} the total number of tracks simulated and reconstructed, respectively. FSET denotes a STAR-{}internal suffix for the set of simulation output files used for analysis in Section~\ref{effcorr_sec_single}.\label{effcorr_tab_requests}\hyperlabel{effcorr_tab_requests}%
}\tabularnewline
\endfirsthead
\caption[]{(continued)}\tabularnewline
\endhead
\hline
\multicolumn{1}{m{11\newtblstarfactor+\arrayrulewidth}}{\centering%
$\sqrt{s_\mathrm{NN}}$ (GeV)
}&\multicolumn{1}{m{11\newtblstarfactor+\arrayrulewidth}}{\centering%
Events (10$^{\text{3}}$)
}&\multicolumn{1}{m{11\newtblstarfactor+\arrayrulewidth}}{\centering%
\emph{MC} (10$^{\text{6}}$)
}&\multicolumn{1}{m{11\newtblstarfactor+\arrayrulewidth}}{\centering%
\emph{Matched} (10$^{\text{6}}$)
}&\multicolumn{1}{m{11\newtblstarfactor+\arrayrulewidth}}{\centering%
FSETs
}&\multicolumn{1}{m{11\newtblstarfactor+\arrayrulewidth}}{\centering%
Id(s) (20\ldots{})
}&\multicolumn{1}{m{11\newtblstarfactor+\arrayrulewidth}}{\centering%
$\vert V_z\vert^\mathrm{max}$ (cm)
}&\multicolumn{1}{m{11\newtblstarfactor+\arrayrulewidth}}{\centering%
\emph{p$_{\text{T}}$} (GeV/c)
}&\multicolumn{1}{m{11\newtblstarfactor+\arrayrulewidth}}{\centering%
STAR Library
}\tabularnewline
\multicolumn{1}{m{11\newtblstarfactor+\arrayrulewidth}}{\centering%
19.6
}&\multicolumn{1}{m{11\newtblstarfactor+\arrayrulewidth}}{\centering%
217
}&\multicolumn{1}{m{11\newtblstarfactor+\arrayrulewidth}}{\centering%
1.43
}&\multicolumn{1}{m{11\newtblstarfactor+\arrayrulewidth}}{\centering%
0.80
}&\multicolumn{1}{m{11\newtblstarfactor+\arrayrulewidth}}{\centering%
200-{}204
}&\multicolumn{1}{m{11\newtblstarfactor+\arrayrulewidth}}{\centering%
121203/4
}&\multicolumn{1}{m{11\newtblstarfactor+\arrayrulewidth}}{\centering%
70
}&\multicolumn{1}{m{11\newtblstarfactor+\arrayrulewidth}}{\centering%
0.2-{}5
}&\multicolumn{1}{m{11\newtblstarfactor+\arrayrulewidth}}{\centering%
P11id
}\tabularnewline
\multicolumn{1}{m{11\newtblstarfactor+\arrayrulewidth}}{\centering%
27
}&\multicolumn{1}{m{11\newtblstarfactor+\arrayrulewidth}}{\centering%
228
}&\multicolumn{1}{m{11\newtblstarfactor+\arrayrulewidth}}{\centering%
1.59
}&\multicolumn{1}{m{11\newtblstarfactor+\arrayrulewidth}}{\centering%
0.94
}&\multicolumn{1}{m{11\newtblstarfactor+\arrayrulewidth}}{\centering%
200-{}204
}&\multicolumn{1}{m{11\newtblstarfactor+\arrayrulewidth}}{\centering%
123601/2
}&\multicolumn{1}{m{11\newtblstarfactor+\arrayrulewidth}}{\centering%
50
}&\multicolumn{1}{m{11\newtblstarfactor+\arrayrulewidth}}{\centering%
0.2-{}4
}&\multicolumn{1}{m{11\newtblstarfactor+\arrayrulewidth}}{\centering%
P11id
}\tabularnewline
\multicolumn{1}{m{11\newtblstarfactor+\arrayrulewidth}}{\centering%
39
}&\multicolumn{1}{m{11\newtblstarfactor+\arrayrulewidth}}{\centering%
141
}&\multicolumn{1}{m{11\newtblstarfactor+\arrayrulewidth}}{\centering%
1.03
}&\multicolumn{1}{m{11\newtblstarfactor+\arrayrulewidth}}{\centering%
0.60
}&\multicolumn{1}{m{11\newtblstarfactor+\arrayrulewidth}}{\centering%
200-{}201
}&\multicolumn{1}{m{11\newtblstarfactor+\arrayrulewidth}}{\centering%
113301
}&\multicolumn{1}{m{11\newtblstarfactor+\arrayrulewidth}}{\centering%
50
}&\multicolumn{1}{m{11\newtblstarfactor+\arrayrulewidth}}{\centering%
<{}2
}&\multicolumn{1}{m{11\newtblstarfactor+\arrayrulewidth}}{\centering%
SL10d
}\tabularnewline
\multicolumn{1}{m{11\newtblstarfactor+\arrayrulewidth}}{\centering%
62.4
}&\multicolumn{1}{m{11\newtblstarfactor+\arrayrulewidth}}{\centering%
103
}&\multicolumn{1}{m{11\newtblstarfactor+\arrayrulewidth}}{\centering%
0.82
}&\multicolumn{1}{m{11\newtblstarfactor+\arrayrulewidth}}{\centering%
0.45
}&\multicolumn{1}{m{11\newtblstarfactor+\arrayrulewidth}}{\centering%
200-{}202
}&\multicolumn{1}{m{11\newtblstarfactor+\arrayrulewidth}}{\centering%
12401
}&\multicolumn{1}{m{11\newtblstarfactor+\arrayrulewidth}}{\centering%
50
}&\multicolumn{1}{m{11\newtblstarfactor+\arrayrulewidth}}{\centering%
<{}2
}&\multicolumn{1}{m{11\newtblstarfactor+\arrayrulewidth}}{\centering%
SL10k
}\tabularnewline
\hline
\end{longtable}\endgroup%

\end{center}

To ascertain the quality of simulated particle samples in general, track
quality and global distance-{}of-{}closest-{}approach (DCA) distributions, for
instance, are compared for tracks from embedding and from experiment
(Section~\ref{effcorr_subsec_det_trackqual_dca}). Hence, sufficiently pure
$e^+/e^-$ and $\pi^+/\pi^-$ samples need to be
generated from measured quantities provided by the TPC alone since TOF hits not
only impose a certain track quality but also restrict the sample to tracks
originating from and using helixes re-{}fitted to the primary vertex.  Both would
distort the according distributions rendering them unsuitable for the use in
quality assurance (QA). In STAR and unlike pions, however, the satisfactory
identification of rarely produced leptons requires the measurement of their
time-{}of-{}flights to achieve a small level of hadronic contamination in the
particle sample (see Section~\ref{ana_sec_pid}). This fundamentally distinguishes the
embedding QA of electron and positron tracks from the rather standard one of
hadrons. Suitable high-{}statistics experimental samples for pions, on the one
hand, can simply be obtained by requiring minimal track qualities of \texttt{nHi\penalty5000 t\penalty5000 s\penalty5000 Fit} >{} 14 and \texttt{nHi\penalty5000 t\penalty5000 s\penalty5000 Fit/\penalty0 nHi\penalty5000 t\penalty5000 s\penalty5000 P\penalty5000 oss} >{} 0.52 for primary tracks with $\vert n\sigma_\pi\vert$ <{} 0.2.  Minimal track quality and the track's energy loss in
the TPC alone, on the other hand, only allows for the selection of
$e^+/e^-$ \emph{candidates} via $\vert n\sigma_e\vert$ <{} 2.

One way to obtain clean experimental $e^+/e^-$ samples is by
combining the unlike-{}sign candidates into dielectrons and employing pair-{}based
cuts to reduce them to the ones originating from Dalitz decays or photon
conversions. The procedure for the electron (positron) sample starts by
requiring the electron (positron) to be associated with a primary track which
is then paired with all global positron (electron) candidates. The physical
helices (\texttt{StP\penalty5000 h\penalty5000 y\penalty5000 s\penalty5000 i\penalty5000 c\penalty5000 a\penalty5000 l\penalty5000 H\penalty5000 e\penalty5000 l\penalty5000 ixD}) of both partner tracks are constructed with the
respective global momenta and used in a minimization algorithm
(\texttt{StH\penalty5000 e\penalty5000 lix:\penalty0 :\penalty0 pat\penalty5000 h\penalty5000 L\penalty5000 e\penalty5000 n\penalty5000 g\penalty5000 t\penalty5000 h\penalty5000 s()}) to determine the pair's DCA (\emph{dca$_{\text{ee}}$}). The
secondary decay vertex is defined midway between the partner helices' closest
points of approach. After rejecting pair combinations with \emph{dca$_{\text{ee}}$} >{} 1 cm
(Figure~\ref{effcorr_fig_elec_sample} upper left) the partners' 4-{}momenta are calculated
at the secondary decay vertex based on the global helices to obtain the correct
"global-{}global" dielectron invariant mass M$_{\text{gg}}$. Dielectrons from physical
sources (Dalitz decay or photon conversion in this context) appear as a peak
structure at very low invariant masses and are separated from pairs of purely
combinatorial origin via M$_{\text{gg}}$ <{} 4 MeV/c$^{\text{2}}$ (Figure~\ref{effcorr_fig_elec_sample} lower
left). The perpendicular decay length L$_{\text{xy}}$ denotes the distance of
secondary to primary vertex in the x-{}y-{}plane and can subsequently be used to
identify dielectrons mostly from photon conversion in the detector material
(3.8 <{} L$_{\text{xy}}$ <{} 6.1 cm) and mostly from Dalitz decays (L$_{\text{xy}}$ <{} 2 cm), see
Figure~\ref{effcorr_fig_elec_sample} upper right. An improved selection of electrons and
positrons attributed to either Dalitz decays or photon conversions can be
achieved using the "primary-{}global" dielectron invariant mass M$_{\text{pg}}$.  As
indicated by the overall M$_{\text{pg}}$ distribution in Figure~\ref{effcorr_fig_elec_sample} (lower right) most of the dielectrons originating from Dalitz decays exhibit
invariant masses M$_{\text{pg}}$ <{} 2.9 MeV/c$^{\text{2}}$ whereas most of the dielectrons caused by
photon conversion appear in 3.0 <{} M$_{\text{pg}}$ <{} 7.0 MeV/c$^{\text{2}}$. This appreciable
separation is due to the invariant mass shift induced on primary tracks that
in fact emanate from off-{}event-{}vertex decays like photon conversion. The
M$_{\text{pg}}$'s distinction power, however, is less than optimal to justify a further
decrease in statistics for the already challengingly small final particle
sample. Electrons and positrons from both sources are hence combined into the
respective final samples used in Section~\ref{effcorr_sec_single}. As soon as a primary
positron (electron) in the $e^+e^-$ ($e^-e^+$)
candidate combinations survives the preceding selection criteria for photon
conversion or Dalitz decay it is attributed to the final experimental positron
(electron) sample.

Figure~\ref{effcorr_fig_elec_sample_dEdx} shows the electron sample's distribution of
TPC $n\sigma_e$ versus its primary momentum. It illustrates the
procedure's power to reduce the candidates sample contaminated by pions, kaons
and protons to the final (almost) pure $e^+/e^-$ sample.

\wrapifneeded{0.50}{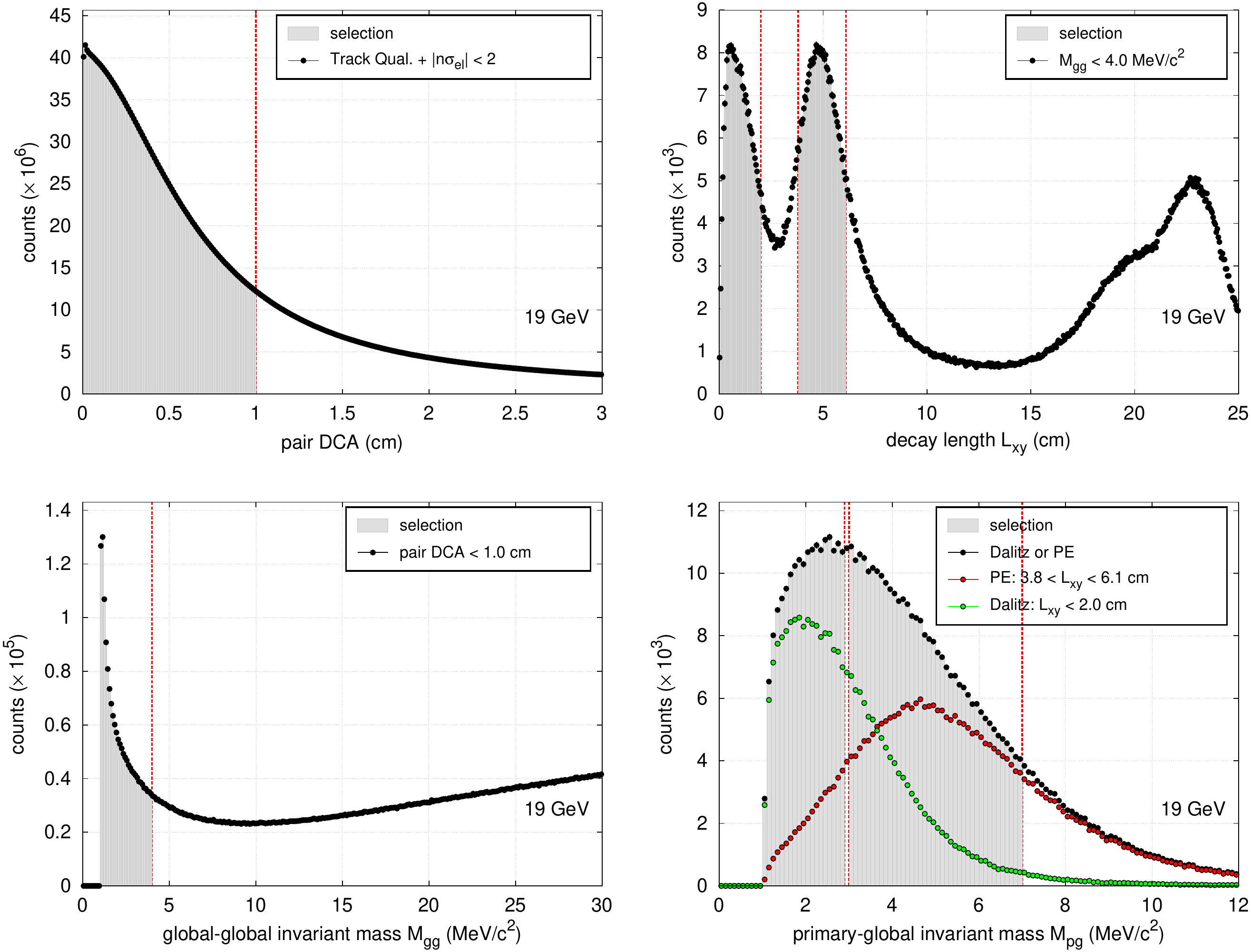}{Overview of the strategy employed to generate pure electron/positron samples from experimental data at 19.6 GeV chosen as representative for all BES-{}I energies. Grey areas denote the part of the respective spectrum selected to generate the distribution for the next criterion in the procedure (directions: top-{}bottom \ensuremath{\rightarrow} left-{}right). The last panel also includes separate contributions based on the selection intervals for Dalitz decays and photon conversions in the perpendicular decay length distribution. See text for details on the selection criteria and associated variable names.}{effcorr_fig_elec_sample}{0.93} %

\wrapifneeded{0.50}{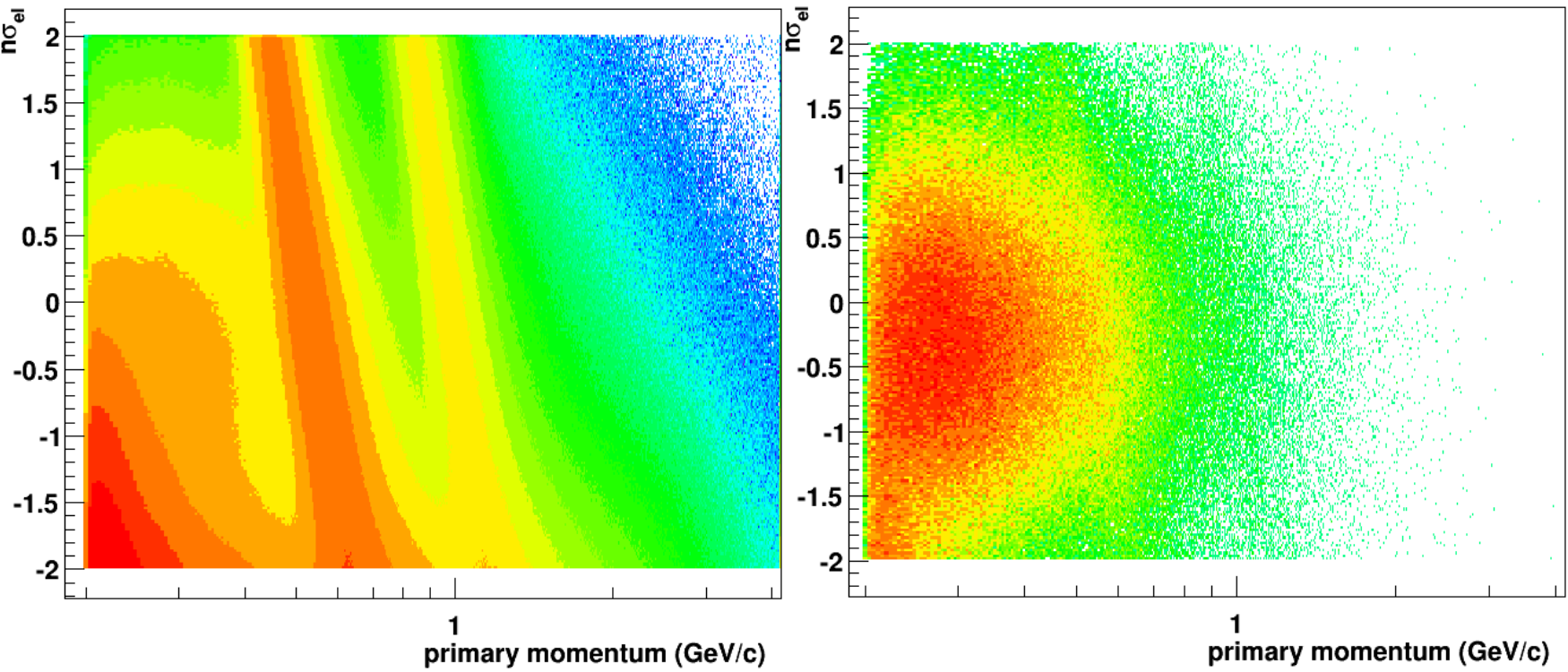}{TPC energy loss distribution versus primary momentum for the electron/positron candidates (left) and the final sample (right). In the candidates sample, the leptons' distribution around zero is hidden beneath hadron contaminations by pions, kaons and protons (narrow bands in increasing momentum). Note that the latter (almost) vanish in the final particle sample.}{effcorr_fig_elec_sample_dEdx}{0.83} %

In addition to their importance in the QA of embedding samples, the final
experimental $e^+/e^-$ samples are used to derive the dominant
systematic uncertainties by comparison of simulated and measured track quality
as well as global DCA distributions. Since a slow simulator for the newly installed
TOF detector as plugin to GSTAR has not been available during the time frame of
this thesis, the final experimental $e^+/e^-$ samples also serve to
calculate TOF matching efficiencies. Moreover, due to its abundant statistics,
the pion samples play a crucial role in the determination of TOF matching
efficiencies (Section~\ref{effcorr_subsec_tofmatch}).

\section{Single-{}Track Efficiencies}
\label{effcorr_sec_single}\hyperlabel{effcorr_sec_single}%

In the following subsections, the particle samples produced in
Section~\ref{effcorr_sec_samples} are used to determine detector-{} and analysis-{}related
efficiencies step by step and separately in particle charges. In general, the
calculations follow the raw data analysis procedure of Chapter~\ref{data_analysis} and
the according selection criteria listed in Table~\ref{ana_tab_dsets_evttrk}. The experimental
as well as simulated particle samples are first reduced to tracks abiding the
criteria for event vertex and minimum reference multiplicity at the respective
energy. For the embedding samples, however, special attention has to be paid to
the removal of electrons and positrons (\emph{i}) emanating from photon conversions
in the detector material (\texttt{Par\penalty5000 e\penalty5000 n\penalty5000 t\penalty5000 G\penalty5000 e\penalty5000 a\penalty5000 n\penalty5000 tId=\penalty0 0}), and (\emph{ii}) sharing a minimum
number of common TPC hits between simulated and associated reconstructed tracks
(\texttt{NCo\penalty5000 m\penalty5000 m\penalty5000 o\penalty5000 n\penalty5000 H\penalty5000 it>{}=\penalty0 10}). Separate histograms for the latter two are subsequently
filled and divided at each step of the analysis chain starting at an
appropriate initial binning in transverse momentum \emph{p$_{\text{T}}$}, pseudo-{}rapidity
\ensuremath{\eta}, and azimuthal angle \ensuremath{\phi}.  The total single track efficiencies
$\epsilon_\mathrm{total}^\mathrm{track}$ (Equation~\ref{effcorr_eq_total}) can
then be obtained through the product of contributions due to detector
inefficiencies (\ensuremath{\epsilon}$_{\text{det}}$), minimum track quality requirements
(\ensuremath{\epsilon}$_{\text{qual}}$), cuts on the global DCA (\ensuremath{\epsilon}$_{\text{glDCA}}$), and track matching to
hits in the TOF detector ($\epsilon_\mathrm{match}^\mathrm{TOF}$),
as well as TOF ($\epsilon_\mathrm{select}^\mathrm{TOF}$) and TPC
($\epsilon_\mathrm{select}^\mathrm{TPC}$) selection.  The
dependencies on \emph{p$_{\text{T}}$}, \ensuremath{\eta}, \ensuremath{\phi}, and momentum \emph{p} indicate the kinematic
variables used to determine and propagate the efficiencies for each of the
contributions. Note that the TOF selection efficiency has been tuned versus
particle momentum in Section~\ref{ana_sec_pid} to a constant value of 3\ensuremath{\sigma}.  Also
shown in Equation~\ref{effcorr_eq_total} are the forms and dependencies of the systematic
uncertainties \ensuremath{\Delta}\ensuremath{\epsilon}$_{\text{syst}}$ for the dominating contributions that are
taken into account in the determination of total systematic uncertainties on
the single track efficiencies. Losses and systematic uncertainties due to
detector inefficiencies as well as track quality and global DCA cuts are
covered in Section~\ref{effcorr_subsec_det_trackqual_dca}, whereas the ones due to TOF
Matching and TPC Selection are addressed in Section~\ref{effcorr_subsec_tofmatch} and
Section~\ref{effcorr_subsec_tpcsel}, respectively. Section~\ref{effcorr_subsec_total} summarizes
the resulting total single track efficiencies for electrons and positrons
including the propagated systematic uncertainties. The full set of efficiencies
and uncertainties are tabularized in Section~\ref{app_effcorr_tables}. For reasons of
clarity and conciseness, most of the figures in the following subsections use
electrons at $\sqrt{s_\mathrm{NN}}$ = 39 GeV as representative
examples for all energies and particles with supporting figures included in
Section~\ref{app_effcorr_figures}.

\begin{dbequation}[H]
\[\epsilon_\mathrm{total}^\mathrm{track}(p_T,\eta,\varphi) = \epsilon_\mathrm{det}(p_T,\eta,\varphi) \times \underbrace{\epsilon_\mathrm{qual}(p_T,\eta,\varphi) \times \epsilon_\mathrm{glDCA}(p_T)}_{\Delta\epsilon_\mathrm{syst}/\epsilon(p_T)} \times \underbrace{\epsilon_\mathrm{match}^\mathrm{TOF}(p_T,\eta,\varphi)}_{\Delta\epsilon_\mathrm{syst}(\eta,\varphi)} \times \underbrace{\epsilon_\mathrm{select}^\mathrm{TOF}}_\mathrm{= 0.9973} \times \underbrace{\epsilon_\mathrm{select}^\mathrm{TPC}(p)}_{\Delta\epsilon_\mathrm{syst}(p)}.\]
\caption{Calculation of total single track efficiencies and according systematic uncertainties.}
\label{effcorr_eq_total}\hyperlabel{effcorr_eq_total}%

\end{dbequation}

\subsection{Detector, Track Quality and Global DCA}
\label{effcorr_subsec_det_trackqual_dca}\hyperlabel{effcorr_subsec_det_trackqual_dca}%

As explained in Section~\ref{effcorr_sec_samples}, the embedding procedure already
emulates the simulation of the detector's response such that the detector
efficiency \ensuremath{\epsilon}$_{\text{det}}$ within STAR acceptance can be calculated as the ratio
of the number of reconstructed over the number of simulated tracks in a chosen
\emph{p$_{\text{T}}$}/\ensuremath{\eta}/\ensuremath{\phi} interval (bin). The fractions \ensuremath{\epsilon}$_{\text{fit}}$ and
\ensuremath{\epsilon}$_{\text{dedx}}$ of tracks that satisfy the minimum track quality requirements in
Table~\ref{ana_tab_dsets_evttrk} for \texttt{nHi\penalty5000 t\penalty5000 s\penalty5000 Fit} and \texttt{nHi\penalty5000 t\penalty5000 s\penalty5000 d\penalty5000 Edx}, respectively, are then based
on the selection of simulated tracks remaining after losses due to the detector
response. The resulting \emph{p$_{\text{T}}$} dependencies of detector and track quality
efficiencies are depicted in Figure~\ref{effcorr_fig_dettrackqual_differential} for
electrons in an examplatory \ensuremath{\eta}/\ensuremath{\phi} window at
$\sqrt{s_\mathrm{NN}}$ = 39 GeV. The \ensuremath{\phi}-{}interval \ensuremath{\Delta}\ensuremath{\phi}
= 15\textdegree{} is chosen according to the detector's sector-{}wise segmentation and
compared to \ensuremath{\Delta}\ensuremath{\phi} = 5\textdegree{}. The \emph{p$_{\text{T}}$}-{}dependency's shape varies
considerably with the narrower azimuthal binning but at the same time suffers
enlarged statistical errors. Similar behavior can be observed when comparing
different choices of \ensuremath{\eta}-{}intervals albeit to much lesser extent.  The two
effects compete when aiming for a better handle on the efficiencies'
differential shapes, on the one hand, as opposed to avoiding sizeable
statistical fluctuations dictated by the limitations of the embedding sample,
on the other hand. The related variations, however, should not be propagated as
part of the systematic uncertainties. Instead, one needs to arrive at a
conscious decision on a binning in \emph{p$_{\text{T}}$}, \ensuremath{\eta}, and \ensuremath{\phi} capable of
reproducing the predominant features of the efficiencies' integrated \emph{p$_{\text{T}}$}-{},
\ensuremath{\eta}, and \ensuremath{\phi} dependencies.

\wrapifneeded{0.50}{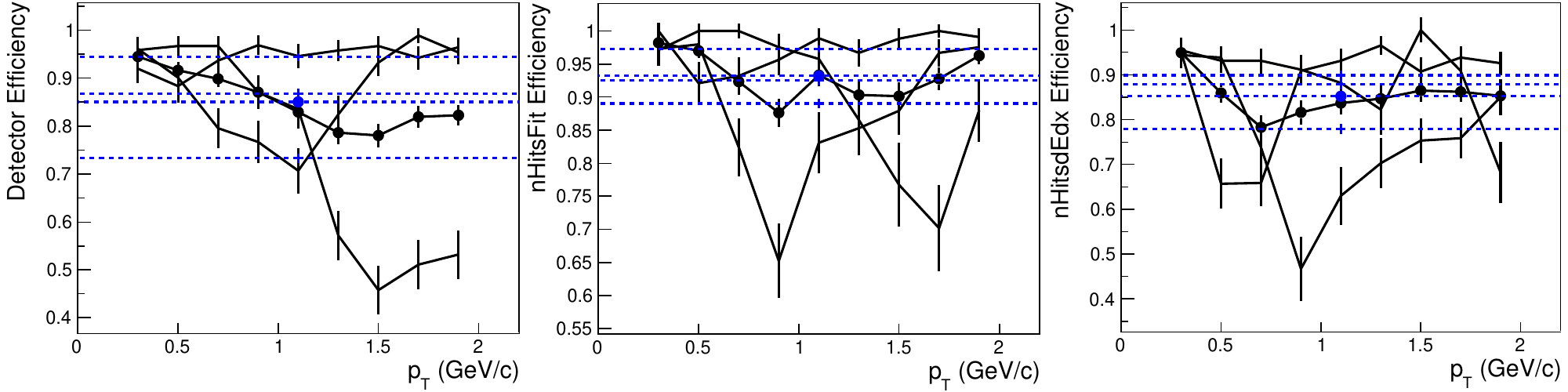}{\emph{e$^{\text{-{}}}$} efficiency comparisons at \ensuremath{\surd}s$_{\text{NN}}$ = 39 GeV for detector response and minimum track quality requirements in 0 <{} \ensuremath{\eta} <{} 0.25 and 45\textdegree{} <{} \ensuremath{\phi} <{} 60\textdegree{} for \ensuremath{\Delta}\ensuremath{\phi} = 15\textdegree{} (lines \& points) versus 5\textdegree{} (lines only). Solid lines show the \emph{p$_{\text{T}}$}-{}dependence for \emph{\ensuremath{\Delta}p$_{\text{T}}$} = 0.2 GeV/c and dashed lines the \emph{p$_{\text{T}}$}-{}integrated efficiencies for the respective \ensuremath{\Delta}\ensuremath{\phi}. All combined efficiencies for \ensuremath{\Delta}\ensuremath{\phi} = 15\textdegree{} are listed in Table~\ref{app_tab_effs_dettrackqual}.}{effcorr_fig_dettrackqual_differential}{1} %

\wrapifneeded{0.50}{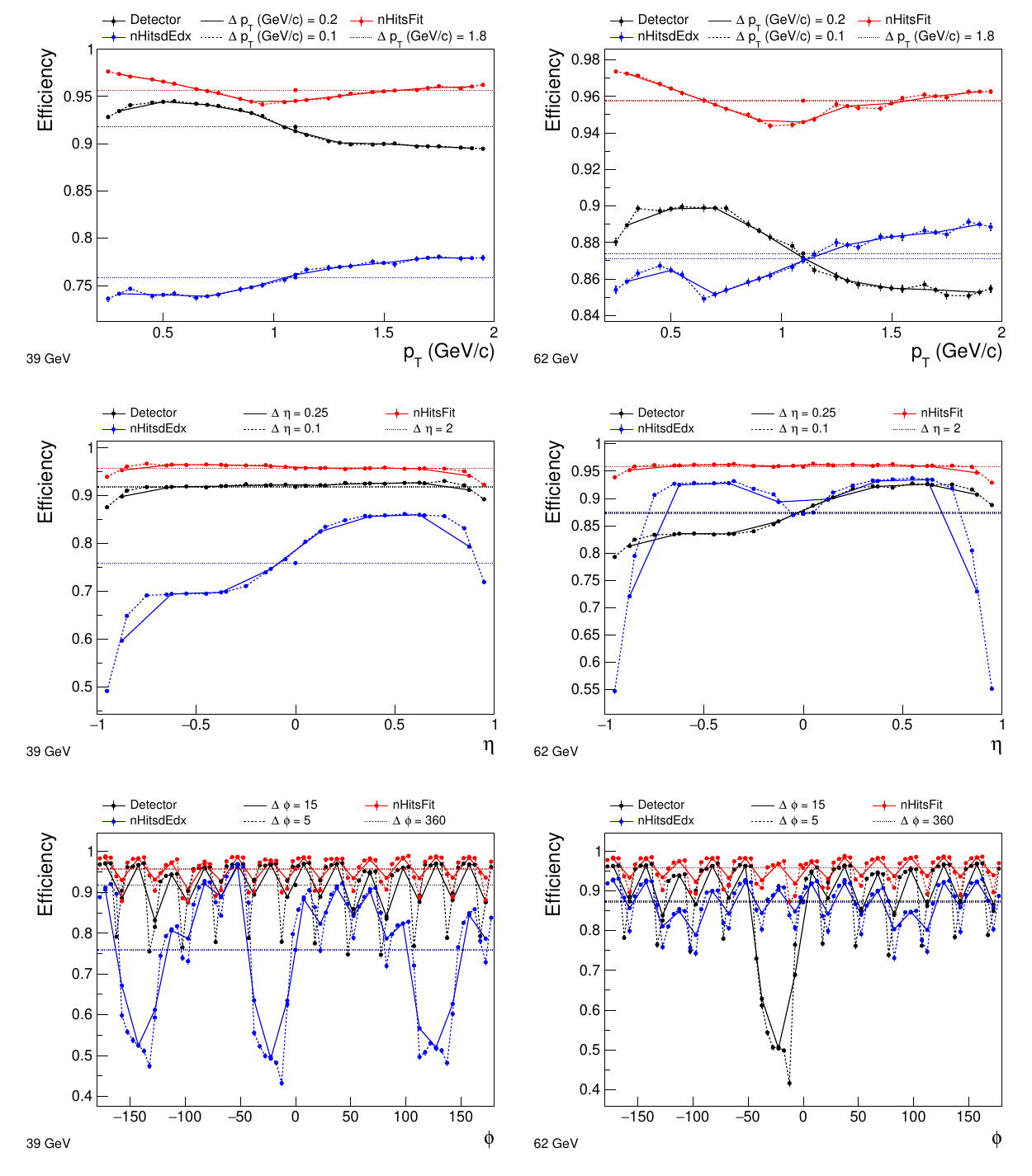}{Overview of calculated electron efficiencies at \ensuremath{\surd}s$_{\text{NN}}$ = 39 GeV (left column) and 62.4 GeV (right column) separately for detector response (black) as well as minimum track quality requirements for \texttt{nHi\penalty5000 t\penalty5000 s\penalty5000 Fit} (red) and \texttt{nHi\penalty5000 t\penalty5000 s\penalty5000 d\penalty5000 Edx} (blue) by means of their respective \emph{p$_{\text{T}}$}-{}, \ensuremath{\eta}-{}, and \ensuremath{\phi}-{}dependencies (top to bottom). For each of the kinematic variables and analysis steps, the dependencies are compared to the fully integrated values (dotted, Table~\ref{effcorr_tab_dettrackqual_integrated}) and to two different suitable choices in binning (dashed/solid lines): \emph{\ensuremath{\Delta}p$_{\text{T}}$} = 0.1/0.2 GeV/c, \ensuremath{\Delta}\ensuremath{\eta} = 0.1/0.25, and \ensuremath{\Delta}\ensuremath{\phi} = 5/15\textdegree{}. See Section~\ref{app_effcorr_figures} for the corresponding figures of other particles and energies.}{effcorr_fig_dettrackqual_overview}{1} %

For this purpose, two reasonable interval choices for the kinematic variables
\emph{p$_{\text{T}}$}, \ensuremath{\eta}, and \ensuremath{\phi} are compared in Figure~\ref{effcorr_fig_dettrackqual_overview} for the according dependencies of detector and track quality efficiencies at
$\sqrt{s_\mathrm{NN}}$ = 39 \& 62.4 GeV. In comparison to their
narrower counterparts, it turns out that the wider choices of \emph{\ensuremath{\Delta}p$_{\text{T}}$} =
0.2 GeV/c, \ensuremath{\Delta}\ensuremath{\eta} = 0.25, and \ensuremath{\Delta}\ensuremath{\phi} = 15\textdegree{} still sufficiently
account not only for the efficiencies' comparably smooth dependence on \emph{p$_{\text{T}}$} but also for their more prominent features in \ensuremath{\eta} and \ensuremath{\phi}. This includes
the differences in positive and negative pseudo-{}rapidities and, in particular,
the azimuthal structures reflecting the lower efficiencies at sector borders
and the overall reduced efficiencies in the less performing sectors.
Figure~\ref{effcorr_fig_dettrackqual_overview} also serves the purpose of giving an
overview comparison across energies allowing for a better understanding of the
detector's performance during the respective beam times and possibly revealing
issues characteristic to a specific energy.  For instance, the number of
missing or mal-{}functioning TPC read-{}out boards (\emph{bad RDOs}) varies considerably
during a run and especially over the time period in which the BES-{}I data has
been taken (\ensuremath{\sim} two years). The reduced efficiencies for some of the sectors
visible in the \ensuremath{\phi}-{}dependencies (c.f. additional figures in
Section~\ref{app_effcorr_figures}) can be correlated with the number of bad RDOs in
different RunID ranges of an energy run. For details, the reader is referred to
the specially developed STAR software module \emph{StBadRdosDb} in Section~\ref{stbadrdos} providing the required databases (Section~\ref{app_stbadrdos_dbfiles}) and interface
utilities. The higher number of bad RDOs during the runs in 2010 as compared to
2011 also explains the lower fully integrated efficiencies for 39 and 62.4 GeV
versus 19.6 and 27 GeV in Table~\ref{effcorr_tab_dettrackqual_integrated}.
\begin{center}
\begingroup%
\setlength{\newtblsparewidth}{\linewidth-2\tabcolsep-2\tabcolsep-2\tabcolsep-2\tabcolsep-2\tabcolsep-2\tabcolsep-2\tabcolsep-2\tabcolsep-2\tabcolsep-2\tabcolsep-2\tabcolsep-2\tabcolsep-2\tabcolsep}%
\setlength{\newtblstarfactor}{\newtblsparewidth / \real{96}}%

\begin{longtable}{llllllllllll}\caption[{Integrated detector (\ensuremath{\epsilon}$_{\text{det}}$), \texttt{nHi\penalty5000 t\penalty5000 s\penalty5000 Fit} (\ensuremath{\epsilon}$_{\text{fit}}$), and \texttt{nHi\penalty5000 t\penalty5000 s\penalty5000 d\penalty5000 Edx} (\ensuremath{\epsilon}$_{\text{dedx}}$) efficiencies.}]{Integrated detector (\ensuremath{\epsilon}$_{\text{det}}$), \texttt{nHi\penalty5000 t\penalty5000 s\penalty5000 Fit} (\ensuremath{\epsilon}$_{\text{fit}}$), and \texttt{nHi\penalty5000 t\penalty5000 s\penalty5000 d\penalty5000 Edx} (\ensuremath{\epsilon}$_{\text{dedx}}$) efficiencies.\label{effcorr_tab_dettrackqual_integrated}\hyperlabel{effcorr_tab_dettrackqual_integrated}%
}\tabularnewline
\endfirsthead
\caption[]{(continued)}\tabularnewline
\endhead
\hline
\multicolumn{3}{m{8\newtblstarfactor+2\tabcolsep+\arrayrulewidth+8\newtblstarfactor+2\tabcolsep+\arrayrulewidth+8\newtblstarfactor+\arrayrulewidth}}{\centering%
19.6 GeV
}&\multicolumn{3}{m{8\newtblstarfactor+2\tabcolsep+\arrayrulewidth+8\newtblstarfactor+2\tabcolsep+\arrayrulewidth+8\newtblstarfactor+\arrayrulewidth}}{\centering%
27 GeV
}&\multicolumn{3}{m{8\newtblstarfactor+2\tabcolsep+\arrayrulewidth+8\newtblstarfactor+2\tabcolsep+\arrayrulewidth+8\newtblstarfactor+\arrayrulewidth}}{\centering%
39 GeV
}&\multicolumn{3}{m{8\newtblstarfactor+2\tabcolsep+\arrayrulewidth+8\newtblstarfactor+2\tabcolsep+\arrayrulewidth+8\newtblstarfactor+\arrayrulewidth}}{\centering%
62.4 GeV
}\tabularnewline
\multicolumn{1}{m{8\newtblstarfactor+\arrayrulewidth}}{\centering%
\ensuremath{\epsilon}$_{\text{det}}$
}&\multicolumn{1}{m{8\newtblstarfactor+\arrayrulewidth}}{\centering%
\ensuremath{\epsilon}$_{\text{fit}}$
}&\multicolumn{1}{m{8\newtblstarfactor+\arrayrulewidth}}{\centering%
\ensuremath{\epsilon}$_{\text{dedx}}$
}&\multicolumn{1}{m{8\newtblstarfactor+\arrayrulewidth}}{\centering%
\ensuremath{\epsilon}$_{\text{det}}$
}&\multicolumn{1}{m{8\newtblstarfactor+\arrayrulewidth}}{\centering%
\ensuremath{\epsilon}$_{\text{fit}}$
}&\multicolumn{1}{m{8\newtblstarfactor+\arrayrulewidth}}{\centering%
\ensuremath{\epsilon}$_{\text{dedx}}$
}&\multicolumn{1}{m{8\newtblstarfactor+\arrayrulewidth}}{\centering%
\ensuremath{\epsilon}$_{\text{det}}$
}&\multicolumn{1}{m{8\newtblstarfactor+\arrayrulewidth}}{\centering%
\ensuremath{\epsilon}$_{\text{fit}}$
}&\multicolumn{1}{m{8\newtblstarfactor+\arrayrulewidth}}{\centering%
\ensuremath{\epsilon}$_{\text{dedx}}$
}&\multicolumn{1}{m{8\newtblstarfactor+\arrayrulewidth}}{\centering%
\ensuremath{\epsilon}$_{\text{det}}$
}&\multicolumn{1}{m{8\newtblstarfactor+\arrayrulewidth}}{\centering%
\ensuremath{\epsilon}$_{\text{fit}}$
}&\multicolumn{1}{m{8\newtblstarfactor+\arrayrulewidth}}{\centering%
\ensuremath{\epsilon}$_{\text{dedx}}$
}\tabularnewline
\multicolumn{1}{m{8\newtblstarfactor+\arrayrulewidth}}{\centering%
92\%
}&\multicolumn{1}{m{8\newtblstarfactor+\arrayrulewidth}}{\centering%
95\%
}&\multicolumn{1}{m{8\newtblstarfactor+\arrayrulewidth}}{\centering%
91\%
}&\multicolumn{1}{m{8\newtblstarfactor+\arrayrulewidth}}{\centering%
92\%
}&\multicolumn{1}{m{8\newtblstarfactor+\arrayrulewidth}}{\centering%
96\%
}&\multicolumn{1}{m{8\newtblstarfactor+\arrayrulewidth}}{\centering%
91\%
}&\multicolumn{1}{m{8\newtblstarfactor+\arrayrulewidth}}{\centering%
92\%
}&\multicolumn{1}{m{8\newtblstarfactor+\arrayrulewidth}}{\centering%
96\%
}&\multicolumn{1}{m{8\newtblstarfactor+\arrayrulewidth}}{\centering%
76\%
}&\multicolumn{1}{m{8\newtblstarfactor+\arrayrulewidth}}{\centering%
87\%
}&\multicolumn{1}{m{8\newtblstarfactor+\arrayrulewidth}}{\centering%
96\%
}&\multicolumn{1}{m{8\newtblstarfactor+\arrayrulewidth}}{\centering%
87\%
}\tabularnewline
\hline
\end{longtable}\endgroup%

\end{center}

A \emph{side note} on the statistical error bars used in
Figure~\ref{effcorr_fig_dettrackqual_differential} and later in
Figure~\ref{effcorr_fig_tofmatch_differential}: Since $\hat{\epsilon}=k/n$ intuitively provides a good estimator for the (unknown) true efficiency,
calculations of the latter boil down to the division of two histogram bin
contents \emph{k} and \emph{n} in the same \emph{p$_{\text{T}}$}/\ensuremath{\eta}/\ensuremath{\phi}-{}interval with \emph{k} (out of
\emph{n}) the number of tracks passing the selection criterion. It proves more
difficult, however, to assign the correct statistical uncertainties on the
calculated efficiencies. As discussed in [184], using poisson or
binomial distributions to derive the mean efficiency
$\bar{\epsilon}$ and its variance $\sigma_\epsilon^2$ either falsely assumes non-{}correlated \emph{k} and \emph{n} or results in unphysical zero
values for the border cases $k=0$ or $k=n$. The correct
probability density function derived in [184], on the other hand,
not only exhibits a most probable value of $\hat{\epsilon}$ but
also dismisses the ill-{}behaving variances in favor of
\[\sigma_\epsilon=\sqrt{\frac{(k+1)(k+2)}{(n+2)(n+3)}-\left(\frac{k+1}{n+2}\right)^2}\]
which yields physical non-{}zero values in the extreme cases for \emph{k}.\newline

For reasons of quality assurance, Figure~\ref{effcorr_fig_trackqualdca_distros} compares
track quality and global DCA distributions produced via embedding to
distributions obtained from the experimental particle samples. All
distributions generally show good agreement between the simulated and measured
data creating confidence in the quality of the embedding samples. The level of
agreement in the \texttt{nHi\penalty5000 t\penalty5000 s\penalty5000 d\penalty5000 Edx} distributions, in particular, is worth
highlighting since it only recently emerged from a STAR-{}internal effort to
improve the simulation of a particle's energy loss in the TPC during the
embedding procedure. As a consequence, \texttt{nHi\penalty5000 t\penalty5000 s\penalty5000 d\penalty5000 Edx} efficiencies can now
confidently be derived from simulated distributions instead of, as previously,
relying on the experimental ones.\newline
 In the case of global DCA, the according embedding distributions in
Figure~\ref{effcorr_fig_trackqualdca_distros} also serve to derive \emph{p$_{\text{T}}$}-{}dependent
efficiencies for the restriction of the particle samples to tracks with global
DCAs (glDCA) of less than 1 cm. The electrons and positrons from physical
sources sought after in this thesis emanate from the primary vertex which is
why the data analysis (Chapter~\ref{data_analysis}) and the detector efficiencies
(Figure~\ref{effcorr_fig_dettrackqual_differential} and Table~\ref{app_tab_effs_dettrackqual})
are based on primary instead of global tracks. However, only tracks with glDCA
<{} 3 cm are tagged as primary and associated with the global partner. Given the
glDCA distribution's shape, the bulk of the tracks appears within the glDCA
<{} 1 cm criterion resulting in \ensuremath{\epsilon}$_{\text{glDCA}}$ close to 1.

\wrapifneeded{0.50}{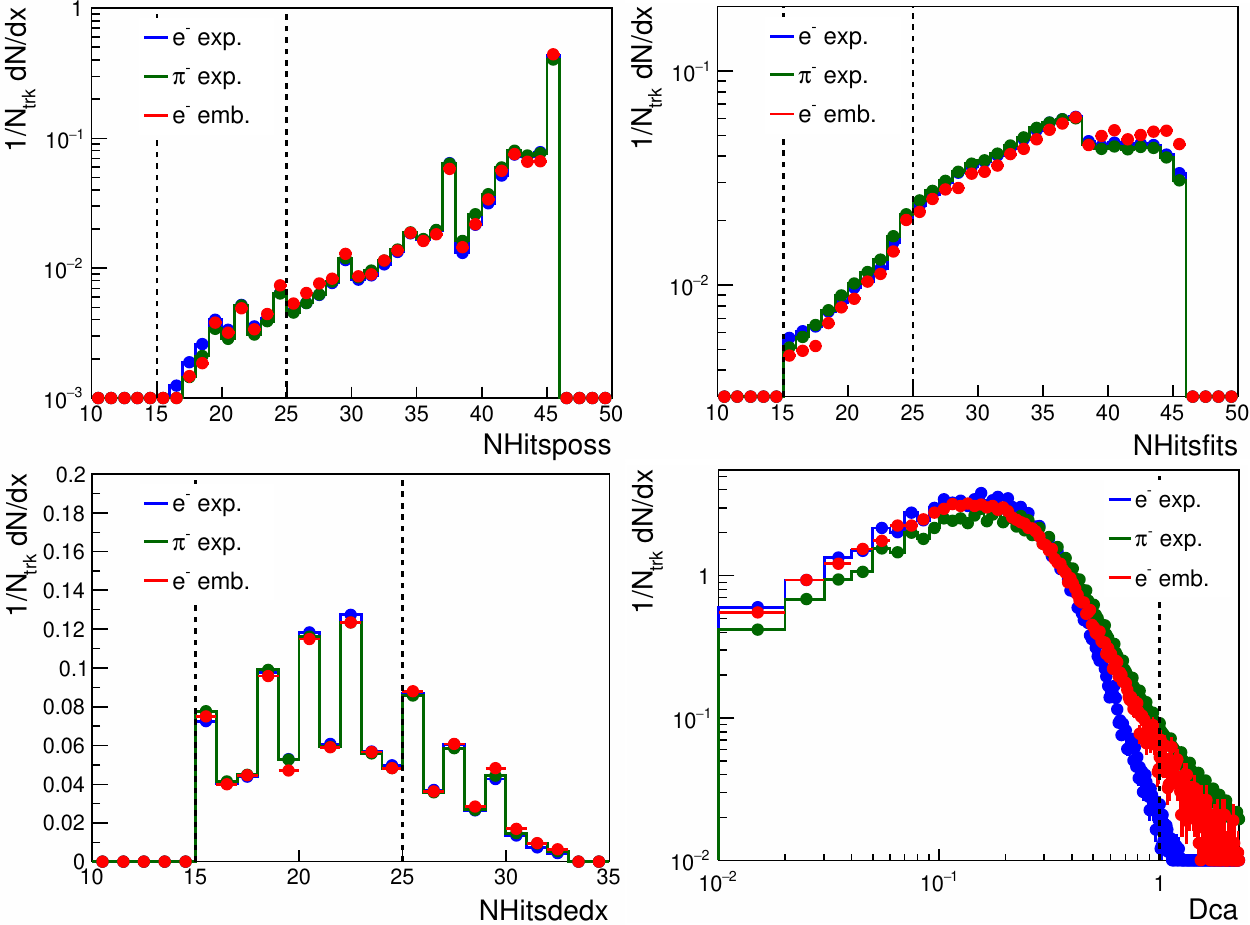}{Comparison of track quality (\texttt{nHi\penalty5000 t\penalty5000 s\penalty5000 P\penalty5000 oss}, \texttt{nHi\penalty5000 t\penalty5000 s\penalty5000 Fit}, \texttt{nHi\penalty5000 t\penalty5000 s\penalty5000 d\penalty5000 Edx}) and global DCA distributions for electrons from embedding (red) as well as electrons (blue) and \ensuremath{\pi}$^{\text{-{}}}$ (green) from experimental data at \ensuremath{\surd}s$_{\text{NN}}$ = 39 GeV for 0.5 <{} \emph{p$_{\text{T}}$} <{} 0.8 GeV/c. Vertical dashed lines denote selection criteria used in the determination of relative systematic uncertainties on the combined track quality and global DCA efficiencies (see text following this figure).}{effcorr_fig_trackqualdca_distros}{0.9} %

Relative systematic uncertainties on the combined track quality and global DCA
efficiencies can be estimated by quantifying how well each of the embedding
distributions reproduces the efficiencies calculated from the according
experimental distribution. Note that, for reasons of simplification during the
raw data analysis, the latter have been produced using the same minimum track
quality requirements as chosen in the data analysis (see Table~\ref{ana_tab_dsets_evttrk})
which also conincide with the lowest values recommended by STAR to define a
good track.  Hence, for the determination of systematic uncertainties on
track quality efficiencies, an alternate minimum requirement of
\texttt{nHi\penalty5000 t\penalty5000 s\penalty5000 P\penalty5000 oss/\penalty0 Fit/\penalty0 dEdx} >{} 24 is employed (c.f. vertical lines in
Figure~\ref{effcorr_fig_trackqualdca_distros}). For each of the distributions, the
respective selection criteria are then used to calculate the relative
efficiencies difference
\[\Delta\epsilon_\mathrm{syst}^\mathrm{rel}= \frac{\vert\epsilon_\mathrm{exp}-\epsilon_\mathrm{emb}\vert}{\epsilon_\mathrm{emb}}\]
with \ensuremath{\epsilon}$_{\text{exp}}$ and \ensuremath{\epsilon}$_{\text{emb}}$ the efficiencies deduced from the
experimental and embedding distributions, respectively. The resulting
$\Delta\epsilon_\mathrm{syst}^\mathrm{rel}$ for each of the track
quality and global DCA distributions are propagated quadratically to obtain the
total relative uncertainties compiled in
Figure~\ref{effcorr_fig_trackqualdca_relsyserr}. For all particles and at all energies,
the \emph{p$_{\text{T}}$}-{}interval 0.5 -{} 0.8 GeV/c exhibits the smallest uncertainties and
hence best overall agreement in the underlying distributions. At the energies
19.6 and 27 GeV, the uncertainties for electrons and positrons in this momentum
bin are about 2\% and well below the ones derived from the comparison to pion
distributions. At 39 and 62.4 GeV, however, these uncertainties range around 5\%
roughly in agreement with the pions except for a 10\% uncertainty for positrons
at 39 GeV. The trend of higher systematic uncertainties at 39 and 62.4 GeV with
respect to 19.6 and 27 GeV is possibly caused by mis-{}matching \texttt{nHi\penalty5000 t\penalty5000 s\penalty5000 d\penalty5000 Edx} distributions due to the larger number of (RunID ranges with) bad RDOs
(Section~\ref{stbadrdos} and Section~\ref{app_stbadrdos_dbfiles}), especially at 39 GeV. For most
energies and particles, one observes systematically larger uncertainties in the
lowest and the higher \emph{p$_{\text{T}}$}-{}intervals which can probably be attributed,
respectively, to the residual pion and kaon/proton contaminations in the
experimental samples (Figure~\ref{effcorr_fig_elec_sample_dEdx}). The combined relative
systematic uncertainties on track quality and global DCA efficiencies are
propagated in Section~\ref{effcorr_subsec_total} along with the systematic uncertainties
of TOF matching (Section~\ref{effcorr_subsec_tofmatch}) and TPC selection
(Section~\ref{effcorr_subsec_tpcsel}) efficiencies.

\wrapifneeded{0.50}{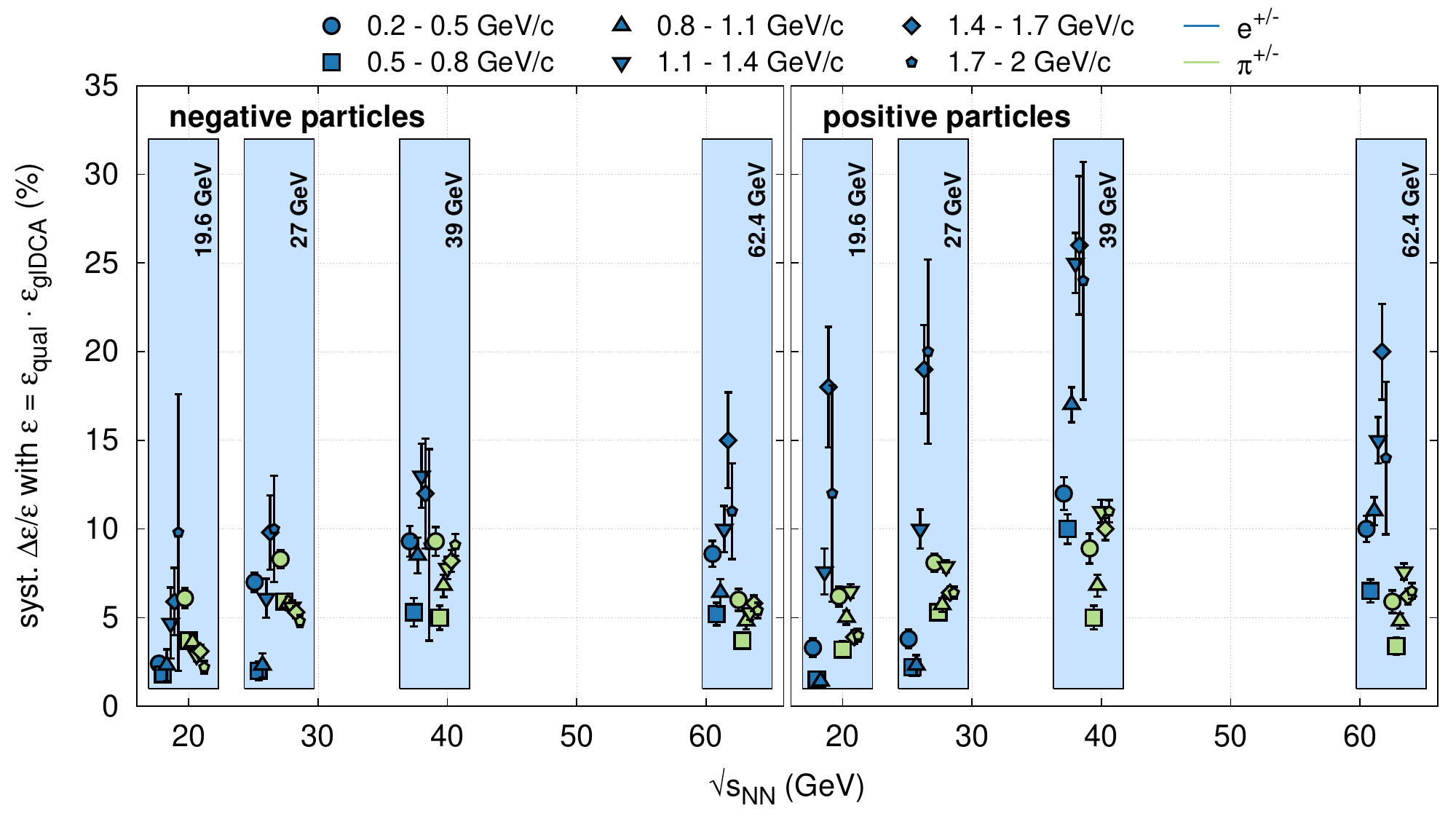}{\ensuremath{\surd}s$_{\text{NN}}$-{} and \emph{p$_{\text{T}}$}-{}dependence of total relative systematic uncertainties on track quality and global DCA efficiencies obtained through the comparison of electrons/positrons from embedding to electrons/positrons (blue symbols) and pions (green symbols) from experimental data, c.f. Figure~\ref{effcorr_fig_trackqualdca_distros}. The \emph{p$_{\text{T}}$}-{}dependence is based on intervals \emph{\ensuremath{\Delta}p$_{\text{T}}$} of 0.3 GeV/c between 0.2 -{} 2 GeV/c. The data points are slightly shifted against each other for better visibility and blue boxes frame the points belonging together at each energy.}{effcorr_fig_trackqualdca_relsyserr}{0.95} %

\subsection{TOF Matching}
\label{effcorr_subsec_tofmatch}\hyperlabel{effcorr_subsec_tofmatch}%

The measurement of a particle's velocity \ensuremath{\beta} via the TOF detector in
addition to its momentum and energy loss in the TPC is crucial for the
successful identification of rare electrons and positrons (Section~\ref{ana_sec_pid}).
Tracks measured in the TPC are therefore matched to TOF detector hits during
the first phase of the event reconstruction analysis chain
(Section~\ref{ana_sec_dsets_evttrk}). The according matching efficiencies are determined by
the selection criteria \ensuremath{\beta} >{} 0 and |yLocal| <{} 1.8 cm
(Table~\ref{ana_tab_dsets_evttrk}), and based on the sample of tracks remaining after detector
response, track quality requirements and global DCA cuts (c.f.
Equation~\ref{effcorr_eq_total}). Two major issues complicate the derivation of TOF
matching efficiencies: (\emph{i}) a slow simulator of the TOF's response for the use
in GSTAR has not been available as of this writing, and (\emph{ii}) the experimental
electron and positron samples generated in Section~\ref{effcorr_sec_samples} do not
provide sufficient statistics to calculate efficiencies in a narrow enough
\emph{p$_{\text{T}}$}/\ensuremath{\eta}/\ensuremath{\phi} binning without being dominated by statistical errors.
Dependence on \emph{p$_{\text{T}}$}/\ensuremath{\eta}/\ensuremath{\phi} during the determination of efficiencies,
however, is imperative since the differential efficiency shapes vary
significantly for different kinematic regions. A possible solution is to
properly scale the TOF matching efficiencies derived from pion samples to the
ones from electrons/positrons given that their shape is satisfactorily
reproduced. Figure~\ref{effcorr_fig_tofmatch_differential} explores this solution by
comparing \emph{p$_{\text{T}}$}-{}dependent electron to \ensuremath{\pi}$^{\text{-{}}}$ TOF matching efficiences for the
different scenarios from \ensuremath{\eta}/\ensuremath{\phi}-{}integrated to \ensuremath{\eta}/\ensuremath{\phi}-{}differential.
For each scenario of \ensuremath{\eta}/\ensuremath{\phi}-{}interval choices the pion efficiencies (red
lines) are independently scaled to the respective electron efficiencies
(blue symbols) using their \emph{p$_{\text{T}}$}-{}averaged ratio as scaling factor.

\wrapifneeded{0.50}{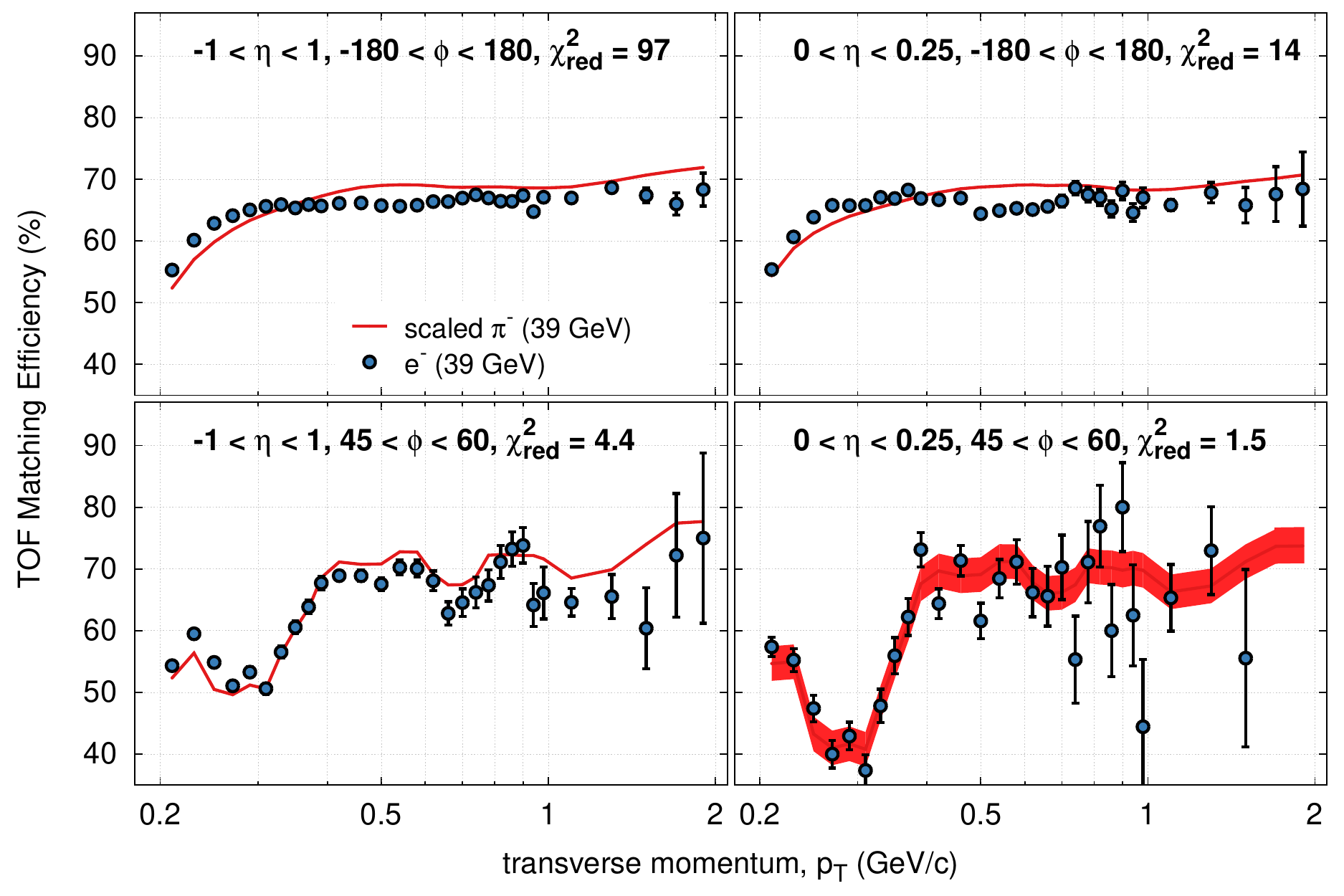}{\emph{p$_{\text{T}}$}-{}dependent comparison of TOF matching efficiencies for electrons (blue symbols) and scaled \ensuremath{\pi}$^{\text{-{}}}$ (red lines) at \ensuremath{\surd}s$_{\text{NN}}$ = 39 GeV using combinations of the interval choices \ensuremath{\Delta}\ensuremath{\eta} = 2 \& 0.25, and \ensuremath{\Delta}\ensuremath{\phi} = 15\textdegree{} \& 360\textdegree{}. The reduced \ensuremath{\chi}$^{\text{2}}$ is a measure for the goodness-{}of-{}fit between electrons and pions. The band around the pion efficiencies in the fully differential scenario indicates \emph{p$_{\text{T}}$}-{}averaged systematic uncertainties. Details see text. All TOF matching efficiencies finally utilized in the efficiency correction are listed in Table~\ref{app_tab_effs_tofmatchtotal0} and Table~\ref{app_tab_effs_tofmatchtotal1} for \ensuremath{\eta} <{} 0 and \ensuremath{\eta} >{} 0, respectively.}{effcorr_fig_tofmatch_differential}{0.9} %

The level of agreement between \emph{p$_{\text{T}}$}-{}dependent electron and pion efficiencies
can be quantified using the reduced \ensuremath{\chi}$^{\text{2}}$ as a measure for the
"goodness-{}of-{}fit". As for its usual parameterization via \ensuremath{\chi}$^{\text{2}}$ minimization,
a value around one indicates a good description of the data (electron
efficiencies) by the model (scaled pion efficiencies) taking into account the
statistical errors on the data. The reduced \ensuremath{\chi}$^{\text{2}}$ values calculated for the
different scenarios in Figure~\ref{effcorr_fig_tofmatch_differential} support the visual
impression that electron efficiencies can only be reproduced well by the pion
efficiencies in the fully differential case (\ensuremath{\Delta}\ensuremath{\eta} = 0.25, \ensuremath{\Delta}\ensuremath{\phi}
= 15\textdegree{}) and not when integrating over one or both of the kinematic
variables \ensuremath{\eta} and \ensuremath{\phi}. The reproducibility in shape also allows for the
\emph{p$_{\text{T}}$}-{}averaged absolute deviation of electron from scaled pion efficiencies to
be taken as estimate for the systematic uncertainty in the respective
\ensuremath{\eta}/\ensuremath{\phi}-{}interval (red band in Figure~\ref{effcorr_fig_tofmatch_differential}).\newline
 To further scrutinize the validity of the procedure,
Figure~\ref{effcorr_fig_tofmatch_syserr_scale_chi2} shows normalized distributions of
the factors \emph{F} involved in the scaling of pion efficiencies as well as of the
reduced \ensuremath{\chi}$^{\text{2}}$ goodness-{}of-{}fit parameters and the resulting systematic
uncertainties on the TOF matching efficiencies. For better comparison and
clarity, all distributions are shifted and scaled by the energy-{}averaged
gaussian means and widths to exhibit $\mu=0$ and
$\sigma=1$, respectively.  The original mean and widths are quoted
in the labels. The first important observation is that 95\% of all reduced
\ensuremath{\chi}$^{\text{2}}$ values for electrons and positrons at all energies lie between 0.5 and
2.1 with a mean of about 1.4 and no significant differences amongst energies or
particles (right column). This strongly supports the previous conclusion that
the pion efficiencies' shapes agree well with the ones of electrons and
positrons. Since, moreover, also 95\% of the scaling factors are in the range
1.01 -{} 1.17 with a mean of about 1.1 (middle column), the scaled pion
efficiencies can confidently be used as a replacement for the statistically
challenged electrons/positrons. The left column in
Figure~\ref{effcorr_fig_tofmatch_syserr_scale_chi2} reveals overall mean systematic
uncertainties of about 5\% at 19.6 GeV and about 3\% at the other energies.

\wrapifneeded{0.50}{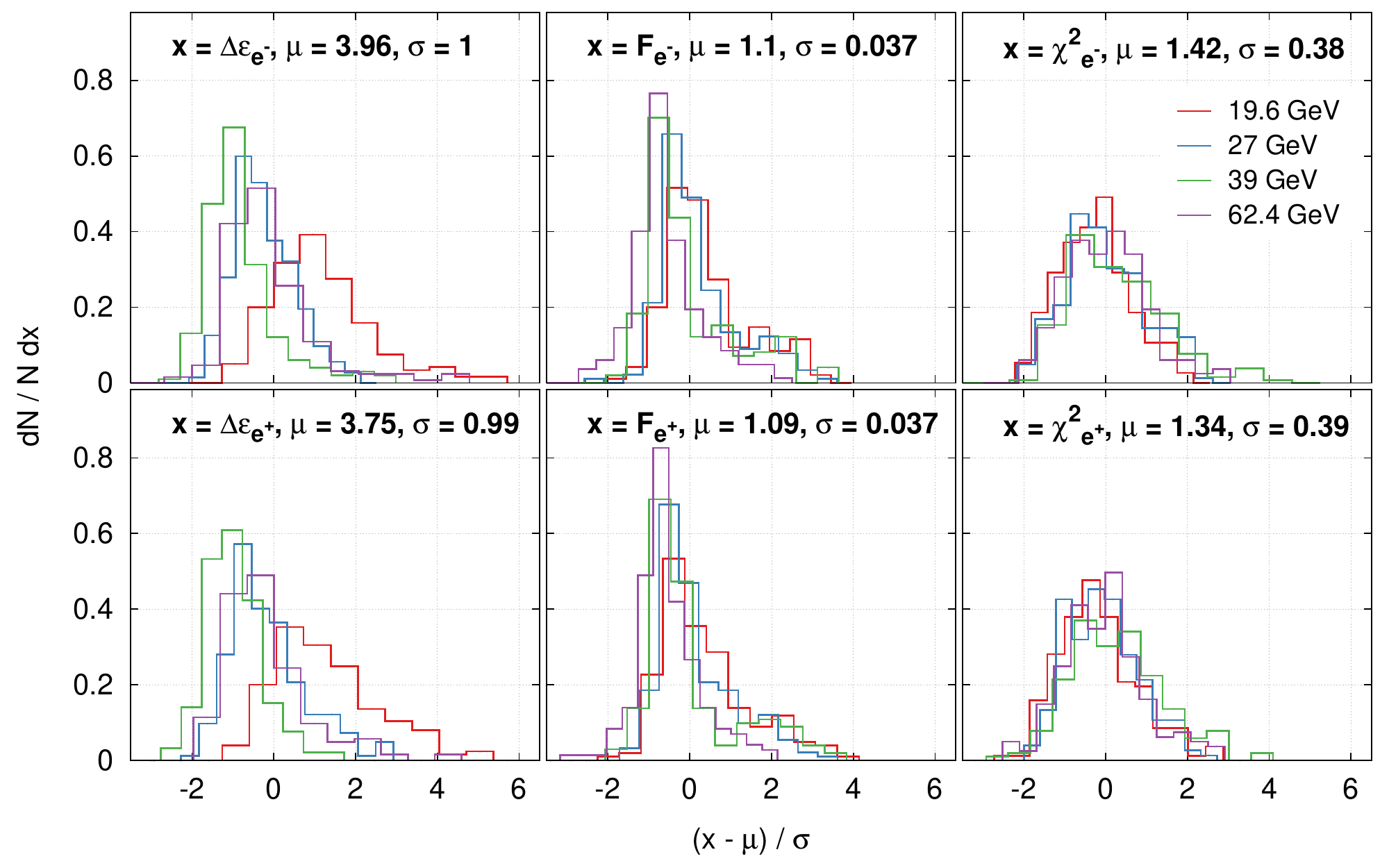}{Distributions of systematic uncertainties \ensuremath{\Delta}\ensuremath{\epsilon} on TOF matching efficiencies (left column), scaling factors \emph{F} for pion efficiencies (middle column) and reduced \ensuremath{\chi}$^{\text{2}}$ (right column) for electrons (top panel) and positrons (bottom panel) at all energies determined in \ensuremath{\eta}-{} and \ensuremath{\phi}-{}intervals of 0.25 and 15\textdegree{}, respectively. Details see text. The data underlying the distributions is listed in Table~\ref{app_tab_effs_tofmatch_extra}.}{effcorr_fig_tofmatch_syserr_scale_chi2}{0.9} %

\subsection{TPC Selection}
\label{effcorr_subsec_tpcsel}\hyperlabel{effcorr_subsec_tpcsel}%

Since the TPC's response and related efficiency losses are already accounted
for in Section~\ref{effcorr_subsec_det_trackqual_dca} (\ensuremath{\epsilon}$_{\text{det}}$), the final step
remaining in the calculation of total single track efficiencies (c.f.
 Equation~\ref{effcorr_eq_total}) is the determination of TPC selection efficiencies
for the momentum-{}dependent particle identification employed in Section~\ref{ana_sec_pid} via the specific energy loss in the TPC. The respective selection cuts are
implemented in a continuous functional form which is why the according
efficiences can be derived solely from the assumedly constant gaussian mean and
width of the underlying n\ensuremath{\sigma}$_{\text{el}}$ distribution. Figure~\ref{effcorr_fig_tpcselect} gives an overview of the procedure described in the following.

\wrapifneeded{0.50}{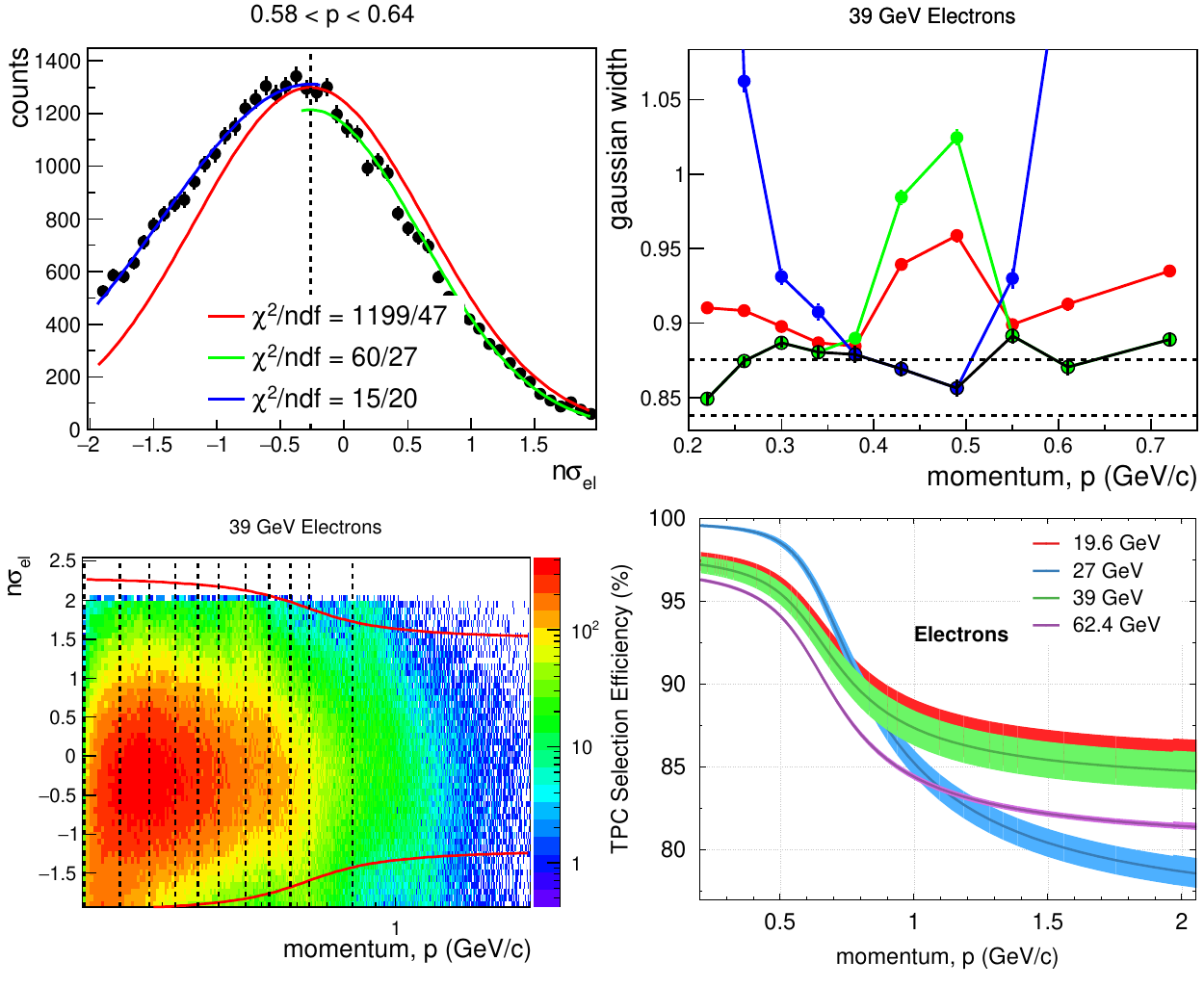}{Procedure for the determination of TPC selection efficiencies for the experimental electron sample at \ensuremath{\surd}s$_{\text{NN}}$ = 39 GeV if not noted otherwise. (upper left) n\ensuremath{\sigma}$_{\text{el}}$ distribution for one of the momentum intervals parameterized via three gaussians with fixed mean (vertical dashed line) over the full range (red) and the two tails (blue, green). (upper right) The resulting gaussian widths versus momentum with the smallest highlighted in open black circles. The upper dashed line denotes the according momentum-{}averaged width whereas the lower indicates the global width of the momentum-{}integrated n\ensuremath{\sigma}$_{\text{el}}$ distribution. (lower left) n\ensuremath{\sigma}$_{\text{el}}$ versus momentum with red lines framing the selection used to identify electrons (see Section~\ref{ana_sec_pid}). Vertical dashed lines indicate the momentum intervals used to generate momentum-{}dependent n\ensuremath{\sigma}$_{\text{el}}$ distributions. (lower right) Resulting continuous TPC selection efficiencies at all energies including systematic uncertainties. The full list of efficiencies and systematic uncertainties can be found in Table~\ref{app_tab_effs_tpcselect}.}{effcorr_fig_tpcselect}{0.95} %

The experimental electron/positron samples of Section~\ref{effcorr_sec_samples} can be
used to calculate global gaussian means \ensuremath{\mu}$_{\text{gl}}$ and widths \ensuremath{\sigma}$_{\text{gl}}$ of the
respective momentum-{}integrated n\ensuremath{\sigma}$_{\text{el}}$ distributions. However, an accurate
gaussian width determination in this manner is impeded by hadron contaminations
in either tail of the n\ensuremath{\sigma}$_{\text{el}}$ distributions at different momenta. Hence,
   each projection of the n\ensuremath{\sigma}$_{\text{el}}$ distributions in the momentum range 0.2
-{} 0.8 GeV/c (Figure~\ref{effcorr_fig_tpcselect} lower left) is first parameterized with
three gaussians independently over the full range and the two tails
(Figure~\ref{effcorr_fig_tpcselect} upper left). The fit functions are initialized with
\ensuremath{\mu}$_{\text{gl}}$ and \ensuremath{\sigma}$_{\text{gl}}$ where the mean is kept fixed and the width is limited
to 10. The widths resulting from the \ensuremath{\chi}$^{\text{2}}$ minimization are studied versus
the momentum and the smallest is selected as the \emph{most probable} width of the
distribution (Figure~\ref{effcorr_fig_tpcselect} upper right). This choice suggests
itself since the contamination bands of pion, kaon and proton run through the
constant electron band (\ensuremath{\langle}n\ensuremath{\sigma}$_{\text{el}}$\ensuremath{\rangle}\ensuremath{\sim}0) and hence effectively
broaden the width preferably on one side of the n\ensuremath{\sigma}$_{\text{el}}$ distribution.
Since the n\ensuremath{\sigma}$_{\text{el}}$ distribution's width is expected to be approximately
constant in momentum (\ensuremath{\sim}1), an optimized width \ensuremath{\sigma}$_{\text{opt}}$ is obtained by
averaging over momentum.  Its systematic uncertainty \ensuremath{\Delta}\ensuremath{\sigma}$_{\text{opt}}$ is
estimated by the larger of three times its standard deviation and
$\vert\sigma_\mathrm{gl}-\sigma_\mathrm{opt}\vert$.
Table~\ref{effcorr_tab_tpcselect} lists the global means as well as the optimized
gaussian widths and their uncertainties.
\begin{center}
\begingroup%
\setlength{\newtblsparewidth}{1\linewidth-2\tabcolsep-2\tabcolsep-2\tabcolsep-2\tabcolsep-2\tabcolsep-2\tabcolsep-2\tabcolsep}%
\setlength{\newtblstarfactor}{\newtblsparewidth / \real{425}}%

\begin{longtable}{llllll}\caption[{Global gaussian means \ensuremath{\mu}$_{\text{gl}}$ and optimized (momentum-{}averaged) gaussian widths \ensuremath{\sigma}$_{\text{opt}}$ with systematic uncertainties \ensuremath{\Delta}\ensuremath{\sigma}$_{\text{opt}}$ of experimental electron/positron n\ensuremath{\sigma}$_{\text{el}}$ distributions.}]{Global gaussian means \ensuremath{\mu}$_{\text{gl}}$ and optimized (momentum-{}averaged) gaussian widths \ensuremath{\sigma}$_{\text{opt}}$ with systematic uncertainties \ensuremath{\Delta}\ensuremath{\sigma}$_{\text{opt}}$ of experimental electron/positron n\ensuremath{\sigma}$_{\text{el}}$ distributions.\label{effcorr_tab_tpcselect}\hyperlabel{effcorr_tab_tpcselect}%
}\tabularnewline
\endfirsthead
\caption[]{(continued)}\tabularnewline
\endhead
\hline
\multicolumn{2}{m{20\newtblstarfactor+2\tabcolsep+\arrayrulewidth+81\newtblstarfactor+\arrayrulewidth}}{\centering%
$\sqrt{s_\mathrm{NN}}$
}&\multicolumn{1}{m{81\newtblstarfactor+\arrayrulewidth}}{\centering%
19.6 GeV
}&\multicolumn{1}{m{81\newtblstarfactor+\arrayrulewidth}}{\centering%
27 GeV
}&\multicolumn{1}{m{81\newtblstarfactor+\arrayrulewidth}}{\centering%
39 GeV
}&\multicolumn{1}{m{81\newtblstarfactor+\arrayrulewidth}}{\centering%
62.4 GeV
}\tabularnewline
\multicolumn{1}{m{20\newtblstarfactor+\arrayrulewidth}}{\centering%
e$^{\text{-{}}}$
}&\multicolumn{1}{m{81\newtblstarfactor+\arrayrulewidth}}{\centering%
\ensuremath{\mu}$_{\text{gl}}$
}&\multicolumn{1}{m{81\newtblstarfactor+\arrayrulewidth}}{\centering%
-{}0.249
}&\multicolumn{1}{m{81\newtblstarfactor+\arrayrulewidth}}{\centering%
-{}0.136
}&\multicolumn{1}{m{81\newtblstarfactor+\arrayrulewidth}}{\centering%
-{}0.264
}&\multicolumn{1}{m{81\newtblstarfactor+\arrayrulewidth}}{\centering%
-{}0.0609
}\tabularnewline
\multicolumn{1}{m{20\newtblstarfactor+\arrayrulewidth}}{\centering%
e$^{\text{-{}}}$
}&\multicolumn{1}{m{81\newtblstarfactor+\arrayrulewidth}}{\centering%
\ensuremath{\sigma}$_{\text{opt}}$\ensuremath{\pm}\ensuremath{\Delta}\ensuremath{\sigma}$_{\text{opt}}$
}&\multicolumn{1}{m{81\newtblstarfactor+\arrayrulewidth}}{\centering%
0.869\ensuremath{\pm}0.043
}&\multicolumn{1}{m{81\newtblstarfactor+\arrayrulewidth}}{\centering%
0.836\ensuremath{\pm}0.023
}&\multicolumn{1}{m{81\newtblstarfactor+\arrayrulewidth}}{\centering%
0.876\ensuremath{\pm}0.037
}&\multicolumn{1}{m{81\newtblstarfactor+\arrayrulewidth}}{\centering%
0.861\ensuremath{\pm}0.006
}\tabularnewline
\multicolumn{1}{m{20\newtblstarfactor+\arrayrulewidth}}{\centering%
e$^{\text{+}}$
}&\multicolumn{1}{m{81\newtblstarfactor+\arrayrulewidth}}{\centering%
\ensuremath{\mu}$_{\text{gl}}$
}&\multicolumn{1}{m{81\newtblstarfactor+\arrayrulewidth}}{\centering%
-{}0.235
}&\multicolumn{1}{m{81\newtblstarfactor+\arrayrulewidth}}{\centering%
-{}0.129
}&\multicolumn{1}{m{81\newtblstarfactor+\arrayrulewidth}}{\centering%
-{}0.257
}&\multicolumn{1}{m{81\newtblstarfactor+\arrayrulewidth}}{\centering%
-{}0.0662
}\tabularnewline
\multicolumn{1}{m{20\newtblstarfactor+\arrayrulewidth}}{\centering%
e$^{\text{+}}$
}&\multicolumn{1}{m{81\newtblstarfactor+\arrayrulewidth}}{\centering%
\ensuremath{\sigma}$_{\text{opt}}$\ensuremath{\pm}\ensuremath{\Delta}\ensuremath{\sigma}$_{\text{opt}}$
}&\multicolumn{1}{m{81\newtblstarfactor+\arrayrulewidth}}{\centering%
0.869\ensuremath{\pm}0.012
}&\multicolumn{1}{m{81\newtblstarfactor+\arrayrulewidth}}{\centering%
0.842\ensuremath{\pm}0.006
}&\multicolumn{1}{m{81\newtblstarfactor+\arrayrulewidth}}{\centering%
0.881\ensuremath{\pm}0.020
}&\multicolumn{1}{m{81\newtblstarfactor+\arrayrulewidth}}{\centering%
0.866\ensuremath{\pm}0.006
}\tabularnewline
\hline
\end{longtable}\endgroup%

\end{center}

The continuous momentum-{}dependent TPC selection efficiencies
$\epsilon_\mathrm{select}^\mathrm{TPC}(p)$ in
Figure~\ref{effcorr_fig_tpcselect} (lower right) can subsequently be calculated
analytically from the integral over a normalized gaussian with mean
$\mu=\mu_\mathrm{gl}$ and width
$\sigma=\sigma_\mathrm{opt}$:
\[\epsilon_\mathrm{select}^\mathrm{TPC}(p)= \left[\int\limits_{x_0^p}^\mu+\int\limits_\mu^{x_1^p}\right] \frac{e^{-t^2}}{\sigma\sqrt{2\pi}}\,\mathrm{d}x= \left[-\int\limits_{t_0^p}^0+\int\limits_0^{t_1^p}\right] \frac{e^{-t^2}}{\cancel{\sigma}\sqrt{\cancel{2}\pi}}\cancel{\sigma}\cancel{\sqrt{2}}\,\mathrm{d}t= \frac{1}{2}\sum\limits_{i=0}^1\mathrm{erf}(t_i^p)\hspace{4mm} \mathrm{with}\hspace{2mm}t=\frac{\vert x-\mu\vert}{\sigma\sqrt{2}}\]
where $\mathrm{erf}(t_i^p)$ the error function evaluated at the integration limits
$t_{0,1}^p$.  For a given momentum \emph{p}, the lower and upper limits
$x_{0,1}^p$ (and indirectly $t_{0,1}^p$) are obtained
from the functions n\ensuremath{\sigma}$_{\text{el}}$(\emph{p}) in Section~\ref{ana_sec_pid} which are used for the
identification of electrons/positrons through their energy loss in the TPC. The
systematic uncertainties \ensuremath{\Delta}\ensuremath{\sigma} = \ensuremath{\Delta}\ensuremath{\sigma}$_{\text{opt}}$ are propagated
quadratically to $\epsilon_\mathrm{select}^\mathrm{TPC}$ via
\[ \Delta\epsilon_\mathrm{select}^\mathrm{TPC}(p)=\sqrt{ \sum\limits_{i=0}^1\left(\Delta t_i^p\frac{\partial\mathrm{erf}(t_i^p)}{2\partial t_i^p}\right)^2 }=\sqrt{ \sum\limits_{i=0}^1\left(\frac{t_i^p\Delta\sigma}{\sigma\sqrt{\pi}}\exp\left[-\left(t_i^p\right)^2\right]\right)^2 }.\]
\subsection{Total Efficiencies and Uncertainties}
\label{effcorr_subsec_total}\hyperlabel{effcorr_subsec_total}%

Concluding this section, the efficiencies derived in the preceding subsections
are combined according to Equation~\ref{effcorr_eq_total} for electrons and positrons
separately using \ensuremath{\Delta}\ensuremath{\eta} = 0.25, \ensuremath{\Delta}\ensuremath{\phi} = 15\textdegree{}, and the variable
\emph{p$_{\text{T}}$} binning defined by the TOF matching efficiencies (c.f.
 Figure~\ref{effcorr_fig_tofmatch_differential}). To account for the different
\emph{p$_{\text{T}}$}-{}intervals used in the determination of efficiencies for detector
response, track quality requirements (\emph{\ensuremath{\Delta}p$_{\text{T}}$} = 0.2 GeV/c) and global DCA
cuts (\emph{\ensuremath{\Delta}p$_{\text{T}}$} = 0.3 GeV/c), efficiencies and associated uncertainties are
interpolated linearly between bin centers. The propagation of efficiencies
beyond the maximum \emph{p$_{\text{T}}$} simulated in embedding is achieved by assuming
constant values and uncertainties evaluated at 2 GeV/c. In the extreme cases of
zero efficiency in which standard propagation of squared relative errors is not
applicable, the Monte Carlo technique is employed to generate a random sample
of combined efficiencies based on sampling the single uncertainties. Mean
and especially width of the sample can then be calculated manually which allows
for the conservation of well-{}defined and physical non-{}zero statistical and
systematic uncertainties. For the application of TPC selection efficiencies and
uncertainties (c.f. Figure~\ref{effcorr_fig_tpcselect}), the center \emph{p$_{\text{T}}$} and \ensuremath{\eta}
given by the binning for TOF matching efficiencies are mapped to the momentum
$p=p_T\cosh(\eta)$. In the case of relative systematic
uncertainties on track quality and global DCA efficiencies
(Figure~\ref{effcorr_fig_trackqualdca_relsyserr}), a conservative estimate is used for
the six \emph{p$_{\text{T}}$} bins by propagating the maximum uncertainty allowed by the
statistical errors. As an example, Figure~\ref{effcorr_fig_totaltrack} shows the
resulting total single track efficiencies in a single \ensuremath{\eta}/\ensuremath{\phi}-{}interval for
electrons and positrons at all energies including their statistical and
systematic uncertainties. It is worth noticing that statistical errors stay
significantly smaller than the systematic uncertainties for almost the entire
\emph{p$_{\text{T}}$} range covered -{} especially at the low momenta where most of the
dielectron production emanates from. The efficiencies range between
40-{}60\% and exhibit appreciable differences in shape for electrons and positrons
which in hindsight motivate the consistently separated treatment of their
respective efficiencies.

\wrapifneeded{0.50}{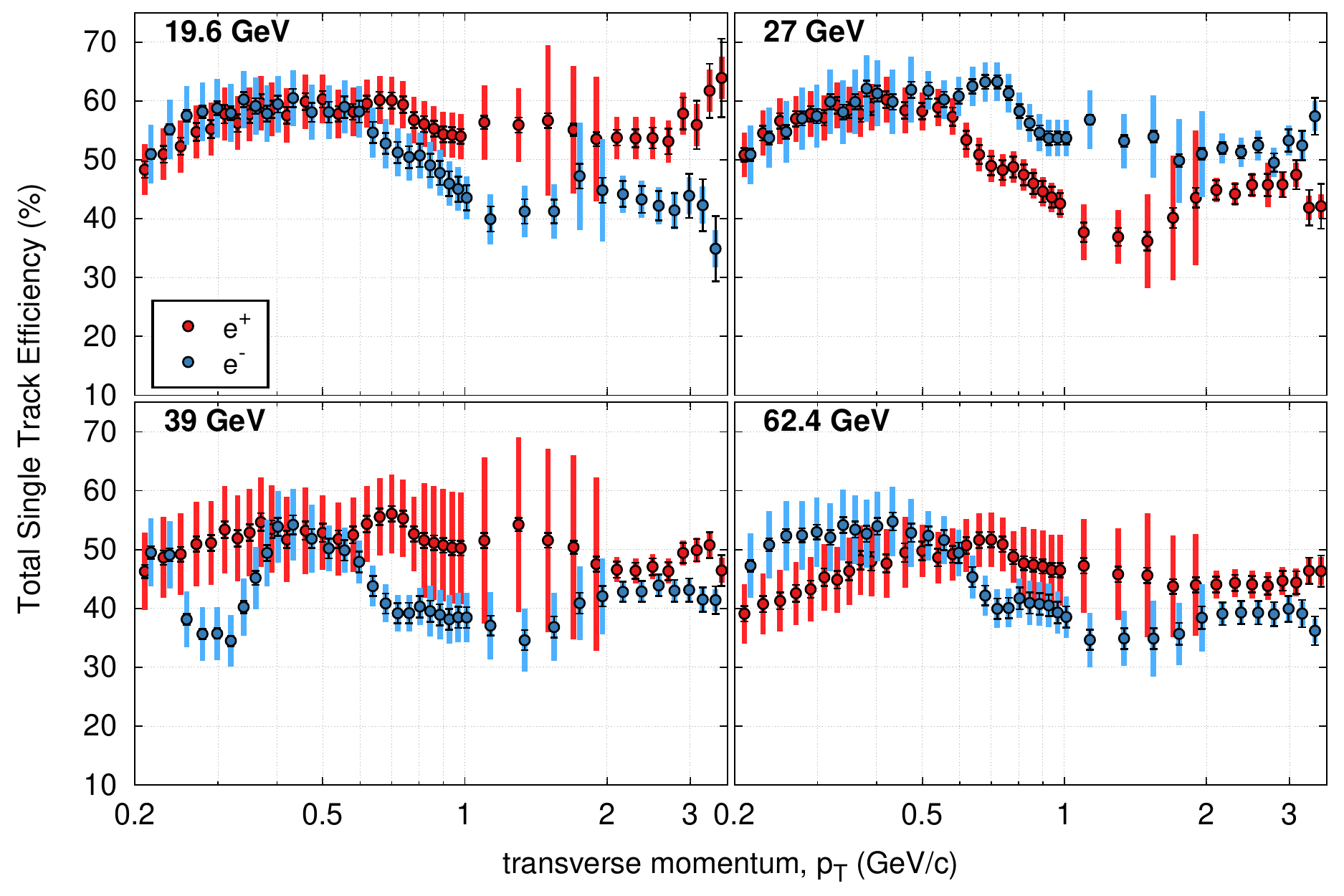}{Total Single Track Efficiencies including statistical and systematic uncertainties in the representative interval of 0 <{} \ensuremath{\eta} <{} 0.25 and 45\textdegree{} <{} \ensuremath{\phi} <{} 60\textdegree{} at all energies with electrons (blue) and positrons (red) separately. For reasons of improved visibility, electrons are shifted upwards by 3\% relative to positrons. Table~\ref{app_tab_effs_tofmatchtotal0} and Table~\ref{app_tab_effs_tofmatchtotal1} list the full set of single track efficiencies and their uncertainties for \ensuremath{\eta} <{} 0 and \ensuremath{\eta} >{} 0, respectively.}{effcorr_fig_totaltrack}{0.95} %

\section{Pair Efficiencies}
\label{effcorr_sec_paireffs}\hyperlabel{effcorr_sec_paireffs}%

The calculation of dielectron (pair) efficiencies is the primary purpose of
this chapter. In the following, they are derived from the total electron and
positron single track efficiencies of Section~\ref{effcorr_subsec_total} via a
Monte Carlo simulation of two-{}body decays. The pair efficiencies need to be
determined in dependence of dielectron invariant mass \emph{M$_{\text{ee}}$} and transverse
momentum \emph{p$_{\text{T}}$} such that they can be applied as an inverse weight on the raw
spectra of Section~\ref{ana_sec_pairrec} and compared directly to cocktail simulations
and model calculations in Chapter~\ref{results}.\newline

The general idea of the simulation is to (\emph{i}) randomly generate the
dielectrons' \emph{M$_{\text{ee}}$}, \emph{p$_{\text{T}}$}, pseudo-{}rapidity \ensuremath{\eta} and azimuthal angle \ensuremath{\phi},
(\emph{ii}) decay them according to two-{}body kinematics, (\emph{iii}) apply STAR
acceptance and single track efficiencies on the two daughters as well as
pair-{}based cuts on the parent, and finally (\emph{iv}) rebin, normalize and divide
appropriate histograms to obtain the pair efficiencies.  As input to this
procedure, each dielectron's \emph{M$_{\text{ee}}$}, \emph{p$_{\text{T}}$} and \ensuremath{\phi} are hence sampled
uniformly in the ranges 2m$_{\text{e}}$ -{} 3.5 GeV/c$^{\text{2}}$, 0 -{} 4 GeV/c and 0 -{} 2\ensuremath{\pi},
respectively. The pair's rapidity \emph{y} is also sampled randomly in [-{}1, 1] and
transformed into \ensuremath{\eta} via
\[\eta=\sinh^{-1}\left(\frac{m_T}{p_T}\sinh(y)\right) \hspace{3mm}\mathrm{with}\hspace{3mm} m_T=\sqrt{M_{ee}^2+p_T^2}.\]
At low invariant masses and transverse momenta, STAR's acceptance misses
dielectrons exhibiting transverse masses \emph{m$_{\text{T}}$} <{} 0.4 GeV/c$^{\text{2}}$. For the lowest
\emph{p$_{\text{T}}$}-{}interval 0 -{} 0.4 GeV/c, this acceptance hole results in quickly falling
but still non-{}zero residual efficiencies at \emph{M$_{\text{ee}}$} <{} 0.4 GeV/c$^{\text{2}}$. Using a
uniform input approach, it is hence very challenging to gather sufficient
statistics in this invariant mass range. To alleviate the situation, sets of
kinematic variables for 10 additional dielectrons are sampled for each
dielectron simulated in the full ranges using identical strategy but separate
random generators. Together with a suitable choice of a variable invariant mass
binning, significantly fewer iterations are consequently required to
sufficiently reduce the statistical errors, i.e. to a level negligible with
respect to the systematic uncertainties and to the statistical errors
on the raw data.\newline

\wrapifneeded{0.50}{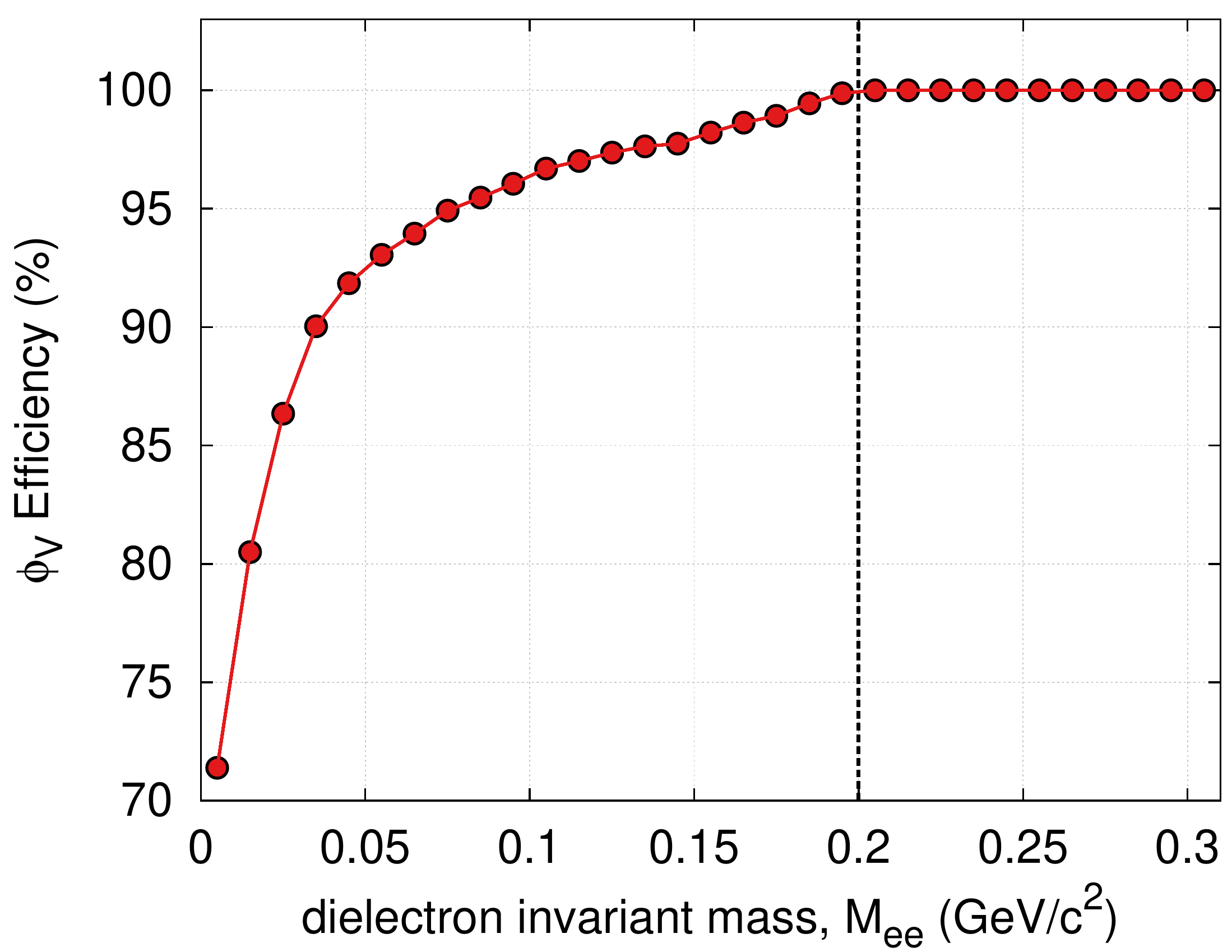}{\ensuremath{\phi}$_{\text{v}}$ efficiency [16]. The vertical dashed line denotes the maximum invariant mass at which the \ensuremath{\phi}$_{\text{V}}$ cut has been applied.}{effcorr_fig_phiv}{0.49} %

The subsequent two-{}body decay of the parent dielectron follows
the kinematics for direct vector meson decays discussed in Section~\ref{sim_sec_cocktail} by calculating the decay daughters' kinematic variables in the dielectron's
center of mass frame and then boosting both into the lab frame. To calculate
the pair efficiencies for the current dielectron, the total single track
efficiencies for each decay leg are obtained for the respective
\ensuremath{\eta}/\ensuremath{\phi}-{}interval and interpolated linearly in \emph{p$_{\text{T}}$} along with the
according statistical and systematic uncertainties. The statistical errors of the efficiencies are
used to randomly vary the single track efficiencies via gaussian sampling.  The
resulting efficiencies are multiplied into pair efficiencies and the according
systematic uncertainties are propagated either through standard or through
Monte Carlo propagation to also cover the extreme cases.
The last step before
filling "output" histograms using the pair efficiencies as weight
\footnote{
allows for the automatic averaging of efficiencies incl.  stat./syst.
uncertainties over the simulated sample.
}, is the application of \ensuremath{\phi}$_{\text{V}}$ cut
efficiencies on the dielectron parent. In Section~\ref{ana_sec_pairrec} photon conversion
is efficiently rejected using a \emph{M$_{\text{ee}}$}-{}dependent selection cut which has been
tuned at top RHIC energy [16] to deliver good \ensuremath{\pi}$^{\text{0}}$-{}Dalitz
selection efficiency at optimal photon conversion rejection efficiency.
Determining the consequential reduction in pair detection efficiency in this
manner, can be achieved through the use of \ensuremath{\pi}$^{\text{0}}$-{}Dalitz and photon conversion
samples simulated via the embedding technique. The resulting \emph{M$_{\text{ee}}$}-{}dependent
\ensuremath{\phi}$_{\text{V}}$ cut efficiencies are depicted in Figure~\ref{effcorr_fig_phiv} and also
interpolated linearly to obtain an efficiency value at arbitrary invariant
mass.\newline

Figure~\ref{effcorr_fig_paireffs} compiles invariant mass and transverse momentum
dependent pair efficiencies for all energies including the respective relative
systematic uncertainties. The former peak at roughly 30\% for 19.6 \& 27 GeV and
at about 20\% for 39 \& 62.4 GeV. At \emph{M$_{\text{ee}}$} <{} 1 GeV/c$^{\text{2}}$, the latter range
between \ensuremath{\sim}10-{}15\% for 19.6 \& 27 GeV, and between \ensuremath{\sim}15-{}20\% at 39 \& 62.4
GeV. As expected, the statistical errors are much smaller than the systematic
uncertainties except for the last bin next to STAR's acceptance hole in the
lowest \emph{p$_{\text{T}}$}-{}interval. The rather appreciable changes in shape with transverse
momentum above \emph{M$_{\text{ee}}$} \ensuremath{\sim} 0.5 GeV/c$^{\text{2}}$ signify the importance of a
two-{}dimensional efficiency correction. Also note that the \emph{p$_{\text{T}}$}-{}intervals
chosen here purposely coincide with the ones in Figure~\ref{results_fig_pT}.

\wrapifneeded{0.50}{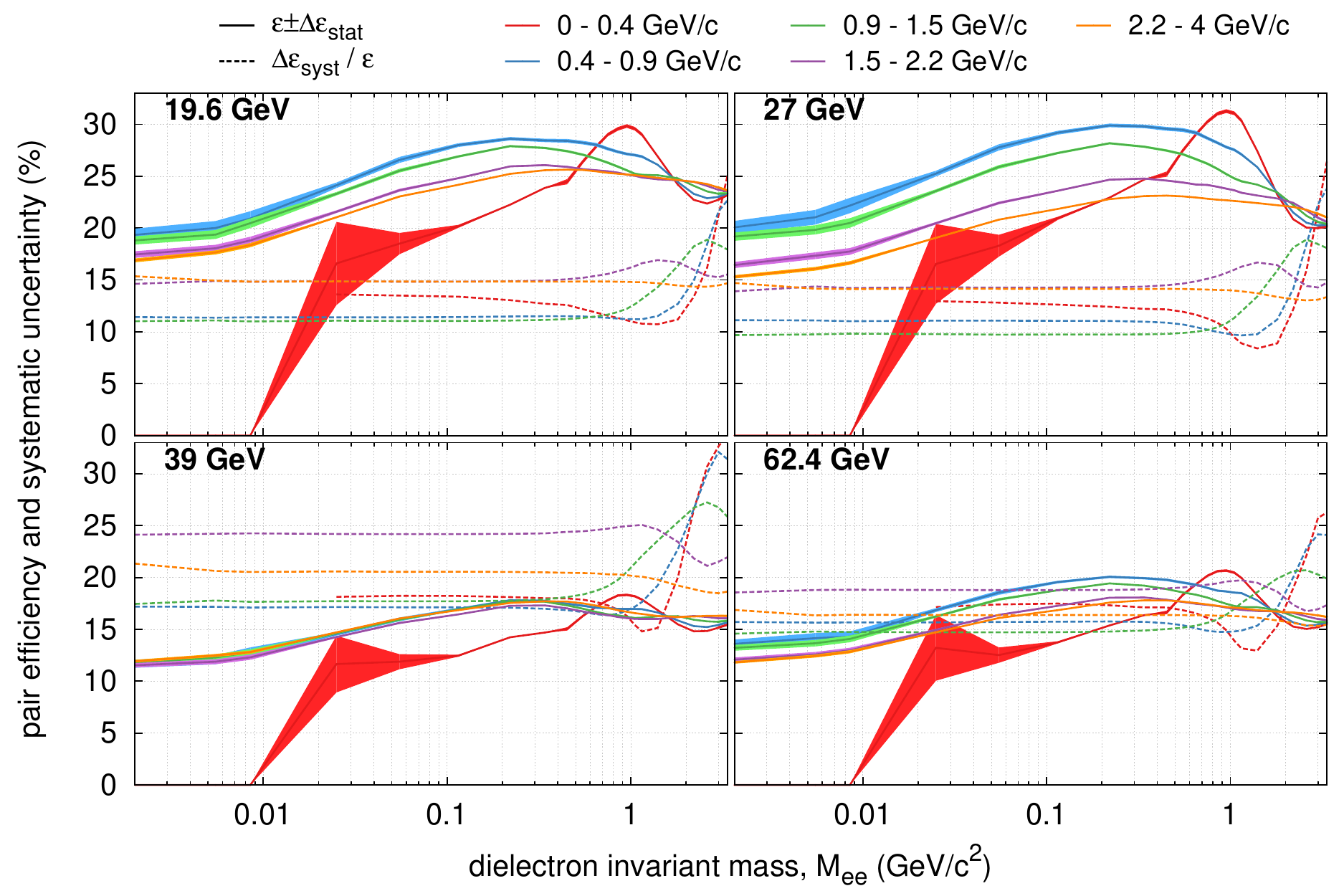}{Pair efficiencies (solid lines) and relative systematic uncertainties (dashed lines) for five \emph{p$_{\text{T}}$}-{}intervals at all BES-{}I energies. The invariant mass range depicted is the full range covered in the simulation. Note that the bands around the efficiency results indicate statistical error bars of the Monte Carlo simulation. The underlying data is listed in Table~\ref{app_tab_effs_pair}.}{effcorr_fig_paireffs}{1} %


\chapter{Simulations and Models}
\label{sim}\hyperlabel{sim}%

The physics interpretation of the experimental results in Chapter~\ref{results} is based on
\emph{(i)} the simulation of hadronic decay channels known to contribute to dielectron
production in elementary collisions, and \emph{(ii)} effective model
calculations that describe the respective in-{}medium effects. The former
constitutes the so-{}called \emph{hadronic cocktail} whose sources and characteristic
mass regions have already been introduced in the context of simulated p+p
collisions in Section~\ref{intro_sec_dielec} along with its according comparison to
previous dielectron measurements in heavy-{}ion collisions. Section~\ref{sim_sec_cocktail} describes the details of the cocktail simulations that are employed in this
thesis for the energy-{}dependent dielectron production in Au+Au collisions at
BES-{}I energies. Section~\ref{intro_sec_dielec} also highlights the excess yields measured
over the cocktail in the Low-{}Mass-{}Region (LMR) of the invariant mass spectra as
well as its possible connection to chiral symmetry restoration through
calculations based on Hadronic Many-{}Body Theory (HMBT). Section~\ref{sim_sec_model} discusses some of the illustrating details and shows the results of these
calculations for the BES-{}I energies studied in this thesis.

\section{Cocktail Simulation}
\label{sim_sec_cocktail}\hyperlabel{sim_sec_cocktail}%

Table~\ref{sim_tab_srcs} lists the $e^+/e^-$ decay channels contributing to
the total cocktails of this thesis and denotes two conceptionally different
categories of dielectron sources: \emph{i}) direct or Dalitz decay channels of the
freeze-{}out pseudo-{}scalar/vector mesons in the mass range of interest, or \emph{ii})
semi-{}leptonic decays of charmed higher-{}mass resonances. A Monte Carlo (MC)
simulation is employed for the meson decays by sampling correctly parameterized
transverse momentum distributions, using appropriate two-{} and three-{}body
kinematics, and accounting for the detector's momentum resolution. The
continuum arising in the Intermediate-{}Mass-{}Range (IMR) from the charmed
resonances decays, on the other hand, is simulated in p+p collisions using
PYTHIA and scaled to Au+Au at the respective energy. At the end of this
section, the results of the simulations for all BES-{}I energies are summarized
in preparation for their comparison to the experimental results.
\begin{center}
\begingroup%
\setlength{\newtblsparewidth}{\linewidth-2\tabcolsep-2\tabcolsep-2\tabcolsep-2\tabcolsep-2\tabcolsep}%
\setlength{\newtblstarfactor}{\newtblsparewidth / \real{100}}%

\begin{longtable}{llll}\caption[{Decay channels from pseudo-{}scalar/vector mesons (PSM/VM) and charmed resonances (CR) included in the cocktail simulations for BES-{}I. Note that the freeze-{}out \ensuremath{\rho}-{}meson is simulated here along with other contributions to allow for comparisons but is not included in the total cocktail (see Chapter~\ref*{results}). Baryonic resonances are not considered for reasons given in Section~\ref*{intro_sec_dielec}.}]{Decay channels from pseudo-{}scalar/vector mesons (PSM/VM) and charmed resonances (CR) included in the cocktail simulations for BES-{}I. Note that the freeze-{}out \ensuremath{\rho}-{}meson is simulated here along with other contributions to allow for comparisons but is not included in the total cocktail (see Chapter~\ref{results}). Baryonic resonances are not considered for reasons given in Section~\ref{intro_sec_dielec}.\label{sim_tab_srcs}\hyperlabel{sim_tab_srcs}%
}\tabularnewline
\endfirsthead
\caption[]{(continued)}\tabularnewline
\endhead
\hline
\multicolumn{1}{m{25\newtblstarfactor+\arrayrulewidth}}{\raggedleft%
Dalitz PSM/VM
}&\multicolumn{1}{m{25\newtblstarfactor+\arrayrulewidth}}{\centering%
$\pi,\eta,\eta' \to e^+e^-\gamma$
}&\multicolumn{1}{m{25\newtblstarfactor+\arrayrulewidth}}{\centering%
$\omega \to e^+e^-\pi^0$
}&\multicolumn{1}{m{25\newtblstarfactor+\arrayrulewidth}}{\centering%
$\phi \to e^+e^-\eta$
}\tabularnewline
\multicolumn{1}{m{25\newtblstarfactor+\arrayrulewidth}}{\raggedleft%
Direct PSM/VM
}&\multicolumn{2}{m{25\newtblstarfactor+2\tabcolsep+\arrayrulewidth+25\newtblstarfactor+\arrayrulewidth}}{\centering%
$\omega,\phi,J/\psi \to e^+e^-$
}&\multicolumn{1}{m{25\newtblstarfactor+\arrayrulewidth}}{\centering%
$(\rho \to e^+e^-)$
}\tabularnewline
\multicolumn{1}{m{25\newtblstarfactor+\arrayrulewidth}}{\raggedleft%
Semi-{}Leptonic CR
}&\multicolumn{3}{m{25\newtblstarfactor+2\tabcolsep+\arrayrulewidth+25\newtblstarfactor+2\tabcolsep+\arrayrulewidth+25\newtblstarfactor+\arrayrulewidth}}{\centering%
$D^0,D^\pm,D_s,\Lambda_c \to e^{+/-}X$
}\tabularnewline
\hline
\end{longtable}\endgroup%

\end{center}

As a reference point for the following discussion, all parameters required as
input to the simulations are listed in Table~\ref{sim_tab_pars}. Many of the details
explained in the remainder of this section can be comprehended to greater depth
by following along in the simulation code published at [185].
\begin{center}
\begingroup%
\setlength{\newtblsparewidth}{\linewidth-2\tabcolsep-2\tabcolsep-2\tabcolsep-2\tabcolsep-2\tabcolsep-2\tabcolsep-2\tabcolsep-2\tabcolsep-2\tabcolsep}%
\setlength{\newtblstarfactor}{\newtblsparewidth / \real{96}}%

\begin{longtable}{llllllll}\caption[{Input parameters for cocktail simulations. See footnotes and text for details.}]{Input parameters for cocktail simulations. See footnotes and text for details.\label{sim_tab_pars}\hyperlabel{sim_tab_pars}%
}\tabularnewline
\endfirsthead
\caption[]{(continued)}\tabularnewline
\endhead
\hline
\multicolumn{8}{m{12\newtblstarfactor+2\tabcolsep+\arrayrulewidth+12\newtblstarfactor+2\tabcolsep+\arrayrulewidth+12\newtblstarfactor+2\tabcolsep+\arrayrulewidth+12\newtblstarfactor+2\tabcolsep+\arrayrulewidth+12\newtblstarfactor+2\tabcolsep+\arrayrulewidth+12\newtblstarfactor+2\tabcolsep+\arrayrulewidth+12\newtblstarfactor+2\tabcolsep+\arrayrulewidth+12\newtblstarfactor+\arrayrulewidth}}{\centering%
\textbf{Particle Properties}
}\tabularnewline
\multicolumn{1}{m{12\newtblstarfactor}|}{\centering%
}&\multicolumn{3}{m{12\newtblstarfactor+2\tabcolsep+\arrayrulewidth+12\newtblstarfactor+2\tabcolsep+\arrayrulewidth+12\newtblstarfactor}|}{\centering%
\emph{pseudo-{}scalar mesons}
}&\multicolumn{4}{m{12\newtblstarfactor+2\tabcolsep+\arrayrulewidth+12\newtblstarfactor+2\tabcolsep+\arrayrulewidth+12\newtblstarfactor+2\tabcolsep+\arrayrulewidth+12\newtblstarfactor+\arrayrulewidth}}{\centering%
\emph{vector mesons}
}\tabularnewline
\multicolumn{1}{m{12\newtblstarfactor}|}{\centering%
particle
}&\multicolumn{1}{m{12\newtblstarfactor}|}{\centering%
\ensuremath{\pi}$^{\text{0}}$
}&\multicolumn{1}{m{12\newtblstarfactor}|}{\centering%
\ensuremath{\eta}
}&\multicolumn{1}{m{12\newtblstarfactor}|}{\centering%
\ensuremath{\eta}'
}&\multicolumn{1}{m{12\newtblstarfactor}|}{\centering%
\ensuremath{\rho}
}&\multicolumn{1}{m{12\newtblstarfactor}|}{\centering%
\ensuremath{\omega}
}&\multicolumn{1}{m{12\newtblstarfactor}|}{\centering%
\ensuremath{\phi}
}&\multicolumn{1}{m{12\newtblstarfactor+\arrayrulewidth}}{\centering%
J/\ensuremath{\psi}
}\tabularnewline
\multicolumn{1}{m{12\newtblstarfactor}|}{\centering%
\emph{m$_{\text{h}}$} \footnotemark{}
}&\multicolumn{1}{m{12\newtblstarfactor}|}{\centering%
0.1349766
}&\multicolumn{1}{m{12\newtblstarfactor}|}{\centering%
0.547853
}&\multicolumn{1}{m{12\newtblstarfactor}|}{\centering%
0.95778
}&\multicolumn{1}{m{12\newtblstarfactor}|}{\centering%
0.77526
}&\multicolumn{1}{m{12\newtblstarfactor}|}{\centering%
0.78265
}&\multicolumn{1}{m{12\newtblstarfactor}|}{\centering%
1.019455
}&\multicolumn{1}{m{12\newtblstarfactor+\arrayrulewidth}}{\centering%
3.096916
}\tabularnewline
\multicolumn{1}{m{12\newtblstarfactor}|}{\centering%
\ensuremath{\Gamma} \footnotemark{}
}&\multicolumn{1}{m{12\newtblstarfactor}|}{\centering%
7.725e-{}9
}&\multicolumn{1}{m{12\newtblstarfactor}|}{\centering%
1.3e-{}6
}&\multicolumn{1}{m{12\newtblstarfactor}|}{\centering%
0.199e-{}3
}&\multicolumn{1}{m{12\newtblstarfactor}|}{\centering%
0.1478
}&\multicolumn{1}{m{12\newtblstarfactor}|}{\centering%
8.49e-{}3
}&\multicolumn{1}{m{12\newtblstarfactor}|}{\centering%
4.26e-{}3
}&\multicolumn{1}{m{12\newtblstarfactor+\arrayrulewidth}}{\centering%
9.29e-{}5
}\tabularnewline
\multicolumn{1}{m{12\newtblstarfactor}|}{\centering%
N$_{\text{FS}}$ \footnotemark{}
}&\multicolumn{3}{m{12\newtblstarfactor+2\tabcolsep+\arrayrulewidth+12\newtblstarfactor+2\tabcolsep+\arrayrulewidth+12\newtblstarfactor}|}{\centering%
2
}&\multicolumn{1}{m{12\newtblstarfactor}|}{\centering%
-{}
}&\multicolumn{2}{m{12\newtblstarfactor+2\tabcolsep+\arrayrulewidth+12\newtblstarfactor}|}{\centering%
1
}&\multicolumn{1}{m{12\newtblstarfactor+\arrayrulewidth}}{\centering%
-{}
}\tabularnewline
\multicolumn{1}{m{12\newtblstarfactor}|}{\centering%
$\Lambda^{-2}$ \footnotemark{}
}&\multicolumn{1}{m{12\newtblstarfactor}|}{\centering%
1.756
}&\multicolumn{1}{m{12\newtblstarfactor}|}{\centering%
1.95
}&\multicolumn{1}{m{12\newtblstarfactor}|}{\centering%
1.84
}&\multicolumn{1}{m{12\newtblstarfactor}|}{\centering%
-{}
}&\multicolumn{1}{m{12\newtblstarfactor}|}{\centering%
2.24
}&\multicolumn{1}{m{12\newtblstarfactor}|}{\centering%
3.23
}&\multicolumn{1}{m{12\newtblstarfactor+\arrayrulewidth}}{\centering%
-{}
}\tabularnewline
\multicolumn{1}{m{12\newtblstarfactor}|}{\centering%
$\Gamma_0^2$ \footnotemark{}
}&\multicolumn{2}{m{12\newtblstarfactor+2\tabcolsep+\arrayrulewidth+12\newtblstarfactor}|}{\centering%
0
}&\multicolumn{1}{m{12\newtblstarfactor}|}{\centering%
0.02
}&\multicolumn{2}{m{12\newtblstarfactor+2\tabcolsep+\arrayrulewidth+12\newtblstarfactor}|}{\centering%
0
}&\multicolumn{1}{m{12\newtblstarfactor}|}{\centering%
0.01
}&\multicolumn{1}{m{12\newtblstarfactor+\arrayrulewidth}}{\centering%
0
}\tabularnewline
\multicolumn{1}{m{12\newtblstarfactor}|}{\centering%
BR$_{\text{ee}}$ \footnotemark{}
}&\multicolumn{3}{m{12\newtblstarfactor+2\tabcolsep+\arrayrulewidth+12\newtblstarfactor+2\tabcolsep+\arrayrulewidth+12\newtblstarfactor}|}{\centering%
-{}
}&\multicolumn{1}{m{12\newtblstarfactor}|}{\centering%
4.72e-{}5
}&\multicolumn{1}{m{12\newtblstarfactor}|}{\centering%
7.28e-{}5
}&\multicolumn{1}{m{12\newtblstarfactor}|}{\centering%
2.954e-{}4
}&\multicolumn{1}{m{12\newtblstarfactor+\arrayrulewidth}}{\centering%
5.94e-{}2
}\tabularnewline
\multicolumn{1}{m{12\newtblstarfactor}|}{\centering%
BR$_{\text{da}}$ \footnotemark{}
}&\multicolumn{1}{m{12\newtblstarfactor}|}{\centering%
1.174e-{}2
}&\multicolumn{1}{m{12\newtblstarfactor}|}{\centering%
6.9e-{}3
}&\multicolumn{1}{m{12\newtblstarfactor}|}{\centering%
9.0e-{}4
}&\multicolumn{1}{m{12\newtblstarfactor}|}{\centering%
-{}
}&\multicolumn{1}{m{12\newtblstarfactor}|}{\centering%
7.7e-{}4
}&\multicolumn{1}{m{12\newtblstarfactor}|}{\centering%
1.15e-{}4
}&\multicolumn{1}{m{12\newtblstarfactor+\arrayrulewidth}}{\centering%
-{}
}\tabularnewline
\multicolumn{1}{m{12\newtblstarfactor}|}{\centering%
M/\ensuremath{\pi} \footnotemark{}
}&\multicolumn{1}{m{12\newtblstarfactor}|}{\centering%
1
}&\multicolumn{1}{m{12\newtblstarfactor}|}{\centering%
0.085
}&\multicolumn{1}{m{12\newtblstarfactor}|}{\centering%
0.0078
}&\multicolumn{1}{m{12\newtblstarfactor}|}{\centering%
0.06
}&\multicolumn{1}{m{12\newtblstarfactor}|}{\centering%
0.069
}&\multicolumn{1}{m{12\newtblstarfactor}|}{\centering%
0.018
}&\multicolumn{1}{m{12\newtblstarfactor+\arrayrulewidth}}{\centering%
6.2e-{}6
}\tabularnewline
\multicolumn{8}{m{12\newtblstarfactor+2\tabcolsep+\arrayrulewidth+12\newtblstarfactor+2\tabcolsep+\arrayrulewidth+12\newtblstarfactor+2\tabcolsep+\arrayrulewidth+12\newtblstarfactor+2\tabcolsep+\arrayrulewidth+12\newtblstarfactor+2\tabcolsep+\arrayrulewidth+12\newtblstarfactor+2\tabcolsep+\arrayrulewidth+12\newtblstarfactor+2\tabcolsep+\arrayrulewidth+12\newtblstarfactor+\arrayrulewidth}}{\centering%
}\tabularnewline
\multicolumn{8}{m{12\newtblstarfactor+2\tabcolsep+\arrayrulewidth+12\newtblstarfactor+2\tabcolsep+\arrayrulewidth+12\newtblstarfactor+2\tabcolsep+\arrayrulewidth+12\newtblstarfactor+2\tabcolsep+\arrayrulewidth+12\newtblstarfactor+2\tabcolsep+\arrayrulewidth+12\newtblstarfactor+2\tabcolsep+\arrayrulewidth+12\newtblstarfactor+2\tabcolsep+\arrayrulewidth+12\newtblstarfactor+\arrayrulewidth}}{\centering%
\textbf{Momentum Smearing}
}\tabularnewline
\multicolumn{2}{m{12\newtblstarfactor+2\tabcolsep+\arrayrulewidth+12\newtblstarfactor}|}{\centering%
\emph{Resolution} \footnotemark{}
}&\multicolumn{6}{m{12\newtblstarfactor+2\tabcolsep+\arrayrulewidth+12\newtblstarfactor+2\tabcolsep+\arrayrulewidth+12\newtblstarfactor+2\tabcolsep+\arrayrulewidth+12\newtblstarfactor+2\tabcolsep+\arrayrulewidth+12\newtblstarfactor+2\tabcolsep+\arrayrulewidth+12\newtblstarfactor+\arrayrulewidth}}{\centering%
\emph{Double-{}Tailed Crystal Ball Function} \footnotemark{}
}\tabularnewline
\multicolumn{1}{m{12\newtblstarfactor}|}{\centering%
a
}&\multicolumn{1}{m{12\newtblstarfactor}|}{\centering%
b
}&\multicolumn{1}{m{12\newtblstarfactor}|}{\centering%
$\bar{x}$
}&\multicolumn{1}{m{12\newtblstarfactor}|}{\centering%
\ensuremath{\sigma}$_{\text{0}}$
}&\multicolumn{1}{m{12\newtblstarfactor}|}{\centering%
n$_{\text{0}}$
}&\multicolumn{1}{m{12\newtblstarfactor}|}{\centering%
\ensuremath{\alpha}$_{\text{0}}$
}&\multicolumn{1}{m{12\newtblstarfactor}|}{\centering%
n$_{\text{1}}$
}&\multicolumn{1}{m{12\newtblstarfactor+\arrayrulewidth}}{\centering%
\ensuremath{\alpha}$_{\text{1}}$
}\tabularnewline
\multicolumn{1}{m{12\newtblstarfactor}|}{\centering%
6e-{}3
}&\multicolumn{1}{m{12\newtblstarfactor}|}{\centering%
8.3e-{}3
}&\multicolumn{1}{m{12\newtblstarfactor}|}{\centering%
-{}0.001
}&\multicolumn{1}{m{12\newtblstarfactor}|}{\centering%
0.01
}&\multicolumn{1}{m{12\newtblstarfactor}|}{\centering%
1.29
}&\multicolumn{1}{m{12\newtblstarfactor}|}{\centering%
-{}1.75
}&\multicolumn{1}{m{12\newtblstarfactor}|}{\centering%
2.92
}&\multicolumn{1}{m{12\newtblstarfactor+\arrayrulewidth}}{\centering%
1.84
}\tabularnewline
\multicolumn{8}{m{12\newtblstarfactor+2\tabcolsep+\arrayrulewidth+12\newtblstarfactor+2\tabcolsep+\arrayrulewidth+12\newtblstarfactor+2\tabcolsep+\arrayrulewidth+12\newtblstarfactor+2\tabcolsep+\arrayrulewidth+12\newtblstarfactor+2\tabcolsep+\arrayrulewidth+12\newtblstarfactor+2\tabcolsep+\arrayrulewidth+12\newtblstarfactor+2\tabcolsep+\arrayrulewidth+12\newtblstarfactor+\arrayrulewidth}}{\centering%
}\tabularnewline
\multicolumn{1}{m{12\newtblstarfactor}|}{\centering%
}&\multicolumn{3}{m{12\newtblstarfactor+2\tabcolsep+\arrayrulewidth+12\newtblstarfactor+2\tabcolsep+\arrayrulewidth+12\newtblstarfactor}|}{\centering%
\textbf{Tsallis-{}Blast-{}Wave Parameters} \footnotemark{}
}&\multicolumn{1}{m{12\newtblstarfactor}|}{\centering%
\textbf{Pion}
}&\multicolumn{3}{m{12\newtblstarfactor+2\tabcolsep+\arrayrulewidth+12\newtblstarfactor+2\tabcolsep+\arrayrulewidth+12\newtblstarfactor+\arrayrulewidth}}{\setlength{\newtblcolwidth}{12\newtblstarfactor+2\tabcolsep+\arrayrulewidth+12\newtblstarfactor+2\tabcolsep+\arrayrulewidth+12\newtblstarfactor}\multirowii[m]{2}{\newtblcolwidth}{\centering%
\textbf{Charm Continuum}
}}\tabularnewline
\multicolumn{1}{m{12\newtblstarfactor}|}{\centering%
}&\multicolumn{3}{m{12\newtblstarfactor+2\tabcolsep+\arrayrulewidth+12\newtblstarfactor+2\tabcolsep+\arrayrulewidth+12\newtblstarfactor}|}{\centering%
(\ensuremath{\beta}$_{\text{S}}$ = 1.5\ensuremath{\beta} for n = 1)
}&\multicolumn{1}{m{12\newtblstarfactor}|}{\centering%
\textbf{Yields}
}&\multicolumn{3}{m{12\newtblstarfactor+2\tabcolsep+\arrayrulewidth+12\newtblstarfactor+2\tabcolsep+\arrayrulewidth+12\newtblstarfactor+\arrayrulewidth}}{\setlength{\newtblcolwidth}{12\newtblstarfactor+2\tabcolsep+\arrayrulewidth+12\newtblstarfactor+2\tabcolsep+\arrayrulewidth+12\newtblstarfactor}\multirowii[m]{2}{-\newtblcolwidth}{\centering%
\textbf{Charm Continuum}
}}\tabularnewline
\multicolumn{1}{m{12\newtblstarfactor}|}{\centering%
$\sqrt{s_{NN}}$
}&\multicolumn{1}{m{12\newtblstarfactor}|}{\centering%
\emph{T} (GeV)
}&\multicolumn{1}{m{12\newtblstarfactor}|}{\centering%
\ensuremath{\beta}
}&\multicolumn{1}{m{12\newtblstarfactor}|}{\centering%
\emph{q}
}&\multicolumn{1}{m{12\newtblstarfactor}|}{\centering%
\emph{dN$_{\text{\ensuremath{\pi}}}$/dy}
}&\multicolumn{1}{m{12\newtblstarfactor}|}{\centering%
$\sigma_{c\bar{c}}^\mathrm{NN}$ (\ensuremath{\mu}b)
}&\multicolumn{1}{m{12\newtblstarfactor}|}{\centering%
$\sigma_\mathrm{mb}^\mathrm{NN}$ (mb)
}&\multicolumn{1}{m{12\newtblstarfactor+\arrayrulewidth}}{\centering%
N$_{\text{coll}}$
}\tabularnewline
\multicolumn{1}{m{12\newtblstarfactor}|}{\centering%
19.6 GeV
}&\multicolumn{1}{m{12\newtblstarfactor}|}{\centering%
0.1189
}&\multicolumn{1}{m{12\newtblstarfactor}|}{\centering%
0.3582
}&\multicolumn{1}{m{12\newtblstarfactor}|}{\centering%
1.0272
}&\multicolumn{1}{m{12\newtblstarfactor}|}{\centering%
51.9
}&\multicolumn{1}{m{12\newtblstarfactor}|}{\centering%
6.1\ensuremath{\pm}3.9
}&\multicolumn{1}{m{12\newtblstarfactor}|}{\centering%
32
}&\multicolumn{1}{m{12\newtblstarfactor+\arrayrulewidth}}{\centering%
231.75
}\tabularnewline
\multicolumn{1}{m{12\newtblstarfactor}|}{\centering%
27 GeV
}&\multicolumn{1}{m{12\newtblstarfactor}|}{\centering%
0.1222
}&\multicolumn{1}{m{12\newtblstarfactor}|}{\centering%
0.3927
}&\multicolumn{1}{m{12\newtblstarfactor}|}{\centering%
1.014
}&\multicolumn{1}{m{12\newtblstarfactor}|}{\centering%
55.8
}&\multicolumn{1}{m{12\newtblstarfactor}|}{\centering%
19.7\ensuremath{\pm}12.6
}&\multicolumn{1}{m{12\newtblstarfactor}|}{\centering%
33
}&\multicolumn{1}{m{12\newtblstarfactor+\arrayrulewidth}}{\centering%
238.7
}\tabularnewline
\multicolumn{1}{m{12\newtblstarfactor}|}{\centering%
39 GeV
}&\multicolumn{1}{m{12\newtblstarfactor}|}{\centering%
0.1229
}&\multicolumn{1}{m{12\newtblstarfactor}|}{\centering%
0.3841
}&\multicolumn{1}{m{12\newtblstarfactor}|}{\centering%
1.02322
}&\multicolumn{1}{m{12\newtblstarfactor}|}{\centering%
58.6
}&\multicolumn{1}{m{12\newtblstarfactor}|}{\centering%
57\ensuremath{\pm}36
}&\multicolumn{1}{m{12\newtblstarfactor}|}{\centering%
34
}&\multicolumn{1}{m{12\newtblstarfactor+\arrayrulewidth}}{\centering%
243.16
}\tabularnewline
\multicolumn{1}{m{12\newtblstarfactor}|}{\centering%
62.4 GeV
}&\multicolumn{1}{m{12\newtblstarfactor}|}{\centering%
0.1168
}&\multicolumn{1}{m{12\newtblstarfactor}|}{\centering%
0.3457
}&\multicolumn{1}{m{12\newtblstarfactor}|}{\centering%
1.049
}&\multicolumn{1}{m{12\newtblstarfactor}|}{\centering%
61.7 [186]
}&\multicolumn{1}{m{12\newtblstarfactor}|}{\centering%
160\ensuremath{\pm}100
}&\multicolumn{1}{m{12\newtblstarfactor}|}{\centering%
36
}&\multicolumn{1}{m{12\newtblstarfactor+\arrayrulewidth}}{\centering%
252.9
}\tabularnewline
\multicolumn{8}{m{12\newtblstarfactor+2\tabcolsep+\arrayrulewidth+12\newtblstarfactor+2\tabcolsep+\arrayrulewidth+12\newtblstarfactor+2\tabcolsep+\arrayrulewidth+12\newtblstarfactor+2\tabcolsep+\arrayrulewidth+12\newtblstarfactor+2\tabcolsep+\arrayrulewidth+12\newtblstarfactor+2\tabcolsep+\arrayrulewidth+12\newtblstarfactor+2\tabcolsep+\arrayrulewidth+12\newtblstarfactor+\arrayrulewidth}}{\centering%
}\tabularnewline
\multicolumn{4}{m{12\newtblstarfactor+2\tabcolsep+\arrayrulewidth+12\newtblstarfactor+2\tabcolsep+\arrayrulewidth+12\newtblstarfactor+2\tabcolsep+\arrayrulewidth+12\newtblstarfactor}|}{\setlength{\newtblcolwidth}{12\newtblstarfactor+2\tabcolsep+\arrayrulewidth+12\newtblstarfactor+2\tabcolsep+\arrayrulewidth+12\newtblstarfactor+2\tabcolsep+\arrayrulewidth+12\newtblstarfactor}\multirowii[m]{2}{\newtblcolwidth}{\centering%
\textbf{Branching Ratios of Charmed Resonances}
}}&\multicolumn{1}{m{12\newtblstarfactor}|}{\centering%
$D^\pm$
}&\multicolumn{1}{m{12\newtblstarfactor}|}{\centering%
\emph{D$^{\text{0}}$}
}&\multicolumn{1}{m{12\newtblstarfactor}|}{\centering%
\emph{D$_{\text{s}}$}
}&\multicolumn{1}{m{12\newtblstarfactor+\arrayrulewidth}}{\centering%
$\Lambda_c$
}\tabularnewline
\multicolumn{4}{m{12\newtblstarfactor+2\tabcolsep+\arrayrulewidth+12\newtblstarfactor+2\tabcolsep+\arrayrulewidth+12\newtblstarfactor+2\tabcolsep+\arrayrulewidth+12\newtblstarfactor}|}{\setlength{\newtblcolwidth}{12\newtblstarfactor+2\tabcolsep+\arrayrulewidth+12\newtblstarfactor+2\tabcolsep+\arrayrulewidth+12\newtblstarfactor+2\tabcolsep+\arrayrulewidth+12\newtblstarfactor}\multirowii[m]{2}{-\newtblcolwidth}{\centering%
\textbf{Branching Ratios of Charmed Resonances}
}}&\multicolumn{1}{m{12\newtblstarfactor}|}{\centering%
0.1607
}&\multicolumn{1}{m{12\newtblstarfactor}|}{\centering%
0.0649
}&\multicolumn{1}{m{12\newtblstarfactor}|}{\centering%
0.065
}&\multicolumn{1}{m{12\newtblstarfactor+\arrayrulewidth}}{\centering%
0.045
}\tabularnewline
\hline
\end{longtable}\endgroup%

\end{center}
\addtocounter{footnote}{-11}\stepcounter{footnote}
\footnotetext{
(hadron) mass of pseudo-{}scalar/vector meson (in GeV/c$^{\text{2}}$).
}\stepcounter{footnote}
\footnotetext{
resonance (or decay) width in Breit-{}Wigner distribution (in GeV).
}\stepcounter{footnote}
\footnotetext{
number of final states in QED part of Kroll-{}Wada formula.
}\stepcounter{footnote}
\footnotetext{
parameter in form factor parameterization of Kroll-{}Wada formula (in GeV$^{\text{-{}2}}$).
}\stepcounter{footnote}
\footnotetext{
additional parameter in form factor parameterization of Kroll-{}Wada formula (in GeV$^{\text{2}}$).
}\stepcounter{footnote}
\footnotetext{
branching ratio for direct decay.
}\stepcounter{footnote}
\footnotetext{
branching ratio for Dalitz decay.
}\stepcounter{footnote}
\footnotetext{
meson-{}to-{}pion ratios of invariant yields.
}\stepcounter{footnote}
\footnotetext{
parameters for p$_{\text{T}}$ dependence of momentum resolution.
}\stepcounter{footnote}
\footnotetext{
left-{}to-{}right: mean and width of gaussian core, and two parameters defining each of the non-{}gaussian tails.
}\stepcounter{footnote}
\footnotetext{
temperature \emph{T}, expansion velocity \ensuremath{\beta}, and exponent \emph{q}.
}
\pagebreak[4]

\subsection{Pseudo-{}Scalar/Vector Meson Sampling and Decays}
\label{_pseudo_scalar_vector_meson_sampling_and_decays}\hyperlabel{_pseudo_scalar_vector_meson_sampling_and_decays}%

One of the important ingredients in the simulation of the cocktails is the
input distributions chosen for the pseudo-{}scalar/vector mesons in Table~\ref{sim_tab_srcs}. The distributions are used to
decay the mesons according to the kinematics of their direct and Dalitz channels.
This step is needed to apply momentum smearing on the daughter particles and
reject those outside the STAR acceptance, i.e. with transverse momenta p$_{\text{T}}$ <{}
0.2 GeV/c and pseudo-{}rapidity |\ensuremath{\eta}| >{} 1. The mesons' rapidity and azimuthal
distributions are sampled uniformly in $y\in(-1,1)$ and
$\phi\in(0,2\pi)$, respectively. Their p$_{\text{T}}$ distributions (i.e. $\mathrm{d}N/m_\mathrm{T}\,\mathrm{d}m_\mathrm{T}$) are
obtained from simultaneous Tsallis-{}Blast-{}Wave parameterizations [82]
to pion, kaon and proton p$_{\text{T}}$ spectra measured at BES-{}I energies
[186-{}188] (Figure~\ref{sim_fig_tbw_fits}) using the
meson mass \emph{m$_{\text{h}}$} to calculate the respective transverse mass
$m_T^2=p_T^2+m_h^2$:
\[ \frac{\mathrm{d}N}{m_\mathrm{T}\,\mathrm{d}m_\mathrm{T}} = \mathcal{F}\,\cdot\,m_\mathrm{T} \int\limits_{-6}^{+6}\cosh(y)\,\mathrm{d}y \int\limits_{-\pi}^{+\pi}\mathrm{d}\phi \int\limits_0^1 r\mathrm{d}r \left( 1+\frac{q-1}{T} \Big\{ m_\mathrm{T}\cosh(y)\cosh(\rho) - p_\mathrm{T}\sinh(\rho)\cos(\phi) \Big\} \right)^{-1/(q-1)} \]
with the proportionality factors $\mathcal{F}$ (Table~\ref{sim_tab_tbw_factors}), temperature \emph{T}, and
$\rho=\tanh^{-1}\left(\beta_S r\right)$ where \ensuremath{\beta}$_{\text{S}}$ the radial
surface expansion velocity of the fireball.

\wrapifneeded{0.50}{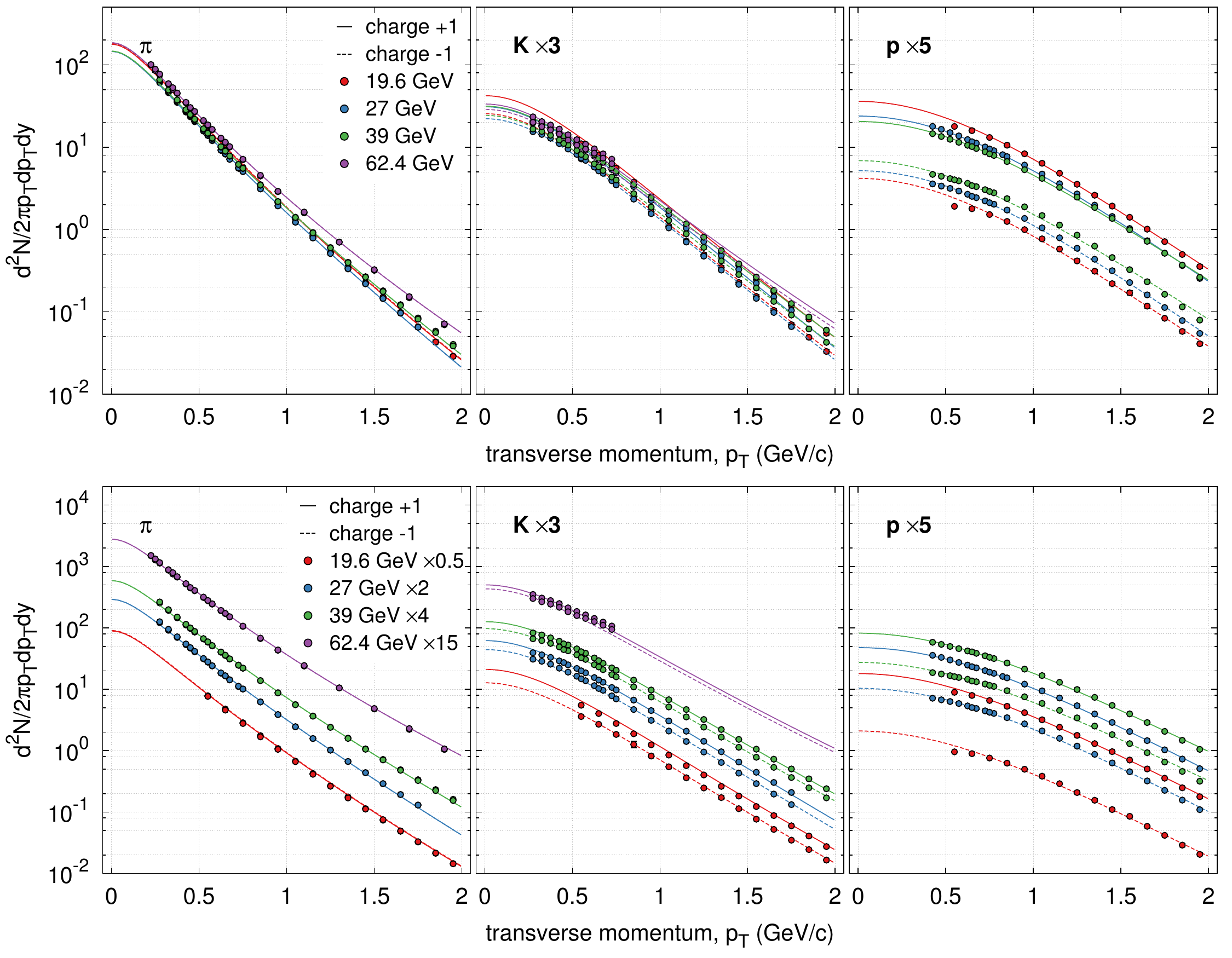}{Simultaneous Tsallis-{}Blast-{}Wave parameterizations of \ensuremath{\pi}/K/p p$_{\text{T}}$-{}spectra at BES-{}I energies. In the lower panel the energies are shifted against each other for better visibility. Solid and dashed lines denote positive and negative particles, respectively [189].}{sim_fig_tbw_fits}{1} %
\begin{center}
\begingroup%
\setlength{\newtblsparewidth}{\linewidth-2\tabcolsep-2\tabcolsep-2\tabcolsep-2\tabcolsep-2\tabcolsep-2\tabcolsep-2\tabcolsep}%
\setlength{\newtblstarfactor}{\newtblsparewidth / \real{96}}%

\begin{longtable}{llllll}\caption[{Tsallis-{}Blast-{}Wave proportionality factors \ensuremath{\mathcal{F}} (in GeV$^{\text{-{}2}}$).}]{Tsallis-{}Blast-{}Wave proportionality factors \ensuremath{\mathcal{F}} (in GeV$^{\text{-{}2}}$).\label{sim_tab_tbw_factors}\hyperlabel{sim_tab_tbw_factors}%
}\tabularnewline
\endfirsthead
\caption[]{(continued)}\tabularnewline
\endhead
\hline
\multicolumn{1}{m{16\newtblstarfactor+\arrayrulewidth}}{\centering%
$\sqrt{s_\mathrm{NN}}$ (GeV)
}&\multicolumn{1}{m{16\newtblstarfactor+\arrayrulewidth}}{\centering%
\ensuremath{\pi}$^{\text{+}}$/\ensuremath{\pi}$^{\text{-{}}}$
}&\multicolumn{1}{m{16\newtblstarfactor+\arrayrulewidth}}{\centering%
K$^{\text{+}}$
}&\multicolumn{1}{m{16\newtblstarfactor+\arrayrulewidth}}{\centering%
K$^{\text{-{}}}$
}&\multicolumn{1}{m{16\newtblstarfactor+\arrayrulewidth}}{\centering%
p
}&\multicolumn{1}{m{16\newtblstarfactor+\arrayrulewidth}}{\centering%
$\bar{\mathrm{p}}$
}\tabularnewline
\multicolumn{1}{m{16\newtblstarfactor+\arrayrulewidth}}{\centering%
19.6
}&\multicolumn{1}{m{16\newtblstarfactor+\arrayrulewidth}}{\centering%
488.7/498.1
}&\multicolumn{1}{m{16\newtblstarfactor+\arrayrulewidth}}{\centering%
448.9
}&\multicolumn{1}{m{16\newtblstarfactor+\arrayrulewidth}}{\centering%
271.7
}&\multicolumn{1}{m{16\newtblstarfactor+\arrayrulewidth}}{\centering%
4970
}&\multicolumn{1}{m{16\newtblstarfactor+\arrayrulewidth}}{\centering%
577.2
}\tabularnewline
\multicolumn{1}{m{16\newtblstarfactor+\arrayrulewidth}}{\centering%
27
}&\multicolumn{1}{m{16\newtblstarfactor+\arrayrulewidth}}{\centering%
404.3
}&\multicolumn{1}{m{16\newtblstarfactor+\arrayrulewidth}}{\centering%
365.1
}&\multicolumn{1}{m{16\newtblstarfactor+\arrayrulewidth}}{\centering%
261.2
}&\multicolumn{1}{m{16\newtblstarfactor+\arrayrulewidth}}{\centering%
4580
}&\multicolumn{1}{m{16\newtblstarfactor+\arrayrulewidth}}{\centering%
994.1
}\tabularnewline
\multicolumn{1}{m{16\newtblstarfactor+\arrayrulewidth}}{\centering%
39
}&\multicolumn{1}{m{16\newtblstarfactor+\arrayrulewidth}}{\centering%
391.7
}&\multicolumn{1}{m{16\newtblstarfactor+\arrayrulewidth}}{\centering%
319.1
}&\multicolumn{1}{m{16\newtblstarfactor+\arrayrulewidth}}{\centering%
246.4
}&\multicolumn{1}{m{16\newtblstarfactor+\arrayrulewidth}}{\centering%
2720
}&\multicolumn{1}{m{16\newtblstarfactor+\arrayrulewidth}}{\centering%
909.4
}\tabularnewline
\multicolumn{1}{m{16\newtblstarfactor+\arrayrulewidth}}{\centering%
62.4
}&\multicolumn{1}{m{16\newtblstarfactor+\arrayrulewidth}}{\centering%
496.3
}&\multicolumn{1}{m{16\newtblstarfactor+\arrayrulewidth}}{\centering%
295.9
}&\multicolumn{1}{m{16\newtblstarfactor+\arrayrulewidth}}{\centering%
254.9
}&\multicolumn{1}{m{16\newtblstarfactor+\arrayrulewidth}}{\centering%
-{}
}&\multicolumn{1}{m{16\newtblstarfactor+\arrayrulewidth}}{\centering%
-{}
}\tabularnewline
\hline
\end{longtable}\endgroup%

\end{center}

For the simple direct two-{}body decays, the resonance's invariant mass (M$_{\text{ee}}$)
shape is sampled based on the Breit-{}Wigner probability distribution
[190]:
\[\frac{dN}{dM_{ee}} = \frac{1}{2\pi}\frac{\Gamma}{(M_{ee}-m_h)^2+\Gamma^2/4}.\]
The daughters' momenta \emph{p} in the pseudo-{}scalar/vector meson's center-{}of-{}mass system are then
given by $p=\sqrt{M_{ee}^2/4-m_e^2}$ with m$_{\text{e}}$ the electron mass
which allows for the derivation of the kinematic variables of $e^-$ (p$_{\text{T}}$, \ensuremath{\eta}, \ensuremath{\phi}) and $e^+$ (p$_{\text{T}}$, -{}\ensuremath{\eta}, \ensuremath{\phi}+\ensuremath{\pi}). In
such back-{}to-{}back decays, polar angle $\theta$ and azimuthal angle
\ensuremath{\phi} are distributed uniformly in (0, \ensuremath{\pi}) and (0, 2\ensuremath{\pi}), respectively.
Given the longitudinal momenta $p_z=p\cos\theta$, the required
\emph{p$_{\text{T}}$} and \ensuremath{\eta} are determined by
\[p_T=\sqrt{p^2-p_z^2}\mbox{\hspace{3mm}and\hspace{3mm}} \eta=0.5\ln\left(\frac{p+p_z}{p-p_z}\right).\]
The resulting $e^+/e^-$ Lorentz vectors are finally boosted into
the lab frame using the PSM/VM's velocity.
For the Dalitz decay channels, the Kroll-{}Wada formula [191] is used
to sample the correct dielectron invariant mass shape within the allowed range
of \emph{2m$_{\text{e}}$} to \emph{m$_{\text{h}}$-{}m$_{\text{d}}$} with m$_{\text{d}}$ the daughter hadron's or photon's (=0) rest
mass:
\[dN/dM_{ee}=2M_{ee} \cdot \mathrm{FF}^2(M_{ee}) \cdot \mathrm{QED}(M_\mathrm{ee}) \cdot \mathrm{PS}(M_\mathrm{ee})\]
where the QED part, the form factors (FF,
[144, 192-{}194]) and the phase
space (PS) are given by
\[\mathrm{QED}(M_{ee}) = N_\mathrm{FS} \frac{\alpha}{3\pi} \frac{1}{M_{ee}^2} \sqrt{1-\frac{4m_e^2}{M_{ee}^2}} \left(1+\frac{2m_e^2}{M_{ee}^2}\right) \hspace{3mm}(N_\mathrm{FS}:\mbox{ number of final states}),\]\[\mathrm{FF}=\begin{cases} \left(1+M_{ee}^2/\Lambda^2\right)^2 & \mbox{ for }\pi^0\\ \left[(1-(M_\mathrm{ee}/\Lambda)^2)^2+(\Gamma_0/\Lambda)^2\right]^{-1} & \mbox{ for }\eta/\eta'/\omega/\phi, \end{cases}\]\[\mathrm{PS}=\begin{cases} \left(1-\left(M_{ee}/m_h\right)^2\right)^3 & \mbox{ for }\pi^0/\eta/\eta'\\ \left( \left(1+M_{ee}^2/\left(m_h^2-m_d^2\right)\right)^2 - 4m_h^2M_{ee}^2/\left(m_h^2-m_d^2\right)^2\right)^{3/2} & \mbox{ for }\omega/\phi. \end{cases}\]
The dielectron invariant mass sampled from the Kroll-{}Wada distribution is first
used to simulate the intermediate virtual photon's back-{}to-{}back decay into
electron and positron in its center-{}of-{}mass system via the two-{}body
dielectron kinematics described above. The underlying two-{}body kinematics can
also be employed to obtain the Lorentz vectors of decay hadron/photon and
lepton pair (virtual photon) assuming uniformly distributed helicity angles.
The required daughter hadron momentum \emph{p} can be calculated in the parent
pseudo-{}scalar/vector meson's center-{}of-{}mass system via
$p^2=E_d^2-m_d^2$ with
$E_d=\left(m_h^2+m_d^2-M_{ee}^2\right)/2m_h$.  The electron's and
positron's Lorentz vectors are boosted first into the lepton pair's rest frame
and subsequently into the parent pseudo-{}scalar/vector meson's lab frame with the latter also
for the daughter decay hadron/photon. In the combined case of allowed Dalitz
and direct decay modes, the simulation randomly switches between the two
kinematics according to the ratio of Dalitz (BR$_{\text{da}}$) and direct (BR$_{\text{ee}}$)
branching ratios.

\subsection{Momentum Smearing and Yield Scaling}
\label{_momentum_smearing_and_yield_scaling}\hyperlabel{_momentum_smearing_and_yield_scaling}%

The final step in the simulation of the pseudo-{}scalar/vector mesons is the smearing of their
decay daughter's momenta, reminiscent of the reconstruction procedure in the
experimental data analysis. Momentum smearing can be achieved by adding a
gaussian-{}like distributed momentum shift $\Delta p_T^\mathrm{rnd}$ to the simulated
\emph{p$_{\text{T}}$} as derived in the following. For the BES-{}I energies, similar momentum resolution as achieved at
$\sqrt{s_{NN}}$ = 200 GeV is assumed which allows to utilize the
same parameterizations tuned to match the measured J/\ensuremath{\psi} signal
[16]. $\Delta p_T^\mathrm{rnd}$ needs to be sampled
according to the decay leptons' (relative) momentum resolution
$\delta p_T\equiv\sigma(p_T)/p_T$ defined by the \emph{p$_{\text{T}}$} dependent
one standard deviation $\sigma(p_T)$ (width) of gaussian fits to
distributions of the relative momentum shift $\Delta p_T^\mathrm{rel}$. This shift caused by the tracking in the TPC with respect
to the true particle momentum can be derived via
\[\Delta p_T^\mathrm{rel}=\left(p_T^\mathrm{rec}-p_T^\mathrm{MC}\right)/p_T^\mathrm{MC}\]
with $p_T^\mathrm{MC}$ the Monte Carlo simulated (true) momentum
and $p_T^\mathrm{rec}$ the according momentum reconstructed in the
STAR-{}specific embedding procedure (see Chapter~\ref{effcorr}). Statistics gathered in the
latter does not allow for reliable fits of the $\Delta p_T^\mathrm{rel}$ distribution's tails on a bin-{}by-{}bin basis in transverse
momentum. However, the influence of the tails on the width of the distribution
is negligible which is why the \emph{p$_{\text{T}}$} dependency of the momentum resolution can
be parameterized using
\[\sigma(p_T)=\sqrt{a^2p_T^2+b^2}\]
and separated from the tail-{}including randomization of $\Delta p_T^\mathrm{rel}$. The $\Delta p_T^\mathrm{rel}$ distributions of
each transverse momentum bin \emph{i} are scaled such that the respective gaussian
widths \ensuremath{\sigma}$_{\text{i}}$ equal \ensuremath{\sigma}$_{\text{0}}$, i.e.
\[\Delta p_{T,i}^\mathrm{rel}=\Delta p_{T,i}^\mathrm{rel}\cdot\sigma_0/\sigma_i.\]
They are subsequently added up to form a universal (\emph{p$_{\text{T}}$} independent)
distribution \emph{dN/dx} parameterized by means of a double-{}tailed Crystal
Ball function [195-{}197]
($x\equiv\Delta p_T^\mathrm{rel}$ and
 $y\equiv\left(x-\bar{x}\right)/\sigma_0$):
\[\frac{dN}{dx}\propto\begin{cases} t_0(x)&\mbox{if }y<\alpha_0\\ \exp\left(-y^2/2\right)&\mbox{if }\alpha_0\leq y\leq\alpha_1\\ t_1(x)&\mbox{if }y>\alpha_1 \end{cases}\]
with
\[t_i(x)=\left[\frac{n_i}{\vert\alpha_i\vert}\middle/ \left(\frac{n_i}{\vert\alpha_i\vert}-\vert\alpha_i\vert+\mbox{sgn}(\alpha_i)\,\cdot\,y\right)\right]^{n_i} \exp\left(-\frac{\alpha_i^2}{2}\right)\]
denoting the two tails of the $\Delta p_T^\mathrm{rel}$ distribution. In summary, it follows for the smeared daughter momenta
$p_T^\mathrm{sm}$, including the proper scaling of the daughter's
functional momentum shape back to the correct width at the specific momentum,
\[p_T^\mathrm{sm}=p_T+\Delta p_T^\mathrm{rnd}=p_T\left(1+\Delta p_T^\mathrm{rel,rnd}\right) =p_T\left(1+\Delta p_{T,\mathrm{CB}}^\mathrm{rel,rnd}\cdot\sigma(p_T)/\sigma_0\right)\]
with $\Delta p_{T,\mathrm{CB}}^\mathrm{rel,rnd}$ sampled randomly
from the above Crystal Ball function.\newline
 Finally, the resulting Lorentz vectors for the decay particles are used to
generate dielectron invariant mass and \emph{p$_{\text{T}}$} spectra at mid-{}rapidity
($|y_{ee}|\sim|y_\mathrm{PSM/VM}|<1$) within STAR acceptance. The
acceptance difference between vector mesons and their daughter dielectrons
caused by the difference in rapidities for Dalitz decays results in an 8\%
overall correction [189] and is included in the scaling
factors listed in Table~\ref{results_tab_scale}.\newline

So far, the pseudo-{}scalar/vector meson's invariant mass spectra obtained from the Monte Carlo
simulations only provide the correct shape but still need to be scaled to the
desired invariant yields before being compiled into the hadronic cocktail. The
previous Tsallis-{}Blast-{}Wave fits also provide the total invariant pion yields in STAR
(\emph{dN$_{\text{\ensuremath{\pi}}}$/dy})  which are used in conjunction with meson-{}to-{}pion (M/\ensuremath{\pi})
ratios from SPS to extract the invariant yields of all other pseudo-{}scalar/vector mesons. The
expected number of dielectrons per event emanating from the decay(s) of a
specific pseudo-{}scalar/vector meson (PSM/VM) within STAR acceptance is calculated from the
according number of simulated dielectrons and folded with the total branching
ratio for the PSM/VM's direct and Dalitz decays, times the number of PSM/VMs per event:
\[\left.\frac{dN^\mathrm{acc}_{ee}}{N_\mathrm{evt}\,dM_{ee}}\right|_\mathrm{PSM/VM}^\mathrm{exp} \hspace{3mm}=\hspace{3mm} \left.\frac{dN_{ee}^\mathrm{acc}}{N_\mathrm{evt}dM_{ee}}\right|_\mathrm{PSM/VM}^\mathrm{sim} \hspace{3mm}\cdot\hspace{3mm} \left.\mathrm{BR}_\mathrm{tot}^{ee(X)}\right|_\mathrm{PSM/VM} \hspace{3mm}\cdot\hspace{3mm} \left.\Delta y\frac{dN}{dy}\right|_\mathrm{PSM/VM} \hspace{5mm}\mbox{ with }\Delta y=2.\]
\wrapifneeded{0.50}{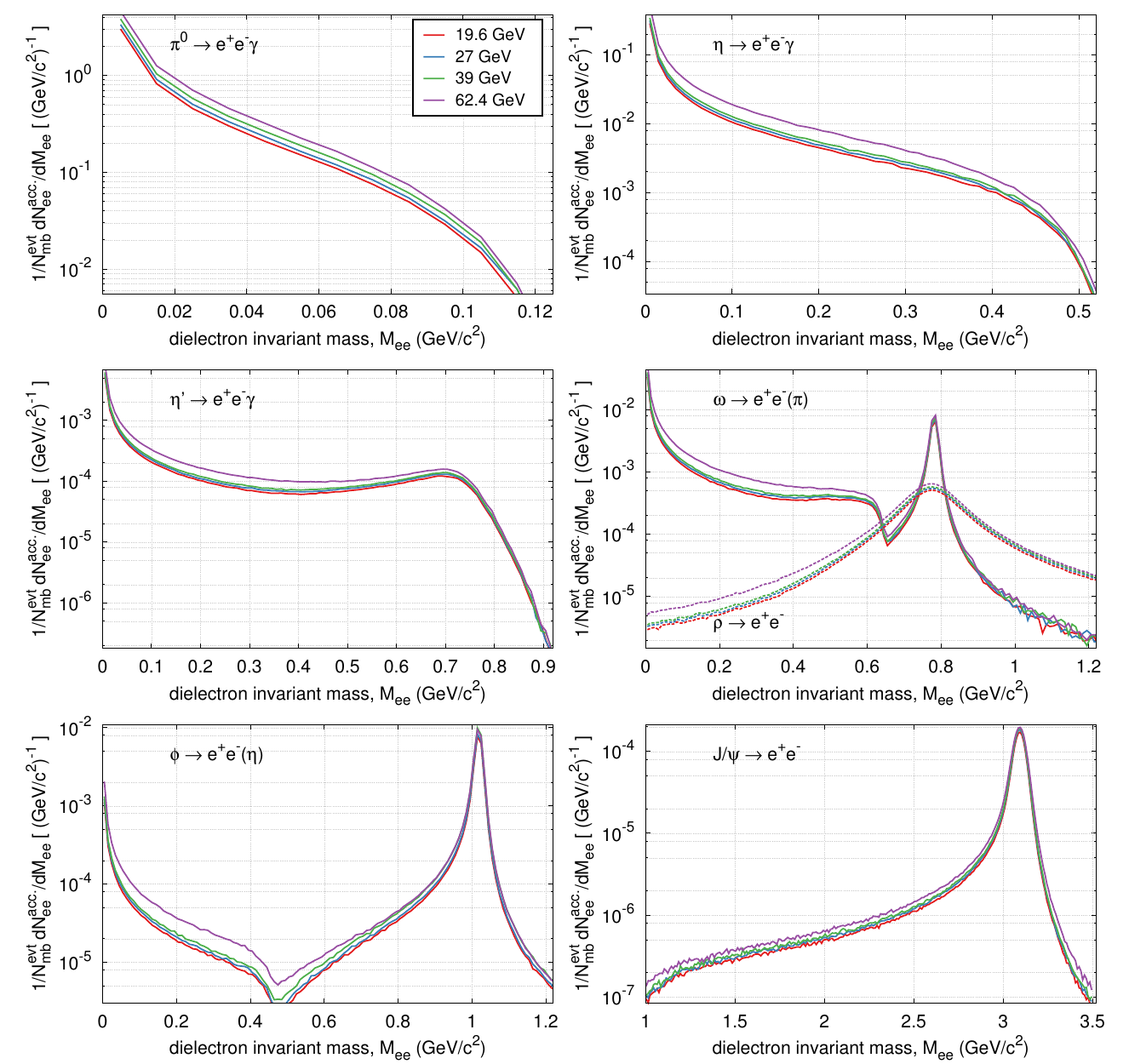}{Simulated invariant mass spectra for all pseudo-{}scalar/vector meson decay channels considered for the cocktail at BES-{}I energies. Though not included in the total cocktail, the \ensuremath{\rho} contribution is depicted, too, since it is used for comparisons in Chapter~\ref{results}. Note that simulated \emph{p$_{\text{T}}$} spectra for six characteristic mass regions have also been generated and are displayed in Figure~\ref{results_fig_pT}.}{sim_fig_vms}{0.94} %

Estimates of systematic uncertainties on the resulting invariant mass spectra
shown in Figure~\ref{sim_fig_vms} are primarily due the pseudo-{}scalar/vector meson yields \emph{dN/dy} and
difficult to obtain at BES-{}I energies since finalized measurements in heavy-{}ion
collisions by STAR and/or PHENIX are only available at top RHIC energy. The
publication of STAR's BES-{}I data available in preliminary status and used for
the pion yields in Table~\ref{sim_tab_pars} is underway at the time of this writing.
Hence, the simulations in this thesis are empirically assigned a lump-{}sum
uncertainty of 30\% on the yields. In the BES-{}I regime, only the rather old
measurements from elementary collisions at the Intersecting Storage Rings are also available at all energies
in addition to a few from Fermilab. Section~\ref{app_isr_spectra} compiles all the data
including references and parameterizes the \emph{p$_{\text{T}}$} spectra via Levy functions.
In the future, this could be used to get a better handle on the
simulation-{}related systematic uncertainties of meson yields using number
of participant scaling from p+p to Au+Au collisions.

\subsection{Charm Continuum}
\label{_charm_continuum}\hyperlabel{_charm_continuum}%

The cocktail simulations also include contributions from correlated
semi-{}leptonic decays of charmed resonances (see Table~\ref{sim_tab_srcs}). The
constituents of a $c\bar{c}$-{}pair generated in the initial
collision hadronize into D-{}mesons and \ensuremath{\Lambda}$_{\text{c}}$-{}baryons by combining with
lighter u-{}, d-{} or s-{}quarks. About 16\% of the hadrons subsequently undergo semi-{}leptonic weak
decays showing up as a continuum of $e^+e^-$ pairs in the
intermediate mass range. The dedicated Monte Carlo event generator PYTHIA
(v6.416 [198]) is used through ROOT's interface class \emph{TPythia6} to
simulate the charm production in center-{}of-{}mass minimum-{}bias p+p collisions.
The resulting distributions are subsequently scaled to Au+Au collisions using
the average number of binary collisions N$_{\text{coll}}$ measured by STAR at the
respective beam energy.

The PYTHIA simulation is initialized in its default minimum-{}bias mode
(\texttt{MSEL=\penalty0 1}) with a predefined subset of hard-{}scattering processes in
hadron-{}hadron collisions, i.e. QCD high-{}p$_{\text{T}}$ processes as well as low-{}p$_{\text{T}}$ production. Two main settings are tuned in Figure~\ref{sim_fig_charm_tune_ccX} (left) to
match STAR's measured charmed meson spectrum in p+p collisions
[199] (also see PYTHIA manual [198]):
(\emph{i}) the current default value of 2 GeV/c for the gaussian width of the
primordial k$_{\text{T}}$ distribution inside a hadron is much larger than explainable by
purely non-{}perturbative terms, hence reset to \texttt{PAR\penalty5000 P\penalty5000 (\penalty5000 91)=\penalty0 1.\penalty0 0}, and
(\emph{ii}) the default Q$^{\text{2}}$=4 scale of the hard 2-{}2 scattering processes is tuned
to high-{}p$_{\text{T}}$ Tevatron studies [200] but heavy-{}flavor production
[201, 202] suggests \texttt{PAR\penalty5000 P\penalty5000 (\penalty5000 67)=\penalty0 1.\penalty0 0}.
In addition, the non-{}electron Individual Decay Channels (IDC) 684-{}735, 755-{}807,
818/824-{}850 and 1097-{}1165 are switched off for the contributing charmed
resonances D$^{\text{+}}$, D$^{\text{0}}$, D$_{\text{s}}$ and \ensuremath{\Lambda}$_{\text{c}}$, respectively.  The mesons
\ensuremath{\eta}$_{\text{c}}$, J/\ensuremath{\psi} and \ensuremath{\chi}$_{\text{2c}}$ are disabled, i.e. IDC=857-{}862 switched off,
since they do not contribute to the charm continuum in the dielectron invariant
mass spectrum.  PYTHIA automatically recalculates the effective branching
ratios for the open decay channels and adjusts the generated cross section
according to the decays simulated. During the simulation run, the hadronization
of the quarks generated in the initial collision is modeled via the Lund string
fragmentation model [203]: The force's color lines are
concentrated in narrow tubes between the quark pairs using appropriate string
tension. The quarks oscillate periodically along the direction of this color
string until it eventually breaks and the pair fragments into new
quark-{}antiquark pairs. For each generated p+p event, the simulation identifies
strings that result from the fragmentation of each of the original c-{} and
$\bar{c}$-{}quarks, checks for the existence of their decay products,
and counts the associated number of $c\bar{c}$-{}pairs
($N_{c\bar{c}}^\mathrm{sim}$).

In the analysis of the simulation results, dielectron invariant mass and p$_{\text{T}}$ spectra of the charm continuum are generated within STAR acceptance.  Since the
PYTHIA settings used here exclusively simulate the semi-{}leptonic decay channels
of the charmed resonances, each of the decay dielectrons needs to be weighted with the
product of the two branching ratios reponsible for $c,\bar{c}\to e^\pm$ when filling the according histograms (denoted below by the
$\otimes$ symbol). The number of dielectrons within STAR acceptance
emanating from charmed resonance decays and expected in a single minimum-{}bias
p+p event is then obtained by scaling the distributions from the relative
$c\bar{c}$ cross section in simulation to the one measured in
minimum-{}bias nucleon-{}nucleon collisions using the factor
\[\frac{\left.\sigma_{c\bar{c}}^\mathrm{NN}\middle/\sigma_\mathrm{mb}^\mathrm{NN}\right.} {\left.\sigma_{c\bar{c}}^\mathrm{sim}\middle/\sigma_\mathrm{mb}^\mathrm{sim}\right.}= \frac{N_\mathrm{evt}^\mathrm{sim}}{N_{c\bar{c}}^\mathrm{sim}}\cdot \frac{\sigma_{c\bar{c}}^\mathrm{NN}}{\sigma_\mathrm{mb}^\mathrm{NN}}\]
which in summary gives
\[\left.\frac{dN_{ee}^\mathrm{acc}}{N_\mathrm{evt}dM_{ee}}\right|_{c\bar{c}}^\mathrm{exp} \hspace{3mm}=\hspace{3mm} \left[\frac{1}{\cancel{N_\mathrm{evt}}} \left(\frac{dN_{ee}^\mathrm{acc}}{dM_{ee}} \otimes\mathrm{BR}^2_{\substack{c\to e^+\\\bar{c}\to e^-}}\right) \right]_{c\bar{c}}^\mathrm{sim} \hspace{3mm}\cdot\hspace{3mm} \frac{\cancel{N_\mathrm{evt}^\mathrm{sim}}}{N_{c\bar{c}}^\mathrm{sim}} \hspace{1.5mm}\cdot\hspace{1.5mm} \frac{\sigma_{c\bar{c}}^\mathrm{NN}}{\sigma_\mathrm{mb}^\mathrm{NN}} \hspace{3mm}\cdot\hspace{3mm} N_\mathrm{coll}.\]
Since to date, many of the according measurements are not available or show
sizeable spreads at BES-{}I energies, a data-{}driven approach is chosen to
determine estimates for the experimental $c\bar{c}$ cross sections
$\sigma_{c\bar{c}}^\mathrm{NN}$. In Figure~\ref{sim_fig_charm_tune_ccX} (right) the double-{}log representation of its measured energy dependence is
parameterized via $y\sim x\ominus\exp(-x)$ in the range from SPS up to
LHC energies. 1.5-{}times the (average) standard deviation of the data points
from the parameterization is taken as the associated systematic uncertainty at
each energy to cover most of the variation in the data.

\wrapifneeded{0.50}{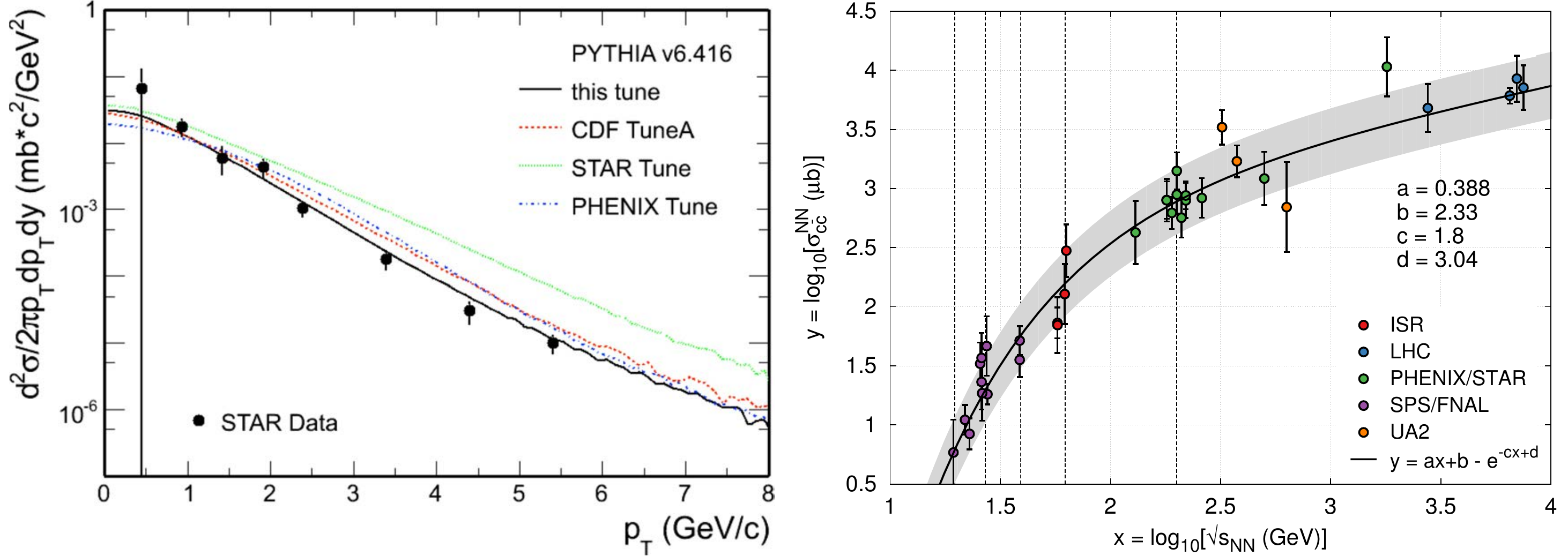}{(left) Parameter tuning for the initialization of the PYTHIA simulation via STAR's measured D$^{\text{0}}$ spectrum in p+p collisions at \ensuremath{\surd}s$_{\text{NN}}$ = 200 GeV [199] in comparison to other or preceding choices. (right) Parameterization of the charm cross section in elementary collisions using data from a variety of experiments and energies. Vertical dashed lines denote BES-{}I and top RHIC energies. The grey band depicts 1.5x the standard deviation chosen as systematic uncertainty.}{sim_fig_charm_tune_ccX}{1} %

\subsection{Summary of Simulation Results}
\label{_summary_of_simulation_results}\hyperlabel{_summary_of_simulation_results}%

In conclusion of this section, the panel in Figure~\ref{sim_fig_panel} compiles the
total simulated dielectron invariant mass spectra for the BES-{}I energies
including a band for its systematic uncertainties. All the single hadronic
contributions to the cocktail are shown as well, with the charm continuum
highlighted in blue. Note that the freeze-{}out \ensuremath{\rho}  is in fact depicted as
dashed lines but not included in the total cocktail since the contributions due
to in-{}medium modified spectral functions are derived from model calculations in
Section~\ref{sim_sec_model} and compared to the measured spectra in Chapter~\ref{results}.\newline
 Propagation of the systematic uncertainties in the pseudo-{}scalar/vector mesons' mid-{}rapidity
yields and the charm cross sections results in total uncertainties of about
20-{}60\% (Figure~\ref{sim_fig_total_overlays} left) on the total cocktail simulation
(Figure~\ref{sim_fig_total_overlays} right).

\wrapifneeded{0.50}{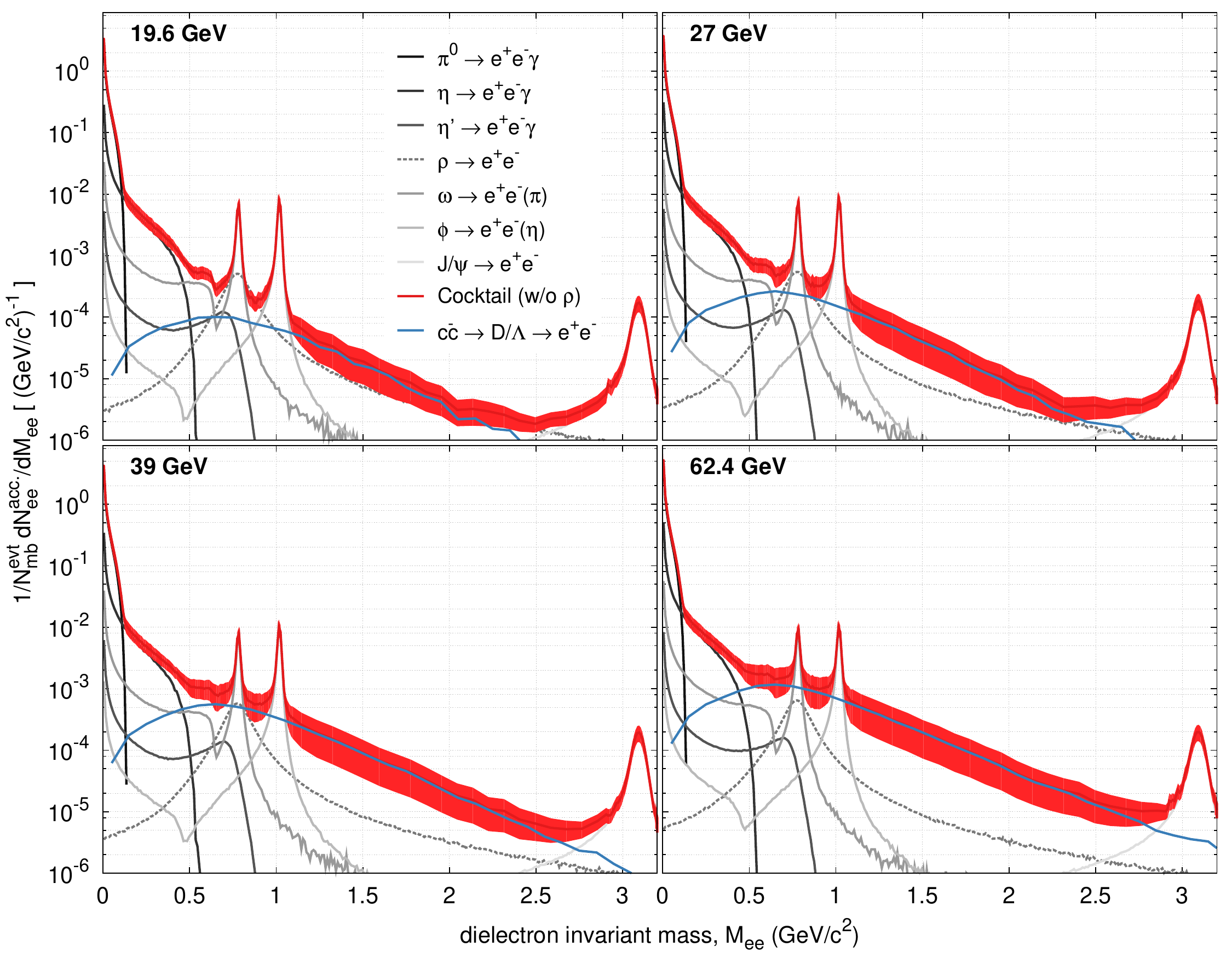}{Compilation of the cocktail simulations at all BES-{}I energies as expected from hadronic sources. The red band denotes the total systematic uncertainty on the cocktail. The freeze-{}out \ensuremath{\rho} contribution is depicted in dashed lines since it is not part of the cocktail here and instead compared to model calculations and data in Chapter~\ref{results}.}{sim_fig_panel}{0.92} %

\wrapifneeded{0.50}{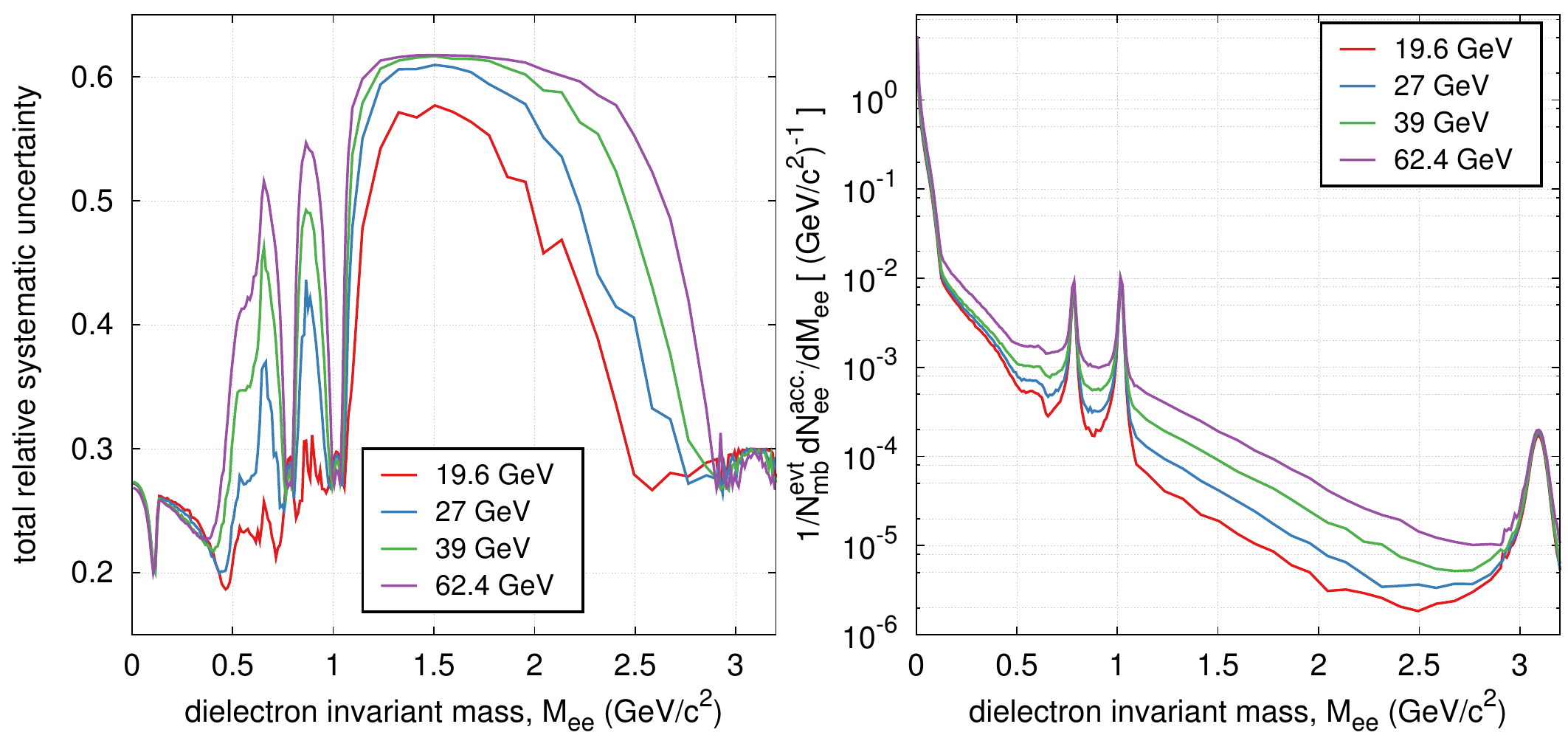}{Simulated invariant mass and energy dependence of the cocktail's total relative uncertainty (left) and total invariant dielectron yields (right). The former is dominated by uncertainties in the mesons' mid-{}rapidity yields and the charm cross sections in elementary collisions.}{sim_fig_total_overlays}{0.85} %

\section{In-{}Medium Modifications}
\label{sim_sec_model}\hyperlabel{sim_sec_model}%

The cocktail simulations of Section~\ref{sim_sec_cocktail} do not include any contributions from the medium, particularly not those
from in-{}medium effects on low-{}mass vector mesons discussed and connected to
chiral symmetry restoration in Chapter~\ref{intro}. In support of the related discussion
and the interpretation of the thesis results in Chapter~\ref{results}, this section is
dedicated to a brief introduction of the most prevalent aspects of the Hadronic
Many-{}Body Theory (HMBT) employed in this thesis to predict the effects of
in-{}medium modifications. The theory addresses in-{}medium spectral functions and
fireball evolution, both necessary to arrive at the integrated dielectron
emission rates measured experimentally via invariant mass spectra. Many and
more of the details selected in the following recapitulation can primarily be
found in [113] but also in [3, 204, 87]. The
conclusion of this section presents the latest model calculations performed by
Rapp [205]. In Chapter~\ref{results}, they are compared to the
efficiency-{}corrected experimental data obtained in Chapter~\ref{data_analysis} and
Chapter~\ref{effcorr}.

\subsection{Dilepton Production Rates}
\label{sim_subsec_prorates}\hyperlabel{sim_subsec_prorates}%

The time-{}reverse process to dielectron production (in the vacuum) is
$e^+e^-$ annihilation into a myriad of hadrons depending on the
center-{}of-{}mass energy $\sqrt{s}=M_{ee}$. The QCD effects can be
separated from the basic QED-{}only processes by taking the ratio
$R(\sqrt{s})$ of the total $e^+e^-$ cross section
$\sigma(e^+e^-\to\mathrm{hadrons})$ to the according
$\sigma(e^+e^-\to\mu^+\mu^-)=4\pi\alpha_s^2(s)/3s$ with the running
coupling constant for the strong force
$\alpha_s^{-1}(s)=\beta_0\ln(s/\Lambda^2)$ where
$\beta_0=(11-2n_f/3)/4\pi$, \ensuremath{\Lambda} = 217 MeV the QCD scale, and
$n_f$ the number of quark flavors [206].

\wrapifneeded{0.50}{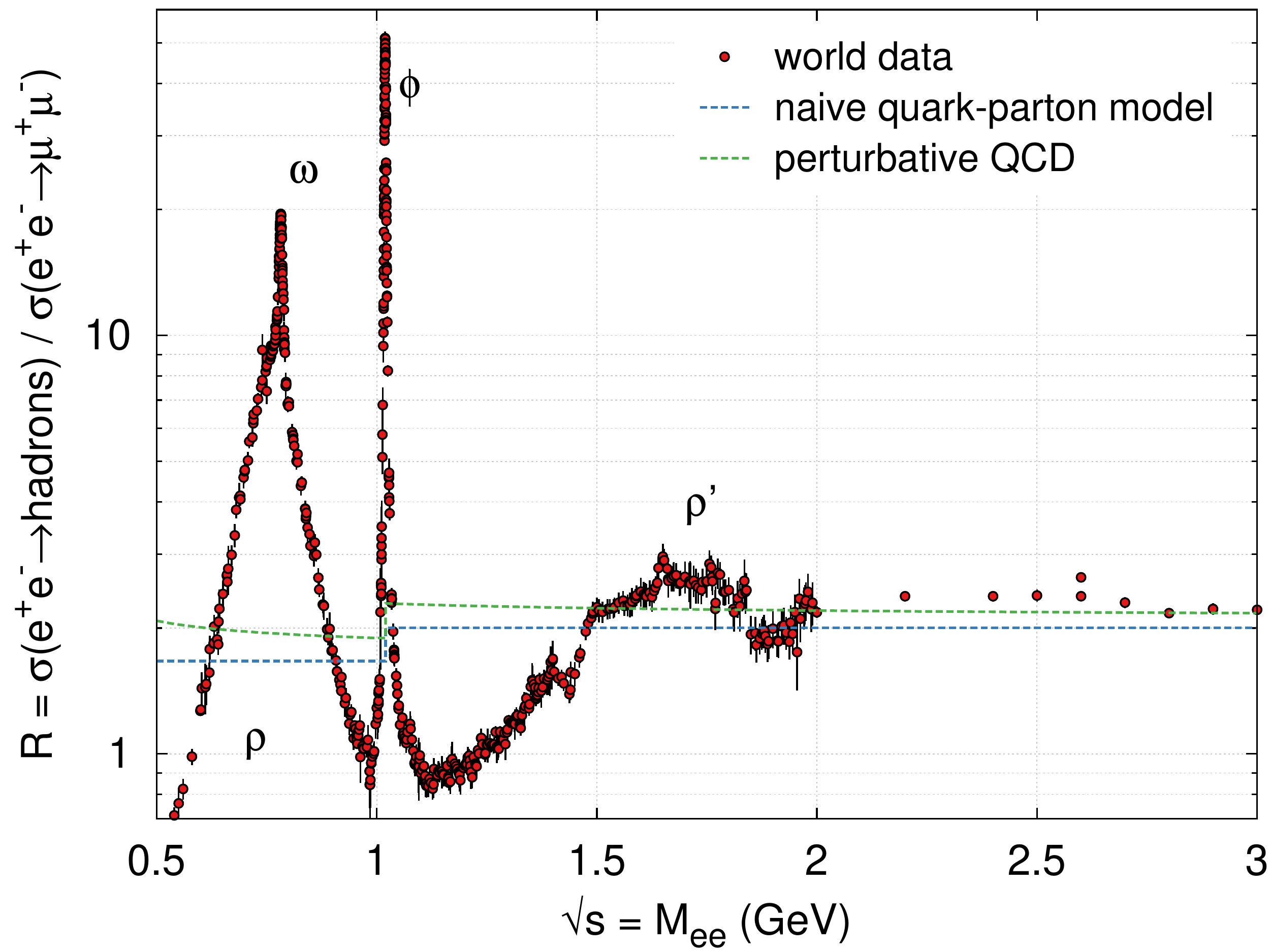}{World data on hadronic annihilation cross section in $e^+e^-$ collisions for ratio to QED-{}only processes. Data from [207]. Details see text.}{ee_hadrons_xsec}{0.49} %

Figure~\ref{ee_hadrons_xsec} shows the available world data for
$R(\sqrt{s})$ below 3 GeV where the light u-{}, d-{} and s-{}quarks
determine the hadronic production dynamics since the collision energy is not
sufficient to produce the bound $c\bar{c}$ state J/\ensuremath{\psi}, yet.  The
data is compared to a naive quark-{}parton model (blue dashed line) in which the
ratio is predicted by purely electro-{}weak considerations to
\[R_\mathrm{EW}(\sqrt{s})=3\sum\limits_q e_q^2=\begin{cases} 5/3 & q\in[u,d],\sqrt{s}<m_\phi\\ 2 & q\in[u,d,s],\sqrt{s}\geq m_\phi \end{cases}\]
with $e_q$ the electric charge of quark \emph{q} and
$m_\phi$ the $\phi$-{}meson rest mass. Also depicted in
green dashed lines is the ratio
\[R(\sqrt{s})=R_\mathrm{EW}(\sqrt{s})\left(1+\delta_\mathrm{QCD}(\sqrt{s})\right)\]
including corrections $\delta_\mathrm{QCD}$ to
$R_\mathrm{EW}$ due to the QCD effects which can be calculated
perturbatively up to $\alpha_s^4$ via [206, 208]
\[\delta_\mathrm{QCD}(\sqrt{s})= \sum\limits_{n=1}^{4}c_n\left(\frac{\alpha_s(s)}{\pi}\right)^n +\mathcal{O}\left(\frac{\Lambda^4}{s^2}\right) \mbox{ with }c_1=1,\begin{cases} c_2=1.76,\,c_3=-9.14,\,c_4=-121&\mbox{for }n_f=2\\ c_2=1.64,\,c_3=-10.3,\,c_4=-107&\mbox{for }n_f=3. \end{cases}\]
Above about $M_\mathrm{dual}$ = 1.5 -{} 2 GeV, the measured
$e^+e^-$ annihiliation cross section approaches and is
satisfactorily accounted for by the calculated perturbative QCD rate and thus
negligibly impacted by hadronization. $M_\mathrm{dual}$ hence
constitutes a \emph{duality threshold} below which the resonance structures can
accurately be described by the light vector mesons \ensuremath{\rho}, \ensuremath{\omega} and \ensuremath{\phi}
through Vector Meson Dominance (VMD, [97]). This allows for the
decomposition of the imaginary part of the vacuum electro-{}magnetic
current-{}current correlator into
\[\mathrm{Im}\,\Pi^\mathrm{vac}_\mathrm{em}= \begin{cases} \sum_Vm_V^4/g_V^2\hspace{2mm}\cdot\hspace{2mm}\mathrm{Im}D_V & \mbox{if }M_{ee}\leq M_\mathrm{dual},\,V=\rho,\omega,\phi\\ -M_{ee}^2/12\pi\hspace{2mm}\cdot\hspace{2mm}R(\sqrt{s}=M_{ee}) & \mbox{if } M_{ee} > M_\mathrm{dual} \end{cases}\]
with the imaginary part of the vector meson propagator
$\mathrm{Im}D_V$ denoting its spectral function.
$\mathrm{Im}\,\Pi_\mathrm{em}$ provides the general framework for
the study of thermal dilepton emission [209, 210] since the
according dilepton production rate can directly be obtained via
\[\frac{\mathrm{d}R_{ee}}{\mathrm{d}^4q} = -\frac{\alpha_\mathrm{em}^2}{\pi^3M_{ee}^2}f^B\,\mathrm{Im}\Pi_\mathrm{em}\]
with \emph{f$^{\text{B}}$} the thermal Bose distribution function.\newline

When embedding low-{}mass vector mesons in an environment of hot hadronic matter
the extension of $\mathrm{Im}\,\Pi_\mathrm{em}$ to finite
temperature and baryo-{}chemical potentials is achieved by calculating
approximations of the respective spectral functions -{} assuming that VMD still
holds in the medium. In HMBT, standard many-{}body techniques are employed to
incorporate medium effects on a microscopic level in the \ensuremath{\rho}-{}meson propagator
$D_\rho$ through self-{}energy corrections \ensuremath{\Sigma} due to
\ensuremath{\pi}-{}cloud polarization and resonant interaction with surrounding mesons \emph{M} and baryons \emph{B}:
\[D_\rho(M_{ee};\mu_B,T) = \left[M_{ee}^2 - m_\rho^2 - \Sigma_{\rho\pi\pi} - \Sigma_{\rho B}-\Sigma_{\rho M} \right]^{-1}\]
with $M=\pi,K,\rho,\dots$ and
$B=N,\Lambda,\Delta,\dots$ from the heat bath. The calculation of
processes involved in the \ensuremath{\Sigma} corrections is represented by the Feynman
diagrams in Figure~\ref{sim_fig_feynm}, for instance, which are allowed to subsequently
occur multiple times and in different combinations.

\wrapifneeded{0.50}{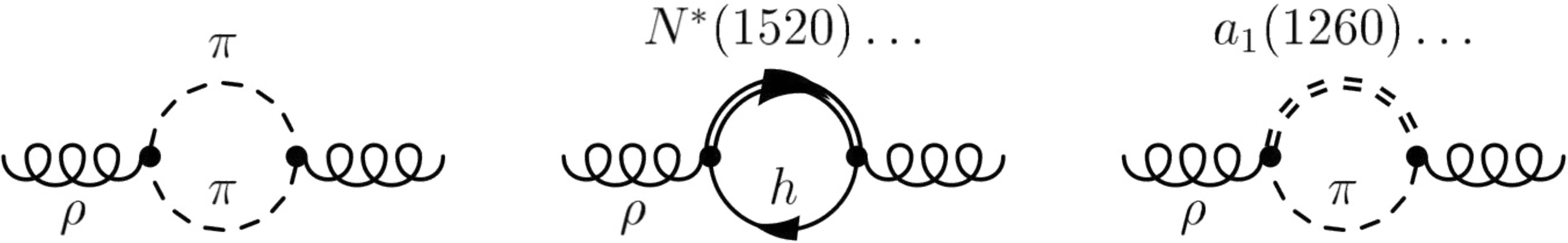}{Feynman diagrams for self-{}energy corrections in the \ensuremath{\rho}-{}meson propagator.}{sim_fig_feynm}{0.6} %

The resulting LMR medium effects for the \ensuremath{\rho}-{}meson turn out to cause the most
prominent contribution to the calculated invariant mass spectra
(Figure~\ref{sim_fig_model_summary}) while being well accessible in experiment
(Chapter~\ref{results}). In comparison, the changes in the spectral functions of \ensuremath{\omega}-{}
and \ensuremath{\phi}-{}meson are mostly hidden beneath the according freeze-{}out decays. For
details on the self-{}energy corrections of \ensuremath{\omega}-{} and \ensuremath{\phi}-{}mesons, the
reader is referred to the literature [113] since the remainder of
this section will concentrate on the discussion of LMR in-{}medium effects caused
by the \ensuremath{\rho}-{}meson.\newline

Figure~\ref{sim_fig_rhosf} (left) shows the resulting temperature dependence for the
\ensuremath{\rho}-{}meson spectral function. A rather strong broadening is observed towards
the phase boundary at \emph{T} \ensuremath{\sim} 180 MeV (with a small mass shift). Since the
\ensuremath{\rho}-{}meson is its own antiparticle with respect to the strong force, an
appreciable additional broadening arises around the phase boundary temperature
from its equal interaction with baryons and antibaryons (Figure~\ref{sim_fig_rhosf} right). In contrary, note that the \ensuremath{\phi}-{}meson is mostly agnostic to these
in-{}medium effects presumably due to the OZI-{}rule which explains the absence of
baryonic resonances with \ensuremath{\phi}N decay channels by the strong suppression of
antistrange quark annihilation needed to supply the excitation energy.

\wrapifneeded{0.50}{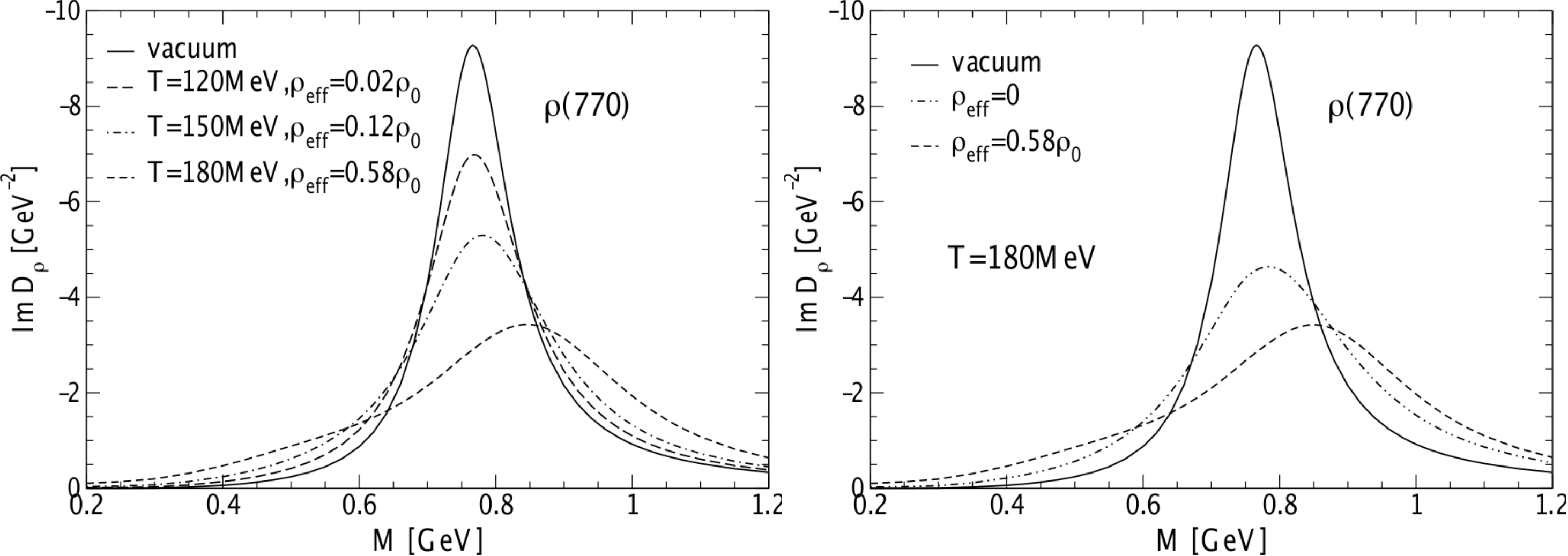}{T-{} (left) and \ensuremath{\mu}$_{\text{B}}$-{}dependence (right) of \ensuremath{\rho}-{}meson spectral function [113]. The effective baryon densities \ensuremath{\rho}$_{\text{eff}}$ correspond to \ensuremath{\mu}$_{\text{B}}$ of 91, 40 and 27 MeV from top to bottom, respectively.}{sim_fig_rhosf}{0.95} %

Above the duality threshold, the free electro-{}magnetic correlator can be
evaluated using the lowest-{}order thermal $q\bar{q}$ annihilation rate. For the QGP phase, this constitutes a lower limit of the
true emission rates since theoretical predictions for in-{}medium corrections are
challenging. Recent progress to calculate QGP dilepton rates non-{}perturbatively
has been achieved using thermal Lattice QCD (lQCD, [211]) which is
used in Section 2.2 of [115] to construct an extension of the QGP
rates to finite 3-{}momentum. When compared to the according HMBT calculations
close to the transition temperature, the "melting" of the \ensuremath{\rho}-{}meson's vacuum
resonance structure in the medium suggests a smooth transition into the
structureless lQCD shape with partonic degrees of freedom.  Since leading-{}order
temperature effects alone already reduce the duality threshold towards the
\ensuremath{\phi}, the QGP rate is extrapolated into the hadronic phase and employed
down to M$_{\text{ee}}$ \ensuremath{\sim} 1.1 GeV/c$^{\text{2}}$.

\subsection{Fireball Evolution}
\label{_fireball_evolution}\hyperlabel{_fireball_evolution}%

\wrapifneeded{0.50}{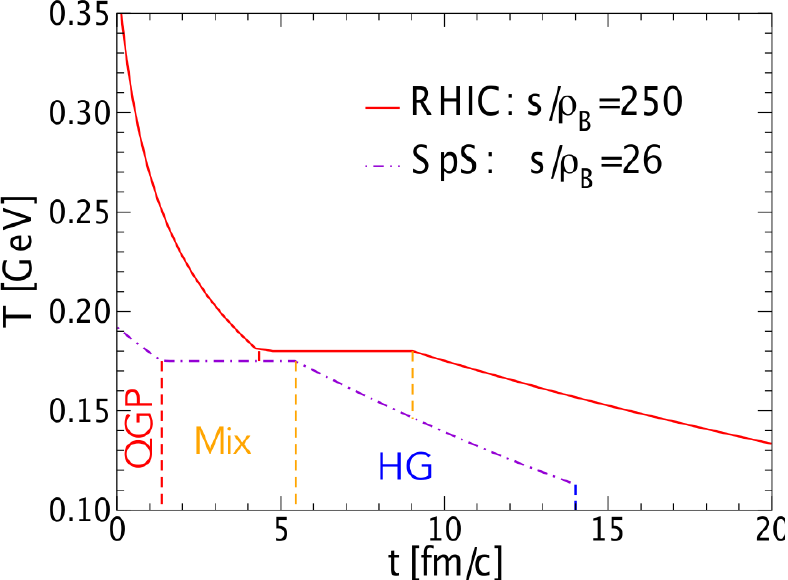}{Fireball temperature profile at SPS and RHIC conditions exhibiting QGP, Mix(ed) and Hadron Gas phase [113].}{sim_fig_fb_profile}{0.47} %

For the calculation of dielectron invariant mass spectra and their comparison
to experimental results, the production rates need to be integrated over the
space-{}time evolution of a heavy-{}ion collision. A simplified fireball expansion
model with time-{}dependent cylindrical volume $V_\mathrm{FC}(t)$ is
employed in [205] using parameters that reflect measured expansion/flow velocities and
particle spectra as most important features of the full hydrodynamic
simulation. Figure~\ref{sim_fig_fb_profile} shows the time-{}dependent temperature profile
of the fireball expansion which is needed for the derivation of the dilepton
production rates. Throughout the entire evolution the system's trajectory in
\emph{(T,\ensuremath{\mu}$_{\text{B}}$)} is determined by isentropic expansion, i.e. constant entropy per
baryon \emph{s/\ensuremath{\rho}$_{\text{B}}$}. At the RHIC energies used in this thesis, chemical freeze-{}out can be assumed to coincide
with the critical temperature for the chiral/deconfinement transition and is
hence fixed at \emph{T$_{\text{c}}$} \ensuremath{\sim} 170 MeV. The hadronic phase's entropy below \emph{T$_{\text{c}}$} develops toward thermal (kinetic) freeze-{}out according to the standard HRG
equation of state (EoS), whereas above \emph{T$_{\text{c}}$}, the QGP phase's entropy is
governed by a lQCD EoS starting at its formation time. Note that
[115] also contains a discussion of this thesis's results in a
preliminary version compared to model calculations using a quasi-{}particle EoS
and \emph{T$_{\text{c}}$} = 180 MeV. However, as already discussed in
Section~\ref{intro_subsec_theo_thermal}, the total yield stays approximately the same due
to the duality of emission rates around \emph{T$_{\text{c}}$} which enhances the QGP but
equally reduces the hadronic yield. During the mixed phase at constant \emph{T$_{\text{c}}$},
the hadronic matter fraction can be inferred from entropy balance since the
entire system is in the QGP phase if its total volumetric entropy density
$S_\mathrm{tot}/V_\mathrm{FC}(t)$ is larger than
$s_\mathrm{QGP}(T_c)$. Table~\ref{sim_tab_T0_dNdy} quotes the charged
particles rapidity densities $dN_\mathrm{ch}/dy$ which are the
remaining necessary input parameters for the model calculations at BES-{}I
energies. Also listed are the initial temperatures \emph{T$_{\text{0}}$} for a chemically
equilibrated QGP resulting from the fireball expansion parameters.
\begin{center}
\begingroup%
\setlength{\newtblsparewidth}{1\linewidth-2\tabcolsep-2\tabcolsep-2\tabcolsep-2\tabcolsep-2\tabcolsep-2\tabcolsep-2\tabcolsep-2\tabcolsep-2\tabcolsep-2\tabcolsep-2\tabcolsep-2\tabcolsep}%
\setlength{\newtblstarfactor}{\newtblsparewidth / \real{429}}%

\begin{longtable}{lllllllllll}\caption[{QGP temperature \emph{T$_{\text{0}}$} and charged particles \emph{dN$_{\text{ch}}$/dy} for HMBT at BES-{}I energies [205].}]{QGP temperature \emph{T$_{\text{0}}$} and charged particles \emph{dN$_{\text{ch}}$/dy} for HMBT at BES-{}I energies [205].\label{sim_tab_T0_dNdy}\hyperlabel{sim_tab_T0_dNdy}%
}\tabularnewline
\endfirsthead
\caption[]{(continued)}\tabularnewline
\endhead
\hline
\multicolumn{1}{m{80\newtblstarfactor+\arrayrulewidth}}{\centering%
$\sqrt{s_\mathrm{NN}}$ (GeV)
}&\multicolumn{1}{m{27\newtblstarfactor+\arrayrulewidth}}{\centering%
19.6
}&\multicolumn{1}{m{27\newtblstarfactor+\arrayrulewidth}}{\centering%
27
}&\multicolumn{1}{m{27\newtblstarfactor+\arrayrulewidth}}{\centering%
39
}&\multicolumn{1}{m{27\newtblstarfactor+\arrayrulewidth}}{\centering%
62.4
}&\multicolumn{1}{m{53\newtblstarfactor+\arrayrulewidth}}{\centering%
}&\multicolumn{1}{m{80\newtblstarfactor+\arrayrulewidth}}{\centering%
$\sqrt{s_\mathrm{NN}}$ (GeV)
}&\multicolumn{1}{m{27\newtblstarfactor+\arrayrulewidth}}{\centering%
19.6
}&\multicolumn{1}{m{27\newtblstarfactor+\arrayrulewidth}}{\centering%
27
}&\multicolumn{1}{m{27\newtblstarfactor+\arrayrulewidth}}{\centering%
39
}&\multicolumn{1}{m{27\newtblstarfactor+\arrayrulewidth}}{\centering%
62.4
}\tabularnewline
\multicolumn{1}{m{80\newtblstarfactor+\arrayrulewidth}}{\centering%
\emph{T$_{\text{0}}$} (MeV)
}&\multicolumn{1}{m{27\newtblstarfactor+\arrayrulewidth}}{\centering%
224
}&\multicolumn{1}{m{27\newtblstarfactor+\arrayrulewidth}}{\centering%
230
}&\multicolumn{1}{m{27\newtblstarfactor+\arrayrulewidth}}{\centering%
237
}&\multicolumn{1}{m{27\newtblstarfactor+\arrayrulewidth}}{\centering%
283
}&\multicolumn{1}{m{53\newtblstarfactor+\arrayrulewidth}}{\centering%
}&\multicolumn{1}{m{80\newtblstarfactor+\arrayrulewidth}}{\centering%
$dN_\mathrm{ch}/dy$
}&\multicolumn{1}{m{27\newtblstarfactor+\arrayrulewidth}}{\centering%
135
}&\multicolumn{1}{m{27\newtblstarfactor+\arrayrulewidth}}{\centering%
140
}&\multicolumn{1}{m{27\newtblstarfactor+\arrayrulewidth}}{\centering%
145
}&\multicolumn{1}{m{27\newtblstarfactor+\arrayrulewidth}}{\centering%
185
}\tabularnewline
\hline
\end{longtable}\endgroup%

\end{center}

\subsection{Summary of Model Calculation Results}
\label{_summary_of_model_calculation_results}\hyperlabel{_summary_of_model_calculation_results}%

The $e^+e^-$ pair yields have been computed for the use in this
thesis [205] uniformly in rapidity $y\in(-1,1)$ employing STAR
acceptance cuts for the individual leptons and neglecting momentum
smearing to account for the detector's mass resolution.
Figure~\ref{sim_fig_model_summary} summarizes the results of the model calculations
described in this section by studying their energy dependence (left) and
discussing the consequences on the dielectron invariant mass spectra using
$\sqrt{s_\mathrm{NN}}$ = 27 GeV as a representative example
(right). The sum of cocktail simulations and model calculations is compared in
invariant mass and transverse momentum to the experimental data for all BES-{}I
energies in Chapter~\ref{results} (c.f.  Figure~\ref{results_fig_panel} and Figure~\ref{results_fig_pT}).\newline

The total of HMBT and QGP calculations in Figure~\ref{sim_fig_model_summary} (left)
exhibits very similar shape with only little structure for all BES-{}I energies.
The single contributions are hence only displayed for
$\sqrt{s_\mathrm{NN}}$ = 19.6 GeV to demonstrate that hadronic
in-{}medium modifications dominate up to M$_{\text{ee}}$ \ensuremath{\sim} 1 GeV/c$^{\text{2}}$ beyond which
thermal radiation from the QGP phase takes over. Since the latter originates
from earlier emission times and higher temperatures, its smaller slope
parameter allows it to reach higher invariant masses than the hadronic sources.
However, in this intermediate mass regime (IMR), the QGP signal competes with
the continuum from correlated semi-{}leptonic charmed decays
(c.f. Figure~\ref{sim_fig_panel}) even though it prevails over Drell-{}Yan
contributions up to the J/\ensuremath{\psi} mass.

Figure~\ref{sim_fig_model_summary} (right) compares the HMBT+QGP in-{}medium invariant
mass distributions at $\sqrt{s_\mathrm{NN}}$ = 27 GeV to the
hadronic cocktail without freeze-{}out \ensuremath{\rho}-{}meson as well as to \ensuremath{\rho}/\ensuremath{\omega}
vacuum spectral functions propagated through the fireball (VacSF+FB). In the
\ensuremath{\rho}/\ensuremath{\omega} and \ensuremath{\phi} mass regions (grey boxes) the medium effects are
unfortunately hidden under the respective freeze-{}out contributions even though
the \ensuremath{\rho}/\ensuremath{\omega} resonance structure experiences a rather strong suppression
(c.f. change from short dashed to solid line versus dash-{}dotted line in left
grey box). Due to its already discussed inertness to in-{}medium modifications,
this is especially true for the \ensuremath{\phi}-{}meson for which such effects are
basically unobservable with the available experimental statistics (omitted in
the figure).

However, the comparison also reveals two promising regions for the medium
effects to be detected in the invariant mass spectra. On the one hand, in the
mass region below the \ensuremath{\rho}/\ensuremath{\omega} resonances from 0.4 to 0.75 GeV/c$^{\text{2}}$ (blue
boxes, identical to Figure~\ref{results_fig_panel} and Figure~\ref{results_fig_diffAbsRel}) as
well as in the region between \ensuremath{\rho}/\ensuremath{\omega} and \ensuremath{\phi}, the dominant
contribution to the total yield is caused by (hadronic) in-{}medium effects
rather than freeze-{}out decays (cocktail).  Moreover, almost the entire effect
in these regions is rooted in the broadening of the \ensuremath{\rho}-{}meson's spectral
function [113] with a contributed yield as much as three times larger
than the cocktail at M$_{\text{ee}}$ = 0.6 GeV/c$^{\text{2}}$ (c.f.  enhancement ratios in
Figure~\ref{results_fig_diffAbsRel} right). In the IMR (M$_{\text{ee}}$ >{} 1.1 GeV/c$^{\text{2}}$), on the
other hand, the in-{}medium contributions from the hadronic and QGP phases
jointly amount to a dielectron yield comparable in size and shape to the
$c\bar{c}$ continuum.\newline

It should be emphasized at this point that the preceding conclusions on
in-{}medium hadronic and QGP effects have been reached without prior knowledge of
their comparison to experimental data in Chapter~\ref{results} and can hence be
considered true predictions with respect to observable consequences on
dielectron spectra.

\wrapifneeded{0.50}{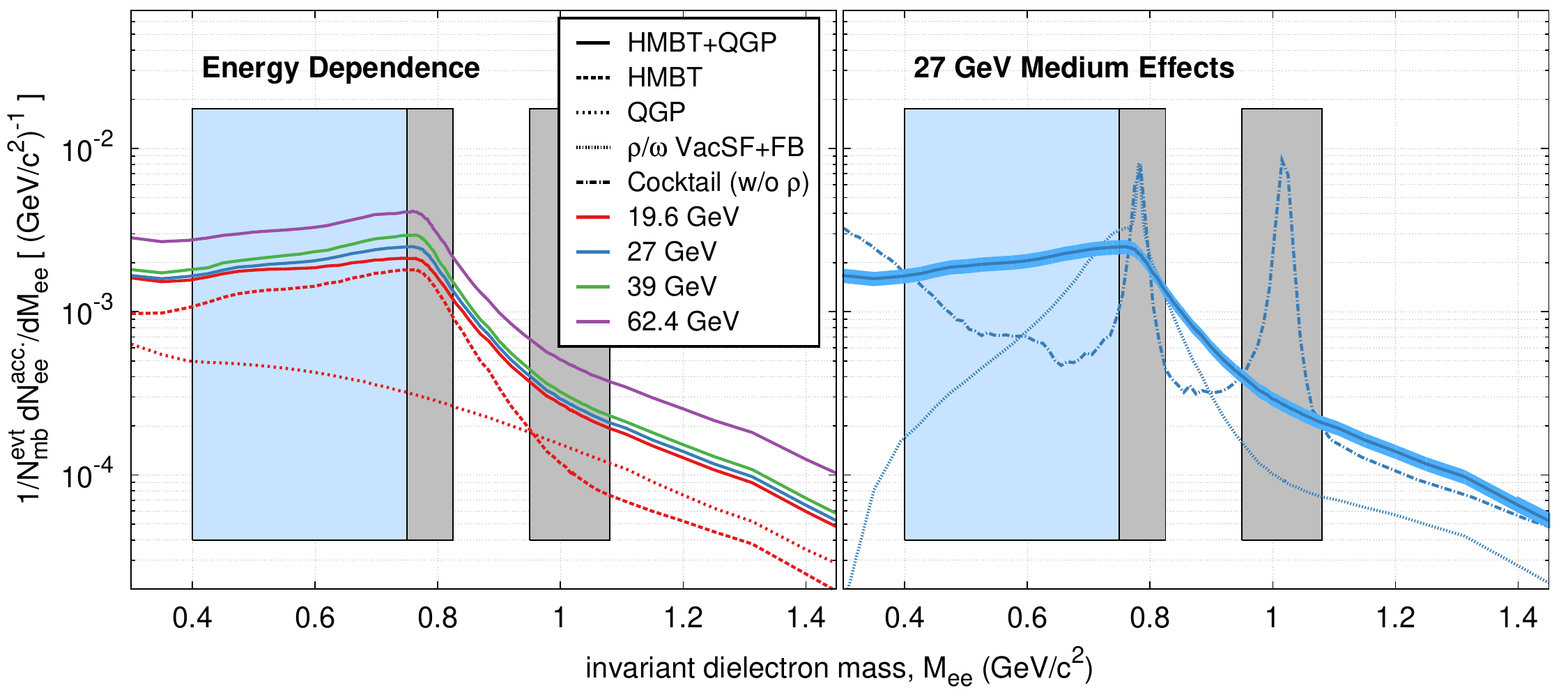}{Dielectron invariant mass spectra of in-{}medium effects in hadronic and QGP phases after fireball evolution. (left) Energy dependence of HMBT+QGP calculations at all BES-{}I energies (solid lines) with single contributions representatively depicted for 19.6 GeV (dashed and dotted lines). (right) Comparison of predicted in-{}medium effects at 27 GeV to cocktail simulation without freeze-{}out \ensuremath{\rho} (dash-{}dotted) and to fireball-{}propagated \ensuremath{\rho}/\ensuremath{\omega} vacuum spectral functions (dense dotted). The blue band denotes a systematic uncertainty of 10\% assigned empirically on the total in-{}medium distribution based on theorists' estimates.}{sim_fig_model_summary}{0.99} %


\chapter{Results and Discussion}
\label{results}\hyperlabel{results}%

This chapter presents the results of this study's energy-{}dependent analysis of
dielectron production in combination with the according published results at
top RHIC energy ($\sqrt{s_\mathrm{NN}}$ = 200 GeV,
[16]). The results cover invariant mass (M$_{\text{ee}}$) and transverse
momentum (p$_{\text{T}}$) spectra of characteristic mass regions as well as their
comparison to simulations and model calculations allowing for the
detailed discussion of their physics interpretation. The Low-{}Mass-{}Region (LMR,
M$_{\text{ee}}$ <{} 1.1 GeV/c$^{\text{2}}$) as one of the particularly interesting regions is
highlighted by studying the energy dependence of observed excess yields and
enhancement ratios over the cocktail with respect to previous measurements,
relation to total baryon density, and expectations for BES-{}II.

Figure~\ref{results_fig_stack} depicts the efficiency-{}corrected invariant mass spectra
of all STAR dielectron measurements in 0-{}80\% minimum bias Au+Au collisions
enabling a systematic study of dielectron production for the wide energy range
from about SPS up to top RHIC energies, namely
$\sqrt{s_\mathrm{NN}}$ = 19.6 -{} 200 GeV. The peak structures of the
\ensuremath{\pi}$^{\text{0}}$ mesons' Dalitz decay and especially the direct decays of the \ensuremath{\omega}-{},
\ensuremath{\phi}-{} and J/\ensuremath{\psi}-{}mesons are nicely identified in the experimental spectra
despite the rather poor signal-{}to-{}background ratio in the \ensuremath{\omega}/\ensuremath{\phi} mass
region (see Section~\ref{ana_sec_pairrec}). The quality of the data allows for the
invariant mass spectra to be measured up to M$_{\text{ee}}$ \ensuremath{\sim} 3.5 GeV/c$^{\text{2}}$. The
scaling factors in Table~\ref{results_tab_scale} are used to scale the cocktail
simulations described in Section~\ref{sim_sec_cocktail} to the experimental
spectra in the \ensuremath{\pi}$^{\text{0}}$ region (M$_{\text{ee}}$ <{} 0.1 GeV/c$^{\text{2}}$).  The choice of employing scaling
factors is justified in light of the difficulties involved in the determination
of cocktail yields
\footnote{
Note: STAR's latest \ensuremath{\pi}/K/p yields are not published and their systematic uncertainties still under evaluation.
} as well as the efficiency correction of the data
(Chapter~\ref{effcorr}). It also allows for a more meaningful comparison to model
calculations above the \ensuremath{\pi}$^{\text{0}}$ mass region since neither conclusions regarding
the absolute excess nor regarding the enhancement factors, for instance, should
be biased by a mass-{}independent overall scaling of the spectra. As for the
magnitude of the scaling factors, it should be noted that even the spectra at
$\sqrt{s_\mathrm{NN}}$ = 200 GeV require an adjustment of about 10\%
despite very high statistics and low(er) systematic uncertainties.  The scaling
factors for BES-{}I energies are thus not only within reach of the respective
systematic uncertainties (see uncertainties in Table~\ref{results_tab_scale}) but also
within acceptable limits as empirically expected from the measurement at top
RHIC energy.
\begin{center}
\begingroup%
\setlength{\newtblsparewidth}{0.8\linewidth-2\tabcolsep-2\tabcolsep-2\tabcolsep-2\tabcolsep-2\tabcolsep-2\tabcolsep-2\tabcolsep}%
\setlength{\newtblstarfactor}{\newtblsparewidth / \real{342}}%

\begin{longtable}{llllll}\caption[{Scaling factors \emph{F} of cocktail to data below 0.1 GeV/c$^{\text{2}}$ of the invariant mass spectrum for all BES-{}I and top RHIC energies. Errors are combined statistical and systematic uncertainties with the former negligible due to the abundant \ensuremath{\pi}$^{\text{0}}$ signal.}]{Scaling factors \emph{F} of cocktail to data below 0.1 GeV/c$^{\text{2}}$ of the invariant mass spectrum for all BES-{}I and top RHIC energies. Errors are combined statistical and systematic uncertainties with the former negligible due to the abundant \ensuremath{\pi}$^{\text{0}}$ signal.\label{results_tab_scale}\hyperlabel{results_tab_scale}%
}\tabularnewline
\endfirsthead
\caption[]{(continued)}\tabularnewline
\endhead
\hline
\multicolumn{1}{m{57\newtblstarfactor+\arrayrulewidth}}{\centering%
$\sqrt{s_\mathrm{NN}}$
}&\multicolumn{1}{m{57\newtblstarfactor+\arrayrulewidth}}{\centering%
19.6 GeV
}&\multicolumn{1}{m{57\newtblstarfactor+\arrayrulewidth}}{\centering%
27 GeV
}&\multicolumn{1}{m{57\newtblstarfactor+\arrayrulewidth}}{\centering%
39 GeV
}&\multicolumn{1}{m{57\newtblstarfactor+\arrayrulewidth}}{\centering%
62.4 GeV
}&\multicolumn{1}{m{57\newtblstarfactor+\arrayrulewidth}}{\centering%
200 GeV
}\tabularnewline
\multicolumn{1}{m{57\newtblstarfactor+\arrayrulewidth}}{\centering%
\emph{F}
}&\multicolumn{1}{m{57\newtblstarfactor+\arrayrulewidth}}{\centering%
0.83\ensuremath{\pm}0.13
}&\multicolumn{1}{m{57\newtblstarfactor+\arrayrulewidth}}{\centering%
0.82\ensuremath{\pm}0.12
}&\multicolumn{1}{m{57\newtblstarfactor+\arrayrulewidth}}{\centering%
0.87\ensuremath{\pm}0.14
}&\multicolumn{1}{m{57\newtblstarfactor+\arrayrulewidth}}{\centering%
0.87\ensuremath{\pm}0.13
}&\multicolumn{1}{m{57\newtblstarfactor+\arrayrulewidth}}{\centering%
0.90\ensuremath{\pm}0.09
}\tabularnewline
\hline
\end{longtable}\endgroup%

\end{center}

\wrapifneeded{0.50}{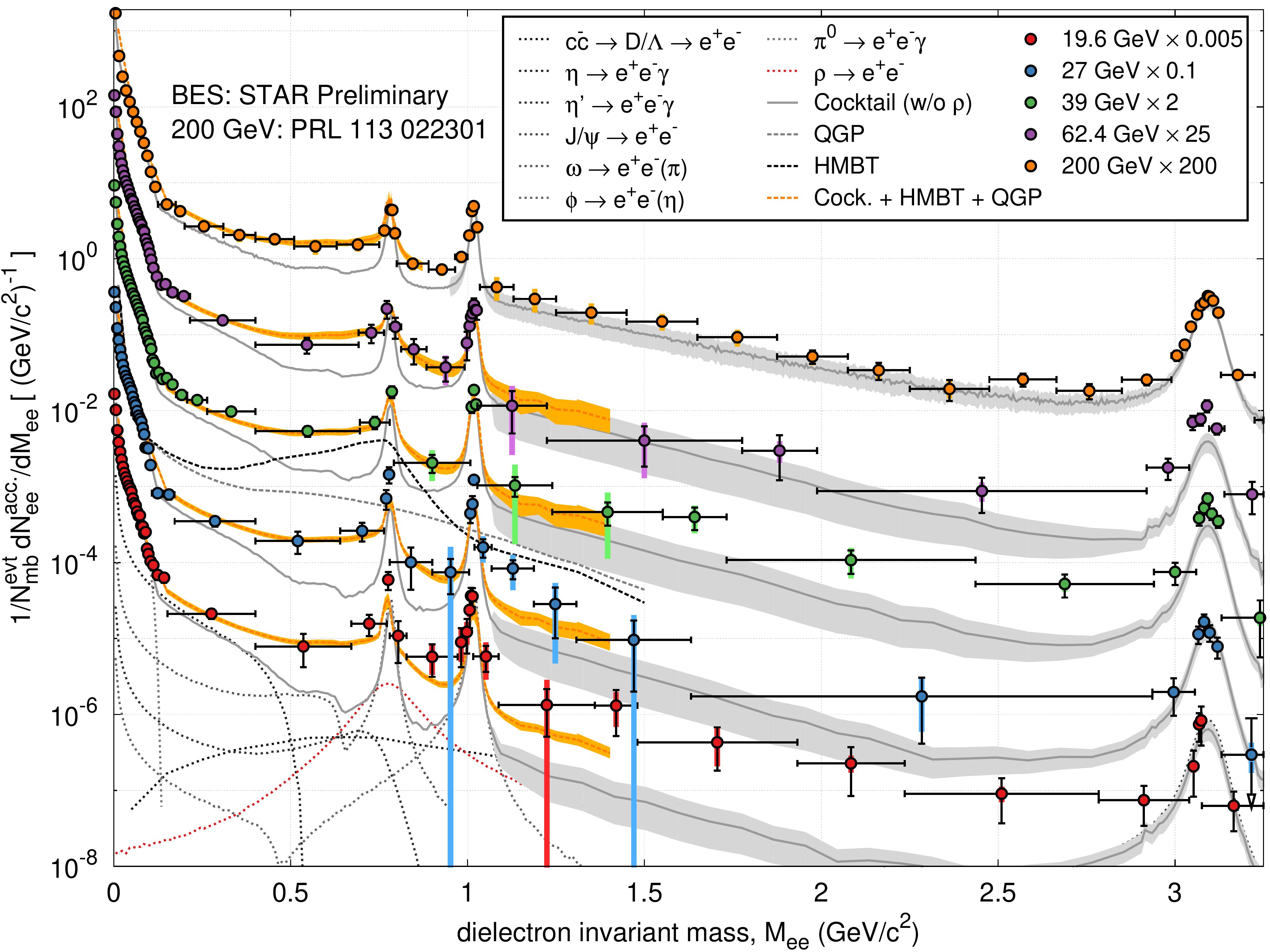}{Efficiency-{}corrected invariant mass dielectron spectra for 0-{}80\% minimum bias Au+Au collisions at BES-{}I and top RHIC energies compared to cocktail simulations and model calculations [3, 205]. In addition to the total cocktail (grey), the single hadronic freeze-{}out contributions are depicted for 19.6 GeV (dotted lines). For 39 GeV, the two model contributions from hadronic gas (HMBT) and QGP phase (dashed lines) are shown next to the total expected yield (orange band).}{results_fig_stack}{0.83} %

The cocktail simulations generally show good agreement with the data over the
entire mass range covered with sufficient statistics during BES-{}I. In the LMR,
though, and as previously reported by other measurements (see
Section~\ref{intro_sec_dielec}), all spectra consistently exhibit a sizeable excess
over the cocktail. Contributions from direct \ensuremath{\rho} decays after freeze-{}out
using a simple Breit-{}Wigner shape are depicted for comparison at
$\sqrt{s_\mathrm{NN}}$ = 19.6 GeV. However, they are not included
in the total cocktails used in this chapter since they by far do not account
for the magnitude of the enhancement [212].  Instead, the
in-{}medium model calculations treat the full evolution of the \ensuremath{\rho}-{}meson
including the appropriate amount of freeze-{}out decays (see Section~\ref{sim_sec_model})
and hence need to be added on top of the cocktail without any \ensuremath{\rho}
contributions. In particular, the scenario of a broadened \ensuremath{\rho} spectral
function resulting from Hadronic Many-{}Body Theory (HMBT) together with QGP
contributions [3, 205], consistently describes the invariant
mass dependence of dielectron LMR yields from SPS to top RHIC energies within
systematic uncertainties. In the IMR, these model calculations also yield a
visible contribution on top of the cocktail resulting in an improved agreement
with the data.  This is possibly the first measurement indicative of IMR
in-{}medium effects emerging in the BES-{}I energy regime. HMBT and QGP
contributions are especially sizeable at lower BES-{}I energies where the
$c\bar{c}$ cross sections in elementary collisions
($\sigma_{c\bar{c}}^\mathrm{NN}$) fall off steeply.

However, substantial conclusions on the possible observation of QGP radiation
in this mass region are not only constrained by the limited statistics in this
region and the large uncertainties on
$\sigma_{c\bar{c}}^\mathrm{NN}$, but also difficult due to the
unknown medium effects on the charm continuum (see Chapter~\ref{summary}) as well as the
correlated background contributions from \ensuremath{\pi}$^{\text{0}}$ decays in jets (details see
Section~\ref{ana_sec_pairrec}). Without adequate statistics in the mass region above
about 1.1 GeV/c$^{\text{2}}$, the latter impedes the precise normalization of mixed event
to same event background distributions, and the former is indistinguishable
from the data even in the simple case of de-{}correlated charm pairs causing
changes in the mass slope. Conclusive measurement of QGP radiation would also
require the study of T$_{\text{eff}}$ versus invariant mass which has proven challenging
already at $\sqrt{s_\mathrm{NN}}$ = 200 GeV with superior
statistics and systematics. The BES-{}I dielectron results for the IMR presented
here, drive most of the motivation for BES-{}II since they reiterate the
importance of measuring the $c\bar{c}$-{}continuum in detail. The
improved beam luminosity hopefully enhances statistics sufficiently enough to
even allow for energy-{}dependent T$_{\text{eff}}$ measurements of the QGP. The present
study not only constitutes the first dielectron results available in heavy-{}ion
collisions at BES-{}I energies but also enables novel comparisons within STAR of
the $\phi\to ee$ and $\phi\to KK$ decay channels, for
instance, or of the J/\ensuremath{\psi} to results obtained using electro-{}magnetic
calorimeters at intermediate \emph{p$_{\text{T}}$}.

\wrapifneeded{0.50}{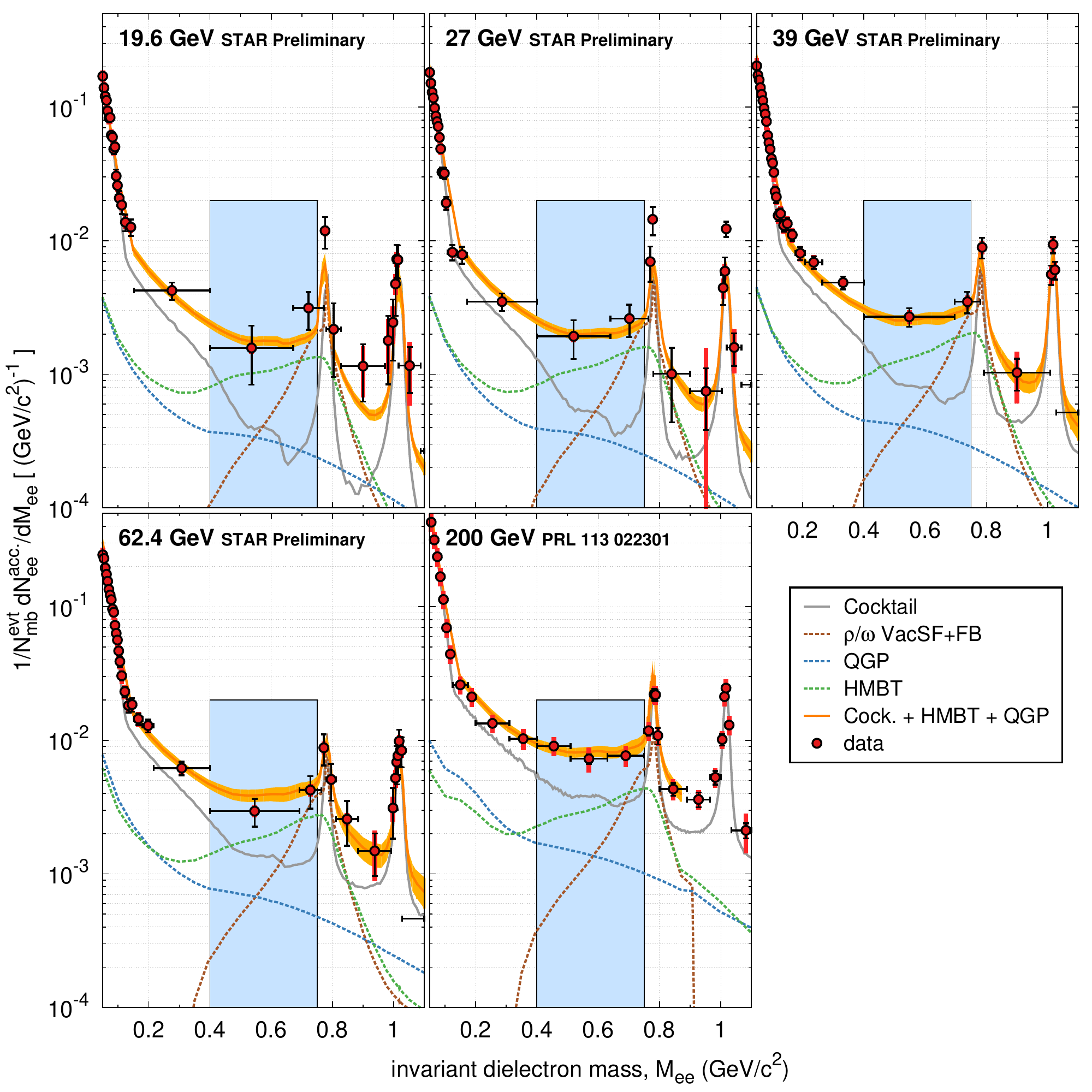}{LMR e$^{\text{+}}$e$^{\text{-{}}}$ spectrum exhibiting excess yields over the entire BES-{}I energy range up to the top RHIC energy. The color code is identical to Figure~\ref{results_fig_stack} showing good agreement of HMBT and QGP model calculations with LMR dielectron yields. Also depicted for comparison is the alternative scenario of \ensuremath{\rho}/\ensuremath{\omega} vacuum spectral functions (VacSF) propagated through the fireball (FB). Note that \ensuremath{\rho} model calculations also include freeze-{}out contributions. Figure~\ref{results_fig_diffAbsRel} makes detailed comparisons of the scenarios to the excess. The blue regions denote the mass range used for the calculation of integrated excess yields and enhancement factors in Figure~\ref{results_fig_excenh} (0.4 <{} M$_{\text{ee}}$ <{} 0.75 GeV/c$^{\text{2}}$).}{results_fig_panel}{0.8} %

Figure~\ref{results_fig_panel} gives more detailed insights into the mass region of the
dielectron spectra between the \ensuremath{\pi}$^{\text{0}}$ tail and beyond the \ensuremath{\phi} resonance,
exhibiting sizeable excess yields over the cocktail of hadronic (freeze-{}out)
sources. In their consistent description over a wide range of energies by the
in-{}medium model calculations, contributions primarily due to the broadened
in-{}medium \ensuremath{\rho} spectral function in HMBT clearly dominate over contributions
from the QGP in the mass region 0.4 to 0.75 GeV/c$^{\text{2}}$ (blue rectangles). Hence,
the measured excess yields are rooted in the \ensuremath{\rho} meson modifications mainly
governed by interactions with baryons in the late-{}stage but hot and dense
hadronic phase. Moreover, the model calculations of the yields from the two
contributing phases in this scenario indicate a smooth increasing evolution
with beam energy and only minor changes in their spectral shape.  An
alternative scenario is to assume a vacuum spectral function for the \ensuremath{\rho} and
\ensuremath{\omega} mesons un-{}modified by the medium. The resulting invariant mass spectra
after integrating the according rates over the fireball evolution and adding
the contributions from freeze-{}out \ensuremath{\rho} decays on top, are included in
Figure~\ref{results_fig_panel} (brown dashed line) for comparison to the broadened \ensuremath{\rho}
spectral function scenario. The respective shape already suggests by eye that
this scenario is insufficient to describe the required yields in the mass
region of the excess. In this context, studying the cocktail-{}subtracted excess
yields (data minus cocktail) and cocktail-{}normalized enhancement ratios (data
divided by cocktail) in Figure~\ref{results_fig_diffAbsRel} draws a clearer picture.

\wrapifneeded{0.50}{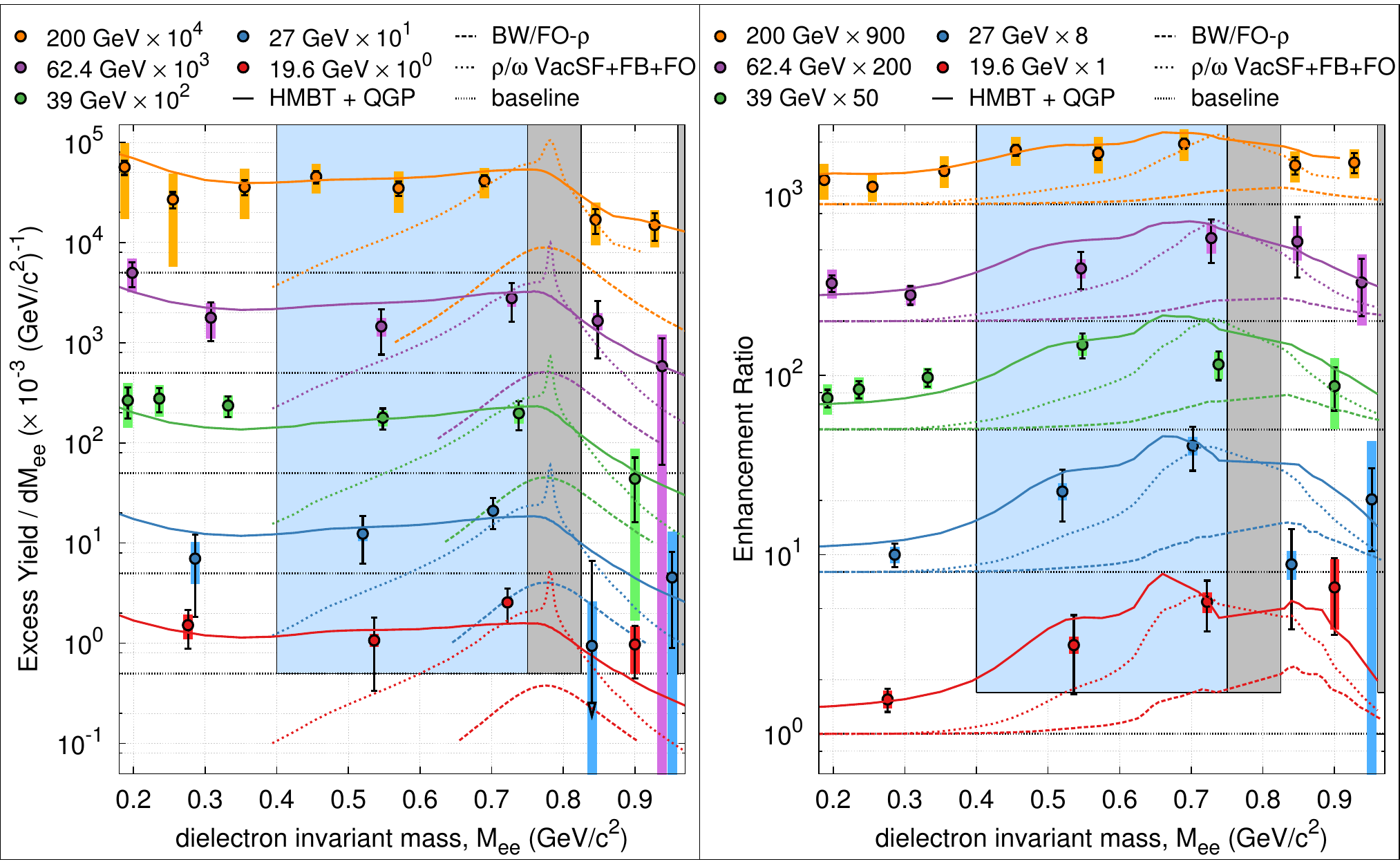}{Invariant mass dependence of absolute LMR excess yields (left) and LMR enhancement ratios (right) over cocktail compared to in-{}medium, vacuum and bare freeze-{}out scenarios. The mass regions in which freeze-{}out \ensuremath{\omega} and \ensuremath{\phi} decays are highly dominant, are masked out for clarity (grey boxes). Also note, that the energies are shifted against each other on a logarithmic scale for better visualization. Black dotted horizontal lines denote the according baselines of 0.5 $\times$ 10$^{\text{-{}3}}$ and 1 for excess yields and enhancement ratios, respectively. All systematic uncertainties are combined into uncertainties on the experimental data points. The regions highlighted in blue are used in Figure~\ref{results_fig_excenh} to derive integrated excess yields and enhancement factors.}{results_fig_diffAbsRel}{1} %

The absolute excess yields below the \ensuremath{\omega} meson evidently
increase smoothly with beam energy (see Table~\ref{results_tab_diffAbs}) and exhibit an
approximately constant dependence on invariant mass. Both observations are
consistently described by the yields predicted in HMBT and QGP calculations,
which are favored at all energies by the data over the scenarios using
fireball-{}propagated vacuum spectral functions or bare freeze-{}out \ensuremath{\rho} decays.
The same conclusion can be reached from the enhancement ratios in this mass
region, which allow for visually improved statistical and systematic
uncertainties but convolute relative differences in the energy dependence of
cocktail and data. Still, the energy-{}dependent measurement of excess yields and
enhancement ratios corroborates the preceding conclusion that the broadened
\ensuremath{\rho} scenario provides a consistent description in good agreement with the
data over a wide energy range.
\begin{center}
\begingroup%
\setlength{\newtblsparewidth}{0.8\linewidth-2\tabcolsep-2\tabcolsep-2\tabcolsep-2\tabcolsep-2\tabcolsep-2\tabcolsep-2\tabcolsep}%
\setlength{\newtblstarfactor}{\newtblsparewidth / \real{342}}%

\begin{longtable}{llllll}\caption[{Excess yields from HMBT+QGP integrated over LMR (blue box in Figure~\ref*{results_fig_diffAbsRel} left).}]{Excess yields from HMBT+QGP integrated over LMR (blue box in Figure~\ref{results_fig_diffAbsRel} left).\label{results_tab_diffAbs}\hyperlabel{results_tab_diffAbs}%
}\tabularnewline
\endfirsthead
\caption[]{(continued)}\tabularnewline
\endhead
\hline
\multicolumn{1}{m{57\newtblstarfactor+\arrayrulewidth}}{\centering%
$\sqrt{s_\mathrm{NN}}$
}&\multicolumn{1}{m{57\newtblstarfactor+\arrayrulewidth}}{\centering%
19.6 GeV
}&\multicolumn{1}{m{57\newtblstarfactor+\arrayrulewidth}}{\centering%
27 GeV
}&\multicolumn{1}{m{57\newtblstarfactor+\arrayrulewidth}}{\centering%
39 GeV
}&\multicolumn{1}{m{57\newtblstarfactor+\arrayrulewidth}}{\centering%
62.4 GeV
}&\multicolumn{1}{m{57\newtblstarfactor+\arrayrulewidth}}{\centering%
200 GeV
}\tabularnewline
\multicolumn{1}{m{57\newtblstarfactor+\arrayrulewidth}}{\centering%
($\times 10^{-3}$)
}&\multicolumn{1}{m{57\newtblstarfactor+\arrayrulewidth}}{\centering%
0.65
}&\multicolumn{1}{m{57\newtblstarfactor+\arrayrulewidth}}{\centering%
0.72
}&\multicolumn{1}{m{57\newtblstarfactor+\arrayrulewidth}}{\centering%
0.82
}&\multicolumn{1}{m{57\newtblstarfactor+\arrayrulewidth}}{\centering%
1.2
}&\multicolumn{1}{m{57\newtblstarfactor+\arrayrulewidth}}{\centering%
1.8
}\tabularnewline
\hline
\end{longtable}\endgroup%

\end{center}

The quality of the STAR data also allows for the systematic measurement of
integrated LMR enhancement factors and excess yields with respect to their
energy dependence in the BES-{}I energy regime. Due to the approximately constant
dependence on invariant mass, these integrated measures are suitable to
quantify the energy dependence of the LMR dielectron production yields with
respect to the cocktail in Figure~\ref{results_fig_excenh}. The mass region 0.4 -{} 0.75
GeV/c$^{\text{2}}$ is chosen as basis to derive the integrated observables since it
avoids STAR's limited acceptance and hence steep decline in detector efficiency
for p$_{\text{T}}$ <{} 0.2 GeV/c, but is still broad enough in range to gather sufficient
statistics while not being influenced by the tail of the \ensuremath{\pi}$^{\text{0}}$ distribution.

\wrapifneeded{0.50}{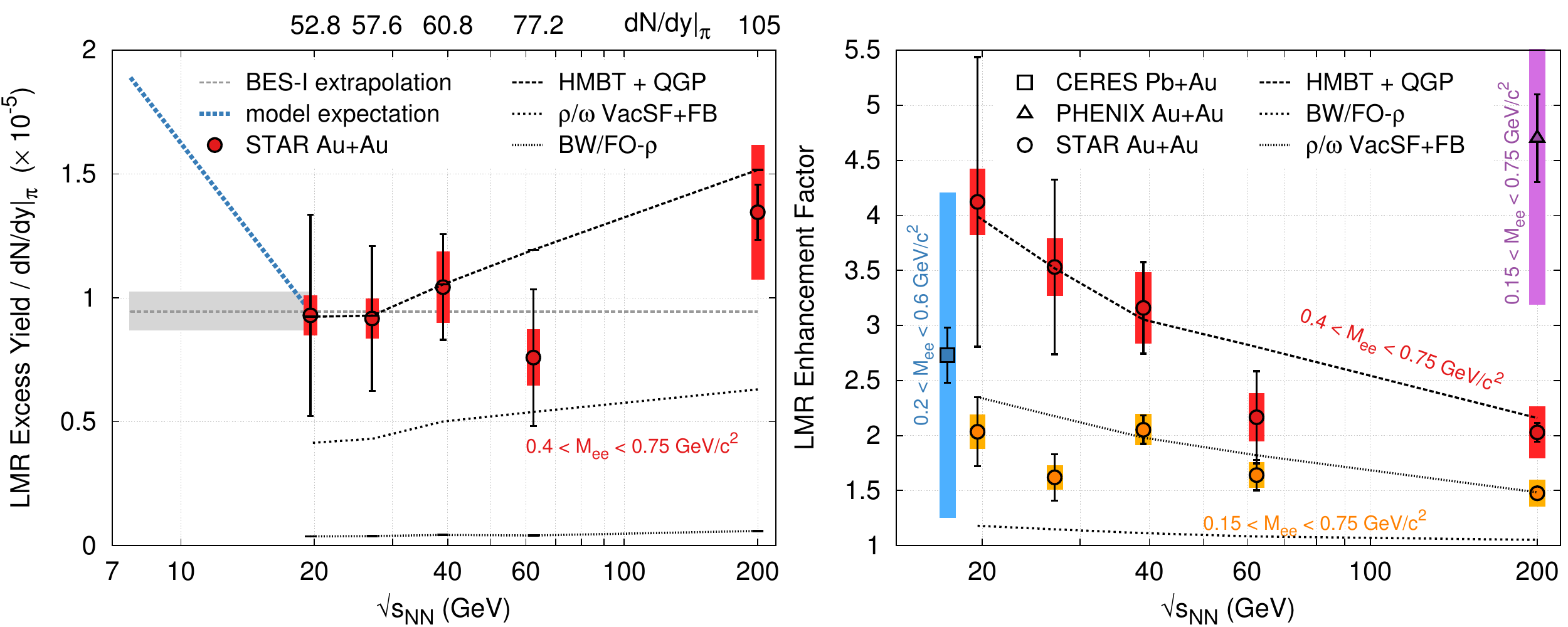}{(Left) Measured BES-{}I and top RHIC energy LMR excess yields normalized to the respective dN/dy|$_{\text{\ensuremath{\pi}}}$ denoted for each energy in the top margin. The grey dashed line and box represent the average excess yield from BES-{}I projected into the BES-{}II regime with the size of the uncertainty taken from 200 GeV. The blue dashed line depicts the expected trend of the LMR excess in the BES-{}II regime as suggested by the according quantity in PHSD calculations (see Figure~\ref{results_fig_totbar}). (Right) LMR enhancement factors measured at STAR's BES-{}I in Au+Au in comparison to Pb+Au measurements by CERES at SPS and Au+Au by PHENIX and STAR at top RHIC energy. The STAR BES-{}I enhancement factors are derived for two different mass regions for comparison. The energy dependence of the different theoretical scenarios explained in the text is depicted, too.}{results_fig_excenh}{1} %

\pagebreak[4]

The measured LMR excess yields in Figure~\ref{results_fig_excenh} (left) are normalized
to the invariant pion yields at the respective energy which allows for the
comparison to the energy dependence of the total baryon density in
Figure~\ref{results_fig_totbar}.

\wrapifneeded{0.50}{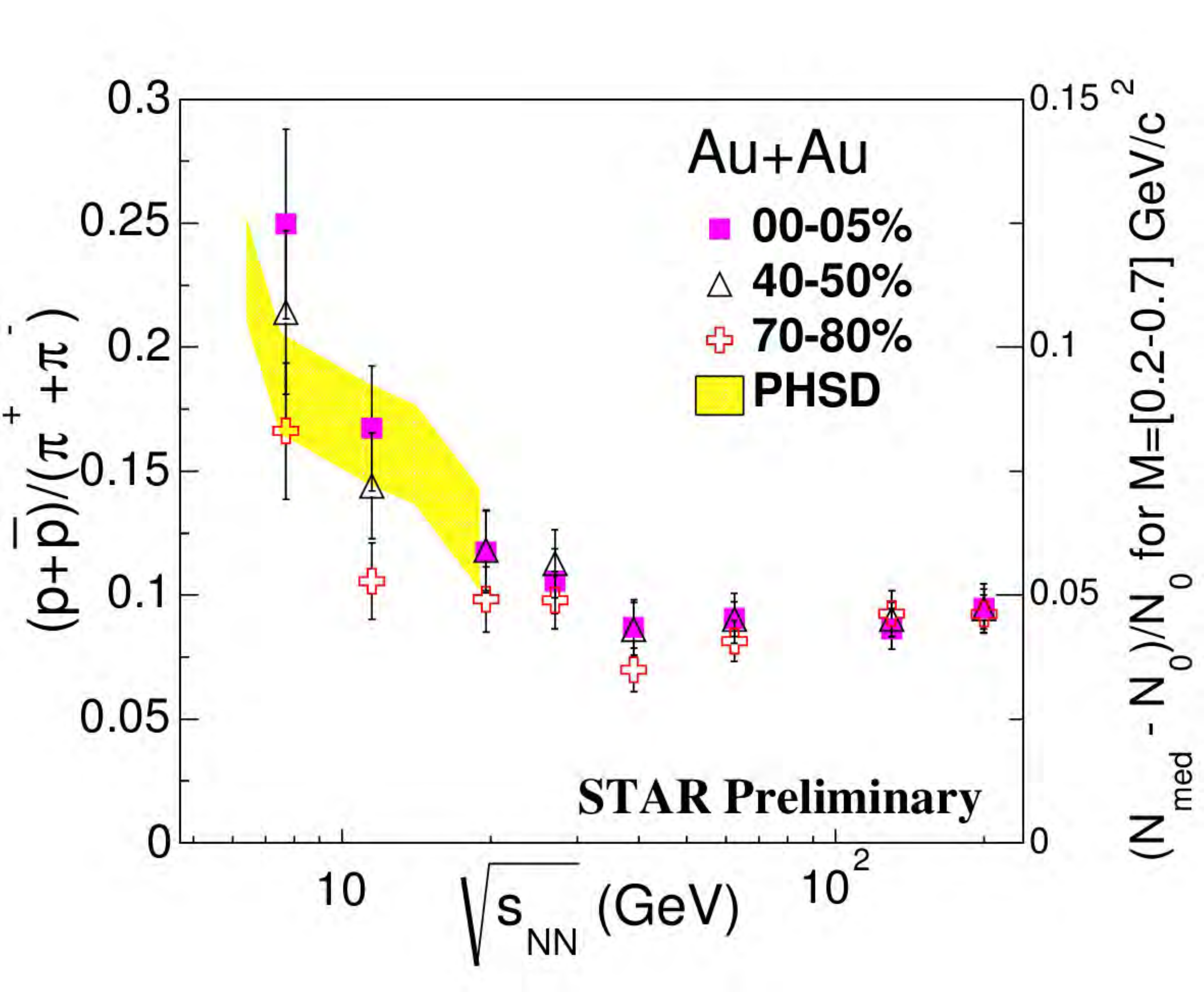}{Energy dependence of total baryon density in Au+Au collisions using the total proton-{}to-{}pion ratio as proxy. The BES-{}II energy regime is overlaid with the expected trend from PHSD for the dielectron excess yield in M$_{\text{ee}}$ = 0.2 -{} 0.7 GeV/c$^{\text{2}}$ [21, 22] (yellow band).}{results_fig_totbar}{0.49} %

In general, protons/pions are the most abundant baryons/mesons emanating from
and routinely measured in a heavy-{}ion collision. The proton-{}to-{}pion ratio
$(p+\bar{p})/(\pi^++\pi^-)$ is about 1/5 at SPS energies and
decreases to about 1/10 at RHIC energies as opposed to a ratio of about 1 at
AGS energies. Hence, at SPS as well as RHIC energies mesons are the dominating
particle species controlling the freeze-{}out conditions [213] which
also makes the proton-{}to-{}pion ratio a valid estimator for the total baryon
density. This is important in the discussion of LMR excess yields since
in-{}medium modifications to the \ensuremath{\rho} spectral function are expected to depend
on total instead of net baryon density due to the seeming CP invariance of the
strong interaction (see Section~\ref{sim_sec_model} and Section 2.7.1 in
[10]). The study in Figure~\ref{results_fig_excenh} (left) does not
reveal a strong energy dependence of the excess yield in line with the
approximately constant total baryon density for BES-{}I energies as observed in
Figure~\ref{results_fig_totbar}. A possible energy-{}dependent enhancement which would
thus be directly related to earlier creation times, seems to be hidden due the
emission of LMR dielectrons during phases driven by hadronic densities and due
to the integration of the according rates over the system's temperature
evolution profile.

The agreement in energy dependence of measured LMR excess
yields and total baryon density in BES-{}I further raises the interesting
question of which trend to expect for the BES-{}II energy regime. \ensuremath{\rho}-{}meson
based PHSD calculations [21, 22] predict a relative
increase of the expected excess yield $(N_\mathrm{med}-N_0)/N_0$ in
M$_{\text{ee}}$ = 0.2 -{} 0.7 GeV/c$^{\text{2}}$ by about a factor of two which is consistent
with the according increase in the measured total baryon densities below
$\sqrt{s_\mathrm{NN}}$ = 20 GeV in Figure~\ref{results_fig_totbar}.  This
trend is translated to Figure~\ref{results_fig_excenh} (left) and compared to the
average excess yield extrapolated from the BES-{}I regime. Systematic
uncertainties from $\sqrt{s_\mathrm{NN}}$ = 200 GeV are associated
with the extrapolation since statistical errors are expected to be small in
BES-{}II with RHIC's improved luminosities at the community's disposal. It shows
that STAR should be able to continue studying the LMR excess yield's trend with
decreasing beam energy and its (possible) correlation with total baryon density
[23].\newline

The extraction of enhancement factors in Figure~\ref{results_fig_excenh} (right) allows
for the comparison to the previously reported results of CERES and PHENIX.
Despite different experimental acceptances, the STAR results reach general
consistency with CERES at $\sqrt{s_\mathrm{NN}}$ = 17 GeV while
providing higher quality measurements. In comparison to PHENIX, however, the
results show a sizeable discrepancy which is not fully resolved yet, since
neither the projection of STAR's results into the PHENIX acceptance, nor the
detailed comparison of cocktail simulations provide satisfactory explanations
([16], also see Section~\ref{intro_sec_dielec}).

In the mass range of 0.4 -{} 0.75 GeV/c$^{\text{2}}$, STAR's measurements of excess yields
as well as enhancements factors are compared to the different theoretical
scenarios used in the context of Figure~\ref{results_fig_panel} and
Figure~\ref{results_fig_diffAbsRel}. Along with the excess yields, one observes an
increasing enhancement factor with decreasing energy in agreement with the HMBT
in-{}medium \ensuremath{\rho}-{}meson calculations.  As a result, the preceding conclusion of a
broadened \ensuremath{\rho} spectral function as preferred scenario for the enhancement, is
further supported in its energy dependence by the agreement of the data over
the wide BES-{}I energy range. Hence, for the first time since the NA60 dimuon
measurements, its findings regarding LMR enhancement are also confirmed in the
dielectron sector with the added bonus of consistency in energy dependence.

\wrapifneeded{0.50}{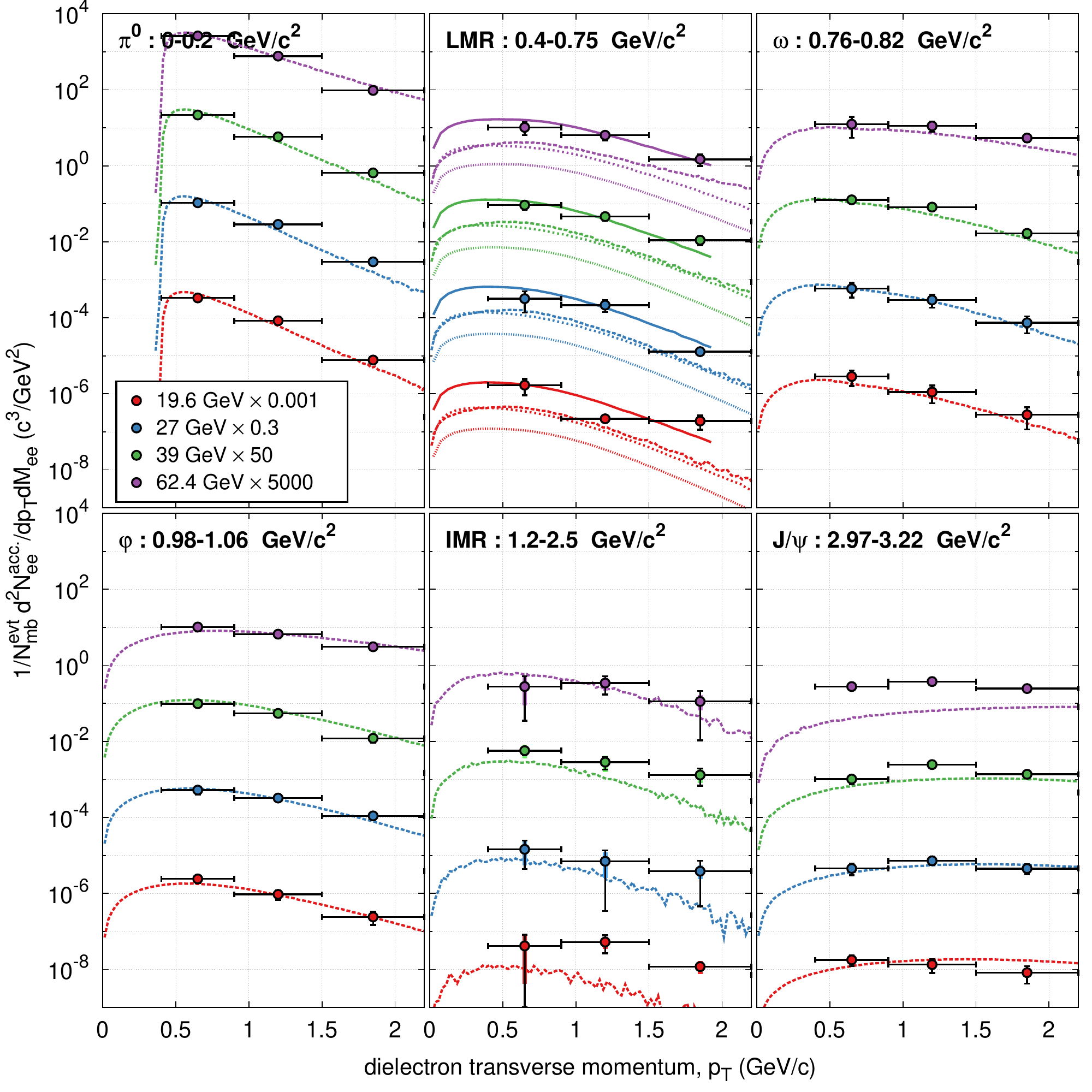}{Efficiency-{}corrected p$_{\text{T}}$ dielectron spectra in 0-{}80\% minimum bias Au+Au collisions measured by STAR at BES-{}I energies in six characteristic mass regions of the invariant mass spectrum. Solid lines in the LMR compare the p$_{\text{T}}$-{}dependence of LMR yields to the expectations from the total of cocktail and model. Dashed and dotted lines in this region represent the respective contributions (cocktail, HMBT, QGP) to the total LMR yield. Dashed lines in the other mass regions denote p$_{\text{T}}$ spectra as expected from cocktail simulations.}{results_fig_pT}{0.92} %

In Figure~\ref{results_fig_pT}, STAR's measurements are extended to the p$_{\text{T}}$ dependence
of dielectron production in six characteristic regions of the invariant mass
spectrum. Note that, though available in Section~\ref{ana_subsec_pair_signal}, the lowest
p$_{\text{T}}$ bin is omitted since it is severely influenced by STAR's acceptance hole
and hence suffers large systematic uncertainties during the efficiency
correction. The comparison of the enhancement's p$_{\text{T}}$ dependence to medium
modifications allows for more resilient conclusions about its source since
transverse momentum and invariant mass together cover the full phase space
available to dielectrons. In line with the invariant mass dependence of Figure~\ref{results_fig_stack}, overall
consistency with the cocktail and also with the scenario of an in-{}medium
broadened \ensuremath{\rho} spectral function in the LMR is observed at all BES-{}I energies.
This further strengthens the conclusion of the latter being the correct
description for the LMR excess at these energies.

In the projected momentum range, the p$_{\text{T}}$ spectra in Figure~\ref{results_fig_pT} show an
approximately exponential shape. When represented as transverse mass
distributions, their inverse slope can be interpreted as the effective
temperature T$_{\text{eff}}$ of the expanding medium. The observed continuous decrease in
slope over the six different invariant mass regions is characteristic for
dielectrons emanating from all stages of the expansion and reveals the
time-{}ordered nature of the different dielectron sources [212].
The slopes (T$_{\text{eff}}$) of the p$_{\text{T}}$ spectra in Figure~\ref{results_fig_pT} also decrease
(increase) faster up to the \ensuremath{\rho}/\ensuremath{\omega} pole mass region than from the \ensuremath{\phi}
to the J/\ensuremath{\psi} mass regions beyond 1 GeV/c$^{\text{2}}$. Though not directly observed
here for the dielectron excess due to insufficient statistics, this is
reminiscent of the observations made for the dimuon excess in
Figure~\ref{intro_fig_ceres_na45_na60}c: T$_{\text{eff}}$ rises linearly with M$_{\text{ee}}$ up to the
\ensuremath{\rho}-{}meson mass but stays below its expected trend. Referring to the same
figure, it is argued in [212] that this is "consistent with radial
flow of an in-{}medium hadronlike source, namely \ensuremath{\pi}\ensuremath{\pi}\ensuremath{\rightarrow}\ensuremath{\rho}, decaying
continuously into lepton pairs". It is further consistent with expectations
that maximal radial flow is reached by the \ensuremath{\rho} due to the its maximal
coupling to pions (c.f. T$_{\text{eff}}$ vs.  \ensuremath{\beta} in Section~\ref{intro_sec_hics}).

All the above observations in the dielectrons' M$_{\text{ee}}$ and p$_{\text{T}}$ distributions
reiterate their unique sensitivity to the space and time evolution dynamics of
a heavy ion collision.

\wrapifneeded{0.50}{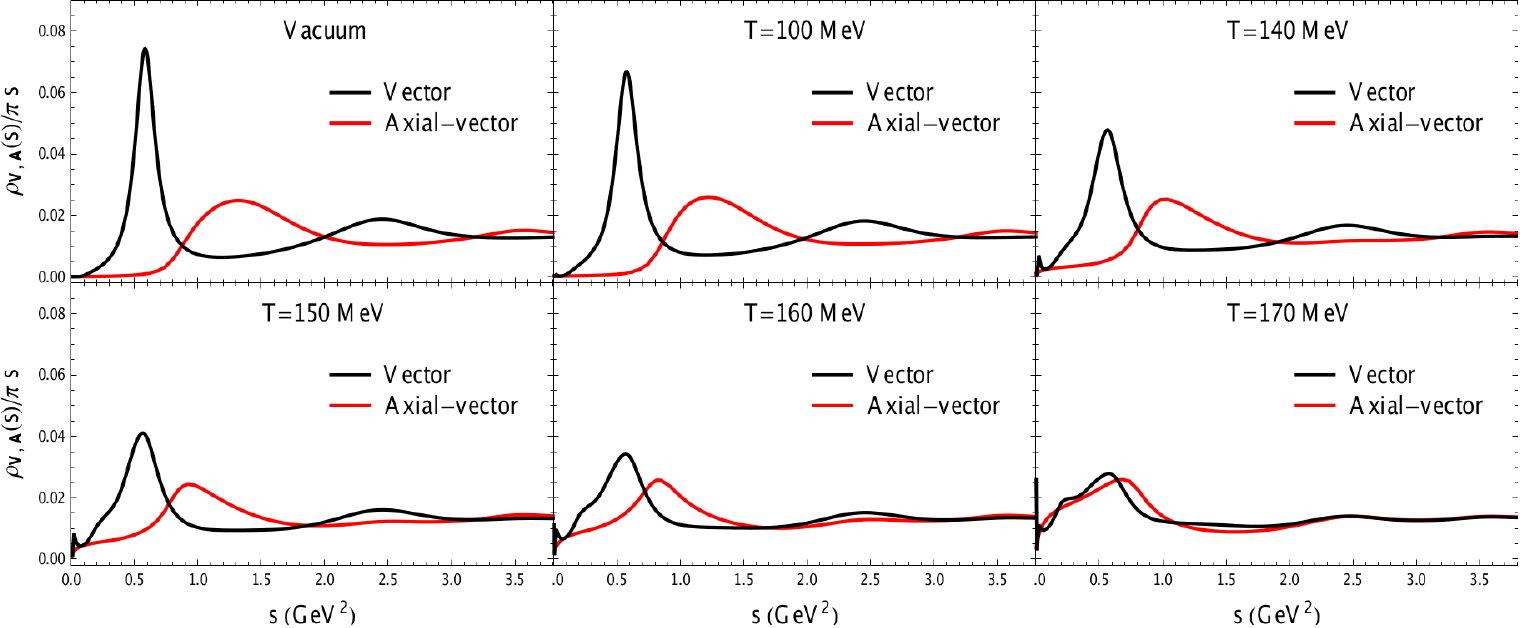}{Vector (\ensuremath{\rho}) and axial-{}vector (a$_{\text{1}}$) in-{}medium spectral functions gradually merging with increasing temperature toward restoration [112, 214].}{results_fig_hohler}{1} %

In summary, the results of this thesis can cumulatively be taken as an
indication for the \emph{universal} nature of the source underlying the observed
enhancement in LMR dielectron production. The theoretical scenario of a
broadened \ensuremath{\rho} spectral function seems to be serving well to describe this
source which itself fuels the required progress on the theory side to
thoroughly connect the measured \ensuremath{\rho} spectral function to its chiral partner
and arrive at the long-{}awaited conclusions on possible chiral symmetry
restoration. For instance, the authors of [112, 214]
subsequently pose the question whether the in-{}medium vector (\ensuremath{\rho}) spectral
functions describing the dilepton data presented in this chapter, are
theoretically compatible with \ensuremath{\chi}SR.  Satisfying theoretical constraints and
Lattice QCD results, they find a smooth reduction of the a$_{\text{1}}$ meson's mass and
a large increase in its width with increasing temperature
(Figure~\ref{results_fig_hohler}). The axial-{}vector spectral function ultimately merges
with the vector spectral function which they argue establishes a direct and compatible
connection of dileptons and \ensuremath{\chi}SR. They admit that
microscopic calculations of the a$_{\text{1}}$ are imperative and efforts in this
direction are ongoing at the time of this writing.


\chapter{Summary and Outlook}
\label{summary}\hyperlabel{summary}%

Quantum Chromo Dynamics (QCD) has emerged [1] as a very successful theory to
predict and confirm the properties of the phases in which nuclear matter exists
at various temperatures and baryon densities (see Section~\ref{intro_sec_qcd_phases}). In
the QCD vacuum, quarks and gluons are confined into color-{}neutral hadrons and
the global chiral symmetry of the QCD Lagrangian is spontaneously broken [2] which
explains the basic features of the experimental hadron spectrum [3]. At
sufficiently high temperatures and baryon densities, nuclear matter is expected
to undergo a transition into the Quark-{}Gluon-{}Plasma (QGP) consisting of
deconfined quarks and gluons as the determining degrees of freedom, and
accompanied by chiral symmetry restoration [4]. Such fundamental properties of QCD
matter can be studied in ultra-{}relativistic heavy-{}ion collisions [5, 6] producing a
relatively large volume of high energy and nucleon densities as existent in the
early universe (see Section~\ref{intro_sec_hics}). During its subsequent expansion and
cooling, the hot and dense matter created in these collisions goes through
multiple stages of QCD phases before hadronizing and freezing out into the
observable particles of the QCD vacuum [7]. Measurement of the elliptic flow of
mesons and baryons, for instance, sheds light on the thermalization time scale
of the medium as well as the degree of collective expansion driven by a
possibly partonic source [8, 9]. It is imperative, however, to observe and
comprehensively measure (smoking-{}gun) signals for the two fundamental QCD
characteristics of deconfinement and chiral symmetry restoration. Annihilation
of $q\bar{q}$ pairs in a thermally equilibrated deconfined QCD
medium would result in thermal electromagnetic radiation (directly from the
QGP), and modifications of vector meson spectral functions in the medium
could reveal precursor effects of the disappearance of the QCD vacuum
structure, respectively [10, 11]. The discussion of the latter
connection in Section~\ref{intro_sec_dielec} identifies the requirement of a coordinated effort from both the
theoretical and experimental side to strengthen the conclusions on (partial)
chiral symmetry restoration based on dielectron measurements of the \ensuremath{\rho}
spectral function.\newline

Dileptons are unique bulk-{}penetrating tools in this regard since they penetrate
through the surrounding medium with negligible interaction and are created
throughout the entire system evolution starting at the early bulk stages (see Figure~\ref{intro_fig_dielec_prod}). The
pairs' invariant mass (M$_{\text{ee}}$) and transverse momentum (p$_{\text{T}}$) distributions are
reminiscent of their time-{}ordered emission and hence sensitive to the system's
evolution dynamics. The Low-{}Mass-{} (LMR) and Intermediate-{}Mass-{}Regions (IMR) of
dielectron spectra, in particular, promise access to the in-{}medium modified
\ensuremath{\rho} spectral function as well as the effective QGP temperature, respectively,
provided for the latter that the contribution by a possibly medium-{}modified
charm continuum is known [12]. A multitude of experiments at SIS18, SPS and RHIC
have taken on the challenging task to measure these rare probes in a heavy-{}ion
environment (see Section~\ref{intro_subsec_dielec_meas}). CERES has first observed a
sizeable and unexplained excess yield in the LMR over a cocktail of expected
hadronic sources [13]. NA60's results from high-{}quality dimuon measurements have
identified the broadened \ensuremath{\rho} spectral function as the favorable scenario to
explain the LMR excess, and partonic sources as dominant in the IMR [14]. PHENIX
followed with dielectron measurements at top RHIC energy exhibiting a
significant LMR enhancement and good agreement of IMR dielectron production
with a charm continuum as expected from N$_{\text{bin}}$-{}scaled p+p collisions [15]. As
opposed to the NA60 results, the former could not be explained in size by the
existing model calculations even when exhausting the available parameter space.
Measurements of dielectron production at the same energy by STAR at RHIC
resulted in a more moderate enhancement consistent with NA60 measurements and
with model calculations based on in-{}medium broadened \ensuremath{\rho} spectral functions [16].
Note that in the low temperature and high baryon density regime, HADES
confirmed the \emph{DLS puzzle} with a measured LMR enhancement factor of about 2 -{} 3 [17, 18].\newline

The two preceding paragraphs set the stage for the extensive study of
dielectron production presented in this thesis. Enabled by the addition of the
TOF detector system completed in 2010, STAR provides excellent PID, low
material budget, full azimuthal acceptance at mid-{}rapidity, and wide p$_{\text{T}}$ coverage [19]. Especially in combination with the first phase of the Beam Energy
Scan (BES-{}I) at RHIC [20], STAR presents the unprecedented opportunity for an
energy-{}dependent study of dielectron production within a homogeneous
experimental environment closing the wide gap in the QCD phase diagram between
SPS and top RHIC energies (see Section~\ref{intro_sec_thesis}). This thesis concentrates
on the analysis of the beam energies 19.6, 27, 39, and 62.4 GeV recorded in
Au+Au collisions during BES-{}I providing sufficient statistics for such a study
(see Chapter~\ref{data_analysis}). In conjunction with the published STAR results at top
RHIC energy, this provides the first comprehensive energy-{}dependent dataset
that will also most likely be the only one available in this energy regime for the forseeable future
given the enormous effort involved in obtaining it
\footnote{
STAR will continue to take high statistics data at \ensuremath{\surd}s$_{\text{NN}}$ <{} 20 GeV during BES Phase-{}II.
}. The results of this thesis
(see Chapter~\ref{results}) include M$_{\text{ee}}$-{} and p$_{\text{T}}$-{}spectra for the four beam energies
measured in 0-{}80\% minimum-{}bias Au+Au collisions with high statistics up
to 3.5 GeV/c$^{\text{2}}$ and 2.2 GeV/c, respectively. Their comparison with cocktail
simulations of hadronic sources reveals a sizeable and steadily increasing
excess yield in the LMR at all beam energies which cannot be described by
\ensuremath{\rho}/\ensuremath{\omega} vacuum spectral functions propagated through the fireball
expansion and much less by the according bare freeze-{}out contributions.  Good
agreement over the entire energy range as well as in M$_{\text{ee}}$ and p$_{\text{T}}$ shape is
instead reached with model calculations within Hadronic Many-{}Body Theory
(HMBT) based on an in-{}medium broadened \ensuremath{\rho} spectral function. This
contribution dominates in the LMR over $q\bar{q}$ contributions
from the QGP and hence universally serves well as the underlying source of
the measured enhancement for the entire RHIC energy regime. It also means that
most of the enhancement is governed by interactions of the \ensuremath{\rho} meson with
thermal resonance excitations in the late(r)-{}stage hot and dense hadronic
phase. This conclusion is supported by the energy-{}dependent measurement of
integrated LMR excess yields normalized to the invariant pion yields as well as
LMR enhancement factors. The former do not exhibit a strong dependence on beam
energy as expected from the approximately constant total baryon density above
20 GeV, and the latter show agreement with the CERES measurement at SPS energy
despite different detector acceptances while providing higher quality data.
The consistency in excess yields and with model calculations measured by STAR
over the wide RHIC energy regime makes a strong case for LMR enhancements on
the order of a factor 2 -{} 3 rather than 5 or more.\newline

The energy-{}dependence of the integrated LMR excess yields allows for an
interesting projection into the second phase of the Beam Energy Scan (BES-{}II).
For the energy regime below 20 GeV, measurements of the total baryon density as
well as \ensuremath{\rho}-{}meson based PHSD calculations suggest an increase of about a
factor of two in LMR excess [21, 22]. High-{}statistics
measurements in BES-{}II should provide enough accuracy to study these
predictions and further strengthen our understanding of the LMR enhancement and
its underlying source [23]. Besides enhanced statistics, BES-{}II will
provide improved tracking due to the proposed iTPC upgrade, and improved
capabilities for the measurement of dimuons, for instance.
In the present study, reliable conclusions on QGP radiation in the IMR are
unfortunately primarily constrained by the available statistics, the
uncertainties in $c\bar{c}$ cross sections in nucleon-{}nucleon
collision, and possible medium-{}modifications of the charm continuum through
decorrelation of the $e^+/e^-$ decay daughters. Measurements during
BES-{}II, however, will enable the study of the charm continuum and hence
indirectly QGP radiation due to the recently completed upgrades of the HFT and
MTD detector subsystems. This is especially important for the LMR since the
$c\bar{c}$ contribution to the total cocktail in the 0.4 -{} 0.7
GeV/c$^{\text{2}}$ mass region increases from about 20\% at 19.6 GeV up to about 60\% at
200 GeV.\newline

In conclusion, STAR's measurements during BES-{}I provide high-{}quality datasets
essential to improve the understanding of the LMR enhancement regarding its
M$_{\text{ee}}$, p$_{\text{T}}$ and energy dependence. The study of its centrality dependence in
the near future would complete the extractable results and could reveal new
information about the energy dependence of the fireball's life time. The
scenario of broadened in-{}medium \ensuremath{\rho} spectral functions has proven to not only
serve well as dominating underlying source but also to be \emph{universal} in nature
since it quantitatively and qualitatively explains the LMR enhancements
measured over the wide range from SPS to top RHIC energies. The extent of the
results presented here enables a more solid discussion of its relation to chiral symmetry restoration
from a theoretical point of view. High-{}statistics measurements at
BES-{}II hold the promise to confirm these conclusions along with the LMR
enhancement's relation to total baryon density with decreasing beam energy.
%
%

\part{Software: STAR and Open-{}Source Projects}
\label{software}\hyperlabel{software}%


\chapter{Contribution of STAR Software Modules}
\label{_contribution_of_star_software_modules}\hyperlabel{_contribution_of_star_software_modules}%

\section{StV0TofCorrection: TOF Correction of Off-{}Vertex Decays}
\label{stv0tofcorr}\hyperlabel{stv0tofcorr}%

\noindent
\begin{description}
\item[{ Abstract
}] \hspace{0em}\\
 The class described in this chapter provides member functions to correct the
time-{}of-{}flight of daughter particles originating from V$_{\text{0}}$ decays.  The
underlying algorithm is based on the reconstruction of mother decay particles.
Detailed studies performed by applying the algorithm to \ensuremath{\Lambda}-{} and
\ensuremath{\Omega}-{}decays, demonstrate its background rejection power and prove it to work
as expected. As an additional feature, a default cut on the particles'
corrected mass is implemented and can be applied by the user to improve the
signal-{}to-{}noise ratio.  The source code of this work has been developed in the
last quarter of 2010 and is part of the \emph{StBTofUtil} [24] module
in the official STAR CVS code repository since January 2011. The latest
revision of the class implementation is from December 2013.  A version of this
chapter from December 2010 has been made available to the STAR collaboration as
an internal report [25]. This work was supported by
the German Exchange Service DAAD.

\end{description}

To identify the species of detected particles, their mass is calculated by
means of their velocity \ensuremath{\beta}. For this purpose, the particles' time-{}of-{}flight
(TOF) from their decay vertex to a matched hit in the TOF detector is required.
In the case of very short-{}lived particles their decay vertex can be assumed to
coincide with the event vertex. However, due to the non-{}negligible decay length
of V$_{\text{0}}$ particles, a TOF correction has to be applied to the emanating daughter
particles which can be achieved by calculating the time-{}of-{}flight for the
reconstructed unstable mother particles. By subsequently applying a cut on the
corrected mass, a higher ratio of those decay particles can be obtained which
actually fit the V$_{\text{0}}$ decay topology. The class \emph{StV0TofCorrection} described
in this chapter provides the necessary tools for such V$_{\text{0}}$ TOF corrections.  It
has been developed for the usage within STAR's analysis framework \emph{StRoot}.
This chapter first explains the implemented algorithm and subsequently tests it
by means of \ensuremath{\Lambda}-{} and \ensuremath{\Omega}-{}decays. It concludes b y demonstrating the
general usage of \emph{StV0TofCorrection} accompanied by examples of its
application.\newline

Figure~\ref{stv0tof_fig_corr} illustrates the TOF correction algorithm. Blue crosses
schematically denote single TPC hits of the proton track. The first TPC hit is
taken as origin for the path length measure. The variables used to correct the
time-{}of-{}flight of a proton produced in the decay chain
$\Omega^-\to\Lambda K^-\to p\pi^- K^-$ are shown as well. Using
this example the general TOF correction algorithm implemented in the
\emph{StV0TofCorrection} class is explained in the following.

\wrapifneeded{0.50}{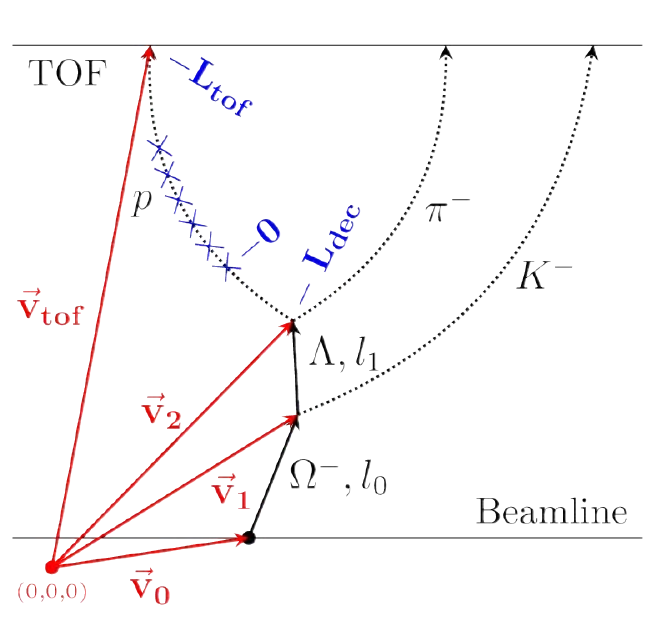}{Sketch for \ensuremath{\beta} correction.}{stv0tof_fig_corr}{0.45} %

Let \emph{v$_{\text{0}}$} be the 3d-{}vector from the global origin of the cartesian coordinate
system to the event vertex, \emph{v$_{\text{1,2}}$} the 3d-{}vectors to the decay vertices of
\ensuremath{\Omega}/\ensuremath{\Lambda} and \emph{v$_{\text{tof}}$} the 3d-{}vector to the TOF hit matched with the
proton track. Then, the decay lengths
$l_i=\left|v_{i+1}-v_i\right|$ of \ensuremath{\Omega}/\ensuremath{\Lambda} and their
velocities
\[\beta_i=\sqrt{1/(1+(M_ic/p_i)^2)}\]
with \emph{M$_{\text{i}}$} the invariant masses and \emph{p$_{\text{i}}$} the reconstructed momenta of
\ensuremath{\Omega}/\ensuremath{\Lambda}, can be used to calculate the \ensuremath{\Omega}/\ensuremath{\Lambda} time-{}of-{}flights
$\Delta t_i=l_i\beta_i c$.  For the decay lengths, the mother
particle tracks are assumed to be straight regardless of their charge.
Deviations caused by this assumption were investigated and found to be
negligible. Consequently, the original proton time-{}of-{}flight TOF$_{\text{p}}$ can be
corrected via TOF$_{\text{p}}$' = TOF$_{\text{p}}$ -{} \ensuremath{\Sigma}\ensuremath{\Delta}'t$_{\text{i}}$'. The use of
$\beta_pc=L_p' \mathrm{TOF}_p'$ and
$m_pc=p\sqrt{\beta_p^{-2}-1}$ provides a way to recalculate the
proton mass with its corrected time-{}of-{}flight if the corresponding path length
$L_p'$ is known.  Latter is deduced by evaluating the points on the
proton helix which correspond to the distance of closest approach (dca) to TOF
hit and $\Lambda$ decay vertex, respectively.  Therefore a function
written by A. Schmah was adapted to the needs of the class.  It iteratively
determines the dca of a point to a helix and returns the corresponding path
lengths. Given the path lengths on the proton helix to the TOF hit
($L_\mathrm{tof}$) and the $\Lambda$ decay vertex
($L_\mathrm{dec}$) the corrected proton path length amounts to
$L_\mathrm{p}' = L_\mathrm{tof} - L_\mathrm{dec}$. Finally, a cut
on the corrected proton mass can be applied to reject protons which do not fit
the given decay topology.  By this means the signal-{}to-{}noise ratio can be
increased significantly. The preceding discussion holds for all types of V$_{\text{0}}$ decays independent of particle species or number of decay steps.  Therefore,
the algorithm has been implemented with unlimited and variable number of
function arguments to assure a most general usage.\newline

In the following the class functionality and its features are investigated
taking the decay $\Lambda\to p\pi^-$ in Au+Au collisions at
7.7 GeV as an example due to its sufficient
statistics. For decay reconstruction the set of cuts listed in the following
table is used resulting in 34 M candidate combinations.
\begin{center}
\begingroup%
\setlength{\newtblsparewidth}{0.8\linewidth-2\tabcolsep-2\tabcolsep-2\tabcolsep}%
\setlength{\newtblstarfactor}{\newtblsparewidth / \real{340}}%

\begin{longtable}{ll}\caption[{Summary of cuts applied for \$\textbackslash{}Lambda\$ candidate selection}]{Summary of cuts applied for $\Lambda$ candidate selection}\tabularnewline
\hline
\multicolumn{1}{m{170\newtblstarfactor+\arrayrulewidth}}{\centering\bfseries%
 \textbf{Event Vertex} %
}&\multicolumn{1}{m{170\newtblstarfactor+\arrayrulewidth}}{\centering\bfseries%
 \textbf{PID TPC dE/dx}%
}\tabularnewline
\endfirsthead
\caption[]{(continued)}\tabularnewline
\hline
\multicolumn{1}{m{170\newtblstarfactor+\arrayrulewidth}}{\centering\bfseries%
 \textbf{Event Vertex} %
}&\multicolumn{1}{m{170\newtblstarfactor+\arrayrulewidth}}{\centering\bfseries%
 \textbf{PID TPC dE/dx}%
}\tabularnewline
\endhead
\multicolumn{1}{m{170\newtblstarfactor+\arrayrulewidth}}{\centering%
Vz <{} 70 cm \& Vr <{} 2 cm
}&\multicolumn{1}{m{170\newtblstarfactor+\arrayrulewidth}}{\centering%
$n\sigma_p<4\,\&\,n\sigma_\pi<4$
}\tabularnewline
\multicolumn{1}{m{170\newtblstarfactor+\arrayrulewidth}}{\centering%
\textbf{Track Properties}
}&\multicolumn{1}{m{170\newtblstarfactor+\arrayrulewidth}}{\centering%
\textbf{Topology}
}\tabularnewline
\multicolumn{1}{m{170\newtblstarfactor+\arrayrulewidth}}{\centering%
nHitsFit >{} 14
}&\multicolumn{1}{m{170\newtblstarfactor+\arrayrulewidth}}{\centering%
decay length >{} 4 cm
}\tabularnewline
\multicolumn{1}{m{170\newtblstarfactor+\arrayrulewidth}}{\centering%
nHitsRatio >{} 0
}&\multicolumn{1}{m{170\newtblstarfactor+\arrayrulewidth}}{\centering%
dca( V$_{\text{0}}$ ) <{} 2.5 cm
}\tabularnewline
\multicolumn{1}{m{170\newtblstarfactor+\arrayrulewidth}}{\centering%
dca( p ) >{} 0.5 cm \& dca( $\pi^-$ ) >{} 1 cm
}&\multicolumn{1}{m{170\newtblstarfactor+\arrayrulewidth}}{\centering%
dca( p $\pi^-$ ) <{} 2.5 cm
}\tabularnewline
\multicolumn{1}{m{170\newtblstarfactor+\arrayrulewidth}}{\centering%
0.1 <{} p <{} 10 GeV/c
}&\multicolumn{1}{m{170\newtblstarfactor+\arrayrulewidth}}{\centering%
1.06 <{} M$_{\text{inv}}$( p $\pi$ ) <{} 1.19 GeV/c$^{\text{2}}$
}\tabularnewline
\hline
\end{longtable}\endgroup%

\end{center}

\pagebreak[4]

\wrapifneeded{0.50}{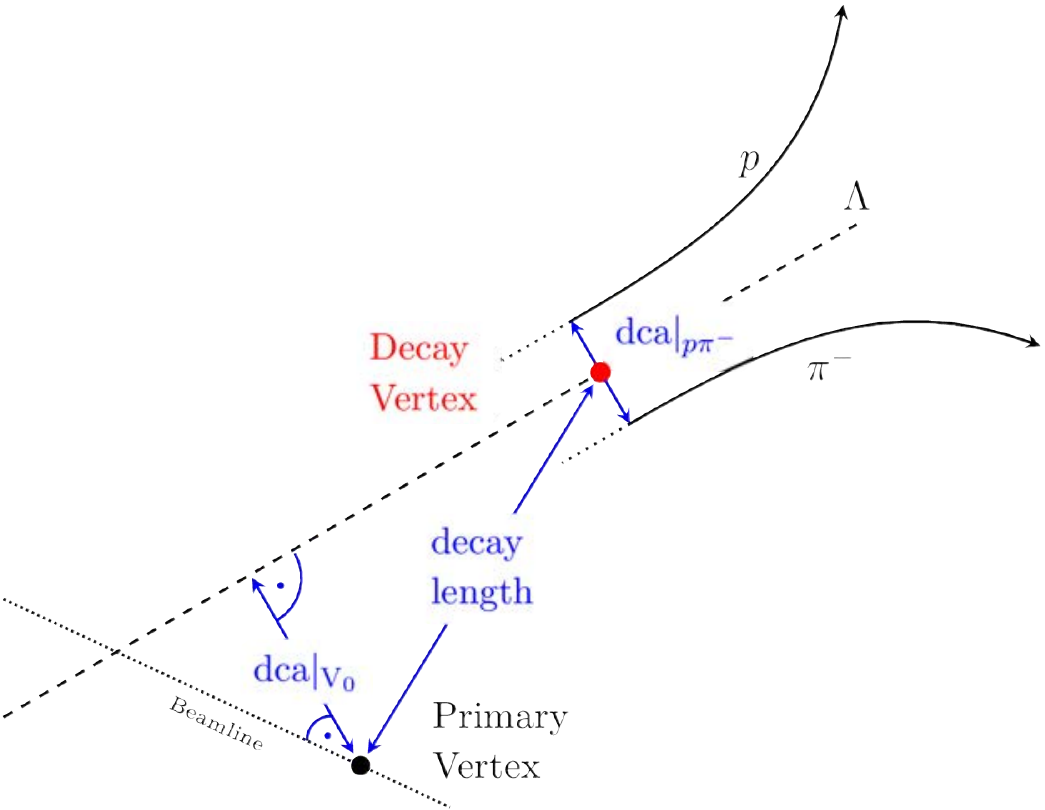}{Illustration of cut variables used for $\Lambda$ candidate selection.}{}{0.49} %

To gain access to the different effects of TOF correction the candidates sample
is divided into three sub-{}samples: In \emph{Sub-{}Sample I} only
$p\pi$ combinations with $\beta$ information for both
particles are selected to check whether the algorithm works and whether the PID
can be improved. In the \emph{Sub-{}Sample I} all other
cases where a TOF correction is possible are gathered to answer the question of
how many candidates can be restored. The last sub-{}sample comprises the residual
$p\pi$ combinations which do not provide any TOF information.  See
Table~\ref{stv0tof_tab_subsamp} for further clarification. The boolean values indicate
whether the specific information is available for the track or not.  Note that
there are cases where TOF information is available for the track but no
$\beta$ information. This is connected to the fact that the timing
information for global tracks is not reliable and therefore omitted on MicroDST
level.
\begin{center}
\begingroup%
\setlength{\newtblsparewidth}{0.75\linewidth-2\tabcolsep-2\tabcolsep-2\tabcolsep-2\tabcolsep}%
\setlength{\newtblstarfactor}{\newtblsparewidth / \real{319}}%

\begin{longtable}{lll}\caption[{List of Sub-{}Samples}]{List of Sub-{}Samples\label{stv0tof_tab_subsamp}\hyperlabel{stv0tof_tab_subsamp}%
}\tabularnewline
\endfirsthead
\caption[]{(continued)}\tabularnewline
\endhead
\hline
\multicolumn{1}{m{35\newtblstarfactor+\arrayrulewidth}}{\centering%
$\boldsymbol{\beta_p\,\beta_\pi}$
}&\multicolumn{1}{m{142\newtblstarfactor+\arrayrulewidth}}{\centering%
$\boldsymbol{\mathrm{TOF}_p\,\mathrm{TOF}_\pi}$
}&\multicolumn{1}{m{142\newtblstarfactor+\arrayrulewidth}}{\centering%
}\tabularnewline
\multicolumn{1}{m{35\newtblstarfactor+\arrayrulewidth}}{\centering%
1 1
}&\multicolumn{1}{m{142\newtblstarfactor+\arrayrulewidth}}{\centering%
1 1
}&\multicolumn{1}{m{142\newtblstarfactor+\arrayrulewidth}}{\centering%
\textbf{Sub-{}Sample I}
}\tabularnewline
\multicolumn{1}{m{35\newtblstarfactor+\arrayrulewidth}}{\centering%
1 0
}&\multicolumn{1}{m{142\newtblstarfactor+\arrayrulewidth}}{\centering%
1 0 \& 1 1
}&\multicolumn{1}{m{142\newtblstarfactor+\arrayrulewidth}}{\setlength{\newtblcolwidth}{142\newtblstarfactor}\multirowii[m]{3}{\newtblcolwidth}{\centering%
\textbf{Sub-{}Sample II}
}}\tabularnewline
\multicolumn{1}{m{35\newtblstarfactor+\arrayrulewidth}}{\centering%
0 1
}&\multicolumn{1}{m{142\newtblstarfactor+\arrayrulewidth}}{\centering%
0 1 \& 1 1
}&\multicolumn{1}{m{142\newtblstarfactor+\arrayrulewidth}}{\setlength{\newtblcolwidth}{142\newtblstarfactor}\multirowii[m]{3}{-\newtblcolwidth}{\centering%
\textbf{Sub-{}Sample II}
}}\tabularnewline
\multicolumn{1}{m{35\newtblstarfactor+\arrayrulewidth}}{\centering%
0 0
}&\multicolumn{1}{m{142\newtblstarfactor+\arrayrulewidth}}{\centering%
1 0 \& 0 1 \& 1 1
}&\multicolumn{1}{m{142\newtblstarfactor+\arrayrulewidth}}{\setlength{\newtblcolwidth}{142\newtblstarfactor}\multirowii[m]{3}{-\newtblcolwidth}{\centering%
\textbf{Sub-{}Sample II}
}}\tabularnewline
\multicolumn{1}{m{35\newtblstarfactor+\arrayrulewidth}}{\centering%
0 0
}&\multicolumn{1}{m{142\newtblstarfactor+\arrayrulewidth}}{\centering%
0 0
}&\multicolumn{1}{m{142\newtblstarfactor+\arrayrulewidth}}{\centering%
\textbf{Sub-{}Sample III}
}\tabularnewline
\hline
\end{longtable}\endgroup%

\end{center}

\wrapifneeded{0.50}{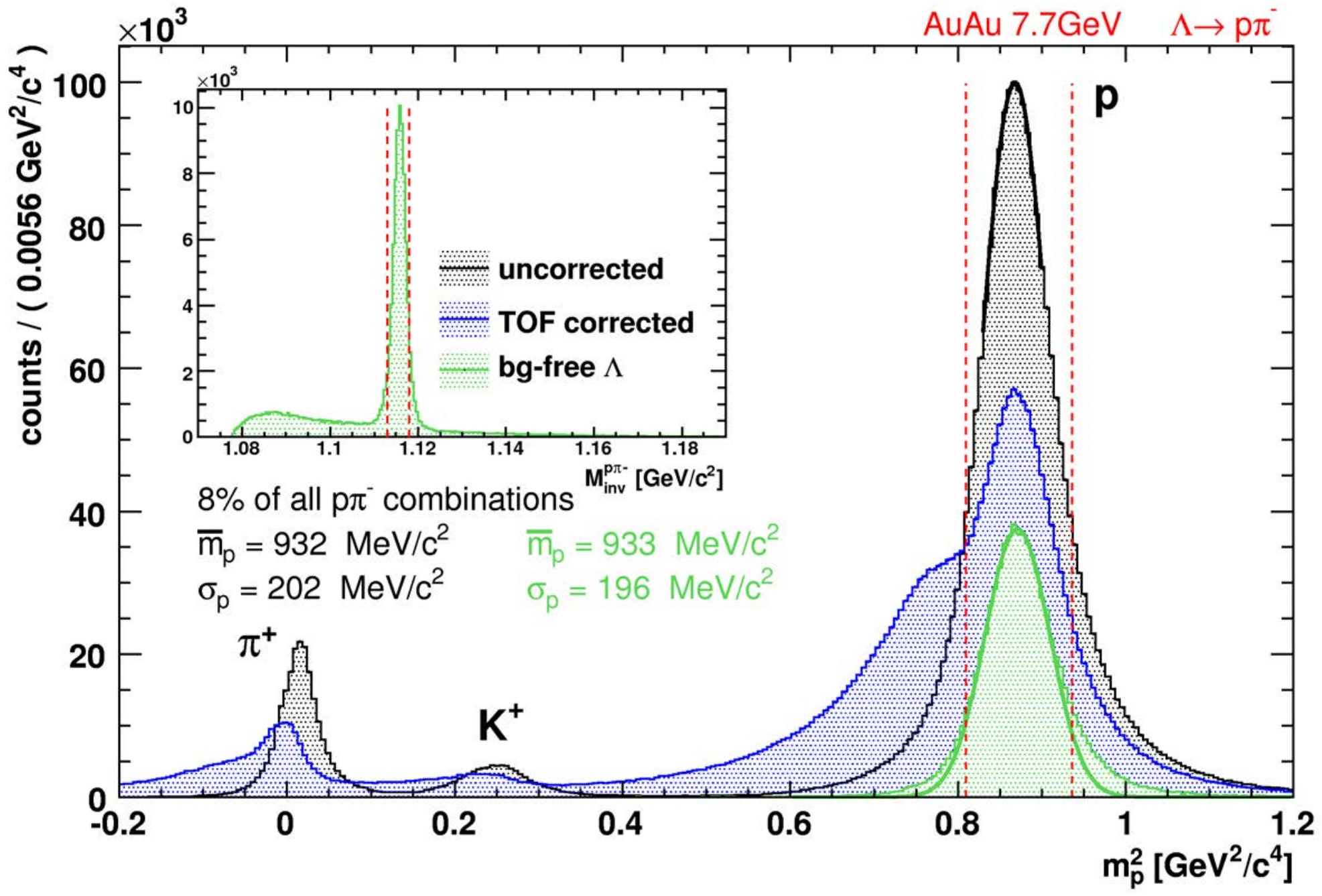}{Proton mass distributions for sub-{}sample I}{stv0tof_fig_protons0}{0.49} %

Figure~\ref{stv0tof_fig_protons0} shows a comparison of mass distributions for protons
steming from $\Lambda\to p\pi^-$ decays. For the sample with
available $\beta$ information for each decay track, distributions
are plotted with (blue) and without (black) applied TOF correction.  A scaled
mass distribution (green) for a background-{}free $\Lambda$ sample
(inlet) is shown as well.  The cut applied on the p $\pi$ invariant
mass is 1.113 -{} 1.118 GeV/c$^{\text{2}}$. Solid lines represent Gaussian functions fitted
in the range $0.83<m^2<0.9\,GeV^2/c^4$. Resulting parameters are
given in corresponding colors.  Red dashed lines indicate the applied mass cuts
for particle identification.  The percentages given in this and the following
figures are based on the total initial number of $p\pi$ combinations. No differentiation has been made regarding the changing momentum
resolution.  As its main feature the TOF correction algorithm changes the mass
distributions with respect to the origin of the particle: the distribution of
protons which do not fit the specific decay topology are shifted to lower
masses and broadened -{} see the shoulder of the blue distribution.  Whereas
\emph{true} protons stay at their original mass. This can be proven by generating a
background-{}free $\Lambda$ sample using tighter and additional cuts
-{} see inlet in Figure~\ref{stv0tof_fig_protons0}. The Gaussian function fitted to the
resulting scaled mass distribution is not shifted and its extrapolation does
not reveal any underlying or broadened structure. Applying a cut on the squared
proton mass inreases the separation power of $\Lambda$-{} and non-{}
$\Lambda$-{}daughters.\newline

\wrapifneeded{0.50}{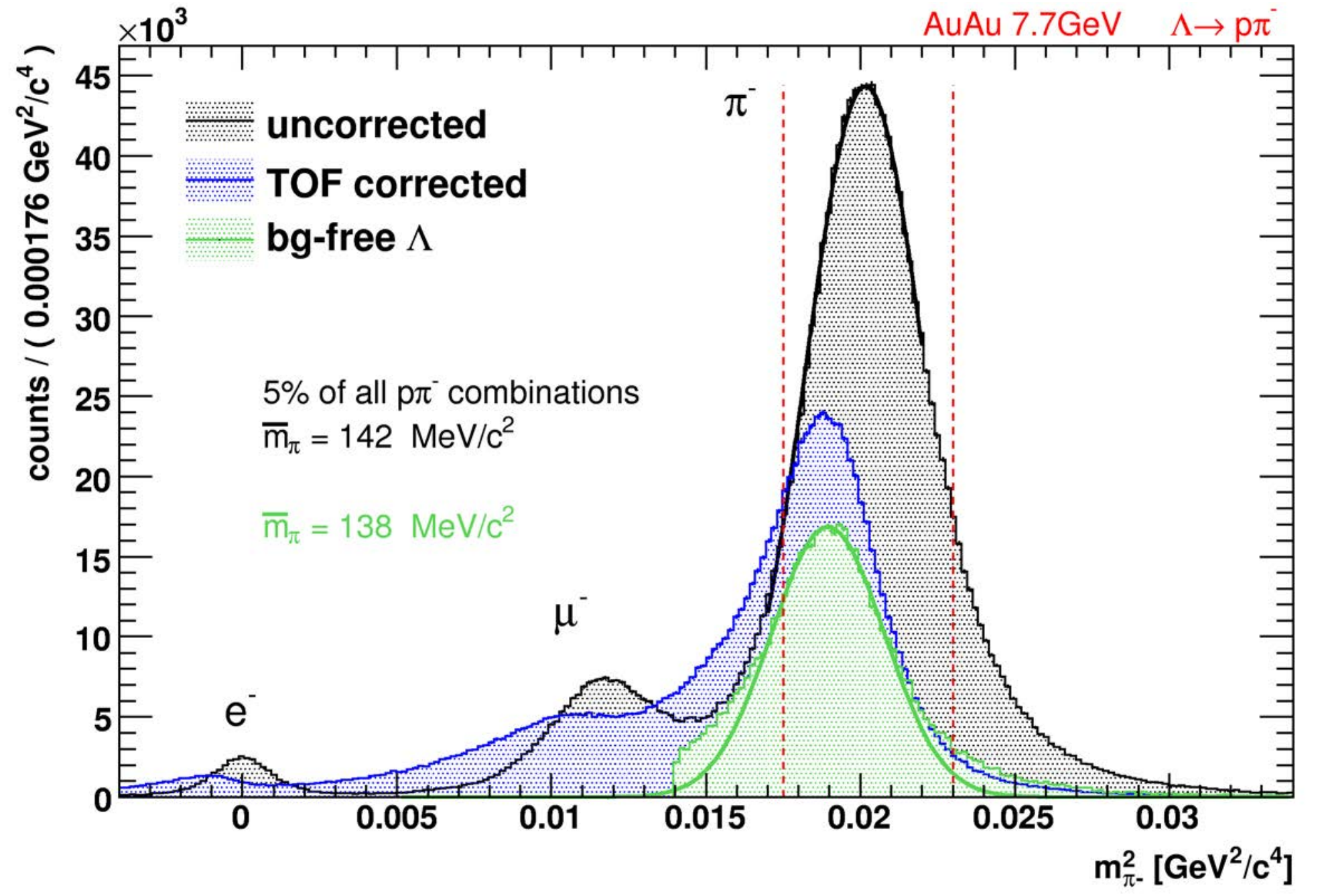}{pion mass distributions}{stv0tof_fig_pions0}{0.49} %

The pion mass distributions after the proton mass cut are shown in
Figure~\ref{stv0tof_fig_pions0} using identical color code. The aforementioned cut on
the proton mass is already applied. Here the non-{}
$\Lambda$-{}daughters can only be seen as tails at the lower mass end
of the pion distribution not allowing a separation as clear as for protons.
However more important, the mean pion mass is shifted by 3 MeV/c$^{\text{2}}$ to its
nominal mass. Thus, using the TOF correction algorithm accounts for the
systematically off pion mass intrinsically.\newline

Figure~\ref{stv0tof_fig_lambdas0} compares the $p\pi$ invariant mass spectra
with (blue) and without (black) applied TOF correction algorithm. The
distributions are fitted with Breit-{}Wigner + higher order polynomial functions
(solid). There are two major improvements to be emphasized:

\wrapifneeded{0.50}{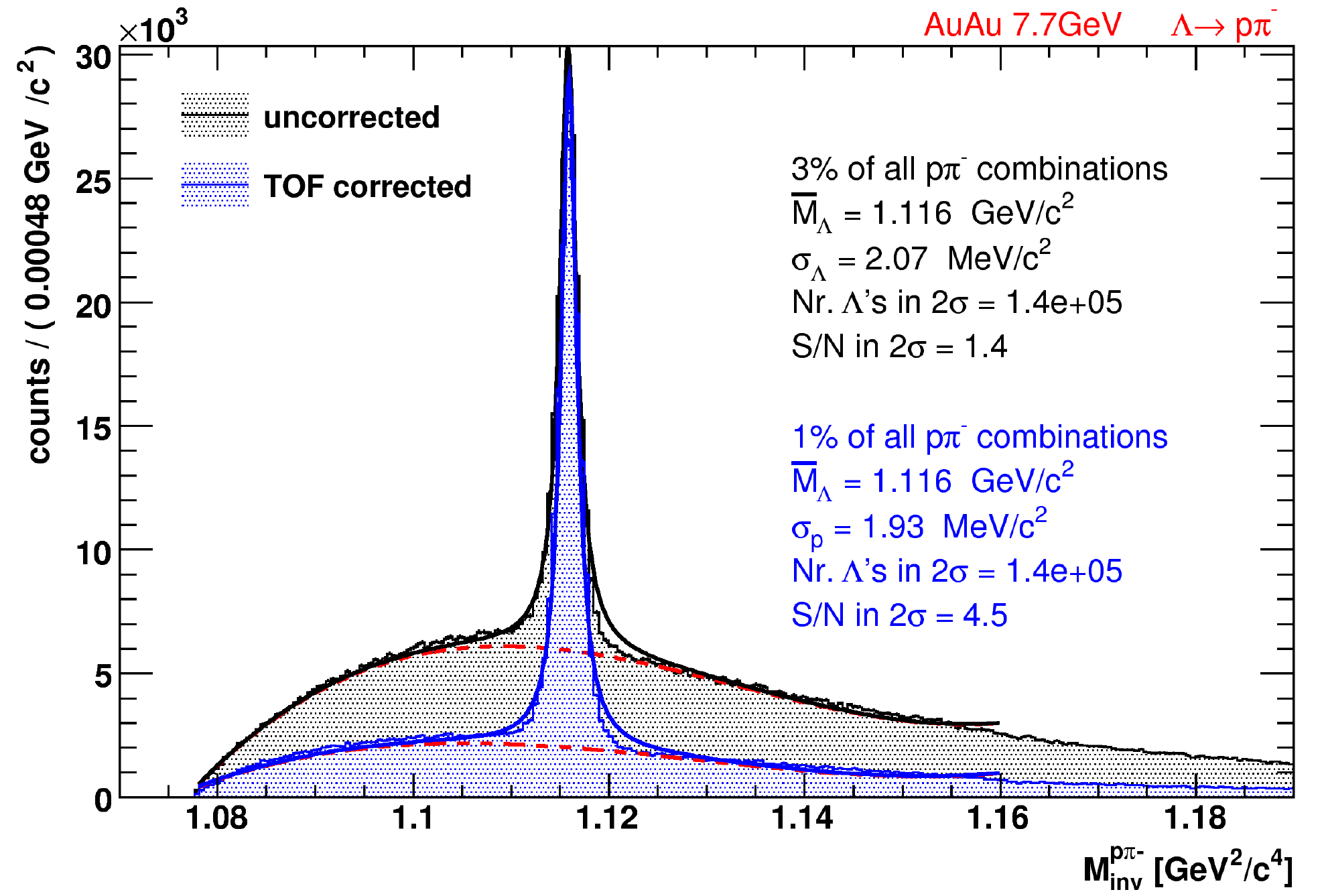}{$p\pi$ invariant mass distributions}{stv0tof_fig_lambdas0}{0.49} %
\begin{itemize}[itemsep=0pt]

\item{} The number of reconstructed $\Lambda$'s stays constant which
  proves that mainly non-{} $\Lambda$-{}daughters have been rejected by
  the new proton and pion particle identification.

\item{} The signal-{}to-{}background ratio improves by more than a factor 3. A comparable
  effect can be seen for the case of incomplete TOF information. This nicely
  shows the background rejection power of the algorithm for the cases of TOF
  information being available.

\end{itemize}

Unfortunately, the current sub-{}sample only covers 8\% of the total number of
$\Lambda$ candidates. The residual sub-{}samples therefore dominate
the signal-{}to-{}background ratio with respect to the overall achievable
improvements.\newline

Figure~\ref{stv0tof_fig_ppi1} shows the mass distributions for sub-{}sample II with (blue)
and without (black) applied TOF correction. Red dashed lines indicate the
mass cuts used for particle identification. In general, the distributions for
both particles indicate the same features as discussed in the previous section,
i.e.  non-{}decay-{}daughter separation and pion pole mass shift. However, in addition
a gain of 35\% can be observed for the pions in a $2\sigma$ range
around the respective mean masses. These are pions whose time-{}of-{}flight was
intrinsically reconstructed by the algorithm using the matched TOF hit but
which originally did not have a $\beta$ information associated
(see Table~\ref{stv0tof_tab_subsamp}). For this sub-{}sample, particle identification
cuts on the proton and pion mass have been applied as well (indicated by red
dashed lines in Figure~\ref{stv0tof_fig_ppi1}).

\wrapifneeded{0.50}{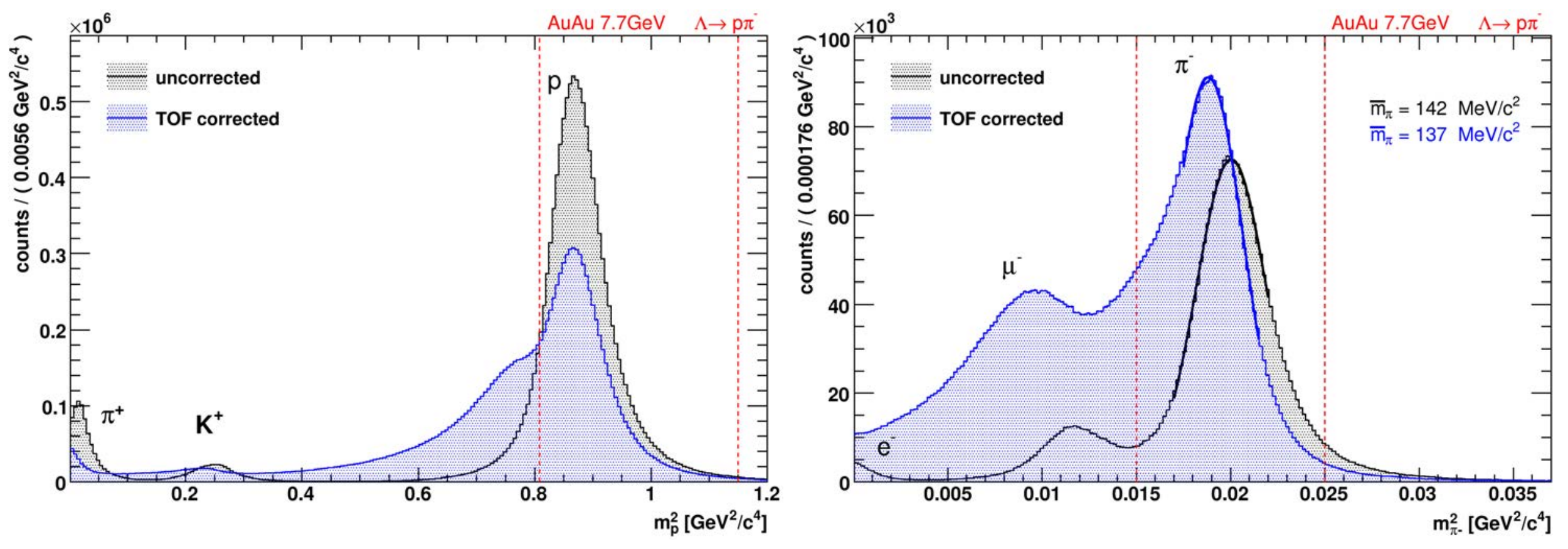}{mass distributions for sub-{}sample II}{stv0tof_fig_ppi1}{1} %

\wrapifneeded{0.50}{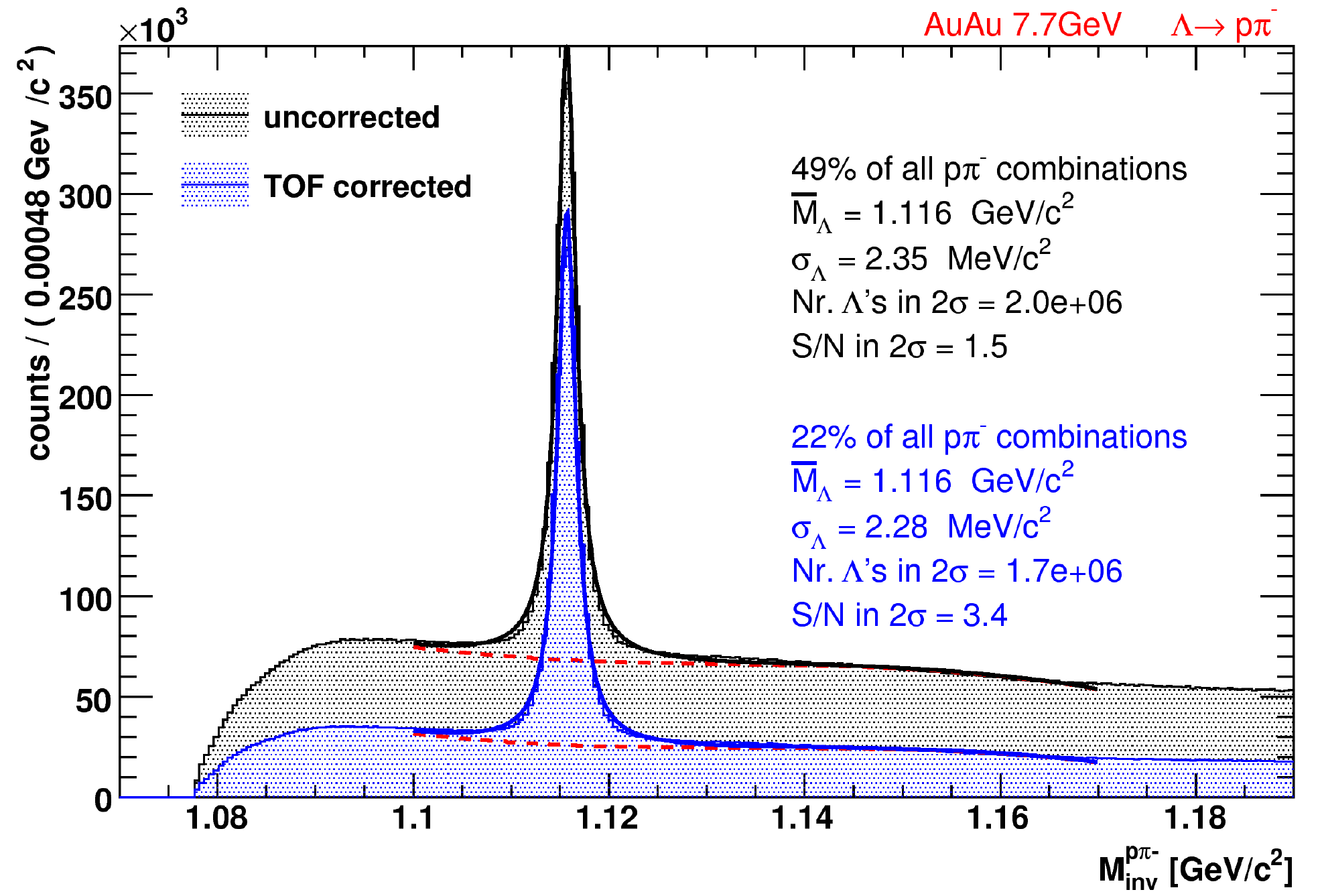}{Comparison of $\Lambda$ invariant mass spectra with and without TOF correction}{stv0tof_fig_lambdas1}{0.49} %

The comparison of the resulting $p\pi^-$ invariant mass
distributions is shown in Figure~\ref{stv0tof_fig_lambdas1}. Even for the current
statistically dominant sub-{}sample (60\%) a significant improvement (factor 2.3)
in signal-{}to-{}background ratio can be observed suffering only a small signal
loss. Latter could mainly be caused by the fact that the applied mass cuts have
rather been chosen in favor of the uncorrected distributions.\newline

The final $p\pi^-$ invariant mass spectrum is shown in
Figure~\ref{stv0tof_fig_overall}.  The corrected spectrum shows a slightly smaller
width and a non-{}significant decrease of 10\% in the number of
$\Lambda$s.

\wrapifneeded{0.50}{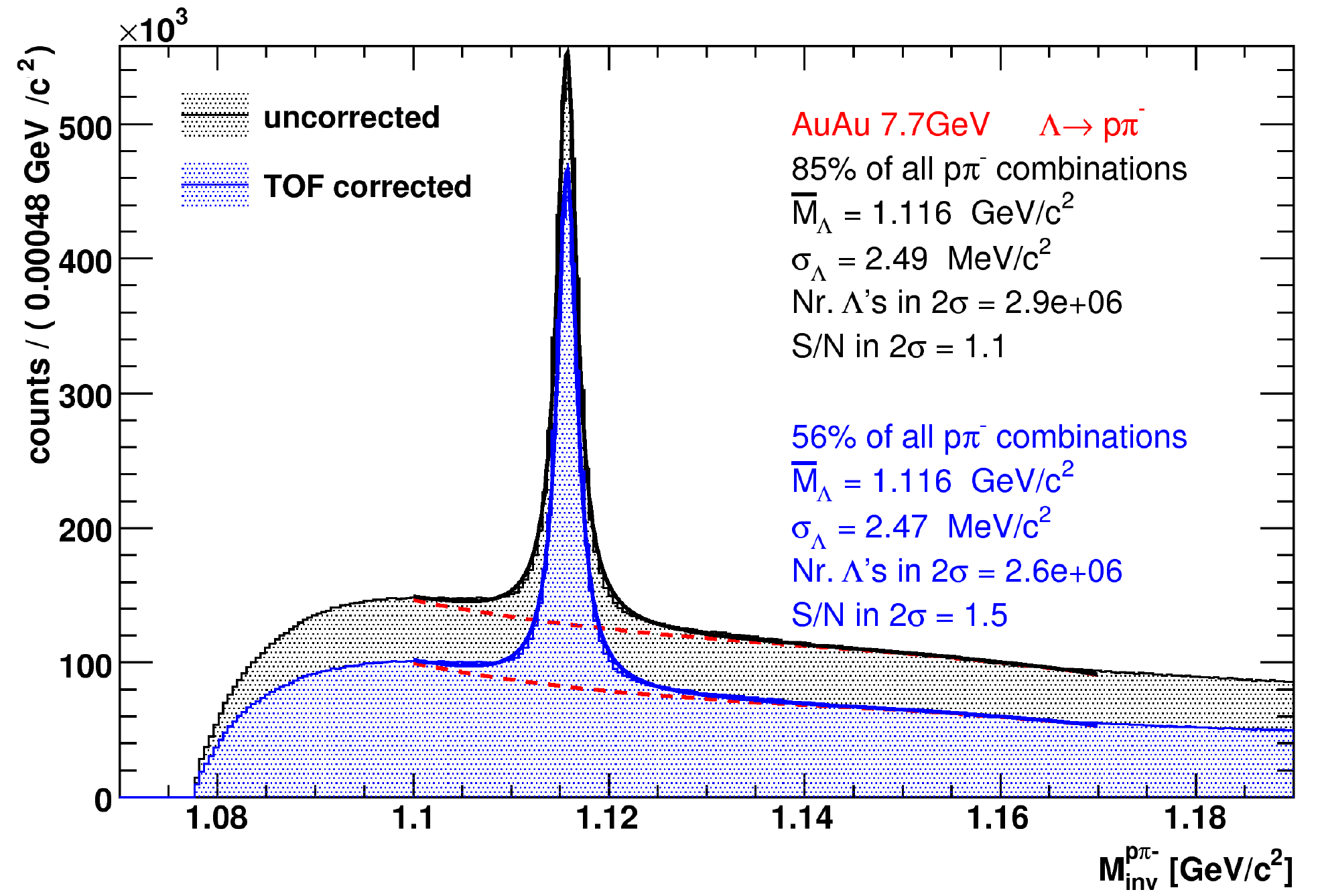}{Overall influence of TOF correction on the $\Lambda$ signal}{stv0tof_fig_overall}{0.49} %

Regarding the signal-{}to-{}background ratio an overall
improvement of approx. 40\% is achieved. The comparably lower gain in contrast
to \emph{Sub-{}Sample I} and \emph{Sub-{}Sample II} is due to the poor
signal-{}to-{}background ratio of the sub-{}sample in which no TOF correction is
possible at all. Nevertheless these improvements show the importance of
applying a TOF correction to the tracks of V$_{\text{0}}$ decay particles regarding
background suppression.\newline

In addition to the preceding $\Lambda$ analysis the TOF
correction algorithm was applied to the decay $\Omega^-\to\Lambda K^-\to p\pi^-K^-$. The corresponding plots can be found in
Figure~\ref{stv0tof_fig_omega}. Applying a TOF correction results in an increase of
signal-{}to-{}noise ratio by a factor of 1.5 in this case.

\wrapifneeded{0.50}{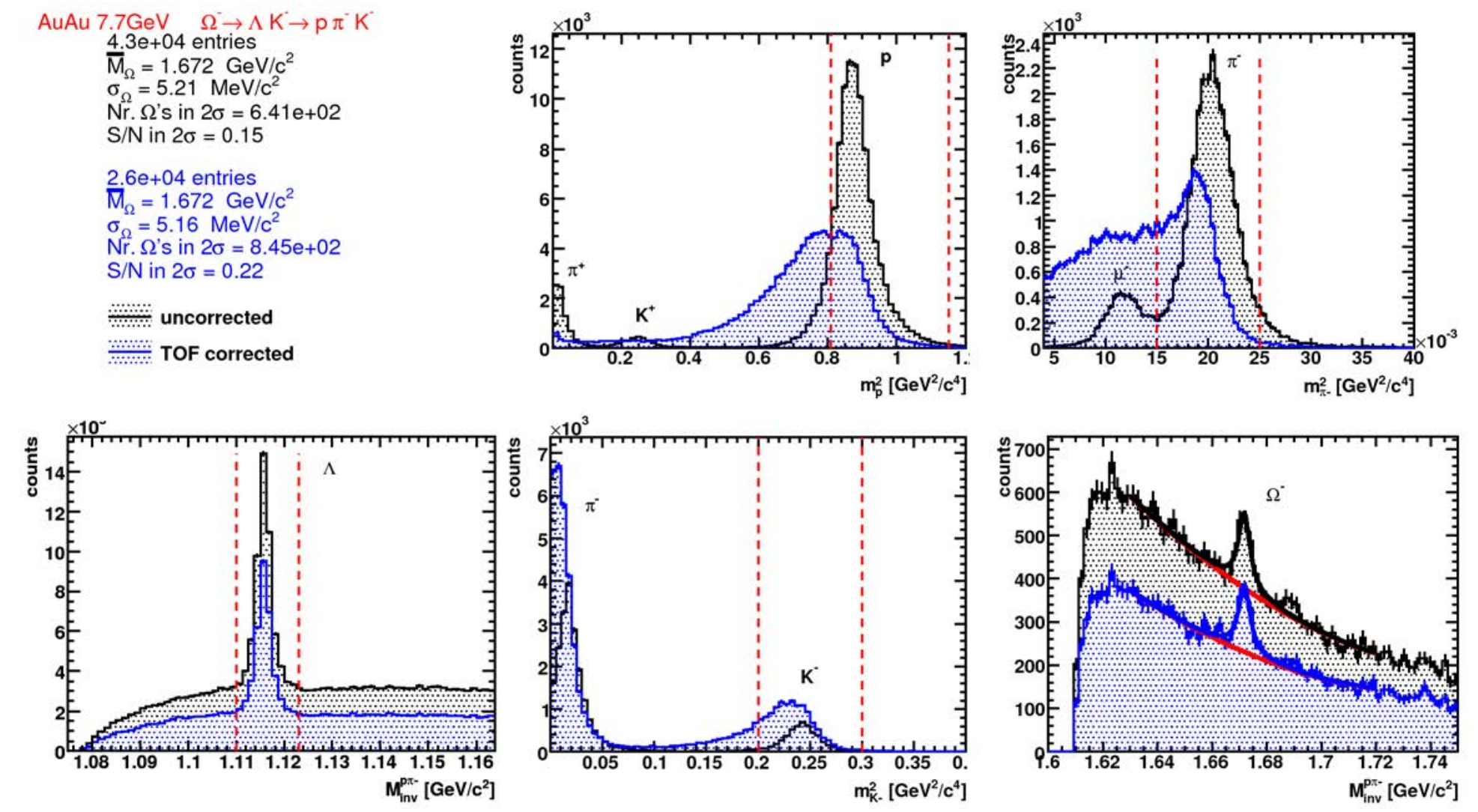}{$\Omega$ Analysis}{stv0tof_fig_omega}{1} %

In general, the class is supposed to be initiated once by calling its
constructor. The member functions \texttt{set\penalty5000 V\penalty5000 e\penalty5000 c\penalty5000 t\penalty5000 o\penalty5000 r\penalty5000 s3D}, \texttt{set\penalty5000 M\penalty5000 o\penalty5000 t\penalty5000 h\penalty5000 e\penalty5000 r\penalty5000 T\penalty5000 r\penalty5000 a\penalty5000 cks} and
\texttt{cor\penalty5000 r\penalty5000 e\penalty5000 c\penalty5000 t\penalty5000 B\penalty5000 eta} are afterwards used multiple times for the TOF correction of each
decay particle. Hence, these member functions have to be invoked for each
combination of daughter particles in the following order.  Remember to call
these functions each time as a block.

\noindent
\begin{description}
\item[{ \texttt{set\penalty5000 V\penalty5000 e\penalty5000 c\penalty5000 t\penalty5000 o\penalty5000 r\penalty5000 s\penalty5000 3\penalty5000 D\penalty5000 (\penalty5000 v\penalty5000 0\penalty5000 )\penalty5000 (\penalty5000 v1).\penalty0 .\penalty0 .\penalty0 (vn\penalty5000 )\penalty5000 (\penalty5000 v\penalty5000 t\penalty5000 of)} }] \hspace{0em}\\
 This function sets all the space vectors needed for TOF correction.  The
function arguments \texttt{v0\ldots{}\penalty5000 v\penalty5000 tof} are all of type \texttt{StT\penalty5000 h\penalty5000 r\penalty5000 e\penalty5000 e\penalty5000 V\penalty5000 e\penalty5000 c\penalty5000 t\penalty5000 orD} and define the
vectors to the following space points.

\begin{itemize}[itemsep=0pt]

\item{} \texttt{v0}: event vertex / primary vertex

\item{} \texttt{v1\ldots{}vn}: vertices in decay chain resulting in the final particle to be corrected

\item{} \texttt{vtof}: TOF hit

\end{itemize}

Special care has to be taken of the vectors' ordering: Start with the event
vertex and add the vertex vectors as you follow the decay chain to the final
particle and its TOF hit. The final argument of \texttt{set\penalty5000 V\penalty5000 e\penalty5000 c\penalty5000 t\penalty5000 o\penalty5000 r\penalty5000 s3D} always corresponds to the
TOF hit vector.\newline
 The event vertex does not change on a candidate-{}by-{}candidate basis, of course.
Nonetheless, it is given to \texttt{set\penalty5000 V\penalty5000 e\penalty5000 c\penalty5000 t\penalty5000 o\penalty5000 r\penalty5000 s3D} as an argument to simplify the
implementation and to minimize the functions to be called by the user.
Make sure to start TOF correction only if a TOF hit for the particle exists!
Carefully check the original time-{}of-{}flights given as an argument to the
\texttt{cor\penalty5000 r\penalty5000 e\penalty5000 c\penalty5000 t\penalty5000 B\penalty5000 eta} function!\newline
 The operator () is overloaded to allow for a variable number of arguments to be
given to the function and pack them into an inline container sequentially.
Thus, make sure to put each function argument inside brackets as indicated in
above.\newline
 Starting gcc 4.3 the so called \emph{variadic templates} were included in the
compiler. These function templates accept a variable number of arguments
(\emph{parameter rack}) which can be accessed and unpacked via the ellipsis (\ldots{})
operator.  This approach would have been a more elegant way to implement the
class.  Unfortunately, variadic templates will only be understood within the
new \emph{C++0x} standard which presumably will be published at the end of 2011.
\item[{ \texttt{set\penalty5000 M\penalty5000 o\penalty5000 t\penalty5000 h\penalty5000 e\penalty5000 r\penalty5000 T\penalty5000 r\penalty5000 a\penalty5000 c\penalty5000 k\penalty5000 s\penalty5000 (\penalty5000 t0).\penalty0 .\penalty0 .\penalty0 (tn-{}\penalty0 1)} }] \hspace{0em}\\
 Call this function to set the reconstructed 4-{}momentum vectors \texttt{t0\ldots{}tn-{}\penalty0 1} of
type \texttt{StL\penalty5000 o\penalty5000 r\penalty5000 e\penalty5000 n\penalty5000 t\penalty5000 z\penalty5000 V\penalty5000 e\penalty5000 c\penalty5000 tor} for all mother particles. Keep the warning in
\texttt{set\penalty5000 V\penalty5000 e\penalty5000 c\penalty5000 t\penalty5000 o\penalty5000 r\penalty5000 s3D} in mind!\newline
 The number of arguments given should be smaller by 2 compared to
\texttt{set\penalty5000 V\penalty5000 e\penalty5000 c\penalty5000 t\penalty5000 o\penalty5000 r\penalty5000 s3D}. If not, the code compiles nevertheless but results in an error
message during runtime.

\item[{ \texttt{cor\penalty5000 r\penalty5000 e\penalty5000 c\penalty5000 t\penalty5000 B\penalty5000 e\penalty5000 t\penalty5000 a\penalty5000 (\penalty5000 h\penalty5000 e\penalty5000 l\penalty5000 i\penalty5000 x\penalty5000 ,\penalty5000 t\penalty5000 o\penalty5000 f\penalty5000 ,\penalty5000 b\penalty5000 e\penalty5000 t\penalty5000 a\penalty5000 c\penalty5000 o\penalty5000 rr)} }] \hspace{0em}\\
 This is the main function of the class. It takes the \texttt{helix} of type
\texttt{StP\penalty5000 h\penalty5000 y\penalty5000 s\penalty5000 i\penalty5000 c\penalty5000 a\penalty5000 l\penalty5000 H\penalty5000 e\penalty5000 l\penalty5000 ixD} and the original time-{}of-{}flight \texttt{tof} of the particle to be
TOF corrected and writes the corrected particle velocity into \texttt{bet\penalty5000 a\penalty5000 c\penalty5000 orr}. The
variable \texttt{bet\penalty5000 a\penalty5000 c\penalty5000 orr} can afterwards be used to calculate a new particle mass and
utilize it for particle identification. Alternatively, \texttt{cor\penalty5000 r\penalty5000 e\penalty5000 c\penalty5000 t\penalty5000 B\penalty5000 eta} can be
called as \texttt{cor\penalty5000 r\penalty5000 e\penalty5000 c\penalty5000 t\penalty5000 B\penalty5000 e\penalty5000 t\penalty5000 a\penalty5000 (\penalty5000 h\penalty5000 e\penalty5000 l\penalty5000 i\penalty5000 x\penalty5000 ,\penalty5000 t\penalty5000 o\penalty5000 f\penalty5000 ,\penalty5000 b\penalty5000 e\penalty5000 t\penalty5000 a\penalty5000 c\penalty5000 o\penalty5000 r\penalty5000 r\penalty5000 ,\penalty5000 p\penalty5000 ,\penalty5000 e\penalty5000 n\penalty5000 um)} with \texttt{p} the particle's
momentum and \texttt{enum} the particle species.  In this form a default hardcoded
m$^{\text{2}}$-{}cut is applied to the particle's mass and the result returned as boolean
value (\texttt{enum} definition: 0 = pi, 1 = K, 2 = p).

\item[{ \texttt{cle\penalty5000 a\penalty5000 r\penalty5000 C\penalty5000 o\penalty5000 n\penalty5000 t\penalty5000 a\penalty5000 i\penalty5000 n\penalty5000 e\penalty5000 r\penalty5000 s()} }] \hspace{0em}\\
 This function needs to be called to clear all inline containers before TOF
correcting the next particle. To assure the filling of a clean container, allocated
memory and pointers are deleted in this function. It could be done
automatically after the TOF correction is finished. However, the user is
encouraged to do it manually to have a more memorizable and controllable block
of functions.

\end{description}

Section~\ref{stv0tof_examples} gives two examples of how to use the \texttt{StV\penalty5000 0\penalty5000 T\penalty5000 o\penalty5000 f\penalty5000 C\penalty5000 o\penalty5000 r\penalty5000 r\penalty5000 e\penalty5000 c\penalty5000 t\penalty5000 ion} class and its member functions. The first example shows the case of correcting
a particle with one mother particle and the second the case of two mother
particles. Both examples are illustrated in Figure~\ref{stv0tof_fig_corr}.

\section{StBadRdosDb: Database Interface for TPC missing RDOs}
\label{stbadrdos}\hyperlabel{stbadrdos}%

The class \emph{StBadRdosDb} has been developed to use information about the TPC's
number of missing read-{}out boards (RDO) during a physics analysis. The data
taken in runs 10 \& 11 are divided into \emph{run ranges} with equal number of
missing RDOs and the result stored in a custom database format. \emph{StBadRdosDb} can then be used to acquire the current run range given a specific run id. This
is especially helpful for analyses which rely on stable detector conditions
throughout a beam time and thus are sensitive to a run-{}range-{}dependent
efficiency correction. Upon request by STAR's physics working group for bulk
correlations, \emph{StBadRdosDb} has been announced to the collaboration
[26] and made available in the STAR CVS
[27] as well as the author's code repository
[215].  The \texttt{mak\penalty5000 e\penalty5000 R\penalty5000 u\penalty5000 n\penalty5000 L\penalty5000 ist.\penalty0 pl} script used to generate official
STAR run lists for each beam time and available in a common area of STAR
scripts [216] has been modified to query the central STAR database
for the number of missing RDOs per subsequent run and the according TPC sectors
and RDO id's. It has been extended to deal with 2010 as well as 2011 data. The
queried dataset is reduced to the list of run id's at which the number of
missing RDOs changes. Given the correct trigger setup name and year the
\texttt{gen\penalty5000 B\penalty5000 a\penalty5000 d\penalty5000 R\penalty5000 d\penalty5000 o\penalty5000 sDb.\penalty0 pl} script is executed as\newline
 \texttt{\$ p\penalty5000 e\penalty5000 r\penalty5000 l g\penalty5000 e\penalty5000 n\penalty5000 B\penalty5000 a\penalty5000 d\penalty5000 R\penalty5000 d\penalty5000 o\penalty5000 sDb.\penalty0 pl A\penalty5000 u\penalty5000 A\penalty5000 u\penalty5000 3\penalty5000 9\penalty5000 \_\penalty5000 p\penalty5000 r\penalty5000 o\penalty5000 d\penalty5000 u\penalty5000 c\penalty5000 t\penalty5000 i\penalty5000 o\penalty5000 n 2\penalty5000 011}\newline
 See the RunLog browser [217] for information on the
trigger setup name. For convenience, the code repository contains a bash script
to (re-{})generate all database files: \texttt{dat\penalty5000 a\penalty5000 b\penalty5000 ase/\penalty0 gen\penalty5000 All.\penalty0 sh}. If database files
for upcoming beam energies are generated one should make sure that the trigger
selection in the \texttt{gen\penalty5000 B\penalty5000 a\penalty5000 d\penalty5000 R\penalty5000 d\penalty5000 o\penalty5000 sDb.\penalty0 pl} script covers all runs needed in the
respective data analysis. The \emph{StBadRdosDb} class does not require any STAR
libraries to be loaded and solely relies on ROOT. It provides the following
member functions to retrieve information about missing RDOs and their
respective run ranges. For their usage see the \texttt{tes\penalty5000 t\penalty5000 S\penalty5000 t\penalty5000 B\penalty5000 a\penalty5000 d\penalty5000 R\penalty5000 d\penalty5000 o\penalty5000 sDb.\penalty0 C} ROOT macro.

\noindent
\begin{description}
\item[{ \texttt{Int\penalty5000 \_\penalty5000 t G\penalty5000 e\penalty5000 t\penalty5000 R\penalty5000 u\penalty5000 n\penalty5000 R\penalty5000 a\penalty5000 n\penalty5000 g\penalty5000 e\penalty5000 (\penalty5000 I\penalty5000 n\penalty5000 t\penalty5000 \_\penalty5000 t r\penalty5000 u\penalty5000 n\penalty5000 id);\penalty0 } }] \hspace{0em}\\
 This function is considered the main purpose of the \emph{StBadRdosDb} implementation. It returns the run range index for a given run id.

\item[{ \texttt{Int\penalty5000 \_\penalty5000 t G\penalty5000 e\penalty5000 t\penalty5000 N\penalty5000 r\penalty5000 R\penalty5000 u\penalty5000 n\penalty5000 R\penalty5000 a\penalty5000 n\penalty5000 g\penalty5000 e\penalty5000 s\penalty5000 (\penalty5000 ) \{ r\penalty5000 e\penalty5000 t\penalty5000 u\penalty5000 r\penalty5000 n N\penalty5000 r\penalty5000 R\penalty5000 a\penalty5000 n\penalty5000 ges;\penalty0  \}} }] \hspace{0em}\\
 A helper function returning the total number of run ranges. It can be used to
define arrays and histogram ranges for run-{}range dependent analyses.

\item[{ \texttt{Int\penalty5000 \_\penalty5000 t G\penalty5000 e\penalty5000 t\penalty5000 N\penalty5000 r\penalty5000 B\penalty5000 a\penalty5000 d\penalty5000 R\penalty5000 d\penalty5000 o\penalty5000 s\penalty5000 (\penalty5000 I\penalty5000 n\penalty5000 t\penalty5000 \_\penalty5000 t r\penalty5000 u\penalty5000 n\penalty5000 \_\penalty5000 r\penalty5000 a\penalty5000 n\penalty5000 ge);\penalty0 } }] \hspace{0em}\\
 An informational function returning the number of missing RDOs for a given run
range.

\item[{ \texttt{Int\penalty5000 \_\penalty5000 t G\penalty5000 e\penalty5000 t\penalty5000 S\penalty5000 e\penalty5000 c\penalty5000 t\penalty5000 o\penalty5000 r\penalty5000 (\penalty5000 I\penalty5000 n\penalty5000 t\penalty5000 \_\penalty5000 t r\penalty5000 u\penalty5000 n\penalty5000 \_\penalty5000 r\penalty5000 a\penalty5000 n\penalty5000 g\penalty5000 e\penalty5000 , I\penalty5000 n\penalty5000 t\penalty5000 \_\penalty5000 t n\penalty5000 r\penalty5000 do);\penalty0 } }] \hspace{0em}\\
 \emph{GetSector} returns the sector to which a specific missing RDO with index
\texttt{nrdo} in run range \texttt{run\penalty5000 \_\penalty5000 r\penalty5000 a\penalty5000 nge} belongs.

\item[{ \texttt{Int\penalty5000 \_\penalty5000 t G\penalty5000 e\penalty5000 t\penalty5000 N\penalty5000 r\penalty5000 B\penalty5000 a\penalty5000 d\penalty5000 R\penalty5000 d\penalty5000 o\penalty5000 s\penalty5000 S\penalty5000 e\penalty5000 c\penalty5000 t\penalty5000 o\penalty5000 r\penalty5000 (\penalty5000 I\penalty5000 n\penalty5000 t\penalty5000 \_\penalty5000 t r\penalty5000 u\penalty5000 n\penalty5000 i\penalty5000 d\penalty5000 , I\penalty5000 n\penalty5000 t\penalty5000 \_\penalty5000 t s\penalty5000 e\penalty5000 c\penalty5000 t\penalty5000 or);\penalty0 } }] \hspace{0em}\\
 retrieves the number of missing RDOs for a specific run id and sector.

\item[{ \texttt{Int\penalty5000 \_\penalty5000 t G\penalty5000 e\penalty5000 t\penalty5000 F\penalty5000 i\penalty5000 r\penalty5000 s\penalty5000 t\penalty5000 I\penalty5000 d\penalty5000 (\penalty5000 I\penalty5000 n\penalty5000 t\penalty5000 \_\penalty5000 t r\penalty5000 r\penalty5000 ) \{ r\penalty5000 e\penalty5000 t\penalty5000 u\penalty5000 r\penalty5000 n p\penalty5000 r\penalty5000 u\penalty5000 n\penalty5000 [\penalty5000 rr];\penalty0  \}} }] \hspace{0em}\\
 Helper function to get the first run id in the database for a specific run
range.

\end{description}

Table~\ref{stbadrdos_tab_dbfile} lists an examplatory database file used by the member
functions of \emph{StBadRdosDb} to provide information about the missing RDOs of a
specific beam time. All energies for the years 2010 and 2011 have been included
in the database (Section~\ref{app_stbadrdos_dbfiles}). Each line in the database files
indicates the start of a run range containing run id's with equal number of
missing RDOs. The second column shows the number of missing RDOs in this range
followed by the according number of RDO \emph{identifiers}. Each identifier encodes
the missing RDO's sector and identification number (format: \texttt{sec\penalty5000 tor.\penalty0 id}). These
identifiers are encoded in \texttt{gen\penalty5000 B\penalty5000 a\penalty5000 d\penalty5000 R\penalty5000 d\penalty5000 o\penalty5000 sDb.\penalty0 pl} and decoded in the class member
functions \texttt{Get\penalty5000 S\penalty5000 e\penalty5000 c\penalty5000 tor} and \texttt{Get\penalty5000 N\penalty5000 r\penalty5000 B\penalty5000 a\penalty5000 d\penalty5000 R\penalty5000 d\penalty5000 o\penalty5000 s\penalty5000 S\penalty5000 e\penalty5000 c\penalty5000 tor}.
\begin{center}
\begingroup%
\setlength{\newtblsparewidth}{1\linewidth-2\tabcolsep-2\tabcolsep-2\tabcolsep-2\tabcolsep}%
\setlength{\newtblstarfactor}{\newtblsparewidth / \real{424}}%

\begin{longtable}{lll}\caption[{\emph{StBadRdosDb} database file for 11 GeV recorded during run 10.}]{\emph{StBadRdosDb} database file for 11 GeV recorded during run 10.\label{stbadrdos_tab_dbfile}\hyperlabel{stbadrdos_tab_dbfile}%
}\tabularnewline
\endfirsthead
\caption[]{(continued)}\tabularnewline
\endhead
\hline
\multicolumn{1}{m{35\newtblstarfactor+\arrayrulewidth}}{\centering%
RunID
}&\multicolumn{1}{m{106\newtblstarfactor+\arrayrulewidth}}{\centering%
\# bad RDOs
}&\multicolumn{1}{m{283\newtblstarfactor+\arrayrulewidth}}{\raggedright%
RDO identifiers (\texttt{sec\penalty5000 tor.\penalty0 id})
}\tabularnewline
\multicolumn{1}{m{35\newtblstarfactor+\arrayrulewidth}}{\centering%
11148001
}&\multicolumn{1}{m{106\newtblstarfactor+\arrayrulewidth}}{\centering%
3
}&\multicolumn{1}{m{283\newtblstarfactor+\arrayrulewidth}}{\raggedright%
17.01 23.02 23.08
}\tabularnewline
\multicolumn{1}{m{35\newtblstarfactor+\arrayrulewidth}}{\centering%
11151010
}&\multicolumn{1}{m{106\newtblstarfactor+\arrayrulewidth}}{\centering%
4
}&\multicolumn{1}{m{283\newtblstarfactor+\arrayrulewidth}}{\raggedright%
11.10 17.01 23.02 23.08
}\tabularnewline
\multicolumn{1}{m{35\newtblstarfactor+\arrayrulewidth}}{\centering%
11151028
}&\multicolumn{1}{m{106\newtblstarfactor+\arrayrulewidth}}{\centering%
3
}&\multicolumn{1}{m{283\newtblstarfactor+\arrayrulewidth}}{\raggedright%
17.01 23.02 23.08
}\tabularnewline
\multicolumn{1}{m{35\newtblstarfactor+\arrayrulewidth}}{\centering%
11151040
}&\multicolumn{1}{m{106\newtblstarfactor+\arrayrulewidth}}{\centering%
4
}&\multicolumn{1}{m{283\newtblstarfactor+\arrayrulewidth}}{\raggedright%
17.01 21.11 23.02 23.08
}\tabularnewline
\multicolumn{1}{m{35\newtblstarfactor+\arrayrulewidth}}{\centering%
11151080
}&\multicolumn{1}{m{106\newtblstarfactor+\arrayrulewidth}}{\centering%
5
}&\multicolumn{1}{m{283\newtblstarfactor+\arrayrulewidth}}{\raggedright%
01.07 17.01 21.11 23.02 23.08
}\tabularnewline
\multicolumn{1}{m{35\newtblstarfactor+\arrayrulewidth}}{\centering%
11152053
}&\multicolumn{1}{m{106\newtblstarfactor+\arrayrulewidth}}{\centering%
6
}&\multicolumn{1}{m{283\newtblstarfactor+\arrayrulewidth}}{\raggedright%
01.07 07.05 17.01 21.11 23.02 23.08
}\tabularnewline
\multicolumn{1}{m{35\newtblstarfactor+\arrayrulewidth}}{\centering%
11153012
}&\multicolumn{1}{m{106\newtblstarfactor+\arrayrulewidth}}{\centering%
7
}&\multicolumn{1}{m{283\newtblstarfactor+\arrayrulewidth}}{\raggedright%
01.07 01.09 07.05 17.01 21.11 23.02 23.08
}\tabularnewline
\multicolumn{1}{m{35\newtblstarfactor+\arrayrulewidth}}{\centering%
11156032
}&\multicolumn{1}{m{106\newtblstarfactor+\arrayrulewidth}}{\centering%
8
}&\multicolumn{1}{m{283\newtblstarfactor+\arrayrulewidth}}{\raggedright%
01.07 01.09 07.05 09.03 17.01 21.11 23.02 23.08
}\tabularnewline
\multicolumn{1}{m{35\newtblstarfactor+\arrayrulewidth}}{\centering%
11158011
}&\multicolumn{1}{m{106\newtblstarfactor+\arrayrulewidth}}{\centering%
9
}&\multicolumn{1}{m{283\newtblstarfactor+\arrayrulewidth}}{\raggedright%
01.07 01.09 07.05 09.03 11.01 17.01 21.11 23.02 23.08
}\tabularnewline
\hline
\end{longtable}\endgroup%

\end{center}

\section{StRunIdEventsDb: Database Interface for Run Id Indexing}
\label{strunideventsdb}\hyperlabel{strunideventsdb}%

\emph{StRunIdEventsDb} has been developed to facilitate handling the run-{}dependent
studies for the multiple beam energies analyzed in Part~\ref{physics} of this
thesis. The class implementation provides a mapping from a specific run id to
its corresponding index for all energies of STAR's beam energy scan during runs
10 \& 11. Moreover, the number of events can be retrieved on a run-{}by-{}run basis
for each trigger at all energies.\newline

Analogous to Section~\ref{stbadrdos} the perl script \texttt{gen\penalty5000 R\penalty5000 u\penalty5000 n\penalty5000 I\penalty5000 d\penalty5000 E\penalty5000 v\penalty5000 e\penalty5000 n\penalty5000 t\penalty5000 sDb.\penalty0 pl} is used to
generate the database files.  The following literal block -{} which also serves
as configuration file for the \emph{StRunIdEventsDb} class -{} lists all triggers for
a specific combination of energy and year.  Figure~\ref{strunidevts_db} shows database
excerpts for 39 and 62 GeV with index, run id, and corresponding number of
events per trigger. The full database and the code are available in the
author's code repository [218].

\noindent
\begin{description}
\item[{ database/config
}] ~
\begin{lstlisting}[language=bash,firstnumber=1,]
7.7 2010 2 mb total
11.5 2010 2 mb total
19.6 2011 6 mb1-fast HLT-tracks vpd-mon bbc-small-mon-narrow zdc-mon-tac total
27. 2011 5 mb1-fast vpd-mon-tac bbc-small-mon-narrow zdc-mon-tac total
39. 2010 4 mb mbslow ht-11 total
62.4 2010 5 min-bias min-bias-slow ht_11_mb central total
200. 2010 7 Central vpd-mb NPE_18 NPE_15 NPE_11 NPEHT_25 total
200. 2011 6 vpd-zdc-mb-protected NPE_18 NPE_15 NPE_11 NPE_18_ftp total
\end{lstlisting}
\end{description}

\wrapifneeded{0.50}{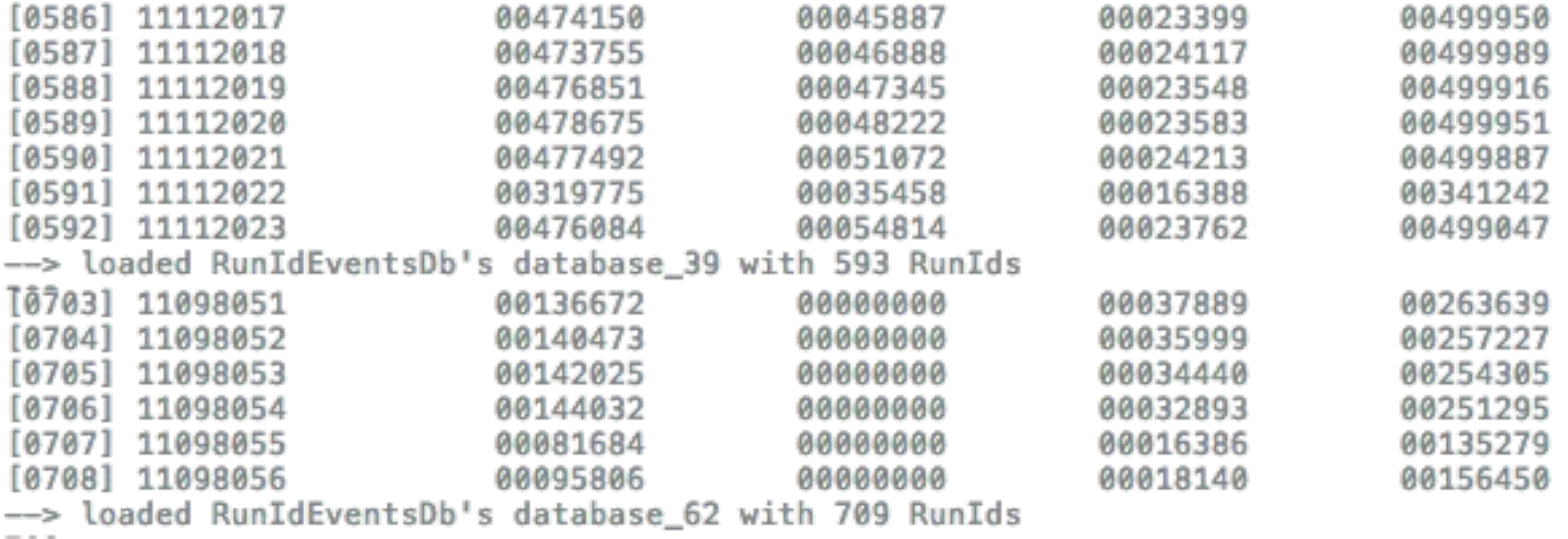}{database excerpts for 39 and 62 GeV}{strunidevts_db}{0.7} %

\emph{StRunIdEventsDb} has been implemented as a singleton class to avoid repeated
loading of the database into memory. Hence, the constructors are declared
private and a public member function is added which decides on returning a
pointer to a newly created instance of \emph{StRunIdEventsDb} or to an already
existing one. To assure singleton behavior this pointer is declared a private
static member. It therefore suffices to initialize the class and load the
database via

\begin{lstlisting}[language=c++,firstnumber=1,]
StRunIdEventsDb* db = StRunIdEventsDb::Instance(mEnergy, mYear);
\end{lstlisting}

The following class member functions are implemented to communicate with the database.

\noindent
\begin{description}
\item[{ \texttt{Int\penalty5000 \_\penalty5000 t g\penalty5000 e\penalty5000 t\penalty5000 T\penalty5000 o\penalty5000 t\penalty5000 a\penalty5000 l\penalty5000 N\penalty5000 r\penalty5000 R\penalty5000 u\penalty5000 n\penalty5000 I\penalty5000 d\penalty5000 s\penalty5000 (\penalty5000 ) \{ r\penalty5000 e\penalty5000 t\penalty5000 u\penalty5000 r\penalty5000 n m\penalty5000 N\penalty5000 r\penalty5000 R\penalty5000 u\penalty5000 n\penalty5000 Ids;\penalty0  \}} }] \hspace{0em}\\
 get the total number of run id's

\item[{ \texttt{Int\penalty5000 \_\penalty5000 t g\penalty5000 e\penalty5000 t\penalty5000 R\penalty5000 u\penalty5000 n\penalty5000 I\penalty5000 d\penalty5000 I\penalty5000 n\penalty5000 d\penalty5000 e\penalty5000 x\penalty5000 (\penalty5000 c\penalty5000 o\penalty5000 n\penalty5000 s\penalty5000 t I\penalty5000 n\penalty5000 t\penalty5000 \_\penalty5000 t\penalty5000 \& r\penalty5000 u\penalty5000 n\penalty5000 id);\penalty0 } }] \hspace{0em}\\
 get the index for a specific run id

\item[{ \texttt{Int\penalty5000 \_\penalty5000 t g\penalty5000 e\penalty5000 t\penalty5000 R\penalty5000 u\penalty5000 n\penalty5000 I\penalty5000 d\penalty5000 (\penalty5000 I\penalty5000 n\penalty5000 t\penalty5000 \_\penalty5000 t i\penalty5000 nd);\penalty0 } }] \hspace{0em}\\
 get the run id for a specific index

\item[{ \texttt{Int\penalty5000 \_\penalty5000 t g\penalty5000 e\penalty5000 t\penalty5000 N\penalty5000 r\penalty5000 E\penalty5000 v\penalty5000 e\penalty5000 n\penalty5000 t\penalty5000 s\penalty5000 (\penalty5000 s\penalty5000 t\penalty5000 r\penalty5000 i\penalty5000 n\penalty5000 g t\penalty5000 n\penalty5000 , I\penalty5000 n\penalty5000 t\penalty5000 \_\penalty5000 t r\penalty5000 u\penalty5000 n\penalty5000 id);\penalty0 } }] \hspace{0em}\\
 get the number of events for a specific runid and trigger name

\item[{ \texttt{voi\penalty5000 d p\penalty5000 r\penalty5000 i\penalty5000 n\penalty5000 t\penalty5000 A\penalty5000 l\penalty5000 l\penalty5000 T\penalty5000 r\penalty5000 i\penalty5000 g\penalty5000 N\penalty5000 a\penalty5000 m\penalty5000 e\penalty5000 s();\penalty0 } }] \hspace{0em}\\
 print all trigger names available for this energy and year

\item[{ \texttt{str\penalty5000 i\penalty5000 n\penalty5000 g g\penalty5000 e\penalty5000 t\penalty5000 T\penalty5000 r\penalty5000 i\penalty5000 g\penalty5000 N\penalty5000 a\penalty5000 m\penalty5000 e\penalty5000 (\penalty5000 I\penalty5000 n\penalty5000 t\penalty5000 \_\penalty5000 t n);\penalty0 } }] \hspace{0em}\\
 get a single trigger name according to the index of \texttt{pri\penalty5000 n\penalty5000 t\penalty5000 A\penalty5000 l\penalty5000 l\penalty5000 T\penalty5000 r\penalty5000 i\penalty5000 g\penalty5000 N\penalty5000 a\penalty5000 m\penalty5000 e\penalty5000 s()} 
\end{description}


\chapter{Transition of STAR CVS to a modern Revision Control System}
\label{cvs2git}\hyperlabel{cvs2git}%

The use of a revision control system to track the changes of its code basis is
essential to the progress of a software development project. STAR is in no way
different to other large technology companies in that many submodules exist for
STAR's detector subsystems with different people responsible for each of them.
Regarding one of the largest experiments in heavy ion physics, however, there
is progress made not only in terms of pure software improvements for data
analyses but also an ever increasing amount of submodules needs to be added
each year accompanying STAR's detector upgrades. Still, along with the data
itself, code used to process the data taken during previous beam times needs to
be available to ensure reproducibility of published results and possible
reproduction of DSTs when new knowledge is acquired or bugs identified. In
addition, running experiments like STAR involves a myriad of custom-{}built tools
needed for tasks such as data acquisition and backup, online monitoring and
analysis, database organization and interfaces, detector alignment and
calibration, track reconstruction and detector matching during data production,
Monte Carlo simulations, ticket/issue submission, documentation, and
publications. Hence, STAR's code organization closely resembles the development
structure and release cycles of other large software products with hierarchical
work flows and a policy which requires all code to be made accessible to all
collaborators. On top, most of the data analysis in STAR is conducted on
multiple remote computing sites at all of which identical code deployments have
to be guaranteed despite differing development environments.

All this requires maintaining a centralized revision control system like the
\emph{Concurrent Versions System} (CVS) [219] which currently is STAR's system
of choice.  Since its beginnings, STAR has collected an immense amount of code
and developed a workflow heavily centered around CVS with many tweaks to
customize it. Even though other centralized successors (Subversion, for
instance) have been released to overcome some of CVS's shortcomings, STAR's
very active data taking and publication cycle have not allowed for a transition
to newer systems.  However, with the rise of distributed revision control
systems like \emph{git} [220] for instance, the many advantages of modern
revision control systems do seem to make a re-{}evaluation of STAR's current
system necessary. Some of these advantages include instant and cheap creation
and deletion of branches, branch merging and rebasing, working without
connection to the remote (central) repository, significantly compressed
repository sizes, fast repository cloning, temporary stashing of changes as
well as selective staging and committing.

There exist many distributed revision control systems but especially \emph{git} has
gained traction and became very popular through the immersion of social coding
and with it an increased number of open source contributions. Judging by the
popularity of it, \emph{git} seems to be best suited to support a developer's
workflow as it encourages frequent commits and work on parallel feature
branches. Thus, likewise in STAR, collaborators seem to prefer the use of \emph{git} for their private analysis code development which starts to increasingly
collide with or at least inhibit the contribution to STAR's central CVS
repository. In general, due to the expensive nature of branching in CVS most of
the code developed by the physicists in STAR seems to stay untracked in their
home directories and unshared through the designated user areas in STAR's CVS.
It therefore seems to be time for an overhaul of STAR's revision control system
and a transition to one of the modern distributed systems.

The following two requirements need to be kept in mind, though. First, STAR's
policy requires \emph{git} as a distributed revision control system to be used in a
centralized but modular manner. A spread of the entire STAR software with each
checkout through many home directories is unfavorable. Second, concerns about
manpower almost dictate the use of already available and established open
source solutions particularly because the transition would have to happen
during ongoing operations. At the same time, these solutions should have a good
chance of long-{}term support or be already stable enough for STAR not to depend
on their fate.

This chapter will not discuss the merits of one revision control system over
the other but give a specific suggestion on how a transition could be carried
out with the best chance of going through smoothly. It is assumed that \emph{git} would become the new revision control system of choice without discussing its
advantages or disadvantages when compared to other distributed systems.  This
chapter concentrates on the execution of a transition from CVS to \emph{git} by
demonstrating that history conversion of all STAR CVS submodules is feasible
via \emph{cvs2git} and by introducing \emph{gerrit} as the new review system for future
code changes and gateway for all sorts of interactions with the central
repository. This includes providing the necessary scripts and instructions as
well as deploying an instance of the review system on a remote server for
demonstration purposes. \emph{cgit} is proposed as a replacement for \emph{cvsweb} as web
service to browse the repository.

Section~\ref{cvs2git_solution} motivates and explains the proposed solution and
Section~\ref{cvs2git_requirements} discusses a list of requirements in the framework of
the proposed solution. In Section~\ref{cvs2git_migration} \emph{gerrit} is set up on a remote
server with \emph{cgit} as external browser, STAR's CVS is converted into multiple
git repositories, synchronized to the remote server and added to \emph{gerrit}.
Finally, Section~\ref{cvs2git_howto} gives instructions on how to interact with the
demonstration deployment as an end user. Note that this chapter is of special
importance to STAR as it introduces a solution for modernizing its revision
control system. It will serve as documentation to accompany the migration to
the new system and provide references to documentation to ease the user's
transition from CVS to \emph{git}.

\emph{Introductory Comment on Git for CVS Users}. Git is commit based with all the
rest being almost exclusively pointers. Each commit has a parent, and each
branch is based on a commit. That's why git is light in committing and
branching. A tag in git, for instance, is only a pointer to a single commit and
as such a direct connection to go back to a certain point in time. In general,
git is fundamentally different from CVS in that its revisions \textbf{are} snapshots
of the directory at a given point in time unlike CVS where everything is file
based. Git for instance, records file renames and moves perfectly fine.

\section{Proposed Solution: Git + Repo + Gerrit + CGit}
\label{cvs2git_solution}\hyperlabel{cvs2git_solution}%

Due to the distributed nature of \emph{git} the concept of submodules is not
equivalent to CVS. Submodules in CVS can be checked out and committed to
separately while the minimum \emph{unit} for a checkout is defined by one \emph{git} repository.  Submodules in \emph{git} [221] are generally used for
foreign libraries which evolve entirely separately but on which the software
itself depends on. They are not checked out together with a \emph{git} repository
but need to be initialized and synchronized separately. The day-{}to-{}day usage of
\emph{git} submodules has some serious trip falls and is rather cumbersome which is
why the community tends to avoid them.  Hence, the "imitation" of a central CVS
repository with \emph{git} is better realized using independent \emph{git} sub-{}repositories whose interplay is controlled via a wrapper script and a
configuration file. This way multiple sub-{}repositories can be checked out and
worked on simultaneously while global tagging and branching is maintained by
the wrapper script.

Google's Android Open Source Project (AOSP) has this issue already resolved and
developed such a wrapper script in python called \emph{repo} [222]. It uses a
so-{}called manifest configuration file in xml format to organize the structural
layout of the project and the remote location of the repositories as well as
coordinated checkouts of specific versions. It does not replace \emph{git} but
operates "on top of git" [223]. AOSP's workflow is directly
applicable to STAR's as it is general enough to serve as a template for any
major collaborative software development project. AOSP's workflow using \emph{repo} is very well documented [224] and with its millions of lines of code
well-{}proven in an heavy software development environment. These are good
credentials for STAR to use it and not to fear its near-{}term obsolescence.

The same counts for \emph{gerrit} [225], a web-{}based code review tool which
controls all the interaction with the single \emph{git} repositories including
amongst others merging patches into the master branch.
\emph{Gerrit} achieves this control mechanism by running ssh and http daemons on two
different non-{}standard ports. Not only does it provide a stream-{}lined,
documented and self-{}sustained code review, its configuration also covers many
aspects of centralized collaboration code development.  For instance, it
provides automated builds via the Jenkins trigger [226] and code
changes will only be accepted and merged if they pass the build process and
according (unit) tests.  It includes email notifications, sign-{}offs etc. and
even a REST API [227].

The standard repository browser \emph{gitweb} comes with every \emph{git} installation by
default and can be configured to be used in \emph{gerrit}. It is suggested, however,
to use the \emph{cgit} browser [228] instead which is more feature-{}rich and
easily exchangeable with \emph{gitweb} in \emph{gerrit}.  \emph{cgit} is written in C and
includes caching if desired, both of which make it extremely fast and very
responsive. \emph{cgit} is straight-{}forward to set up and its development is very
active. In March 2013 it has been selected by the Linux Kernel as its primary
repository browser [229, 230].

In a nutshell, \emph{git + repo + gerrit + cgit} gives STAR the best chances to
succeed in a CVS to Git migration while significantly modernizing its workflow
and code review.

\section{STAR specific Requirements and Requests}
\label{cvs2git_requirements}\hyperlabel{cvs2git_requirements}%

In the following a list of requirements and requests which are specific to STAR
is discussed. These items have been identified as important questions to STAR
through discussion with the project leader of STAR's software \& computing
group, Dr. Jerome Lauret (\href{mailto:jlauret@bnl.gov}{jlauret@bnl.gov}).

\noindent
\begin{description}
\item[{ Centralized Workflow with Feature and Development Branches
}] \hspace{0em}\\
 In the past, the lack of convenient use of feature and temporary branches in
CVS has caused users to maintain copies of entire CVS submodules of \texttt{StR\penalty5000 oot/\penalty0 } in their designated user area at \texttt{off\penalty5000 l\penalty5000 ine/\penalty0 users/\penalty0 }. The user areas do not serve
the purpose of maintaining official STAR code until it is ready to be committed
into the main branch. They exist as an area for users to share code which is
not STAR official but still of interest for the entire collaboration. Thus, the
new system has to enable the use of development and feature branches for the
user to work on while still within the central repository. Also, the new
platform has to take care of the organization of these parallel branches in
terms of sharing, reviewing and merging with the main branch and core
development.\newline
 These concerns and appearances of "offline-{}branching" and
inconvenient parallel development are what caused the development of \emph{git} in
the first place.  Together with \emph{repo} and \emph{gerrit} as tools to organize the
interplay of many git repositories as well as the parallel coding of multiple
developers, the new system proposed in Section~\ref{cvs2git_solution} covers all possible
aspects of such a development environment in a very convenient and well
documented manner. See the quick intro of \emph{gerrit} for a standard "life of a
patch" [225].\newline
 Moreover, even though \emph{git} is of distributed nature all code committing must
be encouraged to go to a central repository. This principle is enforced by STAR
rules and must not be violated when switching to a new revision control system.
Here again, the usage of many git repositories as a joint central entity is the
main purpose for the development of \emph{gerrit} by AOSP as outlined in
Section~\ref{cvs2git_solution}. \emph{Gerrit} serves as the sole gateway to the repositories
and while it is possible to push directly to the git repositories behind
\emph{gerrit} and thus entirely circumvent the review, this is not desired and can
be switched off completely.

\item[{ Authentication and Authorization
}] \hspace{0em}\\
 In contrary to a publicly accessible open source project such as AOSP, strict
authentication and authorization mechanisms have to be enforced in STAR. These
methods have to be consistent with BNL policies. For instance, users without
valid RCF account are not to be allowed write access to repositories served by
the new platform.\newline
 In its default setup \emph{gerrit} manages users and its logins
via the OpenID mechanism. Internally users are organized in groups with
specific access rights globally or on a per project basis. \emph{Gerrit} provides
ACL authorization like CVS does but its capabilities allow very fine grained
and extensive control of the repositories and its branches (see
[225]). For authentication already existing kerberos accounts
can be used instead of OpenID. This is not in the scope of the demonstration deployment of
this chapter as it would only be possible to implement on a dedicated server at
BNL with access to \texttt{krb5.\penalty0 key\penalty5000 tab} [231]. This needs to be
explored after the user's reception of the new system has been reviewed. Hence
for simplicity, the demonstration deployment of the \emph{gerrit} instance in this chapter
will only be based on a single user login (\emph{protected}) and multiple associated
e-{}mail addresses and ssh-{}keys (see Section~\ref{cvs2git_howto}).\newline
 Two aspects of authentication and authorization will have to be evaluated in
the future. First, having an automated mechanism to port the current CVS access
rules (karma) into \emph{gerrit} would be favorable to preserve access rights to
repositories and branches. Second, CVS currently does not have truely public
areas which are not subject to STAR's access control. STAR might profit if such
areas would be realizable with the new platform enabling STAR collaborators to
participate in community work.\newline
 \emph{gitolite} [232] is an alternative
AA system to setup access controlled hosting of \emph{git} repositories. Access on per
project and per branch basis is configured via a dedicated \emph{gitolite-{}admin} repository with a central configuration file and public ssh-{}keys. It integrates very
well with \emph{cgit}, too, and \emph{repo} can still be used to organize multiple \emph{git} repositories at once. Without a \emph{gerrit} instance, however, \emph{repo upload} is
redundant, of course. \emph{Gitolite} is a very good option for the hosting of \emph{git} repositories with a small number of developers who organize the control of a
small amount of development and feature branches themselves and do not need
automated code review nor continuous build integration. Thus, to secure the
long-{}term stability of code organization and review in STAR \emph{gerrit} seems to
be better suited.

\item[{ Demonstration Deployment and Trial Phase
}] \hspace{0em}\\
 For the final migration to \emph{git} and the associated deployment of the \emph{gerrit} platform a phase of daily incremental conversion of CVS to \emph{git} history would
not necessarily be required. During a trial phase with a \emph{gerrit} demonstration
deployment\hspace{0.167em}\textemdash{}\hspace{0.167em}as is the goal of this chapter\hspace{0.167em}\textemdash{}\hspace{0.167em}CVS and \emph{git} could run in
parallel and even diverge. The demonstration deployment of this chapter will already put
in place a full conversion of STAR's CVS repository for users to play with. In
addition, it shall encourage the involvement of the people responsible for each
repository to thoroughly check their converted history. This should help in
identifying failed conversions of single commits as well as possible merging or
splitting of historically linked CVS submodules into \emph{git} repositories.
Following this feedback, these requests can be scripted into the automated
conversion (see Section~\ref{cvs2git_migration}) before a date is scheduled for the final
migration at which the STAR's CVS repository is retired for good.

\item[{ Repository Browser
}] \hspace{0em}\\
 The core of STAR's developer team is long-{}time used to the \emph{cvsweb} experience
of browsing code and its history. The proposed repository browser \emph{cgit} already provides a large set of features and comes very close to the conception
of browsing code in STAR. Contrary to STAR's CVS, the new system will not have
the entire STAR CVS repository and all its submodules mangled into one big
\emph{git} repository but split them up into many smaller ones organized by the
\emph{repo} script. Thus, the \emph{cgit} browser will serve these repositories as
independent unrelated entities. It allows, however, to organize the
repositories into categories which significantly facilitates the browsing.
Pagination can be switched off such that the general layout of the \emph{cgit} browser will look like the Linux Kernel's [230]. \emph{CGit} further
features sorting the repository listing on its landing page based on name,
description, owner and idle time. Together with the very fast search
function, the list of repositories can quickly be reduced to the ones of
interest to the user.\newline
 Still, \emph{cgit} is an open source project and thus helpful features like
"Expand/Collapse Categories" can be suggested to the community or contributed.
As for the representation of a repository itself in \emph{cgit}, functionalities
present in \emph{cvsweb} are mostly implemented in \emph{cgit} [233], too,
but will need some getting used to on the user's side.\newline
 In the author's opinion only \emph{GitLab} [234] would be a serious
alternative candidate to use as repository browser. Rather than only being a
repository browser, however, it aims at being an open source alternative to the
popular \emph{GitHub} social coding platform [235]. Hence, it does not fit
STAR's requirements to interface with \emph{gerrit} as well as \emph{cgit} does.

\end{description}

\section{Migration from CVS to new System}
\label{cvs2git_migration}\hyperlabel{cvs2git_migration}%

This section documents in detail the steps to be taken to first set up a gerrit
instance on a remote server with cgit as an external repository browser
(Section~\ref{cvs2git_webserver}) and then convert the CVS submodules StRoot, kumacs,
pams, StDb, StarDb, asps, QtRoot, StarVMC and mgr into git repositories
(Section~\ref{cvs2git_bnl}). The converted repositories will also be added to the gerrit
review system.

The demonstration deployment of gerrit in this section will only rely on a simple single
sign-{}on via HTTP authorization. The credentials are chosen to be the common
ones for access to usual STAR protected areas. This allows all STAR
collaborators to interact with the system with administrator priviledges given
that they add their ssh-{}key to the \emph{STAR Protected} user on gerrit as described
below. However, it still restricts read access to STAR's code to its
collaborators only. This is by design such that all collaborators contributing to
the STAR code repository can try out the new system before a production
deployment with fine-{}grained access control is started.

The gerrit instance which will be set up in the following will also be used to
serve all scripts and configuration files necessary for the deployment in a
dedicated repository named \emph{git4star}. Thus on all sites and for all users
interacting with the gerrit review system a dedicated ssh-{}key pair has to be
generated and its public key uploaded to the \emph{STAR Protected} user on gerrit
\footnote{
\href{http://gerrit4star.the-huck.com/\#/settings/ssh-keys}{http://gerrit4star.the-{}huck.com/\-\#/\-settings/\-ssh-{}keys}
}. This needs to
be done for end users uploading changes as well as admins converting CVS repos
and creating projects. The source codes for the shell scripts are also appended
to this thesis in Section~\ref{cvs2git_source}.

Last, the following entry needs to be added for the
\emph{gerrit4star} host to \texttt{\textasciitilde{}/\penalty0 .\penalty0 ssh/\penalty0 con\penalty5000 fig}. After that connecting to gerrit via \texttt{ssh
ger\penalty5000 r\penalty5000 i\penalty5000 t\penalty5000 4\penalty5000 s\penalty5000 tar} should present itself with a welcome message and exit. All the
subsections in this section assume that the reader can clone the \emph{git4star} repository via \texttt{git c\penalty5000 l\penalty5000 o\penalty5000 n\penalty5000 e g\penalty5000 e\penalty5000 r\penalty5000 r\penalty5000 i\penalty5000 t\penalty5000 4\penalty5000 s\penalty5000 tar:\penalty0 git\penalty5000 4\penalty5000 s\penalty5000 tar} on all sites.

\begin{lstlisting}[language=bash,firstnumber=1,]
Host=gerrit4star
Hostname=gerrit4star.the-huck.com
User=<replace-with-STAR-Protected-username>
Port=29418
IdentityFile=~/.ssh/<replace-with-ssh-key-name>
IdentitiesOnly=yes
\end{lstlisting}

\subsection{On the Web Server}
\label{cvs2git_webserver}\hyperlabel{cvs2git_webserver}%

\noindent
\begin{description}
\item[{ Assumptions
}] \hspace{0em}\\
 Gerrit runs under a dedicated \emph{gerrit4star} user at the domain
\href{http://gerrit4star.the-huck.com}{http://gerrit4star.the-{}huck.com} with the external cgit browser at
\href{http://cgit4star.the-huck.com}{http://cgit4star.the-{}huck.com}.

\item[{ Apache Configuration
}] \hspace{0em}\\
 As super-{}user, setup a name-{}based virtual host to serve
cgit at \href{http://cgit4star.the-huck.com}{http://cgit4star.the-{}huck.com} [236] and second one with
reverse proxy and HTTP authentication for the gerrit review system at
\href{http://gerrit4star.the-huck.com}{http://gerrit4star.the-{}huck.com} (see [225]). Restrict
access to both by adding the \emph{STAR Protected} user to \texttt{htp\penalty5000 a\penalty5000 s\penalty5000 swd} and including
the appropiate section in the apache configuration. This user will later be
used as single sign-{}on and thus administrator login to gerrit as well as for
all repository interaction through ssh (see above's \texttt{\textasciitilde{}/\penalty0 .\penalty0 ssh/\penalty0 con\penalty5000 fig}). Example
\texttt{cgi\penalty5000 t\penalty5000 4\penalty5000 s\penalty5000 tar.\penalty0 the-{}\penalty0 huck.\penalty0 com} and \texttt{ger\penalty5000 r\penalty5000 i\penalty5000 t\penalty5000 4\penalty5000 s\penalty5000 tar.\penalty0 the-{}\penalty0 huck.\penalty0 com} apache configuration
files are available in the \texttt{git\penalty5000 4\penalty5000 s\penalty5000 tar} script repository (see following step).

\item[{ Get \emph{git4star} script repository
}] \hspace{0em}\\
 As \emph{gerrit4star} user, clone the \emph{git4star} script directory:\newline
 \texttt{git c\penalty5000 l\penalty5000 o\penalty5000 n\penalty5000 e g\penalty5000 e\penalty5000 r\penalty5000 r\penalty5000 i\penalty5000 t\penalty5000 4\penalty5000 s\penalty5000 tar:\penalty0 git\penalty5000 4\penalty5000 s\penalty5000 t\penalty5000 a\penalty5000 r \$\penalty5000 H\penalty5000 OME/\penalty0 git\penalty5000 4\penalty5000 s\penalty5000 tar} 
\item[{ Setup cgit browser
}] \hspace{0em}\\
 The following will clone, init, config and install cgit
[237]. It will configure the \texttt{cgit.\penalty0 cgi} script to be installed in
\texttt{\$CG\penalty5000 I\penalty5000 T\penalty5000 \_\penalty5000 S\penalty5000 C\penalty5000 R\penalty5000 I\penalty5000 P\penalty5000 T\penalty5000 \_\penalty5000 P\penalty5000 ATH} and anything else in \texttt{\$pr\penalty5000 e\penalty5000 fix}. Adjust both in \texttt{cgit.\penalty0 conf} to reflect your web server setup.

\begin{lstlisting}[language=bash,firstnumber=1,]
git clone git://git.zx2c4.com/cgit $HOME/cgit
cd cgit && git submodule init && git submodule update
# copy cgit config file
cp $HOME/git4star/cgit.conf $HOME/cgit/
make -j4
# switch user if 'gerrit4star' doesn't have sudo perms
# go back to 'gerrit4star' user afterwards
sudo make install
# copy cgitrc to $CGIT_CONFIG (see cgit.conf)
cp $HOME/git4star/cgitrc $prefix/etc/cgitrc
# go to http://cgit4star.the-huck.com to see whether cgit runs
# ignore "no repositories found" message
\end{lstlisting}
\item[{ Install gerrit
}] \hspace{0em}\\
 The code below downloads, installs and configures the HTTP
proxy for gerrit [225]. The \texttt{con\penalty5000 f\penalty5000 i\penalty5000 g\penalty5000 u\penalty5000 r\penalty5000 e\penalty5000 H\penalty5000 t\penalty5000 t\penalty5000 p\penalty5000 P\penalty5000 r\penalty5000 oxy.\penalty0 sh} script also
starts the ssh and http daemons. \href{http://cgit4star.the-huck.com}{http://cgit4star.the-{}huck.com} should now list
an \texttt{All\penalty5000 P\penalty5000 r\penalty5000 o\penalty5000 j\penalty5000 e\penalty5000 cts.\penalty0 git} repository which appears to be empty and visiting
\href{http://gerrit4star.the-huck.com}{http://gerrit4star.the-{}huck.com} should now show the gerrit welcome page after
logging in using the \emph{STAR protected} credentials. Fill out "Full Name" as well
as optionally register an e-{}mail and add a ssh-{}key.

\begin{lstlisting}[language=bash,firstnumber=1,]
cd $HOME # for gerrit4star user
wget https://gerrit-releases.storage.googleapis.com/gerrit-2.6.1.war
# the following batch install will raise a FAILED notice but that's ok
java -jar gerrit-2.6.1.war init --batch -d $HOME/gerrit_review
~/git4star/configureHttpProxy.sh
\end{lstlisting}
\item[{ Configure CGit as external browser
}] \hspace{0em}\\
 Gerrit will still show links to its
internally handle gitweb browser. Running \texttt{\$HOME/\penalty0 git\penalty5000 4\penalty5000 s\penalty5000 tar/\penalty0 con\penalty5000 f\penalty5000 i\penalty5000 g\penalty5000 u\penalty5000 r\penalty5000 e\penalty5000 C\penalty5000 Git.\penalty0 sh} configures cgit as external browser instead of gitweb. For the final
\texttt{\$HOME/\penalty0 ger\penalty5000 r\penalty5000 i\penalty5000 t\penalty5000 \_\penalty5000 r\penalty5000 e\penalty5000 v\penalty5000 iew/\penalty0 etc/\penalty0 ger\penalty5000 rit.\penalty0 con\penalty5000 fig} see the \texttt{\$HOME/\penalty0 git\penalty5000 4\penalty5000 s\penalty5000 tar/\penalty0 ger\penalty5000 rit.\penalty0 con\penalty5000 fig} file in the \emph{git4star} script repository. Check the \emph{cgit} links for
\texttt{All\penalty5000 P\penalty5000 r\penalty5000 o\penalty5000 j\penalty5000 e\penalty5000 cts.\penalty0 git} in gerrit after configuration to confirm a correct
configuration.

\item[{ Final Step
}] \hspace{0em}\\
 Set additional access rights "Create Reference" and "Push Merge
Commit" for administrators under \texttt{refs/\penalty0 *} at
\href{http://gerrit4star.the-huck.com/#/admin/projects/All-Projects,access}{http://gerrit4star.the-{}huck.com/\-\#/\-admin/\-projects/\-All-{}Projects,access}. This is
later needed to push the newly converted CVS repositories directly into gerrit.

\end{description}

\subsection{In a Working Directory at BNL with CVSROOT Access}
\label{cvs2git_bnl}\hyperlabel{cvs2git_bnl}%

\noindent
\begin{description}
\item[{ Assumptions
}] \hspace{0em}\\
 All of the following code will be executed within a dedicated
working directory. Access to the \emph{git4star} script repository is assumed as
well.

\begin{lstlisting}[language=bash,firstnumber=1,]
<WORKDIR> = /star/institutions/lbl/huck/star_cvs2git/
\end{lstlisting}
\item[{ Convert STAR CVS
}] \hspace{0em}\\
 The following code downloads and unpacks \emph{cvs2git} [238]. The CVS submodules StRoot, kumacs, pams, StDb, StarDb, asps,
QtRoot, StarVMC and mgr are then converted to git via the \texttt{con\penalty5000 v\penalty5000 e\penalty5000 r\penalty5000 t\penalty5000 S\penalty5000 t\penalty5000 a\penalty5000 r\penalty5000 CVS.\penalty0 sh} script. Make sure to check the environment variables in this script before
executing it!

\begin{lstlisting}[language=bash,firstnumber=1,]
cd <WORKDIR> && git clone gerrit4star:git4star
wget http://cvs2svn.tigris.org/files/documents/1462/49237/cvs2svn-2.4.0.tar.gz
tar -xvzf cvs2svn-2.4.0.tar.gz
cd git4star
# the following converts all repos and takes about 1.5h!
./convertStarCVS.sh | tee cvs2git.log
\end{lstlisting}

Two remarks on the conversion of CVS to git. First, the conversion of STAR's
CVS repository to git for production deployment of gerrit would require a list
of all CVS committers in the format \emph{username} : \emph{Full Name <{}E-{}Mail>{}} to be fed
into the conversion. This list would map usernames to the format in which git
uses them because each git commit requires the committer to provide an e-{}mail
address, too. This is due to the principle that committers should minimally be
reachable by e-{}mail. Such a list could be either generated by using the
\texttt{fin\penalty5000 ger} command or direct access to a database at BNL providing this
information. Second, a general freely choosable "cvs2git" user including name
and e-{}mail would need to be provided during the conversion. This user gets used
for commits for which CVS does not record the original author. More information
about these details can be found in the \texttt{my.\penalty0 cvs\penalty5000 2\penalty5000 git.\penalty0 opt\penalty5000 i\penalty5000 ons} file in the
\emph{git4star} script repository.
\item[{ Create and upload Gerrit Projects
}] \hspace{0em}\\
 Running \texttt{cre\penalty5000 a\penalty5000 t\penalty5000 e\penalty5000 G\penalty5000 e\penalty5000 r\penalty5000 r\penalty5000 i\penalty5000 t\penalty5000 P\penalty5000 r\penalty5000 o\penalty5000 j\penalty5000 e\penalty5000 cts.\penalty0 sh} in
\texttt{\$wo\penalty5000 r\penalty5000 k\penalty5000 dir/\penalty0 git\penalty5000 4\penalty5000 s\penalty5000 tar} creates all projects in gerrit and pushes the according
repositories to gerrit. The script also adds sign-{}off and change-{}id
requirements to all projects in gerrit.

\item[{ Write Manifest for \emph{repo} }] \hspace{0em}\\
 When executing the \texttt{wri\penalty5000 t\penalty5000 e\penalty5000 M\penalty5000 a\penalty5000 n\penalty5000 i\penalty5000 f\penalty5000 est.\penalty0 sh} script a
default manifest named \texttt{def\penalty5000 a\penalty5000 ult.\penalty0 xml} is generated based on the converted CVS
repositories which can later be used to initialize the STAR repository for the
use with the \emph{repo} script (see Section~\ref{cvs2git_howto}).

\end{description}
\begin{center}
\begingroup%
\setlength{\newtblsparewidth}{0.5\linewidth-2\tabcolsep-2\tabcolsep-2\tabcolsep-2\tabcolsep}%
\setlength{\newtblstarfactor}{\newtblsparewidth / \real{213}}%

\begin{longtable}{lll}\caption[{Overview of repository sizes}]{Overview of repository sizes}\tabularnewline
\endfirsthead
\caption[]{(continued)}\tabularnewline
\endhead
\hline
\multicolumn{1}{m{71\newtblstarfactor+\arrayrulewidth}}{\centering%
}&\multicolumn{1}{m{71\newtblstarfactor+\arrayrulewidth}}{\centering%
git (MB)
}&\multicolumn{1}{m{71\newtblstarfactor+\arrayrulewidth}}{\centering%
CVS (MB)
}\tabularnewline
\multicolumn{1}{m{71\newtblstarfactor+\arrayrulewidth}}{\centering%
StRoot
}&\multicolumn{1}{m{71\newtblstarfactor+\arrayrulewidth}}{\centering%
363
}&\multicolumn{1}{m{71\newtblstarfactor+\arrayrulewidth}}{\centering%
480
}\tabularnewline
\multicolumn{1}{m{71\newtblstarfactor+\arrayrulewidth}}{\centering%
kumacs
}&\multicolumn{1}{m{71\newtblstarfactor+\arrayrulewidth}}{\centering%
14.8
}&\multicolumn{1}{m{71\newtblstarfactor+\arrayrulewidth}}{\centering%
1.74
}\tabularnewline
\multicolumn{1}{m{71\newtblstarfactor+\arrayrulewidth}}{\centering%
pams
}&\multicolumn{1}{m{71\newtblstarfactor+\arrayrulewidth}}{\centering%
37.6
}&\multicolumn{1}{m{71\newtblstarfactor+\arrayrulewidth}}{\centering%
128
}\tabularnewline
\multicolumn{1}{m{71\newtblstarfactor+\arrayrulewidth}}{\centering%
StDb
}&\multicolumn{1}{m{71\newtblstarfactor+\arrayrulewidth}}{\centering%
6.47
}&\multicolumn{1}{m{71\newtblstarfactor+\arrayrulewidth}}{\centering%
4.58
}\tabularnewline
\multicolumn{1}{m{71\newtblstarfactor+\arrayrulewidth}}{\centering%
StarDb
}&\multicolumn{1}{m{71\newtblstarfactor+\arrayrulewidth}}{\centering%
199
}&\multicolumn{1}{m{71\newtblstarfactor+\arrayrulewidth}}{\centering%
576
}\tabularnewline
\multicolumn{1}{m{71\newtblstarfactor+\arrayrulewidth}}{\centering%
asps
}&\multicolumn{1}{m{71\newtblstarfactor+\arrayrulewidth}}{\centering%
16.1
}&\multicolumn{1}{m{71\newtblstarfactor+\arrayrulewidth}}{\centering%
47.5
}\tabularnewline
\multicolumn{1}{m{71\newtblstarfactor+\arrayrulewidth}}{\centering%
QtRoot
}&\multicolumn{1}{m{71\newtblstarfactor+\arrayrulewidth}}{\centering%
16.3
}&\multicolumn{1}{m{71\newtblstarfactor+\arrayrulewidth}}{\centering%
12.5
}\tabularnewline
\multicolumn{1}{m{71\newtblstarfactor+\arrayrulewidth}}{\centering%
StarVMC
}&\multicolumn{1}{m{71\newtblstarfactor+\arrayrulewidth}}{\centering%
47.5
}&\multicolumn{1}{m{71\newtblstarfactor+\arrayrulewidth}}{\centering%
116
}\tabularnewline
\multicolumn{1}{m{71\newtblstarfactor+\arrayrulewidth}}{\centering%
mgr
}&\multicolumn{1}{m{71\newtblstarfactor+\arrayrulewidth}}{\centering%
4.66
}&\multicolumn{1}{m{71\newtblstarfactor+\arrayrulewidth}}{\centering%
5.93
}\tabularnewline
\multicolumn{1}{m{71\newtblstarfactor+\arrayrulewidth}}{\centering%
total
}&\multicolumn{1}{m{71\newtblstarfactor+\arrayrulewidth}}{\centering%
705
}&\multicolumn{1}{m{71\newtblstarfactor+\arrayrulewidth}}{\centering%
1373
}\tabularnewline
\hline
\end{longtable}\endgroup%

\end{center}

\section{End User Instructions for Gerrit Demonstration Deployment}
\label{cvs2git_howto}\hyperlabel{cvs2git_howto}%

\noindent
\begin{description}
\item[{ Setup Gerrit Access
}] \hspace{0em}\\
 Add ssh-{}key to \emph{STAR Protected} user in gerrit as
describe in the introduction part of this section. Run \texttt{ssh g\penalty5000 e\penalty5000 r\penalty5000 r\penalty5000 i\penalty5000 t\penalty5000 4\penalty5000 s\penalty5000 tar} to
check whether you see the welcome message and thus are able to connect to
gerrit correctly. Add your git-{}config e-{}mail address to the \emph{STAR Protected} user at \href{http://gerrit4star.the-huck.com/#/settings/contact}{http://gerrit4star.the-{}huck.com/\-\#/\-settings/\-contact} and confirm it via
the link in the e-{}mail sent to you. Put the following into your \texttt{\textasciitilde{}/\penalty0 .\penalty0 git\penalty5000 c\penalty5000 o\penalty5000 n\penalty5000 fig}.

\begin{lstlisting}[language=bash,firstnumber=1,]
[review "http://gerrit4star.the-huck.com"]
   username = protected
\end{lstlisting}
\item[{ Setup, init and start sync'ing with \emph{repo} }] \hspace{0em}\\
 First install the \emph{repo} script as
described in [239]. A good introduction and overview of the usage
of \emph{repo} and the associated workflow can be found in [240, 241].
To get started, create an empty directory to hold your working files.

\begin{lstlisting}[language=bash,firstnumber=1,]
mkdir MyStarProjects && cd MyStarProjects
\end{lstlisting}

The manifest specifies where the various repositories included in the STAR
repository will be placed within your working directory. Init your working
directory using the \texttt{def\penalty5000 a\penalty5000 ult.\penalty0 xml} in the \emph{git4star} script repository:

\begin{lstlisting}[language=bash,firstnumber=1,]
repo init -u gerrit4star:git4star --config-name
\end{lstlisting}

When prompted, configure \emph{repo} with your full name and the email you added
above to the \emph{STAR Protected} user in gerrit. ALL projects can be pulled down
to your working directory as specified in the default manifest by running a
simple \texttt{rep\penalty5000 o s\penalty5000 ync}. However, this is not desired in most cases. The \emph{repo} script allows for only a list of gerrit projects to be synchronized via

\begin{lstlisting}[language=bash,firstnumber=1,]
repo sync StRoot-StAnalysisMaker StRoot-StHbtMaker
\end{lstlisting}
\item[{ Start Work and Upload Changes for Review
}] \hspace{0em}\\
 The workflow of \emph{repo} together with
gerrit is very well documented in [242, 243]. Start a new development branch in all the repos you sync'ed via

\begin{lstlisting}[language=bash,firstnumber=1,]
repo start <branchname> --all
\end{lstlisting}

Change to the git repository of your choice, start playing around, change files and when you are done, commit everything via

\begin{lstlisting}[language=bash,firstnumber=1,]
git commit -a -s -m "commit message"
\end{lstlisting}

Add your ssh-{}key to the ssh-{}agent (\texttt{ssh-{}\penalty0 add \textasciitilde{}/\penalty0 .\penalty0 ssh/\penalty0 <{}ssh-{}\penalty0 key-{}\penalty0 name>{}}) and upload
your changes to gerrit for review via a simple

\begin{lstlisting}[language=bash,firstnumber=1,]
repo upload
\end{lstlisting}

Check \texttt{rep\penalty5000 o h\penalty5000 elp} for more. Screenshots of Gerrit and CGit for STAR are
appended in Section~\ref{app_cvs2git_screen}.
\end{description}


\chapter{Open Source Contribution of Physics Analysis and Documentation Tools}
\label{_open_source_contribution_of_physics_analysis_and_documentation_tools}\hyperlabel{_open_source_contribution_of_physics_analysis_and_documentation_tools}%

\section{ckon: Automatic Build Tool for ROOT Analyses}
\label{ckon}\hyperlabel{ckon}%

ROOT [163] allows for the automatic compilation of shared libraries using
the ACLiC compiler [244]. This is very convenient for the quick use
of compiled libraries in CINT but the organization of code developed for a
physics analysis project would be cumbersome regarding its size and the
resulting intra-{}project library dependencies. On top, ROOT-{}based shared
libraries are mostly invoked through a macro and not a compiled program. The
latter significantly improves the quality of a software project. Most bugs are
detectable during compilation and not runtime and run control can be extended
through command line options and arguments. This avoids constant recompilation
due to hard-{}coding of I/O variables, for instance, and facilitates the use in
shell scripts. Writing Makefiles for each library and program on the other
hand is a difficult, time-{}consuming and repetitive task. This section is
dedicated to a comprehensive solution of this issue.

On all computing sites STAR provides the user with the perl script \emph{cons} to
automatically compile shared libraries which rely on ROOT as well as StRoot
software. However, \emph{cons} is somewhat outdated and does not provide flexible
configuration options. It covers a standard source code layout in which each
subdirectory of \emph{StRoot} is compiled into a library. For custom configuration
like deeper structures which would need \emph{cons} to recursively scan directories,
for instance, the user needs to modify the \emph{cons} script itself. Moreover, it
is not straight-{}forward to set up on a local machine when StRoot software is
not even needed. The author hence uses Autoconf and its simplified
Makefile-{}"language" to automatically generate Makefiles specific to the
underlying operating system.  Autoconf is well established amongst software
looking to be platform independent and provides GNU's known "configure / make /
make install" workflow including distribution package management ("make clean /
make distclean / make dist").

This is a first step towards an easy-{}to-{}maintain project layout but still
requires the setup of Automake-{}Makefiles with library-{}specific naming
conventions. This task is already stream-{}lined
when compared to manually writing Makefiles but still is error-{}prone and very
repetitive. The latter, however, makes the entire process wrapable into the
dedicated software tool \emph{ckon}. On top of relieving the user from any duty to
write Makefiles at all, this tool makes it easy to include foreign code or
third-{}party libraries as well as the user's own core utility and database
classes based on a simple project-{}specific YAML configuration file. All which
is needed to build an entire project is \texttt{cko\penalty5000 n s\penalty5000 e\penalty5000 t\penalty5000 u\penalty5000 p \&\penalty5000 \& c\penalty5000 kon} with the user
deciding on the source directory's layout.

The following first outlines the usage of GNU's autoconf in conjunction with
ROOT which is based on a talk given by the author at the Junior's Day of the
STAR Collaboration Meeting in November 2011 at LBNL. Find the talk titled "An
example-{}driven introduction to using autoconf/automake (\& Doxygen) for the
organization of C++ data analysis software within the CERN ROOT framework"
published on speakerdeck.com [28]. The automatic build tool
\emph{ckon} is subsequently introduced which was developed by the author to automate
the suggested autoconf-{}based workflow for ROOT. Even though doxygen support is
discussed in the presentation the reader is referred to Section~\ref{wp-pdf} in which the
author opts for Asciidoc to cover documentation of a physics analysis project.\newline

The listing below demonstrates a possible layout for a simple ROOT analysis
using Autoconf. The according source code is published at [245].
For references, see
[246-{}251].

\begin{lstlisting}[language=bash,firstnumber=1,escapeinside={<t>}{</t>},moredelim={**[is][\bfseries]{<b>}{</b>}},moredelim={**[is][\itshape]{<i>}{</i>}},]
MySimpleExampleClass/
    configure.ac <t>\coref{1}{CO1-1}</t>
    Makefile.am <t>\coref{2}{CO1-2}</t>
    src/
      runhSimple.cc <t>\coref{3}{CO1-3}</t>
      hSimple/ <t>\coref{4}{CO1-4}</t>
        hSimple.{h,cxx}
        LinkDef.h
      Options/ <t>\coref{5}{CO1-5}</t>
        Options.{h,cxx}
        LinkDef.h
    bin/runhSimple <t>\coref{6}{CO1-6}</t>
    MySimpleExampleClass-11.11.14.tar.gz <t>\coref{7}{CO1-7}</t>
\end{lstlisting}

\begin{description}
[leftmargin=1cm,style=sameline]
\item[{\hyperref[CO1-1]{\conum{1}}}]
 \texttt{con\penalty5000 f\penalty5000 i\penalty5000 g\penalty5000 ure.\penalty0 ac} required for \texttt{.\penalty0 /\penalty0 con\penalty5000 f\penalty5000 i\penalty5000 g\penalty5000 ure} script

\item[{\hyperref[CO1-2]{\conum{2}}}]
 Automake-{}Makefile for generating the main program \texttt{bin/\penalty0 run\penalty5000 h\penalty5000 S\penalty5000 i\penalty5000 m\penalty5000 ple} 
\item[{\hyperref[CO1-3]{\conum{3}}}]
 source code for main program to control the class behavior.\newline
 decodes options and re-{}initializes argc/argv for TRint

\item[{\hyperref[CO1-4]{\conum{4}}}]
 hSimple class based on \texttt{\$RO\penalty5000 O\penalty5000 T\penalty5000 SYS/\penalty0 tut\penalty5000 o\penalty5000 r\penalty5000 i\penalty5000 als/\penalty0 hsi\penalty5000 m\penalty5000 ple.\penalty0 C} [252]

\item[{\hyperref[CO1-5]{\conum{5}}}]
 same for Options class [253].\newline
 for demonstration purposes only (better use Boost's program\_options)

\item[{\hyperref[CO1-6]{\conum{6}}}]
 install program in bin/ and run it:

\begin{lstlisting}[language=bash,firstnumber=1,]
autoreconf -v --force --install && ./configure && make
runhSimple "out/hsimple.root"
runhSimple -N 1000000 "out/hsimple.root"
runhSimple -O "pics/test.png" "out/hsimple.root"
runhSimple -N 20000 -O "pics/test.png" "out/hsimple.root"
\end{lstlisting}
\item[{\hyperref[CO1-7]{\conum{7}}}]
 build \& package distribution management

\begin{lstlisting}[language=bash,firstnumber=1,]
make dist # generate tarball
make clean # remove build files
make distclean # remove everything generated by automake
\end{lstlisting}
\end{description}

\emph{ckon} is a C++ build tool which automatically takes care of compilation,
dictionary generation and linking of programs and libraries developed for data
analyses within the CERN ROOT analysis framework [163]. Given a source
directory it recursively parses header include statements to figure out which
libraries the main programs need to be linked to. It uses automake/autoconf
[254] to be platform independent and GNU install compliant. In
addition, m4 macros [255] are automatically downloaded and the according
compiler flags included based on a list of boost [256] libraries
provided in the config file. For the purpose of YAML database usage, a m4 macro
can be downloaded during setup to link against the yaml-{}cpp library
[257].  The project's source code has been published on github
[258] with the according documentation at [259].  It was written
by the author of this thesis with invaluable contributions by Hiroshi Masui. It
is released under the MIT License [260] and requires the following
list of software already installed on the system: m4/1.4.6, autoconf/2.68,
automake/1.11.4, libtool/2.4, boost/1.50, libcurl/7.27.0. Install \emph{ckon} via

\begin{lstlisting}[language=bash,firstnumber=1,]
git clone https://github.com/tschaume/ckon.git
cd ckon && ./installCkon <install-path>
# replace install-path> with an install path in your PATH
# see `./installCkon -h` for help
# see `./configure --help` for configuration options
\end{lstlisting}

Shown below are the generic command line options which can be given to \emph{ckon}.
The long option \texttt{-{}\penalty0 -{}\penalty0 cko\penalty5000 n\penalty5000 \_\penalty5000 cmd} is implemented as optional positional option to
run the setup, clean all compilation products (i.e. \texttt{mak\penalty5000 e c\penalty5000 l\penalty5000 ean}) and globally
install libraries and programs (i.e. \texttt{mak\penalty5000 e i\penalty5000 n\penalty5000 s\penalty5000 t\penalty5000 all}).

\begin{lstlisting}[language=bash,firstnumber=1,]
Generic Options:
 -h [ --help ]         show this help
 -v [ --verbose ]      verbose output
 -j arg                call make w/ -j <cores>
 --ckon_cmd arg        none | setup | clean | install | dry

Positional Arguments:
ckon setup # init dir, run the setup (mandatory)
ckon # build all
ckon clean # clean all build products
ckon install # install libraries (beta)
ckon dry # only generate automake's pseudo-makefiles
\end{lstlisting}

Run \texttt{cko\penalty5000 n s\penalty5000 e\penalty5000 tup} to generate the files \emph{configure.ac} and \emph{.autom4te.cfg} (both
autoconf specific with no need for modifications) as well as \emph{ckon.cfg}.
Modify the latter to resemble your directory structure and linker options.
Simply remove the lines/options you do not need, thus using the default
options.\newline
 The following options can be set on the command line or preferably in
\emph{ckon.cfg}. Optionally, a file named \emph{ckonignore} with a list of strings to be
ignored during the build process, can be created in the working directory.
Wildcards are not supported (yet). Instead, each path currently processed by
\emph{ckon} will be checked against the strings/lines in \emph{ckonignore}. If one of the
strings in \emph{ckonignore} is contained in the currently scanned path, the path is
ignored/skipped.

\begin{lstlisting}[language=bash,firstnumber=1,]
Configuration:
 -s [ --suffix ] arg    add suffix + in LinkDef.h (bool)
 -y [ --yaml ] arg      use yaml
 --ckon.src_dir arg     source dir
 --ckon.exclSuffix arg  no + suffix
 --ckon.NoRootCint arg  no dictionary
 --ckon.prog_subdir arg progs subdir
 --ckon.build_dir arg   build dir
 --ckon.install_dir arg install dir
 --ckon.cppflags arg    add CPPFLAGS
 --ckon.boost arg       boost libraries
\end{lstlisting}

In addition, unregistered options of the form \texttt{ldadd.\penalty0 pro\penalty5000 g\penalty5000 \_\penalty5000 n\penalty5000 ame} are allowed to
use for adding LDFLAGS to the linker of specific programs. The given
string/value is added verbatim in LDADD.  Unregistered options are only allowed
in \emph{ckon.cfg}. For instance, link the programs \emph{genCharmContrib} and \emph{dedxCut} versus Pythia6 [261] and RooFit [177] by adding the following
to \emph{ckon.cfg}.

\begin{lstlisting}[language=bash,firstnumber=1,]
[ldadd]
genCharmContrib=-lPhysics -lEG -lEGPythia6  # link pythia
dedxCut=-lRooFit -lRooFitCore -lMinuit      # link roofit
\end{lstlisting}

\texttt{ckon.\penalty0 boost} is set during \texttt{cko\penalty5000 n s\penalty5000 e\penalty5000 tup} to use and link against specific boost
libraries. Try not to run rootcint (\texttt{ckon.\penalty0 NoR\penalty5000 o\penalty5000 o\penalty5000 t\penalty5000 C\penalty5000 int}) on the library if
compilation fails.\newline
 Since version 0.4 \emph{ckon} allows for the automatic download
of a \emph{yaml.m4} macro during \texttt{cko\penalty5000 n s\penalty5000 e\penalty5000 tup} to link against the yaml-{}cpp library
[257]. Please submit an issue [262] if the macro does not
find the yaml-{}cpp library after you installed it. This added functionality
should not break anything if you choose not to use YAML during \texttt{cko\penalty5000 n s\penalty5000 e\penalty5000 tup}.\newline
 For the recursive header scan to work, make sure that all include directives
for C++ and ROOT headers are enclosed in \texttt{<{}.\penalty0 .\penalty0 .\penalty0 >{}}! Only your local/private
headers should be enclosed in \texttt{".\penalty0 .\penalty0 .\penalty0 "}. Otherwise \emph{ckon} will fail reporting a
\texttt{bas\penalty5000 i\penalty5000 c\penalty5000 \_\penalty5000 s\penalty5000 t\penalty5000 r\penalty5000 ing:\penalty0 :\penalty0 \_S\_\penalty5000 c\penalty5000 r\penalty5000 e\penalty5000 ate} error.\newline
 Put header and source files for each library into a separate folder in
\texttt{ckon.\penalty0 src\penalty5000 \_\penalty5000 dir}.  Running \emph{ckon} should automagically take the right action for
the current status of your build directory (no need to run \texttt{cko\penalty5000 n c\penalty5000 l\penalty5000 ean} before
re-{}compilation). Makefiles and LinkDef's are generated automatically based on
the contents and timestamps in the \texttt{ckon.\penalty0 src\penalty5000 \_\penalty5000 dir} directory. A typical
directory structure is shown in Section~\ref{ckon_dirlist} using the current defaults for
illustration purposes.

\section{ccsgp: Publication-{}ready Plots with Gnuplot and Python}
\label{get-started-with-ccsgp}\hyperlabel{get-started-with-ccsgp}%

\texttt{ccsgp} [263] is a plotting library based on gnuplot-{}py which wraps the
necessary calls to gnuplot-{}py into one function called \texttt{mak\penalty5000 e\penalty5000 \_\penalty5000 p\penalty5000 lot}. The keyword
arguments to \texttt{mak\penalty5000 e\penalty5000 \_\penalty5000 p\penalty5000 lot} provide easy control over the plot-{}by-{}plot dependent
options while reasonable defaults for legend, grid, borders, font sizes,
terminal etc. are handled internally. By providing the data in a default and
reasonable format, the user does not need to deal with the details of
"gnuplot'ing" nor the internals of the gnuplot-{}py interface library. Every call
of \texttt{mak\penalty5000 e\penalty5000 \_\penalty5000 p\penalty5000 lot} dumps an ascii representation of the plot in the terminal and
generates the eps hardcopy original. The eps figure is also converted
automatically into pdf, png and jpg formats for easy inclusion in presentations
and papers.  In addition, the user can decide to save the data contained in
each image into hdf5 files for easy access via numpy. The function
\texttt{rep\penalty5000 e\penalty5000 a\penalty5000 t\penalty5000 \_\penalty5000 p\penalty5000 lot} allows the user replot a specific graph with different
properties, like axis ranges for instance. The \texttt{mak\penalty5000 e\penalty5000 \_\penalty5000 p\penalty5000 a\penalty5000 nel} user function
facilitates plotting of 1D-{} or 2D-{}panel images with merged axes. The library
allows for the generation of identical plots independent of the data input
source (ROOT, YAML, ascii, pickle, hdf5, etc.) using the full power of python.
The name \texttt{ccsgp} stands for \emph{Carbon Capture and Sequestration GnuPlot} as this
library started off in the context of CCS research [264].  The
package \texttt{ccs\penalty5000 g\penalty5000 p\penalty5000 \_\penalty5000 g\penalty5000 e\penalty5000 t\penalty5000 \_\penalty5000 s\penalty5000 t\penalty5000 a\penalty5000 r\penalty5000 ted} [265, 266] provides a
sample setup to get started with ccsgp. \texttt{ccsgp} is initialized as a module and
its usage demonstrated with dedicated functions in the examples module. Helpful
utility functions are also included to complement the features of \texttt{ccsgp}.
Installation and implementation details for both, \texttt{ccsgp} and
\texttt{ccs\penalty5000 g\penalty5000 p\penalty5000 \_\penalty5000 g\penalty5000 e\penalty5000 t\penalty5000 \_\penalty5000 s\penalty5000 t\penalty5000 a\penalty5000 r\penalty5000 ted}, are appended in Section~\ref{app_ccsgp}.

\section{wp-{}pdf: Setup/Tools for Physics Analysis Homepage and Note}
\label{wp-pdf}\hyperlabel{wp-pdf}%

During the STAR review process of physics analyses for publication detailed
reports have to be provided to the committee in form of a homepage as well as
in pdf-{}format. Usually, this means twice the workload for presenting the same
content. The markup language AsciiDoc and the python script \emph{blogpost} both
developed by S. Rackham are the ideal tools to separate writing and compiling
information from its presentational form [267]. The content only
exists in text files which are pushed to a Wordpress instance for display in
html or which can be converted into a \emph{docbook} [268] to generate an
analysis note. The \emph{a2x} tool-{}chain takes care of this process [269].
Text-{}based source files also make it possible to track progress and changes
with a revision control system. This approach significantly reduces the
workload while at the same time improves the quality of the analysis note
and corresponding homepage by using the tools optimized for the respective task.\newline

In this section the collection of commands required to achieve this goal is
condensed into a small Makefile providing descriptive targets. This makes it very
convenient to update WordPress pages from the command line as well as create
the corresponding pdf versions. The instructions are tuned to the usage of the
Wordpress instance in the context of a high energy physics analysis.

\emph{wp-{}pdf} has been announced on Asciidoc's mailing list [270]
and a public demonstration page which uses the AsciiDoc userguide itself has
been deployed at \href{http://asciidoc.the-huck.com}{http://asciidoc.the-{}huck.com}. The following instructions and
the Makefile are also published on github [271]. On a side note, images
can quite comfortably be organized and served remotely \footnote{
see
\href{http://downloads.the-huck.com}{http://downloads.the-{}huck.com}
} by using git-{}annex
[272, 273].

\noindent
\begin{description}
\item[{ setup wordpress
}] ~\begin{itemize}[itemsep=0pt]

\item{} configure apache [236]

\item{} create mysql database [274]

\item{} install Wordpress [275]

\item{} add upgrade constants to \texttt{wp-{}\penalty0 con\penalty5000 fig.\penalty0 php} [276] (see Section~\ref{wppdf_source})

\item{} change to \texttt{www-{}\penalty0 data} owner: \texttt{sud\penalty5000 o c\penalty5000 h\penalty5000 o\penalty5000 wn -{}\penalty0 R www-{}\penalty0 data:\penalty0 www-{}\penalty0 dat\penalty5000 a p\penalty5000 u\penalty5000 b\penalty5000 lic/\penalty0 }

\item{} install Foghorn theme, adjust \emph{Appearance \ensuremath{\rightarrow} Theme Options}

\item{} install the following wordpress plugins:\newline
   Options Framework, Disable Comments, CMS Page Order, Multi-{}level Navigation
  Plugin, WP Google Fonts (font: Gentium Book Basic), WP PHP widget,
  MathJax-{}Latex (change options: Force Load, Use wp-{}latex syntax?, Use MathJax
  CDN?)

\item{} set up Sidebar in \emph{Appearance \ensuremath{\rightarrow} Widgets}:\newline
   Search, Section Index, PHP Widget (see Section~\ref{wppdf_source})

\item{} go to \emph{Settings \ensuremath{\rightarrow} Permalinks} and use post name as URL

\item{} after pushing front page, go to \emph{Settings \ensuremath{\rightarrow} Reading} and choose it as
  static front page

\end{itemize}
\item[{ install blogpost
}] \hspace{0em}\\
 The official README and source code of blogpost can be found
at [277, 278]. The author of this thesis implemented two
changes on top of blogpost version 0.9.5: First, allow more than the 10 last
most recent posts/pages to show up in info and second, a bugfix for
inline-{}latex via the config macro [279]. This customized version
is available on github [280] along with install instructions in its
README file.

\item[{ prepare for deployment
}] \hspace{0em}\\
 In the working directory adjust docinfo.xml. If you
will be including images, create a symbolic link named \texttt{ima\penalty5000 ges} linking to the
image directory and/or adjust Makefile (see Section~\ref{wppdf_source}).

\item[{ use Makefile for hp, pdf \& latex
}] \hspace{0em}\\
 Let \emph{wp-{}pdf} do the "magic":

\begin{lstlisting}[language=bash,firstnumber=1,]
make
make pdf
make hp
make latex
make out/Introduction/Introduction.txt
make clean
\end{lstlisting}

After publishing a page for the first time, use \emph{Pages \ensuremath{\rightarrow} Page Order} to
arrange pages.
\end{description}

\section{rmrg: Automization Class and Program for ROOTs hadd}
\label{rmrg}\hyperlabel{rmrg}%

\emph{rmrg} is a command line program to merge ROOT Trees and Histograms with
additional features and increased control currently missing in ROOT's hadd
[29]. Primarily,
\begin{itemize}[itemsep=0pt]

\item{} automatic file-{}list generation based on an input directory scan

\item{} step-{}by-{}step verbose TTree and Histogram merging

\item{} options for number of files to merge at once and minimum file size

\end{itemize}

This tool was developed to be included as afterburner in the PicoDst production
at LBNL. Small output files would be merged on a run-{}by-{}run basis which
provides a homogenous landscape of input files to relieve the queueing systems
of the batch farms. The code is published on github [281].
%
%

\part{Appendix}
\label{_appendix}\hyperlabel{_appendix}%


\chapter{Bibliography}
\label{_bibliography}\hyperlabel{_bibliography}%
\begin{enumerate}[label=\arabic*.,itemsep=0pt]

\item{} D. Gross and F. Wilczek. A watershed: the emergence of Quantum-{}Chromo-{}Dynamics. \emph{CERN Courier}. \href{http://cerncourier.com/cws/article/cern/52034}{http://cerncourier.com/\-cws/\-article/\-cern/\-52034}

\item{} K.A. Olive et al. (Particle Data Group). Review of Particle Physics. \emph{Chin. Phys. C38}, 090001. \href{https://doi.org/10.1088/1674-1137/38/9/090001}{https://doi.org/\-10.1088/\-1674-{}1137/\-38/\-9/\-090001}

\item{} R. Rapp and J. Wambach. Chiral symmetry restoration and dileptons in relativistic heavy ion collisions. \emph{Adv. Nucl. Phys. 25}, 1. \href{http://arxiv.org/abs/hep-ph/9909229}{http://arxiv.org/\-abs/\-hep-{}ph/\-9909229}

\item{} E. Shuryak. Quantum-{}Chromo-{}Dynamics and the theory of superdense matter. \emph{Physics Reports 61}, 71\textendash{}158. \href{https://doi.org/10.1016/0370-1573(80)90105-2}{https://doi.org/\-10.1016/\-0370-{}1573(80)90105-{}2}

\item{} National Science Foundation et al. 1975. \emph{Report of the Workshop on BeV/Nucleon Collisions of Heavy Ions -{} How and Why}. Bear Mountain, NY.

\item{} H.G. Baumgardt et al. Shock waves and mach cones in fast nucleus-{}nucleus collisions. \emph{Zeitschrift für Physik A273}, 359\textendash{}371. \href{https://doi.org/10.1007/BF01435578}{https://doi.org/\-10.1007/\-BF01435578}

\item{} U. Heinz and G. Kestin. Jozso's Legacy: Chemical and Kinetic Freeze-{}out in Heavy-{}Ion Collisions. \emph{EPJ 155}, 75. \href{https://arxiv.org/abs/0709.3366}{https://arxiv.org/\-abs/\-0709.3366}

\item{} J.-{}Y. Ollitrault. Anisotropy as a signature of transverse collective flow. \emph{PRD 46}, 229\textendash{}245. \href{https://doi.org/10.1103/PhysRevD.46.229}{https://doi.org/\-10.1103/\-PhysRevD.46.229}

\item{} A. Poskanzer and S. Voloshin. Methods for analyzing anisotropic flow in relativistic nuclear collisions. \emph{PRC 58}, 1671. \href{https://arxiv.org/abs/nucl-ex/9805001}{https://arxiv.org/\-abs/\-nucl-{}ex/\-9805001}

\item{} G. David, R. Rapp, and Z. Xu. Electromagnetic Probes at RHIC-{}II. \emph{Phys.Rept. 462}, 176\textendash{}217. \href{https://arxiv.org/abs/nucl-ex/0611009}{https://arxiv.org/\-abs/\-nucl-{}ex/\-0611009}

\item{} R. Rapp. Update on \ensuremath{\chi}SR in the Context of Dilepton Data. \emph{J.Phys.Conf.Ser. 420}, 012017. \href{https://arxiv.org/abs/1210.3660}{https://arxiv.org/\-abs/\-1210.3660}

\item{} A. Drees. Dileptons and photons at RHIC energies. \emph{Nuclear Physics A830}, 435c\textendash{}442c. \href{https://doi.org/10.1016/j.nuclphysa.2009.10.036}{https://doi.org/\-10.1016/\-j.nuclphysa.2009.10.036}

\item{} CERES Collaboration. Enhanced Production of Low-{}Mass e+e-{} in 200 GeV/N S-{}Au at CERN SPS. \emph{PRL 75}, 1272. \href{https://doi.org/10.1103/PhysRevLett.75.1272}{https://doi.org/\-10.1103/\-PhysRevLett.75.1272}

\item{} NA60 Collaboration. First measurement of the \ensuremath{\rho} SF in high-{}energy N+N. \emph{PRL 96}, 162302. \href{http://arxiv.org/abs/nucl-ex/0605007}{http://arxiv.org/\-abs/\-nucl-{}ex/\-0605007}

\item{} PHENIX Collaboration. Measurement of e+e-{} continuum in pp/AuAu at 200 GeV and implications for direct \ensuremath{\gamma} production. \emph{PRC 81}, 034911. \href{http://arxiv.org/abs/0912.0244}{http://arxiv.org/\-abs/\-0912.0244}

\item{} STAR Collaboration. Dielectron mass spectra in Au+Au at 200 GeV. \emph{PRL 113}, 022301. \href{http://arxiv.org/abs/1312.7397}{http://arxiv.org/\-abs/\-1312.7397}

\item{} HADES Collaboration. Dielectron production in C+C collisions at 2-{}AGeV. \emph{PRL 98}, 052302. \href{https://arxiv.org/abs/nucl-ex/0608031}{https://arxiv.org/\-abs/\-nucl-{}ex/\-0608031}

\item{} HADES Collaboration. Study of dielectron production in collisions at 1A-{}GeV. \emph{PLB 663}, 43. \href{https://doi.org/10.1016/j.physletb.2008.03.062}{https://doi.org/\-10.1016/\-j.physletb.2008.03.062}

\item{} STAR Collaboration. Mid-{}rapidity e+e-{} spectrum in p+p at 200 GeV. \emph{PRC 86}, 024906. \href{https://doi.org/10.1103/PhysRevC.86.024906}{https://doi.org/\-10.1103/\-PhysRevC.86.024906}

\item{} STAR Collaboration. Experimental Exploration of QCD Phase Diagram: Search for Critical Point and Onset of Deconfinement. \href{http://arxiv.org/abs/1007.2613}{http://arxiv.org/\-abs/\-1007.2613}

\item{} W. Cassing et al. Excitation functions of hadronic observables from SIS to RHIC energies. \emph{NPA 674}, 249. \href{http://arxiv.org/abs/nucl-th/0001024}{http://arxiv.org/\-abs/\-nucl-{}th/\-0001024}

\item{} O. Linnyk. 2014. Private Communication.

\item{} STAR Collaboration. STAR Public Note 0598: Studying the phase diagram of QCD matter at RHIC. \href{https://drupal.star.bnl.gov/STAR/starnotes/public/sn0598}{https://drupal.star.bnl.gov/\-STAR/\-starnotes/\-public/\-sn0598}

\item{} StBTofUtil. www.star.bnl.gov/cgi-{}bin/protected/cvsweb.cgi/StRoot/StBTofUtil

\item{} P. Huck. \href{http://drupal.star.bnl.gov/STAR/system/files/stv0tofcorrection_v3.pdf}{http://drupal.star.bnl.gov/\-STAR/\-system/\-files/\-stv0tofcorrection\_v3.pdf}

\item{} P. Huck. \href{http://www.star.bnl.gov/HyperNews-star/protected/get/bulkcorr/1780.html}{http://www.star.bnl.gov/\-HyperNews-{}star/\-protected/\-get/\-bulkcorr/\-1780.html}

\item{} P. Huck. www.star.bnl.gov/cgi-{}bin/protected/cvsweb.cgi/offline/users/huck

\item{} P. Huck. \href{https://speakerdeck.com/tschaume/organize-root-analyses-with-autoconf}{https://speakerdeck.com/\-tschaume/\-organize-{}root-{}analyses-{}with-{}autoconf}

\item{} ROOT's hadd. \href{http://root.cern.ch/root/html/tutorials/io/hadd.C.html}{http://root.cern.ch/\-root/\-html/\-tutorials/\-io/\-hadd.C.html}

\item{} D. Gross and F. Wilczek. Ultraviolet Behavior of Non-{}Abelian Gauge Theories. \emph{PRL 30}, 1343. \href{https://doi.org/10.1103/PhysRevLett.30.1343}{https://doi.org/\-10.1103/\-PhysRevLett.30.1343}

\item{} H. Politzer. Reliable Perturbative Results for Strong Interactions? \emph{PRL 30}, 1346\textendash{}1349. \href{http://doi.org/10.1103/PhysRevLett.30.1346}{http://doi.org/\-10.1103/\-PhysRevLett.30.1346}

\item{} D. Gross and F. Wilczek. Asymptotically Free Gauge Theories I. \emph{Phys. Rev. D8}, 3633. \href{https://doi.org/10.1103/PhysRevD.8.3633}{https://doi.org/\-10.1103/\-PhysRevD.8.3633}

\item{} D. Gross and F. Wilczek. Asymptotically Free Gauge Theories II. \emph{Phys. Rev. D9}, 980. \href{https://doi.org/10.1103/PhysRevD.9.980}{https://doi.org/\-10.1103/\-PhysRevD.9.980}

\item{} H. Georgi and H. Politzer. Electroproduction scaling in asymptotically free theory of strong interactions. \emph{PRD 9}, 416. \href{https://doi.org/10.1103/PhysRevD.9.416}{https://doi.org/\-10.1103/\-PhysRevD.9.416}

\item{} S. Bethke. World summary of the strong coupling constant. \emph{Nucl.Phys.Proc.Suppl. 234}, 229\textendash{}234. \href{https://arxiv.org/abs/1210.0325}{https://arxiv.org/\-abs/\-1210.0325}

\item{} G.F. Sterman. QCD and jets. \href{http://arxiv.org/abs/hep-ph/0412013}{http://arxiv.org/\-abs/\-hep-{}ph/\-0412013}

\item{} M. Peskin and D. Schroeder. \emph{An Introduction To Quantum Field Theory}. Westview Press.

\item{} P. Petreczky. Lattice QCD at T\ensuremath{\neq}0. \emph{J.Phys. G39}, 093002. \href{https://arxiv.org/abs/1203.5320}{https://arxiv.org/\-abs/\-1203.5320}

\item{} K. Liu. Finite Baryon Density Algorithm. \href{http://arxiv.org/abs/hep-lat/0312027}{http://arxiv.org/\-abs/\-hep-{}lat/\-0312027}

\item{} M. Gyulassy and L. McLerran. New forms of QCD matter discovered at RHIC. \emph{NPA 750}, 30. \href{https://arxiv.org/abs/nucl-th/0405013}{https://arxiv.org/\-abs/\-nucl-{}th/\-0405013}

\item{} G. Baym. Relativistic Heavy Ion Collider: From dreams to beams in two decades. \emph{NPA 698}, 23\textendash{}32. \href{https://arxiv.org/abs/hep-ph/0104138}{https://arxiv.org/\-abs/\-hep-{}ph/\-0104138}

\item{} I. Arsene et al. Quark\textendash{}gluon plasma and color glass condensate at RHIC? The BRAHMS perspective. \emph{NPA 757}, 1. \href{https://doi.org/10.1016/j.nuclphysa.2005.02.130}{https://doi.org/\-10.1016/\-j.nuclphysa.2005.02.130}

\item{} B. Back et al. The PHOBOS perspective on discoveries at RHIC. \emph{Nucl. Phys. A757}, 28\textendash{}101. \href{https://doi.org/10.1016/j.nuclphysa.2005.03.084}{https://doi.org/\-10.1016/\-j.nuclphysa.2005.03.084}

\item{} J. Adams et al. Experimental and theoretical challenges in the search for the quark\textendash{}gluon plasma: The STAR Collaboration's critical assessment of the evidence from RHIC collisions. \emph{NPA 757}, 102. \href{https://doi.org/10.1016/j.nuclphysa.2005.03.085}{https://doi.org/\-10.1016/\-j.nuclphysa.2005.03.085}

\item{} K. Adcox et al. Formation of dense partonic matter in relativistic nucleus\textendash{}nucleus collisions at RHIC: Experimental evaluation by the PHENIX Collaboration. \emph{NPA 757}, 184\textendash{}283. \href{https://doi.org/10.1016/j.nuclphysa.2005.03.086}{https://doi.org/\-10.1016/\-j.nuclphysa.2005.03.086}

\item{} L. Cifarelli, L.P. Csernai, and H. Stöcker. The Quark-{}Gluon Plasma, a nearly perfect fluid. \emph{Europhys. News 43}, 29\textendash{}31. \href{https://doi.org/10.1051/epn/2012206}{https://doi.org/\-10.1051/\-epn/\-2012206}

\item{} QM 11. \emph{JPG 38}. \href{http://iopscience.iop.org/0954-3899/38/12}{http://iopscience.iop.org/\-0954-{}3899/\-38/\-12}

\item{} QM 12. \emph{NPA 904}. \href{https://doi.org/10.1016/S0375-9474(13)00372-2}{https://doi.org/\-10.1016/\-S0375-{}9474(13)00372-{}2}

\item{} QM 14. \emph{NPA 931}. www.sciencedirect.com/science/journal/03759474/931

\item{} F. Karsch. Lattice QCD at high temperature and density. \emph{Lect. Notes Phys. 583}, 209\textendash{}249. \href{https://arxiv.org/abs/hep-lat/0106019}{https://arxiv.org/\-abs/\-hep-{}lat/\-0106019}

\item{} I. Bombaci. Quark deconfinement in neutron stars and \ensuremath{\gamma}-{}ray bursts. \emph{J.Phys.Conf.Ser. 50}, 208. \href{https://doi.org/10.1088/1742-6596/50/1/024}{https://doi.org/\-10.1088/\-1742-{}6596/\-50/\-1/\-024}

\item{} R. Hagedorn. New derivation of the statistical theory of particle production with numerical results for p-{}p at 25 GeV. \emph{Nuovo Cimento 15}, 434. \href{http://doi.org/10.1007/BF02902578}{http://doi.org/\-10.1007/\-BF02902578}

\item{} S. Gupta et al. Scale for the Phase Diagram of Quantum Chromodynamics. \emph{Science 332}, 1525. \href{https://doi.org/10.1126/science.1204621}{https://doi.org/\-10.1126/\-science.1204621}

\item{} T. Bhattacharya et al. QCD Phase Transition with Chiral Quarks and Phys. Quark Masses. \emph{PRL 113}, 082001. \href{https://doi.org/10.1103/PhysRevLett.113.082001}{https://doi.org/\-10.1103/\-PhysRevLett.113.082001}

\item{} Y. Nambu and G. Jona-{}Lasinio. Dynamical Model of Elementary Particles Based on Superconductivity I. \emph{Phys. Rev. 122}, 345. \href{https://doi.org/10.1103/PhysRev.122.345}{https://doi.org/\-10.1103/\-PhysRev.122.345}

\item{} Y. Nambu and G. Jona-{}Lasinio. Dynamical Model of Elementary Particles Based on Superconductivity II. \emph{Phys. Rev. 124}, 246. \href{https://doi.org/10.1103/PhysRev.124.246}{https://doi.org/\-10.1103/\-PhysRev.124.246}

\item{} T. Muta. \emph{Foundations of Quantum Chromodynamics. An Introduction to Perturbative Methods in Gauge Theories.} World Scientific Lecture Notes in Physics.

\item{} D. Gross et al. QCD and instantons at finite temperature. \emph{Reviews of Modern Physics 53}, 43. \href{https://doi.org/10.1103/RevModPhys.53.43}{https://doi.org/\-10.1103/\-RevModPhys.53.43}

\item{} R.D. Pisarski and F. Wilczek. Remarks on the chiral phase transition in QCD. \emph{PRD 29}, 338\textendash{}341. \href{https://doi.org/10.1103/PhysRevD.29.338}{https://doi.org/\-10.1103/\-PhysRevD.29.338}

\item{} A. Masayuki and Y. Koichi. Chiral restoration at finite density/temperature. \emph{NPA 504}, 668. \href{https://doi.org/10.1016/0375-9474(89)90002-X}{https://doi.org/\-10.1016/\-0375-{}9474(89)90002-{}X}

\item{} K. Fukushima and T. Hatsuda. Phase diagram of dense QCD. \emph{Rept.Prog.Phys. 74}, 014001. \href{https://arxiv.org/abs/1005.4814}{https://arxiv.org/\-abs/\-1005.4814}

\item{} T. DeGrand et al. Masses and other parameters of the light hadrons. \emph{PRD 12}, 2060\textendash{}2076. \href{https://doi.org/10.1103/PhysRevD.12.2060}{https://doi.org/\-10.1103/\-PhysRevD.12.2060}

\item{} H. Satz. Deconf./percolation. \emph{NPA 642}, 130. \href{https://arxiv.org/abs/hep-ph/9805418}{https://arxiv.org/\-abs/\-hep-{}ph/\-9805418}

\item{} E. Shuryak. Two scales and phase transitions in Quantum-{}Chromo-{}Dynamics. \emph{PLB 107}, 103\textendash{}105. \href{https://doi.org/10.1016/0370-2693(81)91158-8}{https://doi.org/\-10.1016/\-0370-{}2693(81)91158-{}8}

\item{} Y. Aoki et al. The QCD transition temperature: Results with physical masses in the continuum limit. \emph{PLB 643}, 46. \href{https://arxiv.org/abs/hep-lat/0609068}{https://arxiv.org/\-abs/\-hep-{}lat/\-0609068}

\item{} K. Fukushima and C. Sasaki. The phase diagram of nuclear/quark matter at high baryon density. \emph{Prog.Part.Nucl.Phys. 72}, 99. \href{https://arxiv.org/abs/1301.6377}{https://arxiv.org/\-abs/\-1301.6377}

\item{} T. D. Lee and G. C. Wick. Vacuum stability and excitation in a spin-{}0 field theory. \emph{PRD 9}, 2291. \href{https://doi.org/10.1103/PhysRevD.9.2291}{https://doi.org/\-10.1103/\-PhysRevD.9.2291}

\item{} E. Iancu. QCD in heavy ion collisions. \href{http://arxiv.org/abs/1205.0579}{http://arxiv.org/\-abs/\-1205.0579}

\item{} R. Placakyte. Parton Distribution Functions. \href{http://arxiv.org/abs/1111.5452}{http://arxiv.org/\-abs/\-1111.5452}

\item{} L. McLerran. The color glass condensate and small x physics. \emph{Lect. Notes Phys. 583}, 291. \href{https://arxiv.org/abs/hep-ph/0104285}{https://arxiv.org/\-abs/\-hep-{}ph/\-0104285}

\item{} J.L. Albacete and C. Marquet. Gluon saturation and initial conditions for relativistic HICs. \emph{Prog.Part.Nucl.Phys. 76}, 1. \href{https://arxiv.org/abs/1401.4866}{https://arxiv.org/\-abs/\-1401.4866}

\item{} G. Aad et al. Observation of a Centrality-{}Dependent Dijet Asymmetry in Pb+Pb at 2.77 TeV with ATLAS. \emph{PRL 105}, 252303. \href{https://arxiv.org/abs/1011.6182}{https://arxiv.org/\-abs/\-1011.6182}

\item{} CMS. Jet quenching in PbPb at 2.76 TeV. \emph{PRC 84}, 024906. \href{https://arxiv.org/abs/1102.1957}{https://arxiv.org/\-abs/\-1102.1957}

\item{} R. Venugopalan. From Glasma to QGP in heavy ion collisions. \emph{J.Phys.G 35}, 104003. \href{https://doi.org/10.1088/0954-3899/35/10/104003}{https://doi.org/\-10.1088/\-0954-{}3899/\-35/\-10/\-104003}

\item{} U. Heinz. Thermalization at RHIC. \emph{AIP 739}, 163. \href{http://arxiv.org/abs/nucl-th/0407067}{http://arxiv.org/\-abs/\-nucl-{}th/\-0407067}

\item{} F. Gelis. CGC and Glasma. \emph{Int.J.Mod.Phys. A28}, 1330001. \href{https://arxiv.org/abs/1211.3327}{https://arxiv.org/\-abs/\-1211.3327}

\item{} U. Heinz. Strange messages: Chemical/thermal freezeout in nuclear collisions. \emph{JPG 25}, 263. \href{https://arxiv.org/abs/nucl-th/9810056}{https://arxiv.org/\-abs/\-nucl-{}th/\-9810056}

\item{} S. Choi and K.S. Lee. BW model with two freeze-{}out temperatures for hadrons produced in HICs. \emph{PRC 84}, 064905. \href{https://doi.org/10.1103/PhysRevC.84.064905}{https://doi.org/\-10.1103/\-PhysRevC.84.064905}

\item{} P. Braun-{}Munzinger et al. Particle prod. in HICs. \href{https://arxiv.org/abs/nucl-th/0304013}{https://arxiv.org/\-abs/\-nucl-{}th/\-0304013}

\item{} W. Reisdorf and H. Ritter. Collective flow in heavy ion collisions. \emph{Ann.Rev.Nucl.Part.Sci. 47}, 663. \href{https://doi.org/10.1146/annurev.nucl.47.1.663}{https://doi.org/\-10.1146/\-annurev.nucl.47.1.663}

\item{} I. Bearden et al. Collective Expansion in High Energy Heavy Ion Collisions. \emph{PRL 78}, 2080\textendash{}2083. \href{https://doi.org/10.1103/PhysRevLett.78.2080}{https://doi.org/\-10.1103/\-PhysRevLett.78.2080}

\item{} Z. Tang. Spectra and radial flow at RHIC using Tsallis-{}Blast-{}Wave. \emph{PRC 79}, 051901. \href{http://arxiv.org/abs/0812.1609}{http://arxiv.org/\-abs/\-0812.1609}

\item{} E. Schnedermann et al. Thermal phenomenology of hadrons from 200-{}A/GeV S+S collisions. \emph{PRC 48}, 2462. \href{https://arxiv.org/abs/nucl-th/9307020}{https://arxiv.org/\-abs/\-nucl-{}th/\-9307020}

\item{} K. Yagi et al. QGP: From Big Bang to Little Bang. \emph{Cambridge Monographs on Particle Physics, Nuclear Physics and Cosmology 23}. \href{http://utkhii.px.tsukuba.ac.jp/cupbook}{http://utkhii.px.tsukuba.ac.jp/\-cupbook}

\item{} R. Snellings. Elliptic flow. \emph{New J. Phys. 13}, 055008. \href{https://arxiv.org/abs/1102.3010}{https://arxiv.org/\-abs/\-1102.3010}

\item{} B. Zhang, M. Gyulassy, and C.M. Koa. Elliptic flow from a parton cascade. \emph{Physics Letters B455}, 45\textendash{}48. \href{https://arxiv.org/abs/nucl-th/9902016}{https://arxiv.org/\-abs/\-nucl-{}th/\-9902016}

\item{} R. Rapp et al. Chiral Restoration Transition of Quantum-{}Chromo-{}Dynamics and Low Mass Dileptons. \href{http://arxiv.org/abs/0901.3289}{http://arxiv.org/\-abs/\-0901.3289}

\item{} P. Kolb and U. Heinz. Hydrodynamic description of ultrarelativistic heavy ion collisions. \href{http://arxiv.org/abs/nucl-th/0305084}{http://arxiv.org/\-abs/\-nucl-{}th/\-0305084}

\item{} PHENIX Collaboration. Scaling Properties of Azimuthal Anisotropy in Au+Au/Cu+Cu at 200 GeV. \emph{PRL 98}, 162301. \href{https://doi.org/10.1103/PhysRevLett.98.162301}{https://doi.org/\-10.1103/\-PhysRevLett.98.162301}

\item{} STAR Collaboration. Mass, quark-{}number, and s$_{\text{NN}}$ dependence of 2nd/4th flow harmonics in ultrarelativ. NN. \emph{PRC 75}, 054906. \href{https://arxiv.org/abs/nucl-ex/0701010}{https://arxiv.org/\-abs/\-nucl-{}ex/\-0701010}

\item{} R. Barate et al. Measurement of the axial-{}vector \ensuremath{\tau} spectral functions and determination of the strong coupling constant at \ensuremath{\tau} pole mass from hadronic \ensuremath{\tau} decays. \emph{EPJC 4}, 409\textendash{}431. \href{https://doi.org/10.1007/s100529800895}{https://doi.org/\-10.1007/\-s100529800895}

\item{} M. Birse. Chiral symmetry in nuclei: Partial restoration and its consequences. \emph{JPG 20}, 1537. \href{https://arxiv.org/abs/nucl-th/9406029}{https://arxiv.org/\-abs/\-nucl-{}th/\-9406029}

\item{} J. Delorme et al. Chiral Lagrangians and quark condensate in nuclei. \emph{NPA 603}, 239\textendash{}256. \href{https://arxiv.org/abs/nucl-th/9603005}{https://arxiv.org/\-abs/\-nucl-{}th/\-9603005}

\item{} G. Chanfray et al. Quark condensate at finite temperature. \emph{Physics Letters B388}, 673. \href{https://arxiv.org/abs/nucl-th/9607046}{https://arxiv.org/\-abs/\-nucl-{}th/\-9607046}

\item{} R. Rapp and J. Wambach. Low mass dileptons at the SPS: Evidence for \ensuremath{\chi}R? \emph{EPJA 6}, 415. \href{https://arxiv.org/abs/hep-ph/9907502}{https://arxiv.org/\-abs/\-hep-{}ph/\-9907502}

\item{} M. Birse and J. McGovern. \ensuremath{\pi}N \ensuremath{\Sigma}-{}commutator in chiral models of the nucleon. \emph{PLB 292}, 242. \href{https://doi.org/10.1016/0370-2693(92)91169-A}{https://doi.org/\-10.1016/\-0370-{}2693(92)91169-{}A}

\item{} Sakurai. Strong Inter. \emph{Ann.Phys. 11}, 1. \href{https://doi.org/10.1016/0003-4916(60)90126-3}{https://doi.org/\-10.1016/\-0003-{}4916(60)90126-{}3}

\item{} CERES/NA45 Collaboration. Low mass e+e-{} production in 158/AGeV Pb-{}Au at SPS, its dependence on multiplicity/p$_{\text{T}}$. \emph{PLB 422}, 405. \href{http://arxiv.org/abs/nucl-ex/9712008}{http://arxiv.org/\-abs/\-nucl-{}ex/\-9712008}

\item{} CERES Collaboration. e+e-{} pair production in Pb-{}Au collisions at 158/A-{}GeV. \emph{EPJC 41}, 475. \href{http://arxiv.org/abs/nucl-ex/0506002}{http://arxiv.org/\-abs/\-nucl-{}ex/\-0506002}

\item{} CERES Collaboration. \ensuremath{\rho}-{}meson modification detected by low-{}mass e+e-{} in central Pb-{}Au at 158 AGeV/c. \emph{PLB 666}, 425. \href{http://arxiv.org/abs/nucl-ex/0611022}{http://arxiv.org/\-abs/\-nucl-{}ex/\-0611022}

\item{} H. van Hees and R. Rapp. Interpretation of thermal dileptons measured at CERN SPS. \emph{PRL 97}, 102301. \href{https://doi.org/10.1103/PhysRevLett.97.102301}{https://doi.org/\-10.1103/\-PhysRevLett.97.102301}

\item{} HADES Collaboration. Origin of low-{}mass e+e-{} pair excess in light N+N. \emph{PLB 690}, 118. \href{https://doi.org/10.1016/j.physletb.2010.05.010}{https://doi.org/\-10.1016/\-j.physletb.2010.05.010}

\item{} HADES Collaboration. e+e-{} production in Ar+KCl collisions at 1.76AGeV. \emph{PRC 84}, 014902. \href{https://arxiv.org/abs/1103.0876}{https://arxiv.org/\-abs/\-1103.0876}

\item{} D. Sharma. Dielectrons in PHENIX and its implications on heavy flavor. \emph{NPA 932}, 235. \href{https://doi.org/10.1016/j.nuclphysa.2014.10.024}{https://doi.org/\-10.1016/\-j.nuclphysa.2014.10.024}

\item{} E. Atomssa. Dielectron measurements by PHENIX using the HBD. \emph{NPA A904-{}905}, 561c. \href{https://doi.org/10.1016/j.nuclphysa.2013.02.076}{https://doi.org/\-10.1016/\-j.nuclphysa.2013.02.076}

\item{} U. Heinz and K. Lee. \ensuremath{\rho}-{}peak in dimuon spectrum as clock for fireball lifetimes in relativistic N+N. \emph{PLB 259}, 162. \href{https://doi.org/10.1016/0370-2693(91)90152-G}{https://doi.org/\-10.1016/\-0370-{}2693(91)90152-{}G}

\item{} DLS Collaboration. Dielectron cross section measurements in N+N at 1A-{}GeV. \emph{PRL 79}, 1229. \href{https://doi.org/10.1103/PhysRevLett.79.1229}{https://doi.org/\-10.1103/\-PhysRevLett.79.1229}

\item{} C. Ernst et al. Intermediate mass excess of dilepton production in HICs at relativistic energies. \emph{PRC 58}, 447. \href{https://doi.org/10.1103/PhysRevC.58.447}{https://doi.org/\-10.1103/\-PhysRevC.58.447}

\item{} E. Bratkovskaya. Dilepton production and m$_{\text{T}}$-{}scaling at BEVALAC/SIS. \emph{NPA 634}, 168. \href{https://arxiv.org/abs/nucl-th/9710043}{https://arxiv.org/\-abs/\-nucl-{}th/\-9710043}

\item{} R. Rapp. Thermal Electromagnetic Radiation in Heavy-{}Ion Collisions. Talk at TRW '12. \href{http://crunch.ikp.physik.tu-darmstadt.de/erice/2012/sec/talks/thursday/rapp.ppt}{http://crunch.ikp.physik.tu-{}darmstadt.de/\-erice/\-2012/\-sec/\-talks/\-thursday/\-rapp.ppt}

\item{} CLAS Collaboration. Search for medium modifications of the \ensuremath{\rho}-{}meson. \emph{PRL 99}, 262302. \href{https://doi.org/10.1103/PhysRevLett.99.262302}{https://doi.org/\-10.1103/\-PhysRevLett.99.262302}

\item{} P. Hohler and R. Rapp. Realistic implementation of massive Yang-{}Mills theory for \ensuremath{\rho}/a$_{\text{1}}$. \emph{PRD 89}, 125013. \href{https://doi.org/10.1103/PhysRevD.89.125013}{https://doi.org/\-10.1103/\-PhysRevD.89.125013}

\item{} R. Rapp. Signatures of thermal dilepton radiation at ultrarelativistic energies. \emph{PRC 63}, 054907. \href{http://arxiv.org/abs/hep-ph/0010101}{http://arxiv.org/\-abs/\-hep-{}ph/\-0010101}

\item{} X. Dong, F. Geurts, and B. Huang. \emph{Nuclear Physics A 904-{}905}, 19\textendash{}26 \& 217\textendash{}224 \& 565\textendash{}568. \href{https://doi.org/10.1016/S0375-9474(13)00372-2}{https://doi.org/\-10.1016/\-S0375-{}9474(13)00372-{}2}

\item{} R. Rapp. Dilepton spectroscopy of QCD matter at collider energies. \emph{AHEP}, 148253. \href{http://arxiv.org/abs/1304.2309}{http://arxiv.org/\-abs/\-1304.2309}

\item{} STAR Collaboration. Pion, kaon, proton and anti-{}proton p$_{\text{T}}$ distributions from p+p/d+Au at 200 GeV. \emph{PLB 616}, 8. \href{https://doi.org/10.1016/j.physletb.2005.04.041}{https://doi.org/\-10.1016/\-j.physletb.2005.04.041}

\item{} STAR Collaboration. Open charm yields in d+Au at s$_{\text{NN}}$$^{\text{1/2}}$ = 200 GeV. \emph{PRL 94}, 062301. \href{http://arxiv.org/abs/nucl-ex/0407006}{http://arxiv.org/\-abs/\-nucl-{}ex/\-0407006}

\item{} P. Huck. Beam energy dependence of e+e-{} production in Au+Au from STAR. \emph{NPA 931}, 659. \href{https://arxiv.org/abs/1409.5675}{https://arxiv.org/\-abs/\-1409.5675}

\item{} M. Harrison et al. Relativistic Heavy Ion Collider project overview. \emph{NIM A499}, 235\textendash{}244. \href{https://doi.org/10.1016/S0168-9002(02)01937-X}{https://doi.org/\-10.1016/\-S0168-{}9002(02)01937-{}X}

\item{} RHIC \& its Detectors. \href{http://www.sciencedirect.com/science/journal/01689002/499/2}{http://www.sciencedirect.com/\-science/\-journal/\-01689002/\-499/\-2}

\item{} Brookhaven National Laboratory. \href{http://www.bnl.gov}{http://www.bnl.gov}

\item{} J. Benjamin et al. Injecting RHIC from the Brookhaven Tandem Van de Graaff. \emph{PAC 4}, 2277. \href{https://doi.org/10.1109/PAC.1999.792658}{https://doi.org/\-10.1109/\-PAC.1999.792658}

\item{} D. Steski et al. Operation of the RHIC Au-{}ion source. \emph{Rev. Sci. Instrum. 73}, 797. \href{https://doi.org/10.1063/1.1430871}{https://doi.org/\-10.1063/\-1.1430871}

\item{} P. Thieberger et al. Tandem injected relativistic heavy ion facility at BNL. \emph{NIM A268}, 513. \href{https://doi.org/10.1016/0168-9002(88)90570-0}{https://doi.org/\-10.1016/\-0168-{}9002(88)90570-{}0}

\item{} E. Forsyth and Y. Lee. Design and Status of the AGS Booster Accelerator. \emph{PAC Proc.} \href{http://accelconf.web.cern.ch/AccelConf/p87/PDF/PAC1987_0867.PDF}{http://accelconf.web.cern.ch/\-AccelConf/\-p87/\-PDF/\-PAC1987\_0867.PDF}

\item{} D.S. Barton. Heavy Ion Program at Brookhaven National Lab: AGS and RHIC. \emph{PAC Proceedings 1987}. \href{https://accelconf.web.cern.ch/accelconf/p87/PDF/PAC1987_0804.PDF}{https://accelconf.web.cern.ch/\-accelconf/\-p87/\-PDF/\-PAC1987\_0804.PDF}

\item{} \href{http://www.bnl.gov/cad/accelerator/docs/pdf/AGSDesignReport.pdf}{http://www.bnl.gov/\-cad/\-accelerator/\-docs/\-pdf/\-AGSDesignReport.pdf}

\item{} AGS to RHIC Transfer Line. \href{http://www.rhichome.bnl.gov/RHIC/Runs/RhicAtr.pdf}{http://www.rhichome.bnl.gov/\-RHIC/\-Runs/\-RhicAtr.pdf}

\item{} \href{http://www.bnl.gov/cad/accelerator/docs/pdf/RHICConfManual.pdf}{http://www.bnl.gov/\-cad/\-accelerator/\-docs/\-pdf/\-RHICConfManual.pdf}

\item{} T. Ludlam. Overview of experiments and detectors at RHIC. \emph{Nucl. Instr. Meth. A499}, 428. \href{https://doi.org/10.1016/S0168-9002(02)01948-4}{https://doi.org/\-10.1016/\-S0168-{}9002(02)01948-{}4}

\item{} C. Adler et al. The RHIC zero-{}degree calorimeters. \emph{Nucl. Instr. Meth. A461}, 337\textendash{}340. \href{https://doi.org/10.1016/S0168-9002(00)01238-9}{https://doi.org/\-10.1016/\-S0168-{}9002(00)01238-{}9}

\item{} C. Adler et al. The RHIC zero degree calorimeter. \emph{Nucl. Instr. Meth. A470}, 488\textendash{}499. \href{https://arxiv.org/abs/nucl-ex/0008005}{https://arxiv.org/\-abs/\-nucl-{}ex/\-0008005}

\item{} H. Hahn et al. The RHIC design overview. \emph{Nuclear Instrum. Meth. A499}, 245\textendash{}263. \href{https://doi.org/10.1016/S0168-9002(02)01938-1}{https://doi.org/\-10.1016/\-S0168-{}9002(02)01938-{}1}

\item{} M. Beddo et al. STAR CDR. \href{https://drupal.star.bnl.gov/STAR/files/StarCDR.pdf}{https://drupal.star.bnl.gov/\-STAR/\-files/\-StarCDR.pdf}

\item{} K. Ackermann et al. STAR detector overview. \emph{Nucl. Instrum. Meth. A499}, 624\textendash{}632. \href{https://doi.org/10.1016/S0168-9002(02)01960-5}{https://doi.org/\-10.1016/\-S0168-{}9002(02)01960-{}5}

\item{} C. Allgower et al. The STAR endcap electromagnetic calorimeter. \emph{NIMA 499}, 740\textendash{}750. \href{https://doi.org/10.1016/S0168-9002(02)01971-X}{https://doi.org/\-10.1016/\-S0168-{}9002(02)01971-{}X}

\item{} M. Beddo et al. STAR Barrel Electromagnetic Calorimeter. \emph{Nucl. Instr. Meth. A499}, 725. \href{https://doi.org/10.1016/S0168-9002(02)01970-8}{https://doi.org/\-10.1016/\-S0168-{}9002(02)01970-{}8}

\item{} C. Whitten et al. Beam-Beam Counter: A local polarimeter at STAR. \emph{AIP Proc. 980}, 390. \href{https://doi.org/10.1063/1.2888113}{https://doi.org/\-10.1063/\-1.2888113}

\item{} J. Zhou. Construction of a new detector, and calibration strategies, for start timing in STAR at RHIC. \href{https://scholarship.rice.edu/handle/1911/20548}{https://scholarship.rice.edu/\-handle/\-1911/\-20548}

\item{} W.J. Llope et al. STAR VPD. \emph{NIM A759}, 23. \href{https://arxiv.org/abs/1403.6855}{https://arxiv.org/\-abs/\-1403.6855}

\item{} F. Bergsma et al. The STAR detector magnet subsystem. \emph{Nucl. Instr. Meth. A499}, 633\textendash{}639. \href{https://doi.org/10.1016/S0168-9002(02)01961-7}{https://doi.org/\-10.1016/\-S0168-{}9002(02)01961-{}7}

\item{} A. Schmah. 2011. STAR Setup Rendering. Private Communication.

\item{} M. Anderson et al. STAR TPC: unique tool for studying high multiplicity events at RHIC. \emph{NIM A499}, 659. \href{https://doi.org/10.1016/S0168-9002(02)01964-2}{https://doi.org/\-10.1016/\-S0168-{}9002(02)01964-{}2}

\item{} PDG. Particle Phys. Rev. \emph{PRD 86}, 1504. \href{https://doi.org/10.1103/PhysRevD.86.010001}{https://doi.org/\-10.1103/\-PhysRevD.86.010001}

\item{} H. Bichsel. Energy Loss. \href{https://drupal.star.bnl.gov/STAR/starnotes/public/sn0418}{https://drupal.star.bnl.gov/\-STAR/\-starnotes/\-public/\-sn0418}

\item{} H. Bichsel. STAR Public Note 0439: Comparison of Bethe-{}Bloch and Bichsel Functions. \href{https://drupal.star.bnl.gov/STAR/starnotes/public/sn0439}{https://drupal.star.bnl.gov/\-STAR/\-starnotes/\-public/\-sn0439}

\item{} M. Anderson et al. A readout system for the STAR time projection chamber. \emph{NIM A499}, 679. \href{https://doi.org/10.1016/S0168-9002(02)01965-4}{https://doi.org/\-10.1016/\-S0168-{}9002(02)01965-{}4}

\item{} P. Yepes. An algorithm for fast track pattern recognition. \emph{Nucl. Instr. Meth. A380}, 582. \href{https://doi.org/10.1016/0168-9002(96)00726-7}{https://doi.org/\-10.1016/\-0168-{}9002(96)00726-{}7}

\item{} A. Rose. STAR integrated tracker. \emph{CHEP03}. \href{http://arxiv.org/abs/nucl-ex/0307015}{http://arxiv.org/\-abs/\-nucl-{}ex/\-0307015}

\item{} STAR TOF proposal. \href{https://www.star.bnl.gov/public/tof}{https://www.star.bnl.gov/\-public/\-tof}

\item{} W.J. Llope et al. The TOFp/pVPD time-{}of-{}flight system for STAR. \emph{NIM A522}, 252\textendash{}273. \href{https://doi.org/10.1016/j.nima.2003.11.414}{https://doi.org/\-10.1016/\-j.nima.2003.11.414}

\item{} F.S. Bieser et al. The STAR trigger. \emph{Nuclear Instruments and Methods in Physics A499}, 766\textendash{}777. www.star.bnl.gov/public/tpc/NimPapers/trigger/trigger\_nim.pdf

\item{} C. Adler et al. The STAR Level-{}3 trigger system. \emph{Nucl. Instrum. Meth. A499}, 778\textendash{}791. \href{https://doi.org/10.1016/S0168-9002(02)01975-7}{https://doi.org/\-10.1016/\-S0168-{}9002(02)01975-{}7}

\item{} V. Fine et al. OO Model of the STAR offline production "Event Display" and its implementation based on Qt-{}ROOT. \emph{CoRR}. \href{http://arxiv.org/abs/cs.HC/0306087}{http://arxiv.org/\-abs/\-cs.HC/\-0306087}

\item{} J. Landgraf et al. An overview of the STAR DAQ system. \emph{Nucl. Instr. Meth. A499}, 762\textendash{}765. \href{https://doi.org/10.1016/S0168-9002(02)01973-3}{https://doi.org/\-10.1016/\-S0168-{}9002(02)01973-{}3}

\item{} B. Gibbard and T. Throwe. The RHIC computing facility. \emph{Nucl. Instr. Meth. A499}, 814\textendash{}818. \href{https://doi.org/10.1016/S0168-9002(02)01978-2}{https://doi.org/\-10.1016/\-S0168-{}9002(02)01978-{}2}

\item{} High Performance Storage System. \href{http://www.hpss-collaboration.org}{http://www.hpss-{}collaboration.org}

\item{} A.W. Chan et al. The linux farms of the RHIC computing facility. \emph{NIM A499}, 819\textendash{}824. \href{https://doi.org/10.1016/S0168-9002(02)01979-4}{https://doi.org/\-10.1016/\-S0168-{}9002(02)01979-{}4}

\item{} C. Pruneau et al. A new STAR event reconstruction chain. \emph{CERN Document Server}. \href{http://cds.cern.ch/record/688747/files/CERN-2005-002-V1.pdf?version=2}{http://cds.cern.ch/\-record/\-688747/\-files/\-CERN-{}2005-{}002-{}V1.pdf?version=2}

\item{} STAR Reconstruction. \href{https://drupal.star.bnl.gov/STAR/comp/reco}{https://drupal.star.bnl.gov/\-STAR/\-comp/\-reco}

\item{} \href{https://drupal.star.bnl.gov/STAR/comp/sofi/tutorials/stevent-special-documentation}{https://drupal.star.bnl.gov/\-STAR/\-comp/\-sofi/\-tutorials/\-stevent-{}special-{}documentation}

\item{} \href{http://www.star.bnl.gov/public/comp/meet/CM200308/MuDstTutorial.ps.gz}{http://www.star.bnl.gov/\-public/\-comp/\-meet/\-CM200308/\-MuDstTutorial.ps.gz}

\item{} ROOT. \href{http://root.cern.ch}{http://root.cern.ch}

\item{} LBL PicoDst. \href{http://rnc.lbl.gov/~xdong/SoftHadron/picoDst.html}{http://rnc.lbl.gov/\-\textasciitilde{}xdong/\-SoftHadron/\-picoDst.html}

\item{} A. Schmah and R. Reed. 7.7 GeV Background Studies. Private communication.

\item{} M. Miller et al. Glauber modeling in high energy NN collisions. \emph{Ann.Rev.Nucl.Part.Sci. 57}, 205. \href{https://arxiv.org/abs/nucl-ex/0701025}{https://arxiv.org/\-abs/\-nucl-{}ex/\-0701025}

\item{} STAR Collaboration. v$_{\text{2}}$ of identified hadrons in Au+Au at 7.7-{}62.4 GeV. \emph{PRC 88}, 014902. \href{https://arxiv.org/abs/1301.2348}{https://arxiv.org/\-abs/\-1301.2348}

\item{} H. Masui and A. Schmah. StRefMultCorr -{} Correction of Reference Multiplicities in STAR. www.star.bnl.gov/cgi-{}bin/protected/cvsweb.cgi/offline/users/hmasui/StRefMultCorr

\item{} StRefMultCorr. \href{http://cgit.the-huck.com/StRefMultCorr/tree}{http://cgit.the-{}huck.com/\-StRefMultCorr/\-tree}

\item{} F. Grubbs. Sample criteria for testing outlying observations. \emph{Ann.Math.Statist. 21}, 27\textendash{}58. \href{https://doi.org/10.1214/aoms/1177729885}{https://doi.org/\-10.1214/\-aoms/\-1177729885}

\item{} P. Huck. Grubbs test for run QA. \href{http://cgit.the-huck.com/RunQA/tree}{http://cgit.the-{}huck.com/\-RunQA/\-tree}

\item{} S. Voloshin et al. Collect. phenomena in non-{}central NN. \href{http://arxiv.org/abs/0809.2949}{http://arxiv.org/\-abs/\-0809.2949}

\item{} X. Cui. Dielectron v$_{\text{2}}$ in s$_{\text{NN}}$$^{\text{1/2}}$ = 200 GeV Au+Au at STAR. \emph{J.Phys.Conf.Ser. 446}, 012047. \href{https://doi.org/10.1088/1742-6596/446/1/012047}{https://doi.org/\-10.1088/\-1742-{}6596/\-446/\-1/\-012047}

\item{} P. Huck. StLeptonTreeMaker. \href{http://cgit.the-huck.com/StLeptonTreeMaker/tree}{http://cgit.the-{}huck.com/\-StLeptonTreeMaker/\-tree}

\item{} P. Huck. StDielectronMaker. \href{http://cgit.the-huck.com/StDielectronMaker/tree}{http://cgit.the-{}huck.com/\-StDielectronMaker/\-tree}

\item{} P. Huck. ElectronPid. \href{http://cgit.the-huck.com/ElectronPid/tree}{http://cgit.the-{}huck.com/\-ElectronPid/\-tree}

\item{} RooFit. \href{http://root.cern.ch/drupal/content/roofit}{http://root.cern.ch/\-drupal/\-content/\-roofit}

\item{} P. Huck. \href{http://cgit.the-huck.com/pyana/tree/pyana/dielec/dielec.py}{http://cgit.the-{}huck.com/\-pyana/\-tree/\-pyana/\-dielec/\-dielec.py}

\item{} M. Potekhin and O. Barannikova. GSTAR: STAR detector simulation framework using GEANT. \href{http://www.star.bnl.gov/public/comp/simu/gstar/gstar.html}{http://www.star.bnl.gov/\-public/\-comp/\-simu/\-gstar/\-gstar.html}

\item{} S. Agostinelli et al. Geant4 -{} A simulation toolkit. \emph{Nucl. Instr. Meth. A506}, 250\textendash{}303. \href{https://doi.org/10.1016/S0168-9002(03)01368-8}{https://doi.org/\-10.1016/\-S0168-{}9002(03)01368-{}8}

\item{} J. Allison et al. Geant4 developments and applications. \emph{IEEE Trans.Nucl.Sci. 53}, 270\textendash{}278. \href{https://doi.org/10.1109/TNS.2006.869826}{https://doi.org/\-10.1109/\-TNS.2006.869826}

\item{} W.G. Gong. Public STAR Note 0197: Slow Simulator for STAR Time Projection Chamber. \href{http://www.star.bnl.gov/public/comp/simu/TpcRespSim/src/papers/SN0197.pdf}{http://www.star.bnl.gov/\-public/\-comp/\-simu/\-TpcRespSim/\-src/\-papers/\-SN0197.pdf}

\item{} STAR Simulation Requests. \href{https://drupal.star.bnl.gov/STAR/starsimrequest}{https://drupal.star.bnl.gov/\-STAR/\-starsimrequest}

\item{} T. Ullrich and Z. Xu. Efficiency Calculations. \href{http://arxiv.org/abs/physics/0701199}{http://arxiv.org/\-abs/\-physics/\-0701199}

\item{} P. Huck. BesCocktail. \href{https://github.com/tschaume/BesCocktail}{https://github.com/\-tschaume/\-BesCocktail}

\item{} STAR Collaboration. Systematic measurements of identified particle spectra in p+p, d+Au and Au+Au from STAR. \emph{PRC 79}, 034909. \href{https://doi.org/10.1103/PhysRevC.79.034909}{https://doi.org/\-10.1103/\-PhysRevC.79.034909}

\item{} L. Kumar. Systematics of kinetic freeze-{}out properties in NN from STAR. \emph{NPA 931}, 1114. \href{https://arxiv.org/abs/1408.4209}{https://arxiv.org/\-abs/\-1408.4209}

\item{} E. Sangaline. \href{http://nuclear.ucdavis.edu/~sangaline/rcp_www/rawspectra}{http://nuclear.ucdavis.edu/\-\textasciitilde{}sangaline/\-rcp\_www/\-rawspectra}

\item{} J. Butterworth. 2014. Private communication.

\item{} G. Breit and E. Wigner. 1936. Capture of Slow Neutrons. \emph{Physical Reviews 49}, 519\textendash{}531. \href{https://doi.org/10.1103/PhysRev.49.519}{https://doi.org/\-10.1103/\-PhysRev.49.519}

\item{} N. Kroll and W. Wada. Internal pair production associated with the emission of high-{}energy gamma rays. \emph{Phys.Rev. 98}, 1355. \href{https://doi.org/10.1103/PhysRev.98.1355}{https://doi.org/\-10.1103/\-PhysRev.98.1355}

\item{} NA60 Collaboration. Study of the electromagnetic transition form-{}factors in \ensuremath{\eta}\ensuremath{\rightarrow}\ensuremath{\mu}\ensuremath{\mu}\ensuremath{\gamma} and \ensuremath{\omega}\ensuremath{\rightarrow}\ensuremath{\mu}\ensuremath{\mu}\ensuremath{\pi} decays. \emph{PLB 677}, 260\textendash{}266. \href{https://arxiv.org/abs/0902.2547}{https://arxiv.org/\-abs/\-0902.2547}

\item{} L.G. Landsberg. Electromagnetic decays of light mesons. \emph{Physics Reports 128}, 301\textendash{}376. \href{https://doi.org/10.1016/0370-1573(85)90129-2}{https://doi.org/\-10.1016/\-0370-{}1573(85)90129-{}2}

\item{} M.N. Achasov et al. Study of conversion decays \ensuremath{\phi}\ensuremath{\rightarrow}\ensuremath{\eta}e+e-{} \& \ensuremath{\eta}\ensuremath{\rightarrow}\ensuremath{\gamma}e+e-{} with SND Detector at VEPP-{}2M. \emph{PLB 504}, 275\textendash{}281. \href{https://doi.org/10.1016/S0370-2693(01)00320-3}{https://doi.org/\-10.1016/\-S0370-{}2693(01)00320-{}3}

\item{} J. Gaiser. Charmonium spectroscopy from radiative decays of the J/\ensuremath{\psi} and \ensuremath{\psi}' mesons. \href{http://www.slac.stanford.edu/pubs/slacpubs/2750/slac-pub-2899.pdf}{http://www.slac.stanford.edu/\-pubs/\-slacpubs/\-2750/\-slac-{}pub-{}2899.pdf}

\item{} M. Oreglia. \ensuremath{\psi}'\ensuremath{\rightarrow}\ensuremath{\gamma}\ensuremath{\gamma}\ensuremath{\psi} \href{http://www.slac.stanford.edu/cgi-wrap/getdoc/slac-r-236.pdf}{http://www.slac.stanford.edu/\-cgi-{}wrap/\-getdoc/\-slac-{}r-{}236.pdf}

\item{} T. Skwarnicki. Study of radiative CASCADE transitions between the Y' and Y resonances. \href{http://lss.fnal.gov/cgi-bin/find_paper.pl?other/thesis/skwarnicki.pdf}{http://lss.fnal.gov/\-cgi-{}bin/\-find\_paper.pl?other/\-thesis/\-skwarnicki.pdf}

\item{} T. Sjöstrand et al. PYTHIA. \emph{JHEP 05}, 026. \href{https://arxiv.org/abs/hep-ph/0603175}{https://arxiv.org/\-abs/\-hep-{}ph/\-0603175}

\item{} STAR Collaboration. Measurements of D$^{\text{0}}$/D$^{\text{*}}$ production in p+p at 200 GeV. \emph{PRD 86}, 072013. \href{https://doi.org/10.1103/PhysRevD.86.072013}{https://doi.org/\-10.1103/\-PhysRevD.86.072013}

\item{} R. Field and R. Group. PYTHIA tune A at CDF. \href{http://arxiv.org/abs/hep-ph/0510198}{http://arxiv.org/\-abs/\-hep-{}ph/\-0510198}

\item{} E. Norrbin and T. Sjostrand. Production of charm hadrons in the string model. \emph{PLB 442}, 407. \href{https://arxiv.org/abs/hep-ph/9809266}{https://arxiv.org/\-abs/\-hep-{}ph/\-9809266}

\item{} E. Norrbin and T. Sjostrand. Production and hadronization of heavy quarks. \emph{EPJC 17}, 137\textendash{}161. \href{https://arxiv.org/abs/hep-ph/0005110}{https://arxiv.org/\-abs/\-hep-{}ph/\-0005110}

\item{} B. Andersson et al. Parton fragmentation and string dynamics. \emph{Phys.Rep. 97}, 31\textendash{}145. \href{https://doi.org/10.1016/0370-1573(83)90080-7}{https://doi.org/\-10.1016/\-0370-{}1573(83)90080-{}7}

\item{} R. Rapp. Vector probe. \emph{JPG 31}, S217. \href{https://doi.org/10.1088/0954-3899/31/4/027}{https://doi.org/\-10.1088/\-0954-{}3899/\-31/\-4/\-027}

\item{} R. Rapp. 2014. \ensuremath{\rho} in-{}medium model calculations for BES-{}I. Private Communication.

\item{} PDG. QCD Review. \href{http://pdg.lbl.gov/2014/reviews/rpp2014-rev-qcd.pdf}{http://pdg.lbl.gov/\-2014/\-reviews/\-rpp2014-{}rev-{}qcd.pdf}

\item{} PDG. \href{http://pdg.lbl.gov/2014/hadronic-xsections/rpp2014-sigma_R_ee_plots.pdf}{http://pdg.lbl.gov/\-2014/\-hadronic-{}xsections/\-rpp2014-{}sigma\_R\_ee\_plots.pdf}

\item{} P.A. Baikov et al. Adler function, sum rules and Crewther relation of order. \emph{PLB 714}, 62. \href{https://doi.org/10.1016/j.physletb.2012.06.052}{https://doi.org/\-10.1016/\-j.physletb.2012.06.052}

\item{} R. Rapp et al. In-{}medium excitations. \emph{Compressed Baryonic Matter Physics Book 814}, 335\textendash{}529. \href{https://doi.org/10.1007/978-3-642-13293-3_4}{https://doi.org/\-10.1007/\-978-{}3-{}642-{}13293-{}3\_4}

\item{} L. McLerran and T. Toimela. Photon and dilepton emission from the quark-{}gluon plasma: Some general considerations. \emph{PRD 31}, 545. \href{https://doi.org/10.1103/PhysRevD.31.545}{https://doi.org/\-10.1103/\-PhysRevD.31.545}

\item{} H.T. Ding et al. Thermal dilepton rate \& electr. conductivity: Analysis of vector current correlators in quenched lQCD. \emph{PRD 83}, 034504. \href{https://arxiv.org/abs/1012.4963}{https://arxiv.org/\-abs/\-1012.4963}

\item{} NA60 Collaboration. Thermal dimuons. \emph{EPJC 61}, 711. \href{http://arxiv.org/abs/0812.3053}{http://arxiv.org/\-abs/\-0812.3053}

\item{} F.-{}Q. Wang and N. Xu. Baryon phase space density in heavy ion collisions. \emph{PRC 61}, 021904. \href{https://arxiv.org/abs/nucl-th/9909002}{https://arxiv.org/\-abs/\-nucl-{}th/\-9909002}

\item{} P.M. Hohler and R. Rapp. Is \ensuremath{\rho}-{}meson melting compatible with \ensuremath{\chi}R?. \emph{PLB 731}, 103\textendash{}109. \href{https://arxiv.org/abs/1311.2921}{https://arxiv.org/\-abs/\-1311.2921}

\item{} P. Huck. StBadRdosDb. \href{http://cgit.the-huck.com/StBadRdosDb/tree}{http://cgit.the-{}huck.com/\-StBadRdosDb/\-tree}

\item{} \href{http://www.star.bnl.gov/protected/common/common2010/trigger2010/runlists}{http://www.star.bnl.gov/\-protected/\-common/\-common2010/\-trigger2010/\-runlists}

\item{} STAR RunLog Browser. \href{http://online.star.bnl.gov/RunLog}{http://online.star.bnl.gov/\-RunLog}

\item{} P. Huck. StRunIdEventsDb. \href{http://cgit.the-huck.com/cgit.cgi/StRunIdEventsDb}{http://cgit.the-{}huck.com/\-cgit.cgi/\-StRunIdEventsDb}

\item{} Concurrent Versions System. \href{http://ximbiot.com/cvs/manual/cvs-1.12.12/cvs_1.html}{http://ximbiot.com/\-cvs/\-manual/\-cvs-{}1.12.12/\-cvs\_1.html}

\item{} Git Distributed Revision Control System. \href{http://git-scm.com}{http://git-{}scm.com}

\item{} Git Submodules. \href{http://git-scm.com/book/en/Git-Tools-Submodules}{http://git-{}scm.com/\-book/\-en/\-Git-{}Tools-{}Submodules}

\item{} Google Repo Script. \href{https://dl-ssl.google.com/dl/googlesource/git-repo/repo}{https://dl-{}ssl.google.com/\-dl/\-googlesource/\-git-{}repo/\-repo}

\item{} Wikipedia Repo Script. \href{http://en.wikipedia.org/wiki/Repo_%28script%29}{http://en.wikipedia.org/\-wiki/\-Repo\_\%28script\%29}

\item{} Android Open Source Project. \href{http://source.android.com/source/developing.html}{http://source.android.com/\-source/\-developing.html}

\item{} Gerrit. \href{https://www.gerritcodereview.com}{https://www.gerritcodereview.com}

\item{} Jenkins Trigger for Gerrit. \href{https://wiki.jenkins-ci.org/display/JENKINS/Gerrit+Trigger}{https://wiki.jenkins-{}ci.org/\-display/\-JENKINS/\-Gerrit+Trigger}

\item{} Using Gerrit Git Review with Jenkins CI Server. \href{http://vimeo.com/20084957}{http://vimeo.com/\-20084957}

\item{} CGit Web Frontend. \href{http://git.zx2c4.com/cgit/about}{http://git.zx2c4.com/\-cgit/\-about}

\item{} kernel.org announces use of CGit. \href{https://www.kernel.org/pelican.html}{https://www.kernel.org/\-pelican.html}

\item{} kernel.org CGit Browser. \href{https://git.kernel.org/cgit}{https://git.kernel.org/\-cgit}

\item{} Gerrit Kerberos. \href{https://groups.google.com/forum/#!topic/repo-discuss/DF-0moQaJ2I}{https://groups.google.com/\-forum/\-\#!topic/\-repo-{}discuss/\-DF-{}0moQaJ2I}

\item{} Gitolite -{} Git Repository Hosting. \href{http://gitolite.com/gitolite}{http://gitolite.com/\-gitolite}

\item{} CGit Configuration Options. \href{http://git.zx2c4.com/cgit/tree/cgitrc.5.txt}{http://git.zx2c4.com/\-cgit/\-tree/\-cgitrc.5.txt}

\item{} GitLab -{} Self hosted Git Management Software. \href{http://gitlab.org}{http://gitlab.org}

\item{} GitHub -{} Social Coding. \href{https://github.com}{https://github.com}

\item{} \href{https://library.linode.com/hosting-website#sph_configuring-name-based-virtual-hosts}{https://library.linode.com/\-hosting-{}website\#sph\_configuring-{}name-{}based-{}virtual-{}hosts}

\item{} CGit README. \href{http://git.zx2c4.com/cgit/tree/README}{http://git.zx2c4.com/\-cgit/\-tree/\-README}

\item{} cvs2git Documentation. \href{http://cvs2svn.tigris.org/cvs2git.html}{http://cvs2svn.tigris.org/\-cvs2git.html}

\item{} Download and Install Repo. \href{http://source.android.com/source/downloading.html}{http://source.android.com/\-source/\-downloading.html}

\item{} Developing with Repo. \href{http://source.android.com/source/developing.html}{http://source.android.com/\-source/\-developing.html}

\item{} Using Repo. \href{http://source.android.com/source/using-repo.html}{http://source.android.com/\-source/\-using-{}repo.html}

\item{} Gerrit -{} Life of a Patch. \href{http://source.android.com/source/life-of-a-patch.html}{http://source.android.com/\-source/\-life-{}of-{}a-{}patch.html}

\item{} Gerrit -{} Submit Patches. \href{http://source.android.com/source/submit-patches.html}{http://source.android.com/\-source/\-submit-{}patches.html}

\item{} ACLiC. \href{http://root.cern.ch/download/doc/ROOTUsersGuideHTML/ch07s09.html}{http://root.cern.ch/\-download/\-doc/\-ROOTUsersGuideHTML/\-ch07s09.html}

\item{} P. Huck. ROOT and Autoconf. \href{https://github.com/tschaume/autoconf-analysis-root}{https://github.com/\-tschaume/\-autoconf-{}analysis-{}root}

\item{} Automake w/ C++. www.openismus.com/documents/linux/automake/automake

\item{} Doxygen. \href{http://www.stack.nl/~dimitri/doxygen/index.html}{http://www.stack.nl/\-\textasciitilde{}dimitri/\-doxygen/\-index.html}

\item{} Autoconf Manual. \href{http://www.gnu.org/software/hello/manual/autoconf}{http://www.gnu.org/\-software/\-hello/\-manual/\-autoconf}

\item{} Automake Manual. \href{http://www.gnu.org/software/hello/manual/automake}{http://www.gnu.org/\-software/\-hello/\-manual/\-automake}

\item{} Automake Examples. \href{http://realmike.org/blog/2010/07/18/gnu-automake-by-example}{http://realmike.org/\-blog/\-2010/\-07/\-18/\-gnu-{}automake-{}by-{}example}

\item{} \href{http://chris-miceli.blogspot.com/2011/01/integrating-doxygen-with-autotools.html}{http://chris-{}miceli.blogspot.com/\-2011/\-01/\-integrating-{}doxygen-{}with-{}autotools.html}

\item{} hSimple script. \href{http://root.cern.ch/svn/root/tags/v5-24-00b/tutorials/hsimple.C}{http://root.cern.ch/\-svn/\-root/\-tags/\-v5-{}24-{}00b/\-tutorials/\-hsimple.C}

\item{} cmdline. \href{https://github.com/smfr/mactierra/tree/master/Source/cmdline/options}{https://github.com/\-smfr/\-mactierra/\-tree/\-master/\-Source/\-cmdline/\-options}

\item{} autoconf. \href{http://www.gnu.org/software/autoconf}{http://www.gnu.org/\-software/\-autoconf}

\item{} m4. \href{http://www.gnu.org/software/autoconf-archive/The-Macros.html#The-Macros}{http://www.gnu.org/\-software/\-autoconf-{}archive/\-The-{}Macros.html\#The-{}Macros}

\item{} Boost C++ Libraries. \href{http://www.boost.org}{http://www.boost.org}

\item{} A YAML parser and emitter for C++. \href{https://code.google.com/p/yaml-cpp}{https://code.google.com/\-p/\-yaml-{}cpp}

\item{} ckon -{} Automatic Build Tool for ROOT Analyses. \href{https://github.com/tschaume/ckon}{https://github.com/\-tschaume/\-ckon}

\item{} ckon -{} Read the Docs. \href{https://ckon.readthedocs.org}{https://ckon.readthedocs.org}

\item{} MIT License. \href{http://opensource.org/licenses/MIT}{http://opensource.org/\-licenses/\-MIT}

\item{} Pythia6. \href{http://home.thep.lu.se/~torbjorn/Pythia.html}{http://home.thep.lu.se/\-\textasciitilde{}torbjorn/\-Pythia.html}

\item{} ckon -{} Submit Issue. \href{https://github.com/tschaume/ckon/issues}{https://github.com/\-tschaume/\-ckon/\-issues}

\item{} P. Huck. Carbon Capture \& Sequestr. GnuPlot. \href{https://github.com/tschaume/ccsgp}{https://github.com/\-tschaume/\-ccsgp}

\item{} J. Huck et al. Evaluating porous materials for carbon capture. \emph{Energy \& Environmental Science 7}, 4132. \href{https://doi.org/10.1039/C4EE02636E}{https://doi.org/\-10.1039/\-C4EE02636E}

\item{} P. Huck. Get started with ccsgp. \href{https://github.com/tschaume/ccsgp_get_started}{https://github.com/\-tschaume/\-ccsgp\_get\_started}

\item{} P. Huck. ccsgp Documentation. \href{http://ccsgp-get-started.readthedocs.org}{http://ccsgp-{}get-{}started.readthedocs.org}

\item{} AsciiDoc. \href{http://www.methods.co.nz/asciidoc}{http://www.methods.co.nz/\-asciidoc}

\item{} DocBook. \href{http://www.docbook.org}{http://www.docbook.org}

\item{} a2x tool-{}chain. \href{http://www.methods.co.nz/asciidoc/a2x.1.html}{http://www.methods.co.nz/\-asciidoc/\-a2x.1.html}

\item{} \href{https://groups.google.com/forum/#!topic/asciidoc/Sio9vxclWkY/discussion}{https://groups.google.com/\-forum/\-\#!topic/\-asciidoc/\-Sio9vxclWkY/\-discussion}

\item{} wp-{}pdf. \href{https://github.com/tschaume/wp-pdf}{https://github.com/\-tschaume/\-wp-{}pdf}

\item{} git-{}annex. \href{http://git-annex.branchable.com}{http://git-{}annex.branchable.com}

\item{} git-{}annex-{}downloads. \href{https://github.com/tschaume/git-annex-downloads}{https://github.com/\-tschaume/\-git-{}annex-{}downloads}

\item{} Create MySQL DB. \href{https://library.linode.com/hosting-website#sph_creating-a-database}{https://library.linode.com/\-hosting-{}website\#sph\_creating-{}a-{}database}

\item{} Wordpress Install. \href{https://library.linode.com/web-applications/cms-guides/wordpress}{https://library.linode.com/\-web-{}applications/\-cms-{}guides/\-wordpress}

\item{} \href{http://codex.wordpress.org/Editing_wp-config.php#WordPress_Upgrade_Constants}{http://codex.wordpress.org/\-Editing\_wp-{}config.php\#WordPress\_Upgrade\_Constants}

\item{} BlogPost README. \href{http://srackham.wordpress.com/blogpost1}{http://srackham.wordpress.com/\-blogpost1}

\item{} BlogPost Source. \href{https://code.google.com/p/blogpost}{https://code.google.com/\-p/\-blogpost}

\item{} \href{http://www.mail-archive.com/asciidoc@googlegroups.com/msg00104.html}{http://www.mail-{}archive.com/\-asciidoc@googlegroups.com/\-msg00104.html}

\item{} Customized BlogPost. \href{https://github.com/tschaume/blogpost}{https://github.com/\-tschaume/\-blogpost}

\item{} StRootMerger and rmrg. \href{https://github.com/tschaume/rmrg}{https://github.com/\-tschaume/\-rmrg}

\item{} Setup Analysis. \href{http://star.the-huck.com/appendices/setup-analysis}{http://star.the-{}huck.com/\-appendices/\-setup-{}analysis}

\item{} Protected Data Repo. \href{http://cgit.the-huck.com/dielectron_data_protected/tree}{http://cgit.the-{}huck.com/\-dielectron\_data\_protected/\-tree}

\item{} P. Huck. Thesis Supporting Information. \href{http://star.the-huck.com/download}{http://star.the-{}huck.com/\-download}

\item{} hex-{}color scale. \href{http://thadeusb.com/weblog/2010/10/10/python_scale_hex_color}{http://thadeusb.com/\-weblog/\-2010/\-10/\-10/\-python\_scale\_hex\_color} 

\end{enumerate}


\chapter{Xenon Excimer Lamp for HADES RICH}
\label{hardware}\hyperlabel{hardware}%
\begin{itemize}[itemsep=0pt]

\item{} Dec '09 -{} Jul '10, Kruecken, Friese (TU Munich)

\item{} DPG spring meeting 2001, abstract submission
 \footnote{
\href{http://www.dpg-verhandlungen.de/year/2011/conference/muenster/part/hk/session/39/contribution/34}{http://www.dpg-{}verhandlungen.de/\-year/\-2011/\-conference/\-muenster/\-part/\-hk/\-session/\-39/\-contribution/\-34}
}, poster
 \footnote{
\href{http://www.star.bnl.gov/protected/lfspectra/huck/Talks/DPG11_poster_druck.jpg}{http://www.star.bnl.gov/\-protected/\-lfspectra/\-huck/\-Talks/\-DPG11\_poster\_druck.jpg}
}

\item{} new VUV lamp for single photon measurement with HADES RICH

\item{} Hardware development and design project

\item{} 13 years of HADES RICH operation (MWPC w/ CsI photo cathode) and renewal of Front-{}end electronics

\begin{itemize}

\item{} necessitates measurement of cathode pads response on single photons

\item{} develop light source for single cherenkov typical VUV photons (160nm)

\end{itemize}

\item{} requirements:

\begin{itemize}

\item{} trigger for readout electronics

\item{} isotropic and full detector coverage

\item{} externally adjustable light emission/tuning

\end{itemize}

\item{} technique: Xenon Excimer Emission via alpha radiation (2.4cm radiation length)

\item{} R\&D and design

\begin{itemize}

\item{} diaphragm valve

\item{} flange for RICH mounting

\item{} LEMO readout electronics and external control

\item{} getters for Xenon gas purity improvements through heating

\item{} piezo motor for light emission control

\item{} invar inlet for Quarz glass -{} metal connection

\item{} 241Am alpha radiation source

\item{} Si-{}PiN-{}Diode for triggering (fast trigger signal 100ns)

\end{itemize}

\item{} assembly steps in cleanroom (glas-{}metal soldering)

\item{} energy deposition measurements for variable light output and trigger rate

\item{} on-{}site RICH MWPC pad-{}by-{}pad signals / measurements (voltage dependence \ldots{})

\begin{itemize}

\item{} single photon spectra

\item{} homogenous RICH MWPC illumination through reflection on VUV mirror

\item{} photon density, mean quantum efficiency, number of photo-{}electrons, trigger rate

\item{} pad-{}by-{}pad: high statistics, gas amplification, thresholds, detection efficiency

\end{itemize}

\end{itemize}

\wrapifneeded{0.50}{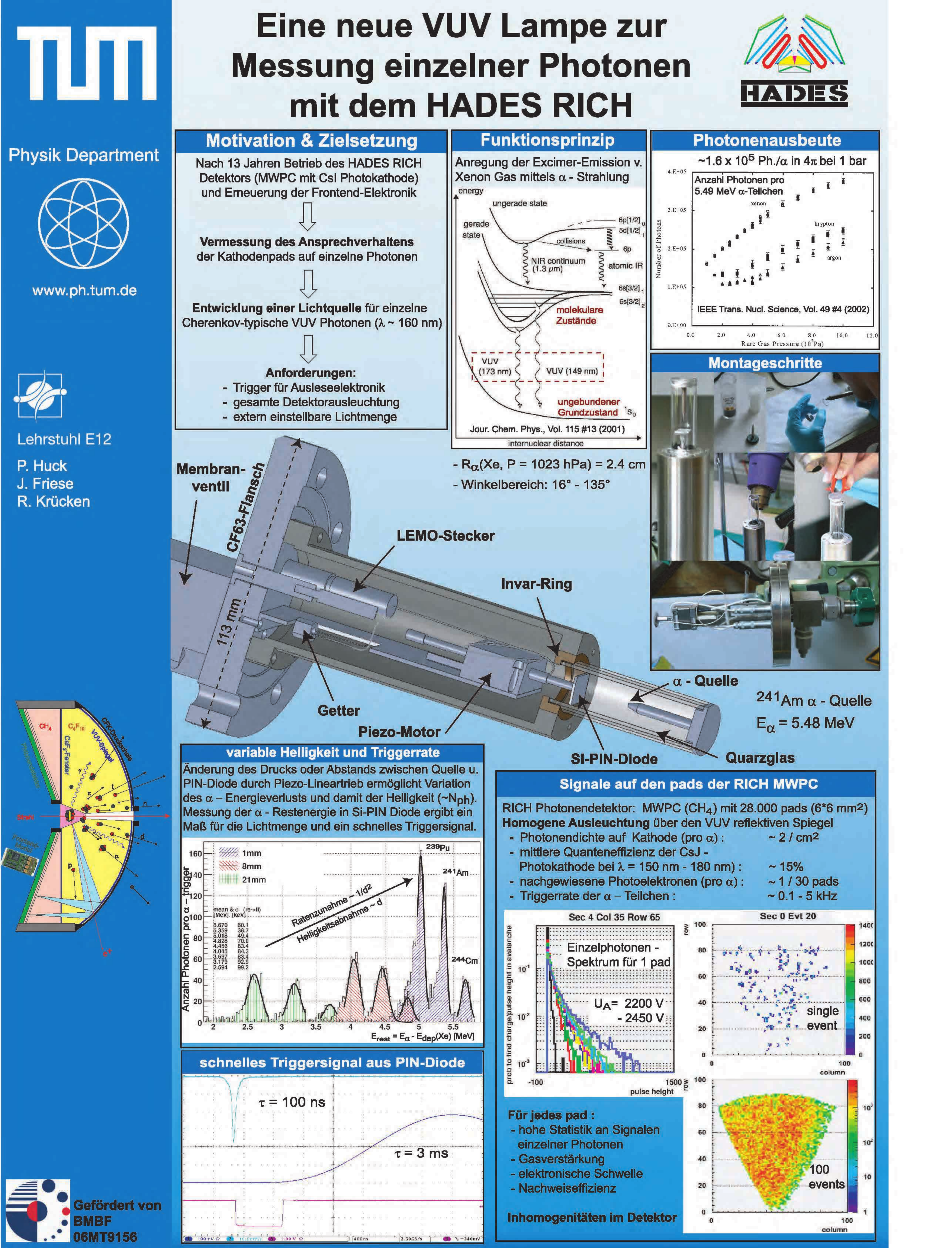}{DPG11 poster}{}{0.8} %
\begin{appendices}

\chapter{Physics}
\label{_physics}\hyperlabel{_physics}%

\section{Technical FAQs}
\label{_technical_faqs}\hyperlabel{_technical_faqs}%

\noindent
\begin{description}
\item[{ Relevant URLs
}] ~\begin{itemize}[itemsep=0pt]

\item{} \href{http://cgit.the-huck.com/}{cgit.the-{}huck.com}: version-{}controlled code \& data repositories
 \footnote{
clone any repo via \texttt{git c\penalty5000 l\penalty5000 o\penalty5000 n\penalty5000 e h\penalty5000 ttp:\penalty0 /\penalty0 /\penalty0 pro\penalty5000 t\penalty5000 e\penalty5000 c\penalty5000 ted@\penalty0 cgit.\penalty0 the-{}\penalty0 huck.\penalty0 com/\penalty0 <{}repo-{}\penalty0 name>{}}
}

\item{} \href{http://downloads.the-huck.com/star/}{downloads.the-{}huck.com/\-star}:
  download page containing all pictures, talks etc

\item{} \href{http://star.the-huck.com/}{star.the-{}huck.com}: central analysis webpage

\item{} Analysis Chain for raw data [282].

\end{itemize}
\item[{ Designated Data Repository
}] \hspace{0em}\\
 The \texttt{die\penalty5000 l\penalty5000 e\penalty5000 c\penalty5000 t\penalty5000 r\penalty5000 o\penalty5000 n\penalty5000 \_\penalty5000 d\penalty5000 a\penalty5000 t\penalty5000 a\penalty5000 \_\penalty5000 p\penalty5000 r\penalty5000 o\penalty5000 t\penalty5000 e\penalty5000 c\penalty5000 ted} repository provided at
[283] contains all necessary text files to go through a
step-{}by-{}step comparison with the raw data analysis and the efficiency
correction as well as the final physics plots. The included README is a good
starting point explaining directory structure and notations.  For the
comparison of background ratio plots at 19 GeV, for instance, the README refers
to \texttt{raw\penalty5000 d\penalty5000 ata/\penalty0 19} sub-{}directory. Same for the single track efficiencies.
M$_{\text{ee}}$-{}dependent ratios are also included (\texttt{rgm\penalty5000 , r\penalty5000 m\penalty5000 m\penalty5000 , rpp}) along with the
acceptance correction factor (\texttt{ac}). Efficiency-{}corrected data is collected in
the input sub-{}directories of \texttt{exa\penalty5000 m\penalty5000 p\penalty5000 les/\penalty0 } (mostly \texttt{gp\_\penalty5000 p\penalty5000 a\penalty5000 nel}).

\end{description}

\pagebreak[4]

\section{Bad Run Tables}
\label{app_badruns}\hyperlabel{app_badruns}%
\begin{footnotesize}
\begin{center}
\begingroup%
\setlength{\newtblsparewidth}{\linewidth-2\tabcolsep-2\tabcolsep-2\tabcolsep-2\tabcolsep-2\tabcolsep-2\tabcolsep-2\tabcolsep-2\tabcolsep-2\tabcolsep}%
\setlength{\newtblstarfactor}{\newtblsparewidth / \real{96}}%

\begin{longtable}{llllllll}\caption[{List of rejected runs at BES-{}I energies in \texttt{StR\penalty5000 e\penalty5000 f\penalty5000 M\penalty5000 u\penalty5000 l\penalty5000 t\penalty5000 C\penalty5000 orr} and refined for this analysis.}]{List of rejected runs at BES-{}I energies in \texttt{StR\penalty5000 e\penalty5000 f\penalty5000 M\penalty5000 u\penalty5000 l\penalty5000 t\penalty5000 C\penalty5000 orr} and refined for this analysis.\label{app_tab_badruns}\hyperlabel{app_tab_badruns}%
}\tabularnewline
\endfirsthead
\caption[]{(continued)}\tabularnewline
\endhead
\hline
\multicolumn{4}{m{12\newtblstarfactor+2\tabcolsep+\arrayrulewidth+12\newtblstarfactor+2\tabcolsep+\arrayrulewidth+12\newtblstarfactor+2\tabcolsep+\arrayrulewidth+12\newtblstarfactor+\arrayrulewidth}}{\centering%
\textbf{\texttt{StR\penalty5000 e\penalty5000 f\penalty5000 M\penalty5000 u\penalty5000 l\penalty5000 t\penalty5000 C\penalty5000 orr}}
}&\multicolumn{4}{m{12\newtblstarfactor+2\tabcolsep+\arrayrulewidth+12\newtblstarfactor+2\tabcolsep+\arrayrulewidth+12\newtblstarfactor+2\tabcolsep+\arrayrulewidth+12\newtblstarfactor+\arrayrulewidth}}{\centering%
\textbf{additional RunIDs rejected}
}\tabularnewline
\multicolumn{2}{m{12\newtblstarfactor+2\tabcolsep+\arrayrulewidth+12\newtblstarfactor+\arrayrulewidth}}{\centering%
\textbf{2010}
}&\multicolumn{2}{m{12\newtblstarfactor+2\tabcolsep+\arrayrulewidth+12\newtblstarfactor+\arrayrulewidth}}{\centering%
\textbf{2011}
}&\multicolumn{1}{m{12\newtblstarfactor+\arrayrulewidth}}{\centering%
\textbf{19.6 GeV}
}&\multicolumn{1}{m{12\newtblstarfactor+\arrayrulewidth}}{\centering%
\textbf{27 GeV}
}&\multicolumn{1}{m{12\newtblstarfactor+\arrayrulewidth}}{\centering%
\textbf{39 GeV}
}&\multicolumn{1}{m{12\newtblstarfactor+\arrayrulewidth}}{\centering%
\textbf{62.4 GeV}
}\tabularnewline
\multicolumn{1}{m{12\newtblstarfactor+\arrayrulewidth}}{\centering%
11080072
}&\multicolumn{1}{m{12\newtblstarfactor+\arrayrulewidth}}{\centering%
11095005
}&\multicolumn{1}{m{12\newtblstarfactor+\arrayrulewidth}}{\centering%
12113091
}&\multicolumn{1}{m{12\newtblstarfactor+\arrayrulewidth}}{\centering%
12136034
}&\multicolumn{1}{m{12\newtblstarfactor+\arrayrulewidth}}{\centering%
12113082
}&\multicolumn{1}{m{12\newtblstarfactor+\arrayrulewidth}}{\centering%
12172044
}&\multicolumn{1}{m{12\newtblstarfactor+\arrayrulewidth}}{\centering%
11102107
}&\multicolumn{1}{m{12\newtblstarfactor+\arrayrulewidth}}{\centering%
11080055
}\tabularnewline
\hline
\end{longtable}\endgroup%

\end{center}
\end{footnotesize}

\pagebreak[4]

\section{Run QA Figures}
\label{app_runqa}\hyperlabel{app_runqa}%

\wrapifneeded{0.50}{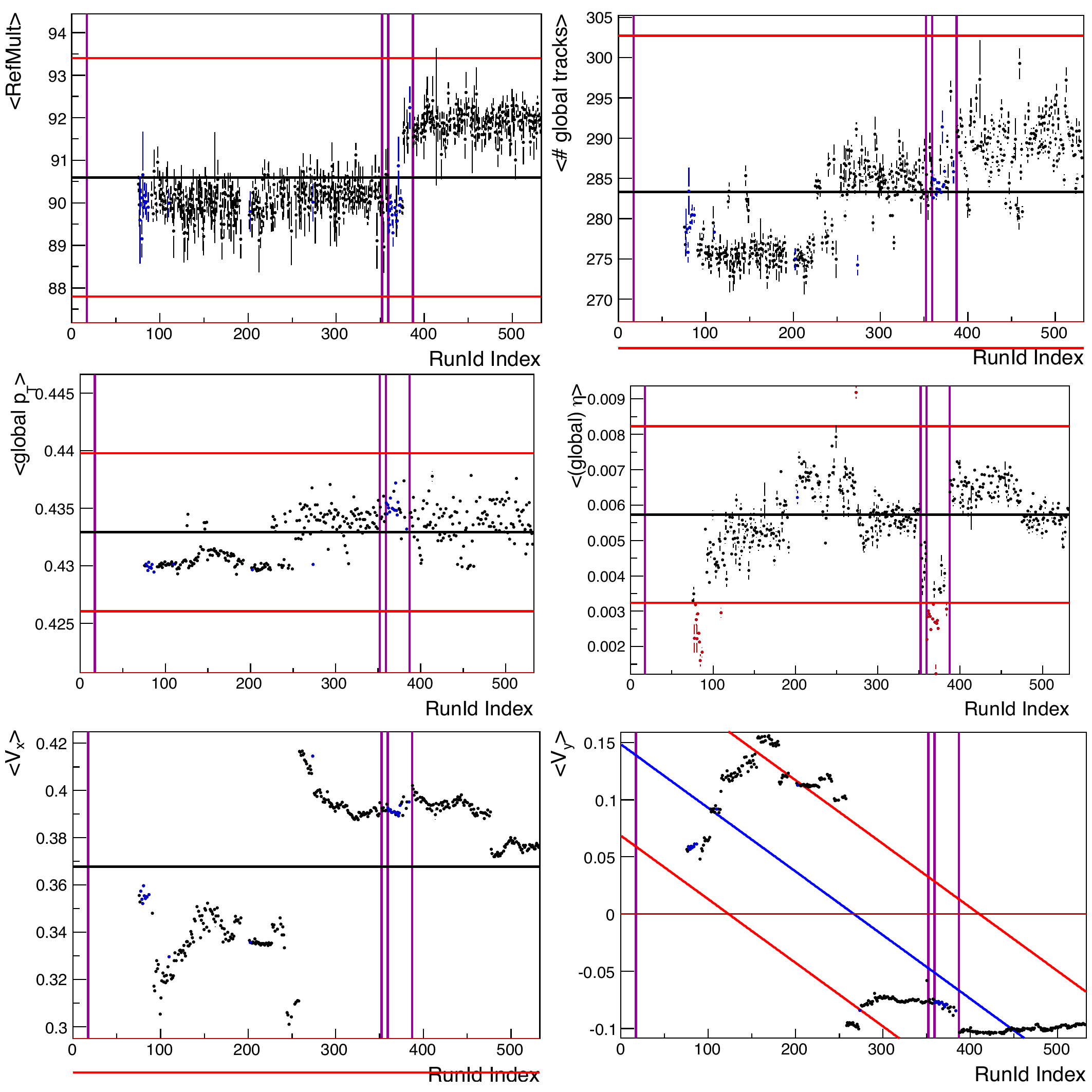}{Run QA 19.6 GeV (Part I)}{}{1} %

\wrapifneeded{0.50}{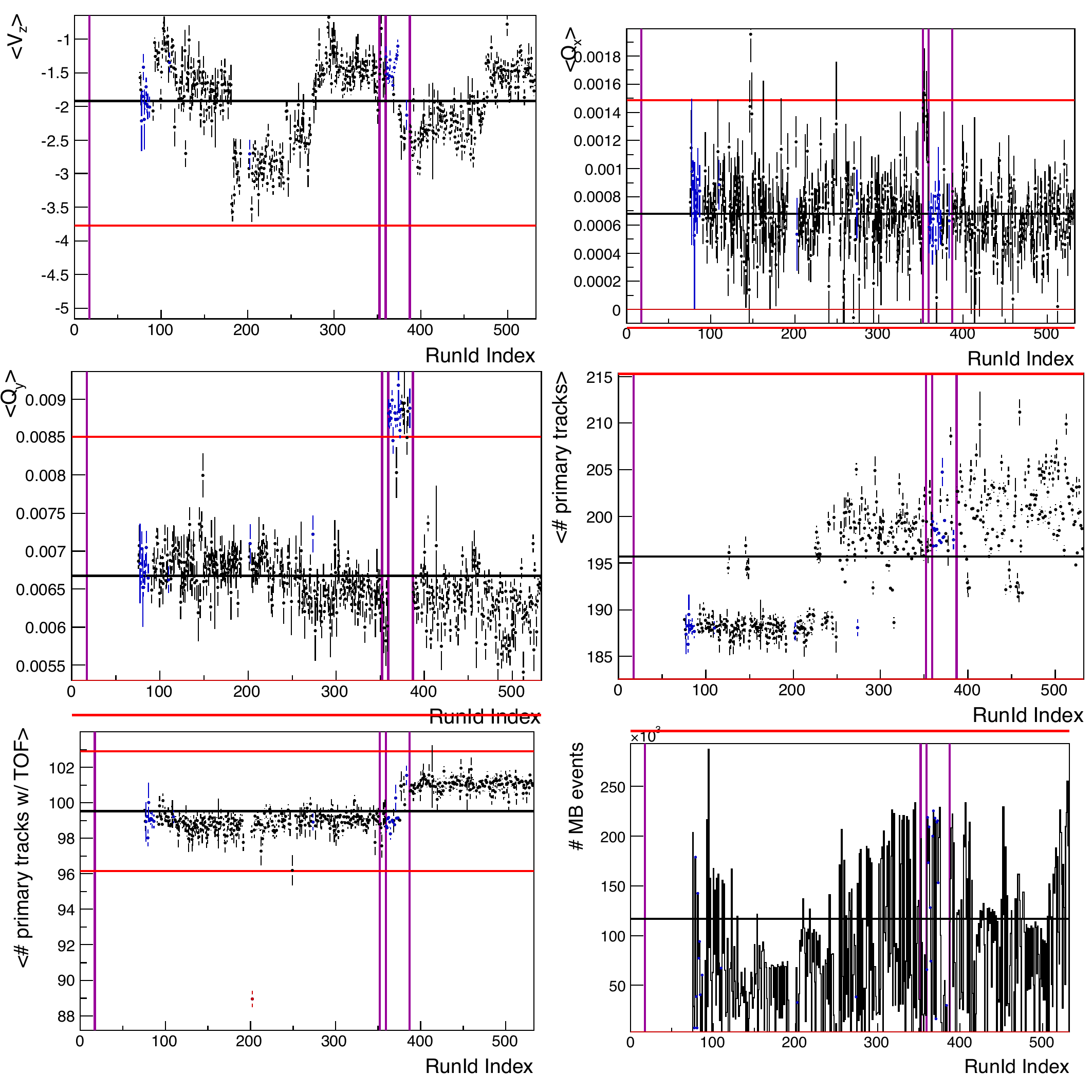}{Run QA 19.6 GeV (Part II)}{}{1} %

\wrapifneeded{0.50}{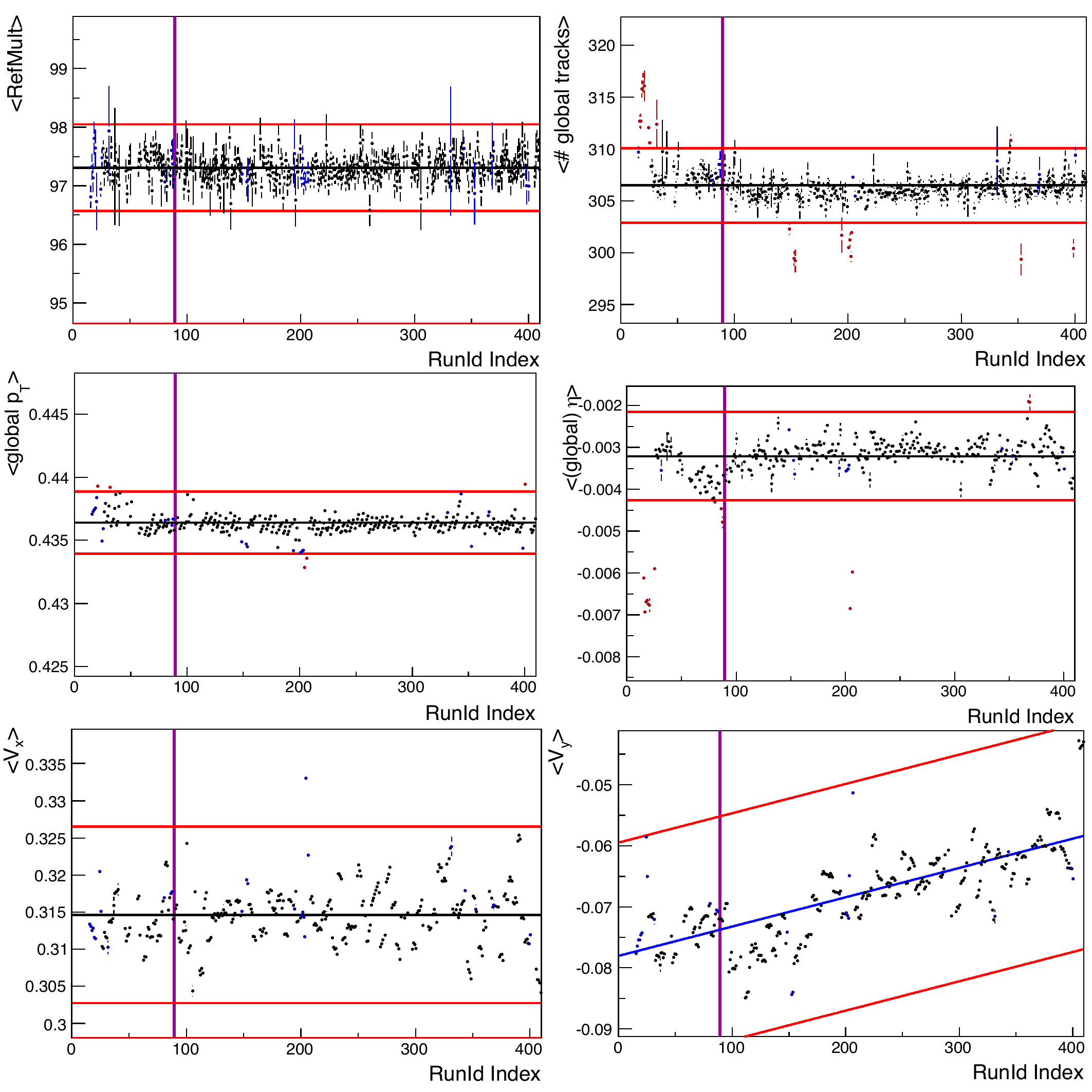}{Run QA 27 GeV (Part I)}{}{1} %

\wrapifneeded{0.50}{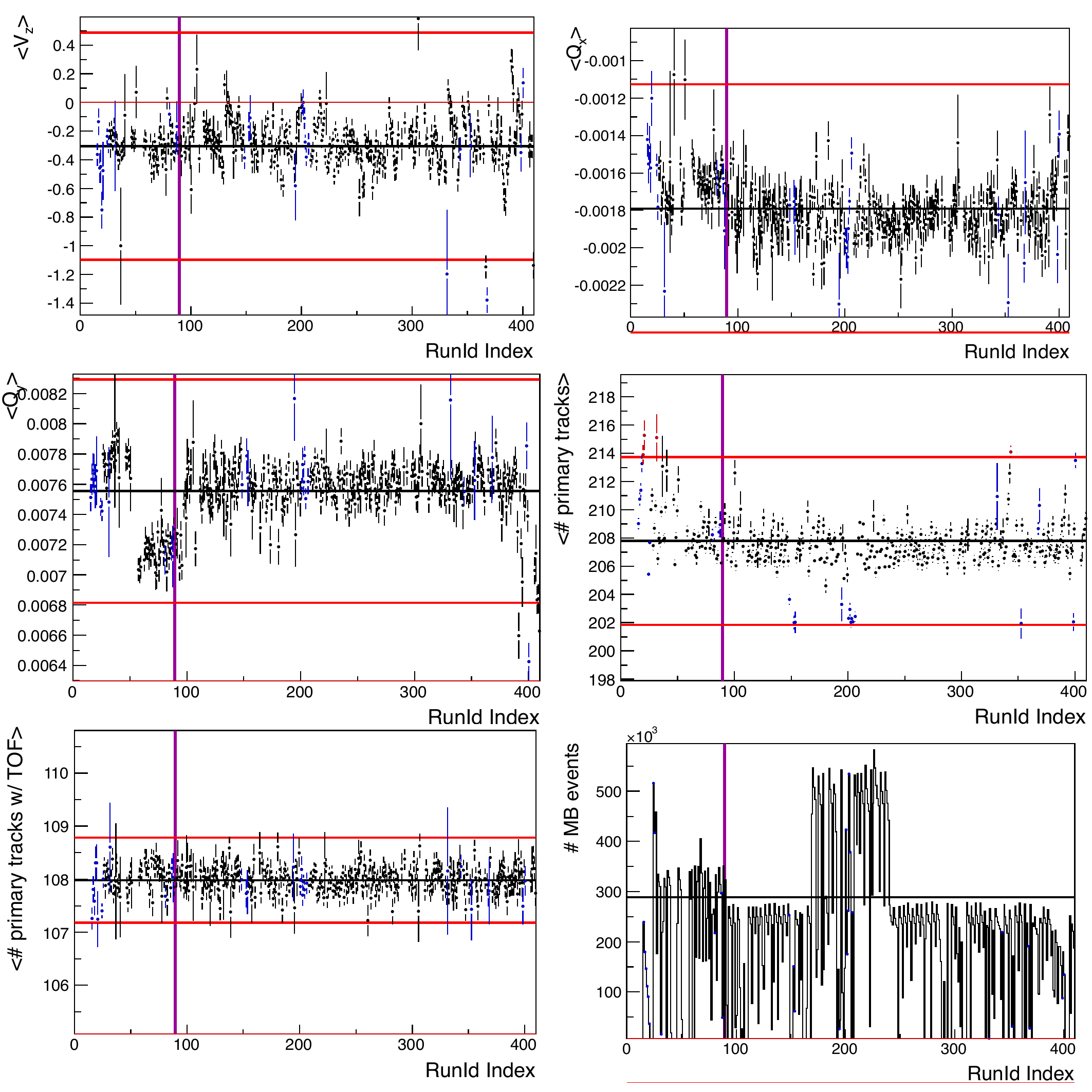}{Run QA 27 GeV (Part II)}{}{1} %

\wrapifneeded{0.50}{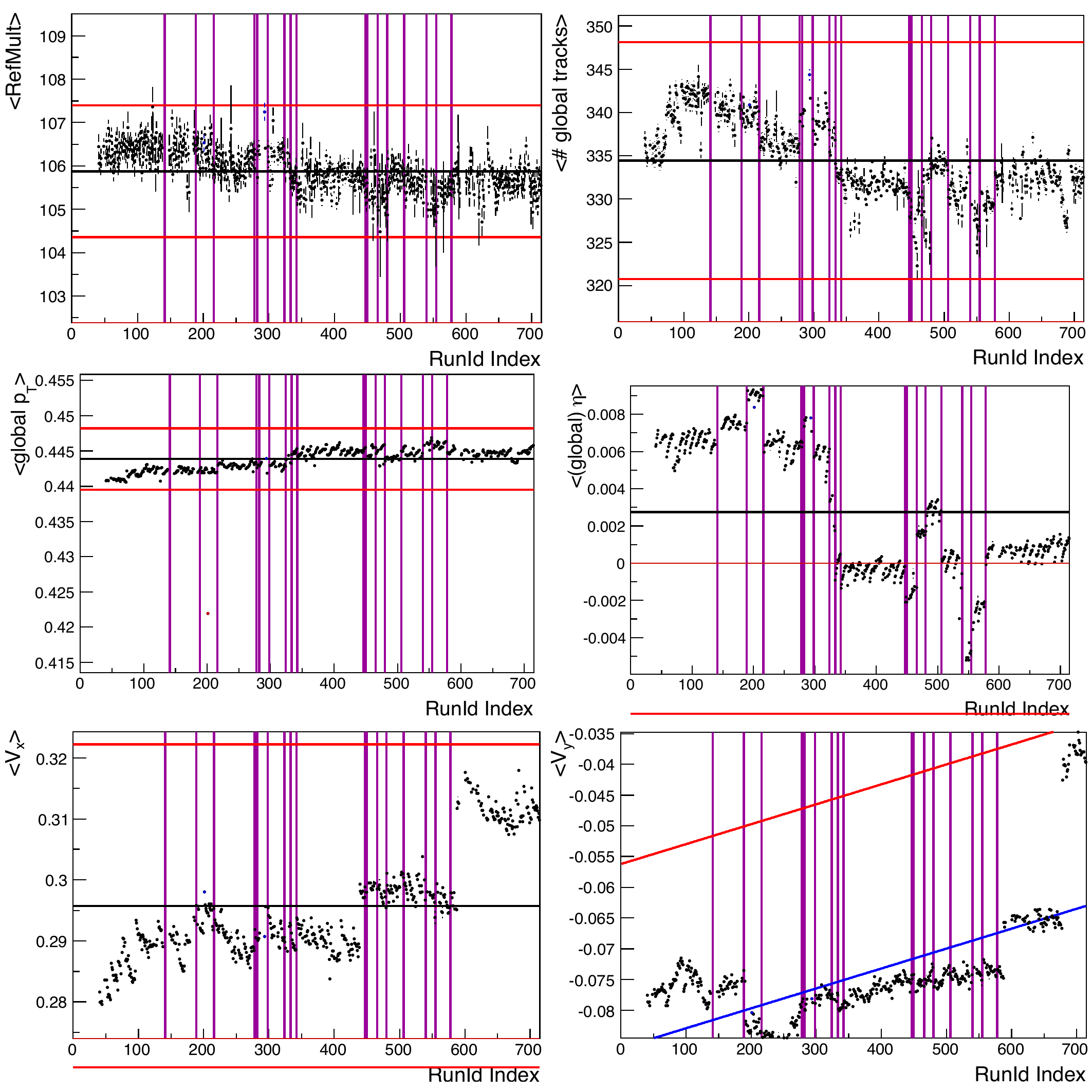}{Run QA 39 GeV (Part I)}{}{1} %

\wrapifneeded{0.50}{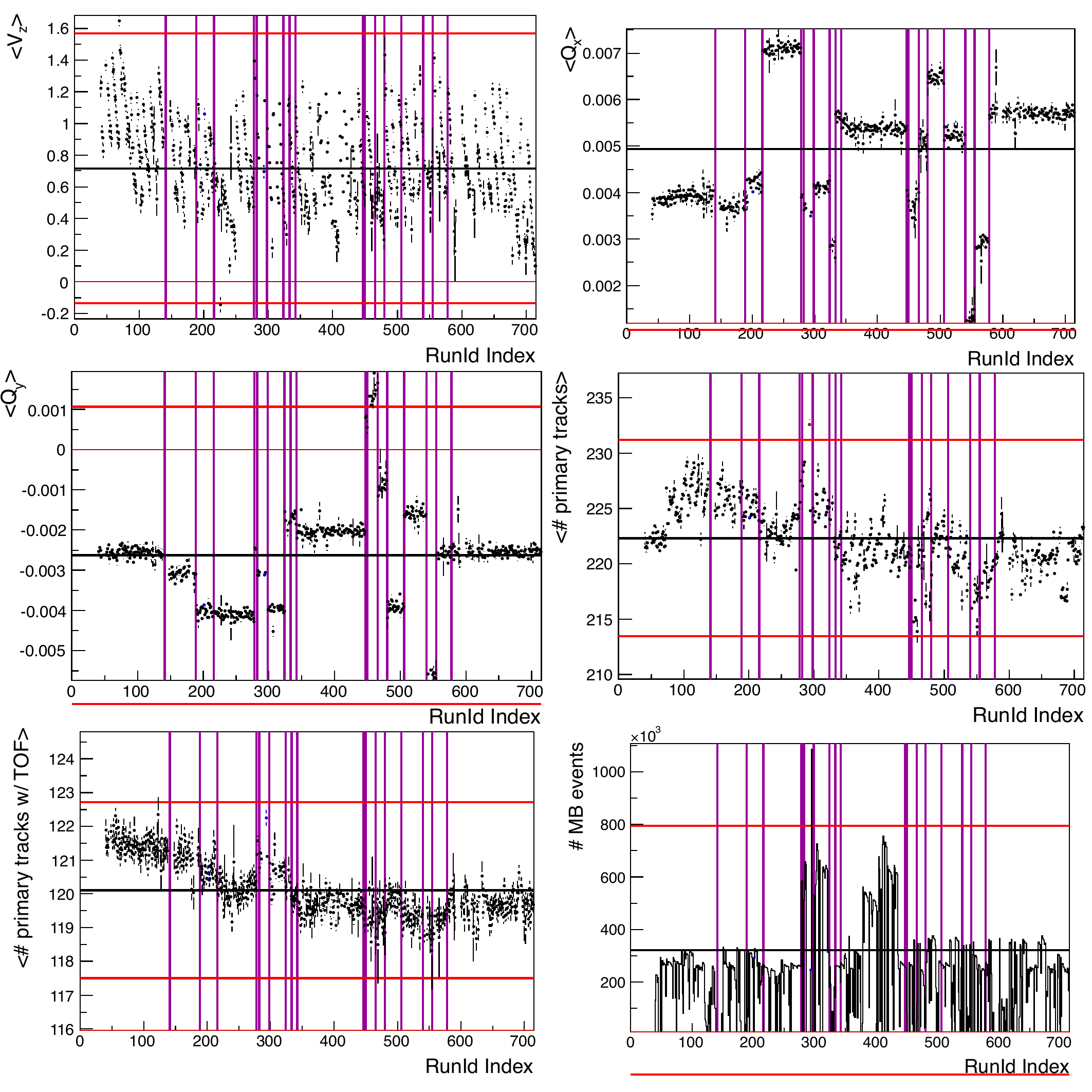}{Run QA 39 GeV (Part II)}{}{1} %

\wrapifneeded{0.50}{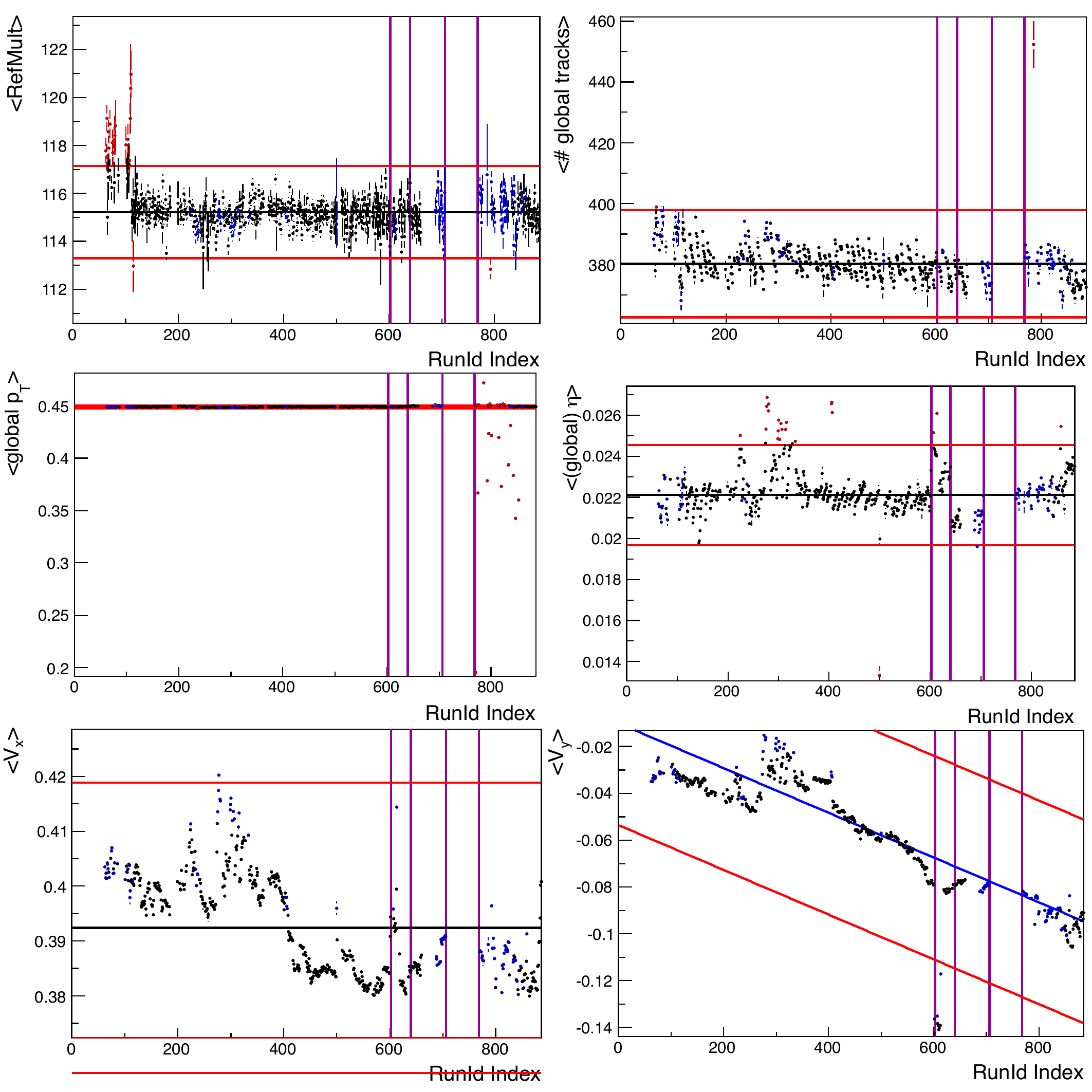}{Run QA 62.4 GeV (Part I)}{}{1} %

\wrapifneeded{0.50}{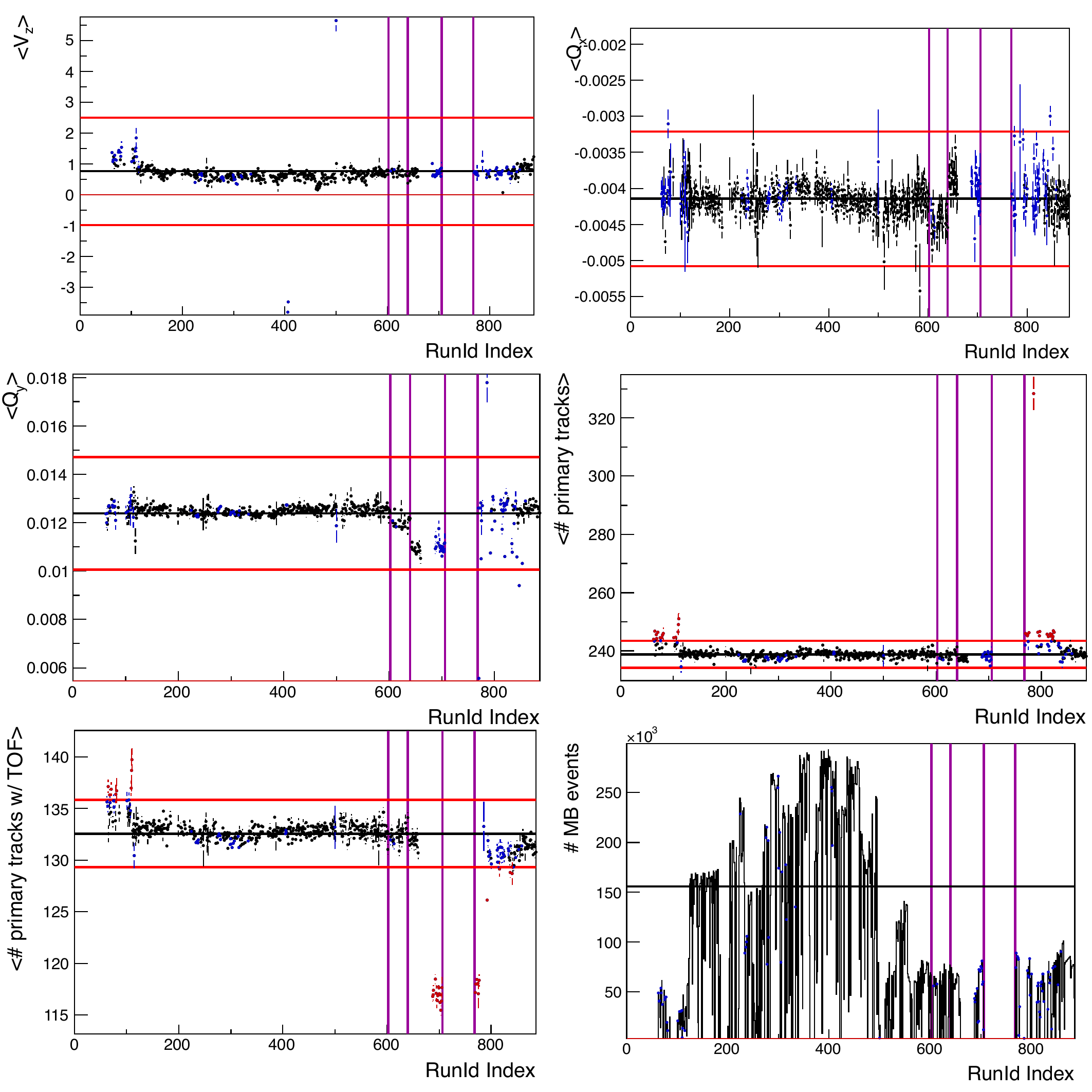}{Run QA 62.4 GeV (Part II)}{}{1} %

\section{Particle Identification Figures}
\label{app_pid_figures}\hyperlabel{app_pid_figures}%

\wrapifneeded{0.50}{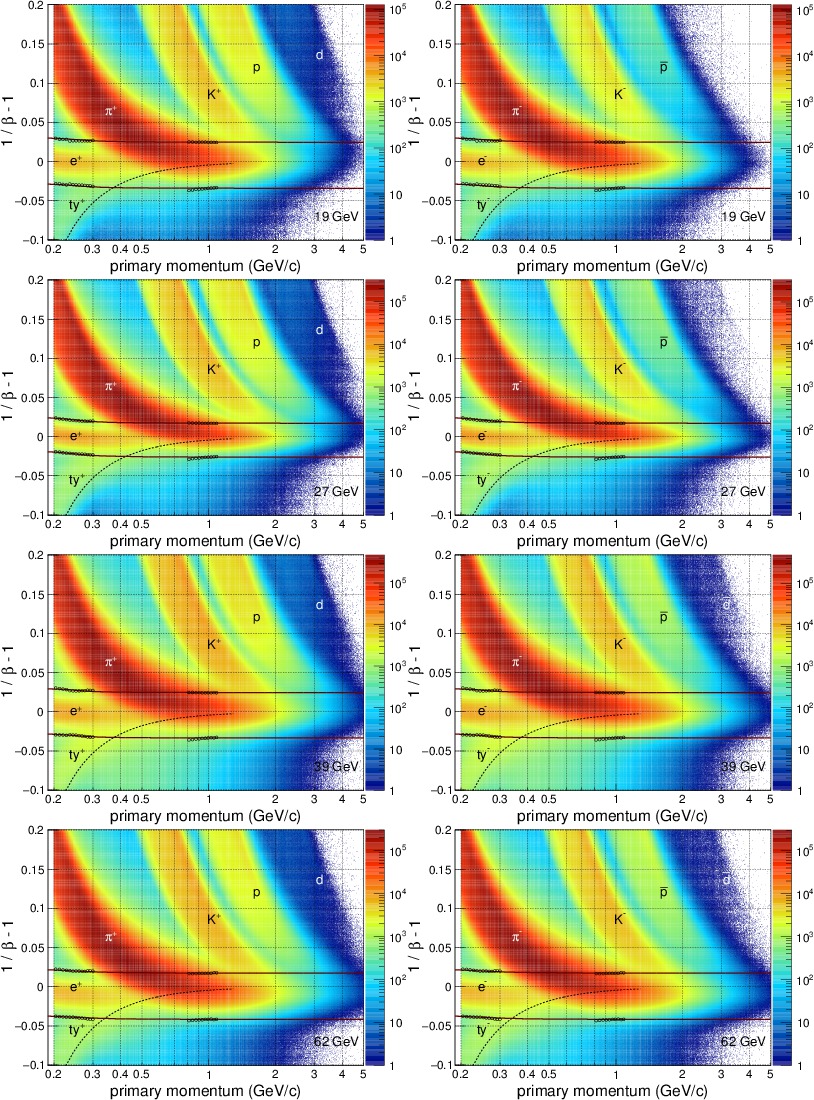}{TOF Selection of positrons (left column) and electrons (right column) for all BES-{}I energies. For a detailed description see Figure~\ref{ana_fig_pid_beta39}.}{app_pid_fig_beta}{0.91} %

\wrapifneeded{0.50}{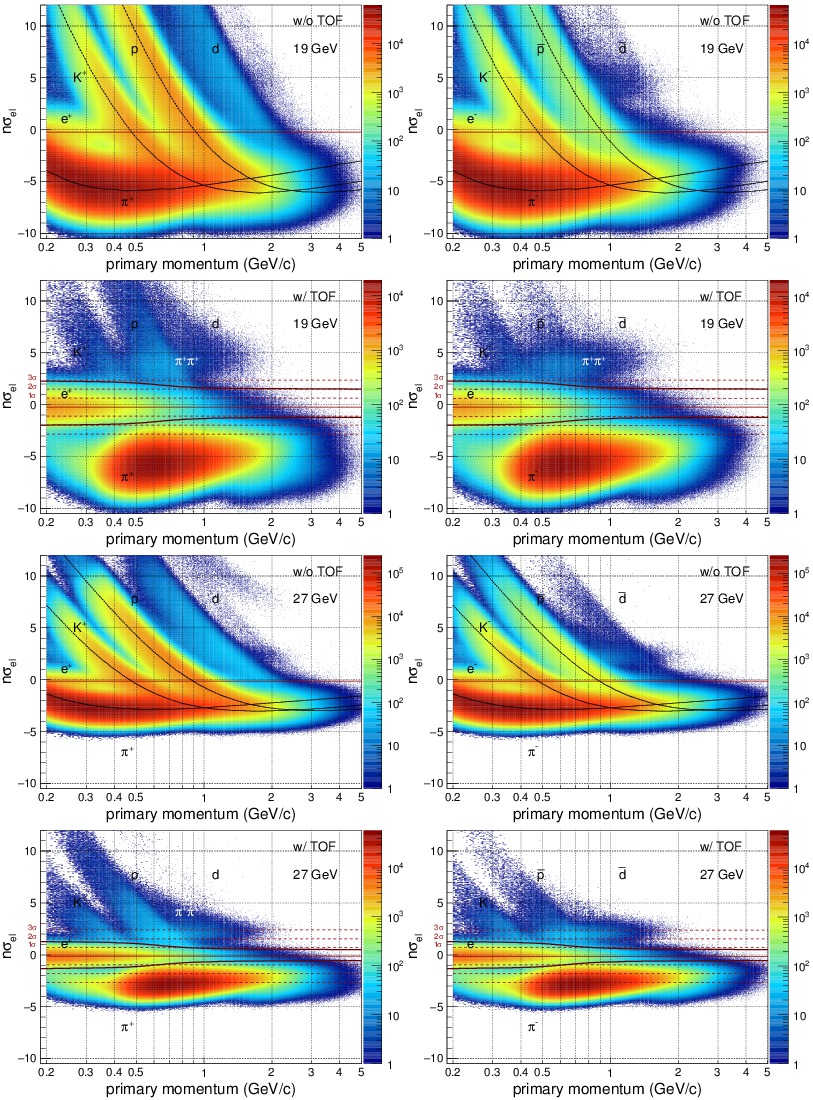}{TPC energy loss and selection of positrons (left column) and electrons (right column) at \ensuremath{\surd}s$_{\text{NN}}$ = 19.6 GeV and 27 GeV with and without TOF selection (Figure~\ref{app_pid_fig_beta}). For a detailed description see Figure~\ref{fig_pidQA}.}{app_pid_fig_dedx1}{0.95} %

\wrapifneeded{0.50}{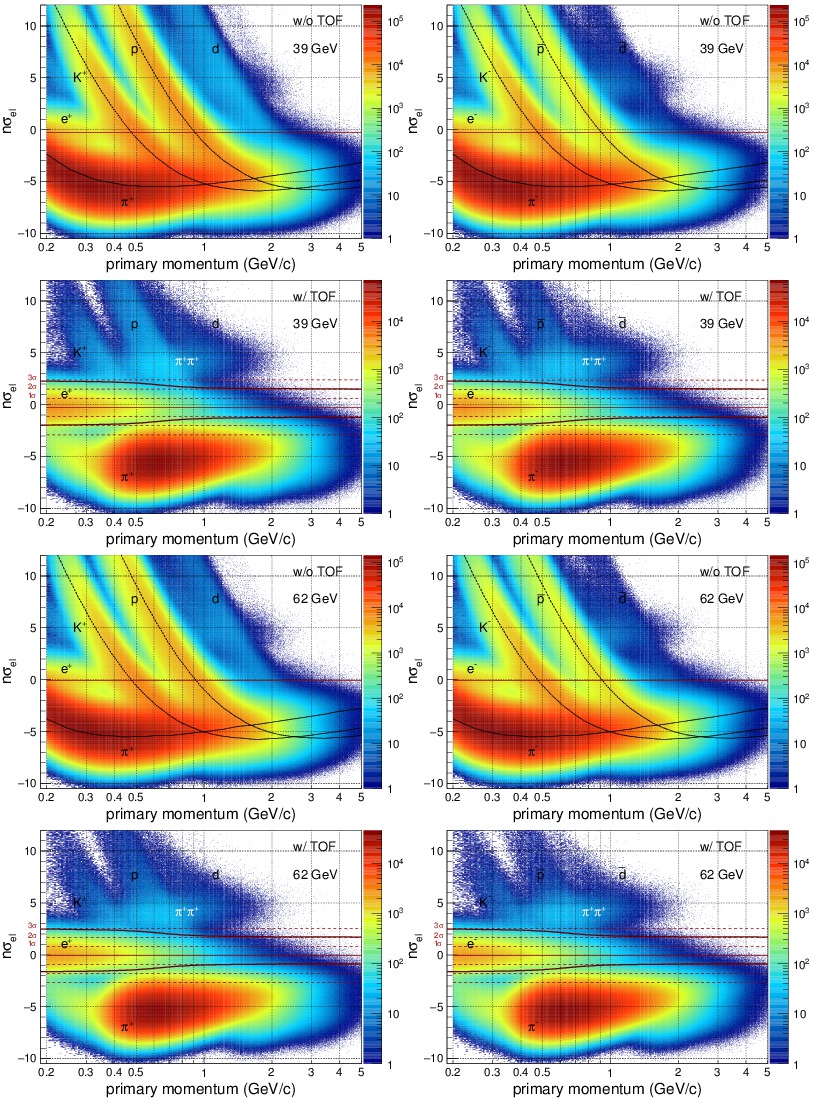}{TPC energy loss and selection of positrons (left column) and electrons (right column) at \ensuremath{\surd}s$_{\text{NN}}$ = 39 GeV and 62.4 GeV with and without TOF selection (Figure~\ref{app_pid_fig_beta}). For a detailed description see Figure~\ref{fig_pidQA}.}{app_pid_fig_dedx2}{0.95} %

\wrapifneeded{0.50}{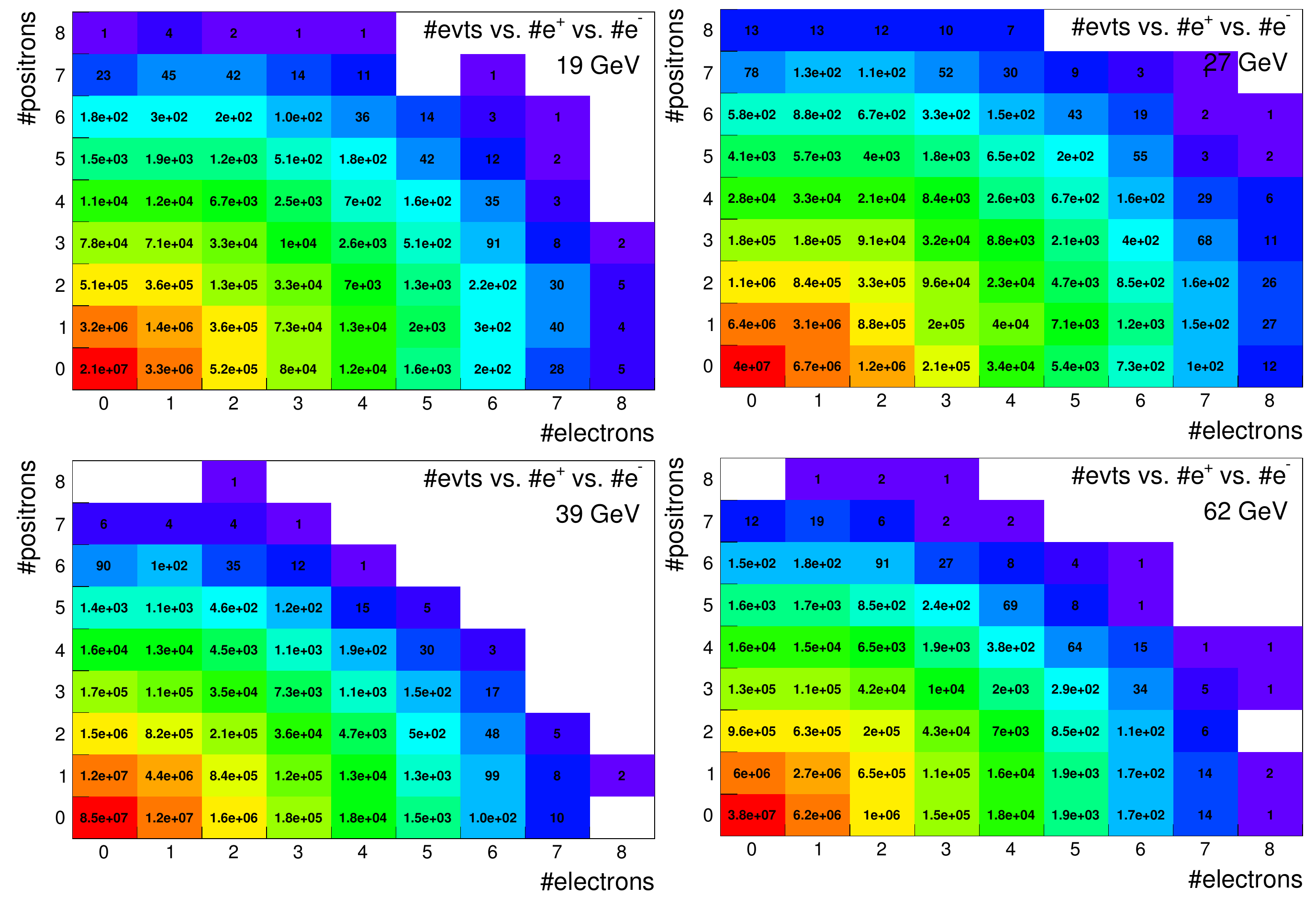}{Number of events for combinations of number of e$^{\text{+}}$/e$^{\text{-{}}}$ at all BES-{}I energies.}{app_pid_fig_nepem}{1} %

\section{Particle Samples Figures}
\label{app_pid_quality_figures}\hyperlabel{app_pid_quality_figures}%

This section is organized in columns for each energy and rows increasing with
momentum. It first shows all multi-{}gaussian fits to pure K, p, merged-{}\ensuremath{\pi}, and
\ensuremath{\pi} samples followed by the simultaneous multi-{}particle fits to n\ensuremath{\sigma}$_{\text{el}}$ distributions. The grey boxes denote the n\ensuremath{\sigma}$_{\text{el}}$ selection criterium used
to identify electrons/positrons. The final page of this section shows the
momentum dependence of the free (but properly initialized and constrained) fit
parameters, namely yield fractions (top) and few correctional shifts/widths
(bottom). Only a representative sample of the fits is included here. For the
full selection see the supporting information [284].

\pagebreak[4]

\begin{center}

\noindent\begin{minipage}[c]{\linewidth}
\begin{center}
\imgexists{puresamp_burst/PureSampleParticleFits_2015-04-16-rotated90_01.jpg}{{\imgevalsize{puresamp_burst/PureSampleParticleFits_2015-04-16-rotated90_01.jpg}{\includegraphics[width=0.95\linewidth,keepaspectratio=true]{puresamp_burst/PureSampleParticleFits_2015-04-16-rotated90_01.jpg}}}}{puresamp\_burst/PureSampleParticleFits\_2015-{}04-{}16-{}rotated90\_01.jpg}\end{center}
\end{minipage}

\end{center}

\begin{center}

\noindent\begin{minipage}[c]{\linewidth}
\begin{center}
\imgexists{nsigmael_burst/nsigmael_2015-04-16-nup_01.pdf}{{\imgevalsize{nsigmael_burst/nsigmael_2015-04-16-nup_01.pdf}{\includegraphics[width=1\linewidth,keepaspectratio=true]{nsigmael_burst/nsigmael_2015-04-16-nup_01.pdf}}}}{nsigmael\_burst/nsigmael\_2015-{}04-{}16-{}nup\_01.pdf}\end{center}
\end{minipage}

\end{center}

\begin{center}

\noindent\begin{minipage}[c]{\linewidth}
\begin{center}
\imgexists{nsigmael_2015-04-16_pars-nup.pdf}{{\imgevalsize{nsigmael_2015-04-16_pars-nup.pdf}{\includegraphics[width=1\linewidth,keepaspectratio=true]{nsigmael_2015-04-16_pars-nup.pdf}}}}{nsigmael\_2015-{}04-{}16\_pars-{}nup.pdf}\end{center}
\end{minipage}

\end{center}

\section{Pair Reconstruction Figures}
\label{app_pairrec_figures}\hyperlabel{app_pairrec_figures}%

\wrapifneeded{0.50}{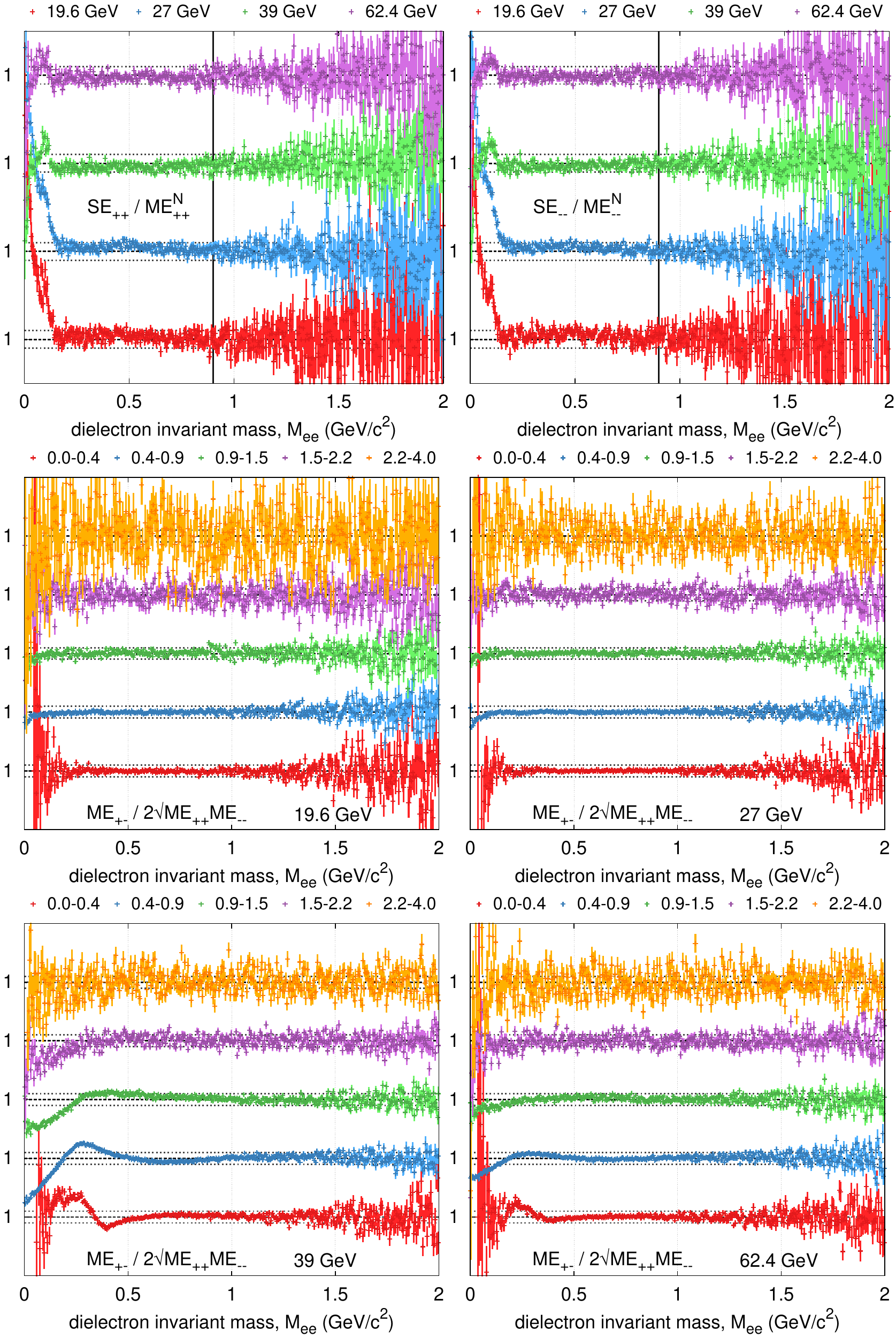}{(top row) \texttt{SE/\penalty0 ME$^{\text{N}}$} like-{}sign ratios for the identification of the normalization range. (remaining rows) p$_{\text{T}}$-{}dependent acceptance correction factors. See Figure~\ref{ana_fig_pair_normrange_accfac}.}{app_pair_fig_normrange_accfac_all}{0.8} %

\pagebreak[4]

\section{ISR pT spectra in p+p}
\label{app_isr_spectra}\hyperlabel{app_isr_spectra}%

data: \href{http://www.star.bnl.gov/protected/lfspectra/huck/html/ppspectra_data/}{http://www.star.bnl.gov/\-protected/\-lfspectra/\-huck/\-html/\-ppspectra\_data/\-}\newline
 code: \href{http://cgit.the-huck.com/cgit.cgi/Obsolete/tree/PPptSpectra}{http://cgit.the-{}huck.com/\-cgit.cgi/\-Obsolete/\-tree/\-PPptSpectra}\newline
 deduce \ensuremath{\eta}/\ensuremath{\pi}0 ratio from ISR/Fermilab data\newline
 make own \ensuremath{\omega} \& \ensuremath{\varphi} pT spectra from dielectron measurement or simply use m$_{\text{T}}$ scaling\newline
 compare to available data in 39 GeV after scaling: \ensuremath{\pi}$^{\text{0}}$, \ensuremath{\phi} \ensuremath{\rightarrow} KK\newline
 (check \ensuremath{\pi}0 yield in AuAu via \# participants scaling from p+p)\newline
 Phys. Lett. B 194 (1987) 4, Nucl. Phys. B 209 (1982) 309-{}320, Nucl. Phys. B 158 (1979) 1-{}10\newline
 Phys. Lett. B 79 (1978), Nucl. Phys. B 124 (1977) 1-{}11, Phys. Lett. B 64 (1976)\newline
 Nucl. Phys. B 116 (1976) 77-{}98, Phys. Rev. Lett 40 (1978) 684, Phys. Rev. Lett. 38 (1977) 112\newline
 Nucl. Phys. B 106 (1976) 1-{}30, Nucl. Phys. B 100 (1975) 237-{}290, Nucl. Phys. B 98 (1975) 49-{}72\newline
 Phys. Lett. B 55 (1975) 232, Nucl. Phys. B 87 (1975) 19-{}40, Phys. Lett. B 47 (1973) 75\newline
 Phys. Lett. B 47 (1973) 275, Phys. Lett. B 46 (1973) 471, Phys. Rev. Lett. 31 (1973) 413\newline
 Phys. Lett. 44B (1973) 521

\wrapifneeded{0.50}{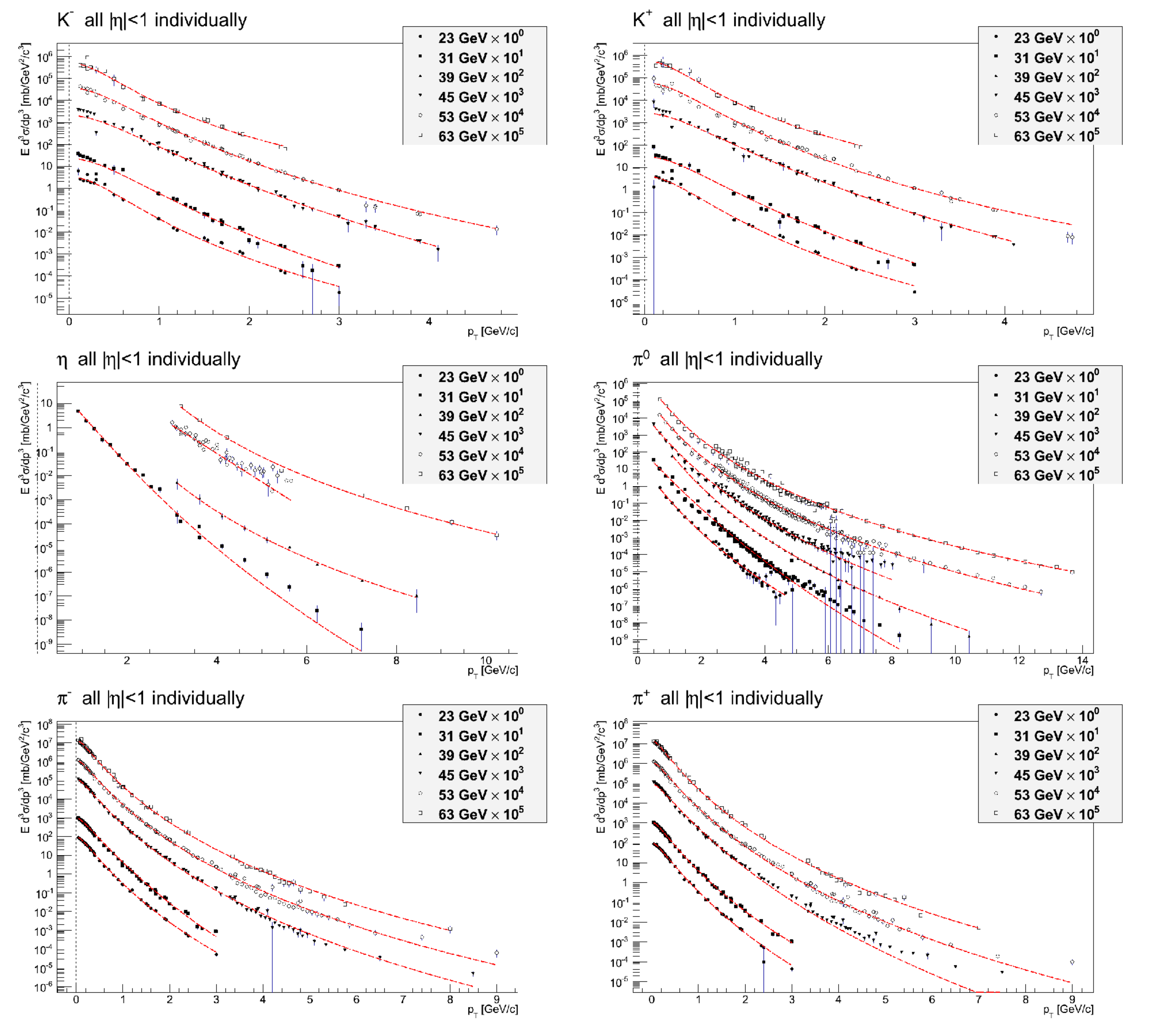}{ISR spectra}{}{0.95} %

\wrapifneeded{0.50}{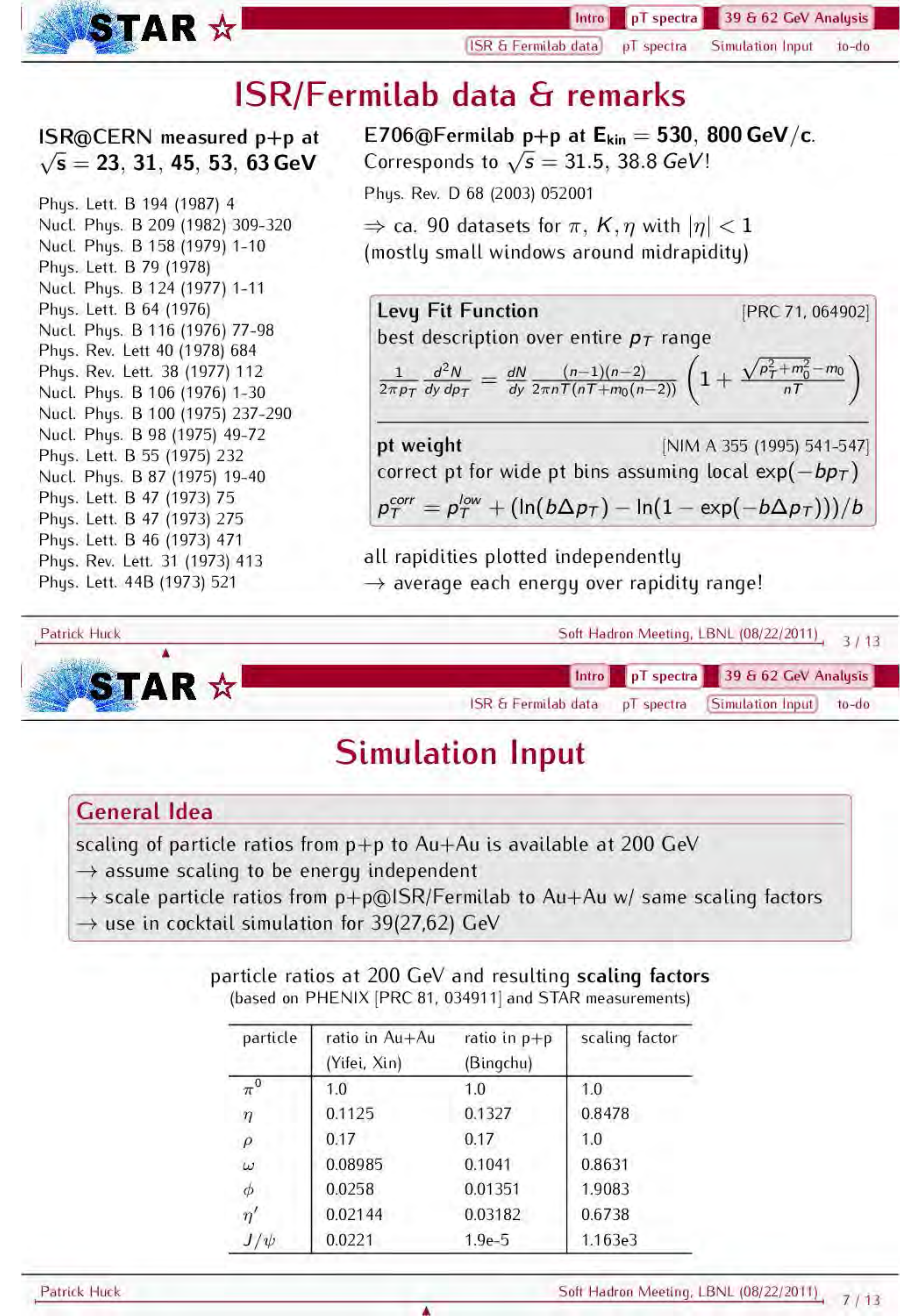}{Notes on ISR Spectra (Levy misses -{}1 exponent)}{}{0.95} %

\pagebreak[4]

\section{Efficiency Tables}
\label{app_effcorr_tables}\hyperlabel{app_effcorr_tables}%
\begin{footnotesize}
\begin{center}
\begingroup%
\setlength{\newtblsparewidth}{\linewidth-2\tabcolsep-2\tabcolsep-2\tabcolsep-2\tabcolsep-2\tabcolsep-2\tabcolsep-2\tabcolsep-2\tabcolsep-2\tabcolsep-2\tabcolsep-2\tabcolsep-2\tabcolsep}%
\setlength{\newtblstarfactor}{\newtblsparewidth / \real{95}}%

\begin{longtable}{lllllllllll}\caption[{p$_{\text{T}}$/\ensuremath{\eta}/\ensuremath{\phi}-{}dependent Combined Detector \& Track Quality         Efficiencies \ensuremath{\epsilon} with statistical uncertainties. Only an excerpt is         listed here with the full table in the supporting information [284].}]{p$_{\text{T}}$/\ensuremath{\eta}/\ensuremath{\phi}-{}dependent Combined Detector \& Track Quality         Efficiencies \ensuremath{\epsilon} with statistical uncertainties. Only an excerpt is         listed here with the full table in the supporting information [284].\label{app_tab_effs_dettrackqual}\hyperlabel{app_tab_effs_dettrackqual}%
}\tabularnewline
\endfirsthead
\caption[]{(continued)}\tabularnewline
\endhead
\hline
\multicolumn{1}{m{5\newtblstarfactor+\arrayrulewidth}}{\setlength{\newtblcolwidth}{5\newtblstarfactor}\multirowii[m]{3}{\newtblcolwidth}{\centering%
p$_{\text{T}}$
}}&\multicolumn{1}{m{10\newtblstarfactor+\arrayrulewidth}}{\setlength{\newtblcolwidth}{10\newtblstarfactor}\multirowii[m]{3}{\newtblcolwidth}{\centering%
\ensuremath{\eta}
}}&\multicolumn{1}{m{10\newtblstarfactor+\arrayrulewidth}}{\setlength{\newtblcolwidth}{10\newtblstarfactor}\multirowii[m]{3}{\newtblcolwidth}{\centering%
\ensuremath{\phi}
}}&\multicolumn{8}{m{5\newtblstarfactor+2\tabcolsep+\arrayrulewidth+5\newtblstarfactor+2\tabcolsep+\arrayrulewidth+5\newtblstarfactor+2\tabcolsep+\arrayrulewidth+5\newtblstarfactor+2\tabcolsep+\arrayrulewidth+15\newtblstarfactor+2\tabcolsep+\arrayrulewidth+15\newtblstarfactor+2\tabcolsep+\arrayrulewidth+10\newtblstarfactor+2\tabcolsep+\arrayrulewidth+10\newtblstarfactor+\arrayrulewidth}}{\centering%
\ensuremath{\epsilon}\ensuremath{\pm}\ensuremath{\Delta}\ensuremath{\epsilon}$_{\text{stat}}$
}\tabularnewline
\hline
\end{longtable}\endgroup%

\end{center}
\end{footnotesize}
\begin{footnotesize}
\begin{center}
\begingroup%
\setlength{\newtblsparewidth}{\linewidth-2\tabcolsep-2\tabcolsep-2\tabcolsep-2\tabcolsep-2\tabcolsep-2\tabcolsep-2\tabcolsep-2\tabcolsep-2\tabcolsep-2\tabcolsep-2\tabcolsep-2\tabcolsep-2\tabcolsep-2\tabcolsep-2\tabcolsep-2\tabcolsep-2\tabcolsep-2\tabcolsep-2\tabcolsep-2\tabcolsep}%
\setlength{\newtblstarfactor}{\newtblsparewidth / \real{98}}%

\begin{longtable}{lllllllllllllllllll}\caption[{TOF Matching (\ensuremath{\epsilon}$_{\text{TOF}}$) and Total Track               (\ensuremath{\epsilon}$_{\text{tot}}$) Efficiencies for \ensuremath{\eta} <{} 0. Only an excerpt is               listed here with the full table in the supporting information               [284].}]{TOF Matching (\ensuremath{\epsilon}$_{\text{TOF}}$) and Total Track               (\ensuremath{\epsilon}$_{\text{tot}}$) Efficiencies for \ensuremath{\eta} <{} 0. Only an excerpt is               listed here with the full table in the supporting information               [284].\label{app_tab_effs_tofmatchtotal0}\hyperlabel{app_tab_effs_tofmatchtotal0}%
}\tabularnewline
\endfirsthead
\caption[]{(continued)}\tabularnewline
\endhead
\hline
\multicolumn{1}{m{2\newtblstarfactor+\arrayrulewidth}}{\setlength{\newtblcolwidth}{2\newtblstarfactor}\multirowii[m]{3}{\newtblcolwidth}{\centering%
p$_{\text{T}}$
}}&\multicolumn{1}{m{10\newtblstarfactor+\arrayrulewidth}}{\setlength{\newtblcolwidth}{10\newtblstarfactor}\multirowii[m]{3}{\newtblcolwidth}{\centering%
\ensuremath{\eta}
}}&\multicolumn{1}{m{10\newtblstarfactor+\arrayrulewidth}}{\setlength{\newtblcolwidth}{10\newtblstarfactor}\multirowii[m]{3}{\newtblcolwidth}{\centering%
\ensuremath{\phi}
}}&\multicolumn{8}{m{1\newtblstarfactor+2\tabcolsep+\arrayrulewidth+1\newtblstarfactor+2\tabcolsep+\arrayrulewidth+1\newtblstarfactor+2\tabcolsep+\arrayrulewidth+1\newtblstarfactor+2\tabcolsep+\arrayrulewidth+1\newtblstarfactor+2\tabcolsep+\arrayrulewidth+1\newtblstarfactor+2\tabcolsep+\arrayrulewidth+1\newtblstarfactor+2\tabcolsep+\arrayrulewidth+1\newtblstarfactor+\arrayrulewidth}}{\centering%
\ensuremath{\epsilon}$_{\text{TOF}}$
}&\multicolumn{8}{m{8\newtblstarfactor+2\tabcolsep+\arrayrulewidth+8\newtblstarfactor+2\tabcolsep+\arrayrulewidth+8\newtblstarfactor+2\tabcolsep+\arrayrulewidth+8\newtblstarfactor+2\tabcolsep+\arrayrulewidth+9\newtblstarfactor+2\tabcolsep+\arrayrulewidth+9\newtblstarfactor+2\tabcolsep+\arrayrulewidth+9\newtblstarfactor+2\tabcolsep+\arrayrulewidth+9\newtblstarfactor+\arrayrulewidth}}{\centering%
\ensuremath{\epsilon}$_{\text{tot}}$\ensuremath{\pm}\ensuremath{\Delta}\ensuremath{\epsilon}$_{\text{stat}}$\ensuremath{\pm}\ensuremath{\Delta}\ensuremath{\epsilon}$_{\text{syst}}$
}\tabularnewline
\hline
\end{longtable}\endgroup%

\end{center}
\end{footnotesize}
\begin{footnotesize}
\begin{center}
\begingroup%
\setlength{\newtblsparewidth}{\linewidth-2\tabcolsep-2\tabcolsep-2\tabcolsep-2\tabcolsep-2\tabcolsep-2\tabcolsep-2\tabcolsep-2\tabcolsep-2\tabcolsep-2\tabcolsep-2\tabcolsep-2\tabcolsep-2\tabcolsep-2\tabcolsep-2\tabcolsep-2\tabcolsep-2\tabcolsep-2\tabcolsep-2\tabcolsep-2\tabcolsep}%
\setlength{\newtblstarfactor}{\newtblsparewidth / \real{98}}%

\begin{longtable}{lllllllllllllllllll}\caption[{TOF Matching (\ensuremath{\epsilon}$_{\text{TOF}}$) and Total Track               (\ensuremath{\epsilon}$_{\text{tot}}$) Efficiencies for \ensuremath{\eta} >{} 0. Only an excerpt is               listed here with the full table in the supporting information               [284].}]{TOF Matching (\ensuremath{\epsilon}$_{\text{TOF}}$) and Total Track               (\ensuremath{\epsilon}$_{\text{tot}}$) Efficiencies for \ensuremath{\eta} >{} 0. Only an excerpt is               listed here with the full table in the supporting information               [284].\label{app_tab_effs_tofmatchtotal1}\hyperlabel{app_tab_effs_tofmatchtotal1}%
}\tabularnewline
\endfirsthead
\caption[]{(continued)}\tabularnewline
\endhead
\hline
\multicolumn{1}{m{2\newtblstarfactor+\arrayrulewidth}}{\setlength{\newtblcolwidth}{2\newtblstarfactor}\multirowii[m]{3}{\newtblcolwidth}{\centering%
p$_{\text{T}}$
}}&\multicolumn{1}{m{10\newtblstarfactor+\arrayrulewidth}}{\setlength{\newtblcolwidth}{10\newtblstarfactor}\multirowii[m]{3}{\newtblcolwidth}{\centering%
\ensuremath{\eta}
}}&\multicolumn{1}{m{10\newtblstarfactor+\arrayrulewidth}}{\setlength{\newtblcolwidth}{10\newtblstarfactor}\multirowii[m]{3}{\newtblcolwidth}{\centering%
\ensuremath{\phi}
}}&\multicolumn{8}{m{1\newtblstarfactor+2\tabcolsep+\arrayrulewidth+1\newtblstarfactor+2\tabcolsep+\arrayrulewidth+1\newtblstarfactor+2\tabcolsep+\arrayrulewidth+1\newtblstarfactor+2\tabcolsep+\arrayrulewidth+1\newtblstarfactor+2\tabcolsep+\arrayrulewidth+1\newtblstarfactor+2\tabcolsep+\arrayrulewidth+1\newtblstarfactor+2\tabcolsep+\arrayrulewidth+1\newtblstarfactor+\arrayrulewidth}}{\centering%
\ensuremath{\epsilon}$_{\text{TOF}}$
}&\multicolumn{8}{m{8\newtblstarfactor+2\tabcolsep+\arrayrulewidth+8\newtblstarfactor+2\tabcolsep+\arrayrulewidth+8\newtblstarfactor+2\tabcolsep+\arrayrulewidth+8\newtblstarfactor+2\tabcolsep+\arrayrulewidth+9\newtblstarfactor+2\tabcolsep+\arrayrulewidth+9\newtblstarfactor+2\tabcolsep+\arrayrulewidth+9\newtblstarfactor+2\tabcolsep+\arrayrulewidth+9\newtblstarfactor+\arrayrulewidth}}{\centering%
\ensuremath{\epsilon}$_{\text{tot}}$\ensuremath{\pm}\ensuremath{\Delta}\ensuremath{\epsilon}$_{\text{stat}}$\ensuremath{\pm}\ensuremath{\Delta}\ensuremath{\epsilon}$_{\text{syst}}$
}\tabularnewline
\hline
\end{longtable}\endgroup%

\end{center}
\end{footnotesize}
\begin{footnotesize}
\begin{center}
\begingroup%
\setlength{\newtblsparewidth}{\linewidth-2\tabcolsep-2\tabcolsep-2\tabcolsep-2\tabcolsep-2\tabcolsep-2\tabcolsep-2\tabcolsep-2\tabcolsep-2\tabcolsep-2\tabcolsep-2\tabcolsep-2\tabcolsep-2\tabcolsep-2\tabcolsep-2\tabcolsep-2\tabcolsep-2\tabcolsep-2\tabcolsep-2\tabcolsep-2\tabcolsep-2\tabcolsep-2\tabcolsep-2\tabcolsep-2\tabcolsep-2\tabcolsep-2\tabcolsep-2\tabcolsep}%
\setlength{\newtblstarfactor}{\newtblsparewidth / \real{84}}%

\begin{longtable}{llllllllllllllllllllllllll}\caption[{\ensuremath{\eta}-{}/\ensuremath{\phi}-{}dependence of systematic uncertainties         \ensuremath{\Delta}\ensuremath{\epsilon}, scaling factors \emph{F}, and reduced \ensuremath{\chi}$^{\text{2}}$ for the         determination of TOF Matching Efficiencies.}]{\ensuremath{\eta}-{}/\ensuremath{\phi}-{}dependence of systematic uncertainties         \ensuremath{\Delta}\ensuremath{\epsilon}, scaling factors \emph{F}, and reduced \ensuremath{\chi}$^{\text{2}}$ for the         determination of TOF Matching Efficiencies.\label{app_tab_effs_tofmatch_extra}\hyperlabel{app_tab_effs_tofmatch_extra}%
}\tabularnewline
\endfirsthead
\caption[]{(continued)}\tabularnewline
\endhead
\hline
\multicolumn{1}{m{14\newtblstarfactor+\arrayrulewidth}}{\setlength{\newtblcolwidth}{14\newtblstarfactor}\multirowii[m]{3}{\newtblcolwidth}{\centering%
\ensuremath{\eta}
}}&\multicolumn{1}{m{14\newtblstarfactor+\arrayrulewidth}}{\setlength{\newtblcolwidth}{14\newtblstarfactor}\multirowii[m]{3}{\newtblcolwidth}{\centering%
\ensuremath{\phi}
}}&\multicolumn{6}{m{1\newtblstarfactor+2\tabcolsep+\arrayrulewidth+4\newtblstarfactor+2\tabcolsep+\arrayrulewidth+2\newtblstarfactor+2\tabcolsep+\arrayrulewidth+1\newtblstarfactor+2\tabcolsep+\arrayrulewidth+4\newtblstarfactor+2\tabcolsep+\arrayrulewidth+2\newtblstarfactor+\arrayrulewidth}}{\centering%
19.6 GeV
}&\multicolumn{6}{m{1\newtblstarfactor+2\tabcolsep+\arrayrulewidth+4\newtblstarfactor+2\tabcolsep+\arrayrulewidth+2\newtblstarfactor+2\tabcolsep+\arrayrulewidth+1\newtblstarfactor+2\tabcolsep+\arrayrulewidth+4\newtblstarfactor+2\tabcolsep+\arrayrulewidth+2\newtblstarfactor+\arrayrulewidth}}{\centering%
27 GeV
}&\multicolumn{6}{m{1\newtblstarfactor+2\tabcolsep+\arrayrulewidth+4\newtblstarfactor+2\tabcolsep+\arrayrulewidth+2\newtblstarfactor+2\tabcolsep+\arrayrulewidth+1\newtblstarfactor+2\tabcolsep+\arrayrulewidth+4\newtblstarfactor+2\tabcolsep+\arrayrulewidth+2\newtblstarfactor+\arrayrulewidth}}{\centering%
39 GeV
}&\multicolumn{6}{m{1\newtblstarfactor+2\tabcolsep+\arrayrulewidth+4\newtblstarfactor+2\tabcolsep+\arrayrulewidth+2\newtblstarfactor+2\tabcolsep+\arrayrulewidth+1\newtblstarfactor+2\tabcolsep+\arrayrulewidth+4\newtblstarfactor+2\tabcolsep+\arrayrulewidth+2\newtblstarfactor+\arrayrulewidth}}{\centering%
62.4 GeV
}\tabularnewline
\hline
\end{longtable}\endgroup%

\end{center}
\end{footnotesize}
\begin{footnotesize}
\begin{center}
\begingroup%
\setlength{\newtblsparewidth}{0.9\linewidth-2\tabcolsep-2\tabcolsep-2\tabcolsep-2\tabcolsep-2\tabcolsep-2\tabcolsep-2\tabcolsep-2\tabcolsep-2\tabcolsep-2\tabcolsep}%
\setlength{\newtblstarfactor}{\newtblsparewidth / \real{387}}%

\begin{longtable}{lllllllll}\caption[{p-{}dependence of TPC Selection Efficiencies}]{p-{}dependence of TPC Selection Efficiencies\label{app_tab_effs_tpcselect}\hyperlabel{app_tab_effs_tpcselect}%
}\tabularnewline
\endfirsthead
\caption[]{(continued)}\tabularnewline
\endhead
\hline
\multicolumn{1}{m{43\newtblstarfactor+\arrayrulewidth}}{\setlength{\newtblcolwidth}{43\newtblstarfactor}\multirowii[m]{3}{\newtblcolwidth}{\centering%
p
}}&\multicolumn{8}{m{43\newtblstarfactor+2\tabcolsep+\arrayrulewidth+43\newtblstarfactor+2\tabcolsep+\arrayrulewidth+43\newtblstarfactor+2\tabcolsep+\arrayrulewidth+43\newtblstarfactor+2\tabcolsep+\arrayrulewidth+43\newtblstarfactor+2\tabcolsep+\arrayrulewidth+43\newtblstarfactor+2\tabcolsep+\arrayrulewidth+43\newtblstarfactor+2\tabcolsep+\arrayrulewidth+43\newtblstarfactor+\arrayrulewidth}}{\centering%
\ensuremath{\epsilon}\ensuremath{\pm}\ensuremath{\Delta}\ensuremath{\epsilon}$_{\text{syst}}$
}\tabularnewline
\multicolumn{1}{m{43\newtblstarfactor+\arrayrulewidth}}{\setlength{\newtblcolwidth}{43\newtblstarfactor}\multirowii[m]{3}{-\newtblcolwidth}{\centering%
p
}}&\multicolumn{2}{m{43\newtblstarfactor+2\tabcolsep+\arrayrulewidth+43\newtblstarfactor+\arrayrulewidth}}{\centering%
19.6 GeV
}&\multicolumn{2}{m{43\newtblstarfactor+2\tabcolsep+\arrayrulewidth+43\newtblstarfactor+\arrayrulewidth}}{\centering%
27 GeV
}&\multicolumn{2}{m{43\newtblstarfactor+2\tabcolsep+\arrayrulewidth+43\newtblstarfactor+\arrayrulewidth}}{\centering%
39 GeV
}&\multicolumn{2}{m{43\newtblstarfactor+2\tabcolsep+\arrayrulewidth+43\newtblstarfactor+\arrayrulewidth}}{\centering%
62.4 GeV
}\tabularnewline
\hline
\end{longtable}\endgroup%

\end{center}
\end{footnotesize}
\begin{footnotesize}
\begin{center}
\begingroup%
\setlength{\newtblsparewidth}{0.7\linewidth-2\tabcolsep-2\tabcolsep-2\tabcolsep-2\tabcolsep-2\tabcolsep-2\tabcolsep-2\tabcolsep}%
\setlength{\newtblstarfactor}{\newtblsparewidth / \real{300}}%

\begin{longtable}{llllll}\caption[{Dielectron (pair) efficiencies.}]{Dielectron (pair) efficiencies.\label{app_tab_effs_pair}\hyperlabel{app_tab_effs_pair}%
}\tabularnewline
\endfirsthead
\caption[]{(continued)}\tabularnewline
\endhead
\hline
\multicolumn{1}{m{40\newtblstarfactor+\arrayrulewidth}}{\setlength{\newtblcolwidth}{40\newtblstarfactor}\multirowii[m]{2}{\newtblcolwidth}{\centering%
\textbf{\emph{M$_{\text{ee}}$}}
}}&\multicolumn{1}{m{20\newtblstarfactor+\arrayrulewidth}}{\setlength{\newtblcolwidth}{20\newtblstarfactor}\multirowii[m]{2}{\newtblcolwidth}{\centering%
\textbf{\emph{p$_{\text{T}}$}}
}}&\multicolumn{4}{m{60\newtblstarfactor+2\tabcolsep+\arrayrulewidth+60\newtblstarfactor+2\tabcolsep+\arrayrulewidth+60\newtblstarfactor+2\tabcolsep+\arrayrulewidth+60\newtblstarfactor+\arrayrulewidth}}{\centering%
\textbf{\ensuremath{\epsilon}\ensuremath{\pm}\ensuremath{\Delta}\ensuremath{\epsilon}$_{\text{syst}}$}
}\tabularnewline
\multicolumn{1}{m{40\newtblstarfactor+\arrayrulewidth}}{\setlength{\newtblcolwidth}{40\newtblstarfactor}\multirowii[m]{2}{-\newtblcolwidth}{\centering%
\textbf{\emph{M$_{\text{ee}}$}}
}}&\multicolumn{1}{m{20\newtblstarfactor+\arrayrulewidth}}{\setlength{\newtblcolwidth}{20\newtblstarfactor}\multirowii[m]{2}{-\newtblcolwidth}{\centering%
\textbf{\emph{p$_{\text{T}}$}}
}}&\multicolumn{1}{m{60\newtblstarfactor+\arrayrulewidth}}{\centering%
\textbf{19.6 GeV}
}&\multicolumn{1}{m{60\newtblstarfactor+\arrayrulewidth}}{\centering%
\textbf{27 GeV}
}&\multicolumn{1}{m{60\newtblstarfactor+\arrayrulewidth}}{\centering%
\textbf{39 GeV}
}&\multicolumn{1}{m{60\newtblstarfactor+\arrayrulewidth}}{\centering%
\textbf{62.4 GeV}
}\tabularnewline
\multicolumn{1}{m{40\newtblstarfactor+\arrayrulewidth}}{\centering%
0.002
}&\multicolumn{1}{m{20\newtblstarfactor+\arrayrulewidth}}{\centering%
0.2
}&\multicolumn{1}{m{60\newtblstarfactor+\arrayrulewidth}}{\centering%
0.0\ensuremath{\pm}0
}&\multicolumn{1}{m{60\newtblstarfactor+\arrayrulewidth}}{\centering%
0.0\ensuremath{\pm}0
}&\multicolumn{1}{m{60\newtblstarfactor+\arrayrulewidth}}{\centering%
0.0\ensuremath{\pm}0
}&\multicolumn{1}{m{60\newtblstarfactor+\arrayrulewidth}}{\centering%
0.0\ensuremath{\pm}0
}\tabularnewline
\hline
\end{longtable}\endgroup%

\end{center}
\end{footnotesize}

\section{Efficiency Figures}
\label{app_effcorr_figures}\hyperlabel{app_effcorr_figures}%

\wrapifneeded{0.50}{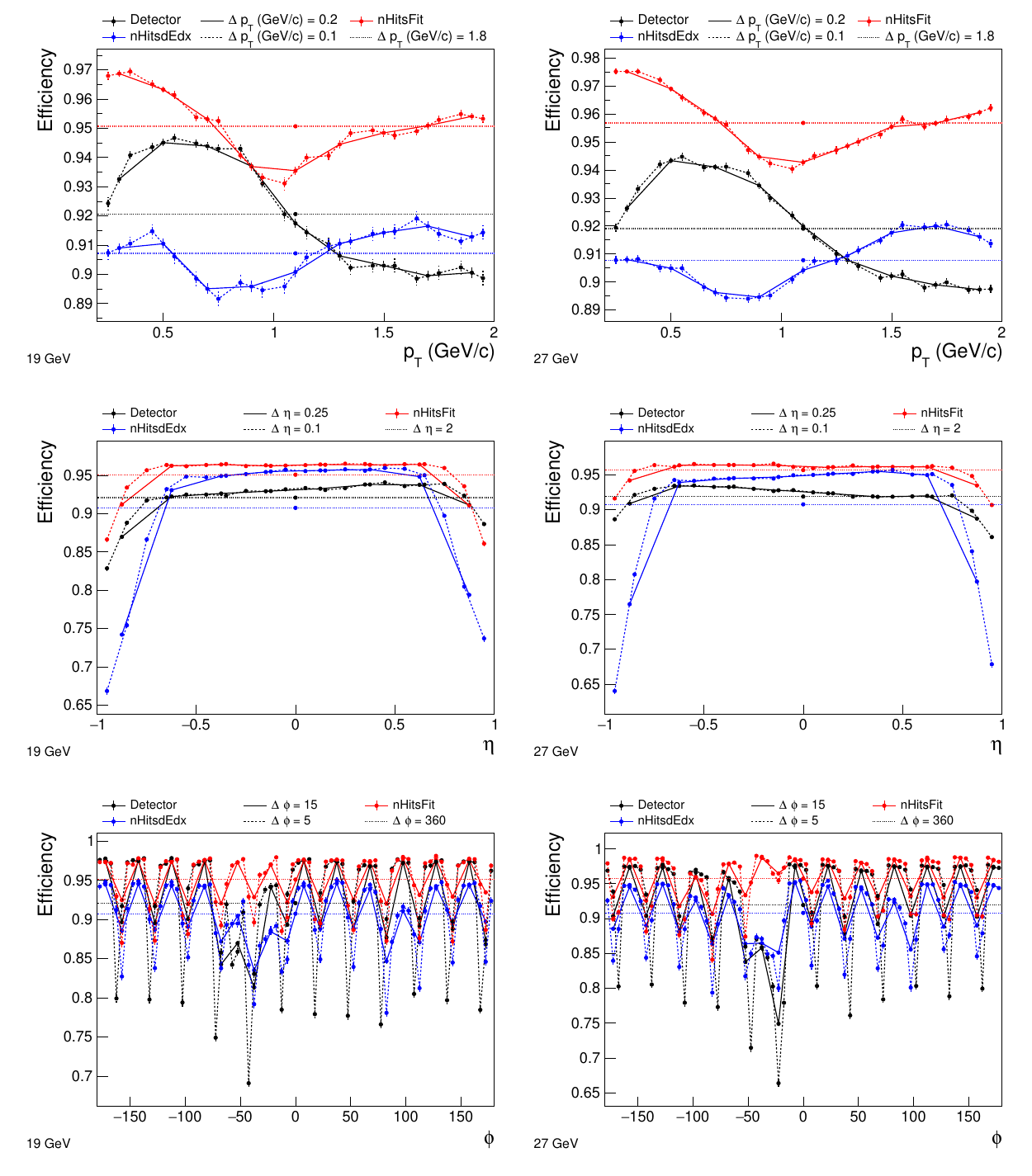}{Overview of calculated electron efficiencies at \ensuremath{\surd}s$_{\text{NN}}$ = 19.6 GeV (left column) \& 27 GeV (right column). For a detailed description see Figure~\ref{effcorr_fig_dettrackqual_overview}.}{effcorr_fig_dettrackqual_overviewA}{1} %

\wrapifneeded{0.50}{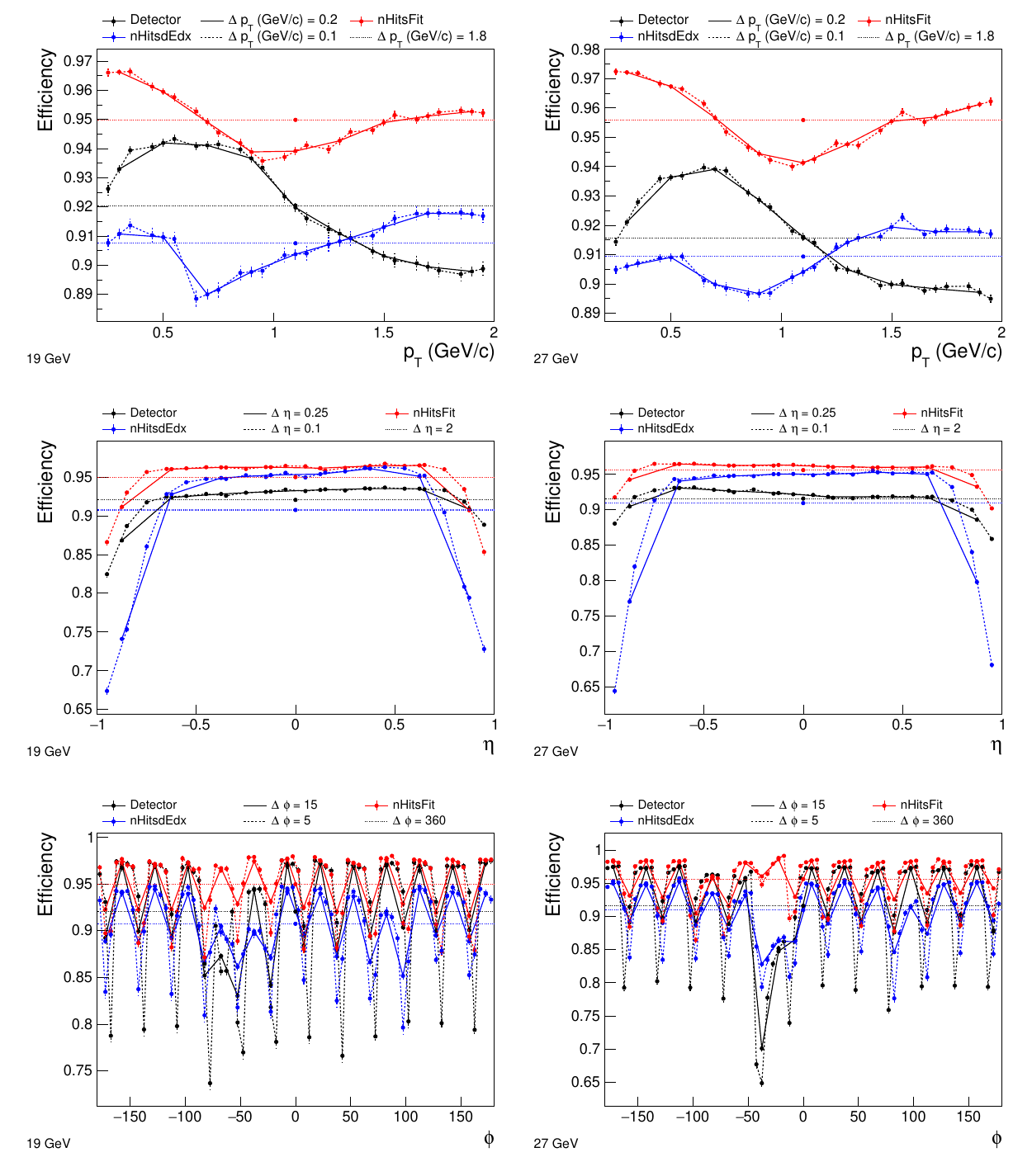}{Overview of calculated positron efficiencies at \ensuremath{\surd}s$_{\text{NN}}$ = 19.6 GeV (left column) \& 27 GeV (right column). For a detailed description see Figure~\ref{effcorr_fig_dettrackqual_overview}.}{effcorr_fig_dettrackqual_overviewB}{1} %

\wrapifneeded{0.50}{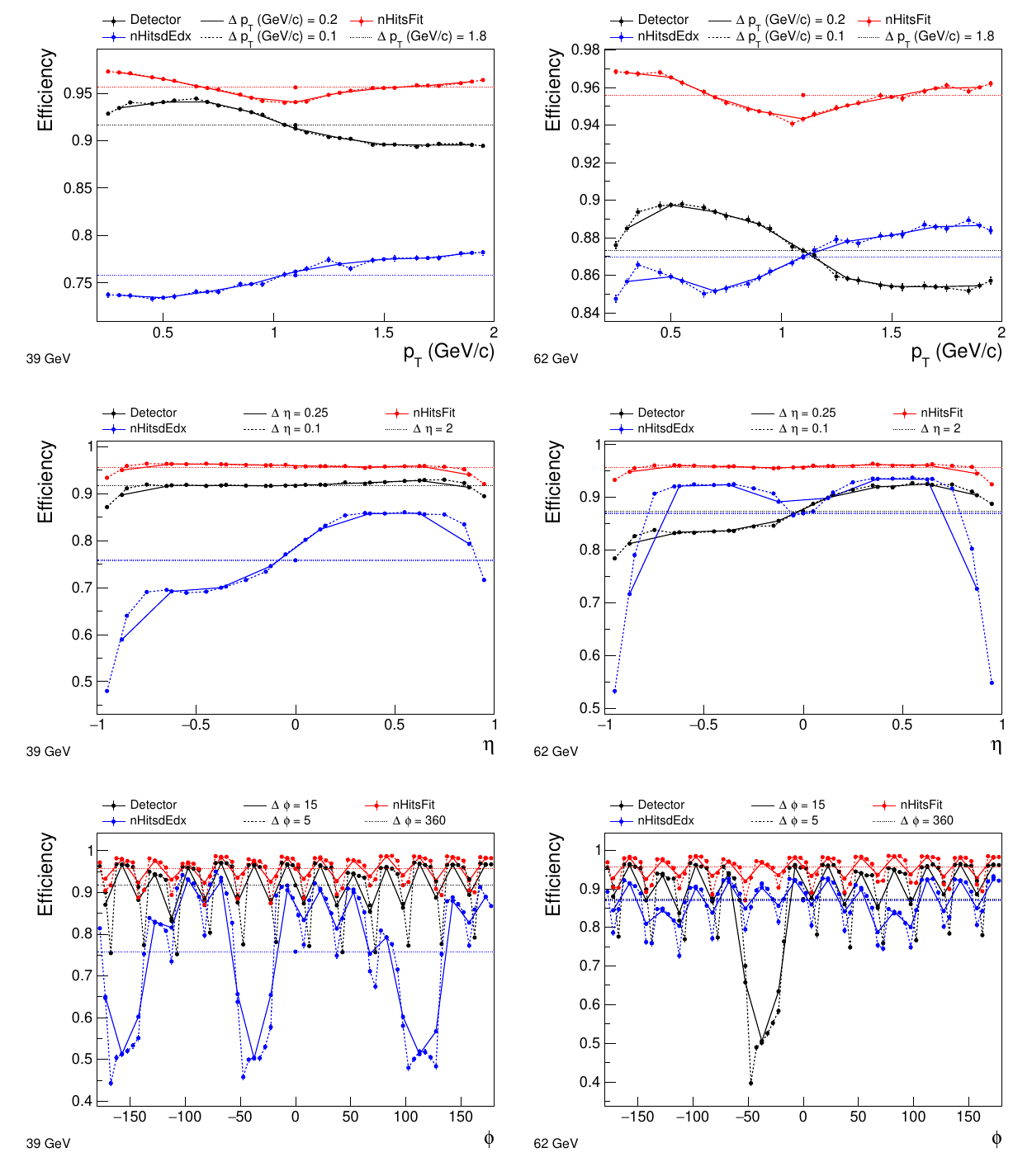}{Overview of calculated positron efficiencies at \ensuremath{\surd}s$_{\text{NN}}$ = 39 GeV (left column) \& 62.4 GeV (right column). For a detailed description see Figure~\ref{effcorr_fig_dettrackqual_overview}.}{effcorr_fig_dettrackqual_overviewC}{1} %

\chapter{Software}
\label{app_software}\hyperlabel{app_software}%

\section{StV0TofCorrection: Examples}
\label{stv0tof_examples}\hyperlabel{stv0tof_examples}%
\begin{enumerate}[label=\arabic*.]

\item{} Kaon from $\Omega^-\to\Lambda~K^-$ (one step case)

\begin{lstlisting}[language=c++,firstnumber=1,]
// init once
StV0TofCorrection *v0tofcorr;
v0tofcorr = new StV0TofCorrection();
// event vertex
StThreeVectorD v0;
// --- for each omega candidate -----------------
// kaon helix
StPhysicalHelixD hel;
// original kaon TOF
Float_t Tof;
// corrected beta set to original
Float_t bCorr = BetaOrig;
// decay vertex of omega
StThreeVectorD v_om;
// kaon TOF hit
StThreeVectorD vtof;
// 4-mom vectors of omega
StLorentzVector t_om;
// kaon momentum
Float_t mom;
// correct beta if TOF hit exists
if ( tofhit ) {
  v0tofcorr->setVectors3D(v0)(v_om)(vtof);
  v0tofcorr->setMotherTracks(t_om);
  // using default cut on m2(K-)
  if (v0tofcorr->correctBeta(hel,Tof,bCorr,mom,1)) {
    // fill ntuple, histos etc.
  }
  v0tofcorr->clearContainers();
}
\end{lstlisting}

\item{} Proton from $\Omega^-\to\Lambda~K^-\to~p\pi^-K^-$ (two step case)

\begin{lstlisting}[language=c++,firstnumber=1,]
// init once
StV0TofCorrection *v0tofcorr;
v0tofcorr = new StV0TofCorrection();
// event vertex
StThreeVectorD v0;
// --- for each omega candidate -----------------
// proton helix
StPhysicalHelixD helix;
// original proton TOF
Float_t Tof;
// corrected beta set to original
Float_t BetaCorr = BetaOrig;
// decay vertices of omega & lambda (ordering!)
StThreeVectorD v_om,v_lam;
// proton TOF hit
StThreeVectorD vtof;
// 4-mom vectors of omega & lambda (ordering!)
StLorentzVector t_om, t_lam;
// correct beta if TOF hit exists
if ( tofhit ) {
  v0tofcorr->setVectors3D(v0)(v_om)(v_lam)(vtof);
  v0tofcorr->setMotherTracks(t_om)(t_lam);
  v0tofcorr->correctBeta(helix,Tof,BetaCorr);
  v0tofcorr->clearContainers();
}
// calc new proton mass m2 by using BetaCorr
Float_t Mass2Corr = p*p*(1./(BetaCorr*BetaCorr)-1.);
// cut on new proton mass
if ( Mass2Corr > ... && Mass2Corr < ... ) {
  // fill ntuple, histos etc.
}
\end{lstlisting}

\end{enumerate}

\section{StBadRdos Database Files}
\label{app_stbadrdos_dbfiles}\hyperlabel{app_stbadrdos_dbfiles}%

{\bf database/dbfiles/database\_11GeV\_2010} 
 
\begin{lstlisting}[firstnumber=1,]
11148001 3 17.01 23.02 23.08
11151010 4 11.10 17.01 23.02 23.08
11151028 3 17.01 23.02 23.08
11151040 4 17.01 21.11 23.02 23.08
11151080 5 01.07 17.01 21.11 23.02 23.08
11152053 6 01.07 07.05 17.01 21.11 23.02 23.08
11153012 7 01.07 01.09 07.05 17.01 21.11 23.02 23.08
11156032 8 01.07 01.09 07.05 09.03 17.01 21.11 23.02 23.08
11158011 9 01.07 01.09 07.05 09.03 11.01 17.01 21.11 23.02 23.08
\end{lstlisting}

{\bf database/dbfiles/database\_19GeV\_2011} 
 
\begin{lstlisting}[firstnumber=1,]
12112008 3 01.06 19.06 19.07
12112060 4 01.06 19.01 19.06 19.07
12119001 5 01.06 05.11 19.01 19.06 19.07
12119008 6 01.06 03.07 05.11 19.01 19.06 19.07
12119049 4 01.06 19.01 19.06 19.07
\end{lstlisting}

{\bf database/dbfiles/database\_200GeV\_2010} 
 
\begin{lstlisting}[firstnumber=1,]
11002120 1 23.08
11003094 2 23.07 23.08
11003101 4 23.07 23.08 23.09 23.10
11004007 5 23.02 23.07 23.08 23.09 23.10
11004040 2 23.02 23.08
11011017 3 09.05 23.02 23.08
11014041 2 23.02 23.08
11021015 8 19.07 19.08 19.09 19.10 19.11 19.12 23.02 23.08
11021034 3 19.09 23.02 23.08
11028018 2 23.02 23.08
11028114 3 23.01 23.02 23.08
11041021 4 17.01 23.01 23.02 23.08
11044005 3 17.01 23.02 23.08
\end{lstlisting}

{\bf database/dbfiles/database\_200GeV\_2011} 
 
\begin{lstlisting}[firstnumber=1,]
12126071 5 01.06 03.07 03.08 19.06 19.07
12129010 3 01.06 19.06 19.07
12129016 4 01.06 03.07 19.06 19.07
12129023 5 01.06 03.07 03.08 19.06 19.07
12138072 7 01.06 03.07 03.08 05.03 19.04 19.06 19.07
12139028 5 01.06 03.07 03.08 19.06 19.07
12158040 11 01.06 03.07 03.08 17.07 17.08 17.09 17.10 17.11 17.12 19.06 19.07
12158051 5 01.06 03.07 03.08 19.06 19.07
12162007 6 01.06 03.07 03.08 05.09 19.06 19.07
12162008 5 01.06 03.07 03.08 19.06 19.07
12162009 6 01.06 03.07 03.08 05.09 19.06 19.07
12162014 5 01.06 03.07 03.08 19.06 19.07
12162029 6 01.06 03.07 03.08 05.09 19.06 19.07
\end{lstlisting}

{\bf database/dbfiles/database\_27GeV\_2011} 
 
\begin{lstlisting}[firstnumber=1,]
12172013 6 01.06 03.07 03.08 05.09 19.06 19.07
12174055 12 01.06 03.07 03.08 05.01 05.02 05.03 05.04 05.05 05.06 05.09 19.06 19.07
12174056 6 01.06 03.07 03.08 05.09 19.06 19.07
\end{lstlisting}

{\bf database/dbfiles/database\_39GeV\_2010} 
 
\begin{lstlisting}[firstnumber=1,]
11099002 3 17.01 23.02 23.08
11101092 4 15.12 17.01 23.02 23.08
11102073 5 15.12 17.01 17.05 23.02 23.08
11103025 6 05.09 15.12 17.01 17.05 23.02 23.08
11105011 3 17.01 23.02 23.08
11105018 4 15.12 17.01 23.02 23.08
11105037 5 01.04 15.12 17.01 23.02 23.08
11106008 6 01.04 03.09 15.12 17.01 23.02 23.08
11106024 7 01.04 03.09 05.09 15.12 17.01 23.02 23.08
11106039 8 01.04 01.12 03.09 05.09 15.12 17.01 23.02 23.08
11108019 9 01.04 01.12 03.09 05.09 09.07 15.12 17.01 23.02 23.08
11108029 10 01.04 01.12 03.09 05.09 09.07 13.12 15.12 17.01 23.02 23.08
11108050 9 01.04 01.12 05.09 09.07 13.12 15.12 17.01 23.02 23.08
11108075 8 01.04 01.12 05.09 13.12 15.12 17.01 23.02 23.08
11109019 9 01.04 01.12 03.09 05.09 13.12 15.12 17.01 23.02 23.08
11109088 12 01.04 01.07 01.12 03.03 03.09 05.09 13.12 15.12 17.01 19.09 23.02 23.08
11109114 11 01.04 01.12 03.03 03.09 05.09 13.12 15.12 17.01 19.09 23.02 23.08
11110025 10 01.04 01.12 03.09 05.09 13.12 15.12 17.01 19.09 23.02 23.08
\end{lstlisting}

{\bf database/dbfiles/database\_62GeV\_2010} 
 
\begin{lstlisting}[firstnumber=1,]
11078018 3 17.01 23.02 23.08
11092009 4 15.12 17.01 23.02 23.08
11092095 5 01.04 15.12 17.01 23.02 23.08
11094007 8 01.04 03.03 03.09 05.09 15.12 17.01 23.02 23.08
11095022 3 17.01 23.02 23.08
\end{lstlisting}

{\bf database/dbfiles/database\_7GeV\_2010} 
 
\begin{lstlisting}[firstnumber=1,]
11114040 4 17.01 19.09 23.02 23.08
11114055 5 03.06 17.01 19.09 23.02 23.08
11116017 4 17.01 19.09 23.02 23.08
11117024 5 03.06 17.01 19.09 23.02 23.08
11118036 6 03.06 09.07 17.01 19.09 23.02 23.08
11119043 7 03.06 05.05 09.07 17.01 19.09 23.02 23.08
11121054 8 03.03 03.06 05.05 09.07 17.01 19.09 23.02 23.08
11122037 9 03.03 03.06 05.05 09.07 11.07 17.01 19.09 23.02 23.08
11122111 10 01.12 03.03 03.06 05.05 09.07 11.07 17.01 19.09 23.02 23.08
11125066 4 01.12 17.01 23.02 23.08
11130034 5 01.12 17.01 17.05 23.02 23.08
11133060 6 01.07 01.12 17.01 17.05 23.02 23.08
11134053 5 01.07 01.12 17.01 23.02 23.08
11136090 6 01.07 01.12 17.01 19.06 23.02 23.08
11136157 10 01.07 01.12 13.09 13.10 13.11 13.12 17.01 19.06 23.02 23.08
11137107 11 01.07 01.12 07.05 13.09 13.10 13.11 13.12 17.01 19.06 23.02 23.08
11138036 12 01.12 09.10 13.09 13.10 13.11 13.12 17.01 17.05 19.06 23.02 23.08 23.09
11139051 4 01.12 17.01 23.02 23.08
11146054 3 17.01 23.02 23.08
\end{lstlisting}

\section{STAR CVS to git Migration}
\label{cvs2git_source}\hyperlabel{cvs2git_source}%

\noindent
\begin{description}
\item[{ configureHttpProxy.sh
}] \hspace{0em}\\
\end{description}

\begin{lstlisting}[language=bash,firstnumber=1,]
#!/bin/bash
cfgfile=$HOME/gerrit_review/etc/gerrit.config
git config --file $cfgfile auth.type HTTP
git config --file $cfgfile --unset auth.httpHeader
git config --file $cfgfile gerrit.canonicalWebUrl http://gerrit4star.the-huck.com/
git config --file $cfgfile httpd.listenUrl proxy-http://127.0.0.1:8081/
$HOME/gerrit_review/bin/gerrit.sh start
\end{lstlisting}

\noindent
\begin{description}
\item[{ configureCGit.sh
}] \hspace{0em}\\
\end{description}

\begin{lstlisting}[language=bash,firstnumber=1,]
#!/bin/bash
cfgfile=$HOME/gerrit_review/etc/gerrit.config
cgipath=/home/patrick/public/cgit4star.the-huck.com/public/cgit.cgi
git config --file $cfgfile gitweb.cgi $cgipath
git config --file $cfgfile gitweb.url http://cgit4star.the-huck.com/
git config --file $cfgfile gitweb.type cgit
$HOME/gerrit_review/bin/gerrit.sh restart
\end{lstlisting}

\noindent
\begin{description}
\item[{ convertStarCVS.sh
}] \hspace{0em}\\
\end{description}

\begin{lstlisting}[language=bash,firstnumber=1,]
#!/bin/bash
#set -xv

#=================================================================
# variables
#=================================================================

workdir="/star/institutions/lbl/huck/star_cvs2git"
gitdir="$workdir/bare"
if [ ! -d $gitdir ]; then mkdir -v $gitdir; fi
cvs2svndir="$workdir/cvs2svn-2.4.0"
optfile="$workdir/git4star/my.cvs2git.options"
tmpopt="$workdir/tmp.cvs2git.options"

splitup="StRoot StarVMC StDb StarDb asps kumacs pams QtRoot"
skip="scripts online Cons conf root root3 group cgi CVSROOT"
skip=$skip" OnlTools root5 .backup offline obsolete Dsv HISTORY"
list=`find $CVSROOT -noleaf -maxdepth 1 -type d ! -name repository`

#=================================================================
# functions
#=================================================================

StrIsInArr() {
  for i in $2; do
    if [[ "$1" == *"$i"* ]]; then return 0; fi
  done
  return 1
}

printRepo() {
  echo "========================================================="
  echo "current repo: $1"
  echo "========================================================="
}

runCvs2Git() {
  sed "s:STARCVS-REPO:$1:" $optfile >> $tmpopt
  cd $cvs2svndir
  cvs2git --options=$tmpopt
  rm $tmpopt
}

genGitRepo() {
  rm -rfv $1; mkdir -pv $1; cd $1
  git init --bare
  tmpdir=$cvs2svndir/cvs2svn-tmp
  cat $tmpdir/git-blob.dat $tmpdir/git-dump.dat | git fast-import
  rm -rfv $tmpdir
  python $cvs2svndir/contrib/git-move-refs.py
  git gc --prune=now
}

convertRepo() {
  repo=$1
  printRepo $repo
  runCvs2Git $repo

  repodir=$2
  repobase=`basename $repo`
  repocat=$repodir
  if [ ! -z "$repodir" ]; then repodir=$repodir"-"; fi
  reponame=$repodir$repobase.git
  outdir="$gitdir/$reponame"
  echo $outdir

  genGitRepo $outdir

  cfgfile="$outdir/config"
  echo "[gitweb]" >> $cfgfile
  echo "        owner = STAR" >> $cfgfile
  echo "        description = $repodir$repobase" >> $cfgfile
  echo "        category = $repocat" >> $cfgfile
}

#=================================================================
# loop repos
# ($(top,sub)repo have to hold absolute urls to afs)
#=================================================================

for toprepo in $list; do

  if StrIsInArr $toprepo "$skip"; then continue; fi # skip dirs

  if StrIsInArr $toprepo "$splitup"; then # descent into dir

    toprepobase=`basename $toprepo`
    sublist=`find $toprepo -noleaf -maxdepth 1 -type d ! -name $toprepobase`
    for subrepo in $sublist; do # loop sub-repositories
      convertRepo $subrepo $toprepobase
    done

  else # convert top-repository
    convertRepo $toprepo
  fi

done
\end{lstlisting}

\noindent
\begin{description}
\item[{ createGerritProjects.sh
}] \hspace{0em}\\
\end{description}

\begin{lstlisting}[language=bash,firstnumber=1,]
#!/bin/bash
workdir="/star/institutions/lbl/huck/star_cvs2git"
gitdir="$workdir/bare"
tmpCloneDir="$workdir/tmp"
mkdir $tmpCloneDir
for dir in `ls $workdir/bare`; do
  cd $workdir
  proj=${dir%.git}
  # create, git clone and push project
  ssh gerrit4star gerrit create-project --name $proj
  git clone $gitdir/$proj.git $tmpCloneDir/$proj
  cd $tmpCloneDir/$proj
  git push ssh://gerrit4star/$proj *:*
  # set require sign-off and change-id
  ssh gerrit4star gerrit set-project $proj --so --id
done
rm -rf $tmpCloneDir
\end{lstlisting}

\noindent
\begin{description}
\item[{ writeManifest.sh
}] \hspace{0em}\\
\end{description}

\begin{lstlisting}[language=bash,firstnumber=1,]
#!/bin/bash
workdir="/star/institutions/lbl/huck/star_cvs2git"
gitdir="$workdir/bare"
manifest=default.xml
echo '<?xml version="1.0" encoding="UTF-8"?>' > $manifest
echo '<manifest>' >> $manifest
echo '  <remote  name="gerrit4star"' >> $manifest
echo '    fetch="ssh://gerrit4star"' >> $manifest
echo '    review="http://gerrit4star.the-huck.com" />' >> $manifest
echo '  <default revision="refs/heads/master"' >> $manifest
echo '    remote="gerrit4star"' >> $manifest
echo '    sync-j="4" />' >> $manifest
for repo in `ls $gitdir`; do
  path=`echo ${repo%.git} | sed 's:-:/:'`
  echo '    <project name="'${repo%.git}'" path="'${path}'"/>' >> $manifest
done
echo '</manifest>' >> $manifest
\end{lstlisting}

\section{Screenshots of Gerrit and CGit for STAR}
\label{app_cvs2git_screen}\hyperlabel{app_cvs2git_screen}%

\wrapifneeded{0.50}{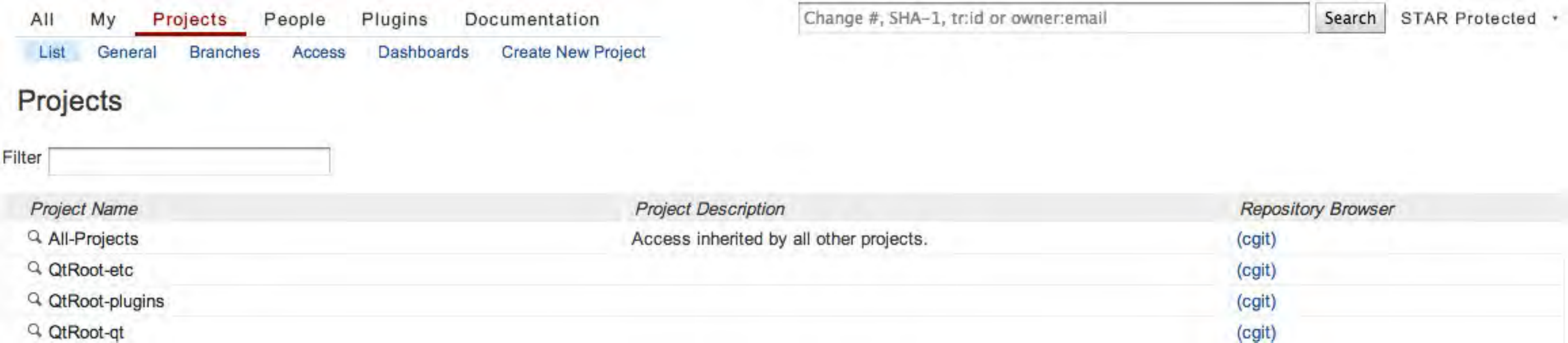}{gerrit4star project list view}{}{0.9} %

\wrapifneeded{0.50}{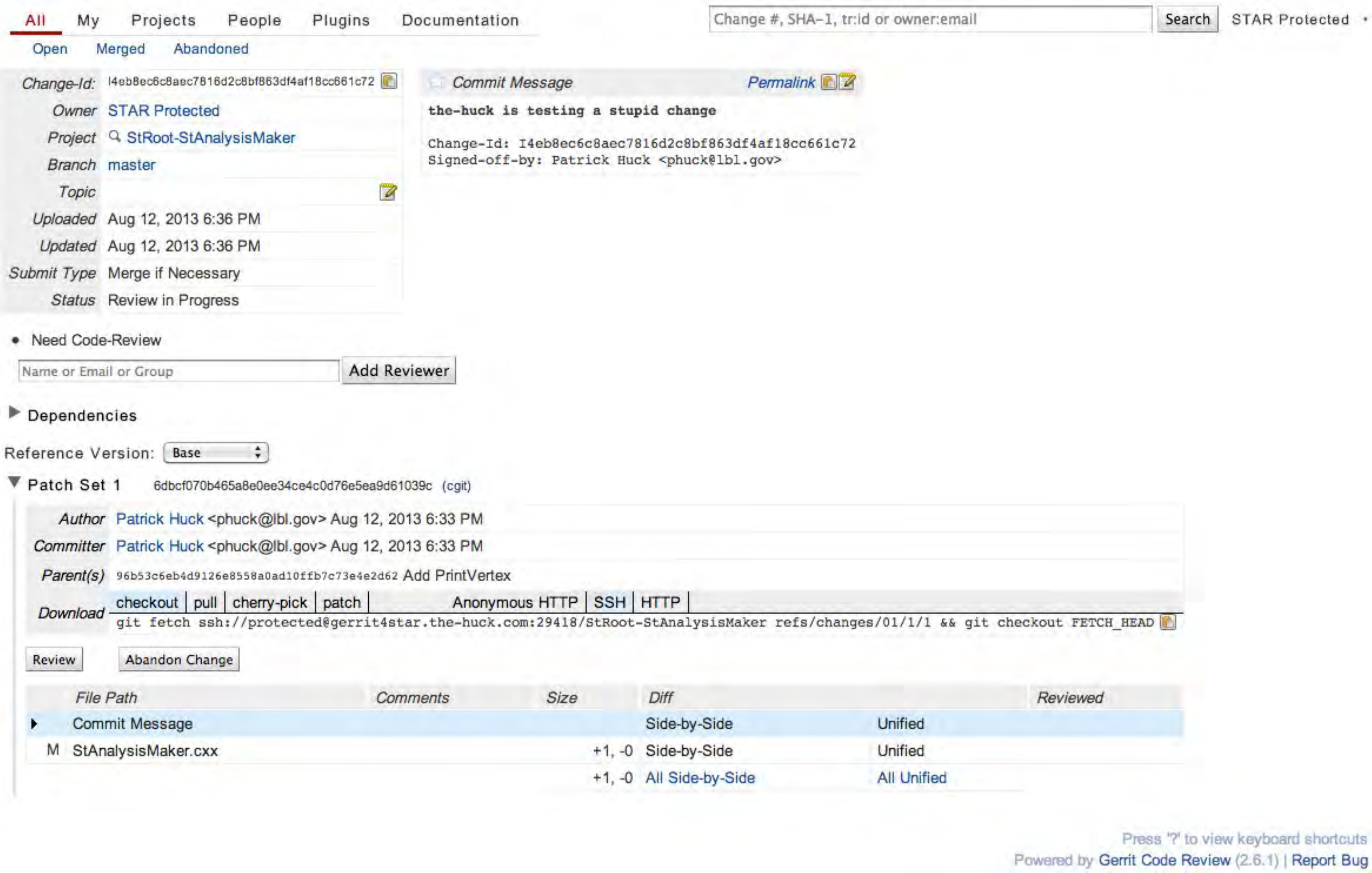}{gerrit4star change view}{}{0.9} %

\wrapifneeded{0.50}{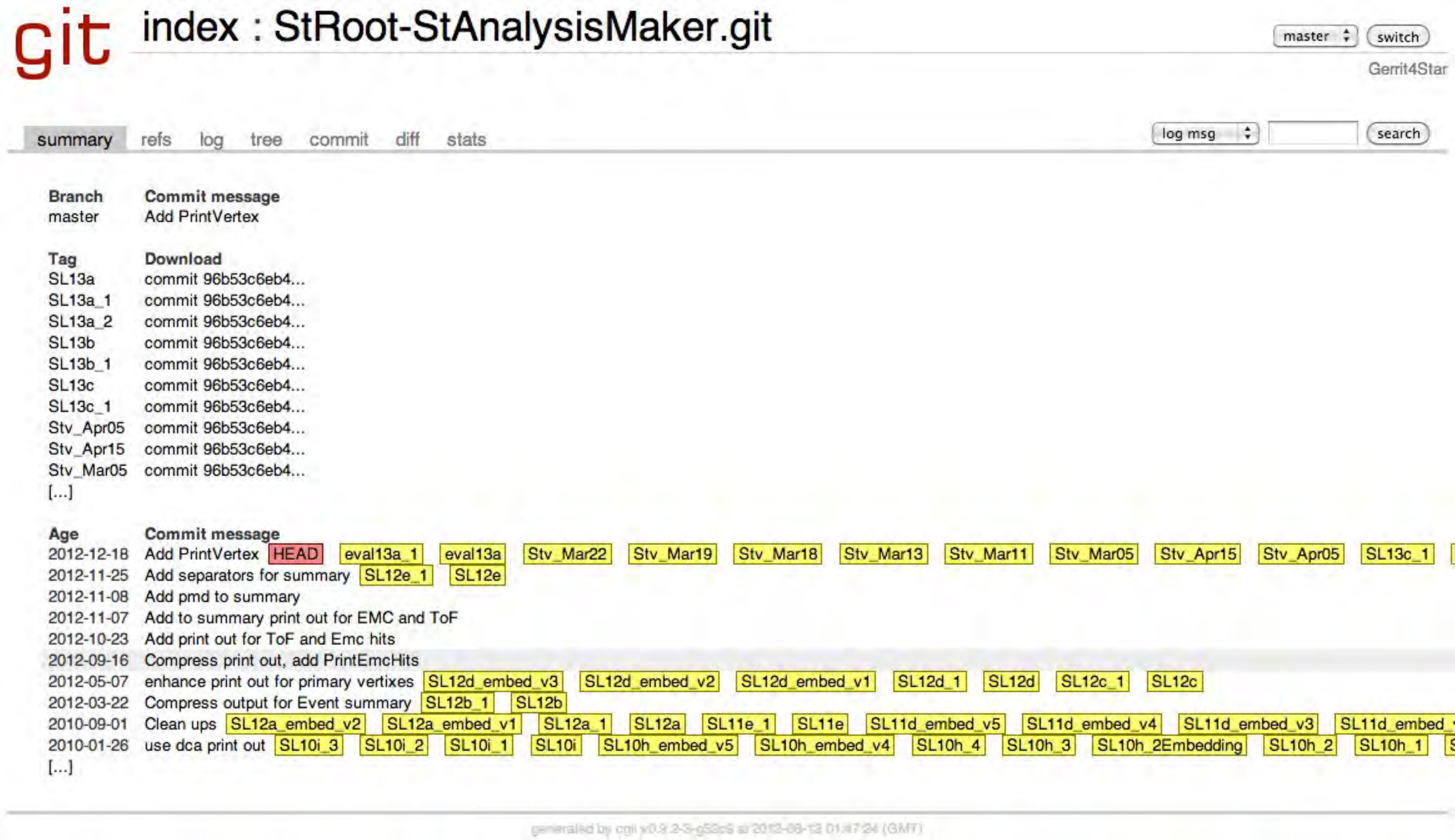}{cgit4star repository view}{}{0.9} %

\section{ckon -{} Typical Directory Structure}
\label{ckon_dirlist}\hyperlabel{ckon_dirlist}%

\begin{lstlisting}[language=bash,firstnumber=1,]
StRoot/
   ElectronPid/
       BetaPanels.cxx
       BetaPanels.h
       PureSampleAnalysis.cxx
       PureSampleAnalysis.h
       SigmaElFitsMaker.cxx
       SigmaElFitsMaker.h
       SigmaElFitsPlotter.cxx
       SigmaElFitsPlotter.h
       SigmaElFitsUtils.cxx
       SigmaElFitsUtils.h
       programs/
           README
           beta3sig.cc
           dedxCut.cc
           nsigparamsGP.cc
           pureSamp.cc
   StBadRdosDb/
       StBadRdosDb.cxx
       StBadRdosDb.h
       database/
           dbfiles
           genAll.sh
           genBadRdosDb.pl
       macros/
           testStBadRdosDb.C
   YamlCfgReader/
       YamlCfgReader.cxx
       YamlCfgReader.h
       config.yml
\end{lstlisting}

\section{ccsgp: Installation and Implementation}
\label{app_ccsgp}\hyperlabel{app_ccsgp}%

\begin{sidebar}
\textbf{NOTE}

This is a reprint of the online documentation in the version at the time of
this thesis. Refer to the public auto-{}generated version at
\href{http://ccsgp-get-started.readthedocs.org/en/latest/}{http://ccsgp-{}get-{}started.readthedocs.org/\-en/\-latest/\-}.
\end{sidebar}

\texttt{ccs\penalty5000 g\penalty5000 p\penalty5000 \_\penalty5000 g\penalty5000 e\penalty5000 t\penalty5000 \_\penalty5000 s\penalty5000 t\penalty5000 a\penalty5000 r\penalty5000 ted} requires gnuplot, git, virtualenv and optionally hdf5.
Install these dependencies via:

\begin{lstlisting}[language=bash,firstnumber=1,]
sudo port install gnuplot git-core py27-virtualenv [hdf5] # MacPorts
sudo apt-get install gnuplot git python-virtualenv [libhdf5-dev] # Debian
\end{lstlisting}

\texttt{hdf5} is optional but if you'd like to save your image data to HDF5, you'll
have to install it. Next, clone the \texttt{ccs\penalty5000 g\penalty5000 p\penalty5000 \_\penalty5000 g\penalty5000 e\penalty5000 t\penalty5000 \_\penalty5000 s\penalty5000 t\penalty5000 a\penalty5000 r\penalty5000 ted} git repository, init
the virtualenv, activate it and install all requirements:

\begin{lstlisting}[language=bash,firstnumber=1,]
git clone https://github.com/tschaume/ccsgp_get_started.git --recursive
cd ccsgp_get_started/
virtualenv-2.7 env
source env/bin/activate
pip install -U numpy
pip install -U -r requirements.txt \
    --allow-external gnuplot-py --allow-unverified gnuplot-py
\end{lstlisting}

Every time you start in a new terminal you have to activate the correct python
environment by sourcing \texttt{env/\penalty0 bin/\penalty0 act\penalty5000 i\penalty5000 v\penalty5000 ate} again or instead use
\texttt{env/\penalty0 bin/\penalty0 pyt\penalty5000 hon} directly!  The h5py package is currently omitted from the
requirements. If you want to use it, uncomment the h5py requirement in
"requirements.txt" and rerun \texttt{pip i\penalty5000 n\penalty5000 s\penalty5000 t\penalty5000 a\penalty5000 ll -{}\penalty0 r r\penalty5000 e\penalty5000 q\penalty5000 u\penalty5000 i\penalty5000 r\penalty5000 e\penalty5000 m\penalty5000 e\penalty5000 nts.\penalty0 txt}.  If you
intend to run the examples, clone the test data repository (somewhere outside
of ccsgp\_get\_started repository) and symlink \texttt{pya\penalty5000 n\penalty5000 a\penalty5000 Dir} to the
\texttt{ccs\penalty5000 g\penalty5000 p\penalty5000 \_\penalty5000 g\penalty5000 e\penalty5000 t\penalty5000 \_\penalty5000 s\penalty5000 t\penalty5000 a\penalty5000 r\penalty5000 t\penalty5000 e\penalty5000 d\penalty5000 \_\penalty5000 d\penalty5000 ata} directory:

\begin{lstlisting}[language=bash,firstnumber=1,]
git clone http://gitlab.the-huck.com/github/ccsgp_get_started_data.git
cd ccsgp_get_started/
ln -s <path/to/ccsgp_get_started_data> pyanaDir
\end{lstlisting}

\texttt{pyana.\penalty0 aux.\penalty0 utils.\penalty0 che\penalty5000 c\penalty5000 k\penalty5000 S\penalty5000 y\penalty5000 m\penalty5000 L\penalty5000 ink} checks for the \texttt{pya\penalty5000 n\penalty5000 a\penalty5000 Dir} symbolic link
and the code won't run without it. Hence you need to generate a symlink
called \texttt{pya\penalty5000 n\penalty5000 a\penalty5000 Dir} either to \texttt{ccs\penalty5000 g\penalty5000 p\penalty5000 \_\penalty5000 g\penalty5000 e\penalty5000 t\penalty5000 \_\penalty5000 s\penalty5000 t\penalty5000 a\penalty5000 r\penalty5000 t\penalty5000 e\penalty5000 d\penalty5000 \_\penalty5000 d\penalty5000 ata} or your own
input/output directory (preferably separated from the code repository).
Pull in public STAR dielectron data into a new branch to use \texttt{gp\_\penalty5000 x\penalty5000 fac} and
\texttt{gp\_\penalty5000 p\penalty5000 a\penalty5000 nel}:

\begin{lstlisting}[language=bash,firstnumber=1,]
cd <path/to/ccsgp_get_started_data>
git remote add dielec_public \
    http://gitlab.the-huck.com/star/dielectron_data_public.git
git checkout -b star_dielec
git pull dielec_public master
\end{lstlisting}

If you are part of the STAR collaboration you can also pull in the
protected STAR dielectron data to include it in \texttt{gp\_\penalty5000 p\penalty5000 a\penalty5000 nel}:

\begin{lstlisting}[language=bash,firstnumber=1,]
git remote add dielec_protect \
    http://cgit.the-huck.com/dielectron_data_protected
git pull -Xtheirs dielec_protect master # STAR protected creds
\end{lstlisting}

The examples are based on a dataset of World Bank Indicators. See the
"genExDat.sh" script in the same directory on how I extracted the data into the
correct format for \texttt{ccsgp}. To generate all example plots based on
\texttt{ccs\penalty5000 g\penalty5000 p\penalty5000 \_\penalty5000 g\penalty5000 e\penalty5000 t\penalty5000 \_\penalty5000 s\penalty5000 t\penalty5000 a\penalty5000 r\penalty5000 t\penalty5000 e\penalty5000 d\penalty5000 \_\penalty5000 d\penalty5000 ata} you can run: \texttt{pyt\penalty5000 h\penalty5000 on -{}\penalty0 m p\penalty5000 y\penalty5000 ana}. Alternatively, you
can run a specific module, for instance, and this way plot specific country
initials:

\begin{lstlisting}[language=bash,firstnumber=1,]
python -m pyana.examples.gp_datdir [--log] <initial> <topN>
\end{lstlisting}

Use \texttt{pdf\penalty5000 nup} to put multiple plots on one page. To start on your own read the
documentation below or the source code and use one of the examples as a
template.

\begin{center}
 \line(1,0){444}
 \end{center}

\noindent
\begin{description}
\item[{ \texttt{pyana.\penalty0 exa\penalty5000 m\penalty5000 p\penalty5000 les.\penalty0 gp\_\penalty5000 d\penalty5000 a\penalty5000 t\penalty5000 dir.\penalty0 gp\_\penalty5000 d\penalty5000 a\penalty5000 t\penalty5000 d\penalty5000 i\penalty5000 r\penalty5000 (\penalty5000 i\penalty5000 n\penalty5000 i\penalty5000 t\penalty5000 i\penalty5000 a\penalty5000 l\penalty5000 , t\penalty5000 o\penalty5000 pN)} }] \hspace{0em}\\
 example for plotting from a text file via \texttt{numpy.\penalty0 loa\penalty5000 d\penalty5000 txt} 

\noindent
\begin{description}
\item[{ Procedure
}] ~\begin{itemize}

\item{} prepare input/output directories

\item{} load the data into an OrderedDict() [adjust axes units]

\item{} sort countries from highest to lowest population

\item{} select the <{}topN>{} most populated countries

\item{} call \texttt{ccsgp.\penalty0 mak\penalty5000 e\penalty5000 \_\penalty5000 p\penalty5000 lot} with data from previous step

\item{} See \texttt{pyt\penalty5000 h\penalty5000 on -{}\penalty0 m p\penalty5000 y\penalty5000 ana.\penalty0 exa\penalty5000 m\penalty5000 p\penalty5000 les.\penalty0 gp\_\penalty5000 d\penalty5000 a\penalty5000 t\penalty5000 d\penalty5000 ir -{}\penalty0 h} for help

\end{itemize}
\item[{ Parameters
}] ~\begin{itemize}

\item{} \textbf{initial} (\emph{str})\hspace{0.167em}\textemdash{}\hspace{0.167em}country initial

\item{} \textbf{topN} (\emph{int})\hspace{0.167em}\textemdash{}\hspace{0.167em}number of most populated countries to plot

\end{itemize}
\item[{ Variables
}] ~\begin{itemize}

\item{} \textbf{inDir}\hspace{0.167em}\textemdash{}\hspace{0.167em}input directory according to package structure and initial

\item{} \textbf{outDir}\hspace{0.167em}\textemdash{}\hspace{0.167em}output directory according to package structure

\item{} \textbf{data}\hspace{0.167em}\textemdash{}\hspace{0.167em}OrderedDict with datasets to plot as separate keys

\item{} \textbf{file}\hspace{0.167em}\textemdash{}\hspace{0.167em}data input file for specific country, format: [x y] OR [x y dx dy]

\item{} \textbf{country}\hspace{0.167em}\textemdash{}\hspace{0.167em}country, filename stem of input file

\item{} \textbf{file\_url}\hspace{0.167em}\textemdash{}\hspace{0.167em}absolute url to input file

\item{} \textbf{nSets}\hspace{0.167em}\textemdash{}\hspace{0.167em}number of datasets

\end{itemize}
\end{description}
\item[{ \texttt{pyana.\penalty0 exa\penalty5000 m\penalty5000 p\penalty5000 les.\penalty0 gp\_\penalty5000 l\penalty5000 c\penalty5000 l\penalty5000 tpt.\penalty0 gp\_\penalty5000 l\penalty5000 c\penalty5000 l\penalty5000 t\penalty5000 p\penalty5000 t()} }] \hspace{0em}\\
 example plot to display linecolors, linetypes and pointtypes

\item[{ \texttt{pyana.\penalty0 exa\penalty5000 m\penalty5000 p\penalty5000 les.\penalty0 gp\_\penalty5000 x\penalty5000 fac.\penalty0 gp\_\penalty5000 x\penalty5000 f\penalty5000 a\penalty5000 c()} }] \hspace{0em}\\
 example using QM12 enhancement factors

\noindent
\begin{description}
\item[{ Procedure
}] ~\begin{itemize}

\item{} uses \texttt{gpc\penalty5000 a\penalty5000 lls} kwarg to reset xtics

\item{} \texttt{numpy.\penalty0 loa\penalty5000 d\penalty5000 txt} needs reshaping for input files with only one datapoint

\item{} according poster presentations see QM12 and NSD review

\end{itemize}
\item[{ Variables
}] ~\begin{itemize}

\item{} \textbf{key}\hspace{0.167em}\textemdash{}\hspace{0.167em}translates filename into legend/key label

\item{} \textbf{shift}\hspace{0.167em}\textemdash{}\hspace{0.167em}slightly shift selected data points

\end{itemize}
\end{description}
\item[{ \texttt{pyana.\penalty0 exa\penalty5000 m\penalty5000 p\penalty5000 les.\penalty0 gp\_\penalty5000 p\penalty5000 a\penalty5000 nel.\penalty0 gp\_\penalty5000 p\penalty5000 a\penalty5000 n\penalty5000 e\penalty5000 l\penalty5000 (\penalty5000 v\penalty5000 e\penalty5000 r\penalty5000 s\penalty5000 i\penalty5000 o\penalty5000 n\penalty5000 , s\penalty5000 k\penalty5000 ip)} }] \hspace{0em}\\
 example for a panel plot using QM12 data (see \texttt{gp\_\penalty5000 x\penalty5000 fac})

\noindent
\begin{description}
\item[{ Parameters
}] \hspace{0em}\\
  \textbf{version} (\emph{str})\hspace{0.167em}\textemdash{}\hspace{0.167em}plot version / input subdir name

\end{description}
\item[{ \texttt{pyana.\penalty0 exa\penalty5000 m\penalty5000 p\penalty5000 les.\penalty0 gp\_\penalty5000 s\penalty5000 t\penalty5000 ack.\penalty0 gp\_\penalty5000 s\penalty5000 t\penalty5000 a\penalty5000 c\penalty5000 k\penalty5000 (\penalty5000 v\penalty5000 e\penalty5000 r\penalty5000 s\penalty5000 i\penalty5000 o\penalty5000 n\penalty5000 , e\penalty5000 n\penalty5000 e\penalty5000 r\penalty5000 g\penalty5000 i\penalty5000 e\penalty5000 s\penalty5000 , i\penalty5000 n\penalty5000 c\penalty5000 l\penalty5000 M\penalty5000 e\penalty5000 d\penalty5000 , i\penalty5000 n\penalty5000 c\penalty5000 l\penalty5000 F\penalty5000 i\penalty5000 ts)} }] \hspace{0em}\\
 example for a plot w/ stacked graphs using QM12 data (see \texttt{gp\_\penalty5000 p\penalty5000 a\penalty5000 nel})

\noindent
\begin{description}
\item[{ Procedure
}] ~\begin{itemize}

\item{} omit keys from the legend

\item{} manually add legend entries

\item{} automatically plot arrows for error bars larger than data point value

\end{itemize}
\item[{ Parameters
}] \hspace{0em}\\
  \textbf{version} (\emph{str})\hspace{0.167em}\textemdash{}\hspace{0.167em}plot version / input subdir name

\end{description}
\item[{ \texttt{pyana.\penalty0 exa\penalty5000 m\penalty5000 p\penalty5000 les.\penalty0 gp\_\penalty5000 r\penalty5000 d\penalty5000 iff.\penalty0 gp\_\penalty5000 r\penalty5000 d\penalty5000 i\penalty5000 f\penalty5000 f\penalty5000 (\penalty5000 v\penalty5000 e\penalty5000 r\penalty5000 s\penalty5000 , n\penalty5000 o\penalty5000 m\penalty5000 e\penalty5000 d\penalty5000 , n\penalty5000 o\penalty5000 x\penalty5000 e\penalty5000 , d\penalty5000 i\penalty5000 f\penalty5000 f\penalty5000 R\penalty5000 e\penalty5000 l\penalty5000 , d\penalty5000 i\penalty5000 v\penalty5000 d\penalty5000 N\penalty5000 dy)} }] \hspace{0em}\\
 example for ratio or difference plots using QM12 data (see \texttt{gp\_\penalty5000 p\penalty5000 a\penalty5000 nel})

\noindent
\begin{description}
\item[{ Procedure
}] ~\begin{itemize}

\item{} uses uncertainties package for easier error propagation and rebinning

\item{} statistical error for medium = 0

\item{} statistical error for cocktail \textasciitilde{} 0

\item{} statistical error bar on data stays the same for diff

\item{} implement ratio

\item{} adjust statistical error on data for ratio

\item{} adjust name and ylabel for ratio

\end{itemize}
\item[{ Parameters
}] ~\begin{itemize}

\item{} \textbf{vers} (\emph{str})\hspace{0.167em}\textemdash{}\hspace{0.167em}plot version

\item{} \textbf{nomed} (\emph{bool})\hspace{0.167em}\textemdash{}\hspace{0.167em}don't plot medium

\item{} \textbf{noxe} (\emph{bool})\hspace{0.167em}\textemdash{}\hspace{0.167em}don't plot x-{}errors

\item{} \textbf{diffRel} (\emph{bool})\hspace{0.167em}\textemdash{}\hspace{0.167em}plot ratio

\item{} \textbf{divdNdy} (\emph{bool})\hspace{0.167em}\textemdash{}\hspace{0.167em}divide by dNdy

\end{itemize}
\end{description}
\item[{ \texttt{pyana.\penalty0 exa\penalty5000 m\penalty5000 p\penalty5000 les.\penalty0 gp\_\penalty5000 p\penalty5000 t\penalty5000 s\penalty5000 pec.\penalty0 gp\_\penalty5000 p\penalty5000 t\penalty5000 s\penalty5000 p\penalty5000 e\penalty5000 c()} }] \hspace{0em}\\
 another example for a 2D-{}panel plot (see \texttt{gp\_\penalty5000 p\penalty5000 a\penalty5000 nel})

\end{description}

\begin{center}
 \line(1,0){444}
 \end{center}

\noindent
\begin{description}
\item[{ \texttt{pyana.\penalty0 ccsgp.\penalty0 ccsgp.\penalty0 mak\penalty5000 e\penalty5000 \_\penalty5000 p\penalty5000 l\penalty5000 o\penalty5000 t\penalty5000 (\penalty5000 d\penalty5000 a\penalty5000 t\penalty5000 a\penalty5000 , p\penalty5000 r\penalty5000 o\penalty5000 p\penalty5000 e\penalty5000 r\penalty5000 t\penalty5000 i\penalty5000 e\penalty5000 s\penalty5000 , t\penalty5000 i\penalty5000 t\penalty5000 l\penalty5000 e\penalty5000 s\penalty5000 , *\penalty5000 *\penalty5000 k\penalty5000 w\penalty5000 a\penalty5000 r\penalty5000 gs)} }] \hspace{0em}\\
 main function to generate a 1D plot

\noindent
\begin{description}
\item[{ Procedure
}] ~\begin{itemize}

\item{} each dataset is represented by a numpy array consisting of data points
in the format "[x, y, dx, dy1, dy2]", dy1 = statistical error, dy2 = systematic uncertainty

\item{} for symbol numbers to use in labels see \href{http://bit.ly/1erBgIk}{http://bit.ly/\-1erBgIk}

\item{} lines format: \texttt{\emph{<{}x/\penalty0 y>{}=\penalty0 <{}va\penalty5000 l\penalty5000 ue>{}}:\penalty0  \emph{<{}gn\penalty5000 u\penalty5000 p\penalty5000 l\penalty5000 o\penalty5000 t o\penalty5000 p\penalty5000 t\penalty5000 i\penalty5000 o\penalty5000 ns>{}}},\newline
   horizontal = (along) x, vertical = (along) y

\item{} labels format: \texttt{\emph{lab\penalty5000 e\penalty5000 l t\penalty5000 ext}:\penalty0  [x\penalty5000 , y\penalty5000 , abs.\penalty0  pl\penalty5000 a\penalty5000 c\penalty5000 e\penalty5000 m\penalty5000 e\penalty5000 n\penalty5000 t t\penalty5000 rue/\penalty0 fal\penalty5000 se]}

\item{} arrows format: \texttt{[<{}x\penalty5000 0\penalty5000 >{}\penalty5000 , <{}\penalty5000 y\penalty5000 0\penalty5000 >{}\penalty5000 ]\penalty5000 , [\penalty5000 <{}\penalty5000 x\penalty5000 1\penalty5000 >{}\penalty5000 , <{}\penalty5000 y\penalty5000 1\penalty5000 >{}\penalty5000 ], \emph{<{}gn\penalty5000 u\penalty5000 p\penalty5000 l\penalty5000 o\penalty5000 t p\penalty5000 r\penalty5000 o\penalty5000 ps>{}}} 

\end{itemize}
\item[{ Parameters
}] ~\begin{itemize}

\item{} \textbf{data} (\emph{list})\hspace{0.167em}\textemdash{}\hspace{0.167em}datasets

\item{} \textbf{properties} (\emph{list})\hspace{0.167em}\textemdash{}\hspace{0.167em}gnuplot property strings for each dataset (lc, lw, pt \ldots{})

\item{} \textbf{titles} (\emph{list})\hspace{0.167em}\textemdash{}\hspace{0.167em}legend/key titles for each dataset

\item{} \textbf{name} (\emph{str})\hspace{0.167em}\textemdash{}\hspace{0.167em}basename of output files

\item{} \textbf{title} (\emph{str})\hspace{0.167em}\textemdash{}\hspace{0.167em}image title

\item{} \textbf{debug} (\emph{bool})\hspace{0.167em}\textemdash{}\hspace{0.167em}flag to switch to debug/verbose mode

\item{} \textbf{key} (\emph{list})\hspace{0.167em}\textemdash{}\hspace{0.167em}legend/key options to be applied on top of default\_key

\item{} \textbf{xlabel} (\emph{str})\hspace{0.167em}\textemdash{}\hspace{0.167em}label for x-{}axis

\item{} \textbf{ylabel} (\emph{str})\hspace{0.167em}\textemdash{}\hspace{0.167em}label for y-{}axis

\item{} \textbf{xr} (\emph{list})\hspace{0.167em}\textemdash{}\hspace{0.167em}x-{}axis range

\item{} \textbf{yr} (\emph{list})\hspace{0.167em}\textemdash{}\hspace{0.167em}y-{}axis range

\item{} \textbf{xlog} (\emph{bool})\hspace{0.167em}\textemdash{}\hspace{0.167em}make x-{}axis logarithmic

\item{} \textbf{ylog} (\emph{bool})\hspace{0.167em}\textemdash{}\hspace{0.167em}make y-{}axis logarithmic

\item{} \textbf{lines} (\emph{dict})\hspace{0.167em}\textemdash{}\hspace{0.167em}vertical and horizontal lines

\item{} \textbf{arrows} (\emph{list})\hspace{0.167em}\textemdash{}\hspace{0.167em}arrows

\item{} \textbf{labels} (\emph{dict})\hspace{0.167em}\textemdash{}\hspace{0.167em}labels

\item{} \textbf{lmargin} (\emph{float})\hspace{0.167em}\textemdash{}\hspace{0.167em}defines left margin size (relative to screen)

\item{} \textbf{bmargin} (\emph{float})\hspace{0.167em}\textemdash{}\hspace{0.167em}defines bottom margin size

\item{} \textbf{rmargin} (\emph{float})\hspace{0.167em}\textemdash{}\hspace{0.167em}defines right margin size

\item{} \textbf{tmargin} (\emph{float})\hspace{0.167em}\textemdash{}\hspace{0.167em}defines top margin size

\item{} \textbf{arrow\_offset} (\emph{float})\hspace{0.167em}\textemdash{}\hspace{0.167em}offset from data point for special error bars (see gp\_panel)

\item{} \textbf{arrow\_length} (\emph{float})\hspace{0.167em}\textemdash{}\hspace{0.167em}length of arrow from data point towards zero for special error bars (see gp\_panel)

\item{} \textbf{arrow\_bar} (\emph{float})\hspace{0.167em}\textemdash{}\hspace{0.167em}width of vertical bar at end of special error bars (see gp\_panel)

\item{} \textbf{gpcalls} (\emph{list})\hspace{0.167em}\textemdash{}\hspace{0.167em}execute arbitrary gnuplot set commands

\end{itemize}
\item[{ Returns
}] \hspace{0em}\\
   MyPlot

\end{description}
\item[{ \texttt{pyana.\penalty0 ccsgp.\penalty0 ccsgp.\penalty0 rep\penalty5000 e\penalty5000 a\penalty5000 t\penalty5000 \_\penalty5000 p\penalty5000 l\penalty5000 o\penalty5000 t\penalty5000 (\penalty5000 p\penalty5000 l\penalty5000 t\penalty5000 , n\penalty5000 a\penalty5000 m\penalty5000 e\penalty5000 , *\penalty5000 *\penalty5000 k\penalty5000 w\penalty5000 a\penalty5000 r\penalty5000 gs)} }] \hspace{0em}\\
 repeat a plot with different properties (kwargs see make\_plot)

\noindent
\begin{description}
\item[{ Parameters
}] ~\begin{itemize}

\item{} \textbf{plt} (\emph{MyPlot})\hspace{0.167em}\textemdash{}\hspace{0.167em}plot to repeat

\item{} \textbf{name} (\emph{str})\hspace{0.167em}\textemdash{}\hspace{0.167em}basename of new output file(s)

\end{itemize}
\item[{ Returns
}] \hspace{0em}\\
   plt

\end{description}
\item[{ \texttt{pyana.\penalty0 ccsgp.\penalty0 ccsgp.\penalty0 mak\penalty5000 e\penalty5000 \_\penalty5000 p\penalty5000 a\penalty5000 n\penalty5000 e\penalty5000 l\penalty5000 (\penalty5000 d\penalty5000 p\penalty5000 t\penalty5000 \_\penalty5000 d\penalty5000 i\penalty5000 c\penalty5000 t\penalty5000 , *\penalty5000 *\penalty5000 k\penalty5000 w\penalty5000 a\penalty5000 r\penalty5000 gs)} }] \hspace{0em}\\
 make a panel plot

\noindent
\begin{description}
\item[{ Parameters
}] ~\begin{itemize}

\item{} \textbf{name/title/debug}\hspace{0.167em}\textemdash{}\hspace{0.167em}are global options used once to initialize the multiplot

\item{} \textbf{x,yr/x,ylog/lines/labels/gpcalls}\hspace{0.167em}\textemdash{}\hspace{0.167em}are applied on each subplot

\item{} same for "r,l,b,tmargin" where "r,lmargin" will be reset to allow for merged y-{}axes

\item{} \textbf{key/ylabel}\hspace{0.167em}\textemdash{}\hspace{0.167em}are only plotted in first subplot

\item{} \textbf{xlabel}\hspace{0.167em}\textemdash{}\hspace{0.167em}is centered over entire panel

\item{} \textbf{layout}\hspace{0.167em}\textemdash{}\hspace{0.167em}\texttt{<{}co\penalty5000 l\penalty5000 s\penalty5000 >{}\penalty5000 x\penalty5000 <{}\penalty5000 r\penalty5000 o\penalty5000 ws>{}}, defaults to horizontal panel if omitted

\item{} \textbf{key\_subplot\_id}\hspace{0.167em}\textemdash{}\hspace{0.167em}sets the desired subplot to put the key in

\item{} \textbf{dpt\_dict} (\emph{dict})\hspace{0.167em}\textemdash{}\hspace{0.167em}\texttt{Ord\penalty5000 e\penalty5000 r\penalty5000 e\penalty5000 d\penalty5000 D\penalty5000 ict} with subplot titles as keys and lists of
\texttt{mak\penalty5000 e\penalty5000 \_\penalty5000 p\penalty5000 lot} "data/properties/titles" as values:

\begin{lstlisting}[language=bash,firstnumber=1,]
OrderedDict('subplot-title': [data, properties, titles], ...)
\end{lstlisting}

\end{itemize}
\end{description}
\end{description}

\begin{center}
 \line(1,0){444}
 \end{center}

\noindent
\begin{description}
\item[{ MyPlot Base Class
}] \hspace{0em}\\
 \texttt{MyP\penalty5000 lot} is the base class of \texttt{ccsgp} for basic gnuplot setup (bars, grid,
title, key, terminal, multiplot) also providing utility functions for general
plotting.

\noindent
\begin{description}
\item[{ Parameters
}] ~\begin{itemize}

\item{} \textbf{title} (\emph{str})\hspace{0.167em}\textemdash{}\hspace{0.167em}image title

\item{} \textbf{name} (\emph{str})\hspace{0.167em}\textemdash{}\hspace{0.167em}basename used for output files

\item{} \textbf{debug} (\emph{bool})\hspace{0.167em}\textemdash{}\hspace{0.167em}debug flag for verbose gnuplot output

\end{itemize}
\item[{ Variables
}] ~\begin{itemize}

\item{} \textbf{name}\hspace{0.167em}\textemdash{}\hspace{0.167em}basename for output files

\item{} \textbf{epsname}\hspace{0.167em}\textemdash{}\hspace{0.167em}basename + \emph{.eps}

\item{} \textbf{gp}\hspace{0.167em}\textemdash{}\hspace{0.167em}Gnuplot.Gnuplot instance

\item{} \textbf{nPanels}\hspace{0.167em}\textemdash{}\hspace{0.167em}number of panels in a multiplot

\item{} \textbf{nVertLines}\hspace{0.167em}\textemdash{}\hspace{0.167em}number of vertical lines

\item{} \textbf{nLabels}\hspace{0.167em}\textemdash{}\hspace{0.167em}number of labels

\item{} \textbf{nArrows}\hspace{0.167em}\textemdash{}\hspace{0.167em}number of arrows

\item{} \textbf{axisLog}\hspace{0.167em}\textemdash{}\hspace{0.167em}flags for logarithmic axes

\item{} \textbf{axisRange}\hspace{0.167em}\textemdash{}\hspace{0.167em}axis range for respective axis (set in setAxisRange)

\end{itemize}
\end{description}
\item[{ \texttt{\_as\penalty5000 c\penalty5000 i\penalty5000 i()} }] \hspace{0em}\\
 write ascii file(s) w/ data contained in plot

\item[{ \texttt{\_cl\penalty5000 a\penalty5000 m\penalty5000 p\penalty5000 (\penalty5000 v\penalty5000 a\penalty5000 l\penalty5000 , m\penalty5000 i\penalty5000 n\penalty5000 i\penalty5000 mum=\penalty0 0, m\penalty5000 a\penalty5000 x\penalty5000 i\penalty5000 mum=\penalty0 255)} }] \hspace{0em}\\
 convenience function to clamp number into min..max range

\item[{ \texttt{\_co\penalty5000 l\penalty5000 o\penalty5000 r\penalty5000 s\penalty5000 c\penalty5000 a\penalty5000 l\penalty5000 e\penalty5000 (\penalty5000 h\penalty5000 e\penalty5000 x\penalty5000 s\penalty5000 t\penalty5000 r\penalty5000 , s\penalty5000 c\penalty5000 a\penalty5000 l\penalty5000 e\penalty5000 f\penalty5000 a\penalty5000 c\penalty5000 tor=\penalty0 1.\penalty0 4)} }] \hspace{0em}\\
 Scales a hex string by "scalefactor". Returns scaled hex string.

\noindent
\begin{description}
\item[{ Notes
}] ~\begin{itemize}

\item{} taken from T. Burgess [285]

\item{} To darken the color, use a float value between 0 and 1.

\item{} To brighten the color, use a float value greater than 1.

\end{itemize}
\item[{ Examples
}] ~
\begin{lstlisting}[language=python,firstnumber=1,]
>>> colorscale("#DF3C3C", .5)
#6F1E1E
>>> colorscale("#52D24F", 1.6)
#83FF7E
>>> colorscale("#4F75D2", 1)
#4F75D2
\end{lstlisting}
\end{description}
\item[{ \texttt{\_co\penalty5000 n\penalty5000 v\penalty5000 e\penalty5000 r\penalty5000 t()} }] \hspace{0em}\\
 convert eps/ps original into pdf, png and jpg format

\item[{ \texttt{\_ge\penalty5000 t\penalty5000 \_\penalty5000 s\penalty5000 t\penalty5000 y\penalty5000 l\penalty5000 e\penalty5000 \_\penalty5000 m\penalty5000 o\penalty5000 d\penalty5000 \_\penalty5000 p\penalty5000 r\penalty5000 o\penalty5000 p\penalty5000 (\penalty5000 p\penalty5000 r\penalty5000 op)} }] \hspace{0em}\\
 get style and modified property string

\item[{ \texttt{\_ha\penalty5000 r\penalty5000 d\penalty5000 c\penalty5000 o\penalty5000 p\penalty5000 y()} }] \hspace{0em}\\
 generate eps, convert to other formats and write data to hdf5

\item[{ \texttt{\_hd\penalty5000 f\penalty5000 5()} }] \hspace{0em}\\
 write data contained in plot to HDF5 file

\noindent
\begin{description}
\item[{ Notes
}] ~\begin{itemize}

\item{} easy numpy import \ensuremath{\rightarrow} (savetxt) \ensuremath{\rightarrow} gnuplot

\item{} export to ROOT objects

\item{} h5py howto (see \href{http://www.h5py.org/docs/intro/quick.html}{http://www.h5py.org/\-docs/\-intro/\-quick.html})

\end{itemize}
\item[{ Raises
}] \hspace{0em}\\
   ImportError

\end{description}
\item[{ \texttt{\_pl\penalty5000 o\penalty5000 t\penalty5000 \_\penalty5000 e\penalty5000 r\penalty5000 r\penalty5000 s\penalty5000 (\penalty5000 d\penalty5000 a\penalty5000 ta)} }] \hspace{0em}\\
 determine whether to plot primary errors separately and plot errorbars if data
has more than two columns which are not all zero

\noindent
\begin{description}
\item[{ Parameters
}] \hspace{0em}\\
  \textbf{data} (\emph{numpy.array})\hspace{0.167em}\textemdash{}\hspace{0.167em}one dataset

\item[{ Variables
}] \hspace{0em}\\
  \textbf{error\_sums}\hspace{0.167em}\textemdash{}\hspace{0.167em}sum of x and y errors

\item[{ Returns
}] \hspace{0em}\\
   True or False

\end{description}
\item[{ \texttt{\_pl\penalty5000 o\penalty5000 t\penalty5000 \_\penalty5000 s\penalty5000 y\penalty5000 s\penalty5000 e\penalty5000 r\penalty5000 r\penalty5000 s\penalty5000 (\penalty5000 d\penalty5000 a\penalty5000 ta)} }] \hspace{0em}\\
 determine whether to plot secondary errors

\noindent
\begin{description}
\item[{ Parameters
}] \hspace{0em}\\
  \textbf{data} (\emph{numpy.array})\hspace{0.167em}\textemdash{}\hspace{0.167em}one dataset

\item[{ Returns
}] \hspace{0em}\\
   True or False

\end{description}
\item[{ \texttt{\_pr\penalty5000 e\penalty5000 t\penalty5000 t\penalty5000 i\penalty5000 f\penalty5000 y\penalty5000 (\penalty5000 s\penalty5000 tr)} }] \hspace{0em}\\
 prettify string, remove special symbols

\item[{ \texttt{\_se\penalty5000 t\penalty5000 t\penalty5000 e\penalty5000 r\penalty5000 (\penalty5000 l\penalty5000 i\penalty5000 st)} }] \hspace{0em}\\
 convenience function to set a list of gnuplot options

\noindent
\begin{description}
\item[{ Parameters
}] \hspace{0em}\\
  \textbf{list} (\emph{list})\hspace{0.167em}\textemdash{}\hspace{0.167em}list of strings given to gnuplot's set command

\end{description}
\item[{ \texttt{\_su\penalty5000 m\penalty5000 \_\penalty5000 e\penalty5000 r\penalty5000 r\penalty5000 s\penalty5000 (\penalty5000 d\penalty5000 a\penalty5000 t\penalty5000 a\penalty5000 , i)} }] \hspace{0em}\\
 convenience function to calculate sum of i-{}th column

\item[{ \texttt{\_us\penalty5000 i\penalty5000 n\penalty5000 g\penalty5000 (\penalty5000 d\penalty5000 a\penalty5000 t\penalty5000 a\penalty5000 , p\penalty5000 rop=\penalty0 None)} }] \hspace{0em}\\
 determine string with columns to use

\noindent
\begin{description}
\item[{ Parameters
}] ~\begin{itemize}

\item{} \textbf{data} (\emph{numpy.array})\hspace{0.167em}\textemdash{}\hspace{0.167em}one dataset

\item{} \textbf{prop} (\emph{str})\hspace{0.167em}\textemdash{}\hspace{0.167em}property string of a dataset

\end{itemize}
\item[{ Returns
}] \hspace{0em}\\
  \emph{1:2:3}, \emph{1:2:4} or \emph{1:2:3:4} 
\end{description}
\item[{ \texttt{\_wi\penalty5000 t\penalty5000 h\penalty5000 \_\penalty5000 e\penalty5000 r\penalty5000 r\penalty5000 s\penalty5000 (\penalty5000 d\penalty5000 a\penalty5000 t\penalty5000 a\penalty5000 , p\penalty5000 r\penalty5000 op)} }] \hspace{0em}\\
 generate special property string for primary errors

\noindent
\begin{description}
\item[{ Notes
}] ~\begin{itemize}

\item{} currently error bars are drawn in black

\item{} use same linewidth as for points

\item{} TODO: give user the option to draw error bars in lighter color
  according to the respective data points

\end{itemize}
\item[{ Parameters
}] ~\begin{itemize}

\item{} \textbf{data} (\emph{numpy.array})\hspace{0.167em}\textemdash{}\hspace{0.167em}one dataset

\item{} \textbf{prop} (\emph{str})\hspace{0.167em}\textemdash{}\hspace{0.167em}property string of a dataset

\end{itemize}
\item[{ Returns
}] \hspace{0em}\\
   property string for primary errors

\end{description}
\item[{ \texttt{\_wi\penalty5000 t\penalty5000 h\penalty5000 \_\penalty5000 m\penalty5000 a\penalty5000 i\penalty5000 n\penalty5000 (\penalty5000 p\penalty5000 r\penalty5000 op)} }] \hspace{0em}\\
 get the correct property string for main data

\item[{ \texttt{\_wi\penalty5000 t\penalty5000 h\penalty5000 \_\penalty5000 s\penalty5000 y\penalty5000 s\penalty5000 e\penalty5000 r\penalty5000 r\penalty5000 s\penalty5000 (\penalty5000 p\penalty5000 r\penalty5000 op)} }] \hspace{0em}\\
 generate special property string for secondary errors

\noindent
\begin{description}
\item[{ Notes
}] ~\begin{itemize}

\item{} draw box in lighter color than point/line color

\item{} does not support integer line colors, only hex

\end{itemize}
\item[{ Parameters
}] \hspace{0em}\\
  \textbf{prop} (\emph{str})\hspace{0.167em}\textemdash{}\hspace{0.167em}property string of a dataset

\item[{ Returns
}] \hspace{0em}\\
   property string for secondary errors

\end{description}
\item[{ \texttt{add\penalty5000 H\penalty5000 o\penalty5000 r\penalty5000 i\penalty5000 z\penalty5000 o\penalty5000 n\penalty5000 t\penalty5000 a\penalty5000 l\penalty5000 L\penalty5000 i\penalty5000 n\penalty5000 e\penalty5000 (\penalty5000 y\penalty5000 , o\penalty5000 p\penalty5000 ts)} }] \hspace{0em}\\
 draw horizontal line

\noindent
\begin{description}
\item[{ Parameters
}] ~\begin{itemize}

\item{} \textbf{y} (\emph{float})\hspace{0.167em}\textemdash{}\hspace{0.167em}y-{}position

\item{} \textbf{opts} (\emph{str})\hspace{0.167em}\textemdash{}\hspace{0.167em}line draw options

\end{itemize}
\end{description}
\item[{ \texttt{ini\penalty5000 t\penalty5000 D\penalty5000 a\penalty5000 t\penalty5000 a\penalty5000 (\penalty5000 d\penalty5000 a\penalty5000 t\penalty5000 a\penalty5000 , p\penalty5000 r\penalty5000 o\penalty5000 p\penalty5000 e\penalty5000 r\penalty5000 t\penalty5000 i\penalty5000 e\penalty5000 s\penalty5000 , t\penalty5000 i\penalty5000 t\penalty5000 l\penalty5000 e\penalty5000 s\penalty5000 , s\penalty5000 u\penalty5000 b\penalty5000 p\penalty5000 l\penalty5000 o\penalty5000 t\penalty5000 \_\penalty5000 t\penalty5000 i\penalty5000 tle=\penalty0 None)} }] \hspace{0em}\\
 initialize the data

\noindent
\begin{description}
\item[{ Notes
}] ~\begin{itemize}

\item{} all lists given as parameters must have the same length.

\item{} each data set is drawn twice to allow for different colors for the errorbars

\item{} error bars use the same linewidth as data points and line color black

\item{} use \emph{boxwidth 0.03 absolute} in gp\_calls to set the width of the uncertainty boxes

\item{} use alternative gnuplot style if "properties" contains a style
specification in the form "with <{}style>{}" and if the style is in
ccsgp.config.supported\_styles (style specification has to be at the
beginning of the property string!)

\end{itemize}
\item[{ Parameters
}] ~\begin{itemize}

\item{} \textbf{data} (\emph{list of numpy arrays})\hspace{0.167em}\textemdash{}\hspace{0.167em}data points w/ format [x, y,
  dx, dy] for each dataset

\item{} \textbf{properties} (\emph{list of str})\hspace{0.167em}\textemdash{}\hspace{0.167em}plot properties for each dataset
  (pt/lw/ps/lc\ldots{})

\item{} \textbf{titles} (\emph{list of strings})\hspace{0.167em}\textemdash{}\hspace{0.167em}key/legend titles for each dataset

\item{} \textbf{subplot\_title} (\emph{str})\hspace{0.167em}\textemdash{}\hspace{0.167em}subplot title for panel plot case

\end{itemize}
\item[{ Variables
}] ~\begin{itemize}

\item{} \textbf{dataSets}\hspace{0.167em}\textemdash{}\hspace{0.167em}zipped titles and data for hdf5/ascii output and
  setAxisRange

\item{} \textbf{data}\hspace{0.167em}\textemdash{}\hspace{0.167em}list of Gnuplot.Data including extra data sets for error
  plotting

\end{itemize}
\end{description}
\item[{ \texttt{plo\penalty5000 t\penalty5000 (\penalty5000 h\penalty5000 a\penalty5000 r\penalty5000 d\penalty5000 c\penalty5000 opy=\penalty0 True)} }] \hspace{0em}\\
 plot and generate output files

\item[{ \texttt{pre\penalty5000 p\penalty5000 a\penalty5000 r\penalty5000 e\penalty5000 \_\penalty5000 p\penalty5000 l\penalty5000 o\penalty5000 t\penalty5000 (\penalty5000 m\penalty5000 a\penalty5000 r\penalty5000 g\penalty5000 ins=\penalty0 Tru\penalty5000 e\penalty5000 , *\penalty5000 *\penalty5000 k\penalty5000 w\penalty5000 a\penalty5000 r\penalty5000 gs)} }] \hspace{0em}\\
 prepare for plotting (calls all members of MyPlot)

\item[{ \texttt{set\penalty5000 A\penalty5000 r\penalty5000 r\penalty5000 o\penalty5000 w\penalty5000 (\penalty5000 p\penalty5000 0\penalty5000 , p\penalty5000 1\penalty5000 , p\penalty5000 r\penalty5000 op)} }] \hspace{0em}\\
 draw an arrow into the figure

\noindent
\begin{description}
\item[{ Parameters
}] ~\begin{itemize}

\item{} \textbf{p0} (\emph{list})\hspace{0.167em}\textemdash{}\hspace{0.167em}start point [x, y]

\item{} \textbf{p1} (\emph{list})\hspace{0.167em}\textemdash{}\hspace{0.167em}end point [x, y]

\item{} \textbf{prop} (\emph{str})\hspace{0.167em}\textemdash{}\hspace{0.167em}gnuplot property string for the arrow

\end{itemize}
\end{description}
\item[{ \texttt{set\penalty5000 A\penalty5000 x\penalty5000 i\penalty5000 s\penalty5000 L\penalty5000 a\penalty5000 b\penalty5000 e\penalty5000 l\penalty5000 (\penalty5000 l\penalty5000 a\penalty5000 b\penalty5000 e\penalty5000 l\penalty5000 , a\penalty5000 xis=\penalty0 \emph{x})} }] \hspace{0em}\\
 set label for specified axis

\noindent
\begin{description}
\item[{ Parameters
}] ~\begin{itemize}

\item{} \textbf{label} (\emph{str})\hspace{0.167em}\textemdash{}\hspace{0.167em}label

\item{} \textbf{axis} (\emph{str})\hspace{0.167em}\textemdash{}\hspace{0.167em}axis which to label

\end{itemize}
\end{description}
\item[{ \texttt{set\penalty5000 A\penalty5000 x\penalty5000 i\penalty5000 s\penalty5000 L\penalty5000 o\penalty5000 g\penalty5000 (\penalty5000 l\penalty5000 o\penalty5000 g\penalty5000 , a\penalty5000 xis=\penalty0 \emph{x})} }] \hspace{0em}\\
 set logarithmic scale for specified axis

\noindent
\begin{description}
\item[{ Parameters
}] ~\begin{itemize}

\item{} \textbf{log} (\emph{bool})\hspace{0.167em}\textemdash{}\hspace{0.167em}whether to set logarithmic

\item{} \textbf{axis} (\emph{str})\hspace{0.167em}\textemdash{}\hspace{0.167em}axis which to set logarithmic

\end{itemize}
\end{description}
\item[{ \texttt{set\penalty5000 A\penalty5000 x\penalty5000 i\penalty5000 s\penalty5000 L\penalty5000 o\penalty5000 g\penalty5000 s\penalty5000 (\penalty5000 *\penalty5000 *\penalty5000 k\penalty5000 w\penalty5000 a\penalty5000 r\penalty5000 gs)} }] \hspace{0em}\\
 set axes logarithmic if requested

\item[{ \texttt{set\penalty5000 A\penalty5000 x\penalty5000 i\penalty5000 s\penalty5000 R\penalty5000 a\penalty5000 n\penalty5000 g\penalty5000 e\penalty5000 (\penalty5000 r\penalty5000 n\penalty5000 g\penalty5000 , a\penalty5000 xis=\penalty0 \emph{x})} }] \hspace{0em}\\
 set range for specified axis

\noindent
\begin{description}
\item[{ Notes
}] ~\begin{itemize}

\item{} automatically determines axis range to include all data points if range is not given.

\item{} logscale and secondary errors taken into account

\item{} y-{}axis range determined for points within given x-{}axis range

\end{itemize}
\item[{ Parameters
}] ~\begin{itemize}

\item{} \textbf{rng} (\emph{list})\hspace{0.167em}\textemdash{}\hspace{0.167em}lower and upper range limits

\item{} \textbf{axis} (\emph{str})\hspace{0.167em}\textemdash{}\hspace{0.167em}axis to which to apply range

\end{itemize}
\end{description}
\item[{ \texttt{set\penalty5000 E\penalty5000 r\penalty5000 r\penalty5000 o\penalty5000 r\penalty5000 A\penalty5000 r\penalty5000 r\penalty5000 o\penalty5000 w\penalty5000 s\penalty5000 (\penalty5000 *\penalty5000 *\penalty5000 k\penalty5000 w\penalty5000 a\penalty5000 r\penalty5000 gs)} }] \hspace{0em}\\
 reset properties of arrows used to plot special errors

\item[{ \texttt{set\penalty5000 K\penalty5000 e\penalty5000 y\penalty5000 O\penalty5000 p\penalty5000 t\penalty5000 i\penalty5000 o\penalty5000 n\penalty5000 s\penalty5000 (\penalty5000 k\penalty5000 e\penalty5000 y\penalty5000 \_\penalty5000 o\penalty5000 p\penalty5000 ts)} }] \hspace{0em}\\
 set key options

\noindent
\begin{description}
\item[{ Parameters
}] \hspace{0em}\\
  \textbf{key\_opts} (\emph{list})\hspace{0.167em}\textemdash{}\hspace{0.167em}strings for key/legend options

\end{description}
\item[{ \texttt{set\penalty5000 L\penalty5000 a\penalty5000 b\penalty5000 e\penalty5000 l\penalty5000 (\penalty5000 l\penalty5000 a\penalty5000 b\penalty5000 e\penalty5000 l\penalty5000 , p\penalty5000 o\penalty5000 s\penalty5000 , a\penalty5000 b\penalty5000 s\penalty5000 \_\penalty5000 p\penalty5000 l\penalty5000 ace=\penalty0 Fal\penalty5000 se)} }] \hspace{0em}\\
 draw a label into the figure

\noindent
\begin{description}
\item[{ Parameters
}] ~\begin{itemize}

\item{} \textbf{label} (\emph{str})\hspace{0.167em}\textemdash{}\hspace{0.167em}label

\item{} \textbf{pos} (\emph{list})\hspace{0.167em}\textemdash{}\hspace{0.167em}x,y -{} position

\item{} \textbf{abs\_place} (\emph{bool})\hspace{0.167em}\textemdash{}\hspace{0.167em}absolute or relative placement

\end{itemize}
\end{description}
\item[{ \texttt{set\penalty5000 M\penalty5000 a\penalty5000 r\penalty5000 g\penalty5000 i\penalty5000 n\penalty5000 s\penalty5000 (\penalty5000 *\penalty5000 *\penalty5000 k\penalty5000 w\penalty5000 a\penalty5000 r\penalty5000 gs)} }] \hspace{0em}\\
 set the margins

\noindent
\begin{description}
\item[{ Notes
}] ~\begin{itemize}

\item{} keys other than l(b,t,r)margin are ignored

\item{} if margin not given leave to gnuplot

\end{itemize}
\end{description}
\item[{ \texttt{set\penalty5000 V\penalty5000 e\penalty5000 r\penalty5000 t\penalty5000 i\penalty5000 c\penalty5000 a\penalty5000 l\penalty5000 L\penalty5000 i\penalty5000 n\penalty5000 e\penalty5000 (\penalty5000 x\penalty5000 , o\penalty5000 p\penalty5000 ts)} }] \hspace{0em}\\
 draw a vertical line

\noindent
\begin{description}
\item[{ Parameters
}] ~\begin{itemize}

\item{} \textbf{x} (\emph{float})\hspace{0.167em}\textemdash{}\hspace{0.167em}position on x-{}axis

\item{} \textbf{opts} (\emph{str})\hspace{0.167em}\textemdash{}\hspace{0.167em}line draw options

\end{itemize}
\end{description}
\end{description}

\begin{center}
 \line(1,0){444}
 \end{center}

\noindent
\begin{description}
\item[{ Configuration Variables
}] ~
\noindent
\begin{description}
\item[{ var default\_key
}] \hspace{0em}\\
   default options for legend/key

\item[{ var basic\_setup
}] \hspace{0em}\\
   bars, grid, terminal and default\_key

\item[{ var default\_margins
}] \hspace{0em}\\
   default margins to define plot area

\item[{ var xPanProps
}] \hspace{0em}\\
   xscale, xsize, xoffset for panel plots

\item[{ var default\_colors
}] \hspace{0em}\\
   provides a reasonable color selection (see palette)

\end{description}
\item[{ \texttt{pyana.\penalty0 ccsgp.\penalty0 utils.\penalty0 get\penalty5000 O\penalty5000 p\penalty5000 t\penalty5000 s\penalty5000 (i)} }] \hspace{0em}\\
 convience function for easy access to gnuplot property string

\item[{ \texttt{pyana.\penalty0 ccsgp.\penalty0 utils.\penalty0 zip\penalty5000 \_\penalty5000 f\penalty5000 l\penalty5000 a\penalty5000 t\penalty5000 (\penalty5000 a\penalty5000 , b\penalty5000 , c=\penalty0 Non\penalty5000 e\penalty5000 , d=\penalty0 None)} }] \hspace{0em}\\
 zips 2-{}4 lists and flattens the result

\item[{ \texttt{pyana.\penalty0 exa\penalty5000 m\penalty5000 p\penalty5000 les.\penalty0 utils.\penalty0 che\penalty5000 c\penalty5000 k\penalty5000 S\penalty5000 y\penalty5000 m\penalty5000 L\penalty5000 i\penalty5000 n\penalty5000 k()} }] \hspace{0em}\\
 check for symbolic link to input directory

\item[{ \texttt{pyana.\penalty0 exa\penalty5000 m\penalty5000 p\penalty5000 les.\penalty0 utils.\penalty0 enu\penalty5000 m\penalty5000 z\penalty5000 i\penalty5000 p\penalty5000 E\penalty5000 d\penalty5000 g\penalty5000 e\penalty5000 s\penalty5000 (\penalty5000 e\penalty5000 A\penalty5000 rr)} }] \hspace{0em}\\
 zip and enumerate edges into pairs of lower and upper limits

\item[{ \texttt{pyana.\penalty0 exa\penalty5000 m\penalty5000 p\penalty5000 les.\penalty0 utils.\penalty0 get\penalty5000 C\penalty5000 o\penalty5000 c\penalty5000 k\penalty5000 t\penalty5000 a\penalty5000 i\penalty5000 l\penalty5000 S\penalty5000 u\penalty5000 m\penalty5000 (\penalty5000 e\penalty5000 0\penalty5000 , e\penalty5000 1\penalty5000 , e\penalty5000 C\penalty5000 o\penalty5000 c\penalty5000 k\penalty5000 t\penalty5000 a\penalty5000 i\penalty5000 l\penalty5000 , u\penalty5000 C\penalty5000 o\penalty5000 c\penalty5000 k\penalty5000 t\penalty5000 a\penalty5000 il)} }] \hspace{0em}\\
 get the cocktail sum for a given data bin range

\item[{ \texttt{pyana.\penalty0 exa\penalty5000 m\penalty5000 p\penalty5000 les.\penalty0 utils.\penalty0 get\penalty5000 E\penalty5000 d\penalty5000 g\penalty5000 e\penalty5000 s\penalty5000 (\penalty5000 n\penalty5000 p\penalty5000 A\penalty5000 rr)} }] \hspace{0em}\\
 get \texttt{numpy.\penalty0 array} of bin edges

\item[{ \texttt{pyana.\penalty0 exa\penalty5000 m\penalty5000 p\penalty5000 les.\penalty0 utils.\penalty0 get\penalty5000 E\penalty5000 r\penalty5000 r\penalty5000 o\penalty5000 r\penalty5000 C\penalty5000 o\penalty5000 m\penalty5000 p\penalty5000 o\penalty5000 n\penalty5000 e\penalty5000 n\penalty5000 t\penalty5000 (\penalty5000 r\penalty5000 e\penalty5000 s\penalty5000 u\penalty5000 l\penalty5000 t\penalty5000 , t\penalty5000 ag)} }] \hspace{0em}\\
 get total error contribution for component with specific tag (stat/syst)

\item[{ \texttt{pyana.\penalty0 exa\penalty5000 m\penalty5000 p\penalty5000 les.\penalty0 utils.\penalty0 get\penalty5000 M\penalty5000 a\penalty5000 s\penalty5000 k\penalty5000 I\penalty5000 n\penalty5000 d\penalty5000 i\penalty5000 c\penalty5000 e\penalty5000 s\penalty5000 (\penalty5000 m\penalty5000 a\penalty5000 sk)} }] \hspace{0em}\\
 get lower and upper index of mask

\item[{ \texttt{pyana.\penalty0 exa\penalty5000 m\penalty5000 p\penalty5000 les.\penalty0 utils.\penalty0 get\penalty5000 U\penalty5000 A\penalty5000 r\penalty5000 r\penalty5000 a\penalty5000 y\penalty5000 (\penalty5000 n\penalty5000 p\penalty5000 A\penalty5000 rr)} }] \hspace{0em}\\
 uncertainty array multiplied by binwidth (col2 = dx)

\item[{ \texttt{pyana.\penalty0 exa\penalty5000 m\penalty5000 p\penalty5000 les.\penalty0 utils.\penalty0 get\penalty5000 W\penalty5000 o\penalty5000 r\penalty5000 k\penalty5000 D\penalty5000 i\penalty5000 r\penalty5000 s()} }] \hspace{0em}\\
 get input/output dirs (same input/output layout as for package)

\end{description}

\section{wp-{}pdf}
\label{wppdf_source}\hyperlabel{wppdf_source}%

\noindent
\begin{description}
\item[{ Upgrade Constant in wp-{}config.php
}] \hspace{0em}\\
\end{description}

\begin{lstlisting}[language=php,firstnumber=1,]
  // forces the filesystem method: "direct", "ssh", "ftpext", or "ftpsockets"
  define('FS_METHOD', 'direct');
  // absolute path to root installation directory
  define('FTP_BASE', '/home/patrick/public/asciidoc.the-huck.com/public/');
  // absolute path to "wp-content" directory
  define('FTP_CONTENT_DIR', '/home/patrick/public/asciidoc.the-huck.com/public/wp-content/');
  // absolute path to "wp-plugins" directory
  define('FTP_PLUGIN_DIR ', '/home/patrick/public/asciidoc.the-huck.com/public/wp-content/plugins/');
  // absolute path to your SSH public key
  define('FTP_PUBKEY', '/home/patrick/.ssh/<your-key>.pub');
  // absolute path to your SSH private key
  define('FTP_PRIVKEY', '/home/patrick/.ssh/<your-key>');
  // either your FTP or SSH username
  define('FTP_USER', '<username>');
  // hostname:port combo for your SSH/FTP server
  define('FTP_HOST', 'the-huck.com');
\end{lstlisting}

\noindent
\begin{description}
\item[{ PHP Widget for Pagination
}] \hspace{0em}\\
\end{description}

\begin{lstlisting}[language=php,firstnumber=1,]
next page: <?php echo next_page_not_post('%title','expand','sort_column=menu_order'); ?><br>
previous page: <?php echo previous_page_not_post(' %title','expand','sort_column=menu_order'); ?>
\end{lstlisting}

\noindent
\begin{description}
\item[{ Makefile
}] \hspace{0em}\\
\end{description}

\begin{lstlisting}[language=make,firstnumber=1,]
# pdf document name
DOCOUT=asciidoc
# your initials
INITIALS=PH

# filelist (needs to be in order!)
# put paths to all asciidoc source files in
# filelist.txt or list them here directly
# you can submit single files via the *.txt target
# use SIMSTR=-n to simulate submission
FILELIST=$(shell cat filelist.txt)

# symbolic links
# this list is checked in the 'check' target
# and requirement for successfull compilation
# remove 'images' if no images included
#    (or adjust to image include in source file)
LINKLIST = #images

###############################
#  NO CHANGES REQUIRED BELOW  #
###############################

# variables
DOCINFO=docinfo.xml
DOCINFOREV=$(DOCOUT)-$(DOCINFO)
TAGLIST=$(shell git tag -l | tr '\n' ' ')

# shell commands
SIMSTR=
BPCMD=blogpost.py $(SIMSTR) -p post
SEDCMD=sed -e '/^:blogpost/d' -e 's:\/\/=:=:'

# phony takes target always as out-of-date
.PHONY: all pdf hp docinfo check clean

# default target if none specified
all: pdf hp latex

# define directive for single revision entry
# argument: tagname
define REVCMD

echo '<revision>' >> $(DOCINFOREV)
echo '  <revnumber>'$(1)'</revnumber>' >> $(DOCINFOREV)
echo '  <date>'$(shell git log -1 $(1) --format=%ad --date=short)'</date>' >> $(DOCINFOREV)
echo '  <authorinitials>$(INITIALS)</authorinitials>' >> $(DOCINFOREV)
echo '  <revremark>'$(shell git log -1 $(1) --format=%B)'</revremark>' >> $(DOCINFOREV)
echo '</revision>' >> $(DOCINFOREV)
endef

# define A2X command
# argument: format (pdf/tex), asciidoc file, outdir
# -a docinfo = include $(DOCOUT)-docinfo.xml
define A2XCMD
a2x -vv -a latexmath -a docinfo -L -D $(3) -f $(1) $(2)
endef

# check whether required symbolic links exist
# abort otherwise
define TESTDEF
$$(if $$(wildcard $(1)),$$(info symlink $(1) ok),$$(error symlink $(1) NOT found))
endef
check:
        $(foreach link, $(LINKLIST), $(eval $(call TESTDEF,$(link))))

# generate $(DOCOUT)-docinfo.xml
docinfo:
        cp $(DOCINFO) $(DOCINFOREV)
        $(foreach tag, $(TAGLIST), $(call REVCMD,$(tag)))
        echo '</revhistory>' >> $(DOCINFOREV)

# generate pdf
pdf: docinfo check
        cp preamb.txt $(DOCOUT).txt
        @$(foreach file, $(FILELIST), $(SEDCMD) $(file) >> $(DOCOUT).txt; )
        $(call A2XCMD,pdf,$(DOCOUT).txt,.)

# generate page-by-page tex files for inclusion in latex document
latex: check
        mkdir tex && ln -s images tex/images
        @$(foreach file, $(FILELIST), $(call A2XCMD,tex,$(file),tex); )

# push all asciidocs to wordpress
hp: check
        @$(foreach file, $(FILELIST), $(BPCMD) $(file); )

# push a single asciidoc to wordpress
FORCE:
%.txt: FORCE
        $(BPCMD) $*.txt

# clean up
clean:
        @if [ -e $(DOCOUT).txt ]; then rm -v $(DOCOUT).txt; fi
        @if [ -e $(DOCOUT).pdf ]; then rm -v $(DOCOUT).pdf; fi
        @if [ -e $(DOCINFOREV) ]; then rm -v $(DOCINFOREV); fi
        @if [ -d tex ]; then rm -rfv tex; fi
\end{lstlisting}

\end{appendices}
\bookmarksetup{startatroot}

\end{document}